\documentclass[12pt]{article}
 
\newcommand{\mrm}[1]{\mathrm{#1}}
\newcommand{\mbf}[1]{\mathbf{#1}}
\newcommand{\mtt}[1]{\mathtt{#1}}
\newcommand{\tsc}[1]{\textsc{#1}}
\newcommand{\tbf}[1]{\textbf{#1}}
\newcommand{\ttt}[1]{\texttt{#1}}
 
\newcommand{\Py}{\tsc{Pythia}}
\newcommand{\Je}{\tsc{Jetset}}
\newcommand{\PyJe}{\tsc{Pythia/Jetset}}
\newcommand{\JePy}{\tsc{Jetset/Pythia}}
 
\newcommand{\alphas}{\alpha_{\mathrm{s}}}
\newcommand{\alphaem}{\alpha_{\mathrm{em}}}
\newcommand{\ssintw}{\sin^2 \! \theta_W}
\newcommand{\scostw}{\cos^2 \! \theta_W}
\newcommand{\pT}{p_{\perp}}
\newcommand{\pTmin}{p_{\perp\mathrm{min}}}
\newcommand{\MSbar}{$\overline{\mrm{MS}}$}
\newcommand{\br}[1]{\overline{#1}}
 
\renewcommand{\a}{\mathrm{a}}
\renewcommand{\b}{\mathrm{b}}
\renewcommand{\c}{\mathrm{c}}
\renewcommand{\d}{\mathrm{d}}
\newcommand{\e}{\mathrm{e}}
\newcommand{\f}{\mathrm{f}}
\newcommand{\g}{\mathrm{g}}
\newcommand{\hrm}{\mathrm{h}}
\newcommand{\lrm}{\mathrm{l}}
\newcommand{\n}{\mathrm{n}}
\newcommand{\p}{\mathrm{p}}
\newcommand{\q}{\mathrm{q}}
\newcommand{\s}{\mathrm{s}}
\renewcommand{\t}{\mathrm{t}}
\renewcommand{\u}{\mathrm{u}}
\newcommand{\A}{\mathrm{A}}
\newcommand{\B}{\mathrm{B}}
\newcommand{\D}{\mathrm{D}}
\newcommand{\F}{\mathrm{F}}
\renewcommand{\H}{\mathrm{H}}
\newcommand{\J}{\mathrm{J}}
\newcommand{\K}{\mathrm{K}}
\renewcommand{\L}{\mathrm{L}}
\newcommand{\Q}{\mathrm{Q}}
\newcommand{\R}{\mathrm{R}}
\newcommand{\T}{\mathrm{T}}
\newcommand{\W}{\mathrm{W}}
\newcommand{\Z}{\mathrm{Z}}
\newcommand{\bbar}{\overline{\mathrm{b}}}
\newcommand{\cbar}{\overline{\mathrm{c}}}
\newcommand{\dbar}{\overline{\mathrm{d}}}
\newcommand{\fbar}{\overline{\mathrm{f}}}
\newcommand{\pbar}{\overline{\mathrm{p}}}
\newcommand{\qbar}{\overline{\mathrm{q}}}
\newcommand{\sbar}{\overline{\mathrm{s}}}
\newcommand{\tbar}{\overline{\mathrm{t}}}
\newcommand{\ubar}{\overline{\mathrm{u}}}
\newcommand{\Fbar}{\overline{\mathrm{F}}}
\newcommand{\Qbar}{\overline{\mathrm{Q}}}
 
\newcommand{\Jpsi}{\mathrm{J}/\psi}
\newcommand{\pomeron}{\mathrm{I}\!\mathrm{P}}
\newcommand{\reggeon}{\mathrm{I}\!\mathrm{R}}
\newcommand{\ee}{\e^+\e^-}
\newcommand{\ep}{\e\p}
\newcommand{\pp}{\p\p}
\newcommand{\ppbar}{\p\pbar}
\newcommand{\gammaZ}{\gamma^* / \Z^0}
\newcommand{\mmin}{\mathrm{min}}
\newcommand{\mmax}{\mathrm{max}}
 
\newenvironment{Itemize}{\begin{list}{$\bullet$}%
{\setlength{\topsep}{0.2mm}\setlength{\partopsep}{0.2mm}%
\setlength{\itemsep}{0.2mm}\setlength{\parsep}{0.2mm}}}%
{\end{list}}
\newcounter{enumct}
\newenvironment{Enumerate}{\begin{list}{\arabic{enumct}.}%
{\usecounter{enumct}\setlength{\topsep}{0.2mm}%
\setlength{\partopsep}{0.2mm}\setlength{\itemsep}{0.2mm}%
\setlength{\parsep}{0.2mm}}}{\end{list}}
 
\newenvironment{entry}%
{\begin{list}{}{\setlength{\topsep}{0mm} \setlength{\itemsep}{0mm}
\setlength{\parskip}{0mm} \setlength{\parsep}{0mm}
\setlength{\leftmargin}{20mm} \setlength{\rightmargin}{0mm}
\setlength{\labelwidth}{18mm} \setlength{\labelsep}{2mm}}}%
{\end{list}}
\newenvironment{subentry}%
{\begin{list}{}{\setlength{\topsep}{0mm} \setlength{\itemsep}{0mm}
\setlength{\parskip}{0mm} \setlength{\parsep}{0mm}
\setlength{\leftmargin}{10mm} \setlength{\rightmargin}{0mm}
\setlength{\labelwidth}{18mm} \setlength{\labelsep}{2mm}}}%
{\end{list}}
\newcommand{\itemc}[1]{\item[\textbf{#1}\hfill]}
\newcommand{\iteme}[1]{\item[\texttt{#1}\hfill]}
\newcommand{\itemn}[1]{\item[{#1}\hfill]}
 
\newcommand{\drawbox}[1]{\vspace{\baselineskip}\noindent%
\fbox{\texttt{#1}}\vspace{0.5\baselineskip}}
\newcommand{\drawboxtwo}[2]{\vspace{\baselineskip}\noindent%
\fbox{\begin{minipage}{150mm}\begin{tabbing}%
{\texttt{#1}}\\{\texttt{#2}}%
\end{tabbing}\end{minipage}}\vspace{0.5\baselineskip}}

\newcommand{\drawboxfour}[4]{\vspace{\baselineskip}\noindent%
\fbox{\begin{minipage}{150mm}\begin{tabbing}%
{\texttt{#1}}\\{\texttt{#2}}\\{\texttt{#3}}\\{\texttt{#4}}%
\end{tabbing}\end{minipage}}\vspace{0.5\baselineskip}}
\newcommand{\boxsep}{\vspace{0.5\baselineskip}} 
 
\newlength{\captivewidth}
\newcommand{\captive}[1]{\rule{5mm}{0mm}%
\begin{minipage}{\captivewidth}%
\caption[small]{#1}\end{minipage}}

\newlength{\tablinsep}
\newlength{\halfpagewid}
\begin{document}
 
\sloppy
 
\pagestyle{empty}
 
\begin{flushright}
LU TP 95--20\\
August 1995\\
(revised version of CERN--TH.7112/93)\\
(first issued December 1993)\\
hep-ph/9508391\\[0.5\baselineskip]
\fbox{W5035/W5044}
\end{flushright}
 
\vspace{\fill}
 
\begin{center}

\tbf{{\Huge P}{\LARGE YTHIA}~~{\Huge 5.7}~~{\LARGE and}~~%
{\Huge J}{\LARGE ETSET}~~{\Huge 7.4}}   \\[5mm]
\tbf{{\LARGE Physics and Manual}} \\[10mm]
\tbf{{\Large Torbj\"orn Sj\"ostrand}} \\[3mm]
{\large Department of Theoretical Physics,} \\[1mm]
{\large University of Lund, S\"olvegatan 14A,} \\[1mm]
{\large S-223 62 \tsc{Lund}, \tsc{Sweden}}\\

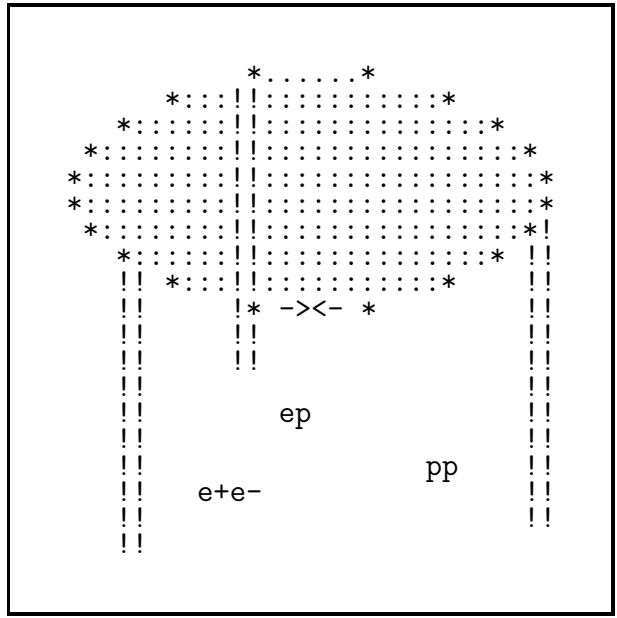
\vspace{\fill}

\fbox{\hspace{1mm}\rule{0mm}{7mm}
\begin{minipage}[t]{150mm}
\begin{tabbing}
{\tt ~~~~~~~~~~~~*......*~~~~~~~~~~~~}\\[-1.1mm]
{\tt ~~~~~~~*:::!!:::::::::::*~~~~~~~}\\[-1.1mm]
{\tt ~~~~*::::::!!::::::::::::::*~~~~}\\[-1.1mm]
{\tt ~~*::::::::!!::::::::::::::::*~~}\\[-1.1mm]
{\tt ~*:::::::::!!:::::::::::::::::*~}\\[-1.1mm]
{\tt ~*:::::::::!!:::::::::::::::::*~}\\[-1.1mm]
{\tt ~~*::::::::!!::::::::::::::::*!~}\\[-1.1mm]
{\tt ~~~~*::::::!!::::::::::::::*~!!~}\\[-1.1mm]
{\tt ~~~~!!~*:::!!:::::::::::*~~~~!!~}\\[-1.1mm]
{\tt ~~~~!!~~~~~!*~-><-~*~~~~~~~~~!!~}\\[-1.1mm]
{\tt ~~~~!!~~~~~!!~~~~~~~~~~~~~~~~!!~}\\[-1.1mm]
{\tt ~~~~!!~~~~~!!~~~~~~~~~~~~~~~~!!~}\\[-1.1mm]
{\tt ~~~~!!~~~~~~~~~~~~~~~~~~~~~~~!!~}\\[-1.1mm]
{\tt ~~~~!!~~~~~~~~ep~~~~~~~~~~~~~!!~}\\[-1.1mm]
{\tt ~~~~!!~~~~~~~~~~~~~~~~~~~~~~~!!~}\\[-1.1mm]
{\tt ~~~~!!~~~~~~~~~~~~~~~~~pp~~~~!!~}\\[-1.1mm]
{\tt ~~~~!!~~~e+e-~~~~~~~~~~~~~~~~!!~}\\[-1.1mm]
{\tt ~~~~!!~~~~~~~~~~~~~~~~~~~~~~~!!~}\\[-1.1mm]
{\tt ~~~~!!~~~~~~~~~~~~~~~~~~~~~~~~~~}\\[-1.1mm]
\end{tabbing}
\end{minipage}
\rule[-6mm]{0mm}{6mm}\hspace{3mm}}

\vspace{\fill}
 
\fbox{\begin{minipage}[t]{126mm}\begin{center}
{\large Important note: this is the long writeup of} \\[1mm]
{\large T. Sj\"ostrand, Computer Physics Commun. {\bf 82} 
(1994) 74.} \\[1mm]
{\large All references should be to the published version.}
\end{center}\end{minipage}}
 
\end{center}
 
\clearpage

\vspace*{\fill}

\noindent
{\hfill\framebox[0.8\textwidth][t]{
\begin{minipage}{0.75\textwidth}

\vspace{\baselineskip}
\begin{center}
Copyright Notice
\end{center}

\vspace{\baselineskip}
{\bf CERNLIB -- CERN Program Library Long writeups}

\vspace{\baselineskip} 
\copyright{} Copyright CERN, Geneva 1993

\vspace{\baselineskip}
Copyright and any other appropriate legal protection of these
computer programs and associated documentation reserved in all
countries of the world.

\vspace{\baselineskip} 
These programs or documentation may not be reproduced by any
method without prior written consent of the Director-General
of CERN or his delegate.

\vspace{\baselineskip} 
Permission for the usage of any programs described herein is
granted apriori to those scientific institutes associated with
the CERN experimental program or with whom CERN has concluded
a scientific collaboration agreement.

\vspace{\baselineskip} 
Requests for information should be addressed to:
\vspace{-0.5\baselineskip}
\begin{center}
\begin{tabular}{l@{\protect\rule{0mm}{\tablinsep}}}
\ttt{CERN Program Library Office }             \\
\ttt{CERN-CN Division }                        \\
\ttt{CH-1211 Geneva 23 }                       \\
\ttt{Switzerland }                             \\
\ttt{Tel.      +41 22 767 4951 }               \\
\ttt{Fax.      +41 22 767 7155 }               \\
\ttt{Bitnet:   CERNLIB@CERNVM  }               \\
\ttt{DECnet:   VXCERN::CERNLIB (node 22.190) } \\
\ttt{Internet: CERNLIB@CERNVM.CERN.CH } 
\end{tabular}
\end{center}
\vspace{0mm}

\end{minipage}}\hfill}

\vspace{2\baselineskip}\noindent
{\hfill\begin{minipage}{0.8\textwidth}
{\bf Trademark notice: All trademarks appearing in this guide are 
acknowledged as such.}
\end{minipage}\hfill}

\vspace*{\fill}

\cleardoublepage

\section*{Preface}

The {\Py} and {\Je} programs are frequently used for event generation
in high-energy physics. The emphasis is on multiparticle production
in collisions between elementary particles. This in particular means
hard interactions in $\ee$, $\pp$ and $\ep$ colliders, although
also other applications are envisaged. The programs are intended to 
generate complete events, in as much detail as 
experimentally observable ones, within the bounds of our current
understanding of the underlying physics. Many of the components of
the programs represent original research, in the sense that models 
have been developed and implemented for a number of aspects not 
covered by standard theory. Although originally conceived
separately, the {\Py} and {\Je} programs today are so often used
together that it makes sense to present them here without too much
distinction.

Both programs have a long history, and several manuals have 
come out. The former round of {\PyJe} program descriptions appeared
in 1987. Meanwhile a large number of additions and 
changes have been made. Recently a new description therefore
appeared in \\[1mm]
T. Sj\"ostrand, Computer Physics Commun. {\bf 82} (1994) 74. \\[1mm]
This is the one and only correct reference to the current versions of
{\Py} and {\Je}. The long writeup that you now have before you is 
an (unpublished) appendix to the publication above, and need not be 
separately cited. Instead remember to cite the original
literature on the physics topics of particular relevance for your
studies. (There is no reason to omit references to good physics 
papers simply because some of their contents have also been
made available as program code.)

Event generators often have a reputation for being `black boxes'; 
if nothing else, this report should provide you with a glimpse of 
what goes on inside the programs. Some such understanding may be of 
special interest for new users, who have no background in the field. 
An attempt has been made to structure the report sufficiently well 
that many of the sections can be read independently of each other,
so you can pick the sections that interest you. I have tried to
keep together the physics and the manual sections on specific 
topics, where practicable, which represents a change of policy
compared with previous manual versions. Any feedback on this and 
other aspects is welcome.

A large number of persons should be thanked for their contributions.
Hans-Uno Bengtsson is the originator of the {\Py} program, and for 
many years we worked in parallel on its further development. Mats
Bengtsson is the main author of the final-state parton-shower 
algorithm. Bo Andersson and G\"osta Gustafson are the originators 
of the Lund model, and strongly influenced the early development of 
the programs. Further comments on the programs have been obtained 
from users too numerous to be mentioned here, but who are all 
gratefully acknowledged. To write programs of this size and 
complexity would be impossible without a strong user feedback. 

The moral responsibility for any remaining errors clearly rests with
me. However, kindly note that this is a `University World' product,
distributed `as is', free of charge, without any binding guarantees. 
And always remember that the programs do not represent a dead
collection of established truths, but rather one of many possible
approaches to the problem of multiparticle production in high-energy
physics, at the frontline of current research. Be critical!

\newpage

\mbox{}

\cleardoublepage

\tableofcontents
 
\cleardoublepage

\pagestyle{plain}
\setcounter{page}{1}
 
\section{Introduction}
 
Multiparticle production is the most characteristic feature
of current high-energy physics. Today, observed particle
multiplicities are typically between ten and a hundred, and
with future machines this range will be extended upwards.
The bulk of the multiplicity is found in jets, i.e. in bunches
of hadrons (or decay products of hadrons) produced by the
hadronization of quarks and gluons.
 
\subsubsection*{The Complexity of High-Energy Processes}
 
To first approximation, all processes have a simple structure at
the level of interactions between the fundamental objects of nature,
i.e. quarks, leptons and gauge bosons. For instance, a lot can
be understood about the structure of hadronic events at LEP just
from the `skeleton' process $\ee \to \Z^0 \to \q \qbar$.
Corrections to this picture can be subdivided, arbitrarily but
conveniently, into three main classes.
 
Firstly, there are bremsstrahlung-type modifications, i.e. the
emission of additional final-state particles by branchings such as
$\e \to \e \gamma$ or $\q \to \q \g$. Because of the largeness of
the strong coupling constant $\alphas$, and because of the presence
of the triple gluon vertex, QCD emission off quarks and gluons is
especially prolific. We therefore speak about `parton showers',
wherein a single initial parton may give rise to a whole bunch of
partons in the final state. Also photon emission may give sizeable
effects in $\ee$ and $\ep$ processes. The bulk of the bremsstrahlung
corrections are universal, i.e. do not depend on the details of
the process studied, but only on one or a few key numbers, such as
the momentum transfer scale of the process. Such universal
corrections may be included to arbitrarily high orders, using a
probabilistic language. Alternatively, exact calculations of
bremsstrahlung corrections may be carried out order by order in
perturbation theory, but rapidly the calculations then become
prohibitively complicated and the answers correspondingly lengthy.
 
Secondly, we have `true' higher-order corrections, which involve a
combination of loop graphs and the soft parts of the
bremsstrahlung graphs above, a combination needed to
cancel some  divergences. In a complete description it is
therefore not possible to consider bremsstrahlung separately,
as assumed here. The
necessary perturbative calculations are usually very difficult;
only rarely have results been presented that include more than one
non-`trivial' order, i.e. more than one loop. As above, answers
are usually very lengthy, but some results are sufficiently simple
to be generally known and used, such as the running of $\alphas$, or
the correction factor $1 + \alphas/\pi + \cdots$ in the partial
widths of $\Z^0 \to \q \qbar$ decay channels. For high-precision
studies it is imperative to take into account the results of
loop calculations, but usually effects are minor for the qualitative
aspects of high-energy processes.
 
Thirdly, quarks and gluons are confined. In the two points above,
we have used a perturbative language to describe the short-distance
interactions of quarks, leptons and gauge bosons. For leptons
and colourless bosons this language is sufficient. However, for
quarks and gluons it must be complemented with a picture for the
hadronization process (which can be subdivided into fragmentation
and decays), wherein the coloured partons are transformed
into jets of colourless hadrons, photons and leptons.
This process is still not yet understood
from first principles, but has to be based on models. In one sense,
hadronization effects are overwhelmingly large, since this is where
the bulk of the multiplicity comes from. In another sense, the
overall energy flow of a high-energy event is mainly determined by
the perturbative processes, with only a minor additional smearing
caused by the hadronization step. One may therefore pick different
levels of ambition, but in general detailed studies require a
detailed modelling of the hadronization process.
 
The simple structure that we started out with has now become
considerably more complex --- instead of maybe two final-state
partons we have a hundred final particles. The original physics
is not gone, but the skeleton process has been dressed up and
is no longer directly visible. A direct comparison between theory
and experiment is therefore complicated at best, and
impossible at worst.
 
\subsubsection*{Event Generators}
 
It is here that event generators come to the rescue. In an event
generator, the objective strived for is to use computers to generate 
events as detailed as could be observed by a perfect detector.
This is not done in one step, but rather by `factorizing' the full
problem into a number of components, each of which can be handled
reasonably accurately. Basically, this means that the hard process
is used as input to generate bremsstrahlung corrections, and that
the result of this exercise is thereafter left to hadronize. This
sounds a bit easier than it really is --- else this report would
be a lot thinner. However, the basic idea is there: if the
full problem is too complicated to be solved in one go, try to
subdivide it into smaller tasks of manageable proportions.
In the actual generation procedure, most steps therefore involve
the branching of one object into two, or at least into a very small
number, each of which being free to branch in its turn. A lot of
bookkeeping is involved, but much is of a repetitive nature, and
can therefore be left for the computer to handle.
 
As the name indicates, the output of an event generator should be
in the form of `events', with the same average behaviour and the
same fluctuations as real data. In the data, fluctuations arise from
the quantum mechanics of the underlying theory. In
generators, Monte Carlo techniques are used to select all relevant
variables according to the desired probability distributions,
and thereby ensure randomness in the final events.
Clearly some loss of information is entailed: quantum mechanics is
based on amplitudes, not probabilities. However, only very rarely
do (known) interference phenomena appear that cannot be cast in a
probabilistic language. This is therefore not a more restraining
approximation than many others.
 
Once there, an event generator can be used in many different ways.
The five main applications are probably the following:
\begin{Itemize}
\item To give physicists a feeling for the kind
of events one may expect/hope to find, and at what rates.
\item As a help in the planning of a new detector, so that detector
performance is optimized, within other constraints, for the
study of interesting physics scenarios.
\item As a tool for devising the analysis strategies that should
be used on real data, so that signal-to-background conditions are
optimized.
\item As a method for estimating detector acceptance corrections
that have to be applied to raw data, in order to extract the
`true' physics signal.
\item As a convenient framework within which to interpret the
observed phenomena in terms of a more fundamental
underlying theory (usually the Standard Model).
\end{Itemize}
 
Where does a generator fit into the overall analysis chain of an
experiment? In `real life', the machine produces interactions.
These events are observed by detectors, and the interesting ones
are written to tape by the
data acquisition system. Afterwards the events may be reconstructed,
i.e. the electronics signals (from wire chambers, calorimeters, and
all the rest) may be
translated into a deduced setup of charged tracks or
neutral energy depositions, in the best of worlds with full knowledge
of momenta and particle species. Based on this cleaned-up
information, one may proceed with the physics analysis.
In the Monte Carlo world, the r\^ole of the machine, namely to produce
events, is taken by the event generators described in this report.
The behaviour of the detectors --- how particles produced by the
event generator traverse the detector, spiral in magnetic
fields, shower in calorimeters, or sneak out through cracks, etc. ---
is simulated in programs such as \tsc{Geant} \cite{Bru89}. 
Traditionally, this latter activity is called event simulation,
which is somewhat unfortunate
since the same words could equally well be applied to what, here, we
call event generation. A more appropriate term is detector
simulation. Ideally, the output of this simulation has exactly the
same format as the real data recorded by the detector, and can
therefore be put through the same event reconstruction and physics
analysis chain, except that here we know what the `right answer'
should be, and so can see how well we are doing.
 
Since the full chain of detector simulation and event
reconstruction is very
time-consuming, one often does `quick and dirty' studies in
which these steps are skipped entirely, or at least replaced by
very simplified procedures which only take into account the geometric
acceptance of the detector and other trivial effects. One may then
use the output of the event generator directly in the physics studies.
 
There are still many holes in our understanding of the full event
structure, despite an impressive amount of work and detailed
calculations. To put together a generator therefore involved making
a choice on what to include, and how to include it. At best, the
spread between generators can be used to give some impression of
the uncertainties involved. A multitude of approximations will
be discussed in the main part of this report, but already here
is should be noted that many major approximations are related to
the almost complete neglect
of the second point above, i.e. of the non-`trivial'
higher-order effects. It can therefore only be hoped that
the `trivial' higher order parts give the bulk of the experimental
behaviour. By and large, this seems to be the case; for $\ee$
annihilation it even turns out to be a very good approximation.
 
The necessity to make compromises has one major implication:
to write a good event generator is an art, not an exact science.
It is therefore essential not to blindly trust the
results of any single event generator, but always to make several
cross-checks. In addition, with computer programs of tens of
thousands of lines, the question is not whether bugs exist, but how
many there are, and how critical their positions.
Further, an event generator cannot be thought of as all-powerful,
or able to give intelligent answers to ill-posed questions;
sound judgement and some understanding of a
generator are necessary prerequisites for successful use. In spite
of these limitations, the event generator approach is the most
powerful tool at our disposal if we wish to gain a detailed and
realistic understanding of physics at current or future high-energy
colliders.
 
\subsubsection*{The Origins of the JETSET and PYTHIA Programs}
 
Over the years, many event generators have appeared. Surveys of
generators for $\ee$ physics in general and LEP in particular
may be found in \cite{Kle89,Sjo89}, for high-energy hadron--hadron
($\pp$) physics in \cite{Ans90,Sjo92,Kno93}, and for $\ep$ physics in
\cite{HER92}. We refer the reader to those for additional details
and references. In this particular report, the two closely
connected programs {\Je} and {\Py} will be described.
 
{\Je} has its roots in the efforts of the Lund
group to understand the hadronization process, starting in the late
seventies \cite{And83}. The so-called string fragmentation model
was developed as an explicit and detailed framework, within which
the long-range confinement forces are allowed to distribute the
energies and flavours of a parton configuration among a collection
of primary hadrons, which subsequently may decay further. This model,
known as the Lund string model, or `Lund' for short, contained a
number of specific predictions, which were confirmed by data from
PETRA and PEP, whence the model gained a widespread
acceptance. The Lund string model is still today the most elaborate
and widely used fragmentation model
at our disposal. It remains at the heart of the
{\JePy} programs.
 
In order to predict the shape of events at PETRA/PEP, and to
study the fragmentation process in detail, it was necessary to start
out from the partonic configurations that were to fragment.
The generation of complete $\ee$ hadronic events was therefore
added, originally based on simple $\gamma$ exchange and
first-order QCD matrix elements, later extended to full $\gammaZ$
exchange with first-order initial-state QED radiation and 
second-order QCD matrix elements. A number of utility routines 
were also provided early on, for everything from event listing 
to jet finding.
 
By the mid-eighties it was clear that the matrix-element approach had
reached the limit of its usefulness, in the sense that it could not
fully describe the multijet topologies of the data. (Later on,
the use of optimized perturbation theory was to lead to a resurgence
of the matrix-element approach, but only for specific applications.)
Therefore a parton-shower description was
developed \cite{Ben87a} as an alternative to the matrix-element one.
The combination of parton showers and string fragmentation has been
very successful, and forms the main approach to the description of
hadronic $\Z^0$ events.
 
In recent years, {\Je} has been a fairly stable product,
covering the four main areas of fragmentation, final-state parton
showers, $\ee$ event generation and general utilities.
 
The successes of string fragmentation in $\ee$ made it interesting
to try to extend this framework to other processes, and explore
possible physics consequences. Therefore a number of
other programs were written, which combined a process-specific
description of the hard interactions with the general fragmentation
framework of {\Je}. The {\Py} program
evolved out of early studies on fixed-target proton--proton
processes, addressed mainly at issues related to string drawing.
 
With time, the interest shifted towards higher energies, first
to the SPS $\ppbar$ collider, and later to SSC and LHC, in the
context of a number of workshops in the USA and Europe. Parton
showers were added, for final-state radiation by making use of
the {\Je} routine, for initial-state one by the development
of the concept of `backwards evolution', specifically for
{\Py} \cite{Sjo85}. Also a framework was developed for
minimum-bias and underlying events \cite{Sjo87a}.
 
Another main change was the introduction of an increasing
number of hard processes, within the Standard Model and beyond.
A special emphasis was put on the search for the Standard Model
Higgs, in different mass ranges and in different channels, with due
respect to possible background processes.
 
The bulk of the machinery developed for hard processes actually
depended little on the choice of initial state, as long as the
appropriate parton distributions were there for the incoming
partons and particles. It therefore made sense to extend the
program from being only a $\pp$ generator to working also for
$\ee$ and $\ep$. This process was only completed in 1991,
again spurred on by physics workshop activities. Currently
{\Py} should therefore work equally well for a selection
of different possible incoming beam particles.
 
The tasks of including new processes, and of improving the simulation
of already present ones, are never-ending. Work therefore continues
apace.
 
While {\Je} still is formally independent of {\Py}, their ties 
have grown much stronger over the years, and the border-line 
between the two programs has become more and more artificial.
It is no coincidence that the two are presented together here;
this way a lot of repetition of common material can be avoided.
The price to be paid is that some differences in philosophy will
have to be discussed.
 
\subsubsection*{About this Report}
 
As we see, {\Je} and {\Py} started out as very
ideologically motivated programs, developed to study specific
physics questions in enough detail that explicit predictions
could be made for experimental quantities. As it was recognized
that experimental imperfections could distort the basic predictions,
the programs were made available for general use by experimentalists.
It thus became feasible to explore the models in more detail
than would otherwise have been possible. As time went by, the
emphasis came to shift somewhat, away from the original strong
coupling to a specific fragmentation model, towards a description of
high-energy multiparticle production processes in general.
Correspondingly, the use expanded from being one of just comparing
data with specific model predictions, to one of extensive use for
the understanding of detector performance, for the derivation of
acceptance correction factors, for the prediction of physics at
future high-energy accelerators, and for the design of related
detectors.
 
While the ideology may be less apparent, it is still there, however.
This is not something unique to the programs discussed here,
but inherent in any event generator, or at least any generator that 
attempts to go beyond the simple parton level skeleton description
of a hard process. Do not accept the myth that everything available
in Monte Carlo form represents ages-old common knowledge, tested
and true. Ideology is present by commissions or omissions
in any number of details. Programs like {\Py} and {\Je} represent
a major amount of original physics research, often on complicated
topics where no simple answers are available.
As a (potential) program user you must be
aware of this, so that you can form your own opinion, not just about
what to trust and what not to trust, but also how much to trust a given
prediction, i.e. how uncertain it is likely to be. {\Je} and
{\Py} are particularly well endowed in this respect, since a
number of publications exist where most of the relevant physics is
explained in considerable detail. In fact, the problem may rather be
the opposite, to find the relevant information among all the possible
places. One main objective of the current report is therefore to
collect much of this information in one single place. Not all the
material found in specialized papers is reproduced, by a wide margin,
but at least enough should be found here to understand the general
picture and to know where to go for details.
 
The current report is therefore intended to replace the previous round 
of published physics descriptions and program manuals
\cite{Sjo86,Sjo87,Ben87}. The formal new standard reference is
\cite{Sjo94}, which is a fairly brief summary of this report
--- for obvious reasons the full description is too long to be published 
in its entirety. Further specification could include a statement of 
the type `We use {\Py} version X.x and {\Je} version Y.y'.
(If you are a {\LaTeX} fan, you may want to know that the program 
names in this report have been generated by the commands
\verb+\textsc{Jetset}+ and \verb+\textsc{Pythia}+.)
Kindly do not refer to {\JePy} as
`unpublished', `private communication' or `in preparation':
such phrases are only creating unnecessary confusion.

In addition, remember that many of the individual physics components
are documented in separate publications. If some of these contain 
ideas that are useful to you, there is every reason to cite them. 
A reasonable selection would vary as a function of the physics you are
studying. The criterion for which to pick should be simple: imagine 
that a Monte Carlo implementation had not been available. Would you
then have cited a given paper on the grounds of its physics contents 
alone? If so, do not punish the extra effort of turning these ideas 
into publicly available software. (Monte Carlo manuals are good for
nothing in the eyes of many theorists, so often only the acceptance 
of `mainstream' publications counts.) Here follows a list of some
main areas where the $\PyJe$ programs contain original research:
\begin{Itemize}
\item The string fragmentation model \cite{And83}.
\item The string effect \cite{And80}.
\item Baryon production (diquark/popcorn) \cite{And82,And85}.
\item Fragmentation of multiparton systems \cite{Sjo84}.
\item Fragmentation effects on $\alphas$ determinations \cite{Sjo84a}.
\item Initial state parton showers \cite{Sjo85}.
\item Final state parton showers \cite{Ben87a}.
\item Photon radiation from quarks \cite{Sjo92c}
\item Deep inelastic scattering \cite{And81a,Ben88}.
\item Photoproduction \cite{Sch93a} and $\gamma\gamma$ physics 
\cite{Sch94a}.
\item Parton distributions of the photon \cite{Sch95}.
\item Colour flow in hard scatterings \cite{Ben84}.
\item Elastic and diffractive cross sections \cite{Sch94}.
\item Minijets (multiple parton--parton interactions) \cite{Sjo87a}.
\item Rapidity gaps \cite{Dok92}.
\item Jet clustering in $k_{\perp}$ \cite{Sjo83}.
\end{Itemize}
 
In addition to a physics survey, the current report also contains a
complete manual for the two programs. Such manuals have always been
updated and distributed jointly with the programs. To a first
approximation, we therefore do not have much new to offer here.
However, an attempt has been made to group the material more logically
according to physics topics than in previous distributions,
to tie it closer to the physics description, and to improve the layout
and therefore the readability. Any feedback is welcome.
 
A word of warning may be in place. The program description is fairly
lengthy, and certainly could not be absorbed in one sitting. This is
not even necessary, since all switches and parameters are provided
with sensible default values, based on our best understanding (of
the physics, and of what you expect to happen if you do not
specify any options). As a new user, you can therefore disregard all
the fancy options, and just run the program with a minimum ado.
Later on, as you gain experience, the options that seem useful can
be tried out. No single user is ever likely to find need for more than
a fraction of the total number of possibilities available,
yet many of them have been added to meet specific user requests.

In some instances, not even this report will provide you with all the 
information you desire. You may wish to find out about recent versions 
of the program, know about related software, pick up a few sample 
main programs to get going, or get hold of related physics papers. 
Some such material can be found if you link to my World Wide Web
homepage:\\
 \ttt{http://thep.lu.se/tf2/staff/torbjorn/Welcome.html}\\
and study the contents there. 
 
\subsubsection*{Disclaimer}
 
At all times it should be remembered that this is not a commercial
product, developed and supported by professionals. Instead it is
a `University World' product, developed by a very few physicists
(mainly the current author)
originally for their own needs, and supplied to other
physicists on an `as-is' basis, free of charge. No guarantees are
therefore given for the proper functioning of the programs, nor for
the validity of physics results. In the end, it is always up to you
to decide for yourself whether to trust a given result or not. Usually
this requires comparison either with analytical results or with
results of other programs, or with both. Even this is not necessarily 
foolproof: for instance, if an error
is made in the calculation of a matrix element for a given process,
this error will be propagated both into the analytical results based
on the original calculation and into all the event generators which
subsequently make use of the published formulae. In the end, there
is no substitute for a sound physics judgement.
 
This does not mean that you are all on your own, with a program
nobody feels responsible for. Attempts are made to check processes as
carefully as possible, to write programs that do not invite
unnecessary errors, and to provide a detailed and accurate
documentation. All of this while maintaining the full power and
flexibility, of course, since the physics must always take precedence
in any conflict of interests. If nevertheless any errors or
unclarities are found, please do communicate them to me,
e.g. on phone +46 -- 46 -- 222 48 16 or e-mail torbjorn@thep.lu.se. 
Every attempt will be made to solve problems as soon as is 
reasonably possible, given that this support is by one person 
alone, who also has other responsibilities.
 
\subsubsection*{Appendix: The Historical Pythia}
 
While the origin and connotations of the `{\Je}' program name
should be commonly known, the `{\Py}' label may need some
explanation.
 
The myth tells how Apollon, the God of Wisdom, killed the powerful
dragon-like monster Python, close to the village of Delphi in Greece.
To commemorate this victory, Apollon founded the Pythic Oracle in
Delphi, on the slopes of Mount Parnassos. Here men could come to
learn the will of the Gods and the course of the future. The oracle
plays an important r\^ole in many of the other Greek myths, such
as those of Heracles and of King Oedipus.
 
Questions were to be put to the Pythia, the `Priestess' or
`Prophetess' of the Oracle. In fact, she was a local woman,
usually a young maiden, of no particular religious schooling.
Seated on a tripod, she inhaled the obnoxious vapours that
seeped up through a crevice in the ground. This brought her
to a trance-like state, in which she would scream seemingly
random words and sounds. It was the task of the professional
priests in Delphi to record those utterings and edit them into
the official Oracle prophecies, which often took the form of
poems in perfect hexameter. In fact, even these edited replies
were often less than easy to interpret. The Pythic oracle
acquired a reputation for ambiguous answers.
 
The Oracle existed already at the beginning of the historical
era in Greece, and was universally recognized as the foremost
religious seat. Individuals and city states came to consult, on
everything from cures for childlessness to matters of war. Lavish
gifts allowed the temple area to be built and decorated. Many
states supplied their own treasury halls, where especially beautiful
gifts were on display. Sideshows included the Omphalos,
a stone reputedly marking the centre of the Earth, and the Pythic
games, second only to the Olympic ones in importance.
 
Strife inside Greece eventually led to a decline in the power of
the Oracle. A serious blow was dealt when the Oracle of Zeus Ammon
(see below) declared Alexander the Great
to be the son of Zeus. The Pythic Oracle lived on, however, and was
only closed by a Roman Imperial decree in 390 \tsc{ad}, at a time 
when Christianity was ruthlessly destroying any religious opposition.
Pythia then had been at the service of man and Gods for a
millenium and a half.
 
The r\^ole of the Pythic Oracle replies on the course of history is
nowhere better described than in `The Histories' by Herodotus
\cite{HerBC}, the classical and captivating description of the
Ancient World at the time of the Great War between Greeks and
Persians. Especially famous is the episode with King Croisus
of Lydia. Contemplating a war against the upstart Persian
Empire, he resolves to ask an oracle what the outcome of a potential
battle would be. However, to have some guarantee for the
veracity of any prophecy, he decides to send embassies to all the
renowned oracles of the known World. The messengers are instructed
to inquire the various divinities, on the hundredth day after
their departure, what King Croisus is doing at that very moment.
From the Pythia the messengers bring back the reply
\begin{em}\begin{verse}
I know the number of grains of sand as well as  the expanse of
the sea, \\
And I comprehend the dumb and hear him who does not speak, \\
There came to my mind the smell of the hard-shelled turtle, \\
Boiled in copper together with the lamb, \\
With copper below and copper above.
\end{verse}\end{em}
The veracity of the Pythia is thus established by the crafty ruler,
who had waited until the appointed day, slaughtered a turtle and a
lamb, and boiled them together in a copper cauldron with a
copper lid. Also the Oracle of Zeus Ammon in the Libyan desert
is able to give a correct reply (lost to posterity), while all
others fail. King Croisus now sends a second embassy to Delphi,
inquiring after the outcome of a battle against the Persians.
The Pythia answers
\begin{em}\begin{verse}
If Croisus passes over the Halys he will dissolve a great Empire.
\end{verse}\end{em}
Taking this to mean he would win, the King collects his army and
crosses the border river, only to suffer a crushing defeat and
see his Kingdom conquered. When the victorious King Cyrus allows
Croisus to send an embassy to upbraid the Oracle, the God Apollon
answers through his Prophetess that he has correctly predicted the
destruction of a great empire --- Croisus' own --- and that he
cannot be held responsible if people choose to interpret the
Oracle answers to their own liking.
 
The history of the {\Py} program is neither as long nor as
dignified as that of its eponym. However, some points of contact
exist. You must be very careful when you formulate the questions:
any ambiguities will corrupt the reply you get. And you must be even
more careful not to misinterpret the answers; in particular not to pick
the interpretation that suits you before considering the alternatives.
Finally, even a perfect God has servants that are only human: a priest
might mishear the screams of the Pythia and therefore produce an
erroneous oracle reply; the current author might unwittingly let a
bug free in the program {\Py}.
 
\clearpage
 
\section{Physics Overview}
 
In this section we will try to give an overview of the main physics
features of {\Je} and {\Py}, and also to introduce some
terminology. The details will be discussed in subsequent sections.
 
For the description of a typical high-energy event, an event
generator should contain a simulation of several physics aspects.
If we try to follow the evolution of an event in some semblance of
a time order, one may arrange these aspects as follows:
\begin{Enumerate}
\item Initially two beam particles are coming in towards each other.
      Normally each particle is characterized by a set of parton
      distribution functions, which defines the partonic substructure 
      in terms of flavour composition and energy sharing.
\item One shower initiator parton from each beam starts off
      a sequence of branchings, such as $\q \to \q \g$, which build up
      an initial-state shower.
\item One incoming parton from each of the two showers
      enters the hard process, where then a number of
      outgoing partons are produced, usually two.
      It is the nature of this process that determines the main
      characteristics of the event.
\item Also the outgoing partons may branch, to build up
      final-state showers.
\item When a shower initiator is taken out of a beam particle,
      a beam remnant is left behind. This remnant may have
      an internal structure, and a net colour charge that relates
      it to the rest of the final state.
\item The QCD confinement mechanism ensures that the outgoing quarks 
      and gluons are not observable, but instead fragment to colour 
      neutral hadrons.
\item Many of the produced hadrons are unstable and decay further.
\end{Enumerate}
 
Conventionally, only quarks and gluons are counted as partons, while
leptons and photons are not. If pushed {\it ad absurdum} this may 
lead to
some unwieldy terminology. We will therefore, where it does not matter,
speak of an electron or a photon in the `partonic' substructure of an
electron, lump branchings $\e \to \e \gamma$ together with other
`parton shower' branchings such as $\q \to \q \g$, and so on. With
this notation, the division into the above seven points applies equally
well to an interaction between two leptons, between a lepton and a
hadron, and between two hadrons.
 
In the following subsections, we will survey the above seven aspects,
not in the same order as given here, but rather in the order in 
which they appear in the program execution, i.e. starting with the 
hard process.
 
\subsection{Hard Processes and Parton Distributions}
 
In {\Je}, only two hard processes are available. The first and
main one is $\ee \to \gammaZ \to \q \qbar$. Here the `$*$' of
$\gamma^*$ is used to denote that the photon must be off the mass shell.
The distinction is of some importance, since a photon on the mass shell
cannot decay. Of course also the $\Z^0$ can be off the mass shell,
but here the  distinction is less relevant (strictly speaking,
a $\Z^0$ is always off the mass shell). In the following we may not
always use `$*$' consistently, but the rule of thumb is to use a `$*$'
only when a process is not kinematically possible for a particle of
nominal mass. The quark $\q$ in the final state of
$\ee \to \gammaZ \to \q \qbar$
may be $\u$, $\d$, $\s$, $\c$, $\b$ or $\t$; the flavour in
each event is picked at random, according to the relative couplings,
evaluated at the hadronic c.m. energy. Also the angular distribution of
the final $\q \qbar$ pair is included. No parton-distribution functions 
are needed.
 
The other {\Je} process is a routine to generate $\g \g \g$ and
$\gamma \g \g$ final states, as expected in onium 1$^{--}$ decays
such as $\Upsilon$. Given the current limits on the top mass,
toponium will decay weakly much too fast for these processes to be of
any interest, so therefore no new applications are expected.
 
{\Py} contains a much richer selection, with close to a
hundred different hard processes. These may be classified in
many different ways.
 
One is according to the number of final-state objects: we speak of
`$2 \to 1$' processes, `$2 \to 2$' ones, `$2 \to 3$' ones, etc.
This aspect is very relevant from a programming point of view:
the more particles in the final state, the more complicated the
phase space and therefore the whole generation procedure. In fact,
{\Py} is optimized for $2 \to 1$ and $2 \to 2$ processes.
There is currently no generic treatment of processes with three or
more particles in the final state, but rather a few different
machineries, each tailored to the pole structure of a specific class of
graphs. This may be seen as a major limitation, and indeed is so
at times. However, often one can come quite far with only one
or two particles in the final state, since showers will add the
required extra activity. The classification may also be misleading
at times, since an $s$-channel resonance is considered as a single
particle, even if it is assumed always to decay into two final-state
particles. Thus the process
$\ee \to \W^+ \W^- \to \q_1 \qbar'_1 \, \q_2 \qbar'_2$ is classified
as $2 \to 2$, although the decay treatment of the $\W$ pair includes
the full $2 \to 4$ matrix elements.
 
Another classification is according to the physics scenario. This will
be the main theme of section \ref{s:pytproc}. The following major
groups may be distinguished:
\begin{Itemize}
\item Hard QCD processes, e.g. $\q \g \to \q \g$.
\item Soft QCD processes, such as diffractive and elastic scattering,
and minimum-bias events.
\item Heavy-flavour production, e.g. $\g \g \to \t \tbar$.
\item Prompt-photon production, e.g. $\q \g \to \q \gamma$.
\item Photon-induced processes, e.g. $\gamma \g \to \q \qbar$.
\item Deep inelastic scattering, e.g. $\q \ell \to \q \ell$.
\item $\W / \Z$ production, such as the $\ee \to \gammaZ$ already
found in {\Je}, or $\q \qbar \to \W^+ \W^-$.
\item Standard model Higgs production, where the Higgs is reasonably
light and narrow, and can therefore still be considered as a resonance.
\item Gauge boson scattering processes, such as $\W \W \to \W \W$,  
when the Standard Model Higgs is so heavy and broad that resonant and
non-resonant contributions have to be considered together.
\item Non-standard Higgs particle production, within the framework
of a two-Higgs-doublet scenario with three neutral and two charged
Higgs states.
\item Production of new gauge bosons, such as a $\Z'$.
\item Production of fourth-generation fermions.
\item Leptoquark production.
\item Deviations from Standard Model processes, e.g. due to contact
interactions or a strongly interacting gauge boson sector. These
scenarios do not always appear as separate processes, but may just
be options to some of the processes above.
\end{Itemize}
This is by no means a survey of all interesting physics.
Most notable is the absence of supersymmetric particle production
and decay, but many other examples could be found. Also, within
the scenarios studied, not all contributing graphs have always been
included, but only the more important and/or more interesting ones.
In many cases, various approximations are involved in the matrix
elements coded.
 
The cross section for a given process $ij \to k$ is given by
\begin{equation}
\sigma_{ij \to k} = \int \d x_1 \int \d x_2 \, f^1_i(x_1) \,
f^2_j(x_2) \, \hat{\sigma}_{ij \to k} ~.
\end{equation}
Here $\hat{\sigma}$ is the cross section for the hard partonic process,
as codified in the matrix elements for each specific process.
For processes with many particles in the final state 
it would be replaced by an
integral over the allowed final-state phase space. The $f^a_i(x)$ are
the parton-distribution functions, which describe the probability to 
find a parton $i$ inside beam particle $a$, with parton $i$ carrying a
fraction $x$ of the total $a$ momentum. Actually, parton distributions
also depend on some momentum scale $Q^2$ that characterizes the hard
process.
 
Parton distributions are most familiar for hadrons, such as the proton.
Hadrons are inherently composite objects, made up of quarks and
gluons. Since we do not understand QCD, a derivation from first 
principles
of hadron parton distributions does not yet exist, although some
progress is being made in lattice QCD studies. It is therefore
necessary to rely on parametrizations, where experimental data are
used in conjunction with the evolution equations for the $Q^2$
dependence, to pin down the parton distributions. Several different
groups have therefore produced their own fits, based on slightly
different sets of data, and with some variation in the theoretical
assumptions.
 
Also for fundamental particles, such as the electron, is it convenient
to introduce parton distributions. The function $f^{\e}_{\e}(x)$ thus
parametrizes the probability that the electron that takes part in the
hard process retains a fraction $x$ of the original energy, the rest
being radiated (into photons) in the initial state. Of course, such
radiation could equally well be made part of the hard interaction,
but the parton-distribution approach usually is much more convenient.
If need be, a description with fundamental electrons is recovered for
the choice $f_{\e}^{\e}(x, Q^2)=\delta(x-1)$. Note that, contrary to
the proton case, electron parton distributions are calculable from first
principles, and reduce to the $\delta$ function above for $Q^2 \to 0$.
 
The electron may also contain photons, and the photon may in its turn
contain quarks and gluons. The internal structure of the
photon is a bit of a problem, since the photon
contains a point-like part, which is perturbatively calculable, and
a vector-meson dominance part, which is not. Normally, the photon
parton distributions are therefore parametrized, just as the hadron
ones. Since the electron ultimately contains quarks and gluons,
hard QCD processes like $\q \g \to \q \g$ therefore not only appear in
$\pp$ collisions, but also in $\ep$ ones (`resolved photoproduction')
and in $\ee$ ones (`doubly resolved 2$\gamma$ events'). The parton 
distribution function approach here makes it much easier to reuse one 
and the same hard process in different contexts.
 
There is also another kind of possible generalization. The two
processes $\q \qbar \to \gammaZ$, studied in hadron colliders,
and $\ee \to \gammaZ$, studied in $\ee$ colliders, are really
special cases of a common process, $\f \fbar \to \gammaZ$,
where $\f$ denotes a fundamental fermion, i.e. a quark, lepton or
neutrino. The whole structure is therefore only coded once, and
then slightly different couplings and colour prefactors are used,
depending on the initial state considered. Usually the
interesting cross section is a sum over several different initial
states, e.g. $\u \ubar \to \gammaZ$ and
$\d \dbar \to \gammaZ$ in a hadron collider. This kind of
summation is always implicitly done, even when not explicitly
mentioned in the text.
 
\subsection{Initial- and Final-State Radiation}
 
In every process that contains coloured and/or charged objects
in the initial or final state, gluon and/or photon radiation
may give large corrections to the overall topology of events.
Starting from a basic $2 \to 2$ process, this kind of corrections
will generate $2 \to 3$, $2 \to 4$, and so on, final-state
topologies. As the available energies are increased, hard
emission of this kind is increasingly important, relative to
fragmentation, in determining the event structure.
 
Two traditional approaches exist to the modelling of perturbative
corrections. One is the matrix-element method, in
which Feynman diagrams are calculated, order by order. In principle,
this is the correct approach, which takes into account exact
kinematics, and the full interference and helicity structure. The
only problem is that calculations become increasingly difficult in
higher orders, in particular for the loop graphs.
Only in exceptional cases have therefore more than one loop been
calculated in full, and often we do not have any loop corrections
at all at our disposal. On the other hand,
we have indirect but strong evidence that, in fact, the emission
of multiple soft gluons plays a significant r\^ole in building up the
event structure, e.g. at LEP, and this sets a limit to the
applicability of matrix elements.
Since the phase space available for gluon emission increases with
the available energy, the matrix-element approach becomes less
relevant for the full structure of events at higher energies.
However, the perturbative expansion by itself is better
behaved at higher energies, owing to the running of $\alphas$.
As a consequence, inclusive measurements, e.g. of the rate of  
well-separated jets, should yield more reliable results.
 
The second possible approach is the parton-shower one. Here an
arbitrary number of branchings of one parton into two (or more)
may be put together, to yield a description of multijet events,
with no explicit upper limit on the number of partons involved.
This is possible since the full matrix-element expressions are
not used, but only approximations derived by simplifying the
kinematics, and the interference and helicity structure. Parton showers
are therefore expected to give a good description of the substructure of
jets, but in principle the shower approach has limited predictive power
for the rate of well-separated jets (i.e. the 2/3/4/5-jet composition).
In practice, shower programs may be patched up to describe the
hard-gluon emission region reasonably well, in particular for the
$\ee$ annihilation process.
Nevertheless, the shower description is not optimal for
absolute $\alphas$ determinations.
 
Thus the two approaches are complementary in many respects,
and both have found use. However, because of its simplicity
and flexibility, the parton-shower option is generally the
first choice, while the matrix elements one is mainly used for
$\alphas$ determinations, angular distribution of jets,
triple-gluon vertex studies, and other specialized studies.
Obviously, the ultimate goal would be to have an approach where the
best aspects of the two worlds are harmoniously married.
 
\subsubsection{Matrix elements}
 
Matrix elements are especially made use of in the {\Je}
implementation of the process $\ee \to \gammaZ \to \q \qbar$.
 
For initial-state QED radiation, a first order (unexponentiated)
description has been adopted. This means that events are subdivided
into two classes, those where a photon is radiated above some
minimum energy, and those without such a photon. In the latter
class, the soft and virtual corrections have been lumped together
to give a total event rate that is correct up to one loop. This 
approach worked fine at PETRA/PEP energies, but does not do so well
for the $\Z^0$ line shape, i.e. in regions where the cross section
is rapidly varying and high precision is strived for.
 
For final-state QCD radiation, several options are available. The
default is the parton-shower one (see below), but the matrix-elements
options are also frequently used. In the definition of 3- or
4-jet events, a cut is introduced whereby it is required that
any two partons have an invariant mass bigger than some fraction
of the c.m. energy. 3-jet events which do not fulfill this
requirement are lumped with the 2-jet ones. The first-order
matrix-element option, which only contains 3- and 2-jet events
therefore involves no ambiguities. In second order, where also
4-jets have to be considered, a main issue is what to do with
4-jet events that fail the cuts. Depending on the choice of
recombination scheme, whereby the two nearby partons are joined
into one, different 3-jet events are produced. Therefore the
second-order differential 3-jet rate has been the subject of
some controversy, and {\Je} actually contains two
different implementations.
 
By contrast, {\Py} does not contain any full higher-order
matrix elements, with loop contributions included. There are a few
cases where higher-order matrix elements are included at the Born
level. Consider e.g. the case of $\W$ production at a hadron collider,
which is contained in the lowest-order process $\q \qbar' \to \W$.
In an inclusive description, additional jets recoiling against the
$\W$ may be generated by parton showers. {\Py} also contains
the two first-order processes $\q \g \to \W \q'$ and
$\q \qbar' \to \W \g$. The cross sections for these processes
are divergent when the $\pT \to 0$. In this region a correct
treatment would therefore have to take into account loop corrections,
which are not available in {\Py}. Depending on the physics
application, one could then use {\Py} in one of two ways.
In the region of small $\pT$, the preferred option is lowest-order 
matrix elements combined with parton showers. For the
production of a $\W$ at large $\pT$, on the other hand, the
shower approach is too imprecise to give the right cross section;
additionally the event selection  machinery is very inefficient.
Here it is advantageous to generate first-order events, and
then add showers only to describe additional softer radiation.
 
\subsubsection{Parton showers}
 
The separation of radiation into initial- and final-state showers is
arbitrary, but very convenient. There are also situations where it
is appropriate: for instance, the process 
$\ee \to \Z^0 \to \q \qbar$ only
contains final-state QCD radiation (QED radiation, however, is
possible both in the initial and final state), while
$\q \qbar \to \Z^0 \to \ee$ only contains initial-state QCD one.
Similarly, the distinction of emission as coming either from the
$\q$ or from the $\qbar$ is arbitrary. In general, the assignment
of radiation to a given mother parton is a good approximation
for an emission close to the direction of motion of that parton,
but not for the wide-angle emission in between two jets, where
interference terms are expected to be important.
 
In both initial- and final-state showers, the structure is given in
terms of branchings $a \to bc$, specifically $\e \to \e \gamma$,
$\q \to \q \g$, $\q \to \q \gamma$, $\g \to \g \g$,
and $\g \to \q \qbar$. Each of these processes is characterized by
a splitting kernel $P_{a \to bc}(z)$. The branching rate is
proportional to the integral $\int P_{a \to bc}(z) \, \d z$.
The $z$ value picked for a branching describes the energy sharing,
with daughter $b$ taking a fraction $z$ and daughter $c$ the
remaining $1-z$ of the $a$ energy. Once formed, the daughters 
$b$ and $c$ may in turn branch, and so on.
 
Each parton is characterized by some virtuality scale $Q^2$,
which gives an approximate sense of time ordering
to the cascade. In the initial-state shower, $Q^2$ values are
gradually increasing as the hard scattering is approached, while
$Q^2$ is decreasing in the final-state showers.
Shower evolution is cut off at some lower
scale $Q_0$, typically around 1 GeV for QCD branchings. The same
cut-off scale is also used to regularize the soft gluon emission
divergences in the splitting kernels. From above,
a maximum scale $Q_{\mmax}$ is introduced, where the showers are
matched to the hard interaction itself. The relation between 
$Q_{\mmax}$ and the kinematics of the hard scattering
is uncertain, and the choice made can strongly affect the
amount of well-separated jets.
 
Despite a number of common traits, the initial- and final-state
radiation machineries are in fact quite different, and are described
separately below. For historical reasons, the final-state shower
is found in {\Je} and the initial-state one in {\Py}.
 
Final-state showers are time-like,
i.e. partons have $m^2 = E^2 - \mbf{p}^2 \geq 0$. The evolution
variable $Q^2$ of the cascade is therefore in {\Je}
associated with the $m^2$ of the
branching parton, but this choice is not unique.
Starting from $Q^2_{\mmax}$, an original parton is evolved
downwards in $Q^2$ until a branching occurs. The selected
$Q^2$ value defines the mass of the branching parton, and the $z$
of the splitting kernel the parton energy division between its
daughters. These daughters may now, in turn, evolve
downwards, in this case with maximum virtuality already defined by
kinematics, and so on down to the $Q_0$ cut-off.
 
In QCD showers, corrections to the leading-log picture, so-called 
coherence effects, lead to an ordering of subsequent emissions in terms
of decreasing angles. This does not follow automatically from the
mass-ordering constraint, but is implemented as an additional
requirement on allowed emissions. Photon emission is not affected
by angular ordering. It is also possible to obtain
non-trivial correlations between azimuthal angles in the various
branchings, some of which are implemented as
options. Finally, the theoretical analysis strongly
suggests the scale choice $\alphas = \alphas(\pT^2) =
\alphas(z(1-z)m^2)$, and this is the default in the program.
 
The final-state radiation machinery is applied in the c.m. frame of the
hard scattering. The total energy and momentum of the hard-scattering
subsystem is preserved, as is the direction of the outgoing partons
(in that frame).
 
In contrast to final-state showers, initial-state ones are space-like.
This means that, in the sequence of branchings $a \to bc$ that lead
up from the shower initiator to the hard interaction,
particles $a$ and $b$ have $m^2 = E^2 - \mbf{p}^2 <0$.
The `side branch' particle $c$, which does not participate
in the hard scattering, may be on the mass shell, or have a time-like
virtuality. In the latter case a time-like shower will evolve off
it, rather like the final-state radiation described above. To first
approximation, the evolution of the space-like main branch
is characterized by the
evolution variable $Q^2 = -m^2$, which is required to be strictly
increasing along the shower, i.e. $Q_b^2 > Q_a^2$. Corrections
to this picture have been calculated,
but are basically absent in {\Py}.
 
Initial-state radiation is handled within the backwards evolution
scheme. In this approach, the
choice of the hard scattering is based on the use of evolved
parton distributions, which means that
the inclusive effects of initial-state radiation are already
included. What remains is therefore to construct the exclusive
showers. This is done starting from the two incoming partons
at the hard interaction, tracing the showers `backwards in time',
back to the two shower initiators. In other words,
given a parton $b$, one tries to find the parton $a$ that branched
into $b$. The evolution in the Monte Carlo is therefore in
terms of a sequence of decreasing space-like virtualities $Q^2$
and increasing momentum fractions $x$. Branchings on
the two sides are interleaved in a common sequence of
decreasing $Q^2$ values.
 
In the above formalism, there is no real distinction between
gluon and photon emission. Some of the details actually do differ,
as will be explained in the full description.
 
The initial- and final-state radiation shifts around the kinematics of
the original hard interaction. In deep inelastic scattering, this
means that the $x$ and $Q^2$ values that can be derived from the 
momentum of the scattered lepton do not agree with the values 
originally picked. In high-$\pT$ processes, it means that one no 
longer has two jets with opposite and compensating $\pT$, but more 
complicated topologies. Effects of any original kinematics selection 
cuts are therefore smeared out, an unfortunate side-effect of the 
parton-shower approach.
 
\subsection{Beam Remnants}
 
In a hadron--hadron collision, the initial-state radiation algorithm
reconstructs one shower initiator in each beam. This initiator only
takes some fraction of the total beam energy, leaving behind a beam
remnant which takes the rest. For a proton beam, a $\u$ quark
initiator would leave behind a $\u \d$ diquark beam remnant, with an
antitriplet colour charge. The remnant is therefore colour-connected
to the hard interaction, and forms part of the same fragmenting
system. It is further customary to assign a primordial transverse
momentum to the shower initiator, to take into account the motion
of quarks inside the original hadron, basically as required by the
uncertainty principle. This primordial $k_{\perp}$ is selected
according to some suitable distribution, and the recoil is assumed
to be taken up by the beam remnant.
 
Often the remnant is more complicated, e.g. a $\g$ initiator
would leave behind a $\u \u \d$ proton remnant system in a colour octet
state, which can conveniently be subdivided into a colour triplet
quark and a colour antitriplet diquark, each of which are 
colour-connected to the hard interaction. The energy sharing between 
these two remnant objects, and their relative transverse momentum,
introduces additional degrees of freedom, which are not understood 
from first principles.
 
Na\"{\i}vely, one would expect an $\ep$ event to have only one beam
remnant, and an $\ee$ event none. This is not always correct, e.g.
a $\gamma \gamma \to \q \qbar$ interaction in an $\ee$ event would
leave behind the $\e^+$ and $\e^-$ as beam remnants, and a
$\q \qbar \to \g \g$ interaction in resolved photoproduction in an
$\ee$ event would leave behind one $\e^{\pm}$ and one $\q / \qbar$
in each remnant. Corresponding complications occur for
photoproduction in $\ep$ events.
 
There is another source of beam remnants. If parton distributions are
used to resolve an electron inside an electron, some of the original
energy is not used in the hard interaction, but is rather associated
with initial-state photon radiation. The initial-state shower is
in principle intended to trace this evolution and reconstruct the
original electron before any radiation at all took place. However,
because of cut-off procedures, some small amount may be left 
unaccounted.
Alternatively the user may have chosen to switch off initial-state
radiation altogether, but still preserved the resolved electron
parton distributions. In either case the remaining energy is given to
a single photon of vanishing transverse momentum, which is then
considered in the same spirit as `true' beam remnants.
 
So far we have assumed that
each event only contains one hard interaction, i.e. that each
incoming particle has only one parton which takes part in hard
processes, and that all other constituents sail through unaffected.
This is appropriate in $\ee$ or $\ep$ events, but not necessarily so in
hadron--hadron collisions. Here each of the beam particles contains a
multitude of partons, and so the probability for several interactions
in one and the same event need not be negligible. In principle these
additional interactions could arise because one single parton from
one beam scatters against several different partons from the other
beam, or because several partons from each beam take place in
separate $2 \to 2$ scatterings. Both are expected, but combinatorics
should favour the latter, which is the mechanism considered in
{\Py}.
 
The dominant $2 \to 2$ QCD cross sections  are
divergent for $\pT \to 0$, and drop rapidly for larger
$\pT$. Probably the lowest-order perturbative cross sections
will be regularized at small $\pT$ by colour coherence effects:
an exchanged gluon of small $\pT$ has a large transverse
wave function and can therefore not resolve the individual colour
charges of the two incoming hadrons; it will only couple to an
average colour charge that vanishes in the limit $\pT \to 0$.
In the program, some effective $\pTmin$ scale is therefore
introduced, below which the perturbative cross section is either
assumed completely vanishing or at least strongly damped.
Phenomenologically, $\pTmin$ comes out to be a number of 
the order of 1.5--2.0 GeV.
 
In a typical `minimum-bias' event one therefore expects to find one
or a few scatterings at scales around or a bit above 
$\pTmin$,
while a high-$\pT$ event also may have additional scatterings
at the $\pTmin$ scale. The probability to have several
high-$\pT$ scatterings in the same event is small, since the
cross section drops so rapidly with $\pT$.
 
The understanding of multiple interaction is still very primitive,
and even the experimental evidence that it exists at all is rather
weak. {\Py} therefore contains several different options,
with a fairly simple one as default. The options differ in particular
on the issue of the `pedestal' effect: is there an increased
probability or not for additional interactions in an event which
is known to contain a hard scattering, compared with one that
contains no hard interactions?
 
\subsection{Fragmentation}
 
QCD perturbation theory, formulated in terms of quarks and
gluons, is valid at short distances. At long distances, QCD
becomes strongly interacting and perturbation theory breaks
down. In this confinement regime, the coloured partons are
transformed into colourless hadrons, a process called either
hadronization or fragmentation. In this paper we reserve the
former term for the combination of fragmentation and the subsequent
decay of unstable particles.
 
The fragmentation process has yet to be understood from first
principles, starting from the QCD Lagrangian. This has left the
way clear for the development of a number of different
phenomenological models. Three main schools are usually
distinguished, string fragmentation (SF), independent
fragmentation (IF) and cluster fragmentation (CF),
but many variants and hybrids exist.
Being models, none of them can lay claims
to being `correct', although some may be better founded than
others. The best that can be aimed for is internal consistency, 
a good representation of existing data, and a predictive power for
properties not yet studied or results at higher energies.
 
{\Je} is intimately connected with string fragmentation,
in the form of the time-honoured `Lund model'. This is the default
for all {\JePy} applications, but independent
fragmentation options also exist, for applications
where one wishes to study the importance of string effects.
 
All current models are of a probabilistic and iterative nature.
This means that the fragmentation process as a whole is described in
terms of one or a few simple underlying branchings, of the type
jet $\to$ hadron + remainder-jet, string $\to$
hadron + remainder-string, and so on. At each branching,
probabilistic rules are given for the production of new flavours,
and for the sharing of energy and momentum between the products.
 
To understand fragmentation models, it is useful to start with
the simplest possible system, a colour-singlet $\q \qbar$ 2-jet
event, as produced in $\ee$ annihilation. Here lattice QCD studies
lend support to a linear confinement picture (in the absence
of dynamical quarks), i.e. the energy stored in the colour
dipole field between a charge and an anticharge increases linearly
with the separation between the charges, if the short-distance
Coulomb term is neglected. This is quite different
from the behaviour in QED, and is related to the presence of a
triple-gluon vertex in QCD. The details are not yet well
understood, however.
 
The assumption of linear confinement provides the starting point for
the string model. As the $\q$ and $\qbar$ partons move apart from
their common production vertex, the physical picture is that of a
colour flux tube (or maybe colour vortex line) being stretched
between the $\q$ and the $\qbar$. The transverse dimensions
of the tube are of typical hadronic sizes, roughly 1 fm. If
the tube is assumed to be uniform along its length, this
automatically leads to a confinement picture with a linearly
rising potential. In order to obtain a Lorentz covariant and causal
description of the energy flow due to this linear confinement,
the most straightforward way is to use the dynamics of the massless
relativistic string with no transverse degrees of freedom.
The mathematical, one-dimensional string can
be thought of as parametrizing the position of the axis of a
cylindrically symmetric flux tube. From
hadron spectroscopy, the string constant, i.e. the amount of
energy per unit length, is deduced to be
$ \kappa \approx 1$ GeV/fm. The
expression `massless' relativistic string is somewhat of a
misnomer: $\kappa$ effectively corresponds to a `mass density' along
the string.
 
Let us now turn to the fragmentation process. As the $\q$ and
$\qbar$ move apart, the potential energy stored in the string
increases, and the string may break by the production of a new
$\q' \qbar'$ pair, so that the system splits into two
colour-singlet systems $\q \qbar'$ and $\q' \qbar$. If the invariant
mass of either of these string pieces is large enough, further
breaks may occur. In the Lund string model, the string break-up
process is assumed to proceed until only on-mass-shell hadrons
remain, each hadron corresponding to a small piece of string
with a quark in one end and an antiquark in the other.
 
In order to generate the quark--antiquark pairs $\q' \qbar'$ which
lead to string break-ups, the Lund model invokes the idea of
quantum mechanical tunnelling. This leads to a flavour-independent
Gaussian spectrum for the $\pT$ of $\q' \qbar'$ pairs.
Since the string is assumed to have no transverse excitations,
this $\pT$ is locally compensated between the quark and the
antiquark of the pair. The total $\pT$ of a hadron is made
up out of the $\pT$ contributions from the quark and
antiquark that together
form the hadron. Some contribution of very soft perturbative gluon
emission may also effectively be included in this description.
 
The tunnelling picture also implies a suppression of heavy-quark
production, $\u : \d : \s : \c \approx 1 : 1 : 0.3 : 10^{-11}$.
Charm and heavier quarks hence are not expected to be produced in
the soft fragmentation, but only in perturbative parton-shower
branchings $\g \to \q \qbar$.
 
When the quark and antiquark from two adjacent string breakings are
combined to form a meson, it is necessary to invoke an algorithm to
choose between the different allowed possibilities, notably
between pseudoscalar and vector mesons.
Here the string model is not particularly predictive. Qualitatively one
expects a $1 : 3$ ratio, from counting the number of spin states,
multiplied by some wave-function normalization factor, which should
disfavour heavier states.
 
A tunnelling mechanism can also be used to explain the production of
baryons. This is still a poorly understood area. In the simplest
possible approach, a diquark in a colour antitriplet state is just
treated like an ordinary antiquark, such that a string can break
either by quark--antiquark or antidiquark--diquark pair production.
A more complex scenario is the `popcorn' one, where
diquarks as such do not exist, but rather quark--antiquark pairs
are produced one after the other. This latter picture gives a less
strong correlation in flavour and momentum space between the
baryon and the antibaryon of a pair.
 
In general, the different string breaks are causally disconnected.
This means that it is possible to describe the breaks in any convenient
order, e.g. from the quark end inwards. One therefore is led to write
down an iterative scheme for the fragmentation, as follows.
Assume an initial quark $\q$ moving out along the $+z$ axis, with the
antiquark going out in the opposite direction.
By the production of a $\q_1 \qbar_1$ pair, a meson $\q \qbar_1$ is
produced, leaving behind an unpaired quark $\q_1$.
A second pair $\q_2 \qbar_2$ may now be produced,
to give a new meson $\q_1 \qbar_2$, etc. At each step the produced
hadron takes some fraction of the available energy and momentum.
This process may be iterated until all energy is used up, with some
modifications close to the $\qbar$ end of the string in order to 
make total energy and momentum come out right.
 
The choice of starting the fragmentation from the quark end is
arbitrary, however. A fragmentation process described in terms of
starting at the $\qbar$ end of the system and fragmenting towards
the $\q$ end should be equivalent.
This `left--right' symmetry constrains the allowed shape of the
fragmentation function $f(z)$, where $z$ is the fraction
of the remaining light-cone momentum $E \pm p_z$ (+ for the $\q$ jet,
$-$ for the $\qbar$ one) taken by each new particle.
The resulting `Lund symmetric fragmentation function' has two free
parameters, which are determined from data.
 
If several partons are moving apart from a common origin, the details
of the string drawing become more complicated. For a $\q \qbar \g$
event, a string is stretched from the
$\q$ end via the $\g$ to the $\qbar$ end, i.e.
the gluon is a kink on the string, carrying energy and momentum.
As a consequence, the gluon has two string pieces attached, and
the ratio of gluon to quark string force is 2, a number which
can be compared with the ratio of colour charge Casimir operators,
$N_C/C_F = 2/(1-1/N_C^2) = 9/4$. In this, as in other
respects, the string model can be viewed as a variant of QCD
where the number of colours $N_C$ is not 3 but infinite.
Note that the factor 2 above does not depend on
the kinematical configuration: a smaller opening angle between
two partons corresponds to a smaller
string length drawn out per unit time, but also to an increased
transverse velocity of the string piece, which gives an exactly
compensating boost factor in the energy density per unit string
length.
 
The $\q \qbar \g$ string will fragment along its length. To first
approximation this means that there is
one fragmenting string piece between
$\q$ and $\g$ and a second one between $\g$ and $\qbar$. One hadron
is straddling both string pieces, i.e. sitting around the gluon
corner. The rest of the particles are produced as in two simple
$\q \qbar$ strings, but strings boosted with respect to the overall
c.m. frame. When considered in detail, the string motion and
fragmentation is more complicated, with the appearance of
additional string regions during the time evolution of the system.
These corrections are especially important for soft and
collinear gluons, since they provide a smooth transition between
events where such radiation took place and events where it did not.
Therefore the string fragmentation scheme is `infrared safe' with
respect to soft or collinear gluon emission.
 
For events that involve many partons, there may be several possible
topologies for their ordering along the string.
An example would be a $\q \qbar \g_1 \g_2$ (the gluon indices are here
used to label two different gluon-momentum vectors), where the
string can connect the partons in either of the sequences
$\q - \g_1 - \g_2 - \qbar$ and $\q - \g_2 - \g_1 - \qbar$.
The matrix elements that are calculable in perturbation theory
contain interference terms between these two possibilities, which
means that the colour flow is not always well-defined. Fortunately,
the interference terms are down in magnitude by a factor
$1/N_C^2$, where $N_C = 3$ is the number of colours, so
approximate recipes can be found. In the leading log shower
description, on the other hand, the rules for the colour flow are
well-defined. A final comment: in the argumentation for the
importance of colour flows there is a tacit assumption that 
soft-gluon exchanges between partons will not normally mess up the 
original colour assignment; this is likely the case but has not 
been proven.
 
\subsection{Decays}
 
A large fraction of the particles produced by fragmentation are
unstable and subsequently decay into the observable stable (or
almost stable) ones. It  is therefore important to include all
particles with their proper mass distributions and decay properties. 
Although involving little deep physics, this is less trivial than 
it may sound: while a lot of experimental information is available, 
there is also very much that is missing. For charm mesons,
it is necessary to put together measured exclusive branching ratios
with some inclusive multiplicity distributions to obtain a consistent
and reasonably complete set of decay channels, a rather delicate
task. For bottom, so far only a rather simple phase-space type of
generator has been used for hadronic decays.
 
Normally it is assumed that decay products are distributed according
to phase space, i.e. that there is no dynamics involved in their
relative distribution. However, in many cases additional assumptions
are necessary, e.g. for semileptonic decays of charm and bottom
hadrons one needs to include the proper weak matrix elements.
Particles may also be produced polarized and impart a non-isotropic
distribution to their decay products. Many of these effects are
not at all treated in the program. In fact, spin information is not
at all carried along, but has to be reconstructed explicitly when
needed.
 
The normal decay treatment is handled by {\Je}, making use
of a set of tables where branching ratios and decay modes are
stored. In {\Py} a separate decay treatment exists, used
exclusively for a specific list of particles:
$\Z^0$, $\W^{\pm}$, $\H^0$, $\Z'^0$, $\W'^{\pm}$, $\H'^0$, $\A^0$,
$\H^{\pm}$, $\eta_{\mrm{tech}}^0$, $\R^0$, $\q^*$, $\ell^*$, and 
the leptoquark $\L_{\Q}$.
Together we call these resonances, and contrast the `particle decay'
treatment of {\Je} with the `resonance decay' one of
{\Py}. Of course, this is just a matter of terminology:
a particle like the $\rho$ could also be called a resonance.
What characterizes a ({\Py}) resonance is that partial widths
and branching ratios are calculated dynamically, as a function of
the actual mass of a particle. Therefore not only do branching ratios
change between an $\H^0$ of nominal mass 100 GeV and one of 200 GeV,
but also for a Higgs of nominal mass 200 GeV, the branching ratios
would change between an actual mass of 190 GeV and 210 GeV, say.
This is particularly relevant for reasonably broad resonances, and
in threshold regions. For an approach like this to work, it is
clearly necessary to have perturbative expressions available for all
partial widths, which is one reason why a corresponding treatment
would not be the same for an ordinary hadronic resonance, like the
$\rho$.
 
The decay products of {\Py} resonances are typically quarks,
leptons, or other resonances, e.g. $\W \to \q \qbar'$ or
$\H^0 \to \W^+ \W^-$. In decays to quarks, parton showers are
automatically added to give a more realistic multijet structure, and
one may also allow photon emission off leptons. If the decay products
in turn are resonances, further decays are necessary. Often
spin information is available in resonance decay matrix elements,
contrary to the normal state of affairs in ordinary particle decays.
This means that the angular orientations in the two decays of a
$\W^+ \W^-$ pair are properly correlated. Occasionally, the
information is not available, and then resonances decay isotropically.
 
The top quark is a special problem. The original machinery is based
on the assumption that the $\t$ is long-lived, so that top hadrons
have time to form in the fragmentation process, and afterwards
these mesons decay weakly. With current `best bet' mass values,
this is not correct, but one should rather consider top decay before
fragmentation. Top should then be handled like one of
the above resonances. Therefore the program now contains an 
alternative along these lines, which is the preferred option. 
 
\clearpage
 
\section{Program Overview}
 
This section contains a diverse collection of information. The first 
part is an overview of previous {\Je} and {\Py} versions.
The second gives instructions for installation of the programs and
describes their philosophy: how they are
constructed and how they are supposed to be used. It also contains
some information on how to read this manual.
The third and final part contains several examples of pieces
of code or short programs, to illustrate the general style of
program usage. The last part is mainly intended as an introduction for
completely new users, and can be skipped by more experienced ones.
 
Since the {\Je} and {\Py} programs today are so closely
connected, and are gradually coalescing, they are presented together
in this report. However, they still appear as separate entities, with
slightly different style and emphasis.
 
{\Je} is the older of the two, and is at the origin of the
whole `Lund' family of event generators. It can be subdivided in two
parts. The larger is a generic package for jet fragmentation,
particle decays, final-state parton showers, event-analysis routines,
and other utitilies. This package can be used in the context of any
hard process, provided one is willing to buy the underlying
assumption of jet universality, i.e. that the fragmentation process
is fundamentally the same whether one is considering an $\ee$ or a
$\pp$ event, and that the only differences are to be found in the
parton-level processes involved. This package is not only used by all
other `Lund' programs, but also by numerous other programs written to
study specific processes. The smaller part of {\Je}
is a generator for $\ee$ annihilation events, according to either
a parton-shower or a matrix-element approach. The
{\Je} program is completely selfcontained.
 
{\Py} is a program made to generate hard or soft processes in
collisions between leptons, hadrons and photons, especially at
$\ee$, $\ep$ and $\pp$ colliders. Where {\Je} is a loose
collection of routines that you can combine as desired,
{\Py} is a more structured program, where you initially
set up what processes you want to study, and thereafter all events
will be generated according to this specification. Included is an
extensive library of hard subprocess differential cross sections,
a library of parton distributions, a process generation machinery,
treatment of initial-state showers and beam remnants, and a few odds
and ends. {\Je} is used for final-state showers, fragmentation
and decay, but no other external libraries are needed.
An interface to external parton-distribution function libraries is 
provided, however.
 
Many programs written by other persons make use of {\Je}, and a few
also of {\Py}. It is not my intention to give a complete list here.
A majority of these programs are specific to given collaborations,
and therefore not publicly distributed. Below we give a list of a
few public programs from the `Lund group', which may have a
somewhat wider application. None of them are supported by the
current author, so any requests should be directed to the persons
mentioned.
\begin{Itemize}
\item \tsc{Ariadne} is a generator for dipole emission, written
mainly by L. L\"onnblad \cite{Pet88}. The dipole provides an
alternative formulation of initial- and final-state showers. {\Je}
or {\Py} can be used to generate the hard process and {\Je} to
do the fragmentation.
\item \tsc{Aroma} is a generator for heavy-flavour processes in
leptoproduction, written by G.~Ingelman and G. Schuler \cite{Ing88}.
It uses {\Je} for fragmentation.
\item \tsc{Fritiof} is a generator for hadron--hadron, hadron--nucleus
and nucleus--nucleus collisions \cite{Nil87}, which makes use of
{\Py} to generate hard QCD scatterings and of {\Je} for fragmentation.
Currently H. Pi is responsible for program development.
\item \tsc{Lepto} is a leptoproduction event generator, written mainly
by G. Ingelman \cite{Ing80}. It can generate parton configurations
in deep inelastic scattering according to a number of possibilities.
It makes use of {\Je} for fragmentation and additionally has a 
parton-shower option based on {\Py}.
\item \tsc{Lucifer} is a photoproduction generator written by
G. Ingelman and A. Weigend \cite{Ing87a}. It is a modification of an
earlier version of {\Py} and makes use of {\Je}.
\item \tsc{Pompyt} is a generator for pomeron interactions written by
P. Bruni and G. Ingelman \cite{Bru93}. This program defines parton
distributions, flux factors and other aspects specific to the pomeron, 
which is combined with the standard {\Py} machinery for process 
generation.  
\item \tsc{Twister} is a generator for higher-twist processes, written
by G. Ingelman \cite{Ing87}. It is a modification of an earlier
version of {\Py} and makes use of {\Je}.
\end{Itemize}
One should also note that a version of {\Py} has been modified to 
include the effects of longitudinally polarized incoming protons. 
This is the work of St. G\"ullenstern et al. \cite{Gul93}.
 
\subsection{Update History}
 
Both {\Je} and {\Py} are by now fairly old and well-established 
programs, but they are still steadily being improved
on. While evolution was especially rapid for {\Je} in the
early days, that program has by now reached a certain level of
maturity, and the pace of change has dropped significantly.
{\Py}, on the other hand, has been continually extended
in recent years, and may still see further growth, although
most of the basic structure should be in place by now.
 
In earlier days, before the advent of electronic mail,
programs were only infrequently distributed, and version numbers
corresponded to distinct new upgrades. Today, the evolutionary
process is more continuous and so is the distribution of new versions.
In particular, the introduction of a new process or feature is often
done on short notice, if no problems of backwards compatibility
are involved. With this distribution, the subversion numbers have
therefore been expanded to three digits, where the last two give 
sub-subversions. For every change made in the public file, the
sub-subversion number is updated, together with the `last date of
change'. In most referencing the shorter `{\Je} version 7.4' 
could still be preferable to e.g. `{\Je} version 7.412'.
 
For the record, in Tables \ref{Jetsetver} and \ref{Pythiaver} we
list the official main versions of {\Je} and {\Py}, respectively, 
with some brief comments.
 
\begin{table}[tp]
\captive{The main versions of {\Je}, with their date of
appearance, published manuals, and main changes from previous
versions. \label{Jetsetver}} \\
\vspace{1ex}
\begin{center}
\begin{tabular}{|c|c|c|l|@{\protect\rule{0mm}{\tablinsep}}}
\hline
No. & Date  & Publ.   &  Main new or improved
features \\[1mm]
\hline
1   & Nov 78 & \cite{Sjo78} &  single-quark jets \\
2   & May 79 & \cite{Sjo79} & heavy-flavour jets \\
3.1 & Aug 79 &   ---   &  2-jets in $\ee$, preliminary 3-jets \\
3.2 & Apr 80 & \cite{Sjo80} &  3-jets in $\ee$ with full matrix
                              elements, \\
    &       &         &  toponium $\to \g \g \g$ decays \\
3.3 & Aug 80 &   ---   &  softer fragmentation spectrum \\
4.1 & Apr 81 &   ---   &  baryon production and diquark
                         fragmentation, \\
    &       &         &  fourth-generation quarks, larger
                         jet systems \\
4.2 & Nov 81 &   ---   &  low-$\pT$ physics \\
4.3 & Mar 82 & \cite{Sjo82} &  4-jets and QFD structure in $\ee$, \\
    & Jul 82 & \cite{Sjo83} &  event-analysis routines \\
5.1 & Apr 83 &   ---   &  improved string fragmentation scheme,
                          symmetric \\
    &       &         &  fragmentation, full 2$^{\mrm{nd}}$ order QCD
                         for $\ee$ \\
5.2 & Nov 83 &   ---   &  momentum-conservation schemes for IF, \\
    &       &         &  initial-state photon radiation in $\ee$ \\
5.3 & May 84 &   ---   &  `popcorn' model for baryon production \\
6.1 & Jan 85 &   ---   &  common blocks restructured, parton showers \\
6.2 & Oct 85 & \cite{Sjo86} &  error detection \\
6.3 & Oct 86 & \cite{Sjo87} &  new parton-shower scheme \\
7.1 & Feb 89 &   ---   &  new particle codes and common block
                         structure, \\
    &       &         &  more mesons, improved decays, vertex
                         information, \\
    &       &         &  Abelian gluon model, Bose--Einstein
                         effects \\
7.2 & Nov 89 &   ---   &  interface to new standard common block, \\
    &         &        &  photon emission in showers \\
7.3 & May 90 & \cite{Sjo92d}  &  expanded support for non-standard
                particles \\
7.4 & Dec 93 & \cite{Sjo94}  &  updated particle data and defaults 
\\[1mm] 
\hline
\end{tabular}
\end{center}
\end{table}
 
\begin{table}[tp]
\captive{The main versions of {\Py}, with their date of
appearance, published manuals, and main changes from previous
versions. \label{Pythiaver}} \\
\vspace{1ex}
\begin{center}
\begin{tabular}{|c|c|c|l|@{\protect\rule{0mm}{\tablinsep}}}
\hline
No. & Date  & Publ.   &  Main new or improved features \\[1mm]
\hline
1   & Dec 82 & \cite{Ben84} &  synthesis of
     predecessors \tsc{Compton}, \tsc{Highpt} and \\
    &       &         &  \tsc{Kassandra} \\
2   &  ---  &         & \\
3.1 &  ---  &         & \\
3.2 &  ---  &         & \\
3.3 & Feb 84 & \cite{Ben84a} & scale-breaking parton distributions \\
3.4 & Sep 84 & \cite{Ben85} & more efficient kinematics selection \\
4.1 & Dec 84 &         &  initial- and final-state parton showers,
      $\W$ and $\Z$ \\
4.2 & Jun 85 &         &  multiple interactions \\
4.3 & Aug 85 &         &  $\W \W$, $\W \Z$, $\Z \Z$ and $\R$
      processes \\
4.4 & Nov 85 &         &  $\gamma \W$, $\gamma \Z$, $\gamma \gamma$
      processes \\
4.5 & Jan 86 &         &  $\H^0$ production, diffractive and
      elastic events \\
4.6 & May 86 &         &  angular correlation in resonance
     pair decays \\
4.7 & May 86 &         &  $\Z'^0$ and $\H^+$ processes \\
4.8 & Jan 87 & \cite{Ben87} &  variable impact parameter
      in multiple interactions \\
4.9 & May 87 &         &  $\g \H^+$ process \\
5.1 & May 87 &         &  massive matrix elements for heavy quarks \\
5.2 & Jun 87 &         &  intermediate boson scattering \\
5.3 & Oct 89 &         &  new particle and subprocess codes,
    new common block \\
    &       &         &  structure, new kinematics selection,
    some \\
    &       &         &  lepton--lepton and lepton--hadron
    interactions, \\
    &       &         &  new subprocesses \\
5.4 & Jun 90 &         &  $s$-dependent widths, resonances not on the
    mass shell, \\
    &       &         &  new processes, new parton distributions \\
5.5 & Jan 91 &         &  improved $\ee$ and $\ep$, several new
    processes \\
5.6 & Sep 91 & \cite{Sjo92d} &  reorganized parton distributions, 
    new processes, \\
    &        &         &  user-defined external processes \\
5.7 & Dec 93 & \cite{Sjo94}  &  new total cross sections, 
    photoproduction, top decay \\[1mm]
\hline
\end{tabular}
\end{center}
\end{table}
 
All versions preceding {\Je} 7.3 and {\Py} 5.6 should now
be considered obsolete, and are no longer supported. For stable
applications, the earlier combination {\Je} 6.3 and {\Py} 4.8
could still be used, however.
 
{\Je} version 7 and {\Py} version 5 have been evolved
in parallel, so some of the processes added in later versions of
{\Py} make use of particle data only found in {\Je}
from that time onwards. Although it would be possible to
combine {\Py} 5.7 with {\Je} 7.3, e.g., it is not recommended. 
From the current versions onwards, checks have therefore
been introduced to detect the use of (potentially) incompatible 
subversions, with warnings issued at initialization if that should 
be the case. 
 
Previous versions of the manuals have contained detailed lists of
modifications from one version to the next, see e.g. \cite{Sjo92d}. 
Below we only reproduce the updates that appear with the most recent 
versions of the programs. Some of them were introduced in later
editions of {\Py} 5.6 with {\Je} 7.3, while others are completely
new. If nothing is explicitly said, these changes do not
affect backwards compatibility, but only add new features.
 
\subsubsection{Updates in JETSET 7.4}
 
Changes from version 7.3 to 7.4 are not so large, although the impact
of the updated particle data and parameter default values may need to
be studied.
\begin{Itemize}
\item Particle data have been updated in accordance with the 1992
Review of Particle Properites \cite{PDG92}. (As usual, with a free 
interpretation of inconsistencies, unclarities and other gaps in 
the knowledge.) Changes are especially drastic for charm and bottom.
In the bottom sector the decay properties are now given individually
for $\B^0$, $\B^+$, $\B_{\s}^0$, $\B_{\c}^+$ and $\Lambda_{\b}^0$, 
i.e. the generic data for `pseudoparticle' 85 are only used for 
other weakly decaying $\B$ baryons.
\item Also a few other Standard Model parameters have been updated,
such as the $\Z^0$ and $\W^{\pm}$ masses and widths, $\ssintw$ and
the CKM matrix elements.
\item Fragmentation and parton shower parameters
have been modified to reflect current LEP knowledge \cite{LEP90}, 
i.e. a minor retuning starting from an average of the `best'
parameter values obtained by the four LEP collaborations. 
Bose-Einstein effects are still left out. Flavour composition is
unchanged, except for a suppression of $\eta'$ production.
Affected by the change are \ttt{MSTJ(11)}, \ttt{PARJ(21)},
\ttt{PARJ(23)}, \ttt{PARJ(24)}, \ttt{PARJ(41)}, \ttt{PARJ(42)},
\ttt{PARJ(54)}--\ttt{PARJ(58)} and \ttt{PARJ(81)}.  
\item Several other default values have been changed for switches and 
parameters in the $\ee$, parton shower and fragmentation parts of 
the programs. These changes are intended to reflect our current 
best knowledge. See \ttt{MSTJ(26)}, \ttt{MSTJ(27)}, \ttt{MSTJ(41)}, 
\ttt{MSTJ(46)}, \ttt{MSTJ(50)}, \ttt{MSTJ(110)}, 
\ttt{PARJ(26)}, and \ttt{PARJ(121)}--\ttt{PARJ(125)}.
\item A common title page for {\Je} and {\Py} has been introduced
with the \ttt{LULOGO} routine. Sub-subversion numbers are also given.
\item Several options have been added for the \ttt{LUSHOW} shower 
routine. See \ttt{MSTJ(41)}, \ttt{MSTJ(47)} and \ttt{MSTJ(50)}.
\item A $\b$ quark produced in the decay of a top hadron is allowed 
to radiate according to the standard parton shower scheme.
\item The scalar gluon option contains the full electroweak angular
distribution of 3-jet events.
\item The \ttt{LUCOMP} routine has been modified. Among other things,
the $\B_{\c}^+$ now appears as a separate compressed code, further
codes for diffractive states have been added to the current list, and
the pomeron (reggeon, $\eta_{\mrm{techni}}$) has been added as particle 
29 (28, 38).
\item A minimum threshold for calorimeter cell energy has been 
introduced for the \ttt{LUCELL} routine.
\item All obsolescent features of the Fortran 90 standard have been
removed, i.e. the program should work well either with a Fortran 77
compiler or with a Fortran 90 one.
\item A few minor errors have been corrected.
\end{Itemize}
 
The following changes have been made since the beginning of 1994,
i.e. since the original distribution 7.400:
\begin{Enumerate}

\item {\Je} version 7.401, 11 February 1994:
\begin{Itemize}
\item Protect against overflow in \ttt{LUZDIS} (needed on some 
machines).  
\end{Itemize}

\item {\Je} version 7.402, 7 April 1994:
\begin{Itemize}
\item New option to suppress either hard or soft radiation in 
\ttt{LUSHOW}, see \ttt{MSTJ(40)}.
\item A generic interface to an external $\tau$ decay library has been
introduced, see \ttt{MSTJ(28)} and \ttt{SUBROUTINE LUTAUD}.
\item In a few places, a dot has been moved from the end of one line 
to the beginning of the next continuation line, or the other way around, 
to keep together tokens such as \ttt{.EQ.} or \ttt{.AND.}, since some 
debuggers may otherwise complain. 
\item A source of (harmless) division by zero in \ttt{LUSHOW} has been 
removed.
\end{Itemize}

\item {\Je} version 7.403, 15 July 1994:
\begin{Itemize}
\item Leptons and photons which are unrelated to the system 
feeling the Bose-Einstein effects do not have their energies and
momenta changed in the global rescaling step of \ttt{LUBOEI}. 
(Example: $\W^+\W^-$ events, where one $\W$ decays leptonically; 
before these lepton momenta could be slightly changed, but now not.)
Further, the \ttt{LUBOEI} routine has been changed to avoid an 
unintentional gap in the limits of the very first bin.
\item The option \ttt{LUEDIT(16)} (used e.g. from \ttt{PYEVNT}) has 
been improved with a more extensive search for missing daughter 
pointers.
\item The \ttt{KLU(I,16)} procedure for finding rank has been rewritten 
to work in the current {\Je} version, which it did not before. However, 
note that it will only work for \ttt{MSTU(16)=2}. As a general comment, 
the options 14--17 of \ttt{KLU} were written at a time when possible 
event histories were less complex, and can not be guaranteed always to 
work today.
\end{Itemize}

\item {\Je} version 7.404, 25 August 1994:
\begin{Itemize}
\item \ttt{LUSHOW} has been corrected, so that if $\t$, $\lrm$ or $\hrm$
quarks (or $\d^*$ or $\u^*$ quarks masked as $\lrm$ or $\hrm$ ones) are 
given with masses that vary from event to event (a Breit-Wigner shape, 
e.g.), the current mass rather than the nominal mass is used to define 
the cut-off scales of parton shower evolution.
\item \ttt{LULOGO} has been modified to take into account that a new 
{\PyJe} description has been published in \\
T. Sj\"ostrand, Computer Phys. Commun. {\bf 82} (1994) 74 \\
and is from now on the standard reference to these two programs.   
\end{Itemize}

\item {\Je} version 7.405, 27 January 1995:
\begin{Itemize}
\item \ttt{LUCELL} has been corrected, in that in the option with 
smearing of energy rather than transverse energy, the conversion 
factor between the two was applied in the wrong direction.
\item \ttt{LUSHOW} has been corrected in one place where the PMTH 
array was addressed with the wrong order of the indices. This affected 
quark mass corrections in the matching to the three-jet matrix 
elements.
\item An additional check has been included in \ttt{LUBOEI} that there 
are at least two particles involved in the Bose-Einstein effects. (No
problem except in some bizarre situations.) 
\end{Itemize}

\item {\Je} version 7.406, 20 February 1995:
\begin{Itemize}
\item A new option has been added for the behaviour of the running
$\alphaem(Q^2)$ in \ttt{ULALEM}. This is not added as a true physics 
scenario, but only to produce results with a given, fixed value for 
the hard events, while still keeping the conventional value in the 
$Q^2=0$ limit. See \ttt{MSTU(101)}, \ttt{PARU(103)}, \ttt{PARU(104)}.  
Additionally, the $G_{\mrm{F}}$ constant has been added to the 
parameter list, see \ttt{PARU(105)}.
\item The LULOGO routine has been updated to reflect my change of
affiliation.
\end{Itemize}

\item {\Je} version 7.407, 21 June 1995:
\begin{Itemize}
\item Header and \ttt{LULOGO} have been updated with respect to phone 
number and WWW access.
\item The \ttt{PHEP} and \ttt{VHEP} variables in the \ttt{/HEPEVT/}
common block are now assumed to be in \ttt{DOUBLE PRECISION}, in 
accord with the proposed LEP 2 workshop addendum to the standard.
\item In \ttt{LUTEST} a missing decimal point on the energy check 
has been reinstated. 
\item In \ttt{LUINDF} an expression has been protected against 
vanishing denominator.
\end{Itemize}

\item {\Je} version 7.408, 23 August 1995:
\begin{Itemize}
\item Check against division by zero in \ttt{LUSHOW}.
\end{Itemize}

\end{Enumerate}
 
\subsubsection{Updates in PYTHIA 5.7}
 
The updates from version 5.6 to 5.7 are all minor, and just about any
program that ran with version 5.6 will also work with {\Py} 5.7.
However, as for {\Je}, it should be noted that some important default 
values have been changed.
\begin{Itemize}
\item New parametrizations of the total cross sections of
hadronic reactions, based in Donnachie--Landshoff \cite{Don92},
which replace the old ones.
\item New parametrizations of elastic and single and double 
diffractive cross sections of hadronic reactions, based on 
Schuler--Sj\"ostrand \cite{Sch94,Sch93a}, which replace the old ones.
Also the slope parameters, the diffractive mass distributions and other
aspects of the event generation have been changed accordingly.
\item A possibility to give own total, elastic and diffractive
cross sections.
\item The single diffractive cross section has been split into its
two constituents, $AB \to XB$ and $AB \to AX$. As a consequence, the 
diffractive subprocess codes 92--94 have received changed meaning.
\item A new common block \ttt{PYINT7} has been added for the expanded
total cross section information, and this information has been partly 
removed from other common blocks.
\item A much extended description of photoproduction physics, with
the possibility to simulate separately VMD, anomalous and direct
processes \cite{Sch93,Sch93a}.
\item The selection of proton parton distributions that come with the 
program has been updated with the CTEQ2 ones, while some others have
been removed. New default is the leading-order fit CTEQ2L.
\item Since the \tsc{Pdflib} library now has been expanded to contain
also parton-distribution functions for the photon, the interfaces to
the \tsc{Pakpdf} and \tsc{Phopdf} libraries have been removed.
In addition, the interface to \tsc{Pdflib} has been modified, and is
now for appropriate for \tsc{Pdflib} version 4. 
\item An extension of hadron parton distributions into the low-$x$ and
low-$Q^2$ region \cite{Sch93a}.
\item The top quark can be made to decay before it has time to 
fragment. In view of the current best estimate for the top mass, 
this is the expected behaviour, and is therefore now default.
Further, a parton shower is allowed to evolve in the top decay.
Also fourth generation quarks are allowed to decay before they 
fragment, and so on.
\item It is possible to call \ttt{PYEVNT} with energies that vary
from one event to the next, without the need to reinitialize.
\item Improved scheme for post-factor conservation of $x$ and $Q^2$ 
in deep inelastic scattering.
\item Processes 15, 19, 30 and 35 have been expanded to cover 
$\gamma^*$ production in addition to the $\Z^0$ one, with full 
interference.
\item New process 80, $\q\gamma \to \q'\pi^{\pm}$.
\item New process 110, $\f\fbar \to \gamma\H^0$.
\item New process 149, $\g\g \to \eta_{\mrm{techni}}$. 
\item New option for initial state radiation to restrict angular
range of emission in accordance with coherence considerations. 
\item Some options have been added or removed, and default values 
have been changed. This includes 
\ttt{KFIN} (top parton distributions off by default),
\ttt{MSTP(7)}, \ttt{MSTP(11)}, \ttt{MSTP(14)},
\ttt{MSTP(23)}, \ttt{MSTP(30)} (removed), 
\ttt{MSTP(31)}, ~\ttt{MSTP(34)}, \ttt{MSTP(45)}, 
\ttt{MSTP(48)}, \ttt{MSTP(49)}, \ttt{MSTP(62)},
\ttt{MSTP(67)}, \ttt{MSTP(101)},
\ttt{PARP(13)}, \ttt{PARP(81)}, \ttt{PARP(82)},
\ttt{PARP(47)} and \ttt{PARP(101)}.
\item All obsolescent features of the Fortran 90 standard have been
removed, i.e. the program should work well either with a Fortran 77
compiler or with a Fortran 90 one.
\item A few minor errors have been corrected.
\end{Itemize}
 
The following changes have been made since the beginning of 1994,
i.e. since the original distribution 5.700:
\begin{Enumerate}

\item {\Py} version 5.701, 27 January 1994:
\begin{Itemize}
\item The machinery to handle $\gamma\gamma$ interactions is expanded
to the level already available for $\gamma\p$. This in particular 
means that a number of new options appear for \ttt{MSTP(14)}. Affected
are also \ttt{MINT(105)}, \ttt{MINT(107)}, \ttt{MINT(108)}, 
\ttt{MINT(109)}, \ttt{VINT(282)} (removed), \ttt{VINT(283)} and
\ttt{VINT(284)}. Parametrizations are introduced for meson--meson
total, elastic and diffractive cross sections, needed for the VMD
part of the photon. The treatment of cross sections for hard processes, 
of initial state radiation, of beam remnants and of other aspects are 
also expanded to cover the new possibilities. A first study of the
relevant physics aspects is found in \cite{Sch94a}.  
\item An option is introduced to modify the $Q^2$ scale of the 
anomalous part of the photon parton distributions, see \ttt{MSTP(59)}
and \ttt{PARP(59)}.
\item Correction of an error, where the generation of jet and 
low-$\pT$ events could give incorrect cross section information with
\ttt{PYSTAT(1)} at low energies. The event generation itself was
correct. (The error was introduced as a consequence of allowing
variable energies.)
\item A rejection is introduced for top events where the top mass 
(selected according to a Breit-Wigner) is too low to allow the decay
into a $\W$ on the mass shell.
\item The correction of a few other minor bugs, probably harmless. 
\end{Itemize}

\item {\Py} version 5.702, 13 February 1994:
\begin{Itemize}
\item The interface to \tsc{Pdflib} has been modified to reflect that 
\ttt{TMAS} should no longer be set except in first \ttt{PDFSET} call. 
(Else a huge amount of irrelevant warning messages are generated by 
\tsc{Pdflib}.)
\item The \ttt{STOP} statement in a few dummy routines has been modifed 
to avoid irrelevant compilation warning messages on IBM mainframes. 
\item A few labels have been renumbered. 
\end{Itemize}

\item {\Py} version 5.703, 22 February 1994:
\begin{Itemize}
\item Removal of a bug in \ttt{PYRESD}, which could give (under some
specific conditions) errors in the colour flow. 
\end{Itemize}

\item {\Py} version 5.704, 7 April 1994:
\begin{Itemize}
\item Process 11 has been corrected, for the part that concerns 
anomalous couplings (contact interactions) in the $\q\q' \to \q\q'$
process. The error was present in the expression for 
$\u\dbar \to \u\dbar$ and obvious permutations, while 
$\u\d \to \u\d$, $\u\ubar \to \u\ubar$ and the others were correct. 
\item The option with post-facto $(x,Q^2)$ 
conservation in deep inelastic scattering can give infinite loops 
when applied to process 83, in particular if one asks for the 
production of a top. (Remember that the standard DIS kinematics 
is defined for massless quarks.) Therefore the switch 
\ttt{MSTP(23)} has been modifed so that by default only process 10 is
affected.
\item \ttt{PYRESD} is modified to ensure isotropic angular distributions 
in the decays of the top or a fourth generation particle, i.e. in
$\t \to \b\W^+$. This may not be the correct distribution but, unless 
explicit knowledge exists for a given process, this should always be 
the default.
\item In processes 16, 20, 31 and 36 the $\W$ propagator has been 
modified to include $s$-dependent widths in the Breit-Wigner shape. 
The most notable effect is a suppression of the low-mass tail of the 
$\W$ mass spectrum.
\item When \tsc{Pdflib} is used, \ttt{PDFSET} is now only called 
whenever a different structure function is requested. For $\p\p$ 
events therefore only one call is made, while $\gamma\p$ interactions 
still involve a call to \ttt{PDFSET} for each \ttt{STRUCTM} one, since 
$\gamma$ and $\p$ structure functions have to be called alternatingly. 
\ttt{MINT(93)} is used to keep track of latest structure function 
called.
\item In a few places, a dot has been moved from the end of one line 
to the beginning of the next continuation line, or the other way around, 
to keep together tokens such as \ttt{.EQ.} or \ttt{.AND.}, since some 
debuggers may otherwise complain. 
\item A number of minor errors have been corrected. 
\end{Itemize}

\item {\Py} version 5.705, 15 July 1994:
\begin{Itemize}
\item A completely new possibility to have {\Py} mix different allowed
processes (direct, VMD and anomalous) in $\gamma\p$ and $\gamma\gamma$
interactions. This option can be accessed with \ttt{MSTP(14)=10}.
The relevant physics description and programming details may be found 
in sections \ref{sss:photoprod} and \ref{sss:photoprodclass}. 
This facility is still not definitive, in that it is hoped to gradually
enhance it with further features. The cross-section output of the 
\ttt{PYSTAT} has been expanded to reflect the further subdivision of the
total cross section.
\item The new facility above has required a major restructuring of some 
of the code: the routine \ttt{PYEVKI} has been removed, new routines 
\ttt{PYINBM}, \ttt{PYINPR} and \ttt{PYSAVE} created, and some material 
has been moved to or from \ttt{PYINIT}, \ttt{PYINRE} and \ttt{PYINKI}.
New variables include \ttt{MSTI(9)}, \ttt{MINT(121)}, \ttt{MINT(122)}, 
\ttt{MINT(123)} and \ttt{VINT(285)}.
\item The GRV leading-order dynamically generated parton distributions
for the $\p$ and $\pi$ have been included as options, see
\ttt{MSTP(51)} and \ttt{MSTP(53)}.
\item A parametrization of the homogeneous solution to the anomalous
photon parton distributions have been added as an option, see 
\ttt{MSTP(56)=3}.
\item The treatment of the anomalous photon component can be modified
with the new switch \ttt{MSTP(15)} and variable \ttt{PARP(17)}; at the 
same time \ttt{MSTP(59)} and \ttt{PARP(59)} have been removed. The new
options are mainly intended for comparative studies and should not
normally be touched.
\item The option \ttt{MSTP(92)=5} for beam remnant treatment 
erroneously missed some statements which now have been inserted.
Further, new options have been added for the beam remnant splitting of 
momentum between a hadron and a quark/diquark jet, where \ttt{MSTP(94)} 
should now be used rather than \ttt{MSTP(92)}.
\item In \ttt{PYDIFF} the recoiling gluon energy is calculated in a 
numerically more stable fashion.
\end{Itemize}

\item {\Py} version 5.706, 25 August 1994:
\begin{Itemize}
\item New processes 167 and 168, $\q \q' \to \q'' \d^*$ and
$\q \q' \to \q'' \d^*$, respectively, have been introduced. 
These contact interaction production processes of excited quarks
complement the quark--gluon fusion ones in processes 147 and 148,
and obey the same general rules, see section \ref{sss:qlstarclass}.
\item The option \ttt{MSTP(57)=3} now also allows a dampening of 
$\pi^{\pm}$ parton distributions.
\item A few minor errors have been corrected.
\end{Itemize}

\item {\Py} version 5.707, 20 October 1994:
\begin{Itemize}
\item A major bug discovered in processes 121 and 122 (and thus also 
affecting 181, 182, 186 and 187), $\g \g (\q \qbar) \to \Q \Qbar \H$: 
the kinematics was incorrectly handed on to the Kunszt matrix elements.
This affected the default option $\Q = \t$, but effects were 
especially dramatic when the alternative $\Q = \b$ was used. 
The choice of appropriate $Q^2$ scale for structure functions 
introduces a further uncertainty in cross sections for the 
processes above. So long as only $\t$ quarks are considered, 
the $\t$ mass is a reasonable choice, but for the $\Q = \b$ 
alternative this is presumably too low. Therefore new 
options have been introduced in \ttt{MSTP(39)}, with the default 
behaviour changed.
\item Another important bug corrected in the calculation of the 
reduction of $\t\tbar$ cross section when decay modes are forced. 
This occured when both $\t$ and $\tbar$ produced a $\W$, and 
$\W^+$ and $\W^-$ decay modes were set differently.

\end{Itemize}

\item {\Py} version 5.708, 25 October 1994:
\begin{Itemize}
\item A few further places changed to make processes 181, 182, 186
and 187 work (see version 5.707 above).

\end{Itemize}

\item {\Py} version 5.709, 26 October 1994:
\begin{Itemize}
\item The matrix element for $\f \fbar \to \W^+ \W^-$ has been 
replaced, using the formulae of \\ 
D. Bardin, M. Bilenky, D. Lehner, A. Olchevski and T. Riemann,
CERN-TH.7295/94, \\
but with the dependence on the $\hat{t}$ variable not integrated 
out (D. Bardin, private communication).
This avoids some problems encountered in the old expressions when 
one or both $\W$'s were far off the mass shell.
\item Change in calls to \tsc{Pdflib}, so that the input $Q$ is 
always at least the $Q_{\mmin}$ of the respective set.
\item Extra protection against infinite loops in PYSSPA. 
\end{Itemize}

\item {\Py} version 5.710, 27 January 1995:
\begin{Itemize}
\item The dimensions of the \ttt{HGZ} array in \ttt{PYRESD} has been 
expanded to avoid accidental writing outside the bounds.
\item \ttt{VINT(41)-VINT(66)} are saved and restored in \ttt{PYSCAT}, 
for use in low-$\pT$ events, when beam remnant treatment has failed 
(with nonzero \ttt{MINT(57)}).
\item The routine \ttt{PYSTGH} has been replaced by the routine 
\ttt{PYSTHG}. This contains an improved parametrization of the
homogeneous evolution of an anomalous photon from some given initial 
scale. The argument \ttt{NF} of the \ttt{PYSTGH} routine has been 
removed; now $\Lambda$ is always automatically converted to the 
relevant $n_f$-flavour value from its 4-flavour one, at flavour 
thresholds. 
\end{Itemize}

\item {\Py} version 5.711, 20 February 1995:
\begin{Itemize}
\item New possibilities have been added to switch between electroweak
couplings being expressed in terms of a running $\alphaem(Q^2)$  or 
in terms of a fixed Fermi constant $G_{\mrm{F}}$. This affects both 
decay widths and process cross sections, in the routines \ttt{PYINRE}, 
\ttt{PYRESD}, \ttt{PYWIDT} and \ttt{PYSIGH}. See \ttt{MSTP(8)} for
details; default corresponds to old behaviour.
\item The option \ttt{MSTP(37)=1}, with running quark masses in 
couplings to Higgs bosons, only works when $\alphas$ is allowed to 
run (so one can define a $\Lambda$ value). Therefore a check has been 
introduced in \ttt{PYWIDT} and \ttt{PYSIGH} that the option 
\ttt{MSTP(37)=1} is only executed if additionally 
\ttt{MSTP(2)}$\geq 1$.
\item Some non-physics changes have been made in the \ttt{RKBBV} 
and \ttt{STRUCTM} codes so as to avoid some (in principle harmless) 
compiler warnings.

\end{Itemize}

\item {\Py} version 5.712, 15 March 1995:
\begin{Itemize}
\item A serious error has been corrected in the \ttt{MSTP(173)=1} 
option, i.e. when the program is run with user-defined weights that 
should compensate for a biased choice of variable beam energies. 
This both affected the relative admixture of low- and high-$\pT$ 
events and the total cross section obtained by Monte Carlo integration.
(\ttt{PYRAND} changed.)
\item In order to improve the flexibility and efficiency of the 
variable-energy option, the user should now set \ttt{PARP(174)} before 
the \ttt{PYINIT} call, and thereafter not change it. This allows 
\ttt{PARP(173)} weights of arbitrary size. (\ttt{PYRAND} and 
\ttt{PYMAXI} changed.)
\item \ttt{MSTI(5)} (and \ttt{MINT(5)}) are now changed so they count 
the number of successfully generated events, rather than the number of 
tries made. This change only affects runs with variable energies, 
\ttt{MSTP(171)=1} and \ttt{MSTP(172)=2}, where \ttt{MSTI(61)=1} 
signals that a user-provided energy has been rejected in the 
weighting. This change also affects \ttt{PARI(2)}, which becomes the 
cross section per fully generated event. (\ttt{PYEVNT} changed.)
\item The option \ttt{MSTP(14)=10} has now been extended so that it 
also works for deep inelastic scattering of an electron off a (real) 
photon, i.e. subprocess 10. What is obtained is a mixture of the 
photon acting as a vector meson and it acting as an anomalous state. 
This should therefore be the sum of what can be obtained with 
\ttt{MSTP(14)=2} and \ttt{=3}. It is distinct from \ttt{MSTP(14)=1} 
in that different sets are used for the parton distributions --- in 
\ttt{MSTP(14)=1} all the contributions to the photon distributions 
are lumped together, while they are split in VMD and anomalous parts 
for \ttt{MSTP(14)=10}. Also the beam remnant treatment is different, 
with a simple Gaussian distribution (at least by default)
for \ttt{MSTP(14)=1} and the VMD part of \ttt{MSTP(14)=10}, but a 
powerlike distribution $\d k_{\perp}^2)/k_{\perp}^2$ between 
\ttt{PARP(15)} and $Q$ for the anomalous part of \ttt{MSTP(14)=10}. 
(\ttt{PYINIT}, \ttt{PYINPR} and \ttt{PYSTAT} changed.)\\
To access this option for $\e$ and $\gamma$ as incoming beams, 
it is only necessary to set \ttt{MSTP(14)=10} and keep \ttt{MSEL} at 
its default value. Unlike the corresponding option for $\gamma\p$ 
and $\gamma\gamma$, no cuts are overwritten, i.e. it is still the 
responsability of the user to set these appropriately. Those 
especially appropriate for DIS usage are \ttt{CKIN(21)-CKIN(22)} or 
\ttt{CKIN(23)-CKIN(24)} for the $x$ range (former or
latter depending on which side is the incoming real photon), 
and \ttt{CKIN(35)-CKIN(36)} for the $Q^2$ range. A further new option 
has been added (in \ttt{PYKLIM}) to set the $W^2$ range as well, see
\ttt{CKIN(39)-CKIN(40)}.\\
A warning about the usage of \tsc{Pdflib} for photons. So long as 
\ttt{MSTP(14)=1}, i.e. the photon is not split up, \tsc{Pdflib} is 
accessed by \ttt{MSTP(56)=2} and \ttt{MSTP(55)} the parton 
distribution set, as described in the manual. However, when the VMD 
and anomalous pieces are split, the VMD part is based on a rescaling 
of pion distributions by VMD factors (except for the SaS sets, that 
already come with a separate VMD piece). Therefore, to access 
\tsc{Pdflib} for \ttt{MSTP(14)=10}, it is not correct to set 
\ttt{MSTP(56)=2} and a photon distribution in \ttt{MSTP(55)}. Instead, 
one should put \ttt{MSTP(56)=2}, \ttt{MSTP(54)=2} and a pion 
distribution code in \ttt{MSTP(53)}, while \ttt{MSTP(55)} has no 
function. The anomalous part is still based on the SaS 
parametrization, with \ttt{PARP(15)} as main free parameter.   
\item A change has been made in \ttt{PYREMN} to reduce the 
possibility of infinite loops. 
\end{Itemize}

\item {\Py} version 5.713, 22 March 1995:
\begin{Itemize}
\item The SaS parton distributions of the photons are now available,
see \cite{Sch95}. There are four new sets. These differ in that two 
use a $Q_0=0.6$~GeV and two a $Q_0=2$~GeV, and in that two use the 
DIS and two the $\overline{\mathrm{MS}}$ conventions 
for the dominant non-leading contributions. (However, the fits are 
formally still leading-order, in that not all next-to-leading
contributions have been included.) New default is the SaS 1D set. 
Furthermore, for the definition of $F_2^{\gamma}$, additional terms 
appear that do not form part of the parton distributions itself. 
To partly take this into account, an additional doubling of the 
possibilities has been included. These eight possibilites can be 
accesed with \ttt{MSTP(55)}. The default value of \ttt{PARP(15)}
has been changed from 0.5 to 0.6~GeV, for consistency with SaS 1D.\\
The generic routine \ttt{PYSTFU} has been rewritten to handle the 
interfacing. The old routines \ttt{PYSTAG}, \ttt{PYSTGS}, \ttt{PYDILN} 
and \ttt{PYSTHG} have been removed. Instead the routines of the 
\tsc{SaSgam} library have been inserted. In order to avoid any 
clashes, the routines \ttt{SAS***} have been renamed \ttt{PYG***}.
Thus new routines are \ttt{PYGGAM}, \ttt{PYGVMD}, \ttt{PYGANO}, 
\ttt{PYGBEH} and \ttt{PYGDIR}. The common block \ttt{SASCOM} is 
renamed \ttt{PYINT8}. If you want to use the parton distributions 
for standalone purposes, you are encouraged to use the original 
\tsc{SaSgam} routines rather than going the way via the
{\Py} adaptations. 
\item \ttt{PYDOCU} has been corrected so that \ttt{PARI(2)} refers 
to the full cross section for $\gamma\p$ and $\gamma\gamma$ processes, 
rather than that of the latest subprocess considered.
\item An additional check has been inserted into PYREMN. 
\end{Itemize}

\item {\Py} version 5.714, 22 March 1995:
\begin{Itemize}
\item Some minor modifications to \ttt{PYSTFU} and \ttt{PYGGAM} in 
the wake of the changes of the previous version.
\end{Itemize}

\item {\Py} version 5.715, 24 April 1995:
\begin{Itemize}
\item An unfortunate choice of default values has been corrected:
the old \ttt{MSTP(3)=2} value implied that $\Lambda_{\mrm{QCD}}$ was 
entirely based on the $\Lambda$ value of the proton structure 
function; also e.g. for $\ee$ annihilation events. Thus the $\Lambda$ 
in \ttt{PARJ(81)} was overwritten, i.e. did not keep the value 
required by standard phenomenology, which typically gave too narrow 
jets. (While switching to \ttt{MSTP(3)=1} it worked fine.) In the 
modified option \ttt{MSTP(3)=2} this has been corrected, to 
better agree with user expectations. Since further changes were made 
in version 5.716, we refer below for additional comments. 
\item The form for \ttt{PTMANO}, the $\pTmin$ for anomalous processes, 
as used in \ttt{PYINPR} when processes are mixed for $\gamma\p$ or 
$\gamma\gamma$ events, has been updated to match (as well as can be 
expected) the SaS 1D photon distributions.
\end{Itemize}

\item {\Py} version 5.716, 30 June 1995:
\begin{Itemize}
\item The strategy for the changes to $\Lambda$ in version 5.715 above 
have been modified for better transparency. Now \ttt{PARJ(81)} is used 
for resonance decays (including e.g. $\Z^0$ decay, from which it is 
determined), and \ttt{PARP(72)} for other timelike showers. 
\ttt{PARJ(81)} is not overwritten \ttt{for MSTP(3)=2}, but only for 
\ttt{=3}. Changes affect \ttt{PYINIT}, \ttt{PYEVNT} and \ttt{PYRESD}.
\item A new multiplicative factor has been introduced for the $Q^2$ 
scale choice of the hard scattering in \ttt{PYSIGH}, affecting
parton distributions and $\alphas$, see \ttt{PARP(34)}.
\item \ttt{PYREMN} has been corrected for occasional too large boost 
factors.
\item An error in \ttt{PYSIGH} for process 148 has been corrected.
\item The \ttt{MSTP(62)=1} option of \ttt{PYSSPA} is modified to 
avoid division by zero.
\item Header has been updated with WWW-information.
\end{Itemize}

\item {\Py} version 5.717, 23 August 1995:
\begin{Itemize}
\item \ttt{MIN1}, \ttt{MIN2}, \ttt{MAX1}, \ttt{MAX2}, \ttt{MINA}
and \ttt{MAXA} in \ttt{PYSIGH} have had an extra \ttt{M} 
prefixed to avoid confusion with Fortran functions.
\item Protect against \ttt{MDCY(0,1)} being accessed in \ttt{PYSIGH}.
\item Protect against \ttt{THB=0} in \ttt{PYRAND}.
\item Protect against \ttt{YSTMAX-YSTMIN = 0} in \ttt{PYSIGH}.
\item Check for moved leptoquark at beginning of \ttt{PYRESD} just like 
for other particles with colour. 	
\end{Itemize}

\end{Enumerate}
 
\subsection{Program Installation}
\label{ss:install}
 
Several `authorized' sources of the programs exist. The `master
copy' of the programs is the one found on my World Wide Web homepage\\
\ttt{http://thep.lu.se/tf2/staff/torbjorn/Welcome.html}\\
There you have:
\begin{tabbing}
~~~ \= \ttt{jetset74.f}~~~~~~~ \= the {\Je} code, \\
\> \ttt{pythia57.f} \> the {\Py} code, \\
\> \ttt{pythia57.tex} \> this common {\PyJe} manual, and \\
\> \ttt{update57.notes} \> plain text update notes to the manual.
\end{tabbing}
In addition to these, one may also find older versions of the
program and manuals, sample main programs and other pieces of
related software, and other physics papers.
 
The lack of stable versions may make it less convenient to rely
on the above files. New versions are introduced in the general 
distribution of the CERN program library, maybe once a year.
These versions are better checked before release, and should be 
useful for most applications. However, clearly, they may be less 
up-to-date. Read the CERN Computer Newsletter for announcements. 
Copies of the programs are also available via anonymous ftp, 
e.g. from the asisftp server at CERN. 
 
The programs are written entirely in standard Fortran 77, and should
run on any machine with such a compiler. To a first approximation,
program compilation should therefore be straightforward.
 
Unfortunately, experience with many different compilers has been
uniform: the options available for obtaining optimized code
actually produce erroneous code (e.g. operations inside \ttt{DO}
loops are moved out before them, where some of the variables have
not yet been properly set). Therefore the general advice is to
use a low optimization level. Note that this is often not the 
default setting.
 
\ttt{SAVE} statements have been included in accordance with the
Fortran standard. Since most ordinary machines take \ttt{SAVE} for
granted, this part is not particularly well tried out, however.
 
All default settings and particle and process data are stored in
\ttt{BLOCK DATA LUDATA} for {\Je} and \ttt{BLOCK DATA PYDATA}
for {\Py}. These subprograms must be linked for a proper
functioning of the other routines. On some machines this is not done
automatically but must be forced by you, in particular if
{\Je} and {\Py} are maintained
as libraries from which routines are to be loaded only when they are
needed. In this connection we note that the library approach does not
give any significant space advantages over a loading of the
packages as a whole, since a normal run will call on most of the
routines anyway, directly or indirectly.
 
Since most machines in current use are 32-bit ones, this is the
precision normally assumed. A few pieces of code have therefore had
to be written in double precision. As a rule of thumb, 
double-precision variables have as first character \ttt{D}, but 
there are a few exceptions.
 
For applications at very high energies, such as LHC, the use of 
single precision for any real variable is a problem. It might
then be necessary to rewrite the program completely, i.e. to have a
declaration \ttt{IMPLICIT DOUBLE PRECISION(A-H,O-Z)} at the beginning 
of each subprogram,
and to change all real constants to double precision. Needless to say,
the latter is a major undertaking. In some cases, shortcuts are
available. On the IBM, for instance, the \ttt{AUTODBL} compiler option 
for automatic precision doubling works fine, provided only that an
even number of integers precede real numbers in common blocks.
In {\Je} you therefore need to introduce an additional
integer variable (\ttt{NPAD}, say) directly after \ttt{N} in the
\ttt{LUJETS} common block, and in {\Py} an additional integer
(\ttt{MSEPAD}) after \ttt{MSEL} in the \ttt{PYSUBS} common block.
Some pieces of code will then actually run in quadruple precision.
 
A test program, \ttt{LUTEST}\label{p:LUTEST}, is included in the 
{\Je} package. It is disguised as a subroutine, so you have to run a 
main program
\begin{verbatim}
      CALL LUTEST(1)
      END
\end{verbatim}
This program will generate six hundred events of different types,
under a variety of conditions. If {\Je} has not been properly
installed, this program is likely to crash, or at least generate a
number of erroneous events. This will then clearly be marked in the
output, which otherwise will just contain a few sample event listings
and a table of the number of different particles produced. To switch
off the output of normal events and final table, use \ttt{LUTEST(0)}
instead of \ttt{LUTEST(1)}. The final tally of errors detected should
read 0.
 
In exactly the same vein, a test program \ttt{PYTEST}\label{p:PYTEST} 
comes with the {\Py} package. You then have to run a program
\begin{verbatim}
      CALL PYTEST(1)
      END
\end{verbatim}
As before the alternative \ttt{PYTEST(0)} will give a less extensive
listing. No errors should appear during execution.
 
\subsection{Program Philosophy}
 
The Monte Carlo programs are built as slave systems, i.e. you,
the user, have to supply the main program. From this the various
subroutines are called on to execute specific tasks, after which
control is returned to the main program. Some of these tasks may
be very trivial, whereas the `high-level' routines by themselves
may make a large number of subroutine calls. Many routines are
not intended to be called directly by you, but only from
higher-level routines such as \ttt{LUEXEC}, \ttt{LUEEVT},
\ttt{PYINIT} or \ttt{PYEVNT}.
 
Basically, this means that there are three ways by which you 
communicate with the
programs. First, by setting common block variables, you specify
the details of how the programs should perform specific tasks,
i.e. which subprocesses should be generated (for {\Py}), which
particle masses should be assumed, which coupling constants used,
which fragmentation scenarios, and so on with hundreds of options
and parameters. Second, by calling subroutines you tell the programs
to generate events according to the rules established above.
Normally there are few subroutine arguments, and those are usually
related to details of the physical situation, such as what
c.m. energy to assume for events. Third, you can either look at the
common block \ttt{LUJETS} to extract information on the generated
event, or you can call on various functions and subroutines
to analyse the event further for you.
 
It should be noted that, while the physics content is obviously at
the centre of attention, the {\JePy} package
also contains a very extensive setup of auxiliary service routines.
The hope is that this will provide a comfortable
working environment, where not only events are generated, but where
you also linger on to perform a lot of the subsequent studies.
Of course, for detailed studies, it may be necessary to interface
the output directly to a detector simulation program.
 
The general rule is that all routines have names that are six 
characters long, beginning with \ttt{LU} for {\Je} routines and 
\ttt{PY} for {\Py} ones. Real-valued functions in {\Je} begin with
\ttt{UL} instead. There are three exceptions to both the length and
the initial character rules: \ttt{KLU}, \ttt{PLU} and \ttt{RLU}. The
former two functions are strongly coupled to the \ttt{K} and \ttt{P}
matrices in the \ttt{LUJETS} common block, the latter uses \ttt{R} to
emphasize the r\^ole as a random-number generator. Also common block
names are six characters long and start with \ttt{LU} or
\ttt{PY}.
 
On the issue of initialization, {\Je} and {\Py} behave
quite differently. Most {\Je} routines work without any
initialization (except for the one implied by the presence of
\ttt{BLOCK DATA LUDATA}, see above), i.e. each event and each task
stand on their own. Current common block values are used to perform
the tasks in specific ways, and those rules can be changed from
one event to the next (or even within the generation of one and the
same event) without any penalty. The random-number generator
is initialized at the first call, but usually this is transparent.
Therefore the two {\Je} routines \ttt{LUEEVT} (and some of
the routines called by it) and \ttt{LUONIA} are basically the only
ones to contain some elements of initialization, where
there are a few advantages if events are generated in a coherent
fashion, but even here the penalty for not doing it is small.
 
In {\Py}, on the other hand, a sizeable amount of
initialization is performed in the \ttt{PYINIT} call, and thereafter
the events generated by \ttt{PYEVNT} all obey the rules established
at that point. Therefore common block variables that specify methods
to be used have to be set before the \ttt{PYINIT} call and then not
be changed afterwards, with few exceptions. Of course, it is possible
to perform several \ttt{PYINIT} calls in the same run, but there is
a significant time overhead involved, so this is not something one
would do for each new event.
 
Apart from writing a title page, giving a brief initialization 
information, printing error messages if need be,
and responding to explicit requests for listings, all tasks of the
programs are performed `silently'. All output is directed to unit
\ttt{MSTU(11)}, by default 6, and it is up to you to set
this unit open for write. The only exceptions are
\ttt{RLUGET}, \ttt{RLUSET} and \ttt{LUUPDA} where, for obvious reasons,
the input/output file number is specified at each call. Here you
again have to see to it that proper read/write access is set.
 
The programs are extremely versatile, but the price to be paid for this 
is having a large number of adjustable parameters and switches for
alternative modes of operation. No single user is ever likely to
need more than a fraction of the available options.
Since all these parameters and switches are assigned sensible default
values, there is no reason to worry about them until the need arises.
 
Unless explicitly stated (or obvious from the context) all switches and
parameters can be changed independently of each other. One should note,
however, that if only a few switches/parameters are changed, this may
result in an artificially bad agreement with data. Many disagreements
can often be cured by a subsequent retuning of some other parameters of
the model, in particular those that were once determined by
a comparison with data in the context of the default scenario.
For example, for $\ee$ annihilation, such a retuning could involve one
QCD parameter ($\alphas$ or $\Lambda$), the longitudinal fragmentation
function, and the average transverse fragmentation momentum.
 
The programs contain a number of checks that requested processes have
been implemented, that flavours specified for jet systems make sense,
that the energy is sufficient to allow hadronization, that the memory 
space in \ttt{LUJETS} is large enough, etc. If anything goes wrong
that the program can catch (obviously this may not always be possible),
an error message will be printed and the treatment of the corresponding
event will be cut short. In serious cases, the program will abort.
As long as no error messages appear on the
output, it may not be worthwhile to look into the rules for error
checking, but if but one message appears, it should be enough cause for
alarm to receive prompt attention. Also warnings are sometimes printed.
These are less serious, and the experienced user might deliberately
do operations which go against the rules, but still can be made to
make sense in their context. Only the first few warnings will be
printed, thereafter the program will be quiet. By default, the program
is set to stop execution after ten errors, after printing
the last erroneous event.
 
It must be emphasized that not all errors will be caught. In particular,
one tricky question is what happens if an integer-valued common block
switch or subroutine/function argument is used with a value that is not
defined. In some subroutine calls, a prompt return will be expedited,
but in most instances the subsequent action is entirely unpredictable,
and often completely haywire. The same goes for real-valued variables
that are assigned values outside the physically sensible range. One 
example will suffice here: if \ttt{PARJ(2)} is defined as the 
$\s / \u$ suppression factor, a value $>1$ will not give more 
profuse production of $\s$ than of $\u$, but actually a spillover 
into $\c$ production. Users, beware!
 
\subsection{Manual Conventions}
 
In the manual parts of this report, some conventions are used.
All names of subprograms, common blocks and variables are given in
upper-case `typewriter' style, e.g. \ttt{MSTP(111)=0}. Also
program examples are given in this style.
 
If a common block variable must have a value set at the beginning of
execution, then a default value is stored in one of the block data
subprograms \ttt{LUDATA} and \ttt{PYDATA}. Such a default value is
usually indicated by a `(D=\ldots)' immediately after the variable
name, e.g.
\begin{entry}
\iteme{MSTJ(1) :} (D=1) choice of fragmentation scheme.
\end{entry}

All variables in the {\Je} common blocks (with very few exceptions,
clearly marked) can be freely changed from one event to the next,
or even within the treatment of one single event. In the {\Py}
common blocks the situation is more complicated. The values of many
switches and parameters are used already in the \ttt{PYINIT} call,
and cannot be changed after that. The problem is mentioned in the
preamble to the afflicted common blocks, which in particular means
\ttt{/PYPARS/} and \ttt{/PYSUBS/}. For the variables which may still 
be changed from one event to the next, a `(C)' is added after
the `(D=\ldots)' statement.

Normally, variables internal to the program are kept in separate
common blocks and arrays, but in a few cases such internal variables
appear among arrays of switches and parameters, mainly for historical
reasons. These are denoted by `(R)' for variables you may want to 
read, because they contain potentially interesting information, and 
by `(I)' for purely internal variables. In neither case may the
variables be changed by you.

In the description of a switch, the alternatives that this
switch may take are often enumerated, e.g.
\begin{entry}
\iteme{MSTJ(1) :} (D=1) choice of fragmentation scheme.
\begin{subentry}
\iteme{= 0 :} no jet fragmentation at all.
\iteme{= 1 :} string fragmentation according to the Lund model.
\iteme{= 2 :} independent fragmentation, according to specification
in \ttt{MSTJ(2)} and \ttt{MSTJ(3)}.
\end{subentry}
\end{entry}
If you then use any value other than 0, 1 or 2, results are 
unpredictable. The action could even be different in different parts 
of the program, depending on the order in which the alternatives are
identified. 

It is also up to you to choose physically sensible values for
parameters: there is no check on the allowed ranges of variables.
We gave an example of this at the end of the preceding section.

Subroutines you are expected to use are enclosed in a box at the
point where they are defined:

\drawbox{CALL LULIST(MLIST)}

\boxsep

This is followed by a description of input or output parameters.
The difference between input and output is not explicitly marked,
but should be obvious from the context. In fact, the event-analysis 
routines of section \ref{ss:evanrout} return values, 
while all the rest only have input variables.  

Routines that are only used internally are not boxed in.
However, we use boxes for all common blocks, so as to 
enhance the readability. 
 
\subsection{Getting Started with JETSET}
\label{ss:JETstarted}
 
As a first example, assume that you want to study the production of
$\u \ubar$ 2-jet systems at 20 GeV energy. To do this, write a main
program
\begin{verbatim}
      CALL LU2ENT(0,2,-2,20.)
      CALL LULIST(1)
      END
\end{verbatim}
and run this program, linked together with {\Je}. The routine
\ttt{LU2ENT}
is specifically intended for storing two entries (jets or particles).
The first argument (0) is a command to perform fragmentation and decay
directly after the entries have been stored, the second and third that
the two entries are $\u$ (2) and $\ubar$ ($-2$), and the last that the 
c.m. energy of the pair is 20 GeV. When this is run, the resulting event 
is stored in the \ttt{LUJETS} common block. This information can then be
read out by you. No output is produced by \ttt{LU2ENT} itself,
except for a title page which appears once for every {\JePy} run.
 
Instead the second command, to \ttt{LULIST}, provides a simple visible
summary of the information stored in \ttt{LUJETS}. The argument (1)
indicates that the short version should be used, which is suitable
for viewing the listing directly on an 80-column terminal screen.
It might look as shown here.
\begin{verbatim}
                        Event listing (summary)
 
  I  particle/jet KS     KF orig   p_x     p_y     p_z      E       m
 
  1  (u)       A  12      2    0   0.000   0.000  10.000  10.000   0.006
  2  (u~)      V  11     -2    0   0.000   0.000 -10.000  10.000   0.006
  3  (string)     11     92    1   0.000   0.000   0.000  20.000  20.000
  4  (rho+)       11    213    3   0.098  -0.154   2.710   2.856   0.885
  5  (rho-)       11   -213    3  -0.227   0.145   6.538   6.590   0.781
  6  pi+           1    211    3   0.125  -0.266   0.097   0.339   0.140
  7  (Sigma0)     11   3212    3  -0.254   0.034  -1.397   1.855   1.193
  8  (K*+)        11    323    3  -0.124   0.709  -2.753   2.968   0.846
  9  p~-           1  -2212    3   0.395  -0.614  -3.806   3.988   0.938
 10  pi-           1   -211    3  -0.013   0.146  -1.389   1.403   0.140
 11  pi+           1    211    4   0.109  -0.456   2.164   2.218   0.140
 12  (pi0)        11    111    4  -0.011   0.301   0.546   0.638   0.135
 13  pi-           1   -211    5   0.089   0.343   2.089   2.124   0.140
 14  (pi0)        11    111    5  -0.316  -0.197   4.449   4.467   0.135
 15  (Lambda0)    11   3122    7  -0.208   0.014  -1.403   1.804   1.116
 16  gamma         1     22    7  -0.046   0.020   0.006   0.050   0.000
 17  K+            1    321    8  -0.084   0.299  -2.139   2.217   0.494
 18  (pi0)        11    111    8  -0.040   0.410  -0.614   0.751   0.135
 19  gamma         1     22   12   0.059   0.146   0.224   0.274   0.000
 20  gamma         1     22   12  -0.070   0.155   0.322   0.364   0.000
 21  gamma         1     22   14  -0.322  -0.162   4.027   4.043   0.000
 22  gamma         1     22   14   0.006  -0.035   0.422   0.423   0.000
 23  p+            1   2212   15  -0.178   0.033  -1.343   1.649   0.938
 24  pi-           1   -211   15  -0.030  -0.018  -0.059   0.156   0.140
 25  gamma         1     22   18  -0.006   0.384  -0.585   0.699   0.000
 26  gamma         1     22   18  -0.034   0.026  -0.029   0.052   0.000
                 sum:  0.00        0.000   0.000   0.000  20.000  20.000
\end{verbatim}
(A few blanks have been removed between the columns to make it fit into
the format of this text.) Look in the particle/jet column and note
that the first two lines are the original $\u$ and $\ubar$, where `bar'
is actually written `$\sim$' to save space in longer names. 
The parentheses enclosing the names, `\ttt{(u)}' and `\ttt{(u\~{})}',
are there as a reminder that these jets actually have been allowed 
to fragment. The jets are still retained so that event histories
can be studied. Also note that the \ttt{KF} (flavour code) column
contains 2 in the first line and $-2$ in the second. These are the
codes actually stored to denote the presence of a $\u$ and a $\ubar$, 
cf. the \ttt{LU2ENT} call, while the names written are just 
conveniences used when producing visible output. The \ttt{A} and 
\ttt{V} near the end of the particle/jet column indicate the beginning 
and end of a string (or cluster, or independent fragmentation) parton 
system; any intermediate entries belonging to the same system would 
have had an \ttt{I} in that column. (This gives a poor man's 
representation of an up-down arrow, $\updownarrow$.)
 
In the \ttt{orig} (origin) column, the zeros indicate that $\u$ and
$\ubar$ are two initial entries. The subsequent line, number 3, denotes
the fragmenting $\u \ubar$ string system as a whole, and has origin 1,
since the first parton of this string system is entry number 1. The
particles in lines 4--10 have origin 3 to denote that they come
directly from the fragmentation of this string. In string
fragmentation it is not meaningful to say that a particle comes from
only the $\u$ quark or only the $\ubar$ one. It is the string system
as a whole that gives a $\rho^+$, a $\rho^-$, a $\pi^+$, a
$\Sigma^0$, a $\K^{*+}$, a $\pbar^-$, and a $\pi^-$. Note that some
of the particle names are again enclosed in parentheses, indicating
that these particles are not present in the final state either, but have 
decayed further. Thus the $\pi^-$ in line 13 and the $\pi^0$ in line
14 have origin 5, as an indication that they come from the decay of the
$\rho^-$ in line 5. Only the names not enclosed in parentheses 
remain at the end of the fragmentation/decay chain, and
are thus experimentally observable. The actual status code used to
distinguish between different classes of entries is given in the
\ttt{KS} column; codes in the range 1--10 correspond to remaining
entries, and those above 10 to those that have fragmented or decayed.
 
The columns with \ttt{p\_x}, \ttt{p\_y}, \ttt{p\_z}, \ttt{E} and
\ttt{m} are quite self-explanatory. All momenta, energies and
masses are given in units of GeV, since the speed of light 
is taken to be $c = 1$.
Note that energy and momentum are conserved at each step of the
fragmentation/decay process (although there exist options where this is
not true). Also note that the $z$ axis plays the r\^ole of preferred
direction, along which the original partons are placed. 
The final line is
intended as a quick check that nothing funny happened. It contains the
summed charge, summed momentum, summed energy and invariant mass of the
final entries at the end of the fragmentation/decay chain, and the
values should agree with the input implied by the \ttt{LU2ENT}
arguments. (In fact, warnings would normally appear on the output if
anything untoward happened, but that is another story.)
 
The above example has illustrated roughly what information is to be had
in the event record, but not so much about how it is stored. This is
better seen by using a 132-column format for listing events. Try e.g.
the following program
\begin{verbatim}
      CALL LU3ENT(0,1,21,-1,30.,0.9,0.7)
      CALL LULIST(2)
      CALL LUEDIT(3)
      CALL LULIST(2)
      END
\end{verbatim}
where a 3-jet $\d \g \dbar$ event is generated in the first line and
listed in the second. This listing will contain the numbers as
directly stored in the common block \ttt{LUJETS}
\begin{verbatim}
      COMMON/LUJETS/N,K(4000,5),P(4000,5),V(4000,5)
\end{verbatim}
For particle \ttt{I}, \ttt{K(I,1)} thus gives information on whether
or not a jet or particle has fragmented or decayed, \ttt{K(I,2)}
gives the particle code, \ttt{K(I,3)} its origin, \ttt{K(I,4)} and
\ttt{K(I,5)} the position of fragmentation/decay products, and
\ttt{P(I,1})--\ttt{P(I,5)} momentum, energy and mass. The number
of lines in current use is given by \ttt{N}, i.e.
1 $\leq$ \ttt{I} $\leq$ \ttt{N}. The \ttt{V} matrix contains decay
vertices; to view those \ttt{LULIST(3)} has to be used. It
is important to learn the rules for how information is stored in
\ttt{LUJETS}.
 
The third line in the program illustrates another important point
about {\Je}: a number of routines are available for
manipulating the event record after the event has been generated.
Thus \ttt{LUEDIT(3)} will remove everything except stable charged
particles, as shown by the result of the second \ttt{LULIST} call.
More advanced possibilities include things like sphericity or
clustering routines.
 
Apart from the input arguments of subroutine calls, control on the
doings of {\Je} may be imposed via the \ttt{LUDAT1},
\ttt{LUDAT2}, \ttt{LUDAT3} and \ttt{LUDAT4} common blocks. Here
sensible default values are always provided. A user might want to
switch off all particle decays by putting \ttt{MSTJ(21)=0} or
increase the $\s / \u$ ratio in fragmentation by putting
\ttt{PARJ(2)=0.40}, to give but two examples. It is by exploring
the possibilities offered here that {\Je} can be turned into an
extremely versatile tool, even if all the nice physics is already
present in the default values.
 
As a final, semirealistic example, assume that the $\pT$
spectrum of $\pi^+$ particles is to be studied in 91.2 GeV
$\ee$ annihilation events, where $\pT$ is to be defined with
respect to the sphericity axis. Using the HBOOK package (version 4,
watch out for version- or installation-specific differences) for
histogramming, a complete program might look like
\begin{verbatim}
C...Common blocks.
      COMMON/LUJETS/N,K(4000,5),P(4000,5),V(4000,5)
      COMMON/PAWC/HMEMOR(10000)

C...Reserve histogram memory and book histograms.
      CALL HLIMIT(10000)
      CALL HBOOK1(1,'pT spectrum of pi+',100,0.,5.,0.)

C...Number of events to generate. Loop over events.
      NEVT=100
      DO 110 IEVT=1,NEVT

C...Generate event. List first one.
      CALL LUEEVT(0,91.2)
      IF(IEVT.EQ.1) CALL LULIST(1)

C...Find sphericity axis and rotate event so sphericity along z axis.
      CALL LUSPHE(SPH,APL)
      CALL LUEDIT(31)

C...Loop over all particles, but skip if not pi+.
      DO 100 I=1,N
      IF(K(I,2).NE.211) GOTO 100

C...Calculate pT and fill in histogram.
      PT=SQRT(P(I,1)**2+P(I,2)**2)
      CALL HF1(1,PT,1.)

C...End of particle and event loops.
  100 CONTINUE
  110 CONTINUE

C...Normalize histogram properly and list it.
      CALL HOPERA(1,'+',1,1,20./NEVT,0.)
      CALL HISTDO

      END
\end{verbatim}
Study this program, try to understand what happens at each step, and
run it to check that it works. You should then be ready to look
at the relevant sections of this report and start writing your own
programs.
 
\subsection{Getting Started with PYTHIA}
\label{ss:PYTstarted}
 
A {\Py} run has to be more strictly organized than
a {\Je} one, in that it is necessary to initialize the
generation before events can be generated, and in that it is
not possible to change switches and parameters freely during
the course of the run. A fairly precise recipe for how a run 
should be structured can therefore be given.
 
Thus, the usage of {\Py} can be subdivided into three steps.
\begin{Enumerate}
\item The initialization step. It is here that all the basic
  characteristics of the coming generation are specified.
  The material in this section includes the following.
  \begin{Itemize}
  \item Common blocks, at least the following, and maybe some more:
    \begin{verbatim}
      COMMON/LUJETS/N,K(4000,5),P(4000,5),V(4000,5)
      COMMON/LUDAT1/MSTU(200),PARU(200),MSTJ(200),PARJ(200)
      COMMON/PYSUBS/MSEL,MSUB(200),KFIN(2,-40:40),CKIN(200)
      COMMON/PYPARS/MSTP(200),PARP(200),MSTI(200),PARI(200)
    \end{verbatim}
    \vspace{-\baselineskip}
  \item Selection of required processes. Some fixed `menus'
     of subprocesses can be selected with different \ttt{MSEL} 
     values, but with {\tt MSEL}=0 it is possible to compose 
     `\`a la carte', using the subprocess numbers.
    To generate processes 14, 18 and 29, for instance, one needs
    \begin{verbatim}
      MSEL=0
      MSUB(14)=1
      MSUB(18)=1
      MSUB(29)=1
    \end{verbatim}
    \vspace{-\baselineskip}
  \item Selection of kinematics cuts in the \ttt{CKIN} array.
  To generate hard scatterings with 5~GeV $\leq \pT \leq$
    10~GeV, for instance, use
    \begin{verbatim}
      CKIN(3)=5.
      CKIN(4)=10.
    \end{verbatim}
    \vspace{-\baselineskip}
  Unfortunately, initial- and final-state radiation will shift
  around the kinematics of the hard scattering, making the effects
  of cuts less predictable. One therefore always has to be very
  careful that no desired event configurations are cut out.
  \item Definition of underlying physics scenario, e.g. top mass.
  \item Selection of parton-distribution sets, $Q^2$ definitions,
    and all other details of the generation.
  \item Switching off of generator parts not needed for toy
    simulations, e.g. fragmentation for parton level studies.
  \item Initialization of the event generation procedure. Here
    kinematics is set up, maxima of differential cross sections
    are found for future Monte Carlo generation, and a number of
    other preparatory tasks carried out. Initialization is performed
    by \ttt{PYINIT}, which should be called only after the switches
    and parameters above have been set to their desired values. The
    frame, the beam particles and the energy have to be specified.
    \begin{verbatim}
      CALL PYINIT('CMS','p','pbar',1800.)
    \end{verbatim}
    \vspace{-\baselineskip}
  \item Any other initial material required by the user, e.g.
    histogram booking.
  \end{Itemize}
\item The generation loop. It is here that events are generated
  and studied. It includes the following tasks:
  \begin{Itemize}
  \item Generation of the next event, with
    \begin{verbatim}
      CALL PYEVNT
    \end{verbatim}
    \vspace{-\baselineskip}
  \item Printing of a few events, to check that everything is
    working as planned, with
    \begin{verbatim}
      CALL LULIST(1)
    \end{verbatim}
    \vspace{-\baselineskip}
  \item An analysis of the event for properties of interest,
    either directly reading out information from the 
    \ttt{LUJETS} common block
    or making use of a number of utility routines in {\Je}.
  \item Saving of events on tape, or interfacing to detector
    simulation.
  \end{Itemize}
\item The finishing step. Here the tasks are:
  \begin{Itemize}
  \item Printing a table of deduced cross sections, obtained as a
  by-product of the Monte Carlo generation activity, with the
  command
    \begin{verbatim}
      CALL PYSTAT(1)
    \end{verbatim}
    \vspace{-\baselineskip}
  \item Printing histograms and other user output.
  \end{Itemize}
\end{Enumerate}
 
To illustrate this structure, imagine a toy example, where one wants
to simulate the production of a 300 GeV Higgs particle. In
{\Py}, a program for this might look something like the
following.
\begin{verbatim}
C...Common blocks.
      COMMON/LUJETS/N,K(4000,5),P(4000,5),V(4000,5)
      COMMON/LUDAT1/MSTU(200),PARU(200),MSTJ(200),PARJ(200)
      COMMON/LUDAT2/KCHG(500,3),PMAS(500,4),PARF(2000),VCKM(4,4)
      COMMON/LUDAT3/MDCY(500,3),MDME(2000,2),BRAT(2000),KFDP(2000,5)
      COMMON/PYSUBS/MSEL,MSUB(200),KFIN(2,-40:40),CKIN(200)
      COMMON/PYPARS/MSTP(200),PARP(200),MSTI(200),PARI(200)
      COMMON/PAWC/HBOOK(10000)
 
C...Number of events to generate. Switch on proper processes.
      NEV=1000
      MSEL=0
      MSUB(102)=1
      MSUB(123)=1
      MSUB(124)=1
 
C...Select t and H masses and kinematics cuts in mass.
      PMAS(6,1)=140.
      PMAS(25,1)=300.
      CKIN(1)=290.
      CKIN(2)=310.
 
C...For simulation of hard process only: cut out unnecessary tasks.
      MSTP(61)=0
      MSTP(71)=0
      MSTP(81)=0
      MSTP(111)=0
 
C...Initialize and list partial widths.
      CALL PYINIT('CMS','p','p',16000.)
      CALL PYSTAT(2)
 
C...Book histograms.
      CALL HLIMIT(10000)
      CALL HBOOK1(1,'Higgs mass',50,275.,325.,0.)
 
C...Generate events. Look at first few.
      DO 200 IEV=1,NEV
      CALL PYEVNT
      IF(IEV.LE.3) CALL LULIST(1)
 
C...Loop over particles to find Higgs and histogram its mass.
      DO 100 I=1,N
  100 IF(K(I,2).EQ.25) HMASS=P(I,5)
      CALL HF1(1,HMASS,1.)
  200 CONTINUE
 
C...Print cross sections and histograms.
      CALL PYSTAT(1)
      CALL HISTDO
 
      END
\end{verbatim}
 
Here 102, 123 and 124 are the three main Higgs production
graphs $\g \g \rightarrow \H$, $\Z \Z \rightarrow \H$, and
$\W \W \rightarrow \H$, and \ttt{MSUB(ISUB)=1} is the command to
switch on process \ttt{ISUB}. Full freedom to combine subprocesses
`\`a la carte' is ensured by \ttt{MSEL=0}; ready-made `menus'
can be ordered with other \ttt{MSEL} numbers.
The \ttt{PMAS} commands set the
masses of the top quark and the Higgs itself, and the \ttt{CKIN}
variables the desired mass range of the Higgs --- a Higgs with a
300 GeV nominal mass actually has a fairly broad Breit--Wigner type
mass distribution. The \ttt{MSTP} switches that come next are there to
modify the generation procedure, in this case to switch off initial-
and final-state radiation, multiple interactions among beam jets,
and fragmentation, to give only the `parton skeleton' of the hard
process. The \ttt{PYINIT} call initializes {\Py}, by finding
maxima of cross sections, recalculating the Higgs decay properties
(which depend on the Higgs mass), etc. The decay properties can
be listed with \ttt{PYSTAT(2)}.
 
Inside the event loop, \ttt{PYEVNT}
is called to generate an event, and \ttt{LULIST(1)} to list the event.
The information used by \ttt{LULIST(1)} is the event record, stored
in the common block \ttt{LUJETS}. Here one finds all produced particles,
both final and intermediate ones, with information on particle
species and event history (\ttt{K} array), particle momenta
(\ttt{P} array) and production vertices (\ttt{V} array).
In the loop over all particles produced, \ttt{1} through \ttt{N},
the Higgs particle is found by its code, \ttt{K(I,2)=25},
and its mass is stored in \ttt{P(I,5)}.
 
After all events have been generated, \ttt{PYSTAT(1)}
gives a summary of the number of events generated in the
various allowed channels, and the inferred cross sections.
 
In the run above, a typical event listing might look like the following.
\begin{verbatim}
                         Event listing (summary)
 
    I  particle/jet     KF    p_x      p_y      p_z       E        m
 
    1  !p+!           2212    0.000    0.000 8000.000 8000.000    0.938
    2  !p+!           2212    0.000    0.000-8000.000 8000.000    0.938
 ======================================================================
    3  !g!              21   -0.505   -0.229   28.553   28.558    0.000
    4  !g!              21    0.224    0.041 -788.073  788.073    0.000
    5  !g!              21   -0.505   -0.229   28.553   28.558    0.000
    6  !g!              21    0.224    0.041 -788.073  788.073    0.000
    7  !H0!             25   -0.281   -0.188 -759.520  816.631  300.027
    8  !W+!             24  120.648   35.239 -397.843  424.829   80.023
    9  !W-!            -24 -120.929  -35.426 -361.677  391.801   82.579
   10  !e+!            -11   12.922   -4.760 -160.940  161.528    0.001
   11  !nu_e!           12  107.726   39.999 -236.903  263.302    0.000
   12  !s!               3  -62.423    7.195 -256.713  264.292    0.199
   13  !c~!             -4  -58.506  -42.621 -104.963  127.509    1.350
 ======================================================================
   14  (H0)             25   -0.281   -0.188 -759.520  816.631  300.027
   15  (W+)             24  120.648   35.239 -397.843  424.829   80.023
   16  (W-)            -24 -120.929  -35.426 -361.677  391.801   82.579
   17  e+              -11   12.922   -4.760 -160.940  161.528    0.001
   18  nu_e             12  107.726   39.999 -236.903  263.302    0.000
   19  s         A       3  -62.423    7.195 -256.713  264.292    0.199
   20  c~        V      -4  -58.506  -42.621 -104.963  127.509    1.350
   21  ud_1      A    2103   -0.101    0.176 7971.328 7971.328    0.771
   22  d         V       1   -0.316    0.001  -87.390   87.390    0.010
   23  u         A       2    0.606    0.052   -0.751    0.967    0.006
   24  uu_1      V    2203    0.092   -0.042-7123.668 7123.668    0.771
 ======================================================================
                sum:  2.00     0.00     0.00     0.00 15999.98 15999.98
\end{verbatim}
The above event listing is abnormally short, in part because some
columns of information were removed to make it fit into this text,
in part because all initial- and final-state QCD radiation, all
non-trivial beam jet structure, and all fragmentation was inhibited
in the generation. Therefore only the skeleton of the process is
visible. In lines 1 and 2 one recognizes the two incoming protons.
In lines 3 and 4 are incoming partons before initial-state radiation
and in 5 and 6 after --- since there is no such radiation they coincide
here. Line 7 shows the Higgs produced by $\g \g$ fusion, 8 and 9 its
decay products and 10--13 the second-step decay products. Up to this
point lines give a summary of the event history, indicated by the
exclamation marks that surround particle names (and also reflected in
the \ttt{K(I,1)} code, not shown). From line 14 onwards come the
particles actually produced in the final states, first in lines 14--16
particles that subsequently decayed, which have their names surrounded
by brackets, and finally the particles and jets left in the end,
including beam remnants.
Here this also includes a number of unfragmented jets, since
fragmentation was inhibited. Ordinarily, the listing would have gone
on for a few hundred more lines, with the particles produced in the
fragmentation and their decay products. The final line gives total
charge and momentum, as a convenient check that nothing unexpected
happened. The first column of the listing
is just a counter, the second gives the particle name and information
on status and string drawing (the \ttt{A} and \ttt{V}), the third
the particle-flavour code (which is used to give the name),
and the subsequent columns give the momentum components.
 
One of the main problems is to select kinematics efficiently. Imagine
for instance that one is interested in the production of a single 
$\Z$ with a
transverse momentum in excess of 50 GeV. If one tries to generate
the inclusive sample of $\Z$ events, by the basic production graphs
$\q \qbar \rightarrow \Z$, then most events will have low transverse
momenta and will have to be discarded. That any of the desired events
are produced at all is due to the initial-state generation machinery, 
which can build up transverse momenta for the incoming
$\q$ and $\qbar$. However, the amount
of initial-state radiation cannot be constrained beforehand. To
increase the efficiency, one may therefore turn to the higher-order
processes $\q \g \rightarrow \Z \q$ and $\q \qbar \rightarrow \Z \g$,
where
already the hard subprocess gives a transverse momentum to the $\Z$.
This transverse momentum can be constrained as one wishes, but again
initial- and final-state radiation will smear the picture. If one were
to set a $\pT$ cut at 50 GeV for the hard-process generation,
those events where the $\Z$ was given only 40 GeV in the hard process 
but got the rest from initial-state radiation would be missed. 
Not only therefore would cross sections
come out wrong, but so might the typical event shapes. In the end,
it is therefore necessary to find some reasonable compromise, by
starting the generation at 30 GeV, say, if one knows that only rarely
do events below this value fluctuate up to 50 GeV. Of course, most 
events will therefore not contain a $\Z$ above 50 GeV, and one will 
have to live with some inefficiency. It is not uncommon that only one 
event out of ten can be used, and occasionally it can be even worse.
 
If it is difficult to set kinematics, it is often easier to set the
flavour content of a process. In a Higgs study, one might wish, for
example, to consider the decay $\H^0 \rightarrow \Z^0 \Z^0$, with each
$\Z^0 \rightarrow \ee$ or $\mu^+ \mu^-$. It is therefore necessary to
inhibit all other $\H^0$ and $\Z^0$ decay channels, and also to adjust
cross sections to take into account this change, all of which is fairly
straightforward. However, if one wanted to consider instead the
decay $\Z^0 \rightarrow \c \cbar$, with a $\D$ meson
producing a lepton, not only
would there then be the problem of different leptonic branching ratios
for different $\D$:s (which means that fragmentation and decay 
treatments would no longer decouple), but also that of additional
$\c \cbar$ pair production in parton-shower evolution, at a rate
that is unknown beforehand. In practice, it is therefore impossible to
force $\D$ decay modes in a consistent manner.
 
\clearpage
 
\section{Monte Carlo Techniques}
 
Quantum mechanics introduces a concept of randomness in the behaviour
of physical processes. The virtue of event generators is that this
randomness can be simulated by the use of Monte Carlo techniques.
In the process, the program authors have to use some ingenuity to 
find the most efficient way to simulate an assumed probability
distribution. A detailed description of possible techniques would 
carry us too far, but
in this section some of the most frequently used approaches are
presented, since they will appear in discussions in
subsequent sections. Further examples may be found e.g. in
\cite{Jam80}.
 
First of all one assumes the existence of a random number generator.
This is a (Fortran) function which, each time it is called, returns 
a number $R$ in the range between 0 and 1, such that the inclusive
distribution of numbers $R$ is flat in the range,
and such that different numbers $R$ are uncorrelated. The random
number generator that comes with {\Je} is described at the
end of this section, and we defer the discussion until then.
 
\subsection{Selection From a Distribution}
\label{ss:MCdistsel}
 
The situation that is probably most common is that we know a 
function $f(x)$ which is non-negative in the allowed $x$ range 
$x_{\mmin} \leq x \leq x_{\mmax}$. We want to select an $x$
`at random' so that the probability for a given $x$ is
proportional to $f(x)$. Here $f(x)$ might be a fragmentation function,
a differential cross section, or any of a number of distributions.
 
One does not
have to assume that the integral of $f(x)$ is explicitly normalized
to unity: by the Monte Carlo procedure of picking exactly one
accepted $x$ value, normalization is implicit in the final result.
Sometimes the integral of $f(x)$ does carry a physics content of
its own, as part of an overall weight factor
we want to keep track of. Consider, for instance, the case when $x$
represents one or several phase-space variables and $f(x)$ a
differential cross section; here the integral has a meaning of total
cross section for the process studied. The task of a Monte Carlo is
then, on the one hand, to generate events one at a time, and, on the
other hand, to estimate the total cross section.
The discussion of this important example is
deferred to section \ref{ss:PYTcrosscalc}.
 
If it is possible to find a primitive function $F(x)$ which
has a known inverse $F^{-1}(x)$, an $x$ can be found as
follows (method 1):
\begin{eqnarray}
& \displaystyle{ \int_{x_{\mmin}}^{x} f(x) \, \d x =
R \int_{x_{\mmin}}^{x_{\mmax}} f(x) \, \d x } & \nonumber \\[1mm]
\Longrightarrow & x = F^{-1}(F(x_{\mmin}) + 
R(F(x_{\mmax}) - F(x_{\mmin}))) ~. &
\end{eqnarray}
The statement of the first line is that a fraction $R$ of the
total area under $f(x)$ should be to the left of $x$.
However, seldom are functions of interest so nice that the
method above works. It is therefore necessary to use more complicated
schemes.
 
Special tricks can sometimes be found. Consider e.g. the generation
of a Gaussian $f(x) = \exp(-x^2)$. This function is not integrable,
but if we combine it with the same Gaussian distribution of a second
variable $y$, it is possible to transform to polar coordinates
\begin{equation}
f(x) \, \d x \, f(y) \, \d y = \exp(-x^2-y^2) \, \d x \, \d y =
r \exp(-r^2) \, \d r \, \d \varphi ~,
\end{equation}
and now the $r$ and $\varphi$ distributions may be easily generated
and recombined to yield $x$. At the same time we get a second number
$y$, which can also be used. For the generation of transverse momenta
in fragmentation, this is very convenient, since in fact we want to
assign two transverse degrees of freedom.
 
If the maximum of $f(x)$ is known, $f(x) \leq f_{\mmax}$ in the 
$x$ range considered, a hit-or-miss method will always yield the
correct answer (method 2):
\begin{Enumerate}
\item select an $x$ with even probability in the allowed range, i.e.
      $x = x_{\mmin} + R(x_{\mmax} - x_{\mmin})$;
\item compare a (new) $R$ with the ratio $f(x)/f_{\mmax}$;
      if $f(x)/f_{\mmax} \le R$, then reject the $x$ value
      and return to point 1 for a new try;
\item otherwise the most recent $x$ value is retained as final answer.
\end{Enumerate}
The probability that $f(x)/f_{\mmax} > R$ is proportional to 
$f(x)$; hence the correct distribution of retained $x$ values.
The efficiency of this method, i.e. the average probability that
an $x$ will be retained, is $(\int \, f(x) \, \d x) 
/ (f_{\mmax}(x_{\mmax} - x_{\mmin}))$.
The method is acceptable if this number is not too low, i.e. if
$f(x)$ does not fluctuate too wildly.
 
Very often $f(x)$ does have narrow spikes, and it may not even be
possible to define an $f_{\mmax}$. An example of the former 
phenomenon is a
function with a singularity just outside the allowed region,
an example of the latter an integrable singularity just at the
$x_{\mmin}$ and/or $x_{\mmax}$ borders.
Variable transformations may then be used to make a function
smoother. Thus a function $f(x)$ which blows up as $1/x$ for
$x \rightarrow 0$, with an $x_{\mmin}$ close to 0, would instead
be roughly constant if transformed to the variable $y = \ln x$.
 
The variable transformation strategy may be seen as a combination
of methods 1 and 2, as follows. Assume the existence of a function
$g(x)$, with $f(x) \leq g(x)$ over the $x$ range of interest.
Here $g(x)$ is picked to be a `simple' function, such that the
primitive function $G(x)$ and its inverse $G^{-1}(x)$ are known.
Then (method 3):
\begin{Enumerate}
\item select an $x$ according to the distribution $g(x)$, using
      method 1;
\item compare a (new) $R$ with the ratio $f(x)/g(x)$;
      if $f(x)/g(x) \le R$, then reject the $x$ value
      and return to point 1 for a new try;
\item otherwise the most recent $x$ value is retained as final answer.
\end{Enumerate}
This works, since the first step will select $x$ with a probability
$g(x) \, \d x = \d G(x)$ and the second retain this choice with
probability $f(x)/g(x)$. The total probability to pick a value $x$
is then just the product of the two, i.e. $f(x) \, \d x$.
 
If $f(x)$ has several spikes, method 3 may work for each spike
separately, but it may not be possible to find a $g(x)$ that
covers all of them at the same time, and which still has an
invertible primitive function.  However, assume that
we can find a function $g(x) = \sum_i g_i(x)$,
such that $f(x) \leq g(x)$ over
the $x$ range considered, and such that the functions $g_i(x)$
each are non-negative and simple, in the sense that we can find
primitive functions and their inverses. In that case (method 4):
\begin{Enumerate}
\item select an $i$ at random, with relative probability given
      by the integrals
      \begin{equation}
      \int_{x_{\mmin}}^{x_{\mmax}} g_i(x) \, \d x =
      G_i(x_{\mmax}) - G_i(x_{\mmin}) ~; \nonumber
      \end{equation}
\item for the $i$ selected, use method 1 to find an $x$, i.e.
      \begin{equation}
      x = G_i^{-1}(G_i(x_{\mmin}) + 
      R(G_i(x_{\mmax})-G_i(x_{\mmin}))) ~;
      \nonumber
      \end{equation}
\item compare a (new) $R$ with the ratio $f(x)/g(x)$;
      if $f(x)/g(x) \le R$, then reject the $x$ value
      and return to point 1 for a new try;
\item otherwise the most recent $x$ value is retained as final answer.
\end{Enumerate}
This is just a trivial extension of method 3, where steps 1 and 2
ensure that, on the average, each $x$ value picked there is
distributed according to $g(x)$: the first step picks $i$
with relative probability $\int g_i(x) \, \d x$, the second $x$ with
absolute probability $g_i(x) / \int g_i(x) \, \d x$ (this is one place
where one must remember to do normalization correctly); the
product of the two is therefore $g_i(x)$ and the sum over all $i$
gives back $g(x)$.
 
We have now arrived at an approach that is sufficiently powerful for
a large selection of problems. In general,
for a function $f(x)$ which is known to have sharp peaks in a few
different places, the generic behaviour at each peak separately
may be covered by one or a few simple functions $g_i(x)$, to which
one adds a few more $g_i(x)$ to cover the basic behaviour away
from the peaks. By a suitable selection
of the relative strengths of the different $g_i$'s, it is
possible to find a function $g(x)$ that matches well the general
behaviour of $f(x)$, and thus achieve a reasonable Monte Carlo 
efficiency.
 
The major additional complication is when $x$ is a multidimensional
variable. Usually the problem is not so much $f(x)$ itself, but
rather that the phase-space boundaries may be very complicated.
If the boundaries factorize it is possible to pick phase-space
points restricted to the desired region. Otherwise the region may
have to be inscribed in a hyper-rectangle, with points picked within
the whole hyper-rectangle but only retained if they are inside the
allowed region. This may lead to a significant loss in efficiency.
Variable transformations may often make the allowed region easier to
handle.
 
There are two main methods to handle several dimensions, each with its
set of variations. The first method is based on a factorized ansatz,
i.e. one attempts to find a function $g(\mbf{x})$ which is
everywhere larger than $f(\mbf{x})$, and which can be factorized
into $g(\mbf{x}) = g^{(1)}(x_1) \, g^{(2)}(x_2) \cdots g^{(n)}(x_n)$,
where $\mbf{x} = (x_1,x_2,\ldots,x_n)$. Here each $g^{(j)}(x_j)$ may
in its turn be a sum of functions $g^{(j)}_i$, as in method 4 above.
First, each $x_j$ is selected independently, and afterwards the ratio
$f(\mbf{x})/g(\mbf{x})$ is used to determine whether to retain the
point.
 
The second method is useful if the boundaries of the allowed region
can be written in a form where the maximum range of $x_1$  is known,
the allowed range of $x_2$ only depends on $x_1$, that of $x_3$ only
on $x_1$ and $x_2$, and so on until $x_n$, whose range may depend on 
all the
preceding variables. In that case it may be possible to find a function
$g(\mbf{x})$ that can be integrated over $x_2$ through $x_n$ to yield
a simple function of $x_1$, according to which $x_1$ is selected.
Having done that, $x_2$ is selected according to a distribution
which now depends on $x_1$, but with $x_3$ through $x_n$ integrated
over. In particular, the allowed range for $x_2$ is known.
The procedure is continued until $x_n$ is reached, where now the
function depends on all the preceding $x_j$ values. In the end, the
ratio $f(\mbf{x})/g(\mbf{x})$ is again used to determine whether to
retain the point.
 
\subsection{The Veto Algorithm}
\label{ss:vetoalg}
 
The `radioactive decay' type of problems is very common,  in particular
in parton showers, but it is also used, e.g. in the multiple
interactions description in {\Py}. In this kind of problems
there is one variable $t$, which may be thought of as giving a kind
of time axis along which different events are ordered. The
probability that `something will happen' (a nucleus decay, a
parton branch) at time $t$ is described by a function $f(t)$, which
is non-negative in the range of $t$ values to be studied. However,
this na\"{\i}ve probability is modified by the additional requirement
that something can only happen at time $t$ if it did not happen
at earlier times $t' < t$. (The original nucleus cannot decay once
again if it already did decay; possibly the decay products may decay
in their turn, but that is another question.)
 
The probability that nothing has happened by time $t$ is expressed by
the function ${\cal N}(t)$ and the differential probability that
something happens at time $t$ by ${\cal P}(t)$. The basic equation
then is
\begin{equation}
 {\cal P}(t) = - \frac{\d {\cal N}}{\d t} = f(t) \, {\cal N}(t) ~.
\end{equation}
For simplicity, we shall assume that the process starts at time
$t = 0$, with ${\cal N}(0) = 1$.
 
The above equation can be solved easily if one notes that
$\d {\cal N} / {\cal N} = \d \ln {\cal N}$:
\begin{equation}
 {\cal N}(t) = {\cal N}(0) \exp \left\{ - \int_0^t f(t') \, \d t'
               \right\}
             = \exp \left\{ - \int_0^t f(t') \, \d t' \right\} ~,
\end{equation}
and thus
\begin{equation}
  {\cal P}(t) = f(t) \exp \left\{ - \int_0^t f(t') \, \d t' \right\} ~.
\label{mc:Pveto}
\end{equation}
With $f(t) = c$ this is nothing but the textbook formulae for
radioactive decay. In particular, at small times the correct
decay probability, ${\cal P}(t)$, agrees well with the input
one, $f(t)$, since the exponential factor is close to unity there.
At larger $t$, the exponential gives a dampening which ensures that
the integral of ${\cal P}(t)$ never can exceed unity, even if the
integral of $f(t)$ does. The exponential can be seen as the
probability that nothing happens between the original time 0 and
the final time $t$. In the parton-shower language, this is (almost)
the so-called Sudakov form factor.
 
If $f(t)$ has a primitive function with a known inverse, it is easy
to select $t$ values correctly:
\begin{equation}
  \int_0^t {\cal P}(t') \, \d t' = {\cal N}(0) - {\cal N}(t) =
  1 -  \exp \left\{ - \int_0^t f(t') \, \d t' \right\} = 1 - R ~,
\end{equation}
which has the solution
\begin{equation}
  F(0) - F(t) = \ln R  ~~~ \Longrightarrow ~~~
  t = F^{-1}(F(0) - \ln R) ~.
\end{equation}
 
If $f(t)$ is not sufficiently nice, one may again try to find a
better function $g(t)$, with $f(t) \leq g(t)$ for all $t \geq 0$.
However to use method 3 with this $g(t)$ would not work, since the
method would not correctly take
into account the effects of the exponential term in ${\cal P}(t)$.
Instead one may use the so-called veto algorithm:
\begin{Enumerate}
\item start with $i = 0$ and $t_0 = 0$;
\item add 1 to $i$ and select $t_i = G^{-1}(G(t_{i-1}) - \ln R)$,
      i.e. according to $g(t)$, but with the constraint that
      $t_i > t_{i-1}$,
\item compare a (new) $R$ with the ratio $f(t_i)/g(t_i)$;
      if $f(t_i)/g(t_i) \le R$, then return to point 2 for a new try;
\item otherwise $t_{i}$ is retained as final answer.
\end{Enumerate}
 
It may not be apparent why this works. Consider, however, the various
ways in which one can select a specific time $t$. The probability that
the first try works, $t = t_1$, i.e. that no intermediate $t$ values
need be rejected, is given by
\begin{equation}
  {\cal P}_0 (t) = \exp \left\{ - \int_0^t g(t') \, \d t' \right\}
  \, g(t) \, \frac{f(t)}{g(t)}
   = f(t) \exp \left\{ - \int_0^t g(t') \, \d t' \right\} ~,
\end{equation}
where the exponential times $g(t)$ comes from eq.~(\ref{mc:Pveto})
applied to $g$, and the ratio $f(t)/g(t)$ is the probability that
$t$ is
accepted. Now consider the case where one intermediate time $t_1$ is
rejected and $t = t_2$ is only accepted in the second step.
This gives
\begin{equation}
  {\cal P}_1 (t) = \int_0^t \d t_1
  \exp \left\{ - \int_0^{t_1} g(t') \, \d t' \right\} g(t_1)
  \left[ 1 - \frac{f(t_1)}{g(t_1)} \right]
  \exp \left\{ - \int_{t_1}^t g(t') \, \d t' \right\} g(t) \,
  \frac{f(t)}{g(t)} ~,
\end{equation}
where the first exponential times $g(t_1)$ gives the probability
that $t_1$ is first selected, the square brackets the probability
that $t_1$
is subsequently rejected, the following piece the probability that
$t = t_2$ is selected when starting from $t_1$, and the final factor
that $t$ is retained. The whole is to be integrated over all
possible intermediate times $t_1$. The exponentials together give an
integral over the range from 0 to $t$, just as in ${\cal P}_0$,
and the factor for the final step being accepted is also the same,
so therefore one finds that
\begin{equation}
  {\cal P}_1 (t) = {\cal P}_0 (t) \int_0^t \d t_1
  \left[ g(t_1) - f(t_1) \right] ~.
\end{equation}
This generalizes.
In ${\cal P}_2$ one has to consider two intermediate times,
$0 \leq t_1 \leq t_2 \leq t_3 = t$, and so
\begin{eqnarray}
{\cal P}_2 (t) & = &  {\cal P}_0 (t)
   \int_0^t \d t_1 \left[ g(t_1) - f(t_1) \right]
   \int_{t_1}^t \d t_2 \left[ g(t_2) - f(t_2) \right]  \nonumber \\
& = & {\cal P}_0 (t) \frac{1}{2} \left(
   \int_0^t \left[ g(t') - f(t') \right] \d t' \right)^2 ~.
\end{eqnarray}
The last equality is most easily seen if one also considers the
alternative region $0 \leq t_2 \leq t_1 \leq t$, where the
r\^oles of $t_1$ and $t_2$ have just been interchanged, and the
integral therefore has the same value as in the region considered.
Adding the two regions, however, the integrals over $t_1$ and
$t_2$ decouple, and become equal. In general, for ${\cal P}_i$,
the $i$ intermediate times can be ordered in $i!$ different ways.
Therefore the total probability to accept $t$, in any step, is
\begin{eqnarray}
{\cal P} (t) & = &
   \displaystyle{ \sum_{i=0}^{\infty} {\cal P}_i (t)
   = {\cal P}_0 (t) \sum_{i=0}^{\infty} \frac{1}{i!}
   \left( \int_0^t \left[ g(t') - f(t') \right] \d t' \right)^i } 
   \nonumber \\
& = & \displaystyle{ f(t) \exp \left\{ - \int_0^t g(t') \, \d t'
   \right\} \exp \left\{ \int_0^t \left[ g(t') - f(t') \right] \d t'
   \right\} }  \nonumber \\
& = & \displaystyle{ f(t) \exp \left\{ - \int_0^t f(t') \, \d t' 
   \right\} ~, }
\end{eqnarray}
which is the desired answer.
 
If the process is to be stopped at some scale $t_{\mmax}$, i.e.
if one would like to remain with a fraction ${\cal N}(t_{\mmax})$
of events where nothing happens at all, this is easy to include
in the veto algorithm: just iterate upwards in $t$ at usual, but
stop the process if no allowed branching is found before 
$t_{\mmax}$.
 
Usually $f(t)$ is a function also of additional variables $x$. The
methods of the preceding subsection are easy to generalize if one
can find a suitable function $g(t,x)$ with $f(t,x) \leq g(t,x)$.
The $g(t)$ used in the veto algorithm is the integral of $g(t,x)$
over $x$. Each time a $t_i$ has been selected also an $x_i$ is
picked, according to $g(t_i,x) \, dx$, and the $(t,x)$ point is
accepted with probability $f(t_i,x_i)/g(t_i,x_i)$.
 
\subsection{The Random Number Generator}
 
The construction of a good, portable (pseudo)random generator is not a
trivial task. Therefore {\Je} has traditionally stayed
away from that area, and just provided the routine \ttt{RLU} as an
interface, which the user could modify to call on an existing routine,
implemented on the actual machine being used.
 
In recent years, progress has been made in constructing portable
generators with large periods and other good properties; see the review
\cite{Jam90}. Therefore the current version contains a random number
generator based on the algorithm proposed by Marsaglia, Zaman and Tsang
\cite{Mar90}. This routine should work on any machine with a mantissa
of at least 24 digits, i.e. all common 32-bit (or more) computers.
Given the same initial state, the sequence will also be identical
on different machines. This need not mean that the same sequence of
events will be generated on an IBM and a VAX, say, since the different
treatments of roundoff errors in numerical operations will lead to
slightly different real numbers being tested against these random
numbers in IF statements. Also code optimization may lead to a
divergence. Apart from nomenclature issues, and the
coding of \ttt{RLU} as a function rather than a subroutine, the only
difference between the {\Je} code and the code given in
\cite{Jam90} is that slightly different algorithms are used to ensure
that the random number is not equal to 0 or 1 within the machine 
precision.
 
The generator has a period of over $10^{43}$, and the
possibility to obtain almost $10^9$
different and disjoint subsequences, selected
by giving an initial integer number. The price to be paid for the
long period is that the state of the generator at a given moment 
cannot be described by a single integer, but requires about 100 words.
Some of these are real numbers, and are thus not correctly represented
in decimal form. The normal procedure, which makes it possible to 
restart the generation from a seed value written to the run output, 
is therefore
not convenient. The CERN library implementation keeps track of the
number of random numbers generated since the start. With this value
saved, in a subsequent run the random generator can be asked to skip
ahead the corresponding number of random numbers. {\Je} 
is a heavy user of random numbers, however: typically 30\% of the full
run time is spent on random number generation. Of this, half is
overhead coming from the function call administration, but the other
half is truly related to the speed of the algorithm. Therefore a
skipping ahead would take place with 15\% of the time cost of the 
original run, i.e. an uncomfortably high figure.
 
Instead a different solution is chosen here. Two special routines are
provided for writing and reading the state of the random number 
generator (plus some initialization information) on a sequential 
file, in a machine-dependent internal representation. The file used 
for this purpose
has to be specified by you, and opened for read and write.
A state is written as a single record, in free format. It is possible
to write an arbitrary number of states on a file, and a record can be
overwritten, if so desired. The event generation loop might then look
something like:
\begin{Enumerate}
\item save the state of the generator on file (using flag set in
point 3 below),
\item generate an event,
\item study the event for errors or other reasons why to regenerate
it later; set flag to overwrite previous generator state if no errors,
otherwise set flag to create new record;
\item loop back to point 1.
\end{Enumerate}
With this procedure, the file will contain the state before each of the
problematical events. An alternative approach might be to save the
state every 100 events or so. If the events are subsequently
processed through a detector simulation, you may
have to save also other sets of seeds, naturally.
 
In addition to the service routines, the common block which contains
the state of the generator is available for manipulation,
if you so desire. In particular, the initial seed value is by
default 19780503, i.e. different from the Marsaglia/CERN default
54217137. It is possible to change this value before any random numbers
have been generated, or to force reinitialization in mid-run with any
desired new seed. Inside {\JePy}, some
initialization may take place in connection with the very first event
generated in a run, so sometimes it may be necessary to generate one
ordinary event before reading in a saved state to generate an
interesting event. In the current {\Py} version, some of the
multiple interaction machinery options contain an element of learning,
which means that the event sequence may be broken.
 
It should be noted that, of course, the appearance of a random
number generator package inside {\Je} does in no way preclude
the use of other routines. You can easily revert to the
old approach, where \ttt{RLU} is nothing but an interface to an
arbitrary external random number generator; e.g. to call a routine
\ttt{RNDM} all you need to have is
\begin{verbatim}
      FUNCTION RLU(IDUMMY)
  100 RLU=RNDM(IDUMMY)
      IF(RLU.LE.0..OR.RLU.GE.1.) GOTO 100
      RETURN
      END
\end{verbatim}
 
The random generator subpackage consists of the following components.
 
\drawbox{R = RLU(IDUMMY)}\label{p:RLU}
\begin{entry}
\itemc{Purpose:} to generate a (pseudo)random number \ttt{R} uniformly
  in the range 0$<$\ttt{R}$<$1, i.e. excluding the endpoints.
\iteme{IDUMMY :} dummy input argument; normally 0.
\end{entry}
 
\drawbox{CALL RLUGET(LFN,MOVE)}\label{p:RLUGET}
\begin{entry}
\itemc{Purpose:} to dump the current state of the random number
  generator on a separate file, using internal representation for
  real and integer numbers. To be precise, the full contents of
  the \ttt{LUDATR} common block are written on the file, with the
  exception of \ttt{MRLU(6)}.
\iteme{LFN :} (logical file number) the file number to which the state
  is dumped. You
  must associate this number with a true file (with a machine-dependent
  name), and see to it that this file is open for write.
\iteme{MOVE :} choice of adding a new record to the file or
  overwriting old record(s). Normally only options 0 or $-$1 should be
  used.
    \begin{subentry}
    \iteme{= 0 (or > 0) :} add a new record to the end of the
      file.
    \iteme{= -1 :} overwrite the last record with a new one (i.e. do
      one \ttt{BACKSPACE} before the new write).
    \iteme{= $-n$ :} back up $n$ records before writing the
      new record. The records following after the new one are lost,
      i.e. the last $n$ old records are lost and one new added.
    \end{subentry}
\end{entry}
 
\drawbox{CALL RLUSET(LFN,MOVE)}\label{p:RLUSET}
\begin{entry}
\itemc{Purpose:} to read in a state for the random number generator,
  from which the subsequent generation can proceed. The state must
  previously have been saved by a \ttt{RLUGET} call. Again the full
  contents of the \ttt{LUDATR} common block are read, with the
  exception of \ttt{MRLU(6)}.
\iteme{LFN :} (logical file number) the file number from which the
  state is read. You
  must associate this number with a true file previously written with
  a \ttt{RLUGET} call, and see to it that this file is open for read.
\iteme{MOVE :} positioning in file before a record is read. With zero
  value, records are read one after the other for each new call, while
  non-zero values may be used to navigate back and forth, and e.g.
  return to the same initial state several times.
    \begin{subentry}
    \iteme{= 0 :} read the next record.
    \iteme{= $+n$ :} skip ahead $n$ records before reading the record
      that sets the state of the random number generator.
    \iteme{= $-n$ :} back up $n$ records before reading the record that
      sets the state of the random number generator.
    \end{subentry}
\end{entry}
 
\drawbox{COMMON/LUDATR/MRLU(6),RRLU(100)}\label{p:LUDATR}
\begin{entry}
\itemc{Purpose:} to contain the state of the random number generator
  at any moment (for communication between \ttt{RLU}, \ttt{RLUGET}
  and \ttt{RLUSET}), and also to provide the user with the
  possibility to initialize different random number sequences, and to
  know how many numbers have been generated.
\iteme{MRLU(1) :}\label{p:MRLU} (D=19780503) the integer number that 
  specifies
  which of the possible subsequences will be initialized in the next
  \ttt{RLU} call for which \ttt{MRLU(2)=0}. Allowed values are
  0$\leq$\ttt{MRLU(1)}$\leq$900\,000\,000, the original Marsaglia
  (and CERN library) seed is 54217137. The \ttt{MRLU(1)} value is not
  changed by any of the {\Je} routines.
\iteme{MRLU(2) :} (D=0) initialization flag, put to 1 in the first
  \ttt{RLU} call of run. A reinitialization of the random number
  generator can be made in mid-run by resetting \ttt{MRLU(2)} to 0
  by hand. In addition, any time the counter \ttt{MRLU(3)} reaches
  1000000000, it is reset to 0 and \ttt{MRLU(2)} is increased by 1.
\iteme{MRLU(3) :} (D=0) counter for the number of random numbers
  generated from the beginning of the run. To avoid overflow when
  very many numbers are generated, \ttt{MRLU(2)} is used as
  described above.
\iteme{MRLU(4), MRLU(5) :} \ttt{I97} and \ttt{J97} of the CERN
  library implementation; part of the state of the generator.
\iteme{MRLU(6) :} (D=0) current position, i.e. how many records after
  beginning, in the file; used by \ttt{RLUGET} and \ttt{RLUSET}.
\iteme{RRLU(1) - RRLU(97) :}\label{p:RRLU} the \ttt{U} array of the 
  CERN library implementation; part of the state of the generator.
\iteme{RRLU(98) - RRLU(100) :} \ttt{C}, \ttt{CD} and \ttt{CM} of the
  CERN library implementation; the first part of the state of the
  generator, the latter two constants calculated at initialization.
\end{entry}
 
\clearpage
 
\section{The Event Record}
 
The event record is the central repository for information about
the particles produced in the current event: flavours, momenta,
event history, and production vertices. It plays a very central
r\^ole: without a proper understanding of what the record is and
how information is stored, it is meaningless to try to use either
{\Je} or {\Py}. The record is stored in the common block 
\ttt{LUJETS}. Almost all the routines thatthe user calls
can be viewed as performing some action on the record: fill a
new event, let partons fragment or particles decay, boost it,
list it, find clusters, etc.
 
In this section we will first describe the KF flavour code,
subsequently the \ttt{LUJETS} common block, and then give a few
comments about the r\^ole of the event record in the programs.
 
To ease the interfacing of different event generators, a
\ttt{HEPEVT} standard common block structure for the event record
has been agreed on. For historical reasons the standard common blocks
are not directly used in {\Je}, but a conversion routine comes with
the program, and is described at the end of this section.
 
\subsection{Particle Codes}
\label{ss:codes}
 
The new particle code now adopted by the Particle Data Group
\cite{PDG88,PDG92} is used consistently throughout the program, and is
referred to as the KF particle code. This code you have to be
thoroughly familiar with. It is described below.
 
Note that a few inconsistencies between the KF and the PDG codes
are known, which stem from differences of interpretation of
the rules agreed on when developing the standard. These rules form
the basis of the PDG tables and (independently) of the {\Je} tables.
(Of course, my private opinion is that I follow the original agreement,
and the PDG deviate from it.) Hopefully, this should have few practical
consequences, since only rarely-produced particles are affected.
Anyway, here is a list of the known discrepancies:
\begin{Enumerate}
\item The PDG has not allowed for the existence of an $\eta_{\b}$, 
which in {\Je} is included with code 551. This code is reserved for
$\chi_{0 \b}$ by the PDG, a particle which appears as 10551 in {\Je}.
(We agree to have $\eta_{\c}$ as 441, which illustrates the basic
difference: I use the additional recurrence figure to refer to a
whole multiplet, whether all particles of that multiplet have been
found or not; the PDG, on the other hand, does not reserve space for 
particles which we know should be there but have not yet been 
discovered, which means that members of a multiplet need not go 
together.)
\item The PDG has not allowed for the existence of an $\hrm_{1 \c}$,
which in {\Je} is represented by 10443. Therefore $\chi_{1 \c}$
is the PDG code 10443 but {\Je} code 20443. Further $\psi'$ is
either 20443 or 30443, and $\Upsilon' = \Upsilon(2S)$ either
20553 or 30553. (Comment as for point 1.)
\item Different conventions for spin $1/2$ baryons with one heavy
flavour (charm, bottom, top), one strange flavour, and one light
($\u$ or $\d$). Here two states exist, e.g. $\Xi_{\c}^+$ and
$\Xi'^+_{\c}$, both with flavour content $\c\s\u$. By analogy with the
$\Lambda^0$--$\Sigma$ pair, {\Je} uses the decreasing order of flavour
content for the heavier state and inversed order of the two lighter
flavours for the lighter state, while the PDG tables use the opposite
convention. Thus in {\Je} $\Xi_{\c}^+$ is 4232 and $\Xi'^+_{\c}$ 4322,
while in PDG it is the other way around.
\end{Enumerate}
There are no plans to change the {\Je} rules to agree with
the PDG ones in either of the cases above.
 
The KF code is not convenient for a direct storing of masses,
decay data, or other particle properties, since the KF
codes are so spread out. Instead a compressed code KC between
1 and 500 is used here, where the most frequently used particles
have a separate code, but many heavy-flavour hadrons are lumped
together in groups. Normally this code is only used at very specific
places in the program, not visible to the user. If need be, the
correspondence can always be obtained by using the function
\ttt{LUCOMP}, \mbox{\ttt{KC = LUCOMP(KF)}}. It is therefore not
intended that you should ever need to know any KC codes at all.
It may be useful to know, however, that for codes smaller than 80,
KF and KC agree.
 
The particle names printed in the tables in this section correspond 
to the ones obtained
with the routine \ttt{LUNAME}, which is used extensively, e.g. in
\ttt{LULIST}. Greek characters are spelt out in full, with a capital
first letter to correspond to a capital Greek letter. Generically the
name of a particle is made up of the following pieces:
\begin{Enumerate}
\item The basic root name. This includes a * for most spin 1
($L = 0$) mesons and spin $3/2$ baryons, and a $'$ for some spin
$1/2$ baryons (where there are two states to be distinguished,
cf. $\Lambda$--$\Sigma^0$). The rules for heavy baryon naming are in
accordance with the 1986 Particle Data Group conventions \cite{PDG86}.
For mesons with one unit of orbital angular momentum, K (D, B,
\ldots) is used for quark-spin 0 and K* (D*, B*, \ldots) for
quark-spin 1 mesons; the convention for `*' may here deviate slightly
from the one used by the PDG.
\item Any lower indices, separated from the root by a \_. For
heavy hadrons, this is the additional heavy-flavour content not
inherent in the root itself. For a diquark, it is the spin.
\item The character $\sim$ (alternatively bar, see \ttt{MSTU(15)})
for an antiparticle, wherever the distinction between particle and
antiparticle is not inherent in the charge information.
\item Charge information: $++$, $+$, $0$, $-$, or $--$.
Charge is not given for quarks or diquarks. Some neutral particles
which are customarily given without a 0 also here lack it,
such as neutrinos, $\g$, $\gamma$,
and flavour-diagonal mesons other than $\pi^0$ and $\rho^0$. Note that
charge is included both for the proton and the neutron. While
non-standard, it is helpful in avoiding misunderstandings when
looking at an event listing.
\end{Enumerate}
 
Below follows a list of KF particle codes. The list is not complete;
a more extensive one may be obtained with \ttt{CALL LULIST(11)}.
Particles are grouped together, and the basic rules are described
for each group. Whenever a distinct antiparticle exists, it is given
the same KF code with a minus sign (whereas KC codes are always
positive).
 
\begin{Enumerate}
 
\item Quarks and leptons, Table \ref{t:codeone}. \\
This group contains the basic building blocks of matter, arranged
according to family, with the lower member of weak isodoublets also
having the smaller code (thus $\d$ precedes $\u$, contrary to the
ordering in previous {\Je} versions). A fourth generation is
included for future reference. The quark codes are used as building
blocks for the diquark, meson and baryon codes below.
 
\begin{table}[ptb]
\captive{Quark and lepton codes.
\protect\label{t:codeone} } \\
\vspace{1ex}
\begin{center}
\begin{tabular}{|c|c|c||c|c|c|@{\protect\rule{0mm}{\tablinsep}}}
\hline
KF & Name & Printed & KF & Name & Printed \\
\hline
    1 & $\d$ & \ttt{d}   &  11 & $\e^-$ &  \ttt{e-}    \\
    2 & $\u$ & \ttt{u}   &  12 & $\nu_{\e}$ &  \ttt{nu\_e}   \\
    3 & $\s$ & \ttt{s}   &  13 & $\mu^-$ &  \ttt{mu-}   \\
    4 & $\c$ & \ttt{c}   &  14 & $\nu_{\mu}$ & \ttt{nu\_mu}   \\
    5 & $\b$ & \ttt{b}   &  15 & $\tau^-$ & \ttt{tau-}   \\
    6 & $\t$ & \ttt{t}   &  16 & $\nu_{\tau}$ &  \ttt{nu\_tau}   \\
    7 & $\lrm$ & \ttt{l}   &  17 & $\chi^-$ & \ttt{chi-}  \\
    8 & $\hrm$ & \ttt{h}   &  18 & $\nu_{\chi}$ & \ttt{nu\_chi}  \\
    9 & & & 19 & & \\
   10 & & & 20 & & \\
\hline
\end{tabular}
\end{center}
\end{table}
 
\item Gauge bosons and other fundamental bosons,
Table \ref{t:codetwo}. \\
This group includes all the gauge and Higgs bosons of the standard
model, as well as some of the bosons appearing in various extensions
of it. The latter are not covered by the standard
PDG codes. They correspond to one extra {\bf U(1)}
group and one extra {\bf SU(2)} one, a further Higgs doublet, a
(scalar, colour octet) techni-$\eta$, a (scalar) leptoquark 
$\L_{\Q}$, and a horizontal gauge boson $\R$
(coupling between families). Additionally, we here include the 
pomeron $\pomeron$ and reggeon $\reggeon$ `particles', which are 
important e.g. in the description of diffractive scattering, but 
have no obvious position anywhere in the classification scheme.
 
\begin{table}[ptb]
\captive{Gauge boson and other fundamental boson codes.
\protect\label{t:codetwo} }  \\
\vspace{1ex}
\begin{center}
\begin{tabular}{|c|c|c||c|c|c|@{\protect\rule{0mm}{\tablinsep}}}
\hline
KF & Name & Printed & KF & Name & Printed \\
\hline
   21 & $\g$ & \ttt{g}             & 31 & &   \\
   22 & $\gamma$ & \ttt{gamma}     & 32 & $\Z'^0$ & \ttt{Z'0}   \\
   23 & $\Z^0$ & \ttt{Z0}          & 33 & $\Z''^0$ & \ttt{Z"0}   \\
   24 & $\W^+$ & \ttt{W+}          & 34 & $\W'^+$ & \ttt{W'+}   \\
   25 & $\H^0$ & \ttt{H0}          & 35 & $\H'^0$ & \ttt{H'0}   \\
   26 & &                          & 36 & $\A^0$ & \ttt{A0}   \\
   27 & &                          & 37 & $\H^+$ & \ttt{H+}   \\
   28 & $\reggeon$ & \ttt{reggeon} &
   38 & $\eta_{\mrm{techni}}$ & \ttt{eta\_tech0}  \\
   29 & $\pomeron$ & \ttt{pomeron} & 39 & $\L_{\Q}$ & \ttt{LQ}   \\
   30 & &                          & 40 & $\R^0$ & \ttt{R0}   \\
\hline
\end{tabular}
\end{center}
\end{table}
 
\item Free space. \\
The positions 41--80 are currently unused. In the future, they might
come to be used, e.g. for supersymmetric partners of the particles 
above, or for some other kind of new physics. At the moment, they are 
at your disposal.
 
\item Various special codes, Table \ref{t:codefour}. \\
In a Monte Carlo, it is always necessary to have codes that do not
correspond to any specific particle, but are used to lump together
groups of similar particles for decay treatment, or to specify generic
decay products. These codes, which again are non-standard, are found
between numbers 81 and 100. Several are not found in the event record,
and therefore properly belong only to the KC group of codes.
 
\begin{table}[ptb]
\captive{Various special codes.
\protect\label{t:codefour} }   \\
\vspace{1ex}
\begin{center}
\begin{tabular}{|c|c|c|@{\protect\rule{0mm}{\tablinsep}}}
\hline
KF & Printed & Meaning \\
\hline
   81 &  \ttt{specflav} & Spectator flavour; used in decay-product
listings  \\
   82 &  \ttt{rndmflav} & A random $\u$, $\d$, or $\s$ flavour;
possible decay product  \\
   83 &  \ttt{phasespa} & Simple isotropic phase-space decay  \\
   84 &  \ttt{c-hadron} & Information on decay of generic charm
hadron  \\
   85 &  \ttt{b-hadron} & Information on decay of generic bottom
hadron  \\
   86 &  \ttt{t-hadron} & Information on decay of generic top
hadron  \\
   87 &  \ttt{l-hadron} & Information on decay of generic low
hadron  \\
   88 &  \ttt{h-hadron} & Information on decay of generic high
hadron  \\
   89 &  \ttt{Wvirt}    & Off-mass-shell $\W$ in weak decays
of $\t$, $\lrm$, $\hrm$ or $\chi$  \\
   90 &  \ttt{diquark}  & Generic code for diquark colour
information  \\
   91 &  \ttt{cluster}  & Parton system in cluster
fragmentation  \\
   92 &  \ttt{string}   & Parton system in string
fragmentation  \\
   93 &  \ttt{indep.}   & Parton system in independent
fragmentation  \\
   94 &  \ttt{CMshower} & Four-momentum of time-like showering
system  \\
   95 &  \ttt{SPHEaxis} & Event axis found with \ttt{LUSPHE}  \\
   96 &  \ttt{THRUaxis} & Event axis found with \ttt{LUTHRU}  \\
   97 &  \ttt{CLUSjet}  & Jet (cluster) found with \ttt{LUCLUS}  \\
   98 &  \ttt{CELLjet}  & Jet (cluster) found with \ttt{LUCELL}  \\
   99 &  \ttt{table}    & Tabular output from \ttt{LUTABU}  \\
  100 &           &   \\
\hline
\end{tabular}
\end{center}
\end{table}
 
\item Diquark codes, Table \ref{t:codefive}. \\
A diquark made up of a quark with code $i$ and another with code $j$,
where $i \geq j$, and with total spin $s$, is given the code
\begin{equation}
\mrm{KF} = 1000 i + 100 j + 2s + 1  ~,
\end{equation}
i.e. the tens position is left empty (cf. the baryon code below).
Some of the most frequently used codes are listed in the table. All 
the lowest-lying spin 0 and 1 diquarks are included in the program.
 
The corresponding KC code is 90, and it is mainly used to store colour
charge.
 
\begin{table}[ptb]
\captive{Diquark codes.
\protect\label{t:codefive} }  \\
\vspace{1ex}
\begin{center}
\begin{tabular}{|c|c|c||c|c|c|@{\protect\rule{0mm}{\tablinsep}}}
\hline
KF & Name & Printed & KF & Name & Printed \\
\hline
      &          &             & 1103 & $\d\d_1$ & \ttt{dd\_1}  \\
 2101 & $\u\d_0$ & \ttt{ud\_0} & 2103 & $\u\d_1$ & \ttt{ud\_1}  \\
      &          &             & 2203 & $\u\u_1$ & \ttt{uu\_1}  \\
 3101 & $\s\d_0$ & \ttt{sd\_0} & 3103 & $\s\d_1$ & \ttt{sd\_1}  \\
 3201 & $\s\u_0$ & \ttt{su\_0} & 3203 & $\s\u_1$ & \ttt{su\_1}  \\
      &          &             & 3303 & $\s\s_1$ & \ttt{ss\_1}  \\
\hline
\end{tabular}
\end{center}
\end{table}
 
\item Meson codes, Tables \ref{t:codesixa} and \ref{t:codesixb}. \\
A meson made up of a quark with code $i$ and an antiquark with
code $-j$, $j \neq i$, and with total spin $s$, is given the code
\begin{equation}
\mrm{KF} = \left\{ 100 \max(i,j) + 10 \min(i,j) + 2s + 1 \right\}
\, \mrm{sign}(i-j) \, (-1)^{\max(i,j)} ~.
\end{equation}
Note the presence of an extra $-$ sign if the heaviest quark is a
down-type one. This is in accordance with the particle--antiparticle
distinction adopted in the 1986 Review of Particle Properties
\cite{PDG86}. It means for example that a $\B$ meson contains a 
$\bbar$ antiquark rather than a $\b$ quark.
 
The flavour-diagonal states are arranged in order of ascending 
mass. The standard rule of having the last digit of the form
$2s+1$ is broken for the $\K_{\mrm{S}}^0$--$\K_{\mrm{L}}^0$ system, 
where it is 0, and this convention should carry over to mixed states 
in the $\B$ meson system. For higher multiplets with the same spin, 
$\pm$10000, $\pm$20000, etc., are added to provide the extra 
distinction needed. Some of the most frequently used codes are given 
below.
 
The full lowest-lying pseudoscalar and vector multiplets are included
in the program, Table \ref{t:codesixa}.
 
\begin{table}[ptb]
\captive{Meson codes, part 1.
\protect\label{t:codesixa} }  \\
\vspace{1ex}
\begin{center}
\begin{tabular}{|c|c|c||c|c|c|@{\protect\rule{0mm}{\tablinsep}}}
\hline
KF & Name & Printed & KF & Name & Printed \\
\hline
  211 & $\pi^+$ & \ttt{pi+}        & 213 & $\rho^+$ & \ttt{rho+}  \\
  311 & $\K^0$ & \ttt{K0}          & 313 & $\K^{*0}$ & \ttt{K*0}  \\
  321 & $\K^+$ & \ttt{K+}          & 323 & $\K^{*+}$ & \ttt{K*+}  \\
  411 & $\D^+$ & \ttt{D+}          & 413 & $\D^{*+}$ & \ttt{D*+}  \\
  421 & $\D^0$ & \ttt{D0}          & 423 & $\D^{*0}$ & \ttt{D*0}  \\
  431 & $\D_s^+$ & \ttt{D\_s+}     &
433 & $\D_{\s}^{*+}$ & \ttt{D*\_s+}  \\
  511 & $\B^0$ & \ttt{B0}          & 513 & $\B^{*0}$ & \ttt{B*0}  \\
  521 & $\B^+$ & \ttt{B+}          & 523 & $\B^{*+}$ & \ttt{B*+}  \\
  531 & $\B_s^0$ & \ttt{B\_s0}     &
533 & $\B_{\s}^{*0}$ & \ttt{B*\_s0}  \\
  541 & $\B_c^+$ & \ttt{B\_c+}     &
543 & $\B_{\c}^{*+}$ & \ttt{B*\_c+}  \\
  111 & $\pi^0$ & \ttt{pi0}        & 113 & $\rho^0$ & \ttt{rho0}  \\
  221 & $\eta$ & \ttt{eta}         & 223 & $\omega$ & \ttt{omega}  \\
  331 & $\eta'$ & \ttt{eta'}       & 333 & $\phi$ & \ttt{phi}  \\
  441 & $\eta_{\c}$ & \ttt{eta\_c} & 443 & $\Jpsi$ & \ttt{J/psi}  \\
  551 & $\eta_{\b}$ & \ttt{eta\_b} &
553 & $\Upsilon$ & \ttt{Upsilon}  \\
  661 & $\eta_{\t}$ & \ttt{eta\_t} & 663 & $\Theta$ & \ttt{Theta}  \\
  130 & $\K_{\mrm{L}}^0$ & \ttt{K\_L0} & & &  \\
  310 & $\K_{\mrm{S}}^0$ & \ttt{K\_S0} & & &  \\
\hline
\end{tabular}
\end{center}
\end{table}
 
Also the lowest-lying orbital angular momentum $L = 1$ mesons are
included, Table \ref{t:codesixb}: one pseudovector multiplet
obtained for total quark-spin 0 ($L = 1, S = 0 \Rightarrow J = 1$)
and one scalar, one pseudovector and one tensor multiplet obtained
for total quark-spin 1 ($L = 1, S = 1 \Rightarrow J = 0, 1$ or 2),
where $J$ is what is conventionally called the spin $s$ of the meson.
Any mixing between the two pseudovector multiplets is
not taken into account. Please note that some members of these
multiplets have still not been found, and are included here only based
on guesswork. Even for known ones, the information on particles 
(mass, width, decay modes) is highly incomplete.

Only two radial excitations are included, the $\psi' = \psi(2S)$ and
$\Upsilon' = \Upsilon(2S)$.
 
\begin{table}[ptb]
\captive{Meson codes, part 2.
\protect\label{t:codesixb} }  \\
\vspace{1ex}
\begin{center}
\begin{tabular}{|c|c|c||c|c|c|@{\protect\rule{0mm}{\tablinsep}}}
\hline
KF & Name & Printed & KF & Name & Printed \\
\hline
10213 & $\b_1$ & \ttt{b\_1+}              &
10211 & $\a_0^+$ & \ttt{a\_0+}  \\
10313 & $\K_1^0$ & \ttt{K\_10}            &
10311 & $\K_0^{*0}$ & \ttt{K*\_00}  \\
10323 & $\K_1^+$ & \ttt{K\_1+}            &
10321 & $\K_0^{*+}$ & \ttt{K*\_0+}  \\
10413 & $\D_1^+$ & \ttt{D\_1+}            &
10411 & $\D_0^{*+}$ & \ttt{D*\_0+}  \\
10423 & $\D_1^0$ & \ttt{D\_10}            &
10421 & $\D_0^{*0}$ & \ttt{D*\_00}  \\
10433 & $\D_{1 \s}^+$ & \ttt{D\_1s+}      &
10431 & $\D_{0 \s}^{*+}$ & \ttt{D*\_0s+}  \\
10113 & $\b_1^0$ & \ttt{b\_10}            &
10111 & $\a_0^0$ & \ttt{a\_00}  \\
10223 & $\hrm_1^0$ & \ttt{h\_10}            &
10221 & $\f_0^0$ & \ttt{f\_00}  \\
10333 & $\hrm'^0_1$ & \ttt{h'\_10}          &
10331 & $\f'^0_0$ & \ttt{f'\_00}  \\
10443 & $\hrm_{1 \c}^0$ & \ttt{h\_1c0}      &
10441 & $\chi_{0 \c}^0$ & \ttt{chi\_0c0}  \\  \hline
20213 & $\a_1^+$ & \ttt{a\_1+}            &
  215 & $\a_2^+$ & \ttt{a\_2+}  \\
20313 & $\K_1^{*0}$ & \ttt{K*\_10}        &
  315 & $\K_2^{*0}$ & \ttt{K*\_20}  \\
20323 & $\K_1^{*+}$ & \ttt{K*\_1+}        &
  325 & $\K_2^{*+}$ & \ttt{K*\_2+}  \\
20413 & $\D_1^{*+}$ & \ttt{D*\_1+}        &
  415 & $\D_2^{*+}$ & \ttt{D*\_2+}  \\
20423 & $\D_1^{*0}$ & \ttt{D*\_10}        &
  425 & $\D_2^{*0}$ & \ttt{D*\_20}  \\
20433 & $\D_{1 \s}^{*+}$ & \ttt{D*\_1s+}  &
  435 & $\D_{2 \s}^{*+}$ & \ttt{D*\_2s+}  \\
20113 & $\a_1^0$ & \ttt{a\_10}            &
  115 & $\a_2^0$ & \ttt{a\_20}  \\
20223 & $\f_1^0$ & \ttt{f\_10}            &
  225 & $\f_2^0$ & \ttt{f\_20}  \\
20333 & $\f'^0_1$ & \ttt{f'\_10}          &
  335 & $\f'^0_2$ & \ttt{f'\_20}  \\
20443 & $\chi_{1 \c}^0$ & \ttt{chi\_1c0}  &
  445 & $\chi_{2 \c}^0$ & \ttt{chi\_2c0}  \\  \hline
30443 & $\psi'$     & \ttt{psi'}     &   &   &  \\
30553 & $\Upsilon'$ & \ttt{Upsilon'} &   &   &  \\
\hline
\end{tabular}
\end{center}
\end{table}
 
The corresponding meson KC codes, used for organizing mass and decay
data, range between 101 and 240.
 
\item Baryon codes, Table \ref{t:codeseven}.  \\
A baryon made up of quarks $i$, $j$ and $k$, with $i \geq j \geq k$,
and total spin $s$, is given the code
\begin{equation}
\mrm{KF} = 1000 i + 100 j + 10 k + 2s + 1  ~.
\end{equation}
An exception is provided by spin $1/2$ baryons made up of three
different types of quarks, where the two lightest quarks form a spin-0
diquark ($\Lambda$-like baryons). Here the order of the $j$ and $k$
quarks is reversed, so as to provide a simple means of distinction
to baryons with the lightest quarks in a spin-1 diquark
($\Sigma$-like baryons).
 
For hadrons with heavy flavours, the root names are Lambda or
Sigma for hadrons with two $\u$ or $\d$ quarks, Xi for those
with one, and Omega for those without $\u$ or $\d$ quarks.
 
Some of the most frequently used codes are given in Table
\ref{t:codeseven}. The full lowest-lying spin $1/2$ and $3/2$
multiplets are included in the program.
 
The corresponding KC codes, used for organizing mass and decay data,
range between 301 and 400, with some slots still free.
 
\begin{table}[ptb]
\captive{Baryon codes.
\protect\label{t:codeseven} }  \\
\vspace{1ex}
\begin{center}
\begin{tabular}{|c|c|c||c|c|c|@{\protect\rule{0mm}{\tablinsep}}}
\hline
KF & Name & Printed & KF & Name & Printed \\
\hline
      &  &                            &
 1114 & $\Delta^-$ & \ttt{Delta-} \\
 2112 & $\n$ & \ttt{n0}                     &
 2114 & $\Delta^0$ & \ttt{Delta0} \\
 2212 & $\p$ & \ttt{p+}                     &
 2214 & $\Delta^+$ & \ttt{Delta+} \\
      &  &                            &
 2224 & $\Delta^{++}$ & \ttt{Delta++} \\
 3112 & $\Sigma^-$ & \ttt{Sigma-}           &
 3114 & $\Sigma^{*-}$ & \ttt{Sigma*-} \\
 3122 & $\Lambda^0$ & \ttt{Lambda0}         & & &  \\
 3212 & $\Sigma^0$ & \ttt{Sigma0}           &
 3214 & $\Sigma^{*0}$ & \ttt{Sigma*0} \\
 3222 & $\Sigma^+$ & \ttt{Sigma+}           &
 3224 & $\Sigma^{*+}$ & \ttt{Sigma*+} \\
 3312 & $\Xi^-$ & \ttt{Xi-}                 &
 3314 & $\Xi^{*-}$ & \ttt{Xi*-} \\
 3322 & $\Xi^0$ & \ttt{Xi0}                 &
 3324 & $\Xi^{*0}$ & \ttt{Xi*0} \\
      &  &                            &
 3334 & $\Omega^-$ & \ttt{Omega-} \\
 4112 & $\Sigma_{\c}^0$ & \ttt{Sigma\_c0}   &
 4114 & $\Sigma_{\c}^{*0}$ & \ttt{Sigma*\_c0} \\
 4122 & $\Lambda_{\c}^+$ & \ttt{Lambda\_c+} & & &  \\
 4212 & $\Sigma_{\c}^+$ & \ttt{Sigma\_c+}   &
 4214 & $\Sigma_{\c}^{*+}$ & \ttt{Sigma*\_c+} \\
 4222 & $\Sigma_{\c}^{++}$ & \ttt{Sigma\_c++} &
 4224 & $\Sigma_{\c}^{*++}$ & \ttt{Sigma*\_c++} \\
 4132 & $\Xi_{\c}^0$ & \ttt{Xi\_c0}         & & &  \\
 4312 & $\Xi'^0_{\c}$ & \ttt{Xi'\_c0}       &
 4314 & $\Xi_{\c}^{*0}$ & \ttt{Xi*\_c0}  \\
 4232 & $\Xi_{\c}^+$ & \ttt{Xi\_c+}         & & &  \\
 4322 & $\Xi'^+_{\c}$ & \ttt{Xi'\_c+}        &
 4324 & $\Xi_{\c}^{*+}$ & \ttt{Xi*\_c+}  \\
 4332 & $\Omega_{\c}^0$ & \ttt{Omega\_c0}   &
 4334 & $\Omega_{\c}^{*0}$ & \ttt{Omega*\_c0}  \\
 5112 & $\Sigma_{\b}^-$ & \ttt{Sigma\_b-}     &
 5114 & $\Sigma_{\b}^{*-}$ & \ttt{Sigma*\_b-}  \\
 5122 & $\Lambda_{\b}^0$ & \ttt{Lambda\_b0} & & &  \\
 5212 & $\Sigma_{\b}^0$ &\ttt{Sigma\_b0}   &
 5214 & $\Sigma_{\b}^{*0}$ & \ttt{Sigma*\_b0}  \\
 5222 & $\Sigma_{\b}^+$ & \ttt{Sigma\_b+}   &
 5224 & $\Sigma_{\b}^{*+}$ &  \ttt{Sigma*\_b+}  \\
\hline
\end{tabular}
\end{center}
\end{table}
 
\item Diffractive states, Table \ref{t:codeeight}.  \\
These codes are not standard ones: they have been defined by analogy
to be used for denoting diffractive states in {\Py},
as part of the event history. The first two or three digits give
flavour content, while the last one is 0, to denote the somewhat
unusual character of the code. Only a few codes have been introduced;
depending on circumstances these also have to double up for other
diffractive states.
 
\begin{table}[ptb]
\captive{Diffractive state codes.
\protect\label{t:codeeight} }  \\
\vspace{1ex}
\begin{center}
\begin{tabular}{|c|c|c|@{\protect\rule{0mm}{\tablinsep}}}
\hline
KF & Printed & Meaning \\
\hline
  110 & \ttt{rho\_diff0} & Diffractive 
$\pi^0 / \rho^0 / \gamma$ state \\
  210 & \ttt{pi\_diffr+} & Diffractive $\pi^+$ state \\
  220 & \ttt{omega\_di0} & Diffractive $\omega$ state \\
  330 & \ttt{phi\_diff0} & Diffractive $\phi$ state \\
  440 & \ttt{J/psi\_di0} & Diffractive $\Jpsi$ state \\
 2110 & \ttt{n\_diffr}   & Diffractive $\n$ state \\
 2210 & \ttt{p\_diffr+}  & Diffractive $\p$ state \\
\hline
\end{tabular}
\end{center}
\end{table}
 
\item Free compressed codes.
The positions 401--500 of mass and decay arrays are left open.
Here a user may map any new kind of particle from the ordinary
KF codes, which probably are above 10000, into a more manageable
KC range for mass and decay data information. The mapping must be
implemented in the \ttt{LUCOMP} function.
 
\end{Enumerate}
 
\subsection{The Event Record}
\label{ss:evrec}
 
Each new event generated is in its entirety stored in the common block
\ttt{LUJETS}, which thus forms the event record. Here each jet or
particle that appears at some stage of the fragmentation or decay
chain will occupy one line in the matrices. The different components
of this line will tell which jet/particle it is, from where it
originates, its present status (fragmented/decayed or not), its
momentum, energy and mass, and the space--time position of its
production vertex. Note that \ttt{K(I,3)}--\ttt{K(I,5)} and the
\ttt{P} and \ttt{V} vectors may take special meaning for some
specific applications (e.g. sphericity or cluster
analysis), as described in those connections.

The event history information stored in \ttt{K(I,3)}--\ttt{K(I,5)}
should not be taken too literally. In the particle decay chains, the
meaning of a mother is well-defined, but the fragmentation description
is more complicated. The primary hadrons produced in string
fragmentation come from the string as a whole, rather than from an
individual parton. Even when the string is not included in the history
(see \ttt{MSTU(16)}), the pointer from hadron to parton is deceptive.
For instance, in a $\q\g\qbar$ event, those hadrons are pointing towards
the $\q$ ($\qbar$) parton that were produced by fragmentation from that
end of the string, according to the random procedure used in the
fragmentation routine. No particles point to the $\g$. This assignment
seldom agrees with the visual impression, and is not intended to. 
 
The common block \ttt{LUJETS} has expanded with time, and can now house
4000 entries. This figure may seem ridiculously large, but actually the
previous limit of 2000 was often reached in studies of high-$\pT$
processes at the LHC and SSC. This is because the event record
contains not only the final particles, but also all intermediate partons
and hadrons, which subsequenty showered, fragmented or decayed. Included
are also a wealth of photons coming from $\pi^0$ decays; the simplest
way of reducing the size of the event record is actually to switch off
$\pi^0$ decays by \ttt{MDCY(LUCOMP(111),1)=0}. Also note that some
routines, such as \ttt{LUCLUS} and \ttt{LUCELL}, use memory after the
event record proper as a working area. Still, to change the size of
the common block, upwards or downwards, is easy: just do a global
substitute in the common block and change the \ttt{MSTU(4)} value to the
new number. If more than 10000 lines are to be used, the packing of
colour information should also be changed, see \ttt{MSTU(5)}.
 
\drawbox{COMMON/LUJETS/N,K(4000,5),P(4000,5),V(4000,5)}\label{p:LUJETS}
 
\begin{entry}
 
\itemc{Purpose:} to contain the event record, i.e. the complete list
of all partons and particles in the current event.
 
\iteme{N :}\label{p:N}  number of lines in the \ttt{K}, \ttt{P} and 
\ttt{V} matrices occupied by the current event. \ttt{N} is continuously
updated as the definition of the original configuration and the
treatment of fragmentation and decay proceed. In the following,
the individual parton/particle number, running between 1 and
\ttt{N}, is called \ttt{I}.
 
\iteme{K(I,1) :}\label{p:K} status code KS, which gives the current 
status of the parton/particle stored in the line. The ground rule is 
that codes 1--10 correspond to currently existing partons/particles, 
while larger codes contain partons/particles which no longer exist, 
or other kinds of event information.
\begin{subentry}
\iteme{= 0 :} empty line.
\iteme{= 1 :} an undecayed particle or an unfragmented jet, the latter
being either a single jet or the last one of a jet system.
\iteme{= 2 :} an unfragmented jet, which is followed by more jets in the
same colour-singlet jet system.
\iteme{= 3 :} an unfragmented jet with special colour flow information
stored in \ttt{K(I,4)} and \ttt{K(I,5)}, such that adjacent partons
along the string need not follow each other in the event record.
\iteme{= 4 :} a particle which could have decayed, but did not
within the allowed volume around the original vertex.
\iteme{= 5 :} a particle which is to be forced to decay in the next
\ttt{LUEXEC} call, in the vertex position given (this code is only
set by user intervention).
\iteme{= 11 :} a decayed particle or a fragmented jet, the latter 
being either a single jet or the last one of a jet system, cf. \ttt{=1}.
\iteme{= 12 :} a fragmented jet, which is followed by more jets in the
same colour-singlet jet system, cf. \ttt{=2}. Further, a $\B$ meson
which decayed as a $\br{\B}$ one, or vice versa, because of
$\B$--$\br{\B}$ mixing, is marked with this code rather than
\ttt{=11}.
\iteme{= 13 :} a jet which has been removed when special colour flow
information has been used to rearrange a jet system, cf. \ttt{=3}.
\iteme{= 14 :} a parton which has branched into further partons, with
special colour-flow information provided, cf. \ttt{=3}.
\iteme{= 15 :} a particle which has been forced to decay (by user
intervention), cf. \ttt{=5}.
\iteme{= 21 :} documentation lines used to give a compressed story of
the event at the beginning of the event record.
\iteme{= 31 :} lines with information on sphericity, thrust or cluster
search.
\iteme{= 32 :} tabular output, as generated by \ttt{LUTABU}.
\iteme{= 41 :} junction (currently not fully implemented).
\iteme{< 0 :} these codes are never used by the program, and are
therefore usually not affected by operations on the record, such as
\ttt{LUROBO}, \ttt{LULIST} and event-analysis routines (the exception
is some \ttt{LUEDIT} calls, where lines are moved but not deleted).
Such codes may therefore be useful in some connections.
\end{subentry}
 
\iteme{K(I,2) :} parton/particle KF code, as described in section
\ref{ss:codes}.
 
\iteme{K(I,3) :} line number of parent particle or jet, where known,
otherwise 0. Note that the assignment of a particle to a given jet in a
jet system is unphysical, and what is given there is only related to
the way the event was generated.
 
\iteme{K(I,4) :} normally the line number of the first daughter;
it is 0 for an undecayed particle or unfragmented jet.
 
For \ttt{K(I,1) = 3, 13} or \ttt{14}, instead, it  contains special
colour-flow information (for internal use only) of the form \\
\ttt{K(I,4)} = 200000000*MCFR + 100000000*MCTO + 10000*ICFR + ICTO, \\
where ICFR and ICTO give the line numbers of the partons from which
the colour comes and to where it goes, respectively; MCFR and
MCTO originally are 0 and are set to 1 when the corresponding
colour connection has been traced in the \ttt{LUPREP} rearrangement
procedure. (The packing may be changed with \ttt{MSTU(5)}.)
The `from' colour position may indicate a parton which branched
to produce the current parton, or a parton created together with
the current parton but with matched anticolour, while the `to'
normally indicates a parton that the current parton branches
into. Thus, for setting up an initial colour configuration, it
is normally only the `from' part that is used, while the `to' part
is added by the program in a subsequent call to parton-shower
evolution (for final-state radiation; it is the other way around
for initial-state radiation).
 
{\bf Note:} normally most users never have to worry about the exact
rules for colour-flow storage, since this is used mainly for
internal purposes. However, when it is necessary to define this
flow, it is recommended to use the \ttt{LUJOIN} routine, since it is
likely that this would reduce the chances of making a mistake.
 
\iteme{K(I,5) :} normally the line number of the last daughter;
it is 0 for an undecayed particle or unfragmented jet.
 
For \ttt{K(I,1) = 3, 13} or \ttt{14}, instead, it contains special
colour-flow information (for internal use only) of the form \\
\ttt{K(I,5)} = 200000000*MCFR + 100000000*MCTO + 10000*ICFR + ICTO, \\
where ICFR and ICTO give the line numbers of the partons from which
the anticolour comes and to where it goes, respectively; MCFR
and MCTO originally are 0 and are set to 1 when the corresponding
colour connection has been traced in the \ttt{LUPREP} rearrangement
procedure. For further discussion, see \ttt{K(I,4)}.
 
\iteme{P(I,1) :}\label{p:P} $p_x$, momentum in the $x$ direction, 
in GeV/$c$.
 
\iteme{P(I,2) :} $p_y$, momentum in the $y$ direction, in GeV/$c$.
 
\iteme{P(I,3) :} $p_z$, momentum in the $z$ direction, in GeV/$c$.
 
\iteme{P(I,4) :} $E$, energy, in GeV.
 
\iteme{P(I,5) :} $m$, mass, in GeV/$c^2$. In parton showers, with
space-like virtualities, i.e. where $Q^2 = - m^2 > 0$,
one puts \ttt{P(I,5)}$ = -Q$.
 
\iteme{V(I,1) :}\label{p:V} $x$ position of production vertex, in mm.
 
\iteme{V(I,2) :} $y$ position of production vertex, in mm.
 
\iteme{V(I,3) :} $z$ position of production vertex, in mm.
 
\iteme{V(I,4) :} time of production, in mm/$c$
($\approx 3.33 \times 10^{-12}$ s).
 
\iteme{V(I,5) :} proper lifetime of particle, in mm/$c$
($\approx 3.33 \times 10^{-12}$ s). If the particle is not expected to
decay, \ttt{V(I,5)=0}. A line with \ttt{K(I,1)=4}, i.e. a
particle that could have decayed, but did not within the
allowed region, has the proper non-zero \ttt{V(I,5)}.
 
In the absence of electric or magnetic fields, or other
disturbances, the decay vertex \ttt{VP} of an unstable particle
may be calculated as \\
\ttt{VP(j) = V(I,j) + V(I,5)*P(I,j)/P(I,5)},
\ttt{j} = 1--4.
 
\end{entry}
 
\subsection{How The Event Record Works}
 
The event record is the main repository for information about an
event. In the generation chain, it is used as a `scoreboard' for
what has already been done and what remains to be done.
This information can be studied by you, to access information
not only about the final state, but also about what came before.
 
\subsubsection{A simple example}
 
The example of section \ref{ss:JETstarted} may help to clarify what
is going on. When \ttt{LU2ENT} is called to generate a $\q\qbar$
pair, the quarks are stored in lines 1 and 2 of the event record,
respectively. Colour information is set to show that they belong
together as a colour singlet. The counter \ttt{N} is also updated
to the value of 2. At no stage is the previously generated event
removed. Lines 1 and 2 are overwritten, but lines 3 
onwards still contain whatever may have been there before. This does
not matter, since \ttt{N} indicates where the `real' record ends.
 
As \ttt{LUEXEC} is called, explicitly by you or indirectly
by \ttt{LU2ENT}, the first entry is considered and found to be
the first jet of a system. Therefore the second entry is also
found, and these two together form a jet system, which may be
allowed to fragment. The `string' that fragments is put in line 3
and the fragmentation products in lines 4 through 10 (in this
particular case). At the same time, the $\q$ and $\qbar$ in the
first two lines are marked as having fragmented, and the same for
the string. At this stage, \ttt{N} is 10. Internally there is
another counter with the value 2, which indicates how far down
in the record the event has been studied.
 
This second counter is gradually increased by one. If the entry in
the corresponding line can fragment or decay, then fragmentation or
decay is perfomed.
The fragmentation/decay products are added at the end of the event
record, and \ttt{N} is updated accordingly. The entry is then also
marked as having been treated. For instance, when line 3 is
considered, the `string' entry of this line is seen to have been
fragmented,
and no action is taken. Line 4, a $\rho^+$, is allowed to decay to
$\pi^+ \pi^0$; the decay products are stored in lines 11 and 12,
and line 4 is marked as having decayed. Next, entry 5 is allowed to
decay. The entry in line 6, $\pi^+$, is a stable particle (by
default) and is therefore passed by without any action being taken.
 
In the beginning of the process, entries are usually unstable, and
\ttt{N} grows faster than the second counter of treated entries.
Later on, an increasing fraction of the entries are stable end
products, and the r\^oles are now reversed, with the second counter
growing faster. When the two coincide, the end of the record has been
reached, and the process can be stopped. All unstable objects have
now been allowed to fragment or decay. They are still present in the
record, so as to simplify the tracing of the history.
 
Notice that \ttt{LUEXEC} could well be called a second time.
The second counter would then start all over from the beginning, but
slide through until the end without causing any action, since
all objects that can be treated already have been.
Unless some of the relevant switches were changed meanwhile, that
is. For instance, if $\pi^0$ decays were switched off the first time
around but on the second, all the $\pi^0$'s found in the record
would be allowed to decay in the second call. A particle once
decayed is not `undecayed', however, so if the $\pi^0$ is put back
stable and \ttt{LUEXEC} is called a third time, nothing will happen.
 
\subsubsection{Applications to PYTHIA}
\label{sss:PYrecord}
 
In a full-blown event generated with {\Py}, the usage of \ttt{LUJETS}
is more complicated, although the general principles survive.
\ttt{LUJETS} is used extensively both by the {\Py} and the {\Je}
routines; indeed it provides the bridge that allows
the general utility routines in {\Je} to be used
also for {\Py} events. The {\Py} event listing begins (optionally)
with a few lines of event summary, specific to the hard process
simulated and thus not described in the overview above. These
specific parts are covered in the following.
 
In most instances, only the partons and particles actually produced
are of interest. For \ttt{MSTP(125)=0}, the event record starts
off with the parton configuration existing after hard interaction,
initial- and final-state radiation, multiple interactions and beam
remnants have been considered. The partons are arranged in colour
singlet clusters, ordered as required for string fragmentation.
Also photons and leptons produced as part of the hard interaction
(e.g. from $\q\qbar \to \g \gamma$ or $\u\ubar \to \Z^0 \to \ee$)
appear in this part of the event record. These original entries
appear with pointer \ttt{K(I,3)=0}, whereas the products of the
subsequent fragmentation and decay have \ttt{K(I,3)} numbers 
pointing back to the line of the parent.
 
The standard documentation, obtained with \ttt{MSTP(125)=1},
includes a few lines at the beginning of the event record, which
contain a brief summary of the process that has taken place. The
number of lines used depends on the nature of the hard process
and is stored in \ttt{MSTI(4)} for the current event. These lines
all have \ttt{K(I,1)=21}. For all processes, lines 1 and 2 give
the two incoming hadrons. When listed with \ttt{LULIST}, these two
lines will be separated from subsequent ones by a sequence of
`\ttt{======}' signs, to improve readability. For diffractive and
elastic events, the two outgoing states in lines 3 and 4 complete the
list. Otherwise, lines 3 and 4 contain the two partons that initiate
the two initial-state parton showers, and 5 and 6 the end products of
these showers, i.e. the partons that enter the hard interaction. With
initial-state radiation switched off, lines 3 and 5 and lines 4 and 6
coincide. For a simple $2 \to 2$ hard scattering, lines 7 and 8 give
the two outgoing partons/particles from the hard  interaction, before
any final-state radiation. For $2 \to 2$ processes proceeding via an
intermediate resonance such as $\gammaZ$, $\W^{\pm}$ or $\H^0$, the
resonance is found in line 7 and the two outgoing partons/particles in
8 and 9. In some cases one of these may be a resonance in its own 
right, or both of them, so that further pairs of lines are added for
subsequent decays. If the decay of a given resonance has been
switched off, then no decay products are listed either in this
initial summary or in the subsequent ordinary listing. Whenever partons
are listed, they are assumed to be on the mass shell for simplicity.
The fact that effective masses may be generated by initial-
and final-state radiation is taken into account in the actual parton
configuration that is allowed to fragment, however. A special case is
provided by $\W^+ \W^-$ or $\Z^0 \Z^0$ fusion to an $\H^0$. Then the
virtual $\W$'s or $\Z$'s are shown in lines 7 and 8, the $\H^0$ in
line 9, and the two recoiling quarks (that emitted the bosons) in 10
and 11, followed by the Higgs decay products. Since the $\W$'s and
$\Z$'s are space-like, what is actually listed as the mass for them
is $-\sqrt{-m^2}$. The listing of the event documentation closes with
another line made up of `\ttt{======}' signs.
 
A few examples may help clarify the picture. For a single diffractive
event $\p \pbar \to \p_{\mrm{diffr}} \pbar$, the event record will start
with \\
\verb& I K(I,1)   K(I,2) K(I,3) &  comment  \\
\verb& 1     21     2212      0 &  incoming $\p$ \\
\verb& 2     21    -2212      0 &  incoming $\pbar$ \\
\verb&========================= &  not part of record; appears in 
listings \\
\verb& 3     21       27      1 &  outgoing $\p_{\mrm{diffr}}$ \\
\verb& 4     21    -2212      2 &  outgoing $\pbar$ \\
\verb&========================= &  again not part of record 
 
The typical QCD $2 \to 2$ process would be \\
\verb& I K(I,1)   K(I,2) K(I,3) &  comment \\
\verb& 1     21     2212      0 &  incoming $\p$ \\
\verb& 2     21    -2212      0 &  incoming $\pbar$ \\
\verb&========================= & \\
\verb& 3     21        2      1 &  $\u$ picked from incoming $\p$ \\
\verb& 4     21       -1      2 &  $\dbar$ picked from incoming 
$\pbar$ \\
\verb& 5     21       21      3 &  $\u$ evolved to $\g$ at hard 
scattering \\
\verb& 6     21       -1      4 &  still $\dbar$ at hard scattering \\
\verb& 7     21       21      0 &  outgoing $\g$ from hard 
scattering \\
\verb& 8     21       -1      0 &  outgoing $\dbar$ from hard 
scattering \\
\verb&========================= & 

Note that, where well defined, the \ttt{K(I,3)} code does contain
information as to which side the different partons come from, e.g.
above the gluon in line 5 points back to the $\u$ in line 3,
which points back to the proton in line 1. In the example above, it
would have been possible to associate the scattered g in line 7
with the incoming one in line 5, but this is not possible in the
general case, consider e.g. $\g \g \to \g \g$.
As a final example, $\W^+\W^-$ fusion to an $\H^0$ in process 8 
(not process 124, which is lengthier) might look like \\
\verb& I K(I,1)   K(I,2) K(I,3) &  comment \\
\verb& 1     21     2212      0 &  first incoming $\p$ \\
\verb& 2     21     2212      0 &  second incoming $\p$ \\
\verb&========================= &  \\
\verb& 3     21        2      1 &  $\u$ picked from first $\p$ \\
\verb& 4     21       21      2 &  $\g$ picked from second $\p$ \\
\verb& 5     21        2      3 &  still $\u$ after initial-state 
radiation \\
\verb& 6     21       -4      4 &  $\g$ evolved to $\cbar$ \\
\verb& 7     21       24      5 &  space-like $\W^+$ emitted by $\u$ 
quark  \\
\verb& 8     21      -24      6 &  space-like $\W^-$ emitted by 
$\cbar$ quark \\
\verb& 9     21       25      0 &  Higgs produced by $\W^+ \W^-$ 
fusion \\
\verb&10     21        1      5 &  $\u$ turned into $\d$ by emission 
of $\W^+$ \\
\verb&11     21       -3      6 &  $\cbar$ turned into $\sbar$ by 
emission of $\W^-$ \\
\verb&12     21       23      9 &  first $\Z^0$ coming from decay 
of $\H^0$ \\
\verb&13     21       23      9 &  second $\Z^0$ coming from decay 
of $\H^0$ \\
\verb&14     21       12     12 &  $\nu_{\e}$ from first $\Z^0$ 
decay \\
\verb&15     21      -12     12 &  $\br{\nu}_{\e}$ from first 
$\Z^0$ decay  \\
\verb&16     21        5     13 &  $\b$ quark from second $\Z^0$ 
decay  \\
\verb&17     21       -5     13 &  $\bbar$ antiquark from second 
$\Z^0$ decay  \\
\verb&========================= &
 
After these lines with the initial information, the event record looks
the same as for \ttt{MSTP(125)=0}, i.e. first comes the parton
configuration to be fragmented and, after another separator line
`\ttt{======}' in the output (but not the event record), the products
of subsequent fragmentation and decay chains. The \ttt{K(I,3)}
pointers for the partons, as well as leptons and photons produced
in the hard interaction, are now pointing towards the documentation
lines above, however. In particular, beam remnants point to 1 or 2,
depending on which side they belong to, and partons emitted in the
initial-state parton showers point to 3 or 4. In the second example
above, the partons produced by final-state radiation will be pointing
back to 7 and 8; as usual, it should be remembered that a specific
assignment to 7 or 8 need not be unique. For the third example, 
final-state radiation partons will come both from partons 10 and 11 
and from partons 16 and 17, and additionally there will be a
neutrino--antineutrino pair pointing to 14 and 15. The extra
pairs of partons that are generated by multiple interactions do not
point back to anything, i.e. they have \ttt{K(I,3)=0}.
 
There exists a third documentation option, \ttt{MSTP(125)=2}. Here
the history of initial- and final-state parton branchings may be traced,
including all details on colour flow. This information has not been
optimized for user-friendliness, and cannot be recommended for
general usage. With this option, the initial documentation lines
are the same. They are followed by blank lines, \ttt{K(I,1)=0}, up to
line 20 (can be changed in \ttt{MSTP(126)}). From line 21 onwards
each parton with \ttt{K(I,1)=} 3, 13 or 14 appears with special
colour-flow information in the \ttt{K(I,4)} and \ttt{K(I,5)}
positions. For an ordinary $2 \to 2$ scattering, the two incoming
partons at the hard scattering are stored in lines 21 and 22, and the
two outgoing in 23 and 24. The colour flow between these partons has
to be chosen according to the proper relative probabilities in
cases when many alternatives are possible, see section 
\ref{sss:QCDjetclass}.
If there is initial-state radiation, the two partons in lines 21 and 
22 are copied down to lines 25 and 26, from which the initial-state 
showers are reconstructed backwards step by step. The branching
history may be read by noting that, for a branching $a \to b c$,
the \ttt{K(I,3)} codes of $b$ and $c$ point towards the line number
of $a$. Since the showers are reconstructed backwards, this actually
means that parton $b$ would appear in the listing before parton
$a$ and $c$, and hence have a pointer to a position below
itself in the list. Associated time-like partons $c$ may initiate
time-like showers, as may the partons of the hard scattering. Again
a showering parton or pair of partons will be copied down towards
the end of the list and allowed to undergo successive branchings
$c \to d e$, with $d$ and $e$ pointing towards $c$. The mass of
time-like partons is properly stored in \ttt{P(I,5)}; for space-like
partons $-\sqrt{-m^2}$ is stored instead. After this
section, containing all the branchings, comes the final parton
configuration, properly arranged in colour, followed by all
subsequent fragmentation and decay products, as usual.
 
\subsection{The HEPEVT Standard}
\label{ss:HEPEVT}
 
A set of common blocks was developed and agreed on within the
framework of the 1989 LEP physics study, see \cite{Sjo89}.
This standard defines an event record structure which should make
the interfacing of different event generators much simpler.
 
It would be a major work to rewrite {\PyJe} to agree with this
standard event record structure. More importantly, the standard
only covers quantities which can be defined unambiguously, i.e.
which are independent of the particular program used. There are
thus no provisions for the need for colour-flow information in
models based on string fragmentation, etc., so the standard
common blocks would anyway have to be supplemented with additional
event information. For the moment, the adopted approach is therefore
to retain the \ttt{LUJETS} event record, but supply a routine
\ttt{LUHEPC} which can convert to or from the standard event record.
Owing to a somewhat different content in the two records, some
ambiguities do exist in the translation procedure. \ttt{LUHEPC}
has therefore to be used with some judgment.
 
In this section, the new standard event structure is first presented,
i.e. the most important points in \cite{Sjo89} are recapitulated.
Thereafter the conversion routine is described, with particular
attention to ambiguities and limitations.
 
The standard event record is stored in two common blocks. The second
of these is specifically intended for spin information. Since {\Je}
never (explicitly) makes use of spin information, this latter
common block is not addressed here. A third common block for colour
flow information has been discussed, but never formalized.
 
In order to make the components of the standard more distinguishable
in user programs, the three characters \ttt{HEP} (for High Energy
Physics) have been chosen to be a part of all names.

Originally it was not specified whether real variables should be in 
single or double precision. At the time, this meant that single 
precision became the default choice, but since then the trend has been
towards increasing precision. In connection with the 1995 LEP~2
workshop, it was therefore agreed to adopt \ttt{DOUBLE PRECISION} 
real variables as part of the standard. 
 
\drawboxfour{~PARAMETER (NMXHEP=2000)}
{~COMMON/HEPEVT/NEVHEP,NHEP,ISTHEP(NMXHEP),IDHEP(NMXHEP),}
{\&JMOHEP(2,NMXHEP),JDAHEP(2,NMXHEP),PHEP(5,NMXHEP),VHEP(4,NMXHEP)}
{~DOUBLE PRECISION PHEP, VHEP}
\label{p:HEPEVT}\begin{entry}
 
\itemc{Purpose:} to contain an event record in a
Monte Carlo-independent format.
 
\iteme{NMXHEP:} maximum numbers of entries (partons/particles) that can
be stored in the common block. The default value of 2000 can be changed
via the parameter construction. In the translation, it is
checked that this value is not exceeded.
 
\iteme{NEVHEP:} is normally the event number, but may have special 
meanings, according to the description below:
\begin{subentry}
\iteme{> 0 :} event number, sequentially increased by 1 for each call 
to the main event generation routine, starting with 1 for the
first event generated.
\iteme{= 0 :} for a program which does not keep track of event numbers,
as {\Je}.
\iteme{= -1 :} special initialization record; not used by {\Je}.
\iteme{= -2 :} special final record; not used by {\Je}.
\end{subentry}
 
\iteme{NHEP:} the actual number of entries stored in the current event.
These are found in the first \ttt{NHEP} positions of the respective
arrays below. Index \ttt{IHEP}, 1$\leq$\ttt{IHEP}$\leq$\ttt{NHEP},
is used below to denote a given entry.
 
\iteme{ISTHEP(IHEP):} status code for entry \ttt{IHEP}, with the 
following meanings:
\begin{subentry}
\iteme{= 0 :} null entry.
\iteme{= 1 :} an existing entry, which has not decayed or fragmented.
This is the main class of entries, which represents the
`final state' given by the generator.
\iteme{= 2 :} an entry which has decayed or fragmented and is 
therefore not appearing in the final state, but is retained for
event history information.
\iteme{= 3 :} a documentation line, defined separately from the event
history. This could include the two incoming reacting particles, etc.
\iteme{= 4 - 10 :} undefined, but reserved for future standards.
\iteme{= 11 - 200 :} at the disposal of each model builder for
constructs specific to his program, but equivalent to a null line
in the context of any other program.
\iteme{= 201 - :} at the disposal of users, in particular for event
tracking in the detector.
\end{subentry}
 
\iteme{IDHEP(IHEP) :} particle identity, according to the PDG
standard. The four additional codes 91--94 have been introduced
to make the event history more legible, see section \ref{ss:codes}
and the \ttt{MSTU(16)} description.
 
\iteme{JMOHEP(1,IHEP) :} pointer to the position where the mother
is stored. The value is 0 for initial entries.
 
\iteme{JMOHEP(2,IHEP) :} pointer to position of second mother.
Normally only one mother exists, in which case the value 0 is to be
used. In {\Je}, entries with codes 91--94 are the only ones to have
two mothers. The flavour contents of these objects, as well as
details of momentum sharing, have to be found by looking at the
mother partons, i.e. the two partons in positions \ttt{JMOHEP(1,IHEP)}
and \ttt{JMOHEP(2,IHEP)} for a cluster or a shower system, and the
range
\ttt{JMOHEP(1,IHEP)}--\ttt{JMOHEP(2,IHEP)} for a string or an
independent fragmentation parton system.
 
\iteme{JDAHEP(1,IHEP) :} pointer to the position of the first daughter.
If an entry has not decayed, this is 0.
 
\iteme{JDAHEP(2,IHEP) :} pointer to the position of the last daughter.
If an entry has not decayed, this is 0. It is assumed that daughters are
stored sequentially, so that the whole range
\ttt{JDAHEP(1,IHEP)}--\ttt{JDAHEP(2,IHEP)} contains daughters. This
variable should be set also when only one daughter is present, as in
$\K^0 \to \K_{\mrm{S}}^0$ decays, so that looping from the first 
daughter to the last one works transparently.
Normally daughters are stored after mothers, but in backwards
evolution of initial-state radiation the opposite may appear,
i.e. that mothers are found below the daughters they branch into.
Also, the two daughters then need not appear one after the other,
but may be separated in the event record.
 
\iteme{PHEP(1,IHEP) :} momentum in the $x$ direction, in GeV/$c$.
 
\iteme{PHEP(2,IHEP) :} momentum in the $y$ direction, in GeV/$c$.
 
\iteme{PHEP(3,IHEP) :} momentum in the $z$ direction, in GeV/$c$.
 
\iteme{PHEP(4,IHEP) :} energy, in GeV.
 
\iteme{PHEP(5,IHEP) :} mass, in GeV/$c^2$. For space-like partons,
it is allowed to use a negative mass, according to
\ttt{PHEP(5,IHEP)}$ = -\sqrt{-m^2}$.
 
\iteme{VHEP(1,IHEP) :} production vertex $x$ position, in mm.
 
\iteme{VHEP(2,IHEP) :} production vertex $y$ position, in mm.
 
\iteme{VHEP(3,IHEP) :} production vertex $z$ position, in mm.
 
\iteme{VHEP(4,IHEP) :} production time, in mm/$c$
($\approx 3.33 \times 10^{-12}$ s).
\end{entry}
\boxsep
 
This completes the brief description of the standard. In {\Je}, the
routine \ttt{LUHEPC} is provided as an interface.
 
\drawbox{CALL LUHEPC(MCONV)}\label{p:LUHEPC}
\begin{entry}
\itemc{Purpose:} to convert between the \ttt{LUJETS} event record and
the \ttt{HEPEVT} event record.
\iteme{MCONV :} direction of conversion.
\begin{subentry}
\iteme{= 1 :} translates the current \ttt{LUJETS} record into the
\ttt{HEPEVT} one, while leaving the original \ttt{LUJETS} one
unaffected.
\iteme{= 2 :} translates the current \ttt{HEPEVT} record into the
\ttt{LUJETS} one, while leaving the original \ttt{HEPEVT} one
unaffected.
\end{subentry}
\end{entry}
\boxsep
 
The conversion of momenta is trivial: it is just a matter of exchanging
the order of the indices. The vertex information is but little more
complicated; the extra fifth component present in \ttt{LUJETS} can be
easily
reconstructed from other information for particles which have decayed.
(Some of the advanced features made possible by this component, such as
the possibility to consider decays within expanding spatial volumes in
subsequent \ttt{LUEXEC} calls, cannot be used if the record is
translated back and forth, however.) Also, the particle codes
\ttt{K(I,2)} and \ttt{IDHEP(I)}
are identical, since they are both based on the PDG codes.
 
The remaining, non-trivial areas deal with the status codes and the 
event history. In moving from \ttt{LUJETS} to \ttt{HEPEVT}, 
information on colour flow is lost. On the other hand, the position 
of a second mother, if any, has to be found; this only affects lines 
with \ttt{K(I,2)=} 91--94. Also, for lines with \ttt{K(I,1)=} 13
or 14, the daughter pointers have to be found. By and large,
however, the translation from \ttt{LUJETS} to \ttt{HEPEVT}
should cause little problem, and there should never be any need for
user intervention. (We assume that {\Je} is run with the default
\ttt{MSTU(16)=1}, otherwise some discrepancies with respect to the 
proposed standard event history description will be present.)
 
In moving from \ttt{HEPEVT} to \ttt{LUJETS}, information on a second
mother is lost. Any codes \ttt{IDHEP(I)} not equal to 1, 2 or 3 are
translated into \ttt{K(I,1)=0}, and so all entries with
\ttt{K(I,1)}$\geq 30$ are effectively lost in a translation back and
forth. All entries with \ttt{IDHEP(I)=2} are translated
into \ttt{K(I,1)=11}, and so entries of type
\ttt{K(I,1) = 12, 13, 14} or \ttt{15} are never found. There is thus
no colour-flow information available for partons which have
fragmented. For partons with \ttt{IDHEP(I)=1},
i.e. which have not fragmented, an attempt is made to subdivide the
partonic system into colour singlets, as required for subsequent
string fragmentation. To this end, it is assumed that partons are
stored sequentially along strings. Normally, a string would then start
at a $\q$ ($\qbar$) or $\qbar\qbar$ ($\q\q$) entry, cover a number
of intermediate gluons, and end at a $\qbar$ ($\q$) or $\q\q$
($\qbar\qbar$) entry. Particles could be interspersed in this list with
no adverse effects, i.e. a $\u-\g-\gamma-\ubar$
sequence would be interpreted as a $\u-\g-\ubar$ string plus an
additional photon. A closed gluon loop would be assumed to be made up
of a sequential listing of the gluons, with the string continuing from
the last gluon up back to the first one. Contrary to the previous, open
string case, the appearance of any particle but a gluon would therefore
signal the end of the gluon loop. For example, a $\g-\g-\g-\g$ sequence
would be interpreted as one single four-gluon loop, while a
$\g-\g-\gamma-\g-\g$ sequence would be seen as composed of two
2-gluon systems.
 
If these interpretations, which are not unique, are not to your liking,
it is up to you to correct them, e.g. by using
\ttt{LUJOIN} to tell exactly which partons should be joined,
in which sequence, to give a string. Calls to \ttt{LUJOIN}
(or the equivalent) are also necessary if \ttt{LUSHOW} is to be used
to have some partons develop a shower.
 
For practical applications, one should note that {\Je} $\ee$ events,
which have been allowed to shower but not to fragment, do have partons
arranged in the order assumed above, so that a translation to
\ttt{HEPEVT} and back does not destroy the possibility to perform
fragmentation by a simple \ttt{LUEXEC} call. Also the hard interactions
in {\Py} fulfil this condition, while problems may appear in the
multiple interaction scenario, where several closed $\g\g$ loops may
appear directly following one another, and thus would be
interpreted as a single multigluon loop after translation back and
forth.
 
\clearpage
 
\section{Hard Processes in JETSET}
\label{s:JETSETproc}
 
{\Je} contains the simulation of two hard processes.
The process of main interest is $\ee \to \gammaZ \to \q \qbar$.
Higher-order QCD corrections can be obtained either with parton
showers or with second-order matrix elements. The details of the
parton-shower evolution are given in section \ref{s:showinfi},
while this section contains the matrix-element description, including
a summary of the {\Je} algorithm for initial-state
photon radiation. Also {\Py} can be used to simulate the
process $\ee \to \gammaZ \to \q \qbar$,
but without the options of using second-order matrix
elements or polarized incoming beams. Some other differences
between the two algorithms are described.
 
The other hard process in {\Je} is $\Upsilon$ decay to
$\g \g \g$ or $\gamma \g \g$, which is briefly commented on.
 
The main sources of information for this chapter are
refs. \cite{Sjo83,Sjo86,Sjo89}.
 
\subsection{Annihilation Events in the Continuum}
\label{ss:eematrix}
 
The description of $\ee$ annihilation into hadronic events involves a
number of components: the $s$ dependence of the total cross section
and flavour composition, multijet matrix elements, angular
orientation of events,  initial-state photon bremsstrahlung
and effects of initial-state electron polarization.
Many of the published formulae
have been derived for the case of massless outgoing quarks. For each
of the components described in the following, we will begin by
discussing the massless case, and then comment on what is done to
accommodate massive quarks.
 
\subsubsection{Electroweak cross sections}
 
In the standard theory, fermions have the following couplings
(illustrated here for the first generation):
\begin{center}
\begin{tabular}{lll@{\protect\rule{0mm}{\tablinsep}}}
$e_{\nu} = 0$, & $v_{\nu} = 1$, & $a_{\nu} = 1$, \\
$e_{\e} = -1$, & $v_{\e} = -1 + 4\ssintw$, & $a_{\e} = -1$, \\
$e_{\u} = 2/3$, & $v_{\u} = 1 - 8\ssintw /3$, & $a_{\nu} = 1$, \\
$e_{\d} = -1/3$, & $v_{\d} = -1 + 4\ssintw /3$, & $a_{\d} = -1$, \\
\end{tabular}
\end{center}
with $e$ the electric charge, and $v$ and $a$ the vector and axial
couplings to the $\Z^0$. The relative energy dependence of the weak
neutral current to the electromagnetic one is given by
\begin{equation}
   \chi(s) = \frac{1}{4\ssintw\scostw} \;
    \frac{s}{s - m_{\Z}^2 + i m_{\Z}\Gamma_{\Z}} ~,
\label{ee:chis}
\end{equation}
where $s = E_{\mrm{cm}}^2$.
In {\Je} the electroweak mixing parameter $\ssintw$ and the
$\Z^0$ mass $m_{\Z}$ and width $\Gamma_{\Z}$ are considered as
constants to be given by you (while
{\Py} itself calculates an $s$-dependent width).
 
Although the incoming $\e^+$ and $\e^-$ beams are normally
unpolarized, we have included the possibility of polarized beams,
following the formalism of \cite{Ols80}. Thus the incoming
$\e^+$ and $\e^-$ are characterized by polarizations
$\mbf{P}^{\pm}$ in the rest frame of the particles:
\begin{equation}
\mbf{P}^{\pm} = P_{\mrm{T}}^{\pm} \hat{\mbf{s}}^{\pm} + 
P_{\mrm{L}}^{\pm} \hat{\mbf{p}}^{\pm} ~,
\end{equation}
where $0 \leq P_{\mrm{T}}^{\pm} \leq 1$ and 
$-1 \leq P_{\mrm{L}}^{\pm} \leq 1$, with the constraint
\begin{equation}
(\mbf{P}^{\pm})^2 = (P_{\mrm{T}}^{\pm})^2 + (P_{\mrm{L}}^{\pm})^2 
\leq 1 ~.
\end{equation}
Here $\hat{\mbf{s}}^{\pm}$ are unit vectors perpendicular to the beam
directions $\hat{\mbf{p}}^{\pm}$. To be specific, we choose a
right-handed coordinate frame with 
$\hat{\mbf{p}}^{\pm} = (0,0, \mp 1)$,
and standard transverse polarization directions (out of the machine
plane for storage rings) $\hat{\mbf{s}}^{\pm} = (0, \pm 1,0)$, the
latter corresponding to azimuthal angles $\varphi^{\pm} = \pm \pi /2$.
As free parameters in the program we choose $P_{\mrm{L}}^+$, 
$P_{\mrm{L}}^-$, $P_{\mrm{T}} = \sqrt{P_{\mrm{T}}^+ P_{\mrm{T}}^-}$ 
and $\Delta \varphi = (\varphi^+ + \varphi^-) /2$.
 
In the massless QED case, the probability to produce a flavour $\f$ is
proportional to $e_{\f}^2$, i.e up-type quarks are four times as likely
as down-type ones. In lowest-order massless QFD the corresponding
relative probabilities are given by \cite{Ols80}
\begin{eqnarray}
   h_{\f}(s) & = & e_{\e}^2 \, (1 - P_{\mrm{L}}^+ P_{\mrm{L}}^-) 
   \, e_{\f}^2 \, + \, 2 e_{\e} \left\{ v_{\e} 
   (1 - P_{\mrm{L}}^+ P_{\mrm{L}}^-) - a_{\e} 
   (P_{\mrm{L}}^- - P_{\mrm{L}}^+)
   \right\} \, \Re\chi(s) \, e_{\f} v_{\f} \, +     \nonumber \\
   & &  + \, \left\{ (v_{\e}^2 + a_{\e}^2) (1 - P_{\mrm{L}}^+ 
   P_{\mrm{L}}^-) - 2 v_{\e} a_{\e} (P_{\mrm{L}}^- - P_{\mrm{L}}^+) 
   \right\} \,
   \left| \chi(s) \right|^2 \, \left\{ v_{\f}^2 + a_{\f}^2 \right\} ~,
\label{ee:hf}
\end{eqnarray}
where $\Re\chi(s)$ denotes the real part of $\chi(s)$.
The $h_{\f}(s)$ expression depends both on the $s$ value and on the 
longitudinal polarization of the $\e^{\pm}$ beams in a non-trivial way.
 
The cross section for the process $\ee \to \gammaZ \to \f \fbar$
may now be written as
\begin{equation}
  \sigma_{\f}(s) = \frac{4 \pi \alphaem^2}{3 s} R_{\f}(s) ~,
\end{equation}
where $R_{\f}$ gives the ratio to the lowest-order QED cross section for
the process $\ee \to \mu^+ \mu^-$,
\begin{equation}
   R_{\f}(s) = N_C \, R_{\mrm{QCD}} \, h_{\f}(s) ~.
\end{equation}
The factor of $N_C = 3$ counts the number of colour states available
for the $\q\qbar$ pair. The $R_{\mrm{QCD}}$
factor takes into account QCD loop corrections to the cross section.
For $n_f$ effective flavours (normally $n_f =5$)
\begin{equation}
 R_{\mrm{QCD}} \approx 1 + \frac{\alphas}{\pi} + (1.986 - 0.115 n_f)
 \left( \frac{\alphas}{\pi} \right)^2 + \cdots  
\label{ee:RQCD}
\end{equation}
in the $\br{\mrm{MS}}$ renormalization scheme \cite{Din79}.
Note that $R_{\mrm{QCD}}$ does not affect the relative quark-flavour 
composition, and so is of peripheral interest in {\Je}.
(For leptons the $N_C$ and $R_{\mrm{QCD}}$ factors would be absent, 
i.e. $N_C \, R_{\mrm{QCD}} = 1$, but leptonic final states are not 
generated in {\Je}.)
 
Neglecting higher-order QCD and QFD effects, the corrections for
massive quarks are given in terms of the velocity $v_{\q}$ of a quark
with mass $m_{\q}$, $v_{\q} = \sqrt{ 1 - 4 m_{\q}^2 /s}$, as follows.
The vector quark current terms in $h_{\f}$ (proportional to
$e_{\f}^2$, $e_{\f} v_{\f}$, or $v_{\f}^2$) are multiplied by a
threshold factor $v_{\q} (3 - v_{\q}^2) /2$, while the axial
vector quark current term (proportional to $a_{\f}^2$) is
multiplied by $v_{\q}^3$. While inclusion of quark masses in the
QFD formulae decreases the total cross section, first-order QCD
corrections tend in the opposite direction \cite{Jer81}. Na\"{\i}vely,
one would expect one factor of $v_{\q}$ to get cancelled. So far,
the available options are either to include threshold factors
in full or not at all.
 
Given that all five quarks are light at
the scale of the $\Z^0$, the issue of quark masses is not really
of interest at LEP. Here, however, purely weak corrections are
important, in particular since they change the $\b$ quark
partial width differently from that of the other ones \cite{Kuh89}.
No such effects are included in the program.
 
\subsubsection{First-order QCD matrix elements}
 
The Born process $\ee \to \q \qbar$ is modified in first-order 
QCD by the probability for the $\q$ or $\qbar$ to
radiate a gluon, i.e. by the process $\ee \to \q \qbar \g$.
The matrix element is conveniently given in terms of scaled energy
variables in the c.m. frame of the event, 
$x_1 = 2E_{\q}/E_{\mrm{cm}}$,
$x_2 = 2E_{\qbar}/E_{\mrm{cm}}$, 
and $x_3 = 2E_{\g}/E_{\mrm{cm}}$,
i.e. $x_1 + x_2 + x_3 = 2$. For massless
quarks the matrix element reads \cite{Ell76}
\begin{equation}
 \frac{1}{\sigma_0} \, \frac{\d \sigma}{\d x_1 \, \d x_2} =
 \frac{\alphas}{2\pi} \, C_F \,
 \frac{x_1^2 + x_2^2}{(1-x_1)(1-x_2)} ~,
  \label{ee:ME3j}
\end{equation}
where $\sigma_0$ is the lowest-order cross section, $C_F = 4/3$ is the
appropriate colour factor, and
the kinematically allowed region is $0 \leq x_i \leq 1, i = 1, 2, 3$.
By kinematics, the $x_k$ variable for parton $k$ is related to the
invariant mass $m_{ij}$ of the other two partons $i$ and $j$ by
$y_{ij} = m_{ij}^2/E_{\mrm{cm}}^2 = 1 - x_k$.
 
The strong coupling constant $\alphas$ is in first order given by
\begin{equation}
 \alphas(Q^2) = \frac{12\pi}{(33-2n_f) \, \ln(Q^2/\Lambda^2)} ~.
 \label{ee:aS3j}
\end{equation}
Conventionally $Q^2 = s = E_{\mrm{cm}}^2$; we will return to this 
issue below.
The number of flavours $n_f$ is 5 for LEP applications, and so the
$\Lambda$ value determined is $\Lambda_5$  (while e.g. most
deep inelastic scattering studies refer to $\Lambda_4$,
the energies for these experiments being below the bottom threshold).
The $\alphas$ values are matched at flavour thresholds, i.e.
as $n_f$ is changed the $\Lambda$ value is also changed. It is
therefore the derivative of $\alphas$ that changes at a
threshold, not $\alphas$ itself.
 
In order to separate 2-jets from 3-jets, it is useful to
introduce jet-resolution parameters. This can be done in several
different ways. Most famous are the $y$ and $(\epsilon, \delta)$
procedures. We will only refer to the $y$ cut, which is the one
used in the program. Here a 3-parton configuration is called
a 2-jet event if
\begin{equation}
\min_{i,j} (y_{ij}) = \min_{i,j} \left( \frac{m_{ij}^2}{E_{\mrm{cm}}^2}
 \right) < y ~.
\end{equation}
 
The cross section in eq.~(\ref{ee:ME3j}) diverges for
$x_1 \rightarrow 1$ or $x_2 \rightarrow 1$ but, when
first-order propagator and vertex corrections are included,
a corresponding singularity with opposite sign appears in the
$\q \qbar$ cross section, so that the total cross section is finite.
In analytical calculations, the average value of any well-behaved
quantity ${\cal Q}$ can therefore be calculated as
\begin{equation}
   \left\langle {\cal Q} \right\rangle =
   \frac{1}{\sigma_{\mrm{tot}}} \lim_{y \rightarrow 0}
   \left( {\cal Q}(\mrm{2parton}) \, \sigma_{\mrm{2parton}}(y) +
   \int_{y_{ij} > y} {\cal Q}(x_1,x_2) \,
   \frac{\d \sigma_{\mrm{3parton}}}{\d x_1 \, \d x_2} \, 
   \d x_1 \, \d x_2 \right) ~,
  \label{ee:Obs}
\end{equation}
where any explicit $y$ dependence disappears in the limit
$y \rightarrow 0$.
 
In a Monte Carlo program, it is not possible to
work with a negative total 2-jet rate, and thus it is necessary to
introduce a fixed non-vanishing $y$ cut in the 3-jet
phase space. Experimentally, there is evidence for the need of a
low $y$ cut, i.e. a large 3-jet rate.
For LEP applications, the recommended value is $y = 0.01$,
which is about as far down as one can go and still retain a positive
2-jet rate. With $\alphas = 0.12$, in full second-order QCD
(see below), the $2:3:4$ jet composition is then approximately
$11 \% : 77 \% : 12 \%$.
 
Note, however, that initial-state QED radiation may
occasionally lower the c.m. energy significantly, i.e. increase
$\alphas$, and thereby bring the 3-jet fraction above unity
if $y$ is kept fixed at 0.01 also in those events. Therefore,
at PETRA/PEP energies, $y$ values slightly above 0.01 are needed.
In addition to the $y$ cut, the program contains a cut on the
invariant mass $m_{ij}$ between any two partons, which is typically 
required to be larger than 2 GeV. This cut corresponds to the
actual merging of two nearby parton jets, i.e. where a treatment with
two separate partons rather than one would be superfluous in view
of the smearing arising from the subsequent fragmentation. Since
the cut-off mass scale $\sqrt{y} E_{\mrm{cm}}$ normally is much larger,
this additional cut only enters for events at low energies.
 
For massive quarks, the amount of QCD radiation is slightly reduced
\cite{Iof78}:
\begin{eqnarray}
  \frac{1}{\sigma_0} \, \frac{\d \sigma}{\d x_1 \, \d x_2} & = &
  \frac{\alphas}{2\pi}
  \, C_F \, \left\{ \frac{x_1^2 + x_2^2}{(1-x_1)(1-x_2)} -
  \frac{4 m_{\q}^2}{s} \left( \frac{1}{1-x_1} + \frac{1}{1-x_2}
  \right) \right.   \nonumber \\[1mm]
  & & - \left. \frac{2 m_{\q}^2}{s} \left( \frac{1}{(1-x_1)^2} +
  \frac{1}{(1-x_2)^2} \right) - \frac{4 m_{\q}^4}{s^2}
  \left( \frac{1}{1-x_1} + \frac{1}{1-x_2} \right)^2 \right\} ~.
\label{ee:threejMEmass}
\end{eqnarray}
In addition, the phase space for emission is reduced by the
requirement
\begin{equation}
\frac{(1-x_1)(1-x_2)(1-x_3)}{x_3^2} \geq \frac{m_{\q}^2}{s} ~.
\end{equation}
For $\b$ quarks at LEP energies, these corrections are fairly small.
 
\subsubsection{4-jet matrix elements}
 
Two new event types are added in second-order QCD,
$\ee \to \q \qbar \g \g$ and $\ee \to \q \qbar \q' \qbar'$.
The 4-jet cross section has been calculated by several
groups \cite{Ali80a,Gae80,Ell81,Dan82}, which agree on the result.
The formulae are too lengthy to be quoted here. In one of the
calculations \cite{Ali80a}, quark masses were explicitly included,
but {\Je} only includes the massless expressions, as taken
from \cite{Ell81}. Here the angular orientation of the event has been
integrated out, so that five independent internal kinematical
variables remain. These may be related to the six $y_{ij}$ and
the four $y_{ijk}$ variables,
$y_{ij} = m_{ij}^2 / s = (p_i + p_j)^2 / s$ and
$y_{ijk} = m_{ijk}^2 / s = (p_i + p_j + p_k)^2 / s$,
in terms of which the matrix elements are given.
 
The original calculations were for the pure $\gamma$-exchange case;
it was recently pointed out \cite{Kni89} that an additional
contribution to the $\ee \to \q \qbar \q' \qbar'$ cross section
arises from the axial part of the $\Z^0$. This term is not included
in the program, but fortunately it is finite and small.
 
Whereas the way the string, i.e. the fragmenting colour flux tube,
is stretched is uniquely given in $\q \qbar \g$ event, for
$\q \qbar \g \g$ events there are two possibilities:
\mbox{$\q - \g_1 - \g_2 - \qbar$} or
\mbox{$\q - \g_2 - \g_1 - \qbar$}.
A knowledge of quark and gluon colours, obtained by perturbation
theory, will uniquely specify the stretching of the string, as long
as the two gluons do not have the same colour. The probability for
the latter is down in magnitude by a factor $1 / N_C^2 = 1 / 9$.
One may either choose to neglect these terms entirely, or to keep
them for the choice of kinematical setup, but then drop them at the
choice of string drawing \cite{Gus82}. We have adopted the latter
procedure. Comparing the two possibilities, differences are
typically 10--20\% for a given kinematical configuration,
and less for the total 4-jet cross section, so from a practical
point of view this is not a major problem.
 
In higher orders, results depend on the renormalization scheme;
we will use $\br{\mrm{MS}}$ throughout. In addition to this choice,
several possible forms can be chosen for $\alphas$,
all of which are equivalent to that order but differ in higher
orders. We have picked the recommended standard \cite{PDG88}
\begin{equation}
\label{ee:aS4j}
 \alphas(Q^2) =
 \frac{12\pi}{(33-2n_f) \, \ln (Q^2 / \Lambda^2_{\br{\mrm{MS}}})}
 \left\{ 1 - 6 \, \frac{153-19n_f}{(33-2n_f)^2} \,
 \frac{\ln (\ln ( Q^2 / \Lambda^2_{\br{\mrm{MS}}}))}
 {\ln ( Q^2 / \Lambda^2_{\br{\mrm{MS}}})}
 \right\} ~.
\end{equation}
 
\subsubsection{Second-order 3-jet matrix elements}
 
As for first order, a full second-order calculation consists both of
real parton emission terms and of vertex and propagator corrections.
These modify the 3-jet and 2-jet cross sections.
Although there was some initial confusion, everybody soon agreed
on the size of the loop corrections \cite{Ell81,Ver81,Fab82}.
In analytic calculations, the procedure of eq.~(\ref{ee:Obs}),
suitably expanded, can therefore be used unambiguously for a
well-behaved variable.
 
For Monte Carlo event simulation, it is again necessary to impose
some finite jet-resolution criterion. This means that four-parton
events which fail the cuts should be reassigned either to the
3-jet or to the 2-jet event class. It is this area that
caused quite a lot of confusion in the past
\cite{Kun81,Got82,Ali82,Zhu83,Gut84,Gut87,Kra88},
and where full agreement does not exist. Most likely, agreement
will never be reached, since there are indeed ambiguous points
in the procedure, related to uncertainties on the theoretical
side, as follows.
 
For the $y$-cut case, any two partons with an invariant mass
$m_{ij}^2 < y E_{\mrm{cm}}^2$ should be recombined into one. If the
four-momenta are simply added, the sum will correspond to a parton
with a positive mass, namely the original $m_{ij}$.
The loop corrections are given in terms of final
massless partons, however. In order to perform the (partial)
cancellation between the four-parton real and the 3-parton
virtual contributions, it is therefore necessary to get rid of
the bothersome mass in the four-parton states. Several
recombinations are used in practice, which go under names such as
`E', `E0', `p' and `p0' \cite{OPA91}. In the `E'-type schemes,
the energy of a recombined parton is given by  $E_{ij} = E_i + E_j$,
and three-momenta may have to be adjusted accordingly. In the
`p'-type schemes, on the other hand, three-momenta are added,
$\mbf{p}_{ij} = \mbf{p}_i + \mbf{p}_j$, and then energies may have
to be adjusted. These procedures result in different 3-jet
topologies, and therefore in different second-order differential
3-jet cross sections.
 
Within each scheme, a number of lesser points remain to be dealt
with, in particular what to do if a recombination of a nearby parton
pair were to give an event with a non-$\q\qbar\g$ flavour structure.
 
{\Je} contains two alternative second-order 3-jet
implementations, GKS and ERT(Zhu). For historical
reasons the former is default, but actually the latter is the
recommended one today. Other parametrizations have also been made
available that run together with {\Je}, see
\cite{Sjo89,Mag89}.
 
The GKS option is based on the GKS \cite{Gut84} calculation, where
some of the original mistakes in FKSS \cite{Fab82} have been
corrected. The GKS formulae have the advantage of giving the
second-order corrections in closed analytic form, as not-too-long
functions of $x_1$, $x_2$, and the $y$ cut. However, it is
today recognized, also by the authors, that important
terms are still missing, and that the matrix elements
should therefore not be taken too seriously. The option is thus
kept mainly for backwards compatibility.
 
The ERT(Zhu) generator \cite{Zhu83} is based on the ERT matrix elements
\cite{Ell81}, with a Monte Carlo recombination procedure suggested
by Kunszt \cite{Kun81} and developed by Ali \cite{Ali82}. It has
the merit of giving corrections in a convenient, parametrized form.
For practical applications, the main limitation is that the
corrections are only given for discrete values of the cut-off
parameter $y$, namely $y$ = 0.01, 0.02, 0.03, 0.04, and 0.05.
 
The basic approach is the following. Without any loss of generality,
the full second-order 3-jet cross section can be written in
terms of the `ratio function' $R(X,Y;y)$, defined by
\begin{equation}
\frac{1}{\sigma_0} \frac{\d \sigma_3^{\mrm{tot}}}{\d X \, \d Y} =
\frac{\alphas}{\pi} A_0(X,Y)
\left\{ 1 + \frac{\alphas}{\pi} R(X,Y;y) \right\} ~,
\label{ee:Zhupar}
\end{equation}
where $X = x_1 - x_2 = x_{\q} - x_{\qbar}$, $Y = x_3 = x_g$,
$\sigma_0$ is the lowest-order hadronic cross section,
and $A_0(X,Y)$ the standard first-order 3-jet cross section,
cf. eq.~(\ref{ee:ME3j}).
By Monte Carlo integration, the value of $R(X,Y;y)$ is
evaluated in bins of $(X,Y)$, and the result parametrized
by a simple function $F(X,Y;y)$.
 
In order to obtain the second-order 3-jet rate, a small cut
$y_0 = 10^{-7}$ was introduced. It was assumed that four-parton events
which fail this cut can be (partly) cancelled analytically against
the virtual 3-jet events, to give a net `regularized virtual'
contribution to the 3-jet rate. For a given choice
of $y$ cut, in the physical range $y \gg y_0$, an additional `soft'
contribution comes from four-parton events which survive the $y_0$ cut
but fail the $y$ one.
 
A large sample (9\,000\,000) of four-parton events was generated inside
the $y_0$ cut region. For events which failed the more stringent $y$
cuts, the parton pair with the smallest invariant mass was recombined
into an effective jet, using the `p0' recombination scheme.
This means that the individual three-momenta were added,
$\mbf{p}_{ij} = \mbf{p}_i + \mbf{p}_j$, the mass
of the recombined pair was set to zero for the calculation of energy,
$E_{ij} = |\mbf{p}_i + \mbf{p}_j|$, and
finally all four-momenta were rescaled by a common factor so as
to preserve the correct c.m. frame energy.
 
In calculating the ${\cal O}(\alphas^2)$ correction functions,
care was taken
to maintain the flavour signature of the jets in the recombination
process. A quark and a gluon were recombined into a quark with the same
flavour as the original quark, two gluons were recombined to form
a gluon, etc. In some cases the
three jets of the final state were not in the
standard $\q \qbar \g$ configuration. The probability for this to happen
corresponded to less than 0.5\% of the total cross section, even
for the most stringent cuts used. For these non-$\q \qbar \g$ final
states, the assignment of $\q$, $\qbar$ and $\g$ was done at random.
 
The sum of `regularized virtual' (1\,000\,000 3-jet events were
generated, with evaluated second-order weights)
and `soft' corrections, normalized to the first-order 3-jet
cross section, was tabulated in the $(X,Y)$ plane, using bins of size
$0.05 \times 0.05$. This estimated $R$-function behaviour was then
fit with a 12-parameter function $F$,
\begin{eqnarray}
   F(X,Y;y)& = & p_1 + p_2 X^2 + p_3 X^4 + (p_4+p_5 X^2)Y+
                 (p_6+p_7X^2)Y^2 +               \nonumber\\
           &   & (p_8+p_9X^2)Y^3 +
                 p_{10}/(X^2-Y^2) + p_{11}/(1-Y) + p_{12}/Y ~.
\label{ee:ZhuFun}
\end{eqnarray}
The parameters $p_i$ are reproduced in \cite{Sjo89}.
 
\subsubsection{The matrix-element event generator scheme}
 
The program contains parametrizations, separately, of the total
first-order 3-jet rate, the total second-order 3-jet rate,
and the total 4-jet rate, all as functions of $y$ (with
$\alphas$ as a separate prefactor).
These parametrizations have been obtained as follows:
\begin{Itemize}
\item
The first-order 3-jet matrix element is almost analytically
integrable; some small finite pieces were obtained by a truncated
series expansion of the relevant integrand.
\item
The  GKS second-order 3-jet matrix elements were integrated for
40 different $y$-cut values, evenly distributed in $\ln y$ between
a smallest value $y = 0.001$ and the kinematical limit $y = 1/3$.
For each $y$ value, 250\,000 phase-space points were generated,
evenly in $\d \ln (1-x_i) = \d x_i/(1-x_i)$, $i = 1,2$, and the
second-order 3-jet rate in the point evaluated. The properly
normalized sum of weights in each of the 40 $y$ points were
then fitted to a polynomial in $\ln(y^{-1}-2)$. For the ERT(Zhu)
matrix elements the parametrizations in eq.~(\ref{ee:ZhuFun})
were used to perform a corresponding Monte Carlo integration for
the five $y$ values available.
\item
The 4-jet rate was integrated numerically, separately for
$\q \qbar \g \g$ and $\q \qbar \q' \qbar'$ events, by generating large
samples of 4-jet phase-space points
within the boundary $y = 0.001$. Each point was classified according
to the actual minimum $y$ between any two partons. The same
events could then be used to update the summed weights for 40
different counters, corresponding to $y$ values evenly distributed
in $\ln y$ between $y = 0.001$ and the kinematical limit $y = 1/6$.
In fact, since
the weight sums for large $y$ values only received contributions
from few phase-space points, extra (smaller) subsamples of events were
generated with larger $y$ cuts. The summed weights,
properly normalized, were then parametrized in terms of
polynomials in $\ln(y^{-1} - 5)$.
Since it turned out to be difficult to obtain one single good fit
over the whole range of $y$ values, different parametrizations are
used above and below $y=0.018$. As originally given, the
$\q \qbar \q' \qbar'$ parametrization only took into account four
$\q'$ flavours, i.e. secondary $\b \bbar$ pairs were not generated,
but this has been corrected for LEP.
\end{Itemize}
 
In the generation stage, each event is treated on its own, which means
that the $\alphas$ and $y$ values may be allowed to vary from event to
event. The main steps are the following.
\begin{Enumerate}
\item
The $y$ value to be used in the current event is determined. If
possible, this is the value given by you, but additional
constraints exist from the validity of the parametrizations
($y \geq 0.001$ for GKS, $0.01 \leq y \leq 0.05$ for ERT(Zhu))
and an extra (user-modifiable) requirement of a minimum absolute
invariant mass between jets (which translates into varying $y$ cuts
due to the effects of initial-state QED radiation).
\item
The $\alphas$ value is calculated.
\item
For the $y$ and $\alphas$ values given, the relative
two/three/4-jet composition is determined. This is achieved by
using the parametrized functions of $y$ for 3- and 4-jet rates,
multiplied by the relevant number of factors of $\alphas$.
In ERT(Zhu), where the second-order 3-jet rate is available
only at a few $y$ values, intermediate results are obtained by linear
interpolation in the ratio of second-order to first-order
3-jet rates. The 3-jet and 4-jet rates are normalized to
the analytically known second-order total event rate, i.e. divided
by $R_{\mrm{QCD}}$ of eq.~(\ref{ee:RQCD}). Finally, the 2-jet rate is 
obtained by conservation of total probability.
\item
If the combination of $y$ and $\alphas$ values is such that the total
3- plus 4-jet fraction is larger than unity, i.e. the remainder
2-jet fraction negative, the $y$-cut value is raised (for that event),
and the process is started over at point 3.
\item
The choice is made between generating a 2-, 3- or 4-jet event,
according to the relative probabilities.
\item
For the generation of 4-jets, it is first necessary to make a choice
between $\q \qbar \g \g$ and $\q \qbar \q' \qbar'$ events, according to
the relative (parametrized) total cross sections. A phase-space point
is then selected, and the differential cross section at this point is
evaluated and compared with a parametrized maximum weight. If the
phase-space point is rejected, a new one is selected, until an
acceptable 4-jet event is found.
\item
For 3-jets, a phase-space point is first chosen according to the
first-order cross section. For this point, the weight
\begin{equation}
W(x_1,x_2;y) = 1 + \frac{\alphas}{\pi} R(x_1,x_2;y)
\label{ee:WTJS}
\end{equation}
is evaluated. Here $R(x_1,x_2;y)$ is analytically given for GKS
\cite{Gut84}, while it is approximated by the parametrization
$F(X,Y;y)$ of eq.~(\ref{ee:ZhuFun}) for ERT(Zhu). Again, linear
interpolation of $F(X,Y;y)$ has to be applied for intermediate $y$
values. The weight $W$ is compared with a maximum weight
\begin{equation}
W_{\mmax}(y) = 1 + \frac{\alphas}{\pi} R_{\mmax}(y) ~,
\end{equation}
which has been numerically determined beforehand and suitably
parametrized. If the phase-space point is rejected, a
new point is generated, etc.
\item
Massive matrix elements are not available in {\Je} for second-order 
QCD (but are in the first-order option). However, if a
3- or 4-jet event determined above falls outside
the phase-space region allowed for massive quarks, the event is
rejected and reassigned to be a 2-jet event. (The way the
$y_{ij}$ and $y_{ijk}$ variables of 4-jet events should be
interpreted for massive quarks is not even unique, so some latitute
has been taken here to provide a reasonable continuity from
3-jet events.) This procedure is known not to give the expected full
mass suppression, but is a reasonable first approximation.
\item
Finally, if the event is classified as a 2-jet event, either
because it was initially so assigned, or because it failed the
massive phase-space cuts for 3- and 4-jets, the
generation of 2-jets is trivial.
\end{Enumerate}
 
\subsubsection{Optimized perturbation theory}
 
Theoretically, it turns out that the second-order corrections to the
3-jet rate are large. It is therefore not unreasonable to expect
large third-order corrections to the 4-jet rate. Indeed, the
experimental 4-jet rate is much larger than second order predicts
(when fragmentation effects have been folded in),
if $\alphas$ is determined based on the 3-jet rate
\cite{Sjo84a,JAD88}.
 
The only consistent way to resolve this issue is to go ahead and
calculate the full next order. This is a tough task, however, so
people have looked at possible shortcuts.
For example, one can try to minimize the higher-order contributions
by a suitable choice of the renormalization scale \cite{Ste81} ---
`optimized perturbation theory'. This
is equivalent to a different choice for the $Q^2$ scale in
$\alphas$, a scale which is not unambiguous anyway. Indeed
the standard value $Q^2 = s = E_{\mrm{cm}}^2$ is larger than the 
natural physical scale of gluon emission in events, given that most 
gluons are fairly soft. One could therefore pick another scale,
$Q^2 = f s$, with $f < 1$. The
${\cal O}(\alphas)$ 3-jet rate would be increased by
such a scale change, and so would the number of 4-jet
events, including those which collapse into 3-jet ones. The loop
corrections depend on the $Q^2$ scale, however,
and compensate the changes above by giving a larger negative
contribution to the 3-jet rate.
 
The possibility of picking an optimized scale $f$ is implemented
as follows \cite{Sjo89}. Assume that the differential 3-jet
rate at scale $Q^2 = s$ is given by the expression
\begin{equation}
R_3 = r_1 \alphas + r_2 \alphas^2 ~,
\end{equation}
where $R_3$, $r_1$ and $r_2$ are functions of the kinematical
variables $x_1$ and $x_2$ and the $y$ cut, as described above.
When the coupling is chosen at a different scale, $Q'^2 = f s$,
the 3-jet rate has to be changed to
\begin{equation}
R_3' = r_1' \alphas' + r_2 \alphas'^2 ~,
\end{equation}
where $r_1' = r_1$,
\begin{equation}
r_2' = r_2 + r_1 \frac{33-2n_f}{12\pi} \ln f ~,
\label{ee:r2optim}
\end{equation}
and $\alphas' = \alphas(fs)$.
Since we only have the Born term for 4-jets, here the effects of a
scale change come only from the change in the coupling constant.
Finally, the 2-jet cross section can still be calculated from the
difference between the total cross section and the 3- and 4-jet
cross sections.
 
If an optimized scale is used in the program, the default value is
$f=0.002$, which is favoured by the studies in ref. \cite{Bet89}. (In
fact, it is also possible to use a correspondingly optimized
$R_{\mrm{QCD}}$ factor, eq.~(\ref{ee:RQCD}), but then the 
corresponding $f$ is chosen independently and much closer to unity.)  
The success of describing the jet rates should not hide the fact that 
one is dabbling in (educated, hopefully) guesswork, and that any
conclusions based on this method have to be taken with a pinch of
salt.
 
One special problem associated with the use of optimized perturbation
theory is that the differential 3-jet rate may become negative
over large regions of the $(x_1, x_2)$ phase space. This problem
already exists, at least in principle, even for a scale $f = 1$,
since $r_2$ is not guaranteed to be positive definite. Indeed,
depending on the choice of $y$ cut, $\alphas$ value and recombination
scheme, one may observe a small region of negative differential
3-jet rate for the full second-order expression. This region
is centred around $\q \qbar \g$ configurations, where the $\q$ and
$\qbar$ are close together in one hemisphere and the $\g$ is alone in
the other, i.e. $x_1 \approx x_2 \approx 1/2$. It is well understood
why second-order corrections should be negative in this region
\cite{Dok89}: the $\q$ and $\qbar$ of a $\q \qbar \g$ state are in a
relative colour octet state, and thus the colour force between them is
repulsive, which translates into a negative second-order term.
 
However, as $f$ is decreased below unity, $r_2'$ receives a negative
contribution from the $\ln f$ term, and the region of negative
differential cross section has a tendency to become larger, also
after taking into account related changes in $\alphas$. In an 
event-generator framework, where all events are supposed to come 
with unit
weight, it is clearly not possible to simulate negative cross sections.
What happens in the program is therefore that no 3-jet events at
all are generated in the regions of negative differential cross section,
and that the 3-jet rate in regions of positive cross sections is
reduced by a constant factor, chosen so that the total number of
3-jet events comes out as it should. This is a consequence of the
way the program works, where it is first decided what kind of event to
generate, based on integrated 3-jet rates in which positive and
negative contributions are added up with sign, and only thereafter
the kinematics is chosen.
 
Based on our physics understanding of the origin of this negative
cross section, the approach adopted is as sensible as any, at least
to that order in perturbation theory (what one might strive for is a
properly exponentiated description of the relevant region). It can
give rise to funny results for low $f$ values, however, as observed
by OPAL \cite{OPA92} for the energy--energy correlation asymmetry.
 
\subsubsection{Angular orientation}
 
While pure $\gamma$ exchange gives a simple $1 + \cos^2\theta$
distribution for the $\q$ (and $\qbar$) direction in $\q \qbar$ events,
$\Z^0$ exchange and $\gammaZ$ interference results in a
forward--backward asymmetry. If one introduces
\begin{eqnarray}
  h'_{\f}(s) & = & 2 e_{\e} \left\{ a_{\e} 
  (1 - P_{\mrm{L}}^+ P_{\mrm{L}}^-) -
  v_{\e} (P_{\mrm{L}}^- - P_{\mrm{L}}^+) \right\} \,  
  \Re\chi(s)  e_{\f} a_{\f}
  \nonumber \\
  & & + \, \left\{ 2 v_{\e} a_{\e} (1 - P_{\mrm{L}}^+ P_{\mrm{L}}^-) -
  (v_{\e}^2 + a_{\e}^2) (P_{\mrm{L}}^- - P_{\mrm{L}}^+) \right\} \,
  |\chi(s)|^2 \, v_{\f} a_{\f} ~,
\end{eqnarray}
then the angular distribution of the quark is given by
\begin{equation}
   \frac{\d \sigma}{\d (\cos\theta_{\f})} \propto
   h_{\f}(s)(1 + \cos^2\theta_{\f}) + 2 h'_{\f}(s) \cos\theta_{\f} ~.
\end{equation}
 
The angular orientation of a 3- or 4-jet event may be described
in terms of three angles $\chi$, $\theta$ and $\varphi$; for 2-jet
events only $\theta$ and $\varphi$ are necessary. From a standard
orientation, with the $\q$ along the $+z$ axis and the $\qbar$ in the
$xz$ plane with $p_x > 0$, an arbitrary orientation may be reached by
the rotations $+\chi$ in azimuthal angle, $+\theta$ in polar angle,
and $+\varphi$ in azimuthal angle, in that order. Differential
cross sections,
including QFD effects and arbitrary beam polarizations have been given
for 2- and 3-jet events in refs. \cite{Ols80,Sch80}. We use the
formalism of ref. \cite{Ols80}, with $\chi \to \pi - \chi$ and
$\varphi^- \to - (\varphi + \pi/2)$. The resulting formulae are
tedious, but
straightforward to apply, once the internal jet configuration has been
chosen. 4-jet events are approximated by 3-jet ones, by joining
the two gluons of a $\q \qbar \g \g$ event and the $\q'$ and $\qbar'$
of a $\q \qbar \q' \qbar'$ event into one effective jet. This means
that some angular asymmetries are neglected \cite{Ali80a}, but that weak
effects are automatically included. It is assumed that the second-order
3-jet events have the same angular orientation as the first-order
ones, some studies on this issue may be found in \cite{Kor85}. Further,
the formulae normally refer to the massless case; only for the QED
2- and 3-jet cases are mass corrections available.
 
The main effect of the angular distribution of multijet events
is to smear the lowest-order result, i.e. to reduce any anisotropies
present in 2-jet systems. In the parton-shower option of the program,
only the initial $\q \qbar$ axis is determined. The subsequent shower
evolution then {\it de facto} leads to a smearing of the jet axis,
although not necessarily in full agreement with the expectations
from multijet matrix-element treatments.
 
\subsubsection{Initial-state radiation}
 
Initial-state photon radiation has been included using the formalism of
ref. \cite{Ber82}. Here each event contains either no photon or one,
i.e. it is a first-order non-exponentiated description.
The main formula for the hard radiative photon
cross section is
\begin{equation}
\frac{\d \sigma}{\d x_{\gamma}} = \frac{\alphaem}{\pi} \,
\left( \ln\frac{s}{m_{\e}^2} -1 \right) \,
\frac{1 + (1-x_{\gamma})^2}{x_{\gamma}} \, \sigma_0 (\hat{s}) ~,
\end{equation}
where $x_{\gamma}$ is the photon energy fraction of the beam energy,
$\hat{s} = (1-x_{\gamma}) s$ is the squared reduced hadronic c.m.
energy, and $\sigma_0$ is the ordinary annihilation cross section at
the reduced energy. In particular, the selection of jet flavours
should be done according to expectations at the reduced
energy. The cross section is divergent both for $x_{\gamma} \to 1$ and
$x_{\gamma} \to 0$. The former is related to the fact that
$\sigma_0$ has a $1/\hat{s}$ singularity (the real photon pole) for
$\hat{s} \to 0$. An upper cut on $x_{\gamma}$ can here be chosen 
to fit the
experimental setup. The latter is a soft photon singularity, which is
to be compensated in the no-radiation cross section. A requirement
$x_{\gamma} > 0.01$ has therefore been chosen so that the hard-photon
fraction is smaller than unity. In the total cross section, effects
from photons with $x_{\gamma} < 0.01$ are taken into account, together
with vertex and vacuum polarization corrections (hadronic vacuum
polarizations using a simple parametrization of the more complicated
formulae of ref. \cite{Ber82}).
 
The hard photon spectrum can be integrated analytically, for the
full $\gammaZ$ structure including interference terms, provided that
no new flavour thresholds are crossed and that the $R_{\mrm{QCD}}$
term in the cross section can be approximated by a constant over the
range of allowed $\hat{s}$ values. In fact, threshold effects can be
taken into account by standard rejection techniques, at the price of
not obtaining the exact cross section analytically, but only by an
effective Monte Carlo integration taking place in parallel with the
ordinary event generation. In addition to $x_{\gamma}$, the polar
angle $\theta_{\gamma}$ and azimuthal angle $\varphi_{\gamma}$ of
the photons are also to be chosen. Further, for the orientation
of the hadronic system, a choice has to be made whether the photon is
to be considered as having been radiated from the $\e^+$ or from the
$\e^-$.
 
Final-state photon radiation, as well as interference between initial-
and final-state radiation, has been left out of this treatment. The
formulae for $\ee \to \mu^+ \mu^-$ cannot be simply taken over for
the case of outgoing quarks, since the quarks as such only live for
a short while before turning into hadrons. Another simplification in
our treatment is that effects of incoming polarized $\e^{\pm}$ beams
have been completely neglected, i.e. neither the effective shift in
azimuthal distribution of photons nor the reduction in polarization
is included. The polarization parameters of the program are to be
thought of as the effective polarization surviving after 
initial-state radiation.
 
\subsubsection{Alternative matrix elements}
 
The program contains two sets of `toy model' matrix elements, one
for an Abelian vector gluon model and one for a scalar gluon model.
Clearly both of these alternatives are already excluded by data,
and are anyway not viable alternatives for a consistent theory of
strong interactions. They are therefore included more as references
to show how well the characteristic features of QCD can be measured
experimentally.
 
Second-order matrix elements are available for the Abelian vector
gluon model. These are easily obtained from the standard QCD matrix
elements by a substitution of the Casimir group factors:
$C_F = 4/3 \to 1$, $N_C = 3 \to 0$, and $T_R = n_{\f}/2 \to 3 n_{\f}$.
First-order matrix elements contain only $C_F$; therefore the
standard first-order QCD results may be recovered by a rescaling
of $\alphas$ by a factor $4/3$. In second order the change of $N_C$
to 0 means that $\g \to \g\g$ couplings are absent from the Abelian
model, while the change of $T_R$ corresponds to an enhancement of the
$\g \to \q'\qbar'$ coupling, i.e. to an enhancement of the
$\q\qbar\q'\qbar'$ 4-jet event rate.
 
The second-order corrections to the 3-jet rate
turn out to be strongly negative --- if
$\alphas$ is fitted to get about the right rate of 4-jet events,
the predicted differential 3-jet rate is negative almost
everywhere in the $(x_1, x_2)$ plane. Whether this unphysical
behaviour would be saved by higher orders is unclear. It has been
pointed out that the rate can be made positive by a suitable choice of
scale, since $\alphas$ runs in opposite directions in an Abelian
model and in QCD \cite{Bet89}. This may be seen directly from eq.
(\ref{ee:r2optim}), where the term $33 = 11 N_C$ is absent in the
Abelian model, and therefore the scale-dependent term changes sign.
In the program, optimized scales have not been implemented for this
toy model. Therefore the alternatives provided for you are either
to generate only 4-jet events, or to neglect second-order
corrections to the 3-jet rate, or to have the total 3-jet
rate set vanishing (so that only 2- and 4-jet events are
generated). Normally we would expect the former to be the one of most
interest, since it is in angular (and flavour) distributions of 4-jet
events that the structure of QCD can be tested.
Also note that the `correct' running of $\alphas$ is not included;
you are expected to use the option where $\alphas$ is just given as
a constant number.
 
The scalar gluon model is even more excluded than the Abelian vector
one, since differences appear already in the 3-jet matrix
element \cite{Lae80}:
\begin{equation}
\frac{\d \sigma}{\d x_1 \, \d x_2} \propto \frac{x_3^2}{(1-x_1)(1-x_2)}
\end{equation}
when only $\gamma$ exchange is included. The axial part of the $\Z^0$
gives a slightly different shape; this is included in the program but
does not make much difference. The angular orientation does include
the full $\gammaZ$ interference \cite{Lae80}, but the main interest is
in the 3-jet topology as such \cite{Ell79}. No higher-order
corrections are included. It is recommended to use the option of a
fixed $\alphas$ also here, since the correct running is not available.
 
\subsection{Decays of Onia Resonances}
\label{ss:oniadecays}
 
Many different possibilities are open for the decay of heavy
$J^{PC} = 1^{--}$ onia resonances. Of special interest are
the decays into three gluons or two gluons plus a photon, since these
offer unique possibilities to study a `pure sample' of gluon jets.
A routine for this purpose is included in the program. It was written
at a time where the expectations were to find toponium at PETRA 
energies.
If, as now seems likely, the top mass is above 100~GeV, weak decays
will dominate, to the extent that the top quark will decay weakly even
before a bound toponium state is formed, and thus the routine will be
of no use for top. The charm system, on the other hand, is far too low
in mass for a jet language to be of any use. The only application is
therefore likely to be for $\Upsilon$, which unfortunately also is on
the low side in mass.
 
The matrix element for $\q \qbar \to \g \g \g$ is (in lowest order)
\cite{Kol78}
\begin{equation}
\frac{1}{\sigma_{\g \g \g}} 
\frac{\d \sigma_{\g \g \g}}{\d x_1 \, \d x_2} =
\frac{1}{\pi^2 - 9} \left\{ \left( \frac{1-x_1}{x_2 x_3} \right)^2 +
\left( \frac{1-x_2}{x_1 x_3} \right)^2 +
\left( \frac{1-x_3}{x_1 x_2} \right)^2 \right\} ~,
\label{ee:Upsilondec}
\end{equation}
where, as before, $x_i = 2 E_i / E_{\mrm{cm}}$ in the c.m. frame of 
the event. This is a well-defined expression, without the kind of
singularities encountered in the $\q \qbar \g$ matrix elements.
In principle, no cuts at all would be necessary, but for reasons
of numerical simplicity we implement a $y$ cut as for continuum
jet production, with all events not fulfilling this cut considered
as (effective) $\g \g$ events. For $\g \g \g$ events, each $\g \g$
invariant mass is required to be at least 2 GeV.
 
Another process is $\q \qbar \to \gamma \g \g$, obtained by replacing
a gluon in $\q \qbar \to \g \g \g$ by a photon. This process has the
same normalized cross section as the one above, if e.g. $x_1$ is taken
to refer to the photon. The relative rate is \cite{Kol78}
\begin{equation}
\frac{\sigma_{\gamma \g \g}}{\sigma_{\g \g \g}} =
\frac{36}{5} \, \frac{e_{\q}^2 \, \alphaem}
{\alphas(Q^2)} ~.
\end{equation}
Here $e_{\q}$ is the charge of the heavy quark, and the scale in
$\alphas$ has been chosen as the mass of the onium state. If the
mass of the recoiling $\g \g$ system is lower than some cut-off
(by default 2 GeV), the event is rejected.
 
In the present implementation the angular orientation of the
$\g \g \g$ and $\gamma \g \g$ events is given for the
$\ee \to \gamma^* \to$ onium case \cite{Kol78} (optionally with
beam polarization effects included), i.e. weak effects have not
been included, since they are negligible at around 10~GeV.
 
It is possible to start a perturbative shower evolution from either
of the two states above. However, for $\Upsilon$ the phase space
for additional evolution is so constrained that not much is to be
gained from that. We therefore do not recommend this possibility.
The shower generation machinery, when starting up from a
$\gamma \g \g$ configuration, is constructed such that the
photon energy is not changed. This means that there is currently no
possibility to use showers to bring the theoretical photon
spectrum in better agreement with the experimental one.
 
In string fragmentation language, a $\g \g \g$ state corresponds to
a closed string triangle with the three gluons at the corners. As
the partons move apart from a common origin, the string triangle
expands. Since the photon does not take part in the fragmentation,
the $\gamma \g \g$ state corresponds to a double string running
between the two gluons.
 
\subsection{Routines and Common Block Variables}
\label{ss:eeroutines}
 
\subsubsection{$\ee$ continuum event generation}
 
The only routine a normal user will call to generate $\ee$ continuum
events is \ttt{LUEEVT}. The other routines listed below, as well as
\ttt{LUSHOW} (see section \ref{ss:showrout}), are called by 
\ttt{LUEEVT}.
 
\drawbox{CALL LUEEVT(KFL,ECM)}\label{p:LUEEVT}
\begin{entry}
\itemc{Purpose:} to generate a complete event
$\ee \to \gammaZ \to \q\qbar \to$ parton shower $\to$ hadrons
according to QFD and QCD cross sections. As an alternative to
parton showers, second-order matrix elements are available for
$\q\qbar + \q\qbar\g + \q\qbar\g\g + \q\qbar\q'\qbar'$ production.
\iteme{KFL :} flavour of events generated.
\begin{subentry}
\iteme{= 0 :} mixture of all allowed flavours according to relevant
probabilities.
\iteme{= 1 - 8 :} primary quarks are only of the specified flavour
\ttt{KFL}.
\end{subentry}
\iteme{ECM :} total c.m. energy of system.
\itemc{Remark:} Each call generates one event, which is independent
of preceding ones, with one exception, as follows. If radiative
corrections are included, the shape of the hard photon spectrum is
recalculated only with each \ttt{LUXTOT} call, which normally is done
only if \ttt{KFL}, \ttt{ECM} or \ttt{MSTJ(102)} is changed. A change
of e.g. the $\Z^0$ mass in mid-run has to be followed either by a user
call to \ttt{LUXTOT} or by an internal call forced e.g. by putting
\ttt{MSTJ(116)=3}.
\end{entry}

\boxsep

\begin{entry} 
\iteme{SUBROUTINE LUXTOT(KFL,ECM,XTOT) :}\label{p:LUXTOT}
to calculate the total hadronic cross section,
including quark thresholds, weak, beam polarization, and QCD effects
and radiative corrections. In the process, variables necessary
for the treatment of hard photon radiation are calculated and
stored.
\begin{subentry}
\iteme{KFL, ECM :} as for \ttt{LUEEVT}.
\iteme{XTOT :} the calculated total cross section in nb.
\end{subentry}
 
\iteme{SUBROUTINE LURADK(ECM,MK,PAK,THEK,PHIK,ALPK) :}\label{p:LURADK}
to describe initial-state hard $\gamma$ radiation.
 
\iteme{SUBROUTINE LUXKFL(KFL,ECM,ECMC,KFLC) :}\label{p:LUXKFL}
to generate the primary quark flavour in case this
is not specified by the user.
 
\iteme{SUBROUTINE LUXJET(ECM,NJET,CUT) :}\label{p:LUXJET}
to determine the number of jets (2, 3 or 4) to be
generated within the kinematically allowed region (characterized by
\ttt{CUT} $= y_{\mrm{cut}}$) in the matrix-element approach; to be 
chosen such that all probabilities are between 0 and 1.
 
\iteme{SUBROUTINE LUX3JT(NJET,CUT,KFL,ECM,X1,X2) :}\label{p:LUX3JT}
to generate the internal momentum variables of a
3-jet event, $\q\qbar\g$, according to first- or second-order
QCD matrix elements.
 
\iteme{SUBROUTINE LUX4JT(NJET,CUT,KFL,ECM,KFLN,X1,X2,X4,X12,X14) :}%
\label{p:LUX4JT} 
to generate the internal momentum variables for a
4-jet event, $\q\qbar\g\g$ or $\q\qbar\q'\qbar'$, according to
second-order QCD matrix elements.
 
\iteme{SUBROUTINE LUXDIF(NC,NJET,KFL,ECM,CHI,THE,PHI) :}\label{p:LUXDIF}
to describe the angular orientation of the jets.
In first-order QCD the complete QED or QFD formulae are used; in
second  order 3-jets are assumed to have the same orientation
as in first, and 4-jets are approximated by 3-jets.

\end{entry}
 
\subsubsection{A routine for onium decay}
 
In \ttt{LUONIA} we have implemented the decays of heavy onia
resonances into three gluons or two gluons plus a photon, which are
the dominant non-background-like decays of $\Upsilon$.
 
\drawbox{CALL LUONIA(KFL,ECM)}\label{p:LUONIA}
\begin{entry}
\itemc{Purpose:} to simulate the process
$\ee \to \gamma^* \to 1^{--}$ onium resonance $\to (\g\g\g$ or
$\g\g\gamma) \to$ shower $\to$ hadrons.
\iteme{KFL :} the flavour of the quark giving rise to the resonance.
\begin{subentry}
\iteme{= 0 :} generate $\g\g\g$ events alone.
\iteme{= 1 - 8 :} generate $\g\g\g$ and $\g\g\gamma$ events in mixture
determined by the squared charge of flavour \ttt{KFL}. Normally
\ttt{KFL=} 5 or 6.
\end{subentry}
\iteme{ECM :} total c.m. energy of system.
\end{entry}
 
\subsubsection{Common block variables}
 
The status codes and parameters relevant for the $\ee$ routines are
found in the common block \ttt{LUDAT1}. This common block also contains
more general status codes and parameters, described elsewhere.
 
\drawbox{COMMON/LUDAT1/MSTU(200),PARU(200),MSTJ(200),PARJ(200)}
\begin{entry}
 
\itemc{Purpose:} to give access to a number of status codes and
parameters regulating the performance of the $\ee$ event generation
routines.
 
\iteme{MSTJ(101) :}\label{p:MSTJ101} (D=5) gives the type of QCD 
corrections used for continuum events.
\begin{subentry}
\iteme{= 0 :} only $\q\qbar$ events are generated.
\iteme{= 1 :} $\q\qbar + \q\qbar\g$ events are generated according
to first-order QCD.
\iteme{= 2 :} $\q\qbar + \q\qbar\g + \q\qbar\g\g + \q\qbar\q'\qbar'$
events are generated according to second-order QCD.
\iteme{= 3 :} $\q\qbar + \q\qbar\g + \q\qbar\g\g + \q\qbar\q'\qbar'$
events are generated, but without second-order corrections to the
3-jet rate.
\iteme{= 5 :} a parton shower is allowed to develop from an original
$\q\qbar$ pair, see \ttt{MSTJ(40) - MSTJ(50)} for details.
\iteme{= -1 :} only $\q\qbar\g$ events are generated (within same
matrix-element cuts as for \ttt{=1}). Since the change in flavour
composition from mass cuts or radiative corrections is not
taken into account, this option is not intended for
quantitative studies.
\iteme{= -2 :} only $\q\qbar\g\g$ and $\q\qbar\q'\qbar'$ events are
generated (as for \ttt{=2}). The same warning as for \ttt{=-1} applies.
\iteme{= -3 :} only $\q\qbar\g\g$ events are generated (as for
\ttt{=2}). The same warning as for \ttt{=-1} applies.
\iteme{= -4 :} only $\q\qbar\q'\qbar'$ events are generated
(as for \ttt{=2}). The same warning as for \ttt{=-1} applies.
\itemc{Note 1:} \ttt{MSTJ(101)} is also used in \ttt{LUONIA}, with
\iteme{$\leq$ 4 :} $\g\g\g + \gamma\g\g$ events are generated
according to lowest-order matrix elements.
\iteme{$\geq$ 5 :} a parton shower is allowed to develop from the
original $\g\g\g$ or $\g\g\gamma$ configuration, see
\ttt{MSTJ(40) - MSTJ(50)} for details.
\itemc{Note 2:} The default values of fragmentation parameters have
been chosen to work well with the default parton-shower approach
above. If any of the other options are used, or if the parton
shower is used in non-default mode, it may be necessary to
retune fragmentation parameters. As an example, we note that
the second-order matrix-element approach (\ttt{MSTJ(101)=2}) at
PETRA/PEP energies gives a better description when the $a$ and
$b$ parameters of the symmetric fragmentation function are set to
$a =$\ttt{PARJ(41)=1}, $b =$\ttt{PARJ(42)=0.7}, and the
width of the transverse momentum distribution to
$\sigma =$\ttt{PARJ(21)=0.40}.
In principle, one also ought to change the joining parameter
to \ttt{PARJ(33)=PARJ(35)=1.1} to preserve a flat rapidity
plateau, but if this should be forgotten, it does not make too
much difference. For applications at TRISTAN or LEP, one
must expect to have to change the matrix-element approach
parameters even more, to make up for additional soft gluon
effects not covered in this approach.
\end{subentry}
 
\iteme{MSTJ(102) :} (D=2) inclusion of weak effects ($\Z^0$ exchange)
for flavour production, angular orientation, cross sections and
initial-state photon radiation in continuum events.
\begin{subentry}
\iteme{= 1 :} QED, i.e. no weak effects are included.
\iteme{= 2 :} QFD, i.e. including weak effects.
\iteme{= 3 :} as \ttt{=2}, but at initialization in \ttt{LUXTOT} the
$\Z^0$ width is calculated from $\ssintw$, $\alphaem$ and 
$\Z^0$ and quark masses (including bottom and top threshold factors for
\ttt{MSTJ(103)} odd), assuming three full generations, and the
result is stored in \ttt{PARJ(124)}.
\end{subentry}
 
\iteme{MSTJ(103) :} (D=7) mass effects in continuum matrix elements,
in the form \ttt{MSTJ(103)} $= M_1 + 2M_2 + 4M_3$, where $M_i = 0$
if no mass effects and $M_i = 1$ if mass effects should be included.
Here;
\begin{subentry}
\iteme{$M_1$ :} threshold factor for new flavour production
according to QFD result;
\iteme{$M_2$ :} gluon emission probability (only applies for
\ttt{|MSTJ(101)|}$\leq 1$, otherwise no mass effects anyhow);
\iteme{$M_3$ :} angular orientation of event (only applies for
\ttt{|MSTJ(101)|}$\leq 1$ and
\ttt{MSTJ(102)=1}, otherwise no mass effects anyhow).
\end{subentry}
 
\iteme{MSTJ(104) :} (D=5) number of allowed flavours, i.e. flavours
that can be produced in a continuum event if the energy is enough.
A change to 6 makes top production allowed above the threshold, etc.
Note that in $\q\qbar\q'\qbar'$ events only the first five flavours
are allowed in the secondary pair, produced by a gluon breakup.
 
\iteme{MSTJ(105) :} (D=1) fragmentation and decay in \ttt{LUEEVT} and
\ttt{LUONIA} calls.
\begin{subentry}
\iteme{= 0 :} no \ttt{LUEXEC} calls, i.e. only matrix-element
and/or parton-shower treatment.
\iteme{= 1 :} \ttt{LUEXEC} calls are made to generate fragmentation
and decay chain.
\iteme{= -1 :} no \ttt{LUEXEC} calls and no collapse of small jet
systems into one or two particles (in \ttt{LUPREP}).
\end{subentry}
 
\iteme{MSTJ(106) :} (D=1) angular orientation in \ttt{LUEEVT} and
\ttt{LUONIA}.
\begin{subentry}
\iteme{= 0 :} standard orientation of events, i.e. $\q$ along $+z$ axis
and $\qbar$ along $-z$ axis or in $xz$ plane with $p_x > 0$ for
continuum events, and $\g_1\g_2\g_3$ or $\gamma\g_2\g_3$ in $xz$ plane
with $\g_1$ or $\gamma$ along the $+z$ axis for onium events.
\iteme{= 1 :} random orientation according to matrix elements.
\end{subentry}
 
\iteme{MSTJ(107) :} (D=0) radiative corrections to continuum events.
\begin{subentry}
\iteme{= 0 :} no radiative corrections.
\iteme{= 1 :} initial-state radiative corrections (including weak
effects for \ttt{MSTJ(102)=} 2 or 3).
\end{subentry}
 
\iteme{MSTJ(108) :} (D=2) calculation of $\alphas$ for matrix-element
alternatives. The \ttt{MSTU(111)} and \ttt{PARU(112)} values are
automatically overwritten in \ttt{LUEEVT} or \ttt{LUONIA} calls
accordingly.
\begin{subentry}
\iteme{= 0 :} fixed $\alphas$ value as given in \ttt{PARU(111)}.
\iteme{= 1 :} first-order formula is always used, with
$\Lambda_{\mrm{QCD}}$ given by \ttt{PARJ(121)}.
\iteme{= 2 :} first- or second-order formula is used, depending on
value of \ttt{MSTJ(101)}, with $\Lambda_{\mrm{QCD}}$ given by
\ttt{PARJ(121)} or \ttt{PARJ(122)}.
\end{subentry}
 
\iteme{MSTJ(109) :} (D=0) gives a possibility to switch from QCD
matrix elements to some alternative toy models. Is not relevant for
shower evolution, \ttt{MSTJ(101)=5}, where one can use
\ttt{MSTJ(49)} instead.
\begin{subentry}
\iteme{= 0 :} standard QCD scenario.
\iteme{= 1 :} a scalar gluon model. Since no second-order corrections
are available in this scenario, one can only use this with
\ttt{MSTJ(101) = 1} or \ttt{-1}. Also note that the event-as-a-whole
angular distribution is for photon exchange only (i.e. no weak
effects), and that no higher-order corrections to the total
cross section are included.
\iteme{= 2 :} an Abelian vector gluon theory, with the colour factors
$C_F = 1$ ($= 4/3$ in QCD), $N_C = 0$ ($= 3$ in QCD) and
$T_R = 3 n_f$ ($= n_f/2$ in QCD). If one selects
$\alpha_{\mrm{Abelian}} = (4/3) \alpha_{\mrm{QCD}}$,
the 3-jet cross section will agree with
the QCD one, and differences are to be found only in 4-jets.
The \ttt{MSTJ(109)=2} option has to be run with
\ttt{MSTJ(110)=1} and \ttt{MSTJ(111)=0}; if need be, the latter
variables will be overwritten by the program. \\
{\bf Warning:} second-order corrections give a large negative
contribution to the 3-jet cross section, so large that
the whole scenario is of doubtful use. In order to make the
second-order options work at all, the 3-jet cross section
is here by hand set exactly equal to zero for \ttt{MSTJ(101)=2}.
It is here probably better to use the option \ttt{MSTJ(101)=3},
although this is not a consistent procedure either.
\end{subentry}
 
\iteme{MSTJ(110) :} (D=2) choice of second-order contributions
to the 3-jet rate.
\begin{subentry}
\iteme{= 1 :} the GKS second-order matrix elements, i.e. the old
{\Je} standard.
\iteme{= 2 :} the Zhu parametrization of the ERT matrix elements,
based on the program of Kunszt and Ali, i.e. in historical sequence
ERT/Kunszt/Ali/Zhu. The parametrization is available for
$y =$ 0.01, 0.02, 0.03, 0.04 and 0.05. Values outside this
range are put at the nearest border, while those inside
it are given by a linear interpolation between the
two nearest points. Since this procedure is rather primitive,
one should try to work at one of the values given above.
Note that no Abelian QCD parametrization is available for
this option.
\end{subentry}
 
\iteme{MSTJ(111) :} (D=0) use of optimized perturbation theory for
second-order matrix elements (it can also be used for first-order
matrix elements, but here it only corresponds to a trivial
rescaling of the $\alphas$ argument).
\begin{subentry}
\iteme{= 0 :} no optimization procedure; i.e. $Q^2 = E_{\mrm{cm}}^2$.
\iteme{= 1 :} an optimized $Q^2$ scale is chosen as
$Q^2 = f E_{\mrm{cm}}^2$, where $f =$\ttt{PARJ(128)} for the total
cross section $R$ factor, while $f =$\ttt{PARJ(129)} for the
3- and 4-jet rates. This $f$ value enters via the
$\alphas$, and also via a term proportional to $\alphas^2 \ln f$.
Some constraints are imposed; thus the optimized `3-jet'
contribution to $R$ is assumed to be positive (for \ttt{PARJ(128)}), 
the total 3-jet rate is not allowed to be negative
(for \ttt{PARJ(129)}), etc.
However, there is no guarantee that the differential 3-jet
cross section is not negative (and truncated to 0) somewhere
(this can also happen with $f = 1$, but is then less frequent).
The actually obtained $f$ values are stored in \ttt{PARJ(168)} and
\ttt{PARJ(169)}, respectively.
If an optimized $Q^2$ scale is used, then the $\Lambda_{\mrm{QCD}}$
(and $\alphas$) should also be changed. With the value $f = 0.002$,
it has been shown \cite{Bet89} that a $\Lambda_{\mrm{QCD}} = 0.100$
GeV gives a reasonable agreement; the parameter to be changed is
\ttt{PARJ(122)} for a second-order running $\alphas$. Note that,
since the optimized $Q^2$ scale is sometimes below the charm
threshold, the effective number of flavours used in $\alphas$ may
well be 4 only. If one feels that it is still appropriate to use 5
flavours (one choice might be as good as the other), it is
necessary to put \ttt{MSTU(113)=5}.
\end{subentry}
 
\iteme{MSTJ(115) :} (D=1) documentation of continuum or onium
events, in increasing order of completeness.
\begin{subentry}
\iteme{= 0 :} only the parton shower, the fragmenting partons and the
generated hadronic system are stored in the \ttt{LUJETS} common block.
\iteme{= 1 :} also a radiative photon is stored (for continuum events).
\iteme{= 2 :} also the original $\ee$ are stored (with 
\ttt{K(I,1)=21}).
\iteme{= 3 :} also the $\gamma$ or $\gammaZ$ exchanged for continuum
events, the onium state for resonance events is stored (with 
\ttt{K(I,1)=21}).
\end{subentry}
 
\iteme{MSTJ(116) :} (D=1) initialization of total cross section and
radiative photon spectrum in \ttt{LUEEVT} calls.
\begin{subentry}
\iteme{= 0 :} never; cannot be used together with radiative
corrections.
\iteme{= 1 :} calculated at first call and then whenever \ttt{KFL}
or \ttt{MSTJ(102)} is changed or \ttt{ECM} is changed by more than
\ttt{PARJ(139)}.
\iteme{= 2 :} calculated at each call.
\iteme{= 3 :} everything is reinitialized in the next call, but
\ttt{MSTJ(116)} is afterwards automatically put \ttt{=1} for use
in subsequent calls.
\end{subentry}
 
\iteme{MSTJ(119) :} (I) check on need to reinitialize \ttt{LUXTOT}.
 
\iteme{MSTJ(120) :} (R) type of continuum event generated with the
matrix-element option (with the shower one, the result is always
\ttt{=1}).
\begin{subentry}
\iteme{= 1 :} $\q\qbar$.
\iteme{= 2 :} $\q\qbar\g$.
\iteme{= 3 :} $\q\qbar\g\g$ from Abelian (QED-like) graphs in
matrix element.
\iteme{= 4 :} $\q\qbar\g\g$ from non-Abelian (i.e. containing
triple-gluon coupling) graphs in matrix element.
\iteme{= 5 :} $\q\qbar\q'\qbar'$.
\end{subentry}
 
\iteme{MSTJ(121) :} (R) flag set if a negative differential
cross section was encountered in the latest \ttt{LUX3JT} call.
Events are still generated, but maybe not quite according to
the distribution one would like (the rate is set to zero in the
regions of negative cross section, and the differential rate
in the regions of positive cross section is rescaled to give
the `correct' total 3-jet rate).

\boxsep 

\iteme{PARJ(121) :}\label{p:PARJ121} (D=1.0 GeV) $\Lambda$ value 
used in first-order
calculation of $\alphas$ in the matrix-element alternative.
 
\iteme{PARJ(122) :} (D=0.25 GeV) $\Lambda$ values used in second-order
calculation of $\alphas$ in the matrix-element alternative.
 
\iteme{PARJ(123) :} (D=91.187 GeV) mass of $\Z^0$ as used in
propagators for the QFD case.
 
\iteme{PARJ(124) :} (D=2.489 GeV) width of $\Z^0$ as used in
propagators for the QFD case. Overwritten at initialization if
\ttt{MSTJ(102)=3}.
 
\iteme{PARJ(125) :} (D=0.01) $y_{\mrm{cut}}$, minimum squared scaled 
invariant mass of any two partons in 3- or 4-jet events; the main
user-controlled matrix-element cut. \ttt{PARJ(126)} provides an
additional constraint. For each new event, it is additionally
checked that the total 3- plus 4-jet fraction does not
exceed unity; if so the effective $y$ cut will be dynamically
increased. The actual $y$-cut value is stored in
\ttt{PARJ(150)}, event by event.
 
\iteme{PARJ(126) :} (D=2. GeV) minimum invariant mass of any two
partons in 3- or 4-jet events; a cut in addition to the one above,
mainly for the case of a radiative photon lowering the hadronic c.m.
energy significantly.
 
\iteme{PARJ(127) :} (D=1. GeV) is used as a safety margin for small
colour-singlet jet systems, cf. \ttt{PARJ(32)}, specifically
$\q\qbar'$ masses in $\q\qbar\q'\qbar'$ 4-jet events and $\g\g$ mass
in onium $\gamma\g\g$ events.
 
\iteme{PARJ(128) :} (D=0.25) optimized $Q^2$ scale for the QCD $R$
(total rate) factor for the \ttt{MSTJ(111)=1} option is given by
$Q^2 = f E_{\mrm{cm}}^2$, where $f =$\ttt{PARJ(128)}. For various 
reasons the actually used $f$ value may be increased compared with 
the nominal one; while \ttt{PARJ(128)} gives the nominal value, 
\ttt{PARJ(168)} gives the actual one for the current event.
 
\iteme{PARJ(129) :} (D=0.002) optimized $Q^2$ scale for the 3-
and 4-jet rate for the \ttt{MSTJ(111)=1} option is given by
$Q^2 = f E_{\mrm{cm}}^2$, where $f =$\ttt{PARJ(129)}. For various 
reasons the actually used $f$ value may be increased compared with 
the nominal one; while \ttt{PARJ(129)} gives the nominal value, 
\ttt{PARJ(169)} gives the actual one for the current event. The 
default value is in agreement with the studies of Bethke \cite{Bet89}.
 
\iteme{PARJ(131), PARJ(132) :} (D=2*0.) longitudinal polarizations
$P_{\mrm{L}}^+$ and $P_{\mrm{L}}^-$ of incoming $\e^+$ and $\e^-$.
 
\iteme{PARJ(133) :} (D=0.) transverse polarization
$P_{\mrm{T}} = \sqrt{P_{\mrm{T}}^+ P_{\mrm{T}}^-}$, with 
$P_{\mrm{T}}^+$ and $P_{\mrm{T}}^-$ transverse
polarizations of incoming $\e^+$ and $\e^-$.
 
\iteme{PARJ(134) :} (D=0.) mean of transverse polarization
directions of incoming $\e^+$ and $\e^-$,
$\Delta \varphi = (\varphi^+ + \varphi^-) /2$, with $\varphi$
the azimuthal angle of polarization, leading to a shift in the
$\varphi$ distribution of jets by $\Delta \varphi$.
 
\iteme{PARJ(135) :} (D=0.01) minimum photon energy fraction
(of beam energy) in initial-state radiation; should normally
never be changed (if lowered too much, the fraction of events
containing a radiative photon will exceed unity, leading to
problems).
 
\iteme{PARJ(136) :} (D=0.99) maximum photon energy fraction
(of beam energy) in initial-state radiation; may be changed
to reflect actual trigger conditions of a detector (but must
always be larger than \ttt{PARJ(135)}).
 
\iteme{PARJ(139) :} (D=0.2 GeV) maximum deviation of $E_{\mrm{cm}}$ 
from the corresponding value at last \ttt{LUXTOT} call, above which 
a new call is made if \ttt{MSTJ(116)=1}.
 
\iteme{PARJ(141) :} (R) value of $R$, the ratio of continuum
cross section to the lowest-order muon pair production cross section,
as given in massless QED (i.e. three times the sum of active
quark squared charges, possibly modified for polarization).
 
\iteme{PARJ(142) :} (R) value of $R$ including quark-mass effects
(for \ttt{MSTJ(102)=1}) and/or weak propagator effects
(for \ttt{MSTJ(102)=2}).
 
\iteme{PARJ(143) :} (R) value of $R$ as \ttt{PARJ(142)}, but
including QCD corrections as given by \ttt{MSTJ(101)}.
 
\iteme{PARJ(144) :} (R) value of $R$ as \ttt{PARJ(143)}, but
additionally including corrections from initial-state photon
radiation (if \ttt{MSTJ(107)=1}). Since the effects of heavy
flavour thresholds are not simply integrable, the initial value
of \ttt{PARJ(144)} is updated during the
course of the run to improve accuracy.
 
\iteme{PARJ(145) - PARJ(148) :} (R) absolute cross sections in nb
as for the cases \ttt{PARJ(141) - PARJ(144)} above.
 
\iteme{PARJ(150) :} (R) current effective matrix element cut-off
$y_{\mrm{cut}}$, as given by \ttt{PARJ(125), PARJ(126)} and the
requirements of having non-negative cross sections for 2-,
3- and 4-jet events. Not used in parton showers.
 
\iteme{PARJ(151) :} (R) value of c.m. energy \ttt{ECM} at last
\ttt{LUXTOT} call.
 
\iteme{PARJ(152) :} (R) current first-order contribution to the
3-jet fraction; modified by mass effects. Not used in parton
showers.
 
\iteme{PARJ(153) :} (R) current second-order contribution to the
3-jet fraction; modified by mass effects. Not used in parton
showers.
 
\iteme{PARJ(154) :} (R) current second-order contribution to the
4-jet fraction; modified by mass effects. Not used in parton
showers.
 
\iteme{PARJ(155) :} (R) current fraction of 4-jet rate
attributable to $\q\qbar\q'\qbar'$ events rather than $\q\qbar\g\g$
ones; modified by mass effects. Not used in parton showers.
 
\iteme{PARJ(156) :} (R) has two functions when using second-order
QCD. For a 3-jet event, it gives the ratio of the second-order
to the total 3-jet cross section in the given kinematical
point. For a 4-jet event, it gives the ratio of the
modified 4-jet cross section, obtained when neglecting interference 
terms whose colour flow is not well defined, to the full
unmodified one, all evaluated in the given kinematical point.
Not used in parton showers.
 
\iteme{PARJ(157) - PARJ(159) :} (I) used for cross-section
calculations to include mass threshold effects to radiative
photon cross section. What is stored is basic cross section,
number of events generated and number that passed cuts.
 
\iteme{PARJ(160) :} (R) nominal fraction of events that should
contain a radiative photon.
\iteme{PARJ(161) - PARJ(164) :} (I) give shape of radiative photon
spectrum including weak effects.
 
\iteme{PARJ(168) :} (R) actual $f$ value of current event in
optimized perturbation theory for $R$; see \ttt{MSTJ(111)} and
\ttt{PARJ(128)}.
 
\iteme{PARJ(169) :} (R) actual $f$ value of current event in
optimized perturbation theory for 3- and 4-jet rate;
see \ttt{MSTJ(111)} and \ttt{PARJ(129)}.
 
\iteme{PARJ(171) :} (R) fraction of cross section corresponding
to the axial coupling of quark pair to the intermediate $\gammaZ$
state; needed for the Abelian gluon model 3-jet matrix
element.
 
\end{entry}
 
\subsection{Examples}
 
An ordinary $\ee$ annihilation event in the continuum, at a c.m.
energy of 40 GeV, may be generated with
\begin{verbatim}
      CALL LUEEVT(0,40.)
\end{verbatim}
In this case a $\q\qbar$ event is generated, including weak effects,
followed by parton-shower evolution and fragmentation/decay treatment.
Before a call to \ttt{LUEEVT}, however, a number of default values
may be changed, e.g. \ttt{MSTJ(101)=2} to use second-order QCD
matrix elements, giving a mixture of $\q\qbar$, $\q\qbar\g$,
$\q\qbar\g\g$, and $\q\qbar\q'\qbar'$ events, \ttt{MSTJ(102)=1} to
have QED only, \ttt{MSTJ(104)=6} to allow $\t\tbar$ production
as well, \ttt{MSTJ(107)=1} to include initial-state photon radiation
(including a treatment of the $\Z^0$ pole), \ttt{PARJ(123)=92.0} to
change the $\Z^0$ mass, \ttt{PARJ(81)=0.3} to change the 
parton-shower $\Lambda$ value, or \ttt{PARJ(82)=1.5} to change the 
parton-shower cut-off. If initial-state photon radiation is used, some
restrictions apply to how one can alternate the generation of
events at different energies or with different $\Z^0$ mass, etc.
These restrictions are not there for
efficiency reasons (the extra time for recalculating the extra
constants every time is small), but because it ties in with the
cross-section calculations (see \ttt{PARJ(144)}).
 
Most parameters can be changed independently of each other. However,
if just one or a few parameters/switches are changed, one should not
be surprised to find a rather bad agreement with the data, like e.g.
a too low or high average hadron multiplicity. It is therefore usually
necessary to retune one parameter related to the perturbative QCD
description, like $\alphas$ or $\Lambda$, one of the two parameters
$a$ and $b$ of the Lund symmetric fragmentation function (since they
are so strongly correlated, it is often not necessary to retune both
of them), and the average fragmentation transverse momentum --- see
Note~2 of the \ttt{MSTJ(101)} description for an example. For very
detailed studies it may be necessary to retune even more parameters.
 
The three-gluon and gluon--gluon--photon decays of $\Upsilon$ may be
simulated by a call
\begin{verbatim}
      CALL LUONIA(5,9.46)
\end{verbatim}
Unfortunately, with present top-mass limits, this routine will not be
of much interest for toponium studies (weak decays will dominate).
 
A typical program for analysis of $\ee$ annihilation events at
100 GeV might look something like
\begin{verbatim}
      COMMON/LUJETS/N,K(4000,5),P(4000,5),V(4000,5)
      COMMON/LUDAT1/MSTU(200),PARU(200),MSTJ(200),PARJ(200)
      COMMON/LUDAT2/KCHG(500,3),PMAS(500,4),PARF(2000),VCKM(4,4)
      COMMON/LUDAT3/MDCY(500,3),MDME(2000,2),BRAT(2000),KFDP(2000,5)
      MDCY(LUCOMP(111),1)=0           ! put pi0 stable
      MSTJ(107)=1                     ! include initial-state radiation
      PARU(41)=1.                     ! use linear sphericity
      .....                           ! other desired changes
      .....                           ! initialize analysis statistics
      DO 100 IEVENT=1,1000            ! loop over events
      CALL LUEEVT(0,100.)             ! generate new event
      IF(IEVENT.EQ.1) CALL LULIST(2)  ! list first event
      CALL LUTABU(11)                 ! save particle composition
                                      !   statistics
      CALL LUEDIT(2)                  ! remove decayed particles
      CALL LUSPHE(SPH,APL)            ! linear sphericity analysis
      IF(SPH.LT.0.) GOTO 100          ! too few particles in event for
                                      !   LUSPHE to work on it (unusual)
      CALL LUEDIT(31)                 ! orient event along axes above
      IF(IEVENT.EQ.1) CALL LULIST(2)  ! list first treated event
      .....                           ! fill analysis statistics
      CALL LUTHRU(THR,OBL)            ! now do thrust analysis
      .....                           ! more analysis statistics
  100 CONTINUE                        !
      CALL LUTABU(12)                 ! print particle composition
                                      !   statistics
      .....                           ! print analysis statistics
      END
\end{verbatim}
 
\clearpage
 
\section{Process Generation in PYTHIA}
\label{s:PYTprocgen}
 
Much can be said about the hard processes in {\Py} and
the way they are generated. Therefore the material has been split
into three sections. In the current one the philo\-sophy underlying
the event generation scheme is presented.  Here we provide a
generic description, where some special cases are swept under the
carpet. In the next section, the existing processes are enumerated,
with some comments about applications and limitations. Finally, in
the third section the generation routines and common block switches
are described.
 
The section starts with a survey of parton distributions, followed
by a detailed description of the simple $2 \to 2$ and $2 \to 1$
hard subprocess generation schemes, including pairs of resonances.
This is followed by a few comments on more complicated configurations.
 
\subsection{Parton Distributions}
\label{ss:structfun}
 
The parton-distribution function $f_i^a(x,Q^2)$ parametrizes the 
probability to find a parton $i$ with a fraction $x$ of the beam energy 
when the beam particle $a$ is probed by a hard scattering at virtuality 
scale $Q^2$. Usually the momentum-weighted combination $x f_i^a(x,Q^2)$ 
is used, for which the normalization condition
$\sum_i \int_0^1 dx \, x f_i^a(x,Q^2) \equiv 1$ normally applies.
The $Q^2$ dependence of parton distributions is perturbatively
calculable, see section \ref{sss:initshowstruc}.
 
The parton distributions in {\Py} come in many shapes, as shown in the
following.
 
\subsubsection{Baryons}
 
For protons, many sets exist on the market. These are obtained by fits
to experimental data, constrained so that the $Q^2$ dependence is in
accordance with the standard QCD evolution equations. The (new) default 
in \tsc{Pythia} is CTEQ2L \cite{Bot93}, a modern leading-order fit. 
Ten other sets are found in \tsc{Pythia}. The complete list is:
\begin{Itemize}
\item EHLQ sets 1 and 2 \cite{Eic84};
\item DO sets 1 and 2 \cite{Duk82};
\item the other CTEQ2 fits, namely CTEQ2M, CTEQ2MS, CTEQ2MF, CTEQ2ML,
and CTEQ2D \cite{Bot93}; and
\item the dynamically generated fit GRV LO (updated version)
\cite{Glu92}.
\end{Itemize}
Of these, EHLQ, DO, CTEQ2L and GRV LO are leading-order parton 
distributions,
while CTEQ2D are in the next-to-leading-order DIS scheme and the rest
in the next-to-leading order $\br{\mrm{MS}}$ scheme. The EHLQ and DO 
sets are by now rather old, and are kept mainly for backwards
compatibility. Since only Born-level matrix elements
are included in the program, there is no particular reason to use
higher-order parton distributions --- the resulting combination is
anyway only good to leading-order accuracy. (Some higher-order 
corrections are effectively included by the parton-shower
treatment, but there is no exact match.)
 
There is a steady flow of new parton-distribution sets on the market.
To keep track of all of them is a major work on its own. Therefore
{\Py} contains an interface to an external library of parton
distribution functions, \tsc{Pdflib} \cite{Plo93}.
This is a truly encyclopedic collection of almost all proton, pion 
and photon parton distributions proposed since the late 70's.
Two dummy routines come with the {\Py} package, so as to avoid 
problems with unresolved external references if \tsc{Pdflib} is not 
linked. One should also note that {\Py} does not check the results, 
but assumes that sensible answers will be returned, also outside the 
nominal $(x, Q^2)$ range of a set. Only the sets that come with 
{\Py} have been suitably modified to provide reasonable answers 
outside their nominal domain of validity.
 
From the proton parton distributions, those of the neutron are obtained
by isospin conjugation, i.e. $f_{\u}^{\n} = f_{\d}^{\p}$ and
 $f_{\d}^{\n} = f_{\u}^{\p}$.
 
The program does allow for incoming beams of a number of hyperons:
$\Lambda^0$, $\Sigma^{-,0,+}$, $\Xi^{-,0}$ and $\Omega^-$. Here
one has essentially no experimental information. One could imagine
to construct models in which valence $\s$ quarks are found at larger
average $x$ values than valence $\u$ and $\d$ ones, because of the
larger $\s$-quark mass. However, hyperon beams is a little-used part
of the program, included only for a few specific studies. Therefore
a simple approach has been taken, in which an average valence
quark distribution is constructed as
$f_{\mrm{val}} = (f_{\u,\mrm{val}}^{\p} + f_{\d,\mrm{val}}^{\p})/3$, 
according to
which each valence quark in a hyperon is assumed to be distributed.
Sea-quark and gluon distributions are taken as in the proton.
Any proton parton distribution set may be used with this procedure.
 
\subsubsection{Mesons and photons}
 
Data on meson parton distributions are scarce, so only very few sets
have been constructed, and only for the $\pi^{\pm}$. {\Py} contains
the Owens set 1 and 2 parton distributions \cite{Owe84}, which for a
long time were essentially the only sets on the market, and the
more recent dynamically generated GRV LO (updated version) 
\cite{Glu92a}. The first one is the default in {\Py}. Further sets are
found in \tsc{Pdflib} and can therefore be used by {\Py}, just as
described above for protons.
 
Sets of photon parton distributions have been obtained as for
hadrons; an additional complication comes from the necessity to
handle the matching of the vector meson dominance (VMD) and the
perturbative pieces in a consistent manner. New sets have been 
produced where this division is explicit and therefore especially
well suited for applications to event generation\cite{Sch95}.
The Schuler and Sj\"ostand set 1D is the default. Although the 
vector-meson philosophy is at the base, the details of the fits do 
not rely on pion data, but only on $F_2^{\gamma}$ data. Here 
follows a brief summary of relevant details.

Photons obey a set of inhomogeneous evolution equations, where the 
inhomogeneous term is induced by $\gamma \to \q\qbar$ branchings.
The solution can be written as the sum of two terms,
\begin{equation}
f_a^{\gamma}(x,Q^2) = f_a^{\gamma,\mrm{NP}}(x,Q^2;Q_0^2) 
+ f_a^{\gamma,\mrm{PT}}(x,Q^2;Q_0^2) ~,
\end{equation}
where the former term is a solution to the homogeneous evolution
with a (non-perturbative) input at $Q=Q_0$ and the latter is a
solution to the full inhomogeneous equation with boundary condition
$f_a^{\gamma,\mrm{PT}}(x,Q_0^2;Q_0^2) \equiv 0$. One possible 
physics interpretation is to let $f_a^{\gamma,\mrm{NP}}$ correspond
to $\gamma \leftrightarrow V$ fluctuations, where 
$V = \rho^0, \omega, \phi, \ldots$ is a set of vector mesons,
and let $f_a^{\gamma,\mrm{PT}}$ correspond to perturbative (`anomalous')
$\gamma \leftrightarrow \q\qbar$ fluctuations. The discrete spectum 
of vector mesons can be combined with the continuous (in virtuality 
$k^2$) spectrum of $\q\qbar$ fluctuations, to give
\begin{equation}
f_a^{\gamma}(x,Q^2) =
\sum_V \frac{4\pi\alphaem}{f_V^2} f_a^{\gamma,V}(x,Q^2) 
+ \frac{\alphaem}{2\pi} \, \sum_{\q} 2 e_{\q}^2 \, 
\int_{Q_0^2}^{Q^2} \frac{{\d} k^2}{k^2} \, 
f_a^{\gamma,\q\qbar}(x,Q^2;k^2) ~,
\end{equation}
where each component $f^{\gamma,V}$ and $f^{\gamma,\q\qbar}$ obeys a
unit momentum sum rule.
  
In sets 1 the $Q_0$ scale is picked at a low value, 0.6~GeV, where
an identification of the non-perturbative component with a set of
low-lying mesons appear natural, while sets 2 use a higher value,
2~GeV, where the validity of perturbation theory is better established.
The data are not good enough to allow a precise determination of
$\Lambda_{\mrm{QCD}}$. Therefore we use a fixed value 
$\Lambda^{(4)} = 200$~MeV, in agreement with conventional 
results for proton distributions. In the VMD component the $\rho^0$ 
and $\omega$ have been added coherently, so that
$\u\ubar : \d\dbar = 4 : 1$ at $Q_0$.

Unlike the $\p$, the $\gamma$ has a direct component where the photon 
acts as an unresolved probe. In the definition of $F_2^{\gamma}$ this 
adds a component $C^{\gamma}$, symbolically
\begin{equation}
F_2^{\gamma}(x,Q^2) = \sum_{\q} e_{\q}^2 \left[ f_{\q}^{\gamma} + 
f_{\qbar}^{\gamma} \right] \otimes C_{\q} + 
f_{\g}^{\gamma} \otimes C_{\g} + C^{\gamma} ~.
\end{equation}
Since $C^{\gamma} \equiv 0$ in leading order, and since we stay with 
leading-order fits, it is permissible to neglect this complication.
Numerically, however, it makes a non-negligible difference. We 
therefore make two kinds of fits, one DIS type with $C^{\gamma} = 0$
and one {\MSbar} type including the universal part
of $C^{\gamma}$.

When jet production is studied for real incoming photons, the standard 
evolution approach is reasonable also for heavy flavours, i.e. 
predominantly the $\c$, but with a lower cut-off $Q_0 \approx m_{\c}$ 
for $\gamma \to \c\cbar$. Moving to deep inelastic scattering, 
$\e\gamma \to \e X$, there is an extra kinematical constraint:
$W^2 = Q^2 (1-x)/x > 4 m_{\c}^2$. It is here better to use the
`Bethe-Heitler' cross section for $\gamma^* \gamma \to \c\cbar$.
Therefore each distribution appears in two variants. For applications
to real $\gamma$'s the parton distributions are calculated as the sum 
of a vector-meson part and an anomalous part including all five flavours.
For applications to DIS, the sum runs over the same vector-meson part, 
an anomalous part and possibly a $C^{\gamma}$ part for the three 
light flavours, and a Bethe-Heitler part for $\c$ and $\b$. 

In addition to the SaS sets, {\Py} also contains the Drees--Grassie set 
of parton distributions \cite{Dre85} and, as for the proton, there is 
an interface to the \tsc{Pdflib} library \cite{Plo93}.
However, these sets do not allow a subdivision of the photon parton
distributions into one VMD part and one anomalous part. This 
subdivision is necessary a sophisticated modelling of $\gamma\p$ and
$\gamma\gamma$ events, see above and section \ref{sss:photoprod}. 
As an alternative, for the VMD part alone, the 
$\rho^0$ parton distribution can be found from the assumed equality
\begin{equation}
f^{\rho^0}_i = f^{\pi^0}_i = \frac{1}{2} \, (f^{\pi^+}_i +
f^{\pi^-}_i) ~.
\end{equation}
Thus any $\pi^+$ parton distribution set, from any library, can be 
turned into a VMD $\rho^0$ set. The $\omega$ parton distribution is 
assumed the same, while the $\phi$ and $\Jpsi$ ones are handled 
in the very crude approximation 
$f^{\phi}_{\s,\mrm{val}} = f^{\pi^+}_{\u,\mrm{val}}$ and
$f^{\phi}_{\mrm{sea}} = f^{\pi^+}_{\mrm{sea}}$.
therefore is default. The VMD part needs to be complemented by an 
anomalous part to make upp a full photon distribution. The latter
is fully perturbatively calculable, given the lower cut-off scale
$Q_0$. The SaS parametrization of the anomalous part is therefore used 
throughout for this purpose. The $Q_0$ scale can be set freely in the
\ttt{PARP(15)} parameter.

The $f_i^{\gamma,\mrm{anom}}$ distribution can be further decomposed, 
by the flavour and the $\pT$ of the original branching 
$\gamma \to \q\qbar$. The flavour is distributed according to squared
charge (plus flavour thresholds for heavy flavours) and the $\pT$
according to $\d \pT^2 / \pT^2$ in the range $Q_0 < \pT < Q$.
At the branching scale, the photon only consists of a $\q\qbar$ pair,
with $x$ distribution $\propto x^2 + (1-x)^2$. A component
$f_a^{\gamma,\q\qbar}(x,Q^2;k^2)$, characterized by its $k \approx \pT$ 
and flavour, then is evolved homogeneously from $\pT$ to $Q$. For 
theoretical studies it is convenient to be able to access a specific 
component of this kind. Therefore also leading-order parametrizations 
of these decomposed distributions are available \cite{Sch95}.    

\subsubsection{Leptons}
\label{sss:estructfun}
 
Contrary to the hadron case, there is no necessity to introduce the
parton-distribution function concept for leptons. A lepton can be 
considered as a point-like particle, with initial-state radiation 
handled by higher-order matrix elements. However, the parton 
distribution function approach offers a slightly simplified but very 
economical description of initial-state radiation effects for any hard 
process, also those for which higher-order corrections are not yet 
calculated.
 
Parton distributions for electrons have been introduced in {\Py},
but not yet for muons, i.e. currently
$f_{\mu}^{\mu}(x, Q^2) = \delta (x-1)$. Also for the electron one
is free to use a simple `unresolved' $\e$, 
$f_{\e}^{\e}(x, Q^2) = \delta(x-1)$, where the $\e$ retains the
full original momentum.
 
Electron parton distributions are calculable entirely from first
principles, but different levels of approximation may be used.
The parton-distribution formulae in {\Py} are based on a
next-to-leading-order exponentiated description, see ref.
\cite{Kle89}, p. 34. The approximate behaviour is
\begin{eqnarray}
& & f_{\e}^{\e}(x,Q^2) \approx \frac{\beta}{2}
(1-x)^{\beta/2-1}   ~;              \nonumber \\
& & \beta = \frac{2 \alphaem}{\pi}
\left( \ln \frac{Q^2}{m_{\e}^2} -1 \right) ~.
\end{eqnarray}
The form is divergent but integrable for $x \to 1$, i.e. the electron
likes to keep most of the energy. To handle the numerical precision
problems for $x$ very close to unity, the parton distribution is set, by
hand, to zero for $x > 0.999999$, and is rescaled upwards in the range
$0.9999 < x < 0.999999$, in such a way that the total area under the
parton distribution is preserved:
\begin{equation}
\left( f_{\e}^{\e}(x,Q^2) \right)_{\mrm{mod}} =
\left\{ \begin{array}{ll}
 f_{\e}^{\e}(x,Q^2) & 0 \leq x \leq 0.9999 \\[2mm]
 \frac{\displaystyle 100^{\beta/2}}{\displaystyle 100^{\beta/2}-1}
\, f_{\e}^{\e}(x,Q^2) & 0.9999 < x \leq 0.999999   \\[4mm]
0 & x > 0.999999 \, ~.
\end{array} \right.
\end{equation}
 
The branchings $\e \to \e \gamma$, which are responsible for the
softening of the $f_{\e}^{\e}$ parton distribution, also gives rise to
a flow of photons. In photon-induced hard processes, the
$f_{\gamma}^{\e}$ parton distribution can be used to describe the
equivalent flow of photons. The formula used in the program is the
simple first-order expression. There is some ambiguity in
the choice of $Q^2$ range over which emissions should be included.
The na\"{\i}ve (default) choice is
\begin{equation}
f_{\gamma}^{\e}(x,Q^2) = \frac{\alphaem}{2 \pi} \,
\frac{1+(1-x)^2}{x} \, \ln \left( \frac{Q^2}{m_{\e}^2} \right) ~.
\end{equation}
Here it is assumed that only one scale enters the problem, namely
that of the hard interaction, and that the scale of the branching
$\e \to \e \gamma$ is bounded from above by the hard interaction scale.
For a pure QCD or pure QED shower this is an appropriate procedure,
cf. section \ref{sss:showermatching}, but in other cases it may not be
optimal. In particular, for photoproduction the alternative that is
probably most appropriate is \cite{Ali88}:
\begin{equation}
f_{\gamma}^{\e}(x,Q^2) =
\frac{\alphaem}{2 \pi} \, \frac{1+(1-x)^2}{x}
\, \ln \left( \frac{Q^2_{\mmax} (1-x)}{m_{\e}^2 \, x^2} \right) ~.
\end{equation}
Here $Q^2_{\mmax}$ is a user-defined cut for the range of
scattered electron kinematics that is counted as
photoproduction. Note that we now deal with two different $Q^2$
scales, one related to the hard subprocess itself, which appears
as the argument of the parton distribution, and the other related to
the scattering of the electron, which is reflected in 
$Q^2_{\mmax}$.
 
In resolved photoproduction or resolved $\gamma\gamma$
interactions, one has to include the parton distributions for quarks
and gluons inside the photon inside the electron. There are no
published sets where results are directly presented in terms of
quark and gluon distributions inside the electron. In the
program, the $f_{\q,\g}^{\e}$ are therefore obtained by a numerical
convolution according to
\begin{equation}
f_{\q,\g}^{\e}(x, Q^2) =
\int_x^1 \frac{dx_{\gamma}}{x_{\gamma}} \, 
f_{\gamma}^{\e}(x_{\gamma}, Q^2) \, f^{\gamma}_{\q,\g} \! 
\left( \frac{x}{x_{\gamma}}, Q^2 \right)   ~,
\label{pg:foldqgine}
\end{equation}
with $f^{\e}_{\gamma}$ as discussed above. The necessity for numerical
convolution makes this parton distribution evaluation rather slow
compared with the others; one should therefore only have it
switched on for resolved photoproduction studies.
 
One can obtain the positron distribution inside an electron,
which is also the electron sea parton distribution, by a convolution
of the two branchings $\e \to \e \gamma$ and $\gamma \to \ee$;
the result is \cite{Che75}
\begin{equation}
f_{\e^+}^{\e^-}(x,Q^2) =
\frac{1}{2} \, \left\{ \frac{\alphaem}{2 \pi} \,
\left( \ln \frac{Q^2}{m_{\e}^2} -1 \right) \right\}^2 \,
\frac{1}{x} \, \left( \frac{4}{3} - x^2 - \frac{4}{3} x^3 +
2x(1+x) \ln x \right)   ~.
\end{equation}
 
Finally, the program also contains the distribution of a
transverse $\W^-$ inside an electron
\begin{equation}
f_{\W}^{\e}(x,Q^2) = \frac{\alphaem}{2 \pi} \,
\frac{1}{4 \ssintw} \, \frac{1+(1-x)^2}{x} \,
\ln \left( 1 + \frac{Q^2}{m_{\W}^2} \right)    ~.
\end{equation}
 
\subsection{Kinematics and Cross section for a $2 \to 2$ Process}
\label{ss:kinemtwo}
 
In this section we begin the description of kinematics selection
and cross-section calculation. The example is for the case of a
$2 \to 2$ process, with final-state masses assumed to be vanishing.
Later on we will expand to finite fixed masses, and to resonances.
 
Consider two incoming beam particles in their c.m. frame, each with
energy $E_{\mrm{beam}}$. The total squared c.m. energy is then
$s = 4 E_{\mrm{beam}}^2$. The two partons that enter the hard 
interaction do not carry the total beam momentum, but only fractions 
$x_1$ and $x_2$, respectively, i.e. they have four-momenta
\begin{eqnarray}
   p_1 & = & E_{\mrm{beam}}(x_1; 0, 0, x_1) ~, \nonumber \\
   p_2 & = & E_{\mrm{beam}}(x_2; 0, 0, -x_2) ~.
\end{eqnarray}
There is no reason to put the incoming partons on the mass shell, 
i.e. to have time-like incoming four-vectors,
since partons inside a particle are always virtual and thus
space-like. These space-like virtualities are introduced as
part of the initial-state parton-shower description, see section
\ref{sss:initshowtrans}, but do not affect the formalism of this
section. The one example where it would be appropriate to put a
parton on the mass shell is for an incoming lepton beam, but even
here the massless kinematics description is adequate as long as
the c.m. energy is correctly calculated with masses.
 
The squared invariant mass of the two partons is defined as
\begin{equation}
  \hat{s} = (p_1 + p_2)^2 = x_1 \, x_2 \, s ~.
\end{equation}
Instead of $x_1$ and $x_2$, it is often customary to use $\tau$
and either $y$ or $x_{\mrm{F}}$:
\begin{eqnarray}
  \tau & = & x_1 x_2 = \frac{\hat{s}}{s} ~;  \\
  y & = & \frac{1}{2} \ln \frac{x_1}{x_2} ~; \\
  x_{\mrm{F}} & = & x_1 - x_2 ~.
\end{eqnarray}
 
In addition to $x_1$ and $x_2$, two additional variables are needed
to describe the kinematics of a scattering $1 + 2 \to 3 + 4$.
One corresponds to the azimuthal angle $\varphi$ of the scattering
plane around the beam axis. This angle is always isotropically
distributed for unpolarized incoming beam particles, and so need not
be considered further. The other variable can be picked as
$\hat{\theta}$, the polar angle of parton 3 in the c.m. frame of the
hard scattering. The conventional choice is to use the variable
\begin{equation}
  \hat{t} = (p_1 - p_3)^2 = (p_2 - p_4)^2 =
  - \frac{\hat{s}}{2} (1 - \cos \hat{\theta}) ~,
\end{equation}
with $\hat{\theta}$ defined as above. In the following, we will make
use of both $\hat{t}$ and $\hat{\theta}$. It is also customary
to define $\hat{u}$,
\begin{equation}
  \hat{u} = (p_1 - p_4)^2 = (p_2 - p_3)^2 =
  - \frac{\hat{s}}{2} (1 + \cos \hat{\theta}) ~,
\end{equation}
but $\hat{u}$ is not an independent variable since
\begin{equation}
  \hat{s} + \hat{t} + \hat{u} = 0 ~.
\end{equation}
 
If the two outgoing particles have masses $m_3$ and $m_4$,
respectively, then the four-momenta in the c.m. frame of the hard
interaction are given by
\begin{equation}
\hat{p}_{3,4} = \left(
\frac{\hat{s} \pm (m_3^2-m_4^2)}{2\sqrt{\hat{s}}} ,
\pm \frac{\sqrt{\hat{s}}}{2} \, \beta_{34} \sin \hat{\theta}, 0,
\pm \frac{\sqrt{\hat{s}}}{2} \, \beta_{34} \cos \hat{\theta}
\right) ~,
\end{equation}
where
\begin{equation}
\beta_{34} = \sqrt{ \left( 1 - \frac{m_3^2}{\hat{s}} -
\frac{m_4^2}{\hat{s}} \right)^2 - 4 \, \frac{m_3^2}{\hat{s}} \,
\frac{m_4^2}{\hat{s}} }    ~.
\end{equation}
Then $\hat{t}$ and $\hat{u}$ are modified to
\begin{equation}
\hat{t}, \hat{u} = - \frac{1}{2} \left\{ ( \hat{s} - m_3^2 - m_4^2 )
\mp \hat{s} \, \beta_{34} \cos \hat{\theta} \right\} ~,
\end{equation}
with
\begin{equation}
\hat{s} + \hat{t} + \hat{u} = m_3^2 + m_4^2 ~.
\end{equation}
 
The cross section for the process $1 + 2 \to 3 + 4$ may be written as
\begin{eqnarray}
\sigma & = & \int \int \int \d x_1 \, \d x_2 \, \d \hat{t} \,
f_1(x_1, Q^2) \, f_2(x_2, Q^2) \, \frac{\d \hat{\sigma}}{\d \hat{t}}
\nonumber \\
& = & \int \int \int \frac{\d \tau}{\tau} \, \d y \, \d \hat{t} \,
x_1 f_1(x_1, Q^2) \, x_2 f_2(x_2, Q^2) \,
\frac{\d \hat{\sigma}}{\d \hat{t}} ~.
\label{pg:sigma}
\end{eqnarray}
 
The choice of $Q^2$ scale is ambiguous, and several alternatives are
available in the program. For massless outgoing particles the default
is the squared transverse momentum
\begin{equation}
Q^2 = \hat{p}_{\perp}^2 = \frac{\hat{s}}{4} \sin^2\hat{\theta} =
    \frac{\hat{t}\hat{u}}{\hat{s}} ~,
\end{equation}
which is modified to
\begin{equation}
Q^2 = \frac{1}{2} (m_{\perp 3}^2 + m_{\perp 4}^2) =
\frac{1}{2} (m_3^2 + m_4^2) + \hat{p}_{\perp}^2 =
\frac{1}{2} (m_3^2 + m_4^2) +
\frac{\hat{t} \hat{u} - m_3^2 m_4^2}{\hat{s}}
\end{equation}
when masses are introduced. The mass term is selected such that,
for $m_3 = m_4 = m$, the expression reduces to the squared transverse
mass, $Q^2 = \hat{m}_{\perp}^2 = m^2 + \hat{p}_{\perp}^2$.
 
The $\d \hat{\sigma}/\d \hat{t}$ expresses the differential
cross section for a scattering, as a function of the kinematical
quantities $\hat{s}$, $\hat{t}$ and $\hat{u}$.
It is in this function that the physics of a given process
resides.
 
The performance of a machine is measured in terms of its
luminosity $\cal L$, which is directly proportional to the
number of particles in each bunch and to the bunch crossing
frequency, and inversely proportional to the area of the bunches at
the collision point. For a process with a $\sigma$ as given by
eq.~(\ref{pg:sigma}), the differential event rate is given
by $\sigma {\cal L}$, and the number of events collected
over a given period of time
\begin{equation}
N = \sigma \, \int {\cal L} \, \d t ~.
\end{equation}
The program does not calculate the number of events, but only the
integrated cross sections.
 
\subsection{Resonance Production}
\label{ss:kinemreson}
 
The simplest way to produce a resonance is by a $2 \to 1$ process.
If the decay of the resonance is not considered, the cross-section
formula does not depend on $\hat{t}$, but takes the form
\begin{equation}
\sigma = \int \int \frac{\d \tau}{\tau} \, \d y \,
x_1 f_1(x_1, Q^2) \, x_2 f_2(x_2, Q^2) \,
\hat{\sigma}(\hat{s})  ~.
\label{pg:sigmares}
\end{equation}
Here the physics is contained in the cross section
$\hat{\sigma}(\hat{s})$. The $Q^2$ scale is usually taken to
be $Q^2 = \hat{s}$.
 
In published formulae, cross sections are often given in
the zero-width approximation, i.e.
$\hat{\sigma}(\hat{s}) \propto \delta (\hat{s} - m_R^2)$,
where $m_R$ is the mass of the resonance. Introducing the
scaled mass $\tau_R = m_R^2/s$, this corresponds to a delta
function $\delta (\tau - \tau_R)$, which can be used to
eliminate the integral over $\tau$.
 
However, what we normally want to do is replace the $\delta$
function by the appropriate Breit--Wigner shape. For a resonance
width $\Gamma_R$ this is achieved by the replacement
\begin{equation}
\delta (\tau - \tau_R) \to \frac{s}{\pi} \,
\frac{m_R \Gamma_R}{(s \tau - m_R^2)^2 + m_R^2 \Gamma_R^2}  ~.
\label{pg:resshapeone}
\end{equation}
In this formula the resonance width $\Gamma_R$ is a constant.
 
An improved description of resonance shapes is obtained if
the width is made $\hat{s}$-dependent (occasionally also 
referred to as mass-dependent width, since $\hat{s}$ is not
always the resonance mass), see e.g. \cite{Ber89}.
To first approximation, this means that the expression
$m_R \Gamma_R$ is to be replaced by $\hat{s} \Gamma_R / m_R$.
To be more precise, in the program the quantity $H_R(\hat{s})$
is introduced, and the Breit--Wigner is written as
\begin{equation}
\delta (\tau - \tau_R) \to \frac{s}{\pi} \,
\frac{H_R(s \tau)}{(s \tau - m_R^2)^2 + H_R^2(s \tau)}  ~.
\label{pg:resshapetwo}
\end{equation}
The $H_R$ factor is evaluated as a sum over all possible final-state
channels, $H_R = \sum_f H_R^{(f)}$. Each decay channel may have its
own $\hat{s}$ dependence, as follows.
 
A decay to a fermion pair, $R \to \f \fbar$, gives no contribution
below threshold, i.e. for $\hat{s} < 4 m_{\f}^2$. Above threshold,
$H_R^{(f)}$ is proportional to $\hat{s}$, multiplied by a threshold
factor $\beta (3 - \beta^2)/2$ for the vector part of a spin 1
resonance, by $\beta^3$ for the axial vector part, and again by
$\beta^3$ for a spin 0 resonance. Here
$\beta = \sqrt{1 - 4m_{\f}^2/\hat{s}}$.
For the decay into unequal masses, e.g. of the $\W^+$, corresponding
but more complicated expressions are used.
 
For decays into a quark pair, the universal first-order strong
correction factor $1 + \alphas(\hat{s}) / \pi$ is included in
$H_R^{(f)}$.
The second-order corrections are often known, but then are specific
to each resonance, and are not included. An option exists for the
$\gamma/\Z^0/\Z'^0$ resonances, where threshold effects due to
$\q\qbar$ bound-state formation are taken into account in a
smeared-out, average sense, see eq.~(\ref{pp:threshenh}).
 
For other decay channels, not into fermion pairs, the $\hat{s}$
dependence is typically more complicated. For instance, the decay
$\H^0 \to \W^+ \W^-$ has a partial width proportional to
$\hat{s}^2$, with a threshold factor $\beta^3$. Since a Higgs
with $m_{\H} < 2 m_{\W}$ could still decay in this channel, it is 
in fact necessary to perform a two-dimensional integral over
the $W^{\pm}$ Breit--Wigner mass distributions to obtain the correct
result (and this has to be done numerically, at least in part).
Fortunately, a Higgs particle lighter than $2 m_{\W}$ is
sufficiently narrow that the integral only needs to be performed
once and for all at initialization (whereas most other partial
widths are recalculated whenever needed). Channels that proceed
via loops, such as $\H \to \g \g$, also display complicated
threshold behaviours.
 
The coupling structure within the electroweak sector is usually
(re)expressed in terms of gauge boson masses, $\alphaem$ and
$\ssintw$, i.e. factors of $G_{\F}$ are replaced according to
\begin{equation}
\sqrt{2} G_{\F} = \frac{\pi \, \alphaem}{\ssintw \, m_{\W}^2} ~.
\end{equation}
Having done that, $\alphaem$ is allowed to run \cite{Kle89}, 
and is evaluated at the $\hat{s}$ scale. Thereby the relevant
electroweak loop correction factors are recovered at the
$m_{\W}/m_{\Z}$ scale. However, the option exists to
go the other way and eliminate $\alpha_em$ in favour of $G_{\F}$.  
Currently $\ssintw$ is not allowed to run.
For the Higgs particle, the couplings to fermions are proportional
to the fermion masses; then also the masses are evaluated at the
$\hat{s}$ scale.
 
In summary, we see that an $\hat{s}$ dependence may enter several
different ways into the $H_R^{(f)}$ expressions from which the
total $H_R$ is built up. Also note that, with the exception of
the term $(\s \tau - m_R^2)^2$ in the denominator of the Breit--Wigner,
no memory remains of the nominal $m_R$ mass: everywhere else, what
enters is the actual resonance mass $\sqrt{\hat{s}}$.
 
When only decays to a specific final state $f$ are considered, the
$H_R$ in the denominator remains the sum over all allowed decay
channels, but the numerator only contains the $H_R^{(f)}$ term
of the final state considered.
 
If the combined production and decay process $i \to R \to f$ is
considered, the same $\hat{s}$ dependence is implicit in the
coupling structure of $i \to R$ as one would have had in
$R \to i$, i.e. to first approximation there is a symmetry between
couplings of a resonance to the initial and to the final state.
The cross section $\hat{\sigma}$
is therefore, in the program, written in the form
\begin{equation}
\hat{\sigma}_{i \to R \to f}(\hat{s}) \propto \frac{\pi}{\hat{s}} \,
\frac{H_R^{(i)}(\hat{s}) \, H_R^{(f)}(\hat{s})}
{(\hat{s} - m_R^2)^2 + H_R^2(\hat{s})} ~.
\label{pg:Hinoutsym}
\end{equation}
As a simple example, the cross section for the process
$\e^- \br{\nu}_{\e} \to \W^- \to \mu^- \br{\nu}_{\mu}$
can be written as
\begin{equation}
\hat{\sigma}(\hat{s}) = 12 \, \frac{\pi}{\hat{s}} \,
\frac{H_{\W}^{(i)}(\hat{s}) \, H_{\W}^{(f)}(\hat{s})}
{(\hat{s} - m_{\W}^2)^2 + H_{\W}^2(\hat{s})} ~,
\end{equation}
where
\begin{equation}
H_{\W}^{(i)}(\hat{s}) = H_{\W}^{(f)}(\hat{s}) =
\frac{\alphaem(\hat{s})}{24 \, \ssintw} \, \hat{s} ~.
\end{equation}
If the effects of several initial and/or final states are studied,
it is straightforward to introduce an appropriate summation in the
numerator.
 
The analogy between the $H_R^{(f)}$ and $H_R^{(i)}$ cannot be pushed
too far, however. The two differ in several important aspects.
Firstly, colour factors appear reversed: the decay $R \to \q\qbar$
contains a colour factor $N_C = 3$ enhancement, while
$\q\qbar \to R$ is instead suppressed by a factor $1/N_C = 1/3$.
Secondly, the $1 + \alphas(\hat{s}) / \pi$ first-order correction
factor for the final state has to be replaced by a more complicated
$K$ factor for the initial state. This factor is not usually known, or 
it is known (to first non-trivial order) but too lengthy to be included
in the program. Thirdly, incoming partons as a rule are space-like.
All the threshold suppression factors of the final state expressions
are therefore irrelevant when production is considered. In sum, the
$H_R^{(f)}$--$H_R^{(i)}$ analogy is mainly useful as a
consistency cross-check, while the two usually are
calculated separately. Exceptions include the rather
messy loop structure involved in $\g\g \to \H^0$ and $\H^0 \to \g\g$,
which is only coded once.
 
It is of some interest to consider the observable resonance shape
when the effects of parton distributions are included. In a
hadron collider, to first approximation, parton distributions tend
to have a behaviour roughly like $f(x) \propto 1/x$ for small $x$
--- this is why $f(x)$ is replaced by $xf(x)$ in eq.
(\ref{pg:sigma}). Instead, the basic parton-distribution behaviour
is shifted into the factor of $1/\tau$ in the integration phase space
$\d \tau/\tau$, cf. eq.~(\ref{pg:sigmares}). When folded with the
Breit--Wigner shape, two effects appear. One is that the overall
resonance is tilted: the low-mass tail is enhanced and the high-mass
one suppressed. The other is that an extremely long tail develops
on the low-mass side of the resonance: when $\tau \to 0$, eq.
(\ref{pg:Hinoutsym}) with $H_R(\hat{s}) \propto \hat{s}$ gives
a $\hat{\sigma}(\hat{s}) \propto \hat{s} \propto \tau$, which
exactly cancels the $1/\tau$ factor mentioned above. Na\"{\i}vely, the
integral over $y$, $\int \d y = - \ln \tau$, therefore gives a
net logarithmic divergence of the resonance shape when $\tau \to 0$.
Clearly, it is then necessary to consider the shape of the 
parton distributions in more detail. At not-too-small $Q^2$, 
the evolution
equations in fact lead to parton distributions more strongly
peaked than $1/x$, typically with $xf(x) \propto x^{-0.3}$, and
therefore a divergence like $\tau^{-0.3}$ in the cross-section
expression. Eventually this divergence is regularized by a closing
of the phase space, i.e. that $H_R(\hat{s})$ vanishes faster than
$\hat{s}$, and by a less drastic small-$x$ parton-distribution
behaviour when $Q^2 \approx \hat{s} \to 0$.
 
The secondary peak at small $\tau$ may give a rather high
cross section, which can even rival that of the ordinary
peak around the nominal mass. This is the case, for instance,
with $\W$ production. Such a peak has never been observed
experimentally, but this is not surprising, since the background
from other processes is overwhelming at low $\hat{s}$.
Thus a lepton of one or a few GeV of transverse momentum is far
more likely to come from the decay of a charm or bottom hadron than
from a `$\W$' of a mass of a few GeV. When resonance production
is studied, it is therefore important to set limits on the mass of
the resonance, so as to agree with the experimental definition,
at least to first approximation. If not, cross-section information
given by the program may be very confusing.

Another problem is that often the matrix elements really are valid
only in the resonance region. The reason is that one usually includes 
only the simplest $s$-channel graph in the calculation. It is this 
`signal' graph that has a peak at the position of the resonance, 
where it (usually) gives much larger cross sections than the other
`background' graphs. Away from the resonance position, `signal' and
`background' may be of comparable order, or the `background' may 
even dominate. There is a quantum mechanical interference when some 
of the `signal' and `background' graphs have the same initial and 
final state, and this interference may be destructive or constructive.
When the interference is non-negligible, it is no longer meaningful 
to speak of a `signal' cross section. As an example, consider the 
scattering of longitudinal $\W$'s, 
$\W^+_{\mrm{L}} \W^-_{\mrm{L}} \to \W^+_{\mrm{L}} \W^-_{\mrm{L}}$,
where the `signal' process is $s$-channel exchange of a Higgs.
This graph by itself is ill-behaved away from the resonance region.
Destructive interference with `background' graphs such as $t$-channel 
exchange of a Higgs and $s$- and $t$-channel exchange of a $\gamma/\Z$ 
is required to save unitarity at large energies.  
 
In $\ee$ colliders, the $f_{\e}^{\e}$ parton distribution is peaked
at $x = 1$ rather than at $x = 0$. The situation therefore is the
opposite, if one considers e.g. $\Z^0$ production in a machine
running at energies above $m_{\Z}$: the tail towards lower masses
is suppressed and the one towards higher masses enhanced, with a 
sharp secondary peak at around the nominal energy of the machine.
Also in this case, an appropriate definition of cross sections  
therefore is necessary --- with additional complications due to the 
interference between $\gamma^*$ and $\Z^0$. When other processes are
considered, problems of interference with background appears also 
here. Numerically the problems may be less pressing, however,
since the secondary peak is occuring in a high-mass region, rather 
than in a more complicated low-mass one. Further, in $\ee$ there is 
little uncertainty from the shape of the parton distributions. 
 
In $2 \to 2$ processes where a pair of resonances are produced, e.g.
$\ee \to \Z^0 \H^0$, cross section are almost always given
in the zero-width approximation for the resonances. Here
two substitutions of the type
\begin{equation}
1 = \int \delta (m^2 - m_R^2) \, dm^2
\to \int \frac{1}{\pi} \,
\frac{m_R \Gamma_R}{(m^2 - m_R^2)^2 + m_R^2 \Gamma_R^2} \, dm^2
\label{pg:resshapethree}
\end{equation}
are used to introduce mass distributions
for the two resonance masses, i.e. $m_3^2$ and $m_4^2$.
In the formula, $m_R$ is the nominal mass and $m$ the actually
selected one. The
phase-space integral over $x_1$, $x_1$ and $\hat{t}$ in eq.
(\ref{pg:sigma}) is then extended to involve also $m_3^2$ and
$m_4^2$. The effects of the mass-dependent width is only partly
taken into account, by replacing the nominal masses $m_3^2$ and
$m_4^2$ in the $\d \hat{\sigma} / \d \hat{t}$ expression by the
actually generated ones (also e.g. in the relation between
$\hat{t}$ and $\cos\hat{\theta}$),
while the widths are evaluated at the nominal masses. This
is the equivalent of a simple replacement of $m_R \Gamma_R$ by
$\hat{s} \Gamma_R / m_R$ in the numerator of eq.
(\ref{pg:resshapeone}), but not in the denominator.
In addition, the full threshold dependence, i.e. the
$\beta$-dependent factors, is not reproduced.
 
There is no particular
reason why the full mass-dependence could not be introduced,
except for the extra work and time consumption needed for each
process. In fact, the matrix elements for several $\gammaZ$
production processes do contain the full expressions.
On the other hand, the matrix elements given in the literature
are often valid only when the resonances are almost on the mass shell,
since some graphs have been omitted. As an example, the process
$\q\qbar \to \e^- \br{\nu}_{\e} \mu^+ \nu_{\mu}$ is dominated by
$\q\qbar \to \W^- \W^+$ when each of the two lepton pairs is close to
$m_{\W}$ in mass, but in general also receives contributions e.g. from
$\q\qbar \to \Z^0 \to \ee$, followed by $\e^+ \to \br{\nu}_{\e} \W^+$
and $\W^+ \to \mu^+ \nu_{\mu}$. The latter contributions are neglected
in cross sections given in the zero-width approximation.
 
Processes with one final-state resonance and another ordinary
final-state product, e.g. $\q \g \to \W^+ \q'$, are treated in
the same spirit as the $2 \to 2$ processes with two resonances,
except that only one mass need be selected according to a
Breit--Wigner.
 
\subsection{Cross-section Calculations}
\label{ss:PYTcrosscalc}
 
In the program, the variables used in the generation of a
$2 \to 2$ process are $\tau$, $y$ and $z = \cos\hat{\theta}$.
For a $2 \to 1$ process, the $z$ variable can be integrated out,
and need therefore not be generated as part of the hard process,
except when the allowed angular range of decays is restricted.
In unresolved lepton beams, i.e. when
$f_{\e}^{\e}(x) = \delta(x-1)$, the variables $\tau$ and/or $y$
may be integrated out. We will cover all these special cases
towards the end of the section, and here concentrate on `standard' 
$2 \to 2$ and $2 \to 1$ processes.
 
\subsubsection{The simple $2 \to 2$ processes}
 
In the spirit of section \ref{ss:MCdistsel}, we want to select simple
functions such that the true $\tau$, $y$ and $z$ dependence of the
cross sections is approximately modelled. In particular, (almost) all
conceivable kinematical peaks should be represented by separate
terms in the approximate formulae. If this can be achieved,
the ratio of the correct to the approximate cross sections will
not fluctuate too much, but allow reasonable Monte Carlo efficiency.
 
Therefore the variables are generated according to the distributions
$h_{\tau}(\tau)$, $h_y(y)$ and $h_z(z)$, where normally
\begin{eqnarray}
h_{\tau}(\tau) & = & \frac{c_1}{{\cal I}_1} \, \frac{1}{\tau} +
\frac{c_2}{{\cal I}_2} \, \frac{1}{\tau^2} +
\frac{c_3}{{\cal I}_3} \, \frac{1}{\tau(\tau + \tau_R)} +
\frac{c_4}{{\cal I}_4} \,
\frac{1}{(s\tau - m_R^2)^2 + m_R^2 \Gamma_R^2} \nonumber \\
& & + \frac{c_5}{{\cal I}_5} \, \frac{1}{\tau(\tau + \tau_{R'})} +
\frac{c_6}{{\cal I}_6} \,
\frac{1}{(s\tau - m_{R'}^2)^2 + m_{R'}^2 \Gamma_{R'}^2} ~,  \\[1mm]
h_y(y) & = & \frac{c_1}{{\cal I}_1} \, (y - y_{\mmin}) +
\frac{c_2}{{\cal I}_2} \, (y_{\mmax} - y) +
\frac{c_3}{{\cal I}_3} \, \frac{1}{\cosh y} ~,  \\[1mm]
h_z(z) & = & \frac{c_1}{{\cal I}_1} +
\frac{c_2}{{\cal I}_2} \, \frac{1}{a-z} +
\frac{c_3}{{\cal I}_3} \, \frac{1}{a+z} +
\frac{c_4}{{\cal I}_4} \, \frac{1}{(a-z)^2} +
\frac{c_5}{{\cal I}_5} \, \frac{1}{(a+z)^2}    ~.
\label{pg:hforms}
\end{eqnarray}
Here each term is separately integrable, with an invertible primitive
function, such that generation of $\tau$, $y$ and $z$ separately
is a standard task, as described in section \ref{ss:MCdistsel}.
In the following we describe the details of the scheme, including
the meaning of the coefficients $c_i$ and ${\cal I}_i$, which are
separate for $\tau$, $y$ and $z$.
 
The first variable to be selected is $\tau$. The range of allowed
values, $\tau_{\mmin} \leq \tau \leq \tau_{\mmax}$, is generally
constrained by a number of user-defined requirements. A cut on the
allowed mass range is directly reflected in $\tau$, a cut on the
$\pT$ range indirectly. The first two terms
of $h_{\tau}$ are intended to represent a smooth $\tau$ dependence,
as generally obtained in processes which do not receive contributions
from $s$-channel resonances. Also $s$-channel exchange of
essentially massless particles ($\gamma$, $\g$, light quarks and
leptons) are accounted for, since these do not produce any separate
peaks at non-vanishing $\tau$. The last four terms of $h_{\tau}$ are
there to catch the peaks in the cross section from resonance
production. These terms are only included when needed. Each resonance
is represented by two pieces, a first to cover the interference with
graphs which peak at $\tau =0$, plus the variation of 
parton distributions, and a second to approximate the
Breit--Wigner shape of the resonance itself. The subscripts $R$ and
$R'$ denote values pertaining to the two resonances, with
$\tau_R = m_R^2/s$. Currently there is only one process where the
full structure with two resonances is used, namely
$\f \fbar \to \gamma^*/\Z^0/\Z'^0$. Otherwise either one or no
resonance peak is taken into account.
 
The kinematically allowed range of $y$ values is
constrained by $\tau$, $|y| \leq - \frac{1}{2} \ln\tau$, and you
may impose additional cuts. Therefore the allowed range
$y_{\mmin} \leq y \leq y_{\mmax}$ is only constructed after 
$\tau$ has been selected. The first two terms of $h_y$ give a fairly 
flat $y$ dependence --- for processes which are symmetric in
$y \leftrightarrow -y$, they will add to give a completely flat $y$
spectrum between the allowed limits. In
principle, the natural subdivision would have been one term flat
in $y$ and one forward--backward asymmetric, i.e. proportional to
$y$. The latter is disallowed by the requirement of positivity,
however. The $y - y_{\mmin}$ and $y_{\mmax} - y$ terms 
actually used give the same amount of freedom, but respect positivity.
The third term is peaked at around $y = 0$, and represents the bias
of parton distributions towards this region.
 
The allowed $z = \cos\hat{\theta}$ range is na\"{\i}vely
$-1 \leq z \leq 1$. However, most cross sections are divergent for
$z \to \pm 1$, so some kind of regularization is necessary. Normally
one requires $\pT \geq \pTmin$, which translates into
$z^2 \leq 1 - 4 \pTmin^2/(\tau s)$ for massless outgoing
particles. Since again the limits depend on $\tau$,
the selection of $z$ is done after that of
$\tau$. Additional requirements may constrain the range further.
In particular, a $p_{\perp\mmax}$ constraint may split the allowed
$z$ range into two, i.e. $z_{-\mmin} \leq z \leq z_{-\mmax}$ or
$z_{+\mmin} \leq z \leq z_{+\mmax}$. An unsplit range is 
represented by $z_{-\mmax} = z_{+\mmin} = 0$. 
For massless outgoing particles
the parameter $a = 1$ in $h_z$, such that the five terms represent
a piece flat in angle and pieces peaked as $1/\hat{t}$, $1/\hat{u}$,
$1/\hat{t}^2$, and $1/\hat{u}^2$, respectively. For non-vanishing
masses one has $a = 1 + 2 m_3^2 m_4^2/\hat{s}^2$.
In this case, the full range $-1 \leq z \leq 1$ is therefore
available --- physically, the standard $\hat{t}$ and $\hat{u}$
singularities are regularized by the masses $m_3$ and $m_4$.
 
For each of the terms, the ${\cal I}_i$ coefficients represent the
integral over the quantity multiplying the coefficient $c_i$; thus,
for instance:
\begin{eqnarray}
h_{\tau}: & & {\cal I}_1 = \int \frac{\d \tau}{\tau} =
     \ln \left( \frac{\tau_{\mmax}}{\tau_{\mmin}} \right) ~, 
     \nonumber \\
     & & {\cal I}_2 = \int \frac{\d \tau}{\tau^2} =
     \frac{1}{\tau_{\mmin}} - \frac{1}{\tau_{\mmax}} ~;  
     \nonumber \\
h_y: & & {\cal I}_1 = \int (y - y_{\mmin}) \, \d y =
     \frac{1}{2} (y_{\mmax} - y_{\mmin})^2 ~; \nonumber \\
h_z: & & {\cal I}_1 = \int \d z = (z_{-\mmax} - z_{-\mmin}) +
     (z_{+\mmax} - z_{+\mmin}) , \nonumber \\
     & & {\cal I}_2 = \int \frac{\d z}{a-z} =
     \ln \left( \frac{(a-z_{-\mmin})(a-z_{+\mmin})}
     {(a-z_{-\mmax})(a-z_{-\mmin})} \right) ~.
\end{eqnarray}
 
The $c_i$ coefficients are normalized to unit sum for $h_{\tau}$, 
$h_y$ and $h_z$ separately. They have a simple interpretation, as
the probability
for each of the terms to be used in the preliminary selection of
$\tau$, $y$ and $z$, respectively. The variation of the cross section
over the allowed phase space is explored in the initialization
procedure of a {\Py} run, and based on this knowledge the $c_i$
are optimized so as to give functions $h_{\tau}$, $h_y$ and $h_z$ that
closely follow the general behaviour of the true cross section.
For instance, the coefficient $c_4$ in $h_{\tau}$ is to be made larger
the more the total cross section is dominated by the region around
the resonance mass.
 
The phase-space points tested at initialization
are put on a grid, with the number of
points in each dimension given by the number of terms in the
respective $h$ expression, and with the position of each point
given by the median value of the distribution of one of the terms.
For instance, the $\d \tau / \tau$ distribution gives a median point at
$\sqrt{\tau_{\mmin}\tau_{\mmax}}$, and $\d \tau / \tau^2$ has the 
median $2 \tau_{\mmin} \tau_{\mmax} 
/ (\tau_{\mmin} + \tau_{\mmax})$.
Since the allowed $y$ and $z$ ranges depend on the $\tau$ value
selected, then so do the median points defined for these two
variables.
 
With only a limited set of phase-space points
studied at the initialization, the `optimal' set of coefficients
is not uniquely defined. To be on the safe side, 40\% of the total
weight is therefore assigned evenly between all allowed $c_i$,
whereas the remaining 60\% are assigned according to the relative
importance surmised, under the constraint that no coefficient is
allowed to receive a negative contribution from this second piece.
 
After a preliminary choice has been made of $\tau$, $y$ and $z$,
it is necessary to find the weight of the event, which is to be used
to determine whether to keep it or generate another one.
Using the relation $\d \hat{t} = \hat{s} \, \beta_{34} \, \d z / 2$, 
eq.~(\ref{pg:sigma}) may be rewritten as
\begin{eqnarray}
\sigma & = & \int \int \int \frac{\d \tau}{\tau} \, \d y \,
\frac{\hat{s} \beta_{34}}{2} \d z \,
x_1 f_1(x_1, Q^2) \, x_2 f_2(x_2, Q^2) \,
\frac{\d \hat{\sigma}}{\d \hat{t}}    \nonumber \\[1mm]
& = & \frac{\pi}{s} \int h_{\tau}(\tau) \, \d \tau
\int h_y(y) \, \d y \int h_z(z) \, \d z \, \beta_{34} \,
\frac{x_1 f_1(x_1, Q^2) \, x_2 f_2(x_2, Q^2)}
{\tau^2 h_{\tau}(\tau) \, h_y(y) \, 2 h_z(z)} \,
\frac{\hat{s}^2}{\pi} \frac{\d \hat{\sigma}}{\d \hat{t}}  
\nonumber \\[1mm]
& = & \left\langle \frac{\pi}{s} \,
\frac{\beta_{34}}{\tau^2 h_{\tau}(\tau) \, h_y(y) \, 2 h_z(z)} \,
x_1 f_1(x_1, Q^2) \, x_2 f_2(x_2, Q^2) \,
\frac{\hat{s}^2}{\pi} \frac{\d \hat{\sigma}}{\d \hat{t}}
\right\rangle ~.
\label{pg:sigmamap}
\end{eqnarray}
In the middle line, a factor of $1 = h_{\tau}/h_{\tau}$
has been introduced to rewrite the $\tau$ integral
in terms of a phase space of unit volume:
$\int h_{\tau}(\tau) \, \d \tau = 1$ according to the relations
above. Correspondingly for the $y$ and $z$ integrals. In addition,
factors of $1 = \hat{s}/ (\tau s)$ and $1 = \pi / \pi$ are used to
isolate the dimensionless cross section
$(\hat{s}^2/\pi) \, \d \hat{\sigma} / \d \hat{t}$.
The content of the last line is that, with $\tau$, $y$ and $z$
selected according to the expressions $h_{\tau}(\tau)$, $h_y(y)$
and $h_z(z)$, respectively, the cross section is obtained as the
average of the final expression over all events. Since the $h$'s
have been picked to give unit volume, there is no need to multiply
by the total phase-space volume.
 
As can be seen, the cross section for a given Monte Carlo event is
given as the product of four factors, as follows:
\begin{Enumerate}
\item The $\pi/s$ factor, which is common to all events, gives the
overall dimensions of the cross section, in GeV$^{-2}$. Since
the final cross section is given in units of mb, the conversion
factor of 1 GeV$^{-2} = 0.3894$ mb is also included here.
\item Next comes the `Jacobian', which compensates for the change
from the original to the final phase-space volume.
\item The parton-distribution function weight is obtained by making 
use of the parton distribution libraries in {\Py} or externally. 
The $x_1$ and $x_2$ values are obtained from $\tau$ and $y$ via the 
relations $x_{1,2} = \sqrt{\tau} \exp(\pm y)$.
\item Finally, the dimensionless cross section
$(\hat{s}^2/\pi) \, \d \hat{\sigma} / \d \hat{t}$
is the quantity that has to be coded for each process separately,
and where the physics content is found.
\end{Enumerate}
 
Of course, the expression in the last line is not strictly necessary
to obtain the cross section by Monte Carlo integration. One could
also have used eq.~(\ref{pg:sigma}) directly, selecting phase-space
points evenly in $\tau$, $y$ and $\hat{t}$, and averaging over those
Monte Carlo weights. Clearly this would be much simpler, but the price
to be paid is that the weights of individual events could fluctuate
wildly. For instance, if the cross section contains a narrow
resonance, the few phase-space points that are generated in the
resonance region obtain large weights, while the rest do not.
With our procedure, a resonance would be included in the
$h_{\tau}(\tau)$ factor, so that more events would be generated
at around the appropriate $\tau_R$ value (owing to the $h_{\tau}$
numerator in the phase-space expression), but with these events
assigned a lower, more normal weight (owing to the factor
$1/h_{\tau}$ in the weight expression).
Since the weights fluctuate less, fewer phase-space points
need be selected to get a reasonable cross-section estimate.
 
In the program, the cross section is obtained as the average over all
phase-space points generated. The events actually handed on to the
user should have unit weight, however (an option with weighted events
exists, but does not represent the mainstream usage). At
initialization, after the $c_i$ coefficients have been determined,
a search inside the allowed phase-space volume is therefore made
to find the maximum of the weight expression in the last line of
eq.~(\ref{pg:sigmamap}). In the subsequent generation of events,
a selected phase-space point is then retained with a probability
equal to the weight in the point divided by the maximum weight.
Only the retained phase-space points are considered further, and
generated as complete events.
 
The search for the maximum is begun by evaluating the weight in the
same grid of points as used to determine the $c_i$ coefficients.
The point with highest weight is used as starting point for a
search towards the maximum. In unfortunate cases, the convergence
could be towards a local maximum which is not the global one.
To somewhat reduce this risk, also the grid point with 
second-highest weight is used for another search. After 
initialization, when events are generated, a warning message 
will be given by default at any time a phase-space
point is selected where the weight is larger than the maximum,
and thereafter the maximum weight is adjusted to reflect the new
knowledge. This means that events generated before this time have
a somewhat erroneous distribution in phase space, but if the
maximum violation is rather modest the effects should be negligible.
The estimation of the cross section is not affected by any of these
considerations, since the maximum weight does not enter into eq.
(\ref{pg:sigmamap}).

For $2 \to 2$ processes with identical final-state particles,
the symmetrization factor of $1/2$ is explicitly included at the
end of the $\d \hat{\sigma} / \d \hat{t}$ calculation. In the final
cross section, a factor of 2 is retrieved because of integration
over the full phase space (rather than only half of it). That
way, no special provisions are needed in the phase-space 
integration machinery.
 
\subsubsection{Resonance production}
 
We have now covered the simple $2 \to 2$ case. In a $2 \to 1$
process, the $\hat{t}$ integral is absent, and the differential
cross section $\d \hat{\sigma} / \d \hat{t}$ is replaced by
$\hat{\sigma}(\hat{s})$. The cross section may now be written as
\begin{eqnarray}
\sigma & = & \int \int \frac{\d \tau}{\tau} \, \d y \,
x_1 f_1(x_1, Q^2) \, x_2 f_2(x_2, Q^2) \,
\hat{\sigma}(\hat{s})    \nonumber \\
& = & \frac{\pi}{s} \int h_{\tau}(\tau) \, \d \tau  
\int h_y(y) \, \d y \,
\frac{x_1 f_1(x_1, Q^2) \, x_2 f_2(x_2, Q^2)}
{\tau^2 h_{\tau}(\tau) \, h_y(y)} \,
\frac{\hat{s}}{\pi} \hat{\sigma}(\hat{s}) \nonumber \\
& = & \left\langle \frac{\pi}{s} \,
\frac{1}{\tau^2 h_{\tau}(\tau) \, h_y(y)} \,
x_1 f_1(x_1, Q^2) \, x_2 f_2(x_2, Q^2) \,
\frac{\hat{s}}{\pi} \hat{\sigma}(\hat{s})
\right\rangle ~.
\label{pg:sigmamapres}
\end{eqnarray}
The structure is thus exactly the same, but the $z$-related pieces
are absent, and the r\^ole of the dimensionless cross section is
played by $(\hat{s}/\pi) \hat{\sigma}(\hat{s})$.
 
If the range of allowed decay angles of the resonance is restricted,
e.g. by requiring the decay products to have a minimum transverse
momentum, effectively this translates into constraints on the
$z = \cos\hat{\theta}$ variable of the $2 \to 2$ process. The
difference is that the angular dependence of a resonance decay is
trivial, and that therefore the $z$-dependent factor can be easily
evaluated. For a spin-0 resonance, which decays isotropically, the
relevant weight is simply
$(z_{-\mmax} - z_{-\mmin})/2 + (z_{+\mmax} - z_{+\mmin})/2$.
For a transversely polarized spin-1 resonance the expression is,
instead,
\begin{equation}
\frac{3}{8}(z_{-\mmax} - z_{-\mmin}) +
\frac{3}{8}(z_{+\mmax} - z_{+\mmin}) +
\frac{1}{8}(z_{-\mmax} - z_{-\mmin})^3 +
\frac{1}{8}(z_{+\mmax} - z_{+\mmin})^3  ~.
\end{equation}
Since the allowed $z$ range could depend on $\tau$ and/or $y$ (it does
for a $\pT$ cut), the factor has to be evaluated for each individual
phase-space point and included in the expression of eq.
(\ref{pg:sigmamapres}).
 
For $2 \to 2$ processes where either of the final-state
particles is a resonance, or both, an additional choice has to 
be made for
each resonance mass, eq.~(\ref{pg:resshapethree}). Since the allowed
$\tau$, $y$ and $z$ ranges depend on $m_3^2$ and $m_4^2$,
the selection of masses have to precede the choice of the other
phase-space variables. Just as for the other variables, masses
are not selected uniformly over the allowed range, but are rather
distributed according to a function $h_m(m^2) \, dm^2$, with a
compensating factor $1/h_m(m^2)$ in the `Jacobian'. The functional
form picked is normally
\begin{equation}
h_m(m^2) = \frac{c_1}{{\cal I}_1} \, \frac{1}{\pi} \,
\frac{m_R \Gamma_R}{(m^2 - m_R^2)^2 + m_R^2 \Gamma_R^2} +
\frac{c_2}{{\cal I}_2} +
\frac{c_3}{{\cal I}_3} \, \frac{1}{m^2} +
\frac{c_4}{{\cal I}_4} \, \frac{1}{m^4}  ~.
\label{pg:reshdist}
\end{equation}
The definition of the ${\cal I}_i$ integrals is analogous to the one
before. The $c_i$ coefficients are not found by optimization, but
predetermined, normally to $c_1 = 0.8$, $c_2 = c_3 =0.1$, $c_4 = 0$.
Clearly, had the phase space and the cross section been independent
of $m_3^2$ and $m_4^2$, the optimal choice would have been to put
$c_1 = 1$ and have all other $c_i$ vanishing --- then the $1/h_m$
factor of the `Jacobian' would exactly have cancelled the
Breit--Wigner of eq.~(\ref{pg:resshapethree}) in the cross section.
The second and the third terms are there to cover the possibility
that the cross section does not die away quite as fast as given by
the na\"{\i}ve Breit--Wigner shape. In particular, the third term covers 
the possibility of a secondary peak at small $m^2$, in a spirit 
slightly similar to the one discussed for resonance production in 
$2 \to 1$ processes.
 
The fourth term is only used for the process
$\f \fbar \to (\gammaZ)(\gammaZ)$, where the $\gamma$ propagator
guarantees that the cross section does have a significant secondary
peak for $m^2 \to 0$. Therefore here the choice is $c_1 = 0.4$,
$c_2 = 0.05$, $c_3 = 0.3$ and $c_4 = 0.25$.
 
A few special tricks have been included to improve efficiency
when the allowed mass range of resonances is constrained by
kinematics or by user cuts. For instance, if a pair of equal
or charge-conjugate resonances are produced, such as in
$\ee \to \W^+ \W^-$, use is made of the constraint that the lighter
of the two has to have a mass smaller than half the c.m. energy.
 
\subsubsection{Lepton beams}
 
Lepton beams have to be handled slightly differently from what has
been described so far. One also has to distinguish between a lepton
for which parton distributions are included and one which is treated
as an unresolved point-like particle. The necessary modifications are
the same for $2 \to 2$ and $2 \to 1$ processes, however, since the
$\hat{t}$ degree of freedom is unaffected.
 
If one incoming beam is an unresolved lepton, the corresponding
parton-distribution piece collapses to a $\delta$ function. This
function can be used to integrate out the $y$ variable:
$\delta(x_{1,2} - 1) = \delta(y \pm (1/2) \ln \tau)$.
It is therefore only necessary to select the $\tau$ and the $z$
variables according to the proper distributions, with compensating
weight factors, and only one set of parton distributions has to be 
evaluated explicitly.
 
If both incoming beams are unresolved leptons, both the $\tau$ and
the $y$ variables are trivially given: $\tau = 1$ and $y = 0$.
Parton-distribution weights disappear completely. For a $2 \to 2$
process, only the $z$ selection remains to be performed, while
a $2 \to 1$ process is completely specified, i.e. the cross section
is a simple number that only depends on the c.m. energy.
 
For a resolved electron, the $f_{\e}^{\e}$ parton distribution is
strongly peaked towards $x = 1$. This affects both the $\tau$
and the $y$ distributions, which are not well described by
either of the pieces in $h_{\tau}(\tau)$ or $h_y(y)$ in processes with
interacting $\e^{\pm}$. (Processes which involve e.g. the $\gamma$
content of the $\e$ are still well simulated, since
$f_{\gamma}^{\e}$ is peaked at small $x$.)
 
If both parton distributions 
are peaked close to 1, the $h_{\tau}(\tau)$ expression in eq.
(\ref{pg:hforms}) is therefore increased with one additional term of
the form $h_{\tau}(\tau) \propto 1 / (1 - \tau)$, with coefficients
$c_7$ and ${\cal I}_7$ determined as before. The divergence when
$\tau \to 1$ is cut off by our regularization procedure for the
$f_{\e}^{\e}$ parton distribution; therefore we only need consider
$\tau < 1 - 2 \times 10^{-6}$.
 
Correspondingly, the $h_y(y)$ expression is expanded with a term
$1/(1 - \exp(y-y_0))$ when incoming beam number 1 consists of a
resolved $\e^{\pm}$, and with a term $1/(1 - \exp(-y-y_0))$
when incoming beam number 2 consists of a resolved $\e^{\pm}$.
Both terms are present for an $\ee$ collider, only one for an
$\ep$ one. The coefficient $y_0 = - (1/2) \ln \tau$ is the na\"{\i}ve
kinematical limit of the $y$ range, $|y| < y_0$. From the
definitions of $y$ and $y_0$ it is easy to see
that the two terms above correspond to $1/(1-x_1)$ and $1/(1-x_2)$,
respectively, and thus are again regularized by our 
parton-distribution function cut-off. Therefore the integration ranges 
are $y < y_0 -10^{-6}$ for the first term and $y > - y_0 + 10^{-6}$
for the second one.
 
\subsubsection{Mixing processes}
\label{sss:mixingproc}
 
In the cross-section formulae given so far, we have deliberately
suppressed a summation over the allowed incoming flavours. For
instance, the process $\f\fbar \to \Z^0$ in a hadron collider
receives contributions from $\u\ubar \to \Z^0$, $\d\dbar \to \Z^0$,
$\s\sbar \to \Z^0$, and so on. These contributions share the
same basic form, but differ in the parton-distribution weights
and (usually) in a few coupling constants in the hard matrix
elements. It it therefore convenient to generate the terms
together, as follows:
\begin{Enumerate}
\item A phase-space point is picked, and all
common factors related to this choice are evaluated, i.e.
the `Jacobian' and the common pieces of the matrix elements
(e.g. for a $\Z^0$ the basic Breit--Wigner shape, excluding
couplings to the initial flavour).
\item The parton-distribution-function library is called to produce 
all the parton distributions, at the relevant $x$ and $Q^2$ values,
for the two incoming beams.
\item A loop is made over the two incoming flavours, one from each
beam particle. For each allowed set of incoming flavours, the full 
matrix-element expression is put together, using the common pieces and 
the flavour-dependent couplings. This is multiplied by the common 
factors and the parton-distribution weights to obtain a cross-section 
weight.
\item Each allowed flavour combination is stored as a separate entry
in a table, together with its weight. In addition, a summed weight
is calculated.
\item The phase-space point is kept or rejected, according to a
comparison of the summed weight with the maximum weight obtained
at initialization. Also the cross-section Monte Carlo integration
is based on the summed weight.
\item If the point is retained, one of the allowed flavour
combinations is picked according to the relative weights stored
in the full table.
\end{Enumerate}
 
Generally, the flavours of the final state are either completely
specified by those of the initial state, e.g. as in $\q\g \to \q\g$,
or completely decoupled from them, e.g. as in
$\f\fbar \to \Z^0 \to \f'\fbar'$. In neither case need therefore the
final-state flavours be specified in the cross-section calculation.
It is only necessary, in the latter case, to include an overall
weight factor, which takes into account the summed contribution of
all final states that are to be simulated. For instance, if only
the process $\Z^0 \to \ee$  is studied, the relevant weight factor
is simply $\Gamma_{\e\e} / \Gamma_{\mrm{tot}}$. Once the kinematics 
and the incoming flavours have been selected, the outgoing flavours
can be picked according to the appropriate relative probabilities.
 
In some processes, such as $\g\g \to \g\g$, several different colour
flows are allowed, each with its own kinematical dependence of the
matrix-element weight, see section \ref{sss:QCDjetclass}. Each colour 
flow is then given as a separate entry in the table mentioned above, 
i.e. in total an entry is characterized by the two incoming flavours, 
a colour-flow index, and the weight. For an accepted phase-space 
point, the colour flow is selected in the same way as the incoming 
flavours.
 
The program can also allow the mixed generation of two or more
completely different processes, such as $\f\fbar \to \Z^0$ and
$\q\qbar \to \g\g$. In that case, each process is initialized
separately, with its own set of coefficients $c_i$ and so on.
The maxima obtained for the individual cross sections are all
expressed in the same units, even when the dimensionality of the
phase space is different. (This is because we always transform to
a phase space of unit volume, 
$\int h_{\tau}(\tau) \, \d \tau \equiv 1$, etc.) The above 
generation scheme need therefore only be generalized as follows:
\begin{Enumerate}
\item One process is selected among the allowed ones, with a relative
probability given by the maximum weight for this process.
\item A phase-space point is found, using the distributions
$h_{\tau}(\tau)$ and so on, optimized for this particular process.
\item The total weight for the phase-space point is evaluated,
again with `Jacobians', matrix elements and allowed incoming flavour
combinations that are specific to the process.
\item The point is retained with a probability given by the ratio of
the actual to the maximum weight of the process. If the point is
rejected, one has to go back to step 1 and pick a new process.
\item Once a phase-space point has been accepted, flavours may be
selected, and the event generated in full.
\end{Enumerate}
It is clear why this works: although phase-space points are selected
among the allowed processes according to relative probabilities given
by the maximum weights, the probability that a point is accepted is
proportional to the ratio of actual to maximum weight. In total,
the probability for a given process to be retained is therefore only
proportional to the average of the actual weights, and any dependence
on the maximum weight is gone.

In $\gamma\p$ and $\gamma\gamma$ physics, the different components
of the photon give different final states, see section 
\ref{sss:photoprod}. Technically, this introduces a further level
of administration, since each event class contains a set of (partly
overlapping) processes. From an ideological point of view, however,
it just represents one more choice to be made, that of event class,
before the selection of process in step 1 above. When a weighting
fails, both class and process have to be picked anew.
 
\subsection{$2 \to 3$ and $2 \to 4$ Processes}
 
The {\Py} machinery to handle $2 \to 1$ and $2 \to 2$ processes
is fairly sophisticated and generic. The same cannot be said about
the generation of hard scattering processes with more than two 
final-state particles. The number of phase-space variables is 
larger, and
it is therefore more difficult to find and transform away all possible
peaks in the cross section by a suitably biased choice of phase-space
points. In addition, matrix-element expressions for $2 \to 3$
processes are typically fairly lengthy. Therefore {\Py} only contains
a very limited number of $2 \to 3$ and $2 \to 4$ processes, and almost
each process is a special case of its own. It is therefore less
interesting to discuss details, and we only give a very generic
overview.
 
If the Higgs mass is not light, interactions among longitudinal
$\W$ and $\Z$ gauge bosons are of interest. In the program,
$2 \to 1$ processes such as $\W_{\mrm{L}}^+ \W_{\mrm{L}}^- \to \H^0$ 
and $2 \to 2$ ones such as 
$\W_{\mrm{L}}^+ \W_{\mrm{L}}^- \to \Z_{\mrm{L}}^0 \Z_{\mrm{L}}^0$ 
are included.
The former are for use when the $\H^0$ still is reasonably narrow,
such that a resonance description is applicable, while the latter
are intended for high energies, where different contributions have
to be added up. Since the program does not contain 
$\W_{\mrm{L}}$ or $\Z_{\mrm{L}}$
distributions inside hadrons, the basic hard scattering has
to be convoluted with the $\q \to \q' \W_{\mrm{L}}$ and 
$\q \to \q \Z_{\mrm{L}}$
branchings, to yield effective $2 \to 3$ and $2 \to 4$ processes.
However, it is possible to integrate out the scattering angles of
the quarks analytically, as well as one energy-sharing variable
\cite{Cha85}. Only after an event has been accepted are these other
kinematical variables selected. This involves further choices of
random variables, according to a separate selection loop.
 
In total, it is therefore only necessary to introduce one additional
variable in the basic phase-space selection, which is chosen to be
$\hat{s}'$, the squared invariant mass of the full $2 \to 3$ or
$2 \to 4$ process, while $\hat{s}$ is used for the squared invariant
mass of the inner $2 \to 1$ or $2 \to 2$ process. The $y$ variable
is coupled to the full process, since parton-distribution weights
have to be given for the original quarks at
$x_{1,2} = \sqrt{\tau'} \exp{(\pm y)}$. The $\hat{t}$ variable is
related to the inner process, and thus not needed for the $2 \to 3$
processes. The selection of the $\tau' = \hat{s}'/s$ variable is
done after $\tau$, but before $y$ has been chosen. To improve the
efficiency, the selection is made according to a weighted phase space
of the form $\int h_{\tau'}(\tau') \, \d \tau'$, where
\begin{equation}
h_{\tau'}(\tau') = \frac{c_1}{{\cal I}_1} \frac{1}{\tau'} +
\frac{c_2}{{\cal I}_2} \, \frac{(1- \tau / \tau')^3}{\tau'^2} +
\frac{c_3}{{\cal I}_3} \, \frac{1}{\tau' (1 - \tau')} ~,
\end{equation}
in conventional notation. The $c_i$ coefficients are optimized
at initialization. The $c_3$ term, peaked at $\tau' \approx 1$,
is only used for $\ee$ collisions.
The choice of $h_{\tau'}$ is roughly matched to the
longitudinal gauge-boson flux factor, which is of the form
\begin{equation}
\left( 1 + \frac{\tau}{\tau'} \right) \,
\ln \left( \frac{\tau}{\tau'} \right) -
2 \left( 1 - \frac{\tau}{\tau'} \right)    ~.
\end{equation}
 
For a light $\H$ the effective $\W$ approximation above breaks down,
and it is necessary to include the full structure of the
$\q \q' \to \q \q' \H^0$ (i.e. $\Z\Z$ fusion) and
$\q \q' \to \q'' \q''' \H^0$ (i.e. $\W\W$ fusion) matrix elements.
The $\tau'$, $\tau$ and $y$ variables are here retained, and selected
according to standard procedures. The Higgs mass is represented by the
$\tau$ choice; normally the $\H^0$ is so narrow that the $\tau$
distribution effectively collapses to a $\delta$ function. In addition,
the three-body final-state phase space is rewritten as
\begin{equation}
\left( \prod_{i=3}^5 \frac{1}{(2 \pi)^3} \frac{\d^3 p_i}{2 E_i}
\right) \, (2 \pi)^4 \delta^{(4)} (p_3 + p_4 + p_5 - p_1 - p_2) =
\frac{1}{(2 \pi)^5} \, \frac{\pi^2}{4 \sqrt{\lambda_{\perp 34}}}
\, \d p_{\perp 3}^2 \, \frac{\d \varphi_3}{2 \pi} \,
\d p_{\perp 4}^2 \, \frac{\d \varphi_4}{2 \pi} \, \d y_5    ~,
\end{equation}
where $\lambda_{\perp 34} = (m_{\perp 34}^2 - m_{\perp 3}^2 -
m_{\perp 4}^2)^2 - 4 m_{\perp 3}^2 m_{\perp 4}^2$.
The outgoing quarks are labelled 3 and 4, and the outgoing Higgs 5.
The $\varphi$ angles are selected isotropically, while the two
transverse momenta are picked, with some foreknowledge of the
shape of the $\W / \Z$ propagators in the cross sections, according
to $h_{\perp} (\pT^2) \, \d \pT^2$, where
\begin{equation}
h_{\perp}(\pT^2) = \frac{c_1}{{\cal I}_1} +
\frac{c_2}{{\cal I}_2} \, \frac{1}{m_R^2 + \pT^2} +
\frac{c_3}{{\cal I}_3} \, \frac{1}{(m_R^2 + \pT^2)^2}    ~,
\end{equation}
with $m_R$ the $\W$ or $\Z$ mass, depending on process, and
$c_1 = c_2 = 0.05$, $c_3 = 0.9$. Within the limits given by the
other variable choices, the rapidity $y_5$ is chosen uniformly.
A final choice remains to be made, which comes from a twofold
ambiguity of exchanging the longitudinal momenta of partons 3
and 4 (with minor modifications if they are massive). Here
the relative weight can be obtained exactly from the form of the
matrix element itself.
 
No good phase-space choice was found for the process
$\g \g \to \Z^0 \b \bbar$. This process is therefore not so easy
to generate with {\Py}. What is currently done is to use the
basic formalism of $2 \to 2$ processes, where the $\b + \bbar$
system is considered as an effective `resonance'. Two masses are
then selected, the $\Z^0$ one according to eq.~(\ref{pg:reshdist})
and the $\b + \bbar$ one according to $\d m^2 / m^2$. Both `decays'
are selected isotropically in the respective rest frame,
to give the final four fermions in
terms of which the matrix element is given. In addition,
$\tau$, $y$ and $z$ are selected according to the standard rules
for $2 \to 2$ processes.
 
\subsection{Resonance Decays}
\label{ss:resdecay}
 
Resonances can be made to decay in two different routines. One is the
standard decay treatment (in \ttt{LUDECY}) that can be used
for any unstable particle, where decay channels
are chosen according to fixed probabilities, and decay angles
usually are picked isotropically in the rest frame of the resonance,
see section \ref{ss:partdecays}.
The more sophisticated treatment (in \ttt{PYRESD}) is the default
one for resonances produced in {\Py}, and is described here.
The following are included in the list of resonances: $\Z^0$,
$\W^{\pm}$, $\H^0$, $\Z'^0$, $\W'^{\pm}$, $\H'^0$, $\A^0$,
$\H^{\pm}$, $\eta_{\mrm{tech}}^0$, $\L_{\Q}$, and $\R^0$.
The top is also considered as a resonance if it is assumed to
decay before it has time to fragment. Likewise for the fourth
generation fermions. If the fourth generation is used to represent
excited quarks and leptons, these are also considered to be 
resonances. 
 
\subsubsection{The decay scheme}
 
In the beginning of the decay treatment, either one or two
resonances may be present, the former represented by processes
such as $\q \qbar' \to \W^+$ and $\q \g \to \W^+ \q'$, the latter
by $\q \qbar \to \W^+ \W^-$. If the latter is the case, the
decay of the two resonances is considered in parallel
(unlike \ttt{LUDECY}, where one particle at a time is made to decay).
 
First the decay channel of each resonance is selected according to
the relative weights $H_R^{(f)}$, as described above, evaluated at
the actual mass of the resonance, rather than at the nominal one.
Threshold factors are therefore fully taken into account, with
channels automatically switched off below the threshold. Normally
the masses of the decay products are well-defined, but e.g. in
decays like $\H^0 \to \W^+ \W^-$ it is also necessary to select
the decay product masses. This is done according to two Breit--Wigners
of the type in eq.~(\ref{pg:resshapethree}), multiplied by the
threshold factor, which depends on both masses.
 
Next the decay angles of the resonance are selected isotropically in
its rest frame. Normally the full range of decay angles is available,
but in $2 \to 1$ processes the decay angles of the original resonance
may be restrained by user cuts, e.g. on the $\pT$ of the decay
products. Based on the angles, the four-momenta of the decay products
are constructed and boosted to the correct frame. As a rule, matrix
elements are given with quark and lepton masses assumed vanishing.
Therefore the four-momentum vectors constructed at this stage are
actually massless for all quarks and leptons.
 
The matrix elements may now be evaluated. For a process such as
$\q \qbar \to \W^+ \W^- \to \e^+ \nu_{\e} \mu^- \br{\nu}_{\mu}$,
the matrix element is a function of the four-momenta of the two
incoming fermions and of the four outgoing ones. An upper limit for
the event weight can be constructed from the cross section for the
basic process $\q \qbar \to \W^+ \W^-$, as already used to select
the two $\W$ momenta. If the weighting fails, new resonance decay
angles are picked and the procedure is iterated until acceptance.
 
Based on the accepted set of angles, the correct decay product
four-momenta are constructed, including previously neglected 
fermion masses. Quarks and, optionally, leptons are allowed to
radiate, using the standard final-state showering machinery, with
maximum virtuality given by the resonance mass.
 
In some decays new resonances are produced, and these are then
subsequently allowed to decay. Only one resonance pair is considered
at a time, i.e. it is not possible to include correlations which
involve the simultaneous decay of three or more resonances. This is
in fact all that is currently needed: in a process like
$\q\qbar \to \Z^0 \H^0 \to \Z^0 \W^+ \W^- \to 6$ fermions,
the spinless nature of the $\H^0$ ensures that the $\W^{\pm}$
decays are decoupled from that of the $\Z^0$ (but not from each other).
 
\subsubsection{Cross-section considerations}
\label{sss:resdecaycross}
 
The cross section for a process which involves the production of one
or several resonances is always reduced to take into account channels
not allowed by user flags. This is trivial for a single $s$-channel
resonance, cf. eq.~(\ref{pg:Hinoutsym}), but can also be included
approximately if several layers of resonance decays are involved.
At initialization, the ratio between the user-allowed width and the
nominally possible one is evaluated and stored, starting from the
lightest resonances and moving upwards. As an example, one first finds
the reduction factors for $\W^+$ and for $\W^-$ decays, which need not
be the same if e.g. $\W^+$ is allowed to decay only to quarks and
$\W^-$ only to leptons. These factors enter together as a weight
for the $\H^0 \to \W^+ \W^-$ channel, which is thus reduced in
importance compared with other possible Higgs decay channels.
This is also reflected
in the weight factor of the $\H^0$ itself, where some channels are open
in full, others completely closed, and finally some (like the one
above) open but with reduced weight. Finally, the weight for the
process $\q\qbar \to \Z^0 \H^0$ is evaluated as the product of the
$\Z^0$ weight factor and the $\H^0$ one. The standard cross section 
of the process is multiplied with this weight.
 
Since the restriction on allowed decay modes is already included in
the hard process cross section, mixing of different event types is
greatly simplified, and the selection of decay channel chains is
straightforward. There is a price to be paid, however. The reduction
factors evaluated at initialization all refer to resonances at their
nominal masses. For instance, the $\W$ reduction factor is evaluated
at the nominal $\W$ mass, even when that factor is used, later on, in
the description of the decay of a 120 GeV Higgs, where at least one
$\W$ would be produced below this mass. We know of no case where this
approximation has any serious consequences, however.
 
The weighting procedure works because the number of resonances to be
produced, directly or in subsequent decays, can be derived
recursively already from the start. It does not work for particles
which could also be produced at later stages, such as the 
parton-shower evolution and the fragmentation. For instance,
$\D^0$ mesons can be produced fairly late in the event generation
chain, in unknown numbers, and so weights could not be introduced
to compensate, e.g. for the forcing of decays only into $\pi^+ \K^-$.
For similar reasons the top is only considered as a resonance if it
is not allowed to hadronize; see discussion in section 
\ref{sss:heavflavclass}.
 
One should note that this reduction factor is separate from the
description of the resonance shape itself, where the full
width of the resonance has to be used. This width is based on the
sum of all possible decay modes, not just the simulated ones.
{\Py} does allow the possibility to change also the underlying
physics scenario, e.g. to include the decay of a $\Z^0$ into a
fourth-generation neutrino.
 
Normally the evaluation of the reduction factors is straightforward.
However, for decays into a pair of equal or charge-conjugate
resonances, such as $\Z^0 \Z^0$ or $\W^+ \W^-$, it is possible to
pick combinations in such a way that the weight of the pair does
not factorize into a product of the weight of each resonance itself.
To be precise, any decay channel can be given seven different status
codes:
\begin{Itemize}
\item $-1$: a non-existent decay mode, completely switched off and of no
concern to us;
\item 0: an existing decay channel, which is switched off;
\item 1: a channel which is switched on;
\item 2: a channel switched on for particles, but off for antiparticles;
\item 3: a channel switched on for antiparticles, but off for particles;
\item 4: a channel switched on for one of the resonances, but not for
   both;
\item 5: a channel switched on for the other of the resonances, but not
  for both.
\end{Itemize}
The meaning of possibilities 4 and 5 is exemplified by the statement
`in a $\W^+\W^-$ pair, one $\W$ decays hadronically and the other
leptonically', which thus covers the cases where either $\W^+$ or
$\W^-$ decays hadronically.
 
Neglecting non-existing channels, each channel belongs to either of the
classes above. If we denote the total branching ratio into channels
of type $i$ by $r_i$, this then translates into the requirement
$r_0 + r_1 + r_2 + r_3 + r_4 + r_5 = 1$. For a single particle the
weight factor is $r_1 + r_2 + r_4$, and for a single antiparticle
$r_1 + r_3 + r_4$. For a pair of identical resonances, the joint weight
is instead
\begin{equation}
(r_1 + r_2)^2 + 2 (r_1 + r_2) (r_4 + r_5) + 2 r_4 r_5 ~,
\end{equation}
and for a resonance--antiresonance pair
\begin{equation}
(r_1 + r_2)(r_1 + r_3) + (2 r_1 + r_2 + r_3) (r_4 + r_5) +
2 r_4 r_5 ~.
\label{eq:WWallchancomb}
\end{equation}
If some channels come with a reduced weight because of restrictions 
on subsequent decay chains, this may be described in terms of
properly reduced $r_i$, so that the sum is less than unity.
For instance, in a $\t\tbar \to \b\W^+ \, \bbar\W^-$ process,
the $\W$ decay modes may be restricted to $\W^+ \to \q\qbar$
and $\W^- \to \e^-\bar{\nu}_{\e}$, in which case
$(\sum r_i)_{\t} \approx 2/3$ and $(\sum r_i)_{\tbar} \approx 1/9$.
With index $\pm$ denoting resonance/antiresonance, 
eq.~(\ref{eq:WWallchancomb}) then generalizes to
\begin{equation}
(r_1 + r_2)^+ (r_1 + r_3)^- + (r_1 + r_2)^+ (r_4 + r_5)^- +
(r_4 + r_5)^+ (r_1 + r_3)^- + r_4^+ r_5^- + r_5^+ r_4^- ~.
\label{eq:WWallchancombgen}
\end{equation}
 
\subsection{Nonperturbative Processes}
\label{ss:nonpertproc}
 
A few processes are not covered by the discussion so far. These are
the ones that depend on the details of hadronic wave functions,
and therefore are not strictly calculable perturbatively
(although perturbation theory may often provide some guidance).
What we have primarily in mind is elastic scattering, diffractive
scattering and low-$\pT$ `minimum-bias' events in hadron--hadron
collisions, but one can also find corresponding processes in
$\gamma \p$ and $\gamma \gamma$ interactions. The description
of these processes is rather differently structured from that of
the other ones, as is explained below. Models for
`minimum-bias' events are discussed in detail in section
\ref{ss:multint}, to which we refer for details on this part
of the program.

\subsubsection{Hadron--hadron interactions}
 
In hadron--hadron interactions, the total hadronic cross section
for $AB \to$ anything, $\sigma^{AB}_{\mrm{tot}}$, is calculated
using the parametrization of Donnachie and Landshoff \cite{Don92}.
In this approach, each cross section appears as the sum of one
pomeron term and one reggeon one
\begin{equation}
\sigma^{AB}_{\mrm{tot}}(s) = X^{AB} \, s^{\epsilon} +  
Y^{AB} \, s^{-\eta} ~,
\label{pg:sigtotpomreg}
\end{equation}
where $s = E_{\mrm{cm}}^2$. The powers $\epsilon = 0.0808$ and
$\eta = 0.4525$ are expected to be universal, whereas the 
coefficients $X^{AB}$ and $Y^{AB}$ are specific to each initial
state. (In fact, the high-energy behaviour given by the pomeron term 
is expected to be the same for particle and antiparticle interactions, 
i.e. $X^{\br{A}B} = X^{AB}$.) Parametrizations not provided
in \cite{Don92} have been calculated in the same spirit, making use
of quark counting rules \cite{Sch93a}.  

The total cross section is subdivided according to
\begin{equation}
\sigma^{AB}_{\mrm{tot}}(s) = \sigma^{AB}_{\mrm{el}}(s) + 
\sigma^{AB}_{\mrm{sd}(XB)}(s) + \sigma^{AB}_{\mrm{sd}(AX)}(s) + 
\sigma^{AB}_{\mrm{dd}}(s) + \sigma^{AB}_{\mrm{nd}}(s)   ~.
\label{pg:sigtotsplit}
\end{equation}
Here `el' is the elastic process $AB \to AB$, `sd$(XB)$' the single 
diffractive $AB \to XB$, `sd$(AX)$' the single diffractive
$AB \to AX$, `dd' the double diffractive $AB \to X_1 X_2$,
and `nd' the non-diffractive ones. Higher diffractive topologies,
such as central diffraction, are currently neglected. 
In the following, the elastic and diffractive cross sections 
and event characteristics are described, as given in the model by
Schuler and Sj\"ostrand \cite{Sch94,Sch93a}. The non-diffractive 
component is identified with the `minimum bias' physics already 
mentioned, a practical but not unambiguous choice. Its cross section 
is given by `whatever is left' according to eq.~(\ref{pg:sigtotsplit}), 
and its properties are discussed in section \ref{ss:multint}.

At not too large squared momentum transfers $t$, the elastic cross
section can be approximated by a simple exponential fall-off. If one
neglects the small real part of the cross section, the optical 
theorem then gives
\begin{equation}
\frac{\d\sigma_{\mrm{el}}}{\d t} = 
\frac{\sigma_{\mrm{tot}}^2}{16 \pi} \, \exp(B_{\mrm{el}} t) ~,
\end{equation}
and $\sigma_{\mrm{el}} = \sigma_{\mrm{tot}}^2 / 16 \pi B_{\mrm{el}}$.
The elastic slope parameter is parametrized by
\begin{equation}
B_{\mrm{el}} = B^{AB}_{\mrm{el}}(s) = 2 b_A + 2 b_B + 
4 s^{\epsilon} -4.2 ~, 
\end{equation}
with $s$ given in units of GeV and $B_{\mrm{el}}$ in GeV$^{-2}$.
The constants $b_{A,B}$ are $b_{\p} = 2.3$, 
$b_{\pi,\rho,\omega,\phi} = 1.4$, $b_{\J/\psi} = 0.23$.
The increase of the slope parameter with c.m. energy is faster
than the logarithmically one conventionally assumed; that way the 
ratio $\sigma_{\mrm{el}} / \sigma_{\mrm{tot}}$ remains well-behaved
at large energies.

The diffractive cross sections are given by
\begin{eqnarray}
\frac{\d\sigma_{\mrm{sd}(XB)}(s)}{\d t \, \d M^2} & = & 
\frac{g_{3\pomeron}}{16\pi} \, \beta_{A\pomeron} \, 
\beta_{B\pomeron}^2 \, \frac{1}{M^2} \, \exp(B_{\mrm{sd}(XB)}t) 
\, F_{\mrm{sd}} ~, \nonumber \\
\frac{\d\sigma_{\mrm{sd}(AX)}(s)}{\d t \, \d M^2} & = & 
\frac{g_{3\pomeron}}{16\pi} \, \beta_{A\pomeron}^2 \, 
\beta_{B\pomeron} \, \frac{1}{M^2} \, \exp(B_{\mrm{sd}(AX)}t)
\, F_{\mrm{sd}} ~, \nonumber \\
\frac{\d\sigma_{\mrm{dd}}(s)}{\d t \, \d M_1^2 \, \d M_2^2} & = &  
\frac{g_{3\pomeron}^2}{16\pi} \, \beta_{A\pomeron} \, 
\beta_{B\pomeron} \, \frac{1}{M_1^2} \, \frac{1}{M_2^2} \, 
\exp(B_{\mrm{dd}}t) \, F_{\mrm{dd}} ~. 
\end{eqnarray}

The couplings $\beta_{A\pomeron}$ are related to the pomeron term
$X^{AB} s^{\epsilon}$ of the total cross section parametrization,
eq.~(\ref{pg:sigtotpomreg}). Picking a reference scale 
$\sqrt{s_{\mrm{ref}}} = 20$ GeV, the couplings are given by 
$\beta_{A\pomeron}\beta_{B\pomeron} = 
X^{AB} \, s_{\mrm{ref}}^{\epsilon}$. The triple-pomeron coupling is
determined from single-diffractive data to be 
$g_{3\pomeron} \approx 0.318$ mb$^{1/2}$; within the context of the 
formulae in this section.

The spectrum of diffractive masses $M$ is taken to begin
0.28 GeV $\approx 2 m_{\pi}$ above the mass of the respective 
incoming particle and extend to the kinematical limit. The simple
$\d M^2 / M^2$ form is modified by the mass-dependence in the 
diffractive slopes and in the $F_{\mrm{sd}}$ and $F_{\mrm{dd}}$ 
factors. 

The slope parameters are assumed to be
\begin{eqnarray}
B_{\mrm{sd}(XB)}(s) & = & 2b_B + 2\alpha' \ln\left(\frac{s}{M^2}\right)
~, \nonumber \\
B_{\mrm{sd}(AX)}(s) & = & 2b_A + 2\alpha' \ln\left(\frac{s}{M^2}\right)
~, \nonumber \\
B_{\mrm{dd}}(s) & = & 2\alpha' \ln\left(e^4 + \frac{s s_0}{M_1^2 M_2^2}
\right) ~. 
\end{eqnarray}
Here $\alpha' = 0.25$ GeV$^{-2}$ and conventionally $s_0$ is picked as
$s_0 = 1 / \alpha'$. The term $e^4$ in $B_{\mrm{dd}}$ is added by hand
to avoid a breakdown of the standard expression for large values of
$M_1^2 M_2^2$. The $b_{A,B}$ terms protect $B_{\mrm{sd}}$ from breaking
down; however a minimum value of 2 is still explicitly required for 
 $B_{\mrm{sd}}$, which comes into play e.g. for a $\Jpsi$ state
(as part of a VMD photon beam).

The kinematical range in $t$ depends on all the masses of the 
problem. In terms of the scaled variables $\mu_1 = m_A^2/s$,
$\mu_2 = m_B^2/s$, $\mu_3 = M_{(1)}^2/s$ ($=m_A^2/s$ when $A$
scatters elastically), $\mu_4 = M_{(2)}^2/s$ ($=m_B^2/s$ when $B$
scatters elastically), and the combinations
\begin{eqnarray}
C_1 & = & 1 - (\mu_1 + \mu_2 + \mu_3 + \mu_4) +
(\mu_1 - \mu_2) (\mu_3 - \mu_4) ~, \nonumber \\
C_2 & = & \sqrt{(1 - \mu_1 -\mu_2)^2 - 4 \mu_1 \mu_2} \,
\sqrt{(1 - \mu_3 - \mu_4)^2 - 4 \mu_3 \mu_4} ~, \nonumber \\
C_3 & = & (\mu_3 - \mu_1) (\mu_4 - \mu_2) +
(\mu_1 + \mu_4 - \mu_2 - \mu_3) (\mu_1 \mu_4 - \mu_2 \mu_3) ~,
\end{eqnarray}
one has $t_{\mmin} < t < t_{\mmax}$ with
\begin{eqnarray}
t_{\mmin} & = & - \frac{s}{2} (C_1 + C_2) ~, \nonumber \\
t_{\mmax} & = & - \frac{s}{2} (C_1 - C_2)  
= - \frac{s}{2} \, \frac{4C_3}{C_1 + C_2}
= \frac{s^2 C_3}{t_{\mmin}} ~.
\end{eqnarray}

The Regge formulae above for single- and double-diffractive events
are supposed to hold in certain asymptotic regions of the total phase
space.  Of course, there will be diffraction also outside these 
restrictive regions. Lacking a theory which predicts differential cross 
sections at arbitrary $t$ and $M^2$ values, the Regge formulae are used
everywhere, but fudge factors are introduced in order to obtain 
`sensible' behaviour in the full phase space. These factors are:  
\begin{eqnarray}
F_{\mrm{sd}} & = & \left( 1 - \frac{M^2}{s} \right)  
\left( 1 + \frac{c_{\mrm{res}} \, M_{\mrm{res}}^2}
{M_{\mrm{res}}^2 + M^2} \right)  ~, \nonumber \\
F_{\mrm{dd}} & = & 
\left( 1 - \frac{\left( M_1 + M_2 \right)^2}{s} \right) 
\left( \frac{s\, m_{\p}^2}{ s\, m_{\p}^2 + M_1^2\, M_2^2} \right) 
\nonumber  \\
& \times & 
\left( 1 + \frac{c_{\mrm{res}} \, M_{\mrm{res}}^2}
{M_{\mrm{res}}^2 + M_1^2} \right)
\left( 1 + \frac{c_{\mrm{res}} \, M_{\mrm{res}}^2}
{M_{\mrm{res}}^2 + M_2^2} \right) ~.
\end{eqnarray}
The first factor in either expression suppresses production close to 
the kinematical limit. The second factor in $F_{dd}$ suppresses 
configurations where the two diffractive systems overlap in rapidity 
space. The final factors give an enhancement of the low-mass region,
where a resonance structure is observed in the data. Clearly a more
detailed modelling would have to be based on a set of exclusive states
rather than on this smeared-out averaging procedure. A reasonable fit
to $\p\p / \pbar\p$ data is obtained for 
$c_{\mrm{res}} = 2$ and $M_{\mrm{res}} = 2$~GeV, 
for an arbitary particle $A$ which is diffractively excited 
we use $M_{\mrm{res}}^A = m_A - m_{\p} + 2$~GeV. 

The diffractive cross-section formulae above have been integrated 
for a set of c.m. energies, starting at 10 GeV, and the results have 
been parametrized. The form of these parametrizations is given in
ref. \cite{Sch94}, with explicit numbers for the $\p\p/\pbar\p$
case. {\Py} also contains similar parametrizations for 
$\pi\p$ (assumed to be same as $\rho\p$ and $\omega\p$),
$\phi\p$, $\Jpsi\p$, $\rho\rho$ ($\pi\pi$ etc.), $\rho\phi$,
$\rho\Jpsi$, $\phi\phi$, $\phi\Jpsi$ and $\Jpsi\Jpsi$.      
 
The processes above do not obey the ordinary event mixing strategy.
First of all, since their total cross sections are known, it is
possible to pick the appropriate process from the start, and then
remain with that choice. In other words, if the selection of
kinematical variables fails, one would not go back and pick a new
process, the way it was done in section \ref{sss:mixingproc}.
Second, it is not possible to impose any cuts or restrain allowed
incoming or outgoing flavours: if not additional information were to
be provided, it would make the whole scenario ill-defined.
Third, it is not recommended to mix generation of these processes
with that of any of the other ones: normally the other processes
have so small cross sections that they would almost never be
generated anyway. (We here exclude the cases of `underlying events'
and `pile-up events', where mixing is provided for, and even is a
central part of the formalism, see sections \ref{ss:multint} and
\ref{ss:pileup}.)

Once the cross-section parametrizations has been used to pick one
of the processes, the variables $t$ and $M$ are selected according
to the formulae given above.

A $\rho^0$ formed by $\gamma \to \rho^0$ in elastic or diffractive
scattering is polarized, and therefore its decay angular distribution 
in $\rho^0 \to \pi^+ \pi^-$ is taken to be proportional to 
$\sin^2 \theta$, where the reference axis is given by the $\rho^0$ 
direction of motion.

A light diffractive system, with a mass less than 1 GeV above the 
mass of the incoming particle, is allowed to decay isotropically into 
a two-body state. Single-resonance diffractive states, such as a 
$\Delta^+$, are therefore not explicily generated, but are assumed 
described in an average, smeared-out sense.

A more massive diffractive system is subsequently treated
as a string with the quantum numbers of the original hadron. Since the
exact nature of the pomeron exchanged between the hadrons is unknown,
two alternatives are included. In the first, the pomeron is assumed to
couple to (valence) quarks, so that the string is stretched directly
between the struck quark and the remnant diquark (antiquark) of the
diffractive state. In the second, the interaction is rather with a
gluon, giving rise to a `hairpin' configuration in which the string
is stretched from a quark to a gluon and then back to a diquark
(antiquark). Both of these scenarios could be present in the data;
the default choice is to mix them in equal proportions.
 
There is experimental support for more complicated scenarios
\cite{Ing85}, wherein the pomeron has a partonic substructure,
which e.g. can lead to high-$\pT$ jet production in the diffractive
system. The full machinery, wherein a pomeron spectrum is folded
with a pomeron-proton hard interaction, is not available in {\Py}.

\subsubsection{Photoproduction and $\gamma\gamma$ physics}
\label{sss:photoprod}
 
The photoproduction part is still under active development. Currently
only interactions between a hadron and a real photon have been studied
in detail. $\gamma\gamma$ physics is under study \cite{Sch94a}, and
is now preliminarily included for real photons. Deep inelastic 
scattering on a real photon is also preliminarily included.
In the future it is hoped to add interactions of mildly virtual photons 
(the transition region between real photons and deep inelastic 
scattering).

The total $\gamma\p$ and $\gamma\gamma$ cross sections can again be 
parametrized in a
form like eq.~(\ref{pg:sigtotpomreg}), which is not so obvious since 
the photon has more complicated structure than an ordinary hadron.
In fact, the structure is still not so well understood. The model we
outline is the one studied by Schuler and Sj\"ostrand 
\cite{Sch93,Sch93a}. In this model the physical photon is represented
by
\begin{equation}
| \gamma \rangle = \sqrt{Z_3} \, | \gamma_B \rangle +
\sum_{V=\rho^0,\omega,\phi,\Jpsi} \frac{e}{f_V} \, | V \rangle +
\frac{e}{f_{\q\qbar}} \, | \q\qbar \rangle + 
\sum_{\ell=\e,\mu,\tau} \frac{e}{f_{\ell\ell}} \, 
| \ell^+ \ell^- \rangle ~.      
\label{pg:gammadecompo}
\end{equation}
 
By virtue of this superposition, one is led to a model of $\gamma\p$ 
interactions, where three different kinds of events may be 
distinguished:
\begin{Itemize}
\item Direct events, wherein the bare photon $| \gamma_B \rangle$
interacts directly with a parton from the proton. The process is 
perturbatively calculable, and no parton distributions of the photon
are involved. The typical event structure is two high-$\pT$ jets and
a proton remnant, while the photon does not leave behind any 
remnant.
\item VMD events, in which the photon fluctuates into a vector meson, 
predominantly a $\rho^0$. All the event classes known from ordinary 
hadron--hadron interactions may thus occur here, such as elastic, 
diffractive, low-$\pT$ and high-$\pT$ events. For the latter, one may 
define (VMD) parton distributions of the photon, and the photon also 
leaves behind a beam remnant. This remnant is smeared in transverse 
momentum by a typical `primordial $k_{\perp}$' of a few hundred MeV.
\item Anomalous events, in which the photon fluctuates into a $\q\qbar$ 
pair of larger virtuality than in the VMD class. This process is 
perturbatively calculable, as is the subsequent QCD evolution. It 
gives rise to the so-called anomalous part of the parton distributions 
ofthe photon, whence the name for the class. It is assumed that only 
high-$\pT$ events may occur. Either the  $\q$ or the $\qbar$ 
plays the r\^ole of a beam remnant, but this remnant has a larger
$\pT$ than in the VMD case, related to the virtuality of the
$\gamma \leftrightarrow \q\qbar$ fluctuation.
\end{Itemize} 
The $| \ell^+ \ell^- \rangle$ states can only interact strongly with 
partons inside the hadron at higher orders, and can therefore be 
neglected. 

In order that the above classification is smooth and free of double
counting, one has to introduce scales that separate the three
components. The main one is $p_0$, which separates the low-mass
vector meson region from the high-mass $| \q\qbar \rangle$ one, 
$p_0 \approx m_{\phi}/2 \approx 0.5$ GeV. Since it is the same 
$\gamma\q\qbar$ vertex that is responsible for the bare $\gamma \p$ 
interactions, $p_0$ is also the lower cut-off of the 
photon--parton cross sections. In addition, a $\pTmin$ cut-off is
needed to separate low-$\pT$ and high-$\pT$ physics; see section
\ref{ss:multint}. As it turns out, somewhat different $\pTmin$ values
are needed for the VMD and anomalous parts; at least qualitatively
this can be understood in terms of different sizes of the 
wave functions. 

The VMD and anomalous events are together called resolved ones.
In terms of high-$\pT$ jet production, the VMD and anomalous
contributions can be combined into a total resolved one, and the
same for parton-distribution functions. However, the two classes
differ in the structure of the underlying event and in the
appearance of soft processes.

In terms of cross sections, eq.\ (\ref{pg:gammadecompo}) corresponds 
to
\begin{equation}
\sigma_{\mrm{tot}}^{\gamma\p}(s) = \sigma_{\mrm{dir}}^{\gamma\p}(s) +
\sigma_{\mrm{VMD}}^{\gamma\p}(s) + \sigma_{\mrm{anom}}^{\gamma\p}(s) ~.
\label{pg:gammacrossdecompo}
\end{equation}

The direct cross section is, to lowest order, the perturbative cross 
section for the two processes $\gamma\q \to \q\g$ and 
$\gamma\g \to \q\qbar$, with a lower cut-off $\pT > p_0$.
Properly speaking, this should be multiplied by the $Z_3$ coefficient, 
\begin{equation}
Z_3 = 1 -  
\sum_{V=\rho^0,\omega,\phi,\Jpsi} \left( \frac{e}{f_V} \right)^2 - 
\left( \frac{e}{f_{\q\qbar}} \right)^2 -
\sum_{\ell=\e,\mu,\tau} \left( \frac{e}{f_{\ell\ell}} \right)^2 ~,
\end{equation}
but normally $Z_3$ is so close to unity as to make no difference.

The VMD factor $(e/f_V)^2 = 4\pi\alphaem/f_V^2$ gives the probability 
for the transition $\gamma \to V$. The coefficients $f_V^2/4\pi$ are
determined from data to be (with a non-negligible amount of 
uncertainty) 2.20 for $\rho^0$, 23.6 for $\omega$, 18.4 for $\phi$ 
and 11.5 for $\Jpsi$. Together these numbers imply that the photon
can be found in a VMD state about 0.4\% of the time, dominated by the
$\rho^0$ contribution. All the properties of the VMD interactions
can be obtained by appropriately scaling down $V\p$ physics 
predictions. Thus the whole machinery developed in the previous 
subsection for hadron--hadron interactions is directly applicable.
Also parton distributions of the VMD component inside the photon 
are obtained by suitable rescaling.

The contribution from the `anomalous' high-mass fluctuations depends
on the typical scale $\mu$ of the interaction
\begin{equation}
\left( \frac{e}{f_{\q\qbar}} \right)^2 \approx 
\frac{\alphaem}{2\pi} \, \frac{N_C}{3} \, 
\left( 2 \sum_{\q} e_{\q}^2 \right) 
\ln\left( \frac{\mu^2}{p_0^2} \right) ~,
\label{anomintfirst}
\end{equation}
where $N_C = 3$ and $\q$ runs over the quarks that can be taken
massless compared with $\mu$. The logarithmic increase with $\mu$
implies that the anomalous contribution to the total photoproduction 
cross section ($\mu \sim m_V$) is less important than 
that to high-$\pT$ jet production ($\mu \sim \pT$). To first 
approximation, therefore only perturbative jet production above some 
$\pTmin$ scale is considered. This includes the standard QCD 
parton--parton scattering processes, with anomalous-photon parton 
distributions that are fully perturbatively calculable \cite{Sch95}. 
In order to satisfy the equality in eq.~(\ref{pg:gammacrossdecompo}), 
with the total cross section known and the direct and VDM 
contributions already fixed, a behaviour roughly like
\begin{equation}
\pTmin^{\mrm{anom}}(s) = 0.70 + 0.17 \log^2(1.+0.05\sqrt{s})
\end{equation}
is needed over the HERA energy range. This is to be seen entirely
as a pragmatic parametrization, not be given any fundamental
interpretation. It is based on SaS set~1D, another set might well 
require a somewhat different form.

In $\gamma\gamma$ physics \cite{Sch94a}, the superposition in 
eq.~(\ref{pg:gammadecompo}) applies separately for each of the two 
incoming photons. In total there are therefore $3 \times 3 = 9$ 
combinations. However, trivial symmetry reduces this to six distinct 
classes, written in terms of the total cross section 
(cf. eq.~(\ref{pg:gammacrossdecompo})) as
\begin{eqnarray}
\sigma_{\mrm{tot}}^{\gamma\gamma}(s) & = &
\sigma_{\mrm{dir}\times\mrm{dir}}^{\gamma\gamma}(s) +
\sigma_{\mrm{VMD}\times\mrm{VMD}}^{\gamma\gamma}(s) + 
\sigma_{\mrm{anom}\times\mrm{anom}}^{\gamma\gamma}(s) \nonumber \\ 
& + & 2 \sigma_{\mrm{dir}\times\mrm{VMD}}^{\gamma\gamma}(s) +
2 \sigma_{\mrm{dir}\times\mrm{anom}}^{\gamma\gamma}(s) + 
2 \sigma_{\mrm{VMD}\times\mrm{anom}}^{\gamma\gamma}(s) ~.
\label{pg:gagacrossdecompo}
\end{eqnarray}
A parametrization of the total $\gamma\gamma$ cross section and 
comments on its subdivision into the six classes is found in
\cite{Sch94a}.

The six different kinds of $\gamma\gamma$ events are thus:
\begin{Itemize}
\item The direct$\times$direct events, which correspond to the 
subprocess $\gamma\gamma \to \q\qbar$ (or $\ell^+\ell^-$). 
The typical event structure is two high-$\pT$ jets and no beam
remnants. The lower cut-off is $\pT > p_0$.
\item The VMD$\times$VMD events, which have the same properties as
the VMD $\gamma\p$ events. There are four by four combinations of
the two incoming vector mesons, with one VMD factor for each meson.
\item The anomalous$\times$anomalous events, wherein each photon
fluctuates into a $\q\qbar$ pair of larger virtuality than in the
VMD class. One parton of each pair gives a beam remnant, whereas
the other (or a daughter parton thereof) participates in a high-$\pT$ 
scattering, with $\pT > \pTmin^{\mrm{anom}}$. 
\item The direct$\times$VMD events, which have the same properties as
the direct $\gamma\p$ events. 
\item The direct$\times$anomalous events, in which a bare photon 
interacts with a parton from the anomalous photon. The lower cut-off
for the hard scattering is given by $\pTmin^{\mrm{anom}}$.
The typical structure is then two high-$\pT$ jets and a beam remnant. 
\item The VMD$\times$anomalous events, which have the same properties 
as the anomalous $\gamma\p$ events. 
\end{Itemize} 
In much of the literature, where a coarser classification us used, 
our direct$\times$direct is called direct, our direct$\times$VMD
and direct$\times$anomalous is called 1-resolved since they both 
involve one resolved photon which gives a beam remnant, 
and the rest are called 2-resolves since both photons are resolved 
and give beam remnants. 
 
\clearpage
 
\section{Physics Processes in PYTHIA}
\label{s:pytproc}
 
In this section we enumerate the physics processes that are available 
in {\Py}, introducing the ISUB code that can be used to select
desired processes. A number of comments are made about the
physics scenarios involved, in particular with respect to
underlying assumptions and domain of validity. The section closes
with a survey of interesting processes by machine.
 
\subsection{The Process Classification Scheme}
\label{ss:ISUBcode}
 
A wide selection of fundamental $2 \to 1$ and $2 \to 2$ tree
processes of the Standard Model (electroweak and strong) has been
included in {\Py}, and slots are provided for many more, not yet
implemented. In addition, a few `minimum-bias'-type processes
(like elastic scattering), loop graphs, box graphs, $2 \to 3$ tree
graphs and some non-Standard Model processes are included. The
classification is not always unique. A process that proceeds only
via an $s$-channel state is classified as a $2 \to 1$ process
(e.g. $\q \qbar \to \gammaZ \to \ee$), but a $2 \to 2$
cross section may well have contributions from $s$-channel diagrams
($\g \g \to \g \g$ obtains contributions from
$\g \g \to \g^* \to \g \g$). Also, in the program, $2 \to 1$ and
$2 \to 2$ graphs may sometimes be folded with two $1 \to 2$
splittings to form effective $2 \to 3$ or $2 \to 4$ processes
($\W^+ \W^- \to \H^0$ is folded with $\q \to \q'' \W^+$ and
$\q' \to \q''' \W^-$ to give $\q \q' \to \q'' \q''' \H^0$).
 
It is possible to select a combination of subprocesses to simulate,
and also afterwards to know which subprocess was actually selected in
each event. For this purpose, all subprocesses are numbered according
to an ISUB code. The list of possible codes is given in Tables
\ref{t:procone}, \ref{t:proctwo}, \ref{t:procthree} and
\ref{t:procfour}. Only processes marked with a `+' sign in the first
column have been implemented in the program to date. Although
ISUB codes were originally designed in a logical fashion,
we must admit that subsequent developments of the program have tended
to obscure the structure. For instance, the process numbers for Higgs
production are spread out, in part as a consequence of the original
classification, in part because further production mechanisms have been
added one at a time, in whatever free slots could be found. At some
future date the subprocess list will therefore be reorganized.
In the thematic descriptions that follow the main tables, the
processes of interest are repeated in a more logical order. If you
want to look for a specific process, it will be easier
to find it there.
 
\begin{table}[pt]
\caption{Subprocess codes, part 1. First column is `+' for processes
implemented and blank for those that are only foreseen. Second is
the subprocess number \ttt{ISUB}, and third the description of the
process. The final column gives references from which the
cross sections have been obtained. See text for further information.
\protect\label{t:procone} }
\begin{center}
\begin{tabular}{|c|r|l|l|@{\protect\rule{0mm}{\tablinsep}}}
\hline
In & No. & Subprocess & Reference \\
\hline
  &     & a) $2 \to 1$, tree &    \\
+ &   1 & $\f_i \fbar_i \to \gammaZ$ & \cite{Eic84} \\
+ &   2 & $\f_i \fbar_j \to \W^+$ & \cite{Eic84} \\
+ &   3 & $\f_i \fbar_i \to \H^0$ & \cite{Eic84} \\
  &   4 & $\gamma \W^+ \to \W^+$ &  \\
+ &   5 & $\Z^0 \Z^0 \to \H^0$ & \cite{Eic84,Cha85} \\
  &   6 & $\Z^0 \W^+ \to \W^+$ &  \\
  &   7 & $\W^+ \W^- \to \Z^0$ &  \\
+ &   8 & $\W^+ \W^- \to \H^0$ & \cite{Eic84,Cha85} \\
\hline
  &     & b) $2 \to 2$, tree &    \\
+ &  10 & $\f_i \f_j \to \f_i \f_j$ (QFD) & \cite{Ing87b}  \\
+ &  11 & $\f_i \f_j \to \f_i \f_j$ (QCD) & 
\cite{Com77,Ben84,Eic84,Chi90} \\
+ &  12 & $\f_i \fbar_i \to \f_k \fbar_k$ & 
\cite{Com77,Ben84,Eic84,Chi90} \\
+ &  13 & $\f_i \fbar_i \to \g \g$ & \cite{Com77,Ben84}  \\
+ &  14 & $\f_i \fbar_i \to \g \gamma$ & \cite{Hal78,Ben84}  \\
+ &  15 & $\f_i \fbar_i \to \g \Z^0$ & \cite{Eic84}  \\
+ &  16 & $\f_i \fbar_j \to \g \W^+$ & \cite{Eic84}  \\
  &  17 & $\f_i \fbar_i \to \g \H^0$ &   \\
+ &  18 & $\f_i \fbar_i \to \gamma \gamma$ & \cite{Ber84}  \\
+ &  19 & $\f_i \fbar_i \to \gamma \Z^0$ & \cite{Eic84}  \\
+ &  20 & $\f_i \fbar_j \to \gamma \W^+$ & \cite{Eic84,Sam91}  \\
  &  21 & $\f_i \fbar_i \to \gamma \H^0$ &   \\
+ &  22 & $\f_i \fbar_i \to \Z^0 \Z^0$ & \cite{Eic84,Gun86}  \\
+ &  23 & $\f_i \fbar_j \to \Z^0 \W^+$ & \cite{Eic84,Gun86}  \\
+ &  24 & $\f_i \fbar_i \to \Z^0 \H^0$ & \cite{Ber85}  \\
+ &  25 & $\f_i \fbar_i \to \W^+ \W^-$ & \cite{Bar94,Gun86}  \\
+ &  26 & $\f_i \fbar_j \to \W^+ \H^0$ & \cite{Eic84}  \\
  &  27 & $\f_i \fbar_i \to \H^0 \H^0$ &   \\
+ &  28 & $\f_i \g \to \f_i \g$ & \cite{Com77,Ben84}  \\
+ &  29 & $\f_i \g \to \f_i \gamma$ & \cite{Hal78,Ben84}  \\
+ &  30 & $\f_i \g \to \f_i \Z^0$ & \cite{Eic84}  \\
+ &  31 & $\f_i \g \to \f_k \W^+$ & \cite{Eic84}  \\
  &  32 & $\f_i \g \to \f_i \H^0$ &   \\
+ &  33 & $\f_i \gamma \to \f_i \g$ & \cite{Duk82}  \\
+ &  34 & $\f_i \gamma \to \f_i \gamma$ & \cite{Duk82}  \\
+ &  35 & $\f_i \gamma \to \f_i \Z^0$ & \cite{Gab86}  \\
+ &  36 & $\f_i \gamma \to \f_k \W^+$ & \cite{Gab86}  \\
  &  37 & $\f_i \gamma \to \f_i \H^0$ &   \\
  &  38 & $\f_i \Z^0 \to \f_i \g$ &   \\
  &  39 & $\f_i \Z^0 \to \f_i \gamma$ &   \\
  &  40 & $\f_i \Z^0 \to \f_i \Z^0$ &   \\
\hline
\end{tabular}
\end{center}
\end{table}
 
\begin{table}[pt]
\caption{Subprocess codes, part 2. First column is `+' for processes
implemented and blank for those that are only foreseen. Second is
the subprocess number \ttt{ISUB}, and third the description of the
process. The final column gives references from which the
cross sections have been obtained. See text for further information.
\protect\label{t:proctwo} }
\begin{center}
\begin{tabular}{|c|r|l|l|@{\protect\rule{0mm}{\tablinsep}}}
\hline
In & No. & Subprocess & Reference \\
\hline
  &     & b) $2 \to 2$, tree (cont'd) &    \\
  &  41 & $\f_i \Z^0 \to \f_k \W^+$ &   \\
  &  42 & $\f_i \Z^0 \to \f_i \H^0$ &   \\
  &  43 & $\f_i \W^+ \to \f_k \g$ &   \\
  &  44 & $\f_i \W^+ \to \f_k \gamma$ &   \\
  &  45 & $\f_i \W^+ \to \f_k \Z^0$ &   \\
  &  46 & $\f_i \W^+ \to \f_k \W^+$ &   \\
  &  47 & $\f_i \W^+ \to \f_k \H^0$ &   \\
  &  48 & $\f_i \H^0 \to \f_i \g$ &   \\
  &  49 & $\f_i \H^0 \to \f_i \gamma$ &   \\
  &  50 & $\f_i \H^0 \to \f_i \Z^0$ &   \\
  &  51 & $\f_i \H^0 \to \f_k \W^+$ &   \\
  &  52 & $\f_i \H^0 \to \f_i \H^0$ &   \\
+ &  53 & $\g \g \to \f_k \fbar_k$ & \cite{Com77,Ben84}  \\
+ &  54 & $\g \gamma \to \f_k \fbar_k$ & \cite{Duk82}  \\
  &  55 & $\g \Z^0 \to \f_k \fbar_k$ &   \\
  &  56 & $\g \W^+ \to \f_k \fbar_l$ &   \\
  &  57 & $\g \H^0 \to \f_k \fbar_l$ &   \\
+ &  58 & $\gamma \gamma \to \f_k \fbar_k$ & \cite{Bar90}  \\
  &  59 & $\gamma \Z^0 \to \f_k \fbar_k$ &   \\
  &  60 & $\gamma \W^+ \to \f_k \fbar_l$ &   \\
  &  61 & $\gamma \H^0 \to \f_k \fbar_k$ &   \\
  &  62 & $\Z^0 \Z^0 \to \f_k \fbar_k$ &   \\
  &  63 & $\Z^0 \W^+ \to \f_k \fbar_l$ &   \\
  &  64 & $\Z^0 \H^0 \to \f_k \fbar_k$ &   \\
  &  65 & $\W^+ \W^- \to \f_k \fbar_k$ &   \\
  &  66 & $\W^+ \H^0 \to \f_k \fbar_l$ &   \\
  &  67 & $\H^0 \H^0 \to \f_k \fbar_k$ &   \\
+ &  68 & $\g \g \to \g \g$ &  \cite{Com77,Ben84} \\
+ &  69 & $\gamma \gamma \to \W^+ \W^-$ & \cite{Kat83}   \\
+ &  70 & $\gamma \W^+ \to \Z^0 \W^+$ & \cite{Kun87}  \\
+ &  71 & $\Z^0 \Z^0 \to \Z^0 \Z^0$ (longitudinal) &  \cite{Abb87} \\
+ &  72 & $\Z^0 \Z^0 \to \W^+ \W^-$ (longitudinal) &  \cite{Abb87} \\
+ &  73 & $\Z^0 \W^+ \to \Z^0 \W^+$ (longitudinal) &  \cite{Dob91} \\
  &  74 & $\Z^0 \H^0 \to \Z^0 \H^0$ &   \\
  &  75 & $\W^+ \W^- \to \gamma \gamma$ &   \\
+ &  76 & $\W^+ \W^- \to \Z^0 \Z^0$ (longitudinal) & \cite{Ben87b} \\
+ &  77 & $\W^+ \W^{\pm} \to \W^+ \W^{\pm}$ (longitudinal) & 
\cite{Dun86,Bar90a}  \\
  &  78 & $\W^+ \H^0 \to \W^+ \H^0$ &   \\
  &  79 & $\H^0 \H^0 \to \H^0 \H^0$ &   \\
+ &  80 & $\q_i \gamma \to \q_k \pi^{\pm}$ & \cite{Bag82} \\
\hline
\end{tabular}
\end{center}
\end{table}
 
\begin{table}[pt]
\caption{Subprocess codes, part 3. First column is `+' for processes
implemented and blank for those that are only foreseen. Second is
the subprocess number \ttt{ISUB}, and third the description of the
process. The final column gives references from which the
cross sections have been obtained. See text for further information.
\protect\label{t:procthree} }
\begin{center}
\begin{tabular}{|c|r|l|l|@{\protect\rule{0mm}{\tablinsep}}}
\hline
In & No. & Subprocess & Reference \\
\hline
  &     & c) $2 \to 2$, tree, massive final quarks &    \\
+ &  81 & $\f_i \fbar_i \to \Q_k \Qbar_k$ & \cite{Com79}  \\
+ &  82 & $\g \g \to \Q_k \Qbar_k$ & \cite{Com79}  \\
+ &  83 & $\q_i \f_j \to \Q_k \f_l$ & \cite{Dic86}  \\
+ &  84 & $\g \gamma \to \Q_k \Qbar_k$ & \cite{Fon81}  \\
+ &  85 & $\gamma \gamma \to \F_k \Fbar_k$ & \cite{Bar90}  \\
+ &  86 & $\g \g \to \Jpsi \g$ & \cite{Bai83}  \\
+ &  87 & $\g \g \to \chi_{0 \c} \g$ & \cite{Gas87}  \\
+ &  88 & $\g \g \to \chi_{1 \c} \g$ & \cite{Gas87}  \\
+ &  89 & $\g \g \to \chi_{2 \c} \g$ & \cite{Gas87}  \\
\hline
  &     & d) `minimum bias' &   \\
+ &  91 & elastic scattering               & \cite{Sch94}  \\
+ &  92 & single diffraction ($AB \to XB$) & \cite{Sch94}  \\
+ &  93 & single diffraction ($AB \to AX$) & \cite{Sch94}  \\
+ &  94 & double diffraction               & \cite{Sch94}  \\
+ &  95 & low-$\pT$ production & \cite{Sjo87}  \\
\hline
  &    & e) $2 \to 1$, loop &   \\
  & 101 & $\g \g \to \Z^0$ &   \\
+ & 102 & $\g \g \to \H^0$ & \cite{Eic84}  \\
+ & 103 & $\gamma \gamma \to \H^0$ & \cite{Dre89}  \\
\hline
  &     & f) $2 \to 2$, box &   \\
+ & 110 & $\f_i \fbar_i \to \gamma \H^0$ & \cite{Ber85a} \\
+ & 111 & $\f_i \fbar_i \to \g \H^0$ & \cite{Ell88}  \\
+ & 112 & $\f_i \g \to \f_i \H^0$ & \cite{Ell88}  \\
+ & 113 & $\g \g \to \g \H^0$ & \cite{Ell88}  \\
+ & 114 & $\g \g \to \gamma \gamma$ & \cite{Con71,Ber84,Dic88}  \\
+ & 115 & $\g \g \to \g \gamma$ & \cite{Con71,Ber84,Dic88}  \\
  & 116 & $\g \g \to \gamma \Z^0$ &   \\
  & 117 & $\g \g \to \Z^0 \Z^0$ &   \\
  & 118 & $\g \g \to \W^+ \W^-$ &   \\
  & 119 & $\gamma \gamma \to \g \g$ &   \\
\hline
  &     & g) $2 \to 3$, tree &   \\
+ & 121 & $\g \g \to \Q_k \Qbar_k \H^0$ & \cite{Kun84}  \\
+ & 122 & $\q_i \qbar_i \to \Q_k \Qbar_k \H^0$ & \cite{Kun84}  \\
+ & 123 & $\f_i \f_j \to \f_i \f_j \H^0$ ($\Z \Z$ fusion) & 
\cite{Cah84}  \\
+ & 124 & $\f_i \f_j \to \f_k \f_l \H^0$ ($\W^+\W^-$ fusion) & 
\cite{Cah84}  \\
+ & 131 & $\g \g \to \Z^0 \Q_k \br{\Q}_k$ & \cite{Eij90}  \\
\hline
\end{tabular}
\end{center}
\end{table}
 
\begin{table}[pt]
\caption{Subprocess codes, part 4. First column is `+' for processes
implemented and blank for those that are only foreseen. Second is
the subprocess number \ttt{ISUB}, and third the description of the
process. The final column gives references from which the
cross sections have been obtained. See text for further information.
\protect\label{t:procfour} }
\begin{center}
\begin{tabular}{|c|r|l|l|@{\protect\rule{0mm}{\tablinsep}}}
\hline
In & No. & Subprocess & Reference \\
\hline
  &     & h) non-Standard Model, $2 \to 1$ &    \\
+ & 141 & $\f_i \fbar_i \to \gamma/\Z^0/\Z'^0$ & \cite{Alt89}  \\
+ & 142 & $\f_i \fbar_j \to \W'^+$ & \cite{Alt89}  \\
+ & 143 & $\f_i \fbar_j \to \H^+$ & \cite{Gun87}  \\
+ & 144 & $\f_i \fbar_j \to \R$ & \cite{Ben85a}  \\
+ & 145 & $\q_i \ell_j \to \L_{\Q}$ & \cite{Wud86}  \\
+ & 147 & $\d \g \to \d^*$ & \cite{Bau90}  \\
+ & 148 & $\u \g \to \u^*$ & \cite{Bau90}  \\
+ & 149 & $\g \g \to \eta_{\mrm{techni}}$ & \cite{Eic84,App92}  \\
+ & 151 & $\f_i \fbar_i \to \H'^0$ & \cite{Eic84}  \\
+ & 152 & $\g \g \to \H'^0$ & \cite{Eic84}  \\
+ & 153 & $\gamma \gamma \to \H'^0$ & \cite{Dre89}  \\
+ & 156 & $\f_i \fbar_i \to \A^0$ & \cite{Eic84}  \\
+ & 157 & $\g \g \to \A^0$ & \cite{Eic84}  \\
+ & 158 & $\gamma \gamma \to \A^0$ & \cite{Dre89}  \\
\hline
  &     & i) non-Standard Model, $2 \to 2$ and $2 \to 3$ &    \\
+ & 161 & $\f_i \g \to \f_k \H^+$ & \cite{Bar88}  \\
+ & 162 & $\q \g \to \ell \L_{\Q}$ & \cite{Hew88}  \\
+ & 163 & $\g \g \to \L_{\Q} \br{\L}_{\Q}$ & 
\cite{Hew88,Eic84} \\
+ & 164 & $\q_i \qbar_i \to \L_{\Q} \br{\L}_{\Q}$ & 
\cite{Hew88}  \\
+ & 165 & $\f_i \fbar_i \to \f_k \fbar_k$ (via $\gammaZ$) & 
\cite{Eic84,Lan91}  \\
+ & 166 & $\f_i \fbar_j \to \f_k \fbar_l$ (via $\W^{\pm}$) & 
\cite{Eic84,Lan91}  \\
+ & 167 & $\q \q' \to \q'' \d^*$ & \cite{Bau90}  \\
+ & 168 & $\q \q' \to \q'' \u^*$ & \cite{Bau90}  \\
+ & 171 & $\f_i \fbar_i \to \Z^0 \H'^0$ & \cite{Eic84}  \\
+ & 172 & $\f_i \fbar_j \to \W^+ \H'^0$ & \cite{Eic84}  \\
+ & 173 & $\f_i \f_j \to \f_i \f_j \H'^0$ ($\Z \Z$ fusion) & 
\cite{Cah84}  \\
+ & 174 & $\f_i \f_j \to \f_k \f_l \H'^0$ ($\W^+\W^-$ fusion) & 
\cite{Cah84}  \\
+ & 176 & $\f_i \fbar_i \to \Z^0 \A^0$ & \cite{Eic84}  \\
+ & 177 & $\f_i \fbar_j \to \W^+ \A^0$ & \cite{Eic84}  \\
+ & 178 & $\f_i \f_j \to \f_i \f_j \A^0$ ($\Z \Z$ fusion) & 
\cite{Cah84}  \\
+ & 179 & $\f_i \f_j \to \f_k \f_l \A^0$ ($\W^+\W^-$ fusion) & 
\cite{Cah84}  \\
+ & 181 & $\g \g \to \Q_k \Qbar_k \H'^0$ & \cite{Kun84}  \\
+ & 182 & $\q_i \qbar_i \to \Q_k \Qbar_k \H'^0$ & \cite{Kun84}  \\
+ & 186 & $\g \g \to \Q_k \Qbar_k \A^0$ & \cite{Kun84}  \\
+ & 187 & $\q_i \qbar_i \to \Q_k \Qbar_k \A^0$ & \cite{Kun84}  \\
\hline
\end{tabular}
\end{center}
\end{table}
 
In the following, $\f_i$ represents a fundamental fermion of flavour
$i$, i.e. $\d$, $\u$, $\s$, $\c$, $\b$, $\t$, $\lrm$,
$\hrm$, $\e^-$, $\nu_{\e}$, $\mu^-$, $\nu_{\mu}$, $\tau^-$,
$\nu_{\tau}$, $\chi^-$ or $\nu_{\chi}$. A corresponding antifermion
is denoted by $\fbar_i$. In several cases, some classes of fermions 
are explicitly excluded, since they do not couple to the $\g$ or
$\gamma$ (no $\ee \to \g \g$, e.g.). When processes have only been
included for quarks, while leptons might also have been possible,
the notation $\q_i$ is used. A lepton is denoted by $\ell$; in a few
cases neutrinos are also lumped under this heading.
In processes where fermion masses are
explicitly included in the matrix elements, an $\F$ is used to denote
an arbitrary fermion and a $\Q$ a quark. Flavours appearing already in
the initial state are denoted by indices $i$ and $j$, whereas new 
flavours in the final state are denoted by $k$ and $l$.
 
Charge-conjugate channels are always assumed included as well (where
separate), and processes involving a $\W^+$ also imply those involving
a $\W^-$. Wherever $\Z^0$ is written, it is understood that $\gamma^*$
and $\gammaZ$ interference should be included as well (with
possibilities to switch off either, if so desired). In some cases
this is not fully implemented, see further below.
Correspondingly, $\Z'^0$ denotes the complete set
$\gamma^*/\Z^0/\Z'^0$ (or some subset of it). Thus the notation 
$\gamma$ is only used for a photon on the mass shell.
 
In the last column of the tables below, references are given to works 
from which formulae have been taken. Sometimes these references are to 
the original works on the subject, sometimes only to the place where 
the formulae are given in the most convenient or accessible form, or
where chance lead us. Apologies to all matrix-element calculators who
are not mentioned. However, remember that this is not a review article
on physics processes, but only a way for readers to know what is
actually found in the program, for better or worse. In several
instances, errata have been obtained from the authors. Often
the formulae given in the literature have been generalized to include
trivial radiative corrections, Breit--Wigner line shapes with
$\hat{s}$-dependent widths (see section \ref{ss:kinemreson}), etc.
 
The following sections contain some useful comments on the processes
included in the program, grouped by physics interest rather than
sequentially by ISUB or \ttt{MSEL} code (see \ref{ss:PYswitchkin} 
for further information on the \ttt{MSEL} code). The different 
ISUB and \ttt{MSEL} codes
that can be used to simulate the different groups are given. ISUB
codes within brackets indicate the kind of processes that indirectly
involve the given physics topic, although only as part of a larger
whole. Some obvious examples, such as the possibility to produce jets 
in just about any process, are not spelled out in detail.
 
The text at times contains information on which special switches or
parameters are of particular interest to a given process. All these
switches are described in detail in section \ref{ss:PYswitchpar}, but 
are alluded to here so as to provide a more complete picture of the
possibilities available for the different subprocesses. However, the
list of possibilities is certainly not exhausted by the text below.
 
\subsection{QCD Processes}

In this section we discuss scatterings exclusively between coloured 
partons --- a process like $\ee \to \gammaZ \to \q\qbar$ is also
traditionally called a QCD event, but is here book-kept as 
$\gammaZ$ production.

\subsubsection{QCD jets} 
\label{sss:QCDjetclass}
  
\ttt{MSEL} = 1, 2 \\
ISUB = \\
\begin{tabular}{rl@{\protect\rule{0mm}{\tablinsep}}}
11 & $\q_i \q_j \to \q_i \q_j$ \\
12 & $\q_i \qbar_i \to \q_k \qbar_k$ \\
13 & $\q_i \qbar_i \to \g \g$ \\
28 & $\q_i \g \to \q_i \g$ \\ 
53 & $\g \g \to \q_k \qbar_k$ \\
68 & $\g \g \to \g \g$ \\
\end{tabular}
 
No higher-order processes are explicitly included,
nor any higher-order loop corrections to the $2 \to 2$ processes.
However, by initial- and final-state QCD radiation, multijet events 
are being generated, starting from the above processes. The shower 
rate of multijet production is clearly uncertain
by some amount, especially for well-separated jets. 

A string-based fragmentation scheme such as the Lund model needs 
cross sections for the different colour flows; these have been 
calculated in \cite{Ben84} and differ from the usual calculations by 
interference terms of the order $1/N_C^2$. By default, the standard 
QCD expressions for the differential cross sections are used. In this 
case, the interference terms are distributed on the various colour 
flows according to the pole structure of the terms. However, the
interference terms can be excluded, by changing \ttt{MSTP(34)}
 
As an example, consider subprocess 28, $\q \g \to \q \g$. The total 
cross section for this process, obtained by summing and squaring the 
Feynman $\hat{s}$-, $\hat{t}$-, and $\hat{u}$-channel graphs, is
\cite{Com77}
\begin{equation} 
2 \left( 1 - \frac{\hat{u}\hat{s}}{\hat{t}^2} \right) - 
\frac{4}{9} \left( \frac{\hat{s}}{\hat{u}} + 
\frac{\hat{u}}{\hat{s}} \right) - 1 ~. 
\end{equation}
(An overall factor $\pi \alphas^2/\hat{s}^2$ is ignored.) 
Using the identity of the Mandelstam variables 
for the massless case, $\hat{s} + \hat{t} + \hat{u} = 0$,  
this can be rewritten as 
\begin{equation}
\frac{\hat{s}^2 + \hat{u}^2}{\hat{t}^2} -
\frac{4}{9} \left( \frac{\hat{s}}{\hat{u}} + 
\frac{\hat{u}}{\hat{s}} \right) ~. 
\end{equation}

On the other hand, the cross sections for the two possible colour 
flows of this subprocess are \cite{Ben84}
\begin{eqnarray} 
   A: & & \frac{4}{9} \left( 2 \frac{\hat{u}^2}{\hat{t}^2} -
\frac{\hat{u}}{\hat{s}} \right) ~;   \nonumber \\
   B: & & \frac{4}{9} \left( 2 \frac{\hat{s}^2}{\hat{t}^2} -
\frac{\hat{s}}{\hat{u}} \right) ~.
\end{eqnarray}
Colour configuration $A$ is one in which the original colour
of the $\q$ annihilates with the anticolour of the $\g$,
the $\g$ colour flows through, and a new colour--anticolour is
created between the final $\q$ and $\g$. In colour configuration
$B$, the gluon anticolour flows through, but the $\q$ and $\g$
colours are interchanged. Note that these two colour
configurations have different kinematics dependence.
For \ttt{MSTP(34)=0}, these are the cross sections actually 
used.  

The sum of the $A$ and $B$ contributions is
\begin{equation} 
\frac{8}{9} \frac{\hat{s}^2 + \hat{u}^2}{\hat{t}^2} -
\frac{4}{9} \left( \frac{\hat{s}}{\hat{u}} + 
\frac{\hat{u}}{\hat{s}} \right) ~. 
\end{equation} 
The difference between this expression and that of \cite{Com77}, 
corresponding to the interference between the two colour-flow 
configurations, is then 
\begin{equation} 
\frac{1}{9} \frac{\hat{s}^2 + \hat{u}^2}{\hat{t}^2} ~,
\end{equation}  
which can be naturally divided between colour flows $A$ and $B$: 
\begin{eqnarray}
   A: & & \frac{1}{9} \frac{\hat{u}^2}{\hat{t}^2} ~; \nonumber \\
   B: & & \frac{1}{9} \frac{\hat{s}^2}{\hat{t}^2} ~.
\end{eqnarray} 
For \ttt{MSTP(34)=1}, the standard QCD matrix element is therefore
used, with the same relative importance of the two colour configurations 
as above. Similar procedures are followed also for the other QCD 
subprocesses. 
  
All the matrix elements in this group are for massless quarks 
(although final-state quarks are of course put on the mass shell). 
As a consequence, cross sections are divergent for $\pT \to 0$,
and some kind of regularization is required. Normally you 
are expected to set the desired $\pTmin$ value in 
\ttt{CKIN(3)}. 
  
The new flavour produced in the annihilation processes (ISUB = 12 and
53) is determined by the flavours allowed for gluon splitting into 
quark--antiquark; see switch \ttt{MDME}. 
    
\subsubsection{Heavy flavours}
\label{sss:heavflavclass} 
  
\ttt{MSEL} = 4, 5, 6, 7, 8 \\
ISUB = \\
\begin{tabular}{rl@{\protect\rule{0mm}{\tablinsep}}}
81 & $\q_i \qbar_i \to \Q_k \Qbar_k$  \\
82 & $\g \g \to \Q_k \Qbar_k$  \\
(83) & $\q_i \f_j \to Q_k f_l$  \\
\end{tabular}
  
The matrix elements in this group differ from the corresponding ones 
in the group above in that they correctly take into account the quark 
masses. As a consequence, the cross sections are finite for 
$\pT \to 0$. It is therefore not necessary to introduce any special 
cuts. 
  
The two first processes that appear here are the dominant lowest-order
graphs in hadron colliders --- a few other graphs will be mentioned
later, such as process 83, which is important for a heavy top.

The choice of flavour to produce is according to a hierarchy of options:
\begin{Enumerate}
\item if \ttt{MSEL=4-8} then the flavour is set by the \ttt{MSEL} value;
\item else if \ttt{MSTP(7)=1-8} then the flavour is set by the
\ttt{MSTP(7)} value;
\item else the flavour is determined by the heaviest flavour allowed for 
gluon splitting into quark--antiquark; see switch \ttt{MDME}.
\end{Enumerate} 
Note that only one heavy flavour is allowed at a time; if more than one 
is turned on in \ttt{MDME}, only the heaviest will be produced (as 
opposed to the case for ISUB = 12 and 53 above, where more than one 
flavour is allowed simultaneously). 
  
The lowest-order processes listed above just represent one source of 
heavy-flavour production. Heavy quarks can also be present in the 
parton distributions at the $Q^2$ scale of the hard interaction, 
leading to processes like $\Q \g \to \Q \g$, 
so-called flavour excitation, or they can be 
created by gluon splittings $\g \to \Q \Qbar$ in initial- or final-state 
shower evolution. In fact, as the c.m. energy is increased, these other 
processes gain in importance relative to the lowest-order production 
graphs above. As as example, only 10\% of the $\b$ production at LHC 
energies come from the lowest-order graphs. The figure is even smaller 
for charm, while it is at or above 50\% for top. At LHC energies, 
the specialized treatment described in this subsection is therefore 
only of interest for top (and potential fourth-generation quarks) --- 
the higher-order corrections can here be approximated by an effective 
$K$ factor, except possibly in some rare corners of phase space. For 
charm and bottom, on the other hand, it is necessary to simulate the 
full event sample (within the desired kinematics cuts), and then only 
keep those events with $\b/\c$ either from lowest-order production, or 
flavour excitation, or gluon splitting. Obviously this may be a 
time-consuming enterprise --- although the probability for a high-$\pT$ 
event to contain (at least) one charm or bottom pair is fairly large, 
most of these heavy flavours are carrying a small fraction of the total 
$\pT$ flow of the jets, and therefore do not survive normal experimental 
cuts. 
  
As an aside, it is not only for the lowest-order graphs that events 
may be generated with a guaranteed heavy-flavour content. One may also 
generate the flavour excitation process by itself, in the massless 
approximation, using ISUB = 28 and setting the \ttt{KFIN} array 
appropriately. No trick exists to force the gluon splittings without 
introducing undesirable biases, however. 
  
The cross section for a heavy quark pair close to threshold can be 
modified according to the formulae of \cite{Fad90}, see \ttt{MSTP(35)}. 
Here threshold effects due to $\Q\Qbar$ bound-state formation are taken 
into account in a smeared-out, average sense. Then the na\"{\i}ve 
cross section is multiplied by the squared wave function at the origin. 
In a colour-singlet channel this gives a net enhancement of the form
\begin{equation}
|\Psi^{(s)}(0)|^2 = \frac{X_{(s)}}{1 - \exp(- X_{(s)})}  ~ ,
~~~ \mrm{where}~ X_{(s)} = \frac{4}{3} 
\frac{\pi \alphas}{\beta}  ~,
\label{pp:threshenh}
\end{equation}
while in a colour octet channel there is a net suppression given by
\begin{equation}
|\Psi^{(8)}(0)|^2 = \frac{X_{(8)}}{\exp(- X_{(8)}) -1}  ~ ,
~~~ \mrm{where}~ X_{(8)} = \frac{1}{6} 
\frac{\pi \alphas}{\beta}  ~.
\label{pp:threshsup}
\end{equation}
The $\alphas$ factor in this expression is related to the energy
scale of bound-state formation; it is selected independently from
the one of the standard production cross section.
The presence of a threshold factor affects the total rate and also 
kinematical distributions. 

Heavy flavours, i.e. top and fourth generation, are assumed to be so
short-lived that they decay before they have time to hadronize. This
means that the light quark in the decay $\Q \to \W^{\pm} \q$ 
inherits the colour of the heavy one.
The new {\Py} description represents a change of philosophy compared
to previous versions, formulated at a time when the top was thought 
to be much lighter than is believed currently. However, optionally
the old description may still be used, where top hadrons are formed
and these subsequently allowed to decay; see \ttt{MSTP(48)} and 
\ttt{MSTP(49)}. For event shapes the difference between the two 
time orderings normally has only marginal effects \cite{Sjo92a}.

It should be noted that cross section calculations are different in 
the two cases. If the top (or the fourth generation fermion) is 
assumed short-lived, then it is treated like a resonance in the sense 
of section \ref{sss:resdecaycross}, i.e. the cross-section is reduced 
so as only to correspond to the channels left open by the user.
This also includes the restrictions on secondary decays, i.e. on the
decays of a $\W^+$ or a $\H^+$ produced in the top decay.
If the top is allowed to form hadrons, no such reduction takes place. 
Branching ratios then have to be folded in by hand to get the correct 
cross sections. The logic behind this difference is that if 
hadronization takes place, one would be allowed e.g. to decay the 
$\T^0$ and $\T^+$ meson according to different branching ratios.
But which $\T$ mesons are to be formed is not known at the top quark
creation, so one could not weight for that. For a $\t$ quark which
decays rapidly this ambiguity does not exist, and so a reduction
factor can be introduced directly coupled to the $\t$ quark production
process.

This rule about cross-section calculations applies to all the 
processes explicitly set up to handle heavy flavour creation.
In addition to the ones above, this means all the ones in Tables
\ref{t:procone}--\ref{t:procfour} where the fermion final state is 
given as capital letters (`$\Q$' and `$\F$') and also flavours produced
in resonance decays ($\Z^0$, $\W^{\pm}$, $\H^0$, etc., including
processes 165 and 166). However, heavy flavours can also be produced
in a process such as 31, $\q_i \g \to \q_k \W^{\pm}$, where $\q_k$
could be a top quark. In this case, the thrust of the description is
clearly on light flavours --- the kinematics of the process is 
formulated in the massless fermion limit --- so any top production
is purely incidental. Since here the choice of scattered flavour is
only done at a later stage, the top branching ratios are not
correctly folded in to the hard scattering cross section. So, for 
applications like these, it is not recommended to restrict the allowed
top decay modes. Often one might like to get rid of the possibility
of producing top together with light flavours. This can be 
done by switching off (i.e. setting \ttt{MDME(I,1)=0}) the 
`channels' $\d \to \W^- \t$, $\s \to \W^- \t$, $\b \to \W^- \t$,
$\g \to \t\tbar$ and $\gamma \to \t\tbar$. Also any heavy flavours 
produced by parton shower evolution would not be correctly weighted
into the cross section. However, currently top production is switched
off in both initial (see \ttt{KFIN} array) and final (see 
\ttt{MSTJ(45)}) state radiation.

\subsubsection{J/$\psi$}
\label{sss:Jpsiclass}

ISUB = \\
\begin{tabular}{rl@{\protect\rule{0mm}{\tablinsep}}}
 86 & $\g \g \to \Jpsi \g$ \\
 87 & $\g \g \to \chi_{0 \c} \g$ \\
 88 & $\g \g \to \chi_{1 \c} \g$ \\
 89 & $\g \g \to \chi_{2 \c} \g$ \\
\end{tabular}
  
One may distinguish three main sources of $\Jpsi$ production.
\begin{Enumerate}
\item Decays of $\B$ mesons and baryons.
\item Parton-shower evolution, wherein a $\c$ and a $\cbar$ quark
produced in adjacent branchings 
(e.g. $\g \to \g \g \to \c \cbar \c \cbar$)
turn out to have so small an invariant mass that the pair collapses 
to a single particle.
\item Direct production, where a $\c$ quark loop gives a coupling 
between a set of gluons and a $\c\cbar$ bound state. Higher-lying 
states, like the $\chi_c$ ones, may subsequently decay to $\Jpsi$.
\end{Enumerate}

In this section are given the main processes for the third source,
intended for applications at hadron colliders at non-vanishing
transverse momenta --- in the limit of $\pT \to 0$ it is necessary
to include a number of $2 \to 1$ processes and to regularize 
divergences in the $2 \to 2$ graphs above. The cross sections depend
on wave function values at the origin, see \ttt{PARP(38)} and
\ttt{PARP(39)}. A review of the physics issues involved may be found 
in \cite{Glo88} (note, however, that the choice of $Q^2$ scale is 
different in {\Py}).

\subsubsection{Minimum bias}
\label{sss:minbiasclass} 
  
\ttt{MSEL} = 1, 2 \\
ISUB = \\
\begin{tabular}{rl@{\protect\rule{0mm}{\tablinsep}}}
91 & elastic scattering  \\
92 & single diffraction ($AB \to XB$) \\
93 & single diffraction ($AB \to AX$) \\
94 & double diffraction  \\
95 & low-$\pT$ production  \\
\end{tabular}
  
These processes are briefly discussed in section 
\ref{ss:nonpertproc}.
Currently they are mainly intended for interactions between hadrons,
although one may also consider $\gamma \p$ interactions in the 
option where the incoming photon is assumed resolved, 
\ttt{MSTP(14)=1} or \ttt{=2}. A possible extension to 
$\gamma \gamma$ interactions is not yet available.

Uncertainties come from a number of sources, e.g. from the 
parametrizations of the various cross sections and slope parameters.

In diffractive scattering, the structure of the selected hadronic 
system may be regulated with \ttt{MSTP(101)}. No high-$\pT$ 
jet production in diffractive events is included so far. 
  
The subprocess 95, low-$\pT$ events, is somewhat unique in 
that no meaningful physical border-line to high-$\pT$ events can be 
defined. Even if the QCD $2 \to 2$ high-$\pT$ processes are formally 
switched off, some of the generated events will be classified as 
belonging to this group, with a $\pT$ spectrum of interactions to 
match the `minimum-bias' event sample. Only with the option 
\ttt{MSTP(82)=0} will subprocess 95 yield strictly low-$\pT$  
events, events which will then probably not be compatible with any 
experimental data. A number of options exist for the detailed 
structure of low-$\pT$ events, see in particular \ttt{MSTP(81)} and 
\ttt{MSTP(82)}. Further details on the model(s) for minimum-bias
events are found in section \ref{ss:multint}.
  
\subsection{Electroweak Gauge Bosons}

This section covers the production and/or exchange of $\gamma$, 
$\Z^0$ and $\W^{\pm}$ gauge bosons, singly and in pairs. The topic 
of longitudinal gauge-boson scattering at high energies is deferred 
to the Higgs section, since the presence or absence of a Higgs here 
makes a big difference.

\subsubsection{Prompt photon production}
\label{sss:promptgammaclass} 
  
\ttt{MSEL} = 10 \\ 
ISUB = \\
\begin{tabular}{rl@{\protect\rule{0mm}{\tablinsep}}}
 14 & $\q_i \qbar_i \to \g \gamma$  \\
 18 & $\f_i \fbar_i \to \gamma \gamma$ \\
 29 & $\q_i \g \to \q_i \gamma$ \\
114 & $\g \g \to \gamma \gamma$  \\
115 & $\g \g \to \g \gamma$ \\
\end{tabular}

In hadron colliders, processes ISUB = 14 and 29 give the main source 
of single-$\gamma$ production, with ISUB = 115 giving an additional 
contribution which, in some kinematics regions, may become important. 
For $\gamma$-pair production, the process ISUB = 18 is often 
overshadowed in importance by ISUB = 114. 
  
Another source of photons is bremsstrahlung off incoming or outgoing 
quarks. This has to be treated on an equal footing with QCD parton 
showering. For time-like parton-shower evolution, i.e. in the 
final-state showering and in the side branches of the initial-state 
showering, photon emission may be switched on or off with 
\ttt{MSTJ(41)}. Photon radiation off the space-like incoming 
quark legs is not yet included, but should be of lesser importance 
for production at reasonably large $\pT$ values. Radiation off an 
incoming electron is included in a leading-log approximation. 
  
{\bf Warning:} the cross sections for the box graphs 114 and 115 become 
very complicated, numerically unstable and slow when the 
full quark mass dependence is included. For quark masses much 
below the $\hat{s}$ scale, the simplified massless expressions are 
therefore used --- a fairly accurate approximation. However, there 
is another set of subtle numerical cancellations between different 
terms in the massive matrix elements in the region of small-angle 
scattering. The associated problems have not been sorted out yet. 
There are therefore two possible solutions. One is to use the 
massless formulae throughout. The program then becomes faster and 
numerically stable, but does not give, for example, the characteristic 
dip (due to destructive interference) at top threshold. This is the 
current default procedure, with five flavours assumed, but this 
number can be changed in \ttt{MSTP(38)}. The other possibility is 
to impose cuts on the scattering angle of the hard process, see 
\ttt{CKIN(27)} and \ttt{CKIN(28)}, since the numerically unstable 
regions are when $|\cos\hat{\theta}|$ is close to unity. It is then 
also necessary to change \ttt{MSTP(38)} to 0. 

\subsubsection{Photoproduction and $\gamma\gamma$ physics}
\label{sss:photoprodclass}

\ttt{MSEL} = 1, 2, 4, 5, 6, 7, 8 \\
ISUB = \\
\begin{tabular}{rl@{\protect\rule{0mm}{\tablinsep}}}
33 & $\q_i \gamma \to \q_i \g$ \\
34 & $\f_i \gamma \to \f_i \gamma$ \\
54 & $\g \gamma \to \q_k \qbar_k$ \\
58 & $\gamma \gamma \to \f_k \fbar_k$ \\
80 & $\q_i \gamma \to \q_k \pi^{\pm}$ \\
84 & $\g \gamma \to \Q_k \Qbar_k$ \\
85 & $\gamma \gamma \to \F_k \Fbar_k$ \\
\end{tabular}

An (almost) real photon has both a point-like component and a 
hadron-like one. This means that several classes of processes 
may be distinguished, see section \ref{sss:photoprod}.
\begin{Enumerate}
\item The processes listed above are possible when the photon 
interacts as a point-like particle, i.e. couples directly to 
quarks and leptons.
\item When the photon acts like a hadron, i.e. is resolved in a 
partonic substructure, then high-$\pT$  parton--parton interactions
are possible, as already described in sections \ref{sss:QCDjetclass} 
and \ref{sss:promptgammaclass}. These interactions may be further
subdivided into VMD and anomalous ones \cite{Sch93,Sch93a}.
\item A hadron-like photon can also produce the equivalent of the 
minimum bias processes of section \ref{sss:minbiasclass}. 
\end{Enumerate}

For $\gamma\p$ events, we believe that the best description can be 
obtained when three separate event classes are combined, one for 
direct, one for VMD and one for anomalous events, see the detailed 
description in \cite{Sch93,Sch93a}.
These correspond to \ttt{MSTP(14)} being 0, 2 and 3, respectively.
The direct and anomalous components are high-$\pT$ only, while
VMD contains both high-$\pT$ and low-$\pT$ events. The option
\ttt{MSTP(14)=1} combines the VMD and anomalous parts of the photon
into one single resolved photon concept, which therefore is less
precise than the full subdivision.

When combining three runs to obtain the totality of $\gamma\p$
interactions, to the best of our knowledge, it is necessary to choose
the $\pT$ cut-offs with some care, so as to represent the expected
total cross section. 
\begin{Itemize}
\item The direct processes only depend on the 
\ttt{CKIN(3)} cut-off of the generation, with preferred value
0.5 GeV \cite{Sch93,Sch93a}. Since this value is so low, one must 
remember to reduce a few other defaults values: 
\ttt{CKIN(1)=2.*CKIN(3)}, \ttt{CKIN(5)=CKIN(6)=0.5*CKIN(3)}. 
For the same reason it is recommended to include a dampening of 
proton parton distributions, \ttt{MSTP(57)=2}. 
\item The VMD processes work as ordinary
hadron--hadron ones, i.e. one obtains both low- and high-$\pT$ 
events by default, with dividing line set by \ttt{PARP(81)}
(or \ttt{PARP(82)}, depending on minijet unitarization scheme).
\item For the anomalous, finally, the minimal $\pT$ of the 
$\gamma \to \q\qbar$ branching is set in \ttt{PARP(15)}. The
default is 0.5 GeV, in agreement with the recommended cutoff for
the same vertex in direct proccesses. In addition, a lower 
\ttt{CKIN(3)} cut-off should be selected for the hard interactions.
This needs some fine-tuning, which in principle should be
done separately for each c.m. energy. A good first approximation
in the HERA energy range (but not beyond 300 GeV) is 
\ttt{CKIN(3)}$=1.50 + 0.0035 \, E_{\mrm{cm}}$.
\end{Itemize} 

The processes in points 1 and 2 can be simulated either with a photon
beam or with an electron beam. For a photon beam it is necessary to
use option \ttt{MSTP(14)} to switch between a point-like and a resolved
photon --- it is not possible to simulate the two sets of processes
in a single run. An electron by default is assumed to contain photons,
but this can be switched off by \ttt{MSTP(11)=0}.  To have quark and 
gluon distributions inside the photon (itself inside the electron),
\ttt{MSTP(12)=1} must be used. For the electron, the two kinds of 
processes may be generated together, unlike for the photon. It is
not possible to have also the low-$\pT$ physics (including multiple
interactions in high-$\pT$ events) for an electron beam. Kindly note
that subprocess 34 contains both the scattering of an electron off
a photon and the scattering of a quark (inside a photon inside an 
electron) off a photon; the former can be switched off with the
help of the \ttt{KFIN} array.

If you are only concerned with standard QCD physics, the option 
\ttt{MSTP(14)=10} gives an automatic mixture of the VMD, direct and
anomalous event classes. The mixture is properly given according to
the relative cross sections. Whenever possible, this option is therefore
preferrable in terms of user-friendliness. However, it can only work
because of a completely new layer of administration, not found anywhere 
else in {\Py}. For instance, a subprocess like $\q\g \to \q\g$ is 
allowed in several of the classes, but appears with different sets of 
parton distributions and different $\pT$ cut-offs in each of these,
so that it is necessary to switch gears between each event in the 
generation. It is therefore not possible to avoid a number of 
restrictions on what you can do in this case:
\begin{Itemize}
\item The \ttt{MSTP(14)=10} option can only be used for 
incoming photon beams, i.e. when \ttt{'gamma'} is the argument in the 
\ttt{PYINIT} call. A convolution with the bremsstrahlung photon 
spectrum in an electron beam may come one day, but not in the immediate 
future.
\item The machinery has only been set up to generate standard
QCD physics, specifically either `minimum-bias' one or high-$\pT$ jets.
For minimum bias, you are not allowed to use the \ttt{CKIN} variables 
at all. This is not a major limitation, since it is in the spirit of 
minimum-bias physics not to impose any contraints on allowed jet 
production. (If you still do, these cuts will be ineffective for the 
VMD processes but take effect for the other ones, giving 
inconsistencies.) 
The minimum-bias physics option is obtained by default; by switching 
from \ttt{MSEL=1} to \ttt{MSEL=2} also the elastic and diffractive 
components of the VMD part are included. High-$\pT$ jet production is 
obtained by setting the \ttt{CKIN(3)} cut-off larger than each of 
the (energy-dependent) cut-off scales for the VMD, direct 
and  anomalous components; typically this means at least 3 GeV. For 
lower input \ttt{CKIN(3)} values the program will automatically switch 
back to minimum-bias physics.
\item Some variables are internally recalculated and reset: 
\ttt{CKIN(1)}, \ttt{CKIN(3)}, \ttt{CKIN(5)}, \ttt{CKIN(6)}, 
\ttt{MSTP(57)}, \ttt{MSTP(85)}, \ttt{PARP(2)}, \ttt{PARP(81)}, 
\ttt{PARP(82)}, \ttt{PARU(115)} and \ttt{MDME(22,J)}. This is because 
they must have values that depend on the component studied.
These variables can therefore not be modified without changing 
\ttt{PYINPR} and recompiling the program, which obviously is a major 
exercise.  
\item Pileup events are not at all allowed.  
\end{Itemize}

Also, a warning about the usage of \tsc{Pdflib} for photons. So long 
as \ttt{MSTP(14)=1}, i.e. the photon is not split up, \tsc{Pdflib} 
is accessed by \ttt{MSTP(56)=2} and \ttt{MSTP(55)} as the parton 
distribution set. However, when the VMD and anomalous pieces are split, 
the VMD part is based on a rescaling of pion distributions by VMD factors
(except for the SaS sets, that already come with a separate VMD piece).
Therefore, to access \tsc{Pdflib} for \ttt{MSTP(14)=10}, it is not 
correct to set \ttt{MSTP(56)=2} and a photon distribution in 
\ttt{MSTP(55)}. Instead, one should put \ttt{MSTP(56)=2}, 
\ttt{MSTP(54)=2} and a pion distribution code in \ttt{MSTP(53)},
while \ttt{MSTP(55)} has no function. The anomalous part is still based on
the SaS parametrization, with \ttt{PARP(15)} as main free parameter.

Currently, hadrons are not defined with any photonic content. None
of the processes are therefore relevant in hadron--hadron collisions.
In $\e\p$ collisions, the electron can emit an almost real photon,
which may interact directly or be resolved. In $\ee$ collisions, 
one may have direct, singly-resolved or doubly-resolved processes.

The $\gamma\gamma$ equivalent to the $\gamma\p$ description involves 
six different event classes, see section \ref{sss:photoprod}.
These classes can be obtained by setting \ttt{MSTP(14)} to 0, 2, 3, 
5, 6 and 7, respectively. If one combines the VMD and anomalous
parts of the parton distributions of the photon, in a more coarse 
description,
it is enough to use the \ttt{MSTP(14)} options 0, 1 and 4. The cut-off
procedures follows from the ones used for the $\gamma\p$ ones above.
Thus the direct$\times$direct and direct$\times$VMD processes require 
the same cut-offs as used for direct $\gamma\p$ events, the 
VMD$\times$VMD ones the same as used for VMD $\gamma\p$ events, 
and the rest (anomalous$\times$anomalous, direct$\times$anomalous
and VMD$\times$anomalous) the same as used for anomalous $\gamma\p$
events.

As with $\gamma\p$ events, the option \ttt{MSTP(14)=10} gives a mixture 
of the six possible $\gamma\gamma$ event classes. The same complications
and restrictions exist here as already listed above. For normal use
the advantages should outweight the disadvantages.

It is hoped to extend the formalism also to mildly virtual photons.
Currently this is not done. The interaction of a highly virtual 
photon with a real photon is included in the deep inelastic scattering 
formalism below, however.

Process 54 generates a mixture of quark flavours; allowed flavours
are set by the gluon \ttt{MDME values}. Process 58 can generate both 
quark and lepton pairs, according to the \ttt{MDME} values of the 
photon. Processes 84 and 85 are variants of these matrix elements, 
with fermion masses included in the matrix elements, but where only 
one flavour can be generated at a time. This flavour is selected as
described for processes 81 and 82 in section \ref{sss:heavflavclass},
with the exception that for process 85 the `heaviest' flavour allowed
for photon splitting takes to place of the heaviest flavour allowed 
for gluon splitting. Since lepton KF codes come after quark ones, 
they are counted as being `heavier', and thus take precedence if 
they have been allowed.

Process 80 is a higher twist one. The theory for such processes 
is rather shaky, so results should not be taken too literally. 
The messy formulae given in \cite{Bag82} have not been programmed
in full, instead the pion form factor has been parametrized as
$Q^2 F_{\pi}(Q^Q) \approx 0.55 / \ln Q^2$, with $Q$ in GeV. 

\subsubsection{Deep inelastic scattering}
\label{sss:DISclass}

\ttt{MSEL} = 1, 2, 35, 36, 37, 38 \\
ISUB = \\
\begin{tabular}{rl@{\protect\rule{0mm}{\tablinsep}}}
 10 & $\f_i \f_j \to \f_k \f_l$ \\
 83 & $\q_i \f_j \to \Q_k \f_l$ \\
\end{tabular}

The `deep inelastic scattering' (DIS) processes, i.e. $t$-channel 
electroweak gauge boson exchange, are traditionally associated
with interactions between a lepton or neutrino and a hadron, but 
processes 10  and 83 can equally well be applied for $\q\q$ scattering 
in hadron colliders (with a cross section much smaller than 
corresponding QCD processes, however). If applied to incoming $\ee$
beams, process 10 corresponds to Bhabha scattering.

For process 10 both $\gamma$, $\Z^0$ and $\W^{\pm}$ exchange 
contribute, including interference between $\gamma$ and $\Z^0$.
The switch \ttt{MSTP(21)} may be used to restrict to only some
of these, e.g. neutral or charged current only.

The option \ttt{MSTP(14)=10} (see previous section) has now been 
extended so that it also works for deep inelastic sacattering of 
an electron off a (real) photon, i.e. process 10. What is obtained 
is a mixture of the photon acting as a vector meson and it acting 
as an anomalous state. This should therefore be the sum of what can 
be obtained with \ttt{MSTP(14)=2} and \ttt{=3}. It is distinct from 
\ttt{MSTP(14)=1} in that different sets are used for the parton 
distributions --- in \ttt{MSTP(14)=1} all the contributions to the
photon distributions are lumped together, while they are split in 
VMD and anomalous parts for \ttt{MSTP(14)=10}. Also the beam remnant 
treatment is different, with a simple Gaussian distribution (at least 
by default) for \ttt{MSTP(14)=1} and the VMD part of \ttt{MSTP(14)=10}, 
but a powerlike distribution $\d k_{\perp}^2 / k_{\perp}^2$ between 
\ttt{PARP(15)} and $Q$ for the anomalous part of \ttt{MSTP(14)=10}. 

To access this option for $\e$ and $\gamma$ as incoming beams, it is 
only necessary to set \ttt{MSTP(14)=10} and keep \ttt{MSEL} at its 
default value. Unlike the corresponding option for $\gamma\p$ and 
$\gamma\gamma$, no cuts are overwritten, i.e. it is still the 
responsability of the user to set these appropriately. 

Cuts especially appropriate for DIS usage include either
\ttt{CKIN(21)-CKIN(22)} or 
\ttt{CKIN(23)-CKIN(24)} for the $x$ range (former or latter depending 
on which side is the incoming real photon), \ttt{CKIN(35)-CKIN(36)} for 
the $Q^2$ range, and \ttt{CKIN(39)-CKIN(40)} for the $W^2$ range. 

In principle, the DIS $x$ variable of an event corresponds to the
$x$ value stored in \ttt{PARI(33)} or \ttt{PARI(34)}, depending
on which side the incoming hadron is on, while the DIS
$Q^2 = -\hat{t} = $\ttt{-PARI(15)}. However, just like initial- and 
final-state radiation can shift jet momenta, they can modify
the momentum of the scattered lepton. Therefore the DIS $x$ and 
$Q^2$ variables are not automatically conserved. An option, on by
default, exists in \ttt{MSTP(23)}, where the event can be `modified 
back' so as to conserve $x$ and $Q^2$, but this option is still rather 
primitive and should not be taken too literally.

Process 83 is the equivalent of process 10 for $\W^{\pm}$ exchange
only, but with the heavy-quark mass included in the matrix element.
In hadron colliders it is mainly of interest for the production of 
very heavy flavours, where the possibility of producing just one 
heavy quark is kinematically favoured over pair production. The
selection of the heavy flavour is already discussed in section
\ref{sss:heavflavclass}.
  
\subsubsection{Single $\W/\Z$ production}
\label{sss:WZclass} 
  
\ttt{MSEL} = 11, 12, 13, 14, 15, (21) \\
ISUB = \\
\begin{tabular}{rl@{\protect\rule{0mm}{\tablinsep}}}
  1 & $\f_i \fbar_i \to \gammaZ$ \\
  2 & $\f_i \fbar_j \to \W^+$ \\
 15 & $\f_i \fbar_i \to \g (\gammaZ)$ \\
 16 & $\f_i \fbar_j \to \g \W^+$ \\
 19 & $\f_i \fbar_i \to \gamma (\gammaZ)$ \\
 20 & $\f_i \fbar_j \to \gamma \W^+$ \\
 30 & $\f_i \g \to \f_i (\gammaZ)$ \\
 31 & $\f_i \g \to \f_k \W^+$ \\
 35 & $\f_i \gamma \to \f_i (\gammaZ)$ \\
 36 & $\f_i \gamma \to \f_k \W^+$ \\
131 & $\g \g \to \Z^0 \Q_k \Qbar_k$ \\
(141) & $\f_i \fbar_i \to \gamma/\Z^0/\Z'^0$ \\
\end{tabular}
  
This group consists of $2 \to 1$ processes, i.e. production of a 
single resonance, and $2 \to 2$ processes, where the resonance 
is recoiling against a jet or a photon. The process 141, which 
also is listed here, is described further elsewhere.

With initial-state showers turned on, the $2 \to 1$ processes also 
generate additional jets; in order to avoid double-counting, the 
corresponding $2 \to 2$ processes should therefore not be turned 
on simultaneously. The basic rule is to use the $2 \to 1$ processes 
for inclusive generation of $\W/\Z$, i.e. where the bulk of the
events studied have $\pT \ll m_{\W/\Z}$, which is where parton showers
may be expected to do a good job. For dedicated studies of $\W/\Z$
production at larger transverse momenta, the parton showers tend 
to underestimate the event rates. It is here better to start from
the $2 \to 2$ matrix elements and add showers to these. However, the 
$2 \to 2$ matrix elements are divergent for $\pT \to 0$, and should
not be used down to the low-$\pT$ region, or one may get unphysical
cross sections. The problem of double-counting applies not only 
to $\W/\Z$ production in hadron colliders, but also to a process
like $\ee \to \Z^0 \gamma$, which clearly is part of the 
initial-state radiation corrections to $\ee \to \Z^0$ obtained for 
\ttt{MSTP(11)=1}. As is the case for $\Z$ production in association 
with jets, the $2 \to 2$ process should therefore only be used for the 
high-$\pT$ region. 
  
The $\Z^0$ of subprocess 1 includes the full interference structure 
$\gammaZ$; via \ttt{MSTP(43)} you can select to produce only 
$\gamma^*$, only $\Z^0$, or the full $\gammaZ$. The same holds true 
for the $\Z'^0$ of subprocess 141; via \ttt{MSTP(44)} any combination 
of $\gamma^*$, $\Z^0$ and $\Z'^0$ can be selected. Thus, subprocess 
141 with \ttt{MSTP(44)=4} is essentially equivalent to subprocess 
1 with \ttt{MSTP(43)=3}; however, process 141 also includes the 
possibility of a decay into Higgses. Also processes 15, 19, 30 and 35
contain the full mixture of $\gammaZ$, with \ttt{MSTP(43)} available
to change this. Only the $\Z^0$ that appears in process 131 does not 
contain the $\gamma^*$ contribution. 

Note that process 1, with only 
$\q\qbar \to \gamma^* \to \ell^+ \ell^-$ allowed,
and studied in the region well below the $\Z^0$ mass, is what is 
conventionally called Drell--Yan. This latter process therefore does 
not appear under a separate heading, but can be obtained by a
suitable setting of switches and parameters.    
  
A process like $\f_i \fbar_j \to \gamma \W^+$ is only included in
the limit that the $\gamma$ is emitted in the `initial state',
while the possibility of a final-state radiation off the $\W^+$
decay products is not explicitly included (but can be obtained
implicitly by the parton-shower machinery) and various interference
terms are not at all present. Some caution must therefore be
exercised; see also section \ref{sss:WZpairclass} for related 
comments.

For the $2 \to 1$ processes, the Breit--Wigner includes an 
$\hat{s}$-dependent width, which should provide an improved 
description of line shapes. In fact, from a line-shape point of view, 
process 1 should provide a more accurate simulation of $\ee$
annihilation events than the dedicated $\ee$ generation scheme of 
{\Je} (see section \ref{ss:eematrix}). However, the $\pT$ 
distribution of
radiated initial-state photons is probably still better modelled in
the {\Je} routines. Another difference is that {\Je} only allows
the generation of $\gammaZ \to \q\qbar$, while process 1 additionally
contains $\gammaZ \to \ell^+ \ell^-$ and $\gammaZ \to \nu \br{\nu}$.
The parton-shower and fragmentation descriptions are the same,
but the {\Py} implementation has not been interfaced with the
first- and second-order matrix-element options available in {\Je}. 
  
Almost all processes in this group have been included with the 
correct angular distribution in the subsequent $\W/\Z \to \f \fbar$
decays. The exception is process 36, where currently
the $\W$ decays isotropically.  

The process $\ee \to \e^+ \e^- \Z^0$ can be simulated in two 
different ways. One is to make use of the $\e$ `sea' distribution 
inside $\e$, i.e. have splittings $\e \to \gamma \to \e$. 
This can be obtained, together with ordinary $\Z^0$ production, by 
using subprocess 1, with \ttt{MSTP(11)=1} and \ttt{MSTP(12)=1}. Then 
the contribution of the type above is 5.0 pb for a 500 GeV $\ee$ 
collider, compared with the correct 6.2 pb \cite{Hag91}. Alternatively 
one may use process 35, with \ttt{MSTP(11)=1} and \ttt{MSTP(12)=0}. 
To catch the singularity in the forward direction, regularized by 
the electron mass, it is necessary to set 
\ttt{CKIN(3)=CKIN(5)=0.01} --- using lower values will only slow 
down execution, not significantly increasing the cross section. One 
then obtains 5.1 pb, i.e. again 20\% below the correct value, but now 
also generates a $\pT$ distribution for the $\Z^0$; this is therefore 
to be preferred. 
  
Process 36, $\f \gamma \to \f' \W^{\pm}$ may have corresponding 
problems; except that in $\ee$ the forward scattering amplitude for 
$\e \gamma \to \nu \W$ is killed (radiation zero), which means 
that the differential cross section is vanishing for $\pT \to 0$.
It is therefore feasible to use the default \ttt{CKIN(3)} and 
\ttt{CKIN(5)} values in $\ee$, and one also comes closer to the 
correct cross section. 
  
One single true $2 \to 3$ process is included in this class as well; 
namely $\g \g \to \Z^0 \Q \Qbar$, with full massive matrix elements. 
The more complicated phase space and the lengthy matrix-element 
evaluations make this process extremely slow. With the quark flavour
picked to be $\b$, it may form an important background to intermediate 
mass Higgs searches in the multilepton channel. The quark flavour is 
stored in \ttt{KFPR(131,2)}; the default is $5 = \b$. The kinematics 
is set up in terms of a $\Z^0$ recoiling against the $\Q \Qbar$ 
system, and all ordinary kinematics cut for a $2 \to 2$ process can 
be used on this level, including \ttt{CKIN(43)} and \ttt{CKIN(44)} to 
restrict the range of the $\Q \Qbar$ invariant mass. In addition, for 
this process alone, \ttt{CKIN(51) - CKIN(54)} can be used to set the 
$\pT$ range of the two quarks; as is to be expected, that of the 
$\Z^0$ is set by \ttt{CKIN(3) - CKIN(4)}. Since the optimization 
procedure is not set up to probe the full multidimensional phase space 
allowed in this process, maximum violations may be quite large. It may 
then be useful to make a preliminary run to find how big the violations 
are in total, and then use the \ttt{MSTP(121)=1} option in the full 
run. 

\subsubsection{$\W/\Z$ pair production}
\label{sss:WZpairclass} 
  
\ttt{MSEL} = 15 \\
ISUB = \\
\begin{tabular}{rl@{\protect\rule{0mm}{\tablinsep}}}
 22 & $\f_i \fbar_i \to (\gammaZ) (\gammaZ)$ \\
 23 & $\f_i \fbar_j \to \Z^0 \W^+$ \\
 25 & $\f_i \fbar_i \to \W^+ \W^-$ \\
 69 & $\gamma \gamma \to \W^+ \W^-$ \\
 70 & $\gamma \W^+ \to \Z^0 \W^+$ \\
\end{tabular}
  
In this section we mainly consider the production of $\W/\Z$ pairs
by fermion--antifermion annihilation, but also include two processes
which involve $\gamma/\W$ beams. Scatterings between gauge-boson 
pairs, i.e. processes like $\W^+ \W^- \to \Z^0 \Z^0$, depend so 
crucially on the assumed Higgs scenario that they are considered 
separately in section \ref{sss:heavySMHclass}.

The cross sections used for the above processes are those derived
in the narrow-width limit, but have been extended to include 
Breit--Wigner shapes with mass-dependent widths. However, one
should realize that other graphs, not included here, can contribute
in regions away from the $\W/\Z$ mass. This problem is especially
important if several flavours coincide in the four-fermion final 
state. Consider, as an example, 
$\ee \to \mu^+ \mu^- \nu_{\mu} \br{\nu}_{\mu}$.
Not only would such a final state receive contributions from 
intermediate $\Z^0\Z^0$ and $\W^+\W^-$ states, but also 
from processes $\ee \to \Z^0 \to \mu^+ \mu^-$, followed
either by $\mu^+ \to \mu^+ \Z^0 \to \mu^+ \nu_{\mu} \br{\nu}_{\mu}$, 
or by 
$\mu^+ \to \br{\nu}_{\mu} \W^+ \to \br{\nu}_{\mu} \mu^+ \nu_{\mu}$.
In addition, all possible interferences should be considered.
Since this is not done, the processes have to be used with some 
sound judgement. Very often, one may wish to constrain a
lepton pair mass to be close to $m_{\Z}$, in which case a number
of the possible `other' processes are negligible.

Of the above processes, the first contains the full
$\f_i \fbar_i \to (\gammaZ)(\gammaZ)$ structure, obtained by a 
straightforward generalization of the formulae in ref. \cite{Gun86} 
(done by the present author). Of course, the possibility of
there being significant contributions from graphs that are not 
included is increased, in particular 
if one $\gamma^*$ is very light and therefore could be a 
bremsstrahlung-type photon. It is possible to use \ttt{MSTP(43)} to 
recover the pure $\Z^0$ case, i.e. $\f_i \fbar_i \to \Z^0 \Z^0$
exclusively. In processes 23 and 70, only the pure $\Z^0$ contribution
is included.

Full angular correlations are included for the first three processes,
i.e. the full $2 \to 2 \to 4$ matrix elements are included in the
resonance decays, including the appropriate $\gammaZ$ interference
in process 22. In the latter two processes no spin 
information is currently preserved, i.e. the $\W/\Z$ bosons are 
allowed to decay isotropically.

We remind you that the mass ranges of the two resonances may be set 
with the \ttt{CKIN(41) - CKIN(44)} parameters; this is particularly 
convenient, for instance, to pick one resonance almost on the mass 
shell and the other not. 

\subsection{Higgs Production}

A fair fraction of all the processes in {\Py} deal with Higgs 
production in one form or another. This multiplication is caused by 
the need to consider production by several different processes, 
depending on Higgs mass and machine type. Further, the program 
contains a full two-Higgs-multiplet scenario, as predicted for example
in the Minimal Supersymmetric extension of the Standard Model
(MSSM). Therefore the continued discussion is, somewhat arbitrarily,
subdivided into a few different scenarios.

\subsubsection{Light Standard Model Higgs}
\label{sss:lightSMHclass}
  
\ttt{MSEL} = 16, 17, 18 \\
ISUB = \\
\begin{tabular}{rl@{\protect\rule{0mm}{\tablinsep}}}
  3 & $\f_i \fbar_i \to \H^0$ \\
 24 & $\f_i \fbar_i \to \Z^0 \H^0$ \\
 26 & $\f_i \fbar_j \to \W^+ \H^0$ \\
102 & $\g \g \to \H^0$ \\
103 & $\gamma \gamma \to \H^0$ \\
110 & $\f_i \fbar_i \to \gamma \H^0$ \\
111 & $\f_i \fbar_i \to \g \H^0$ \\
112 & $\f_i \g \to \f_i \H^0$ \\
113 & $\g \g \to \g \H^0$ \\
121 & $\g \g \to \Q_k \Qbar_k \H^0$ \\
122 & $\q_i \qbar_i \to \Q_k \Qbar_k \H^0$ \\
123 & $\f_i \f_j \to \f_i \f_j \H^0$ ($\Z^0 \Z^0$ fusion) \\
124 & $\f_i \f_j \to \f_k \f_l \H^0$ ($\W^+ \W^-$ fusion) \\
\end{tabular}
  
In this section we discuss the production of a reasonably light 
Standard Model Higgs, below 700 GeV, say, so that the narrow
width approximation can be used with some confidence. Below 400 GeV
there would certainly be no trouble, while above that the narrow
width approximation is gradually starting to break down.

In a hadron collider, the main production processes are 102, 123 
and 124, i.e. $\g\g$, $\Z^0 \Z^0$ and $\W^+ \W^-$ fusion. In the
latter two processes, it is also necessary to take into account
the emission of the space-like $\W/\Z$ bosons off quarks, which
in total gives the $2 \to 3$ processes above. 

Further processes of lower cross sections may be of interest
because of easier signals. For instance, processes 24 and 26 give 
associated production of a $\Z$ or a $\W$ together with the $\H^0$.
There is also the processes 3, 121 and 122, which involve production
of heavy flavours. 

Process 3 contains contributions from all flavours, but is 
completely dominated by the subprocess $\t \tbar \to \H^0$,
i.e. by the contribution from the top sea distributions.
This process is by now known to overestimate the cross section 
for Higgs production as compared with a more careful calculation 
based on the subprocess $\g \g \to \t \tbar \H^0$ (121). The
difference between the two is that in process 3 the
$\t$ and $\tbar$ are added by the initial-state shower, while in 
121 the full matrix element is used. The price to be paid is that
the complicated multibody phase space in process 121 makes the
program run slower than with most other processes. One should 
therefore think twice before using it. As usual, it would be 
double-counting to include both 3 and 121. Process 122 is 
similar in structure to 121, but is less important. In both
process 121 and 122 the produced quark is assumed to be a $\t$;
this can be changed in \ttt{KFPR(121,2)} and \ttt{KFPR(122,2)} 
before initialization, however.
  
A subprocess like 113, with a Higgs recoiling against a gluon jet, is 
also effectively generated by initial-state corrections to subprocess 
102; thus, in order to avoid double-counting, just as for the case 
of $\Z^0/\W^+$ production, section \ref{sss:WZclass}, these subprocesses 
should not be switched on simultaneously. Process 102 should
be used for inclusive production of Higgs, and 111--113 for the study 
of the Higgs subsample with high transverse momentum. 
  
In $\ee$ annihilation, associated production of an $\H^0$ with a 
$\Z^0$, process 24, is usually the dominant one close to threshold,
while the $\Z^0 \Z^0$ and $\W^+ \W^-$ fusion processes 123 and 124
win out at high energies. Process 103, $\gamma\gamma$ fusion, may
also be of interest, in particular when the possibilities of 
beamstrahlung photons and backscattered photons are included. 
Process 110, which gives an $\H^0$ in association with a $\gamma$,
is a loop process and is therefore suppressed in rate. Only for a 
rather massive $\H^0$ (mass above 60 GeV at LEP 1) can it start to
compete with the associated production of a $\Z^0$, since phase space
suppression is less severe for the former than for the latter.
   
The branching ratios of the Higgs are very strongly dependent on 
the mass. In principle, the program is set up to calculate these 
correctly, as a function of the actual Higgs mass, i.e. not just
at the nominal mass. However, higher-order corrections may at 
times be important and not fully unambiguous; see for instance 
\ttt{MSTP(37)}. 

Since the Higgs is a spin-0 particle it decays isotropically. In decay
processes such as $\H^0 \to \W^+ \W^- \to 4$ fermions angular 
correlations are included. Also in processes 24 and 26, $\Z^0$ 
and $\W^{\pm}$ decay angular distributions are correctly taken into 
account.

\subsubsection{Heavy Standard Model Higgs}
\label{sss:heavySMHclass}

ISUB = \\
\begin{tabular}{rl@{\protect\rule{0mm}{\tablinsep}}}
  5 & $\Z^0 \Z^0 \to \H^0$ \\
  8 & $\W^+ \W^- \to \H^0$ \\
 71 & $\Z^0 \Z^0 \to \Z^0 \Z^0$ (longitudinal) \\
 72 & $\Z^0 \Z^0 \to \W^+ \W^-$ (longitudinal) \\
 73 & $\Z^0 \W^+ \to \Z^0 \W^+$ (longitudinal) \\
 76 & $\W^+ \W^- \to \Z^0 \Z^0$ (longitudinal) \\
 77 & $\W^+ \W^{\pm} \to \W^+ \W^{\pm}$ (longitudinal) \\
\end{tabular}

Processes 5 and 8 are the simple $2 \to 1$ versions of what is now
available in 123 and 124 with the full $2 \to 3$ kinematics.
For low Higgs masses processes 5 and 8 overestimate the correct
cross sections and should not be used, whereas good agreement between
the $2 \to 1$ and $2 \to 3$ descriptions is observed when heavy 
Higgs production is studied. 

The subprocesses 5 and 8, $V V \to \H^0$, which contribute to the 
processes $V V \to V' V'$, show a bad high-energy behaviour. Here 
$V$ denotes a longitudinal intermediate gauge boson, $\Z^0$ or 
$\W^{\pm}$. This  can be cured only by the inclusion of all 
$V V \to V' V'$ graphs, as is done in subprocesses 71, 72, 73, 76 
and 77. In particular, subprocesses 5 and 8 give rise to a fictitious 
high-mass tail of the Higgs. If this tail is thrown away, however, 
the agreement between the $s$-channel graphs only (subprocesses 
5 and 8) and the full set of graphs (subprocesses 71 etc.) is very 
good: for a Higgs of nominal mass 300 (800) GeV, a cut at 600 (1200) 
GeV retains 95\% (84\%) of the total cross section, and differs from 
the exact calculation, cut at the same values, by only 2\% (11\%) 
(numbers for SSC energies). With this prescription there is 
therefore no need to use subprocesses 71 etc. rather than 
subprocesses 5 and 8. 

For subprocess 77, there is an option, see \ttt{MSTP(45)}, to select
the charge combination of the scattering $\W$'s: like-sign, 
opposite-sign (relevant for Higgs), or both.

Process 77 contains a divergence for $\pT \to 0$ due to 
$\gamma$-exchange contributions. This leads to an infinite total 
cross section, which is entirely fictitious, since the simple 
parton-distribution function approach to the longitudinal $\W$ flux 
is not appropriate in this limit. For this process, it is therefore 
necessary to make use of a cut, e.g. $\pT > m_{\W}$.
  
For subprocesses 71, 72, 76 and 77, an option is included (see 
\ttt{MSTP(46)}) whereby the user can select only the $s$-channel 
Higgs graph; this will then be essentially equivalent to running 
subprocess 5 or 8 with the proper decay channels (i.e. $\Z^0\Z^0$ or 
$\W^+\W^-$) set via \ttt{MDME}. The difference is that the 
Breit--Wigners in subprocesses 5 and 8 contain a mass-dependent 
width, whereas the width in subprocesses 71--77 is calculated at 
the nominal Higgs mass; also, higher-order corrections to the widths 
are treated more accurately in subprocesses 5 and 8. Further, 
processes 71--77 assume the incoming $\W/\Z$ to be on the mass shell, 
with associated kinematics factors, while processes 5 and 8 have 
$\W/\Z$ correctly space-like. All this leads to differences in the 
cross sections by up to a factor of 1.5. 
  
In the absence of a Higgs, the sector of longitudinal $\Z$ and $\W$ 
scattering will become strongly interacting at energies above 1 TeV. 
The models proposed by Dobado, Herrero and Terron \cite{Dob91} to 
describe this kind of physics have been included as alternative matrix 
elements for subprocesses 71, 72, 73, 76 and 77, selectable by 
\ttt{MSTP(46)}. From the point of view of the general classification 
scheme for subprocesses, this kind of models should appropriately be 
included as separate subprocesses with numbers above 100, but the 
current solution allows a more efficient reuse of existing code. 
By a proper choice of parameters, it is also here possible to 
simulate the production of a techni-$\rho$. 

Currently, the scattering of transverse gauge bosons has not been 
included, neither that of mixed transverse--longitudinal scatterings.
These are expected to be less important at high energies, and do not
contain an $\H^0$ resonance peak, but need not be
entirely negligible in magnitude. As a rule of thumb, processes 
71--77 should not be used for $VV$ invariant masses below 500 GeV.

The decay products of the longitudinal gauge bosons are correctly 
distributed in angle.

\subsubsection{Extended neutral Higgs sector}
\label{sss:extneutHclass} 

\ttt{MSEL} = 19 \\  
ISUB = \\
\begin{tabular}{rrrl@{\protect\rule{0mm}{\tablinsep}}}
$\H^0$ & $\H'^0$ & $\A^0$ &  \\
  3 & 151 & 156 & $\f_i \fbar_i \to X$ \\
102 & 152 & 157 & $\g \g \to X$ \\
103 & 153 & 158 & $\gamma \gamma \to X$ \\
 24 & 171 & 176 & $\f_i \fbar_i \to \Z^0 X$ \\
 26 & 172 & 177 & $\f_i \fbar_j \to \W^+ X$ \\
123 & 173 & 178 & $\f_i \f_j \to \f_i \f_j X$ ($\Z \Z$ fusion) \\
124 & 174 & 179 & $\f_i \f_j \to \f_k \f_l X$ ($\W^+\W^-$ fusion) \\
121 & 181 & 186 & $\g \g \to \Q_k \Qbar_k X$ \\
122 & 182 & 187 & $\q_i \qbar_i \to \Q_k \Qbar_k X$ \\
\end{tabular} \\
ISUB = \\
\begin{tabular}{rl@{\protect\rule{0mm}{\tablinsep}}}
(141) & $\f_i \fbar_i \to \gamma/\Z^0/\Z'^0$ \\
\end{tabular}
  
In {\Py}, the particle content of a two-Higgs-doublet scenario is 
included: two neutral scalar particles, 25 and 35, one pseudoscalar 
one, 36, and a charged doublet, $\pm 37$. (Of course, these particles 
may also be associated with corresponding Higgs states in larger 
multiplets.) By convention, we choose to call the lighter scalar 
Higgs $\H^0$ and the heavier $\H'^0$ --- this differs from the 
convention in the MSSM, where the lighter is called $\hrm^0$ and the 
heavier 
$\H^0$, but allows us to call the Higgs of the one-Higgs scenario 
$\H^0$. The pseudoscalar is called $\A^0$ and the charged $\H^{\pm}$.
Charged-Higgs production is covered in section \ref{sss:chHclass}.
   
A number of $\H^0$ processes have been duplicated for $\H'^0$ and 
$\A^0$. The correspondence between ISUB numbers is shown in the table
above: the first column of ISUB numbers corresponds to
$X = \H^0$, the second to $X = \H'^0$, and the third to $X = \A^0$. 
Note that several of these processes are not expected to take 
place at all, owing to vanishing Born term couplings. We have still 
included them for flexibility in simulating arbitrary couplings at 
the Born or loop level. 
  
A few Standard Model Higgs processes have no correspondence in the 
scheme above. These include
\begin{Itemize}
\item 5 and 8, which anyway have been superseded by 123 and 124; 
\item 71, 72, 73, 76 and 77, which deal with what happens if there 
is no light Higgs, and so is a scenario complementary to the one 
above, where several light Higgses are assumed;
\item 110, which is mainly of interest in Standard Model Higgs 
searches; and  
\item 111, 112 and 113, which describe the high-$\pT$ tail of the 
Higgs production, and are less interesting for most Higgs studies.
\end{Itemize} 
  
In processes 121, 122, 181, 182, 186 and 187 the recoiling heavy 
flavour is assumed to be top, which is the only one of interest in 
the Standard Model, and the one where the parton-distribution-function 
approach invoked in processes 3, 151 and 156 is least reliable. 
However, it is possible to change the quark flavour in 121 etc.; 
for each process ISUB this flavour is given by \ttt{KFPR(ISUB,2)}. 
This may become relevant if couplings to $\b\bbar$ states are
enhanced, e.g. if $\tan\beta \gg 1$ in the MSSM.
  
By default, the $\H^0$ has the couplings of the Standard Model 
Higgs, while the $\H'^0$ and $\A^0$ have couplings set in 
\ttt{PARU(171) - PARU(178)} and \ttt{PARU(181) - PARU(190)}, 
respectively. The default values for the $\H'^0$ and $\A^0$ have no 
deep physics motivation, but are set just so that the program will 
not crash due to the absence of any couplings whatsoever. You 
should therefore set the above couplings to your desired values if 
you want to simulate either $\H'^0$ or $\A^0$. Also the couplings 
of the $\H^0$ particle can be modified, in 
\ttt{PARU(161) - PARU(165)}, provided that \ttt{MSTP(4)} is set to 
1. 
  
For \ttt{MSTP(4)=2}, the mass of the $\H^0$ (in \ttt{PMAS(25,1)}) 
and the $\tan\beta$ value (in \ttt{PARU(141)}) are used to derive 
the masses of the other Higgses, as well as all Higgs couplings. 
\ttt{PMAS(35,1) - PMAS(37,1)} and \ttt{PARU(161) - PARU(195)} are 
overwritten accordingly. The relations used are the ones of the 
Born-level MSSM \cite{Gun90}. Today, loop corrections to those
expressions have been calculated, and are known to have 
non-negligible effects on the resulting phenomenology. Eventually
the modified relations will be included as an additional option, but
this has not yet been done.
 
Note that not all combinations of $m_{\H}$ and $\tan\beta$ are 
allowed; the requirement of a finite $\A^0$ mass imposes the 
constraint 
\begin{equation}
m_{\H} < m_{\Z} \, \frac{\tan^2\beta - 1}{\tan^2\beta + 1}, 
\end{equation}
or, equivalently, 
\begin{equation}
\tan^2\beta > \frac{m_{\Z} + m_{\H}}{m_{\Z} - m_{\H}}. 
\end{equation}
If this condition is not fulfilled, the program will crash.
  
Process 141 can also be used to simulate $\Z^0 \to \H^0 \A^0$ and 
$\Z^0 \to \H'^0 \A^0$ for associated neutral Higgs production. 
The fact that we here make use of the $\Z'^0$ can easily be 
discounted, either by letting the relevant couplings vanish,
or by the option \ttt{MSTP(44)=4}.

Finally, heavier Higgses may decay into lighter ones, if 
kinematically allowed, in processes like $\A^0 \to \Z^0 \H^0$ or 
$\H^+ \to \W^+ \H^0$. Such modes are included as part of the
general mixture of decay channels, but they can be enhanced if 
the uninteresting channels are switched off.

\subsubsection{Charged Higgs sector}
\label{sss:chHclass}

\ttt{MSEL} = 23 \\
ISUB = \\
\begin{tabular}{rl@{\protect\rule{0mm}{\tablinsep}}}
143 & $\f_i \fbar_j \to \H^+$ \\
161 & $\f_i \g \to \f_k \H^+$ \\
(141) & $\f_i \fbar_i \to \gamma/\Z^0/\Z'^0$ \\
\end{tabular}

A charged Higgs doublet, $\H^{\pm}$, is included in the program. 
This doublet may be the one predicted in the MSSM scenario,
see section \ref{sss:extneutHclass}, or in any other scenario.
The $\tan\beta$ parameter, which is relevant also 
for charged Higgs couplings, is set via \ttt{PARU(141)}. 
  
The basic subprocess for charged Higgs production in hadron
colliders is ISUB = 143. However, this process is dominated 
by $\t \bbar \to \H^+$, and so depends on the choice of $\t$ 
parton distribution. A better representation is provided by 
subprocess 161, $\f \g \to \f' \H^+$; i.e. actually 
$\bbar \g \to \tbar \H^+$. It is therefore recommended to use 
161 and not 143; to use both would be double-counting. 
  
In subprocess 141, the decay $\gamma^*/\Z^0/\Z'^0 \to \H^+ \H^-$
allows the production of a pair of charged Higgs particles.
This process is especially important in $\ee$ colliders.
The coupling of the $\gamma^*$ to $\H^+ \H^-$ is determined by
the charge alone, while the $\Z^0$ coupling is regulated 
by \ttt{PARU(142)}, and that of the $\Z'^0$ by \ttt{PARU(143)}. 
The $\Z'^0$ piece can be switched off, e.g. by \ttt{MSTP(44)=4}.
An ordinary $\Z^0$, i.e. particle code 23, cannot be made to 
decay into $\H^+ \H^-$, however.

A major potential source of charged Higgs production is top decay.
When the top is treated as a resonance (the default option), it is
possible to switch on the decay channel $\t \to \b \H^+$. Top will 
then decay to $\H^+$ a fraction of the time, whichever way it is 
produced. The branching ratio is automatically calculated, based
on the $\tan\beta$ value and masses.
It is possible to only have the $\H^+$ decay mode switched on,  
in which case the cross section is reduced accordingly. If one 
instead assumes that top hadrons are formed, branching ratios
are not automatically calculated. However, you can set, for 
the generic top hadron 86, the branching ratios for the two main
channels $\t \to \b \H^+$ and $\t \to \b \W^+$. In this option
the cross section for top production will not be reduced if only 
the $\t \to \b \H^+$ decay is switched on, cf. section 
\ref{sss:resdecaycross}.

\subsection{Non-Standard Physics}

The number of possible non-Standard Model scenarios is essentially
infinite, but many of the studied scenarios still share a lot of
aspects. For instance, new $\W'$ and $\Z'$ gauge bosons can
arise in a number of different ways. Therefore it still makes sense
to try to cover a few basic classes of particles, with enough
freedom in couplings that many kinds of detailed scenarios can be
accommodated by suitable parameter choices. We have already seen one
example of this, in the extended Higgs sector above. In this section
a few other kinds of non-standard generic physics is discussed.
Clearly many others could have been included, but there is probably
only one glaring omission: currently no supersymmetric particle
production has been included. One main reason for this is the large
number of particles, processes, possible mass hierarchies and decay 
chains.  

\subsubsection{Fourth-generation fermions}
  
\ttt{MSEL} = 7, 8, 37, 38  \\
ISUB = \\
\begin{tabular}{rl@{\protect\rule{0mm}{\tablinsep}}}
 1 & $\f_i \fbar_i \to \gammaZ$ \\
 2 & $\f_i \fbar_j \to \W^+$ \\
81 & $\q_i \qbar_i \to \Q_k \Qbar_k$  \\
82 & $\g \g \to \Q_k \Qbar_k$  \\
83 & $\q_i \f_j \to \Q_k \f_l$ \\
84 & $\g \gamma \to \Q_k \Qbar_k$ \\
85 & $\gamma \gamma \to \F_k \Fbar_k$ \\
141 & $\f_i \fbar_i \to \gamma/\Z^0/\Z'^0$ \\
142 & $\f_i \fbar_j \to \W'^+$ \\
\end{tabular}

The prospects of a fourth generation currently seem rather dim, but 
the appropriate flavour content is still found in the program. In 
fact, the fourth generation is included on an equal basis with the
first three, provided \ttt{MSTP(1)=4}. Also processes other than 
the ones above can therefore be used, e.g. all other processes with
gauge bosons, including non-standard ones such as the $\Z'^0$. We
therefore do not repeat the descriptions found elsewhere, e.g. how
to set only the desired flavour in processes 81--85. Note that it
may be convenient to set \ttt{CKIN(1)} and other cuts such that the
mass of produced gauge bosons is enough for the wanted particle
production --- in principle the program will cope even without that,
but possibly at the expense of very slow execution.
  
\subsubsection{New gauge bosons} 
  
\ttt{MSEL} = 21, 22, 24 \\
ISUB = \\
\begin{tabular}{rl@{\protect\rule{0mm}{\tablinsep}}}
141 & $\f_i \fbar_i \to \gamma/\Z^0/\Z'^0$ \\
142 & $\f_i \fbar_j \to \W'^+$ \\
144 & $\f_i \fbar_j \to \R$ \\
\end{tabular}
  
The $\Z'^0$ of subprocess 141 contains the full $\gamma^*/\Z^0/\Z'^0$ 
interference structure for couplings to fermion pairs. With 
\ttt{MSTP(44)} it is possible to pick only a subset, e.g. only the
pure $\Z'^0$ piece. The couplings of the $\Z'^0$ to quarks and leptons 
can be set via \ttt{PARU(121) - PARU(128)}. The eight numbers 
correspond to the vector and axial couplings of down-type quarks,
up-type quarks, leptons and neutrinos, respectively. The default 
corresponds to the same couplings as that of the Standard Model
$\Z^0$, with axial couplings $a_{\f} = \pm 1$ and vector couplings
$v_{\f} = a_{\f} - 4 e_{\f} \ssintw$. This implies a resonance width
that increases linearly with the mass. By a suitable choice of the
parameters, it is possible to simulate just about any imaginable
$\Z'^0$ scenario, with full interference effects in cross sections
and decay angular distributions.

The coupling to the decay channel $\Z'^0 \to \W^+ \W^-$ is regulated
by \ttt{PARU(129) - PARU(130)}. The former gives the strength of the
coupling, which determines the rate. The default, \ttt{PARU(129)=1.},
corresponds to the `extended gauge model' of \cite{Alt89}, wherein
the $\Z^0 \to \W^+ \W^-$ coupling is used, scaled down by a factor
$m_{\W}^2/m_{\Z'}^2$, to give a $\Z'^0$ partial width into this channel
that again increases linearly. If this factor is cancelled, by having
\ttt{PARU(129)} proportional to $m_{\Z'}^2/m_{\W}^2$, one obtains a
partial width that goes like the fifth power of the $\Z'^0$ mass, the
`reference model' of \cite{Alt89}. In the decay angular distribution
one could imagine a much richer structure than is given by the one
parameter \ttt{PARU(130)}.

Other decay modes include $\Z'^0 \to \Z^0 \H^0$, predicted in
left--right symmetric models (see \ttt{PARU(145)} and ref. 
\cite{Coc91}), and a number of other Higgs decay channels, see 
sections \ref{sss:extneutHclass} and \ref{sss:chHclass}.
  
The $\W'^{\pm}$ of subprocess 142 so far does not contain 
interference with the Standard Model $\W^{\pm}$ --- in practice this 
should not be a major limitation. The couplings of the $\W'$ to 
quarks and leptons are set via \ttt{PARU(131) - PARU(134)}.
Again one may set vector and axial couplings freely, separately
for the $\q \qbar'$ and the $\ell \nu_{\ell}$ decay channels.
The defaults correspond to the $V-A$ structure of the Standard Model
$\W$, but can be changed to simulate a wide selection of models.
One possible limitation is that the same Cabibbo--Kobayashi--Maskawa
quark mixing matrix is assumed as for the standard $\W$.

The coupling $\W' \to \Z^0 \W$ can be set via 
\ttt{PARU(135) - PARU(136)}. Further comments on this channel as for 
$\Z'$; in particular, default couplings again agree with 
the `extended gauge model' of \cite{Alt89}. A $\W' \to \W \H^0$
channel is also included, in analogy with the $\Z'^0 \to \Z^0 \H^0$
one, see \ttt{PARU(146)}.
  
The $\R$ boson (particle code 40) of subprocess 144 represents one 
possible scenario for a horizontal gauge boson, i.e. a gauge boson 
that couples between the generations, inducing processes like 
$\s \dbar \to \R^0 \to \mu^- \e^+$. Experimental limits on 
flavour-changing neutral currents forces such a boson to be fairly 
heavy. The model implemented is the one described in \cite{Ben85a}. 
  
\subsubsection{Leptoquarks}
\label{sss:LQclass} 
  
\ttt{MSEL} = 25 \\
ISUB = \\
\begin{tabular}{rl@{\protect\rule{0mm}{\tablinsep}}}
145 & $\q_i \ell_j \to \L_{\Q}$ \\
162 & $\q \g \to \ell \L_{\Q}$ \\
163 & $\g \g \to \L_{\Q} \br{\L}_{\Q}$ \\
164 & $\q_i \qbar_i \to \L_{\Q} \br{\L}_{\Q}$ \\
\end{tabular}
  
Several processes that can generate a leptoquark have been included. 
Currently only one leptoquark has been implemented, as particle 39,
denoted $\L_{\Q}$. The leptoquark is assumed to carry specific quark 
and lepton quantum numbers, by default $\u$ quark plus electron. 
These flavour numbers are conserved, i.e. a process such as 
$\u \e^- \to \L_{\Q} \to \d \nu_{\e}$ is not allowed. This may be a 
bit restrictive, but it represents one of many leptoquark possibilities.
The spin of the leptoquark is assumed to be zero, i.e. its decay is
isotropical. 
  
Although only one leptoquark is implemented, its flavours may be 
changed arbitrarily to study the different possibilities. The 
flavours of the leptoquark are defined by the quark and lepton 
flavours in the decay mode list. Since only one decay channel is 
allowed, this means that the quark flavour is stored in 
\ttt{KFDP(MDCY(39,2),1)} and the lepton one in 
\ttt{KFDP(MDCY(39,2),2)}. The former must always be a quark, while 
the latter could be a lepton or an antilepton; a charge-conjugate 
partner is automatically defined by the program. At initialization, 
the charge is recalculated as a function of the flavours defined; 
also the leptoquark name is redefined to be of the type 
\ttt{'LQ\_(q)(l)'}, where actual quark \ttt{(q)} and lepton \ttt{(l)} 
flavours are displayed. 
  
The $\L_{\Q} \to \q \ell$ vertex contains an undetermined Yukawa 
coupling strength, which affects both the width of the leptoquark 
and the cross section for many of the production graphs. This 
strength may be changed in \ttt{PARU(151)}. The definition of 
\ttt{PARU(151)} corresponds to the $k$ factor of \cite{Hew88}, i.e. 
to $\lambda^2/(4\pi\alphaem)$, where $\lambda$ is the Yukawa 
coupling strength of \cite{Wud86}. Note that \ttt{PARU(151)} 
is thus quadratic in the coupling. 
  
The leptoquark is likely to be fairly long-lived, in which case it 
has time to fragment into a mesonic- or baryonic-type state, which 
would decay later on. This is a bit tedious to handle; therefore 
the leptoquark is always assumed to decay before fragmentation has 
to be considered. This may give some imperfections in the event 
generation, but should not be off by much in the final analysis. 
  
Inside the program, the leptoquark is treated as a resonance. 
Since it carries colour, some extra care is required. 
In particular, it is not allowed to put the leptoquark stable, 
by modifying either \ttt{MDCY(39,1)} or \ttt{MSTP(41)}: then the 
leptoquark would be handed undecayed to {\Je}, which would try 
to fragment it (as it does with any other coloured object), and 
most likely crash. 
  
\subsubsection{Compositeness and anomalous couplings} 
\label{sss:ancoupclass}
  
ISUB = \\
\begin{tabular}{rl@{\protect\rule{0mm}{\tablinsep}}}
 11 & $\f_i \f_j \to \f_i \f_j$ (QCD) \\
 12 & $\f_i \fbar_i \to \f_k \fbar_k$  \\
 20 & $\f_i \fbar_j \to \gamma \W^+$  \\
165 & $\f_i \fbar_i \to \f_k \fbar_k$ (via $\gammaZ$) \\
166 & $\f_i \fbar_j \to \f_k \fbar_l$ (via $\W^{\pm}$) \\
\end{tabular}
  
Some processes have been set up to allow anomalous coupling to be 
introduced, in addition to the Standard Model ones. These can be 
switched on by \ttt{MSTP(5)}$\geq 1$; the default \ttt{MSTP(5)=0} 
corresponds to the Standard Model behaviour. 
  
In processes 11 and 12, the quark substructure is included in the 
left--left isoscalar model \cite{Eic84,Chi90} for 
\ttt{MSTP(5)=1}, with compositeness 
scale $\Lambda$ given in \ttt{PARU(155)} (default 1000~GeV) and sign 
$\eta$ of interference term in \ttt{PARU(156)} (default $+1$; only 
other alternative $-1$). The above model assumes that only $\u$ and 
$\d$ quarks are composite (at least at the scale studied); with 
\ttt{MSTP(5)=2} compositeness terms are included in the 
interactions between all quarks. 
  
The processes 165 and 166 are basically equivalent to 1 and 2, i.e. 
$\gammaZ$ and $\W^{\pm}$ exchange, respectively, but a bit less 
fancy (no mass-dependent width etc.). The reason for this duplication 
is that the resonance treatment formalism of processes 1 and 2 could
not easily be extended to include other than $s$-channel graphs.
In processes 165 and 166, only one final-state flavour 
is generated at the time; this flavour should be set in 
\ttt{KFPR(165,1)} and \ttt{KFPR(166,1)}, respectively. For process 
166 one gives the down-type flavour, and the program will associate 
the up-type flavour of the same generation. Defaults are 11 in both 
cases, i.e. $\ee$ and $\e^+ \nu_{\e}$ ($\e^- \br{\nu}_{\e}$) final 
states. While \ttt{MSTP(5)=0} gives the Standard Model results, 
\ttt{MSTP(5)=1} contains the left--left isoscalar model (which 
does not affect process 166), and \ttt{MSTP(5)=3} the 
helicity-non-conserving model (which affects both) \cite{Eic84,Lan91}. 
Both models above assume that only $\u$ and $\d$ quarks are 
composite; with \ttt{MSTP(5)=} 2 or 4, respectively, contact terms 
are included for all quarks in the initial state. Parameters are 
\ttt{PARU(155)} and \ttt{PARU(156)}, as above. 
  
Note that processes 165 and 166 are bookkept as $2 \to 2$ processes, 
while 1 and 2 are $2 \to 1$ ones. This means that the default $\Q^2$
scale in parton distributions is $\pT^2$ for the former and $\hat{s}$ 
for the latter. To make contact between the two, it is recommended to 
set \ttt{MSTP(32)=4}, so as to use $\hat{s}$ as scale also for 
processes 165 and 166. 
  
In process 20, for $\W \gamma$ pair production, it is possible to set 
an anomalous magnetic moment for the $\W$ in \ttt{PARU(153)}
($= \eta = \kappa-1$; where $\kappa = 1$ is the Standard Model 
value). The production process is affected according to the formulae 
of \cite{Sam91}, while $\W$ decay currently remains 
unaffected. It is necessary to set \ttt{MSTP(5)=1} to enable this 
extension. 
  
\subsubsection{Excited fermions}
\label{sss:qlstarclass} 
  
ISUB = \\
\begin{tabular}{rl@{\protect\rule{0mm}{\tablinsep}}}
147 & $\d \g \to \d^*$ \\
148 & $\u \g \to \u^*$ \\
167 & $\q \q' \to \q'' \d^*$ \\
168 & $\q \q' \to \q'' \u^*$ \\
\end{tabular}
  
Compositeness scenarios may also give rise to sharp resonances of 
excited quarks and leptons. If \ttt{MSTP(6)=1}, then at 
initialization the standard fourth generation of fermions will be 
overwritten, and made to correspond to an excited copy of the first 
generation, consisting of spin $1/2$ particles $\d^*$ (code 7), 
$\u^*$ (8), $\e^*$ (17) and $\nu^*_{\e}$ (18). Since the original 
fourth-generation information is lost, it is then not possible to 
generate fourth-generation particles in the same run. 
  
The current implementation contains gauge interaction production
by quark--gluon fusion (processes 147 and 148) and contact interaction 
production by quark--quark or quark--antiquark scattering (processes
167 and 168) . The couplings $f$, $f'$ and $f_s$ to the 
{\bf SU(2)}, {\bf U(1)} and {\bf SU(3)} groups are stored in 
\ttt{PARU(157) - PARU(159)}, the scale parameter $\Lambda$ in 
\ttt{PARU(155)}; you are also expected to change the $\f^*$ masses 
in accordance with what is desired --- see \cite{Bau90} 
for details on conventions. Decay processes are of 
the types $\q^* \to \q \g$, $\q^* \to \q \gamma$, 
$\q^* \to \q \Z^0$ or $\q^* \to \q' \W^{\pm}$. 
A non-trivial angular dependence is included in the $\q^*$
decay for processes 147 and 148, but has not been included for 
processes 167 and 168.

\subsubsection{Technicolor}
\label{sss:technicolorclass}
  
ISUB = \\
\begin{tabular}{rl@{\protect\rule{0mm}{\tablinsep}}}
149 & $\g \g \to \eta_{\mrm{techni}}$ \\
\end{tabular}

The technicolor scenario offers an alternative to the ordinary Higgs
mechanism for giving masses to the $\W$ and $\Z$. The technicolor 
gauge group is an analogue of QCD, with a rich spectrum of 
technimesons made out of techniquarks. Three of the technipions
assume the role of the longitudinal components of the $\W$ and $\Z$ 
bosons, but many other states remain as separate particles.
No fully realistic model has been found so far, however, so any
phenomenology has to be taken as indicative only.

In section \ref{sss:heavySMHclass} it is discussed how processes
71--77, in some of its options, can be used to simulate a scenario 
with techni-$\rho$ resonances in longitudinal gauge boson scattering.

Here we present another process, that of the production of a 
techni-$\eta$. This particle has zero spin, is a singlet under 
electroweak {\bf SU(2)}$\times${\bf U(1)}, but carries octet colour 
charge. It is one of the possible techni-$\pi$ particles; the name 
techni-$\eta$ is part of a subclassification not used by all authors.

The techni-$\eta$ couples to ordinary fermions according to the fermion
squared mass. The dominant decay mode is therefore $\t \tbar$, if
allowed. The coupling to a $\g \g$ state is roughly comparable with
that to $\b \bbar$. Production at hadron colliders is therefore 
predominantly through $\g \g$ fusion, as implemented in process 149.

The two main free parameters are the techni-$\eta$ mass and the decay
constant $F_{\pi}$. The latter appears inversely quadratically in all
the partial widths. Also the total cross section is affected, since 
the cross section is proportional to the $\g \g$ partial width. 
$F_{\pi}$ is stored in \ttt{PARP(46)} and has the default value 123 
GeV, which is the number predicted in some models.

\subsection{Main Processes by Machine}

In the previous section we have already commented on which processes
have limited validity, or have different meanings (according to
conventional terminology) in different contexts. Let us just repeat 
a few of the main points to be remembered for different machines.

\subsubsection{$\ee$ collisions}

The main annihilation process is number 1, $\ee \to \Z^0$,
where in fact the full $\gamma^*/\Z^0$ interference structure is
included. This process can be used, with some confidence, for
c.m. energies from about 4 GeV upwards, i.e. at DORIS/CESR,
PETRA/PEP, TRISTAN, LEP, and any future linear colliders.
(To get below 10 GeV, you have to change \ttt{PARP(2)}, however.) 
This is the default process obtained when \ttt{MSEL=1}, i.e. 
when you do not change anything yourself. 

Process 141 contains a $\Z'^0$, including 
full interference with the standard $\gammaZ$. With the value 
\ttt{MSTP(44)=4} in fact one is back at the standard $\gammaZ$
structure, i.e. the $\Z'^0$ piece has been switched off. Even so,
this process may be useful, since it can simulate e.g.
$\ee \to \H^0 \A^0$. Since the $\H^0$ may in its turn decay to
$\Z^0 \Z^0$, a decay channel of the ordinary $\Z^0$ to 
$\H^0 \A^0$, although physically correct, would be technically
confusing. In particular, it would be messy to set the original 
$\Z^0$ to decay one way and the subsequent ones another. So, in 
this sense, the $\Z'^0$ could be used as a copy of the ordinary
$\Z^0$, but with a distinguishable label.

The process $\ee \to \Upsilon$ does not exist as a separate process
in {\Py}, but can be simulated by using \ttt{LUONIA}, see section
\ref{ss:oniadecays}.

At LEP 2 and even higher energy machines, the simple $s$-channel
process 1 will lose out to other processes, such as 
$\ee \to \Z^0 \Z^0$ and $\ee \to \W^+ \W^-$, i.e. processes 
22 and 25. The former process in fact includes the structure
$\ee \to (\gammaZ)(\gammaZ)$, which means that the cross section 
is singular if either of the two $\gammaZ$ masses is allowed to 
vanish. A mass cut therefore needs to be introduced, and is 
actually also used in other processes, such as $\ee \to \W^+ \W^-$.
  
For practical
applications, both with respect to cross sections and to event
shapes, it is imperative to include initial-state radiation effects. 
Therefore \ttt{MSTP(11)=1} is the default, wherein exponentiated
electron-inside-electron distributions are used to give the
momentum of the actually interacting electron. By radiative 
corrections to process 1, such processes as $\ee \to \gamma \Z^0$
are therefore automatically generated. If process 19 were to be
used at the same time, this would mean that radiation were to be
double-counted. In the alternative \ttt{MSTP(11)=0}, electrons are 
assumed to deposit their full energy in 
the hard process, i.e. initial-state QED radiation is not included.
This option is very useful, since it often corresponds to the
`ideal' events that one wants to correct back to. 

Resolved electrons also means that one may have interactions
between photons. This opens up the whole field of $\gamma\gamma$
processes, which is described in section \ref{sss:photoprodclass}.
In particular, with \ttt{MSTP(12)=1} photons may be resolved, i.e.
photons need not only interact point-like, but can also interact
like a hadron with a partonic substructure. The whole menagerie
of hadron--hadron collider processes can then be accessed. However,
it is not yet possible to include the low-$\pT$ processes with a
variable photon energy spectrum. That is, to generate the `total'
$\gamma\gamma$ spectrum, the program also has to be initialized for
a $\gamma\gamma$ collider.

The thrust of the {\PyJe} programs is towards processes that involve
hadron production, one way or another. Because of generalizations
from other areas, also a few completely
non-hadronic processes are available. These include Bhabha
scattering, $\ee \to \ee$ in process 10, and photon pair production,
$\ee \to \gamma \gamma$ in process 18. However, note that the 
precision that could be expected in a {\Py} simulation of those
processes is certainly far less than that of dedicated programs.
For one thing, electroweak loop effects are not included.
For another, nowhere is the electron mass taken into account,
which means that explicit cut-offs at some minimum $\pT$ are always
necessary. 

\subsubsection{Lepton--hadron collisions}

The issue of applications to $\e\p$ colliders has been covered in a
recent report \cite{Sjo92b}.

The default process for a lepton--hadron collider is deep inelastic
scattering, $\ell \q \to \ell' \q'$, of process 10. This includes
$\gamma^0/\Z^0/\W^{\pm}$ exchange, with full interference, as
described in section \ref{sss:DISclass}. Radiation off the incoming
lepton leg is included by \ttt{MSTP(11)=1} and off the outgoing
one by \ttt{MSTJ(41)=2} (both are default). Note that both 
QED and QCD radiation (off the $\e$ and the $\q$ legs, respectively) 
are allowed to modify the $x$ and $Q^2$ values of the process, while 
the conventional approach in the literature is to allow only the 
former. Therefore an option (on by default) has been added to preserve 
these values by a post-facto rescaling, \ttt{MSTP(23)=1}.  

In terms of cross sections, a more important set of processes are
related to photoproduction, either with a point-like or with a 
resolved photon, see section \ref{sss:photoprodclass}. A complete
description of photoproduction is available \cite{Sch93,Sch93a}, but 
needs three separate runs for the three distinct behaviours of a 
photon: point-like, VMD resolved and anomalous resolved. 

\subsubsection{Hadron--hadron collisions}

The default is to include QCD jet production by $2 \to 2$ processes,
see section \ref{sss:QCDjetclass}. Since the differential 
cross section is divergent for $\pT \to 0$, a lower cut-off has to
be introduced. Normally that cut-off is given by the user-set 
$\pTmin$ value in \ttt{CKIN(3)}. If \ttt{CKIN(3)} is chosen
smaller than a given value of the order of 2 GeV (see \ttt{PARP(81)}
and \ttt{PARP(82)}), then low-$\pT$ events are also switched on. 
The jet cross section is regularized at low $\pT$, so as to 
obtain a smooth joining between the high-$\pT$ and the low-$\pT$
descriptions, see further section \ref{ss:multint}. 
As \ttt{CKIN(3)} is varied, the jump from one 
scenario to another is abrupt, in terms of cross section: in a 
high-energy hadron collider, the cross section for jets down to 
a $\pTmin$ scale of a few GeV can well reach values much 
larger than the total inelastic, non-diffractive cross section.
Clearly this is nonsense; therefore either $\pTmin$ should 
be picked so large that the jet cross section be only a fraction
of the total one, or else one should select $\pTmin = 0$ and 
make use of the full description.

If one switches to \ttt{MSEL=2}, also elastic and diffractive 
processes are switched on, see section \ref{sss:minbiasclass}.
However, the simulation of these processes is fairly primitive,
and should not be used for dedicated studies,
but only to estimate how much they may contaminate the class of
non-diffractive minimum bias events.

Most processes can be simulated in hadron colliders, since the 
bulk of {\Py} processes can be initiated by quarks or gluons.
However, there are limits. Currently we include no photon
or lepton parton distributions, which means that a process like
$\gamma \q \to \gamma \q$ is not accessible. Further, the
possibility of having $\Z^0$ and $\W^{\pm}$ interacting in 
processes such as 71--77 has been hardwired process by process,
and does not mean that there is a generic treatment of $\Z^0$ and 
$\W^{\pm}$ distributions.

The emphasis in the hadron--hadron process description is on high
energy hadron colliders. The program can be used also at 
fixed-target energies, but the multiple interaction model for underlying
events then breaks down and should not be used. The limit of 
applicability is somewhere at around 100 GeV. Below that, one is also
recommended to change to \ttt{MSTP(92)=3}, to obtain a reasonable 
amount of beam remnant particle production in the absence of 
multiple interactions.
 
\clearpage
 
\section{The PYTHIA Program Elements}
 
In the previous two sections, the physics processes and the 
event-generation schemes of {\Py} have been presented. Here, finally,
the event-generation routines and the common block variables are
described. However, routines and variables related to initial- and
final-state showers, beam remnants and underlying events, and
fragmentation and decay are relegated to subsequent sections on
these topics. Further, for historical reasons, many adjustable 
coupling constants are found in the \ttt{LUDAT1} common block in 
{\Je}, rather than somewhere in the {\Py} common blocks;
these parameters are described in section \ref{ss:coupcons}. 
 
In the presentation in this section, information less important for
an efficient use of {\Py} has been put closer to the end.
We therefore begin with the main event generation routines, and
follow this by the main common block variables.
 
It is useful to distinguish three phases in a normal run with {\Py}.
In the first phase, the initialization, the general character of the
run is determined. At a minimum, this requires the specification of the
incoming hadrons and the energies involved. At the discretion of the
user, it is also possible to select specific final states, and to make
a number of decisions about details in the subsequent generation.
This step is finished by a \ttt{PYINIT} call, at which time several
variables are initialized in accordance with the values set. The
second phase consists of the main loop over the number of events,
with each new event being generated by a \ttt{PYEVNT} call. This event
may then be analysed, using information stored in some common blocks,
and the statistics accumulated. In the final phase, results are
presented. This may often be done without the invocation of any {\Py}
routines. From \ttt{PYSTAT}, however, it is possible to obtain a
useful list of cross sections for the different subprocesses.
 
\subsection{The Main Subroutines}
\label{ss:PYTmainroutines}
 
There are two routines that you must know: \ttt{PYINIT} for
initialization and \ttt{PYEVNT} for the subsequent generation of
each new event. In addition, the cross section and other kinds of
information available with \ttt{PYSTAT} are frequently useful. The
other three routines described here, \ttt{PYFRAM}, \ttt{PYKCUT},
and \ttt{PYEVWT}, are of more specialized interest.
 
\drawbox{CALL PYINIT(FRAME,BEAM,TARGET,WIN)}\label{p:PYINIT}
\begin{entry}
\itemc{Purpose:} to initialize the generation procedure.
 
\iteme{FRAME :} a character variable used to specify the frame of the
experiment. Upper-case and lower-case letters may be freely mixed.
\begin{subentry}
\iteme{= 'CMS' :} colliding beam experiment in c.m. frame, with beam
momentum in $+z$ direction and target momentum in $-z$ direction.
\iteme{= 'FIXT' :} fixed-target experiment, with beam particle
momentum pointing in $+z$ direction.
\iteme{= 'USER' :} full freedom to specify frame by giving beam
momentum in \ttt{P(1,1)}, \ttt{P(1,2)} and \ttt{P(1,3)} and target
momentum in \ttt{P(2,1)}, \ttt{P(2,2)} and \ttt{P(2,3)} in
common block \ttt{LUJETS}. Particles are assumed on the mass shell,
and energies are calculated accordingly.
\iteme{= 'FOUR' :} as \ttt{'USER'}, except also energies should be
specified, in \ttt{P(1,4)} and \ttt{P(2,4)}, respectively. The
particles need not be on the mass shell; effective masses are 
calculated from energy and momentum. (But note that numerical
precision may suffer; if you know the masses the option \ttt{'FIVE'}
below is preferrable.)
\iteme{= 'FIVE' :} as \ttt{'USER'}, except also energies and masses
should be specified, i.e the full momentum information in
\ttt{P(1,1) - P(1,5)} and \ttt{P(2,1) - P(2,5)} should be given for
beam and target, respectively. Particles need not be on the mass
shell. Space-like virtualities should be stored as $-\sqrt{-m^2}$.
Four-momentum and mass information must match. 
\iteme{= 'NONE' :} there will be no initialization of any
processes, but only of resonance widths and a few other
process-independent variables. Subsequent to such a call,
\ttt{PYEVNT} cannot be used to generate events, so this option
is mainly intended for those who will want to construct their own
events afterwards, but still want to have access to some of the
{\Py} facilities. In this option, the \ttt{BEAM}, \ttt{TARGET} and
\ttt{WIN} arguments are dummy.
\end{subentry}
 
\iteme{BEAM, TARGET :} character variables to specify beam and
target particles. Upper-case and lower-case letters may be freely
mixed. An antiparticle may be denoted either by `$\sim$' or
`bar' at the end of the name. It is also possible to leave out the
underscore (`\_') directly after `nu' in neutrino names, and the
charge for proton and neutron.
\begin{subentry}
\iteme{= 'e-' :} electron.
\iteme{= 'e+' :} positron.
\iteme{= 'nu\_e' :} $\nu_{\e}$.
\iteme{= 'nu\_e$\sim$' :} $\br{\nu}_{\e}$.
\iteme{= 'mu-' :} $\mu^-$.
\iteme{= 'mu+' :} $\mu^+$.
\iteme{= 'nu\_mu' :} $\nu_{\mu}$.
\iteme{= 'nu\_mu$\sim$' :} $\br{\nu}_{\mu}$.
\iteme{= 'tau-' :} $\tau^-$.
\iteme{= 'tau+' :} $\tau^+$.
\iteme{= 'nu\_tau' :} $\nu_{\tau}$.
\iteme{= 'nu\_tau$\sim$' :} $\br{\nu}_{\tau}$.
\iteme{= 'gamma' :} photon (real, i.e. on the mass shell).
\iteme{= 'pi0' :} $\pi^0$.
\iteme{= 'pi+' :} $\pi^+$.
\iteme{= 'pi-' :} $\pi^-$.
\iteme{= 'n0' :} neutron.
\iteme{= 'n$\sim$0' :} antineutron.
\iteme{= 'p+' :} proton.
\iteme{= 'p$\sim$-' :} antiproton.
\iteme{= 'Lambda0' :} $\Lambda$ baryon.
\iteme{= 'Sigma-' :} $\Sigma^-$ baryon.
\iteme{= 'Sigma0' :} $\Sigma^0$ baryon.
\iteme{= 'Sigma+' :} $\Sigma^+$ baryon.
\iteme{= 'Xi-' :} $\Xi^-$ baryon.
\iteme{= 'Xi0' :} $\Xi^0$ baryon.
\iteme{= 'Omega-' :} $\Omega^-$ baryon.
\iteme{= 'pomeron' :} the pomeron $\pomeron$; since pomeron parton
distribution functions have not been defined this option can not
be used currently.
\iteme{= 'reggeon' :} the reggeon $\reggeon$, with comments as for 
the pomeron above.
\end{subentry}
 
\iteme{WIN :} related to energy of system, exact meaning depends on
\ttt{FRAME}.
\begin{subentry}
\iteme{FRAME='CMS' :} total energy of system (in GeV).
\iteme{FRAME='FIXT' :} momentum of beam particle (in GeV/$c$).
\iteme{FRAME='USER' :} dummy (information is taken from the \ttt{P}
vectors, see above).
\end{subentry}
\end{entry}
 
\drawbox{CALL PYEVNT}\label{p:PYEVNT}
\begin{entry}
\itemc{Purpose:} to generate one event of the type specified by the
\ttt{PYINIT} call. (This is the main routine, which calls a number
of other routines for specific tasks.)
\end{entry}
 
\drawbox{CALL PYSTAT(MSTAT)}\label{p:PYSTAT}
\begin{entry}
\itemc{Purpose:} to print out cross-sections statistics, decay
widths, branching ratios, status codes and parameter values.
\ttt{PYSTAT} may be called at any time, after the \ttt{PYINIT} call,
e.g. at the end of the run, or not at all.
 
\iteme{MSTAT :} specification of desired information.
\begin{subentry}
\iteme{= 1 :} prints a table of how many events of the different
kinds that have been generated and the corresponding
cross sections. All numbers already include the effects of cuts
required by the user in \ttt{PYKCUT}.
\iteme{= 2 :} prints a table of the resonances defined in the
program, with their particle codes (KF), and all allowed decay
channels. (If the number of generations in \ttt{MSTP(1)} is 3,
however, channels involving fourth-generation particles are not
displayed.) For each decay channel is shown the sequential channel
number (IDC) of the {\Je} decay tables, the partial decay width,
branching ratio and effective branching ratio (in the event
some channels have been excluded by the user).
\iteme{= 3 :} prints a table with the allowed hard interaction
flavours \ttt{KFIN(I,J)} for beam and target particles.
\iteme{= 4 :} prints a table of the kinematical cuts \ttt{CKIN(I)}
set by the user in the current run.
\iteme{= 5 :} prints a table with all the values of the status codes
\ttt{MSTP(I)} and the parameters \ttt{PARP(I)} used in the current
run.
\end{subentry}
\end{entry}
 
\drawbox{CALL PYFRAM(IFRAME)}\label{p:PYFRAM}
\begin{entry}
\itemc{Purpose:} to transform an event listing between different 
reference frames, if so desired.
\iteme{IFRAME :} specification of frame the event is to be
boosted to.
\begin{subentry}
\iteme{= 1 :} frame specified by user in the \ttt{PYINIT} call.
\iteme{= 2 :} c.m. frame of incoming particles.
\iteme{= 3 :} hadronic c.m. frame of lepton--hadron interaction 
events. Mainly intended for deep inelastic scattering, but can also
be used in photoproduction. Note that both the lepton and any 
photons radiated off the lepton remain in the event listing, 
and have to be removed separately if you only want to study the
hadronic subsystem.
\end{subentry}
\end{entry}
 
\drawbox{CALL PYKCUT(MCUT)}\label{p:PYKCUT}
\begin{entry}
\itemc{Purpose:} to enable you to reject a given set of
kinematic variables at an early stage of the generation procedure
(before evaluation of cross sections), so as not to spend unnecessary
time on the generation of events that are not wanted.
The routine will not be called unless you require is by
setting \ttt{MSTP(141)=1}, and never if `minimum-bias'-type events
(including elastic and diffractive scattering) are to be generated
as well. A dummy routine \ttt{PYKCUT} is included in the program file,
so as to avoid unresolved external references when the routine is
not used.
\iteme{MCUT :} flag to signal effect of user-defined cuts.
\begin{subentry}
\iteme{= 0 :} event is to be retained and generated in full.
\iteme{= 1 :} event is to be rejected and a new one generated.
\end{subentry}
\itemc{Remark :} at the time of selection, several variables in the
\ttt{MINT} and \ttt{VINT} arrays in the \ttt{PYINT1} common block
contain information that can be used to make the decision. The routine
provided in the program file explicitly reads the variables that
have been defined at the time \ttt{PYKCUT} is called, and also
calculates some derived quantities. The information available
includes subprocess type ISUB, $E_{\mrm{cm}}$, $\hat{s}$, $\hat{t}$,
$\hat{u}$, $\hat{p}_{\perp}$, $x_1$, $x_2$, $x_{\mrm{F}}$, $\tau$, $y$,
$\tau'$, $\cos\hat{\theta}$, and a few more. Some of these may not
be relevant for the process under study, and are then set to zero.
\end{entry}
 
\drawbox{CALL PYEVWT(WTXS)}\label{p:PYEVWT}
\begin{entry}
\itemc{Purpose:} to allow you to reweight event cross sections,
by process type and kinematics of the hard scattering. There exists
two separate modes of usage, described in the following. \\
For \ttt{MSTP(142)=1}, it is assumed that the cross section of the
process is correctly given by default in {\Py}, but that one
wishes to generate events biased to a specific region of phase
space. While the \ttt{WTXS} factor therefore multiplies the na\"{\i}ve
cross section in the choice of subprocess type and kinematics,
the produced event comes with a compensating weight
\ttt{PARI(10)=1./WTXS}, which should be used when filling histograms
etc. In the \ttt{PYSTAT(1)} table, the cross sections are unchanged
(up to statistical errors) compared with the standard cross sections,
but the relative composition of events may be changed and need
no longer be in proportion to relative cross sections. A typical
example of this usage is if one wishes to enhance the production
of high-$\pT$ events; then a weight like
\ttt{WTXS}$=(\pT/p_{\perp 0})^2$ (with $p_{\perp 0}$ some fixed
number) might be appropriate. \\
For \ttt{MSTP(142)=2}, on the other hand, it is assumed that the true
cross section is really to be modifed by the multiplicative factor
\ttt{WTXS}. The generated events therefore come with unit weight, just
as usual. This option is really equivalent to replacing the basic
cross sections coded in {\Py}, but allows more flexibility: no
need to recompile the whole of {\Py}. \\
The routine will not be called unless \ttt{MSTP(142)}$\geq 1$, and
never if `minimum-bias'-type events (including elastic and
diffractive scattering) are to be generated as well. Further,
cross sections for additional multiple interactions or pile-up events
are never affected. A dummy routine \ttt{PYEVWT} is included in the
program file, so as to avoid unresolved external references when the
routine is not used.
\iteme{WTXS:} multiplication factor to ordinary event cross section;
to be set (by you) in \ttt{PYEVWT} call.
\itemc{Remark :} at the time of selection, several variables in the
\ttt{MINT} and \ttt{VINT} arrays in the \ttt{PYINT1} common block
contain information that can be used to make the decision. The routine
provided in the program file explicitly reads the variables that
have been defined at the time \ttt{PYEVWT} is called, and also
calculates some derived quantities. The given list of information
includes subprocess type ISUB, $E_{\mrm{cm}}$, $\hat{s}$, $\hat{t}$,
$\hat{u}$, $\hat{p}_{\perp}$, $x_1$, $x_2$, $x_{\mrm{F}}$, $\tau$, $y$,
$\tau'$, $\cos\hat{\theta}$, and a few more. Some of these may
not be relevant for the process under study, and are then set to
zero.
\itemc{Warning:} the weights only apply to the hard scattering
subprocesses. There is no way to reweight the shape of initial- and
final-state showers, fragmentation, or other aspects of the event.
\end{entry}
 
\subsection{Switches for Event Type and Kinematics Selection}
\label{ss:PYswitchkin}
 
By default, if {\Py} is run for a hadron collider,
only QCD $2 \to 2$ processes are generated,
composed of hard interactions above $\pTmin=$\ttt{PARP(81)},
with low-$\pT$ processes added on so as to give the full
(parametrized) inelastic, non-diffractive cross section.
In an $\ee$ collider, $\gammaZ$ production is the default, and in
an $\ep$ one it is deep inelastic scattering. With the help of the
common block \ttt{PYSUBS}, it is possible to select the generation
of another process, or combination of processes. It is also allowed
to restrict the generation to specific incoming partons/particles
at the hard interaction. This often automatically also restricts
final-state flavours but, in processes such as resonance production
or QCD/QED production of new flavours, switches in the {\Je}
program may be used to this end; see section \ref{ss:parapartdat}.
 
The \ttt{CKIN} array may be used to impose specific kinematics cuts.
You should here be warned that, if kinematical variables are
too strongly restricted, the generation time per event may become
very long. In extreme cases, where the cuts effectively close the
full phase space, the event generation may run into an infinite
loop. The generation of $2 \to 1$ resonance production is performed
in terms of the $\hat{s}$ and $y$ variables, and so the ranges
\ttt{CKIN(1) - CKIN(2)} and \ttt{CKIN(7) - CKIN(8)} may be
arbitrarily restricted without a significant loss of speed.
For $2 \to 2$ processes, $\cos\hat{\theta}$ is added as a third
generation variable, and so additionally the range
\ttt{CKIN(27) - CKIN(28)} may be restricted without any danger.

Effects from initial- and final-state radiation
are not included, since they are not known at the time the 
kinematics at the hard interaction is selected. The sharp 
kinematical cut-offs that can be imposed on the generation
process are therefore smeared, both by QCD radiation and by
fragmentation. A few examples of such effects follow.
\begin{Itemize}
\item Initial-state radiation implies that each of the two incoming
partons has a non-vanishing $\pT$ when they interact. The hard
scattering subsystem thus receives a net transverse boost,
and is rotated with respect to the beam directions.
In a $2 \to 2$ process, what typically happens is that one of the
scattered partons receives an increased $\pT$, while the $\pT$
of the other parton is reduced. 
\item Since the initial-state radiation
machinery assigns space-like virtualities to the incoming partons,
the definitions of $x$ in terms of energy fractions and in terms of
momentum fractions no longer coincide, and so the interacting
subsystem may receive a net longitudinal boost compared with 
na\"{\i}ve expectations, as part of the parton-shower machinery. 
\item Initial-state radiation gives rise to 
additional jets, which in extreme cases may be mistaken for either
of the jets of the hard interaction.
\item Final-state radiation gives rise to additional jets, which 
smears the meaning of the basic $2 \to 2$ scattering. The assignment
of soft jets is not unique. The energy of a jet becomes dependent
on the way it is identified, e.g. what jet cone size is used. 
\item The beam remnant description assigns primordial $k_{\perp}$
values, which also gives a net $\pT$ shift of the hard-interaction 
subsystem; except at low energies this effect is overshadowed by
initial-state radiation, however. Beam remnants may also add 
further activity under the `perturbative' event.
\item Fragmentation will further broaden jet profiles, and make
jet assignments and energy determinations even more uncertain.   
\end{Itemize}
In a study of events within a given window of
experimentally defined variables, it is up to you to leave
such liberal margins that no events are missed. In other words, cuts
have to be chosen such that a negligible fraction of events migrate
from outside the simulated region to inside the interesting region.
Often this may lead to low efficiency in terms of what fraction of
the generated events are actually of interest to you. 
See also section \ref{ss:PYTstarted}.
 
In addition to the variables found in \ttt{PYSUBS}, also those in the
\ttt{PYPARS} common block may be used to select exactly what one wants
to have simulated. These possibilities will be described in the
following subsection.
 
The notation used above and in the following is that `$\hat{~}$'
denotes internal variables in the hard scattering subsystem,
while `$^*$' is for variables in the c.m. frame of the
event as a whole. 
 
\drawbox{COMMON/PYSUBS/MSEL,MSUB(200),KFIN(2,-40:40),CKIN(200)}%
\label{p:PYSUBS}
\begin{entry}
\itemc{Purpose:} to allow the user to run the program with any desired
subset of processes, or restrict flavours or kinematics. If the
default values, denoted below by (D=\ldots), are not satisfactory,
they must be changed before the \ttt{PYINIT} call.
\boxsep
 
\iteme{MSEL :}\label{p:MSEL} (D=1) a switch to select between full 
user control and some preprogrammed alternatives.
\begin{subentry}
\iteme{= 0 :} desired subprocesses have to be switched on in
\ttt{MSUB}, i.e. full user control.
\iteme{= 1 :} depending on incoming particles, different
alternatives are used. \\
Lepton--lepton: $\Z$ or $\W$ production (ISUB = 1 or 2).  \\
Lepton--hadron: deep inelastic scattering (ISUB = 10). \\
Hadron--hadron: QCD high-$\pT$ processes (ISUB = 11, 12, 13, 28,
53, 68); additionally low-$\pT$ production if
\ttt{CKIN(3)}$<$\ttt{PARP(81)} or \ttt{PARP(82)}, depending on
\ttt{MSTP(82)} (ISUB = 95). If low-$\pT$ is switched on, the other
\ttt{CKIN} cuts are not used. \\
A resolved photon counts as hadron, except that an anomalous photon 
cannot have low-$\pT$ interactions. When the photon is not resolved, 
the following cases are possible. \\
Photon--lepton: Compton scattering (ISUB = 34). \\
Photon--hadron: photon-parton scattering (ISUB = 33, 34, 54). \\
Photon--photon: fermion pair production (ISUB = 58).
\iteme{= 2 :} as \ttt{MSEL = 1} for lepton--lepton, lepton--hadron
and unresolved photons. For hadron--hadron (including resolved 
photons) all QCD processes, including low-$\pT$, single and
double diffractive and elastic scattering, are included (ISUB =
11, 12, 13, 28, 53, 68, 91, 92, 93, 94, 95). The \ttt{CKIN} cuts are
here not used.
\iteme{= 4 :} charm ($\c\cbar$) production with massive matrix
elements (ISUB = 81, 82, 84, 85).
\iteme{= 5 :} bottom ($\b\bbar$) production with massive matrix
elements (ISUB = 81, 82, 84, 85).
\iteme{= 6 :} top ($\t\tbar$) production with massive matrix
elements (ISUB = 81, 82, 84, 85).
\iteme{= 7 :} low ($\lrm\br{\lrm}$) production with
massive matrix elements (ISUB = 81, 82, 84, 85).
\iteme{= 8 :} high ($\hrm\br{\hrm}$) production with massive matrix
elements (ISUB = 81, 82, 84, 85).
\iteme{= 10 :} prompt photons (ISUB = 14, 18, 29).
\iteme{= 11 :} $\Z^0$ production (ISUB = 1).
\iteme{= 12 :} $\W^{\pm}$ production (ISUB = 2).
\iteme{= 13 :} $\Z^0$ + jet production (ISUB = 15, 30).
\iteme{= 14 :} $\W^{\pm}$ + jet production (ISUB = 16, 31).
\iteme{= 15 :} pair production of different combinations of $\gamma$,
$\Z^0$ and $\W^{\pm}$ (except $\gamma\gamma$; see \ttt{MSEL} = 10)
(ISUB = 19, 20, 22, 23, 25).
\iteme{= 16 :} $\H^0$ production (ISUB = 3, 102, 103, 123, 124).
\iteme{= 17 :} $\H^0 \Z^0$ or $\H^0 \W^{\pm}$ (ISUB = 24, 26).
\iteme{= 18 :} $\H^0$ production, combination relevant for $\ee$
annihilation (ISUB = 24, 103, 123, 124).
\iteme{= 19 :} $\H^0$, $\H'^0$ and $\A^0$ production, excepting pair
production (ISUB = 24, 103, 123, 124, 153, 158, 171, 173, 174, 176,
178, 179).
\iteme{= 21 :} $\Z'^0$ production (ISUB = 141).
\iteme{= 22 :} $\W'^{\pm}$ production (ISUB = 142).
\iteme{= 23 :} $\H^{\pm}$ production (ISUB = 143).
\iteme{= 24 :} $\R^0$ production (ISUB = 144).
\iteme{= 25 :} $\L_{\Q}$ (leptoquark) production (ISUB = 145, 162,
163, 164).
\iteme{= 35:} single bottom production by $\W$ exchange (ISUB = 83).
\iteme{= 36:} single top production by $\W$ exchange (ISUB = 83).
\iteme{= 37:} single low production by $\W$ exchange (ISUB = 83).
\iteme{= 38:} single high production by $\W$ exchange (ISUB = 83).
\end{subentry}
\boxsep
 
\iteme{MSUB :}\label{p:MSUB} (D=200*0) array to be set when 
\ttt{MSEL=0} (for \ttt{MSEL}$\geq 1$ relevant entries are set in 
\ttt{PYINIT}) to choose which subset of subprocesses to include 
in the generation. The ordering follows the ISUB code given in 
section \ref{ss:ISUBcode} (with comments as given there).
\begin{subentry}
\iteme{MSUB(ISUB) = 0 :} the subprocess is excluded.
\iteme{MSUB(ISUB) = 1 :} the subprocess is included.
\itemc{Note:} when \ttt{MSEL=0}, the \ttt{MSUB} values set by the
user are never changed by {\Py}. If you want to combine several
different `subruns', each with its own \ttt{PYINIT} call, into one
single run, it is up to you to remember not only to switch on
the new processes before each new \ttt{PYINIT} call, but also to
switch off the old ones that are no longer desired.
\end{subentry}
\boxsep
 
\iteme{KFIN(I,J) :}\label{p:KFIN} provides an option to selectively 
switch on and
off contributions to the cross sections from the different incoming
partons/particles at the hard interaction. In combination with the
{\Je} resonance decay switches, this also allows you to set
restrictions on flavours appearing in the final state.
\begin{subentry}
\iteme{I :} is 1 for beam side of event and 2 for target side.
\iteme{J :} enumerates flavours according to the KF code; see section
\ref{ss:codes}.
\iteme{KFIN(I,J) = 0 :} the parton/particle is forbidden.
\iteme{KFIN(I,J) = 1 :} the parton/particle is allowed.
\itemc{Note:} By default, the following are switched on: $\d$, $\u$, 
$\s$, $\c$, $\b$, $\e^-$, $\nu_{\e}$, $\mu^-$, $\nu_{\mu}$, $\tau^-$,
$\nu_{\tau}$, $\g$, $\gamma$, $\Z^0$, $\W^+$ and their antiparticles.
In particular, top is off, and has to be switched on explicitly if
needed. 

\end{subentry}
\boxsep
 
\iteme{CKIN :}\label{p:CKIN} kinematics cuts that can be set by you 
before the \ttt{PYINIT} call, and that affect the region of phase space
within which events are generated. Some cuts are `hardwired' while
most are `softwired'. The hardwired ones are directly related to the
kinematical variables used in the event selection procedure,
and therefore have negligible effects on program efficiency.
The most important of these are \ttt{CKIN(1) - CKIN(8)},
\ttt{CKIN(27) - CKIN(28)}, and \ttt{CKIN(31) - CKIN(32)}.
The softwired ones are most of the remaining ones, that cannot
be fully taken into account
in the kinematical variable selection, so that generation in
constrained regions of phase space may be slow. In extreme
cases the phase space may be so small that the maximization
procedure fails to find any allowed points at all (although some
small region might still exist somewhere), and therefore switches
off some subprocesses, or aborts altogether.
 
\iteme{CKIN(1), CKIN(2) :} (D=2.,-1. GeV) range of allowed
$\hat{m} = \sqrt{\hat{s}}$ values. If \ttt{CKIN(2)}$ < 0.$, the upper
limit is inactive.
 
\iteme{CKIN(3), CKIN(4) :} (D=0.,-1. GeV) range of allowed
$\hat{p}_{\perp}$ values for hard $2 \to 2$ processes, with
transverse momentum $\hat{p}_{\perp}$ defined in the rest frame of
the hard interaction. If \ttt{CKIN(4)}$ < 0.$, the upper limit is
inactive. For processes that are singular in the limit
$\hat{p}_{\perp} \to 0$
(see \ttt{CKIN(6)}), \ttt{CKIN(5)} provides an additional constraint.
The \ttt{CKIN(3)} and \ttt{CKIN(4)} limits can also be used in
$2 \to 1 \to 2$ processes. Here, however, the product
masses are not known and hence are assumed to be vanishing in the event
selection. The actual $\pT$ range for massive products is thus
shifted downwards with respect to the nominal one.
 
\iteme{CKIN(5) :} (D=1. GeV) lower cut-off on $\hat{p}_{\perp}$ values,
in addition to the \ttt{CKIN(3)} cut above, for processes that are
singular in the limit $\hat{p}_{\perp} \to 0$ (see \ttt{CKIN(6)}).
 
\iteme{CKIN(6) :} (D=1. GeV) hard $2 \to 2$ processes, which do not
proceed only via an intermediate resonance (i.e. are $2 \to 1 \to 2$
processes), are classified as singular in the limit
$\hat{p}_{\perp} \to 0$ if either or both of the two final-state
products has a mass $m < $\ttt{CKIN(6)}.
 
\iteme{CKIN(7), CKIN(8) :} (D=-10.,10.) range of allowed scattering
subsystem rapidities $y = y^*$ in the c.m. frame of the event,
where $y = (1/2)  \ln(x_1/x_2)$. (Following the notation of this
section, the variable should be given as $y^*$, but because of its
frequent use, it was called $y$ in section \ref{ss:kinemtwo}.)
 
\iteme{CKIN(9), CKIN(10) :} (D=-10.,10.) range of allowed (true)
rapidities for the product with largest rapidity in a $2 \to 2$ or a
$2 \to 1 \to 2$ process, defined in the c.m. frame of the event,
i.e. $\max(y^*_3, y^*_4)$. Note that rapidities are counted with sign,
i.e. if $y^*_3 = 1$ and $y^*_4 = -2$ then $\max(y^*_3, y^*_4) = 1$.

\iteme{CKIN(11), CKIN(12) :} (D=-10.,10.) range of allowed (true)
rapidities for the product with smallest rapidity in a $2 \to 2$ or a
$2 \to 1 \to 2$ process, defined in the c.m. frame of the event,
i.e. $\min(y^*_3, y^*_4)$. Consistency thus requires
\ttt{CKIN(11)}$\leq$\ttt{CKIN(9)} and
\ttt{CKIN(12)}$\leq$\ttt{CKIN(10)}.
 
\iteme{CKIN(13), CKIN(14) :} (D=-10.,10.) range of allowed
pseudorapidities for the product with largest pseudorapidity
in a $2 \to 2$ or a $2 \to 1 \to 2$ process, defined in the c.m.
frame of the event, i.e. $\max(\eta^*_3, \eta^*_4)$. Note that 
pseudorapidities are counted with sign, i.e. if $\eta^*_3 = 1$ and 
$\eta^*_4 = -2$ then $\max(\eta^*_3, \eta^*_4) = 1$.
 
\iteme{CKIN(15), CKIN(16) :} (D=-10.,10.) range of allowed
pseudorapidities for the product with smallest pseudorapidity
in a $2 \to 2$ or a $2 \to 1 \to 2$ process, defined in the c.m.
frame of the event, i.e. $\min(\eta^*_3, \eta^*_4)$. Consistency
thus requires \ttt{CKIN(15)}$\leq$\ttt{CKIN(13)} and
\ttt{CKIN(16)}$\leq$\ttt{CKIN(14)}.
 
\iteme{CKIN(17), CKIN(18) :} (D=-1.,1.) range of allowed
$\cos\theta^*$ values for the product with largest $\cos\theta^*$
value in a $2 \to 2$ or a $2 \to 1 \to 2$ process, defined in the
c.m. frame of the event, i.e. $\max(\cos\theta^*_3,\cos\theta^*_4)$.
 
\iteme{CKIN(19), CKIN(20) :} (D=-1.,1.) range of allowed
$\cos\theta^*$ values for the product with smallest $\cos\theta^*$
value in a $2 \to 2$ or a $2 \to 1 \to 2$ process, defined in the
c.m. frame of the event, i.e. $\min(\cos\theta^*_3,\cos\theta^*_4)$.
Consistency thus requires \ttt{CKIN(19)}$\leq$\ttt{CKIN(17)} and
\ttt{CKIN(20)}$\leq$\ttt{CKIN(18)}.
 
\iteme{CKIN(21), CKIN(22) :} (D=0.,1.) range of allowed $x_1$ values
for the parton on side 1 that enters the hard interaction.
 
\iteme{CKIN(23), CKIN(24) :} (D=0.,1.) range of allowed $x_2$ values
for the parton on side 2 that enters the hard interaction.
 
\iteme{CKIN(25), CKIN(26) :} (D=-1.,1.) range of allowed Feynman-$x$
values, where $x_{\mrm{F}} = x_1 - x_2$.
 
\iteme{CKIN(27), CKIN(28) :} (D=-1.,1.) range of allowed
$\cos\hat{\theta}$ values in a hard $2 \to 2$ scattering, where
$\hat{\theta}$ is the scattering angle in the rest frame of the
hard interaction.
 
\iteme{CKIN(31), CKIN(32) :} (D=2.,-1. GeV) range of allowed
$\hat{m}' = \sqrt{\hat{s}'}$ values, where $\hat{m}'$ is the mass of
the complete three- or four-body final state in $2 \to 3$ or
$2 \to 4$ processes (while $\hat{m}$, constrained in \ttt{CKIN(1)}
and \ttt{CKIN(2)}, here corresponds to the one- or two-body central
system). If \ttt{CKIN(32)}$ < 0.$, the upper limit is inactive.
 
\iteme{CKIN(35), CKIN(36) :} (D=0.,-1. GeV$^2$) range of allowed
$|\hat{t}| = - \hat{t}$ values in $2 \to 2$ processes. Note that
for deep inelastic scattering this is nothing but the $Q^2$ scale,
in the limit that initial- and final-state radiation is neglected.
If \ttt{CKIN(36)}$ < 0.$, the upper limit is inactive.
 
\iteme{CKIN(37), CKIN(38) :} (D=0.,-1. GeV$^2$) range of allowed
$|\hat{u}| = - \hat{u}$ values in $2 \to 2$ processes. If
\ttt{CKIN(38)}$ < 0.$, the upper limit is inactive.

\iteme{CKIN(39), CKIN(40) :} (D=4., -1. GeV$^2$) the $W^2$ range 
allowed in DIS processes, i.e. subprocess number 10. If 
\ttt{CKIN(40)}$ < 0.$, the upper limit is inactive. Here $W^2$ is 
defined in terms of $W^2 = Q^2 (1-x)/x$. This formula is not quite 
correct, in that \textit{(i)} it neglects the target mass (for a 
proton), and \textit{(ii)} it neglects initial-state photon radiation 
off the incoming electron. It should be good enough for loose cuts, 
however. 
 
\iteme{CKIN(41) - CKIN(44) :} (D=12.,-1.,12.,-1. GeV) range of allowed
mass values of the two (or one) resonances produced in a `true'
$2 \to 2$ process, i.e. one not (only) proceeding through a single
$s$-channel resonance ($2 \to 1 \to 2$). (These are the ones listed
as $2 \to 2$ in the tables in section \ref{ss:ISUBcode}.)
Only particles with a width above \ttt{PARP(41)} are considered as
bona fide resonances and tested against the \ttt{CKIN}
limits; particles with a smaller width are put on the mass shell
without applying any cuts. The exact interpretation of the \ttt{CKIN}
variables depends on the flavours of the two produced resonances. \\
For two resonances like $\Z^0 \W^+$ (produced from
$\f \f' \to \Z^0\W^+$), which are not identical and which are not
each other's antiparticles, one has \\
\ttt{CKIN(41)}$ < m_1 < $\ttt{CKIN(42)}, and \\
\ttt{CKIN(43)}$ < m_2 < $\ttt{CKIN(44)}, \\
where $m_1$ and $m_2$ are the actually generated masses of the two
resonances, and 1 and 2 are defined by the order in which they are
given in the production process specification.  \\
For two resonances like $\Z^0 \Z^0$, which are identical, or
$\W^+ \W^-$, which are each other's antiparticles, one instead
has \\
\ttt{CKIN(41)}$ < \min(m_1,m_2) < $\ttt{CKIN(42)}, and \\
\ttt{CKIN(43)}$ < \max(m_1,m_2) < $\ttt{CKIN(44)}. \\
In addition, whatever limits are set on \ttt{CKIN(1)} and, in
particular, on \ttt{CKIN(2)} obviously affect the masses actually
selected.
\begin{subentry}
\itemc{Note 1:} If \ttt{MSTP(42)=0}, so that no mass smearing is
allowed, the \ttt{CKIN} values have no effect (the same as for
particles with too narrow a width).
\itemc{Note 2:} If \ttt{CKIN(42)}$ < $\ttt{CKIN(41)} it means that 
the \ttt{CKIN(42)} limit is inactive; correspondingly, if
\ttt{CKIN(44)}$ < $\ttt{CKIN(43)} then \ttt{CKIN(44)} is inactive.
\itemc{Note 3:} If limits are active and the resonances are
identical, it is up to you to ensure that
\ttt{CKIN(41)}$\leq$\ttt{CKIN(43)} and
\ttt{CKIN(42)}$\leq$\ttt{CKIN(44)}.
\itemc{Note 4:} For identical resonances, it is not possible to
preselect which of the resonances is the lighter one; if, for
instance, one
$\Z^0$ is to decay to leptons and the other to quarks, there is no
mechanism to guarantee that the lepton pair has a mass smaller
than the quark one.
\itemc{Note 5:} The \ttt{CKIN} values are applied to all relevant
$2 \to 2$ processes equally, which may not be what one desires if
several processes are generated simultaneously. Some caution is
therefore urged in the use of the \ttt{CKIN(41) - CKIN(44)} values.
Also in other respects, users are recommended to take proper
care: if a $\Z^0$ is only allowed to decay into $\b\bbar$, for example,
setting its mass range to be 2--8 GeV is obviously not a
good idea.
\itemc{Note 6:} In principle, the machinery should work for any
$2 \to 2$ process with resonances in the final state, but so far
it has only been checked for processes 22--26, so also from this
point some caution is urged.
\end{subentry}
 
\iteme{CKIN(45) - CKIN(48) :} (D=12.,-1.,12.,-1. GeV) range of allowed
mass values of the two (or one) secondary resonances produced in
a $2 \to 1 \to 2$ process (like $\g \g \to \H^0 \to \Z^0 \Z^0$) or
even a $2 \to 2 \to 4$ (or 3) process (like
$\q \qbar \to \Z^0 \H^0 \to \Z^0 \W^+ \W^-$). Note that these
\ttt{CKIN} values only affect the secondary resonances; the
primary ones are constrained by \ttt{CKIN(1)}, \ttt{CKIN(2)} and
\ttt{CKIN(41) - CKIN(44)} (indirectly, of course, the choice of
primary resonance masses affects the allowed mass range for the
secondary ones). What is considered to be a resonance is defined
by \ttt{PARP(41)}; particles with a width smaller than this are
automatically put on the mass shell. The description closely
parallels the one given for \ttt{CKIN(41) - CKIN(44)}. Thus, for
two resonances that are not identical or each other's
antiparticles, one has \\
\ttt{CKIN(45)}$ < m_1 < $\ttt{CKIN(46)}, and \\
\ttt{CKIN(47)}$ < m_2 < $\ttt{CKIN(48)}, \\
where $m_1$ and $m_2$ are the actually generated masses of the two
resonances, and 1 and 2 are defined by the order in which they
given in the decay channel specification in the program (see e.g.
output from \ttt{PYSTAT(2)} or \ttt{LULIST(12)}). For two
resonances that are identical or each other's antiparticles,
one instead has \\
\ttt{CKIN(45)}$ < \min(m_1,m_2) < $\ttt{CKIN(46)}, and \\
\ttt{CKIN(47)}$ < \max(m_1,m_2) < $\ttt{CKIN(48)}.
\begin{subentry}
\itemc{Notes 1 - 5:} as for \ttt{CKIN(41) - CKIN(44)}, with trivial
modifications.
\itemc{Note 6:} Setting limits on secondary resonance masses is
possible in any of the channels of the allowed types (see above).
However, so far only $\H^0 \to \Z^0 \Z^0$ and $\H^0 \to \W^+ \W^-$
have been fully implemented, such that an arbitrary mass range
below the na\"{\i}ve mass threshold may be picked. For other possible
resonances, any restrictions made on the allowed mass range
are not reflected in the cross section; and further it is not
recommendable to pick mass windows that make a decay on the
mass shell impossible. These limitations will be relaxed in future
versions.
\end{subentry}
 
\iteme{CKIN(51) - CKIN(56) :} (D=0.,-1.,0.,-1.,0.,-1. GeV) range of
allowed transverse momenta in a true $2 \to 3$ process. Currently two
different alternatives are around. For subprocess 131, the $\pT$
of the first product (the $\Z^0$) is set by \ttt{CKIN(3)} and
\ttt{CKIN(4)}, while for the quark and antiquark $\pT$'s one has \\
\ttt{CKIN(51)}$<\min(p_{\perp \q},p_{\perp \qbar})<$\ttt{CKIN(52)},
and \\
\ttt{CKIN(53)}$<\max(p_{\perp \q},p_{\perp \qbar})<$\ttt{CKIN(54)}. \\
Negative \ttt{CKIN(52)} and \ttt{CKIN(54)} values means that the
corresponding limits are inactive. For subprocesses 121--124, and
their $\H'^0$ and $\A^0$ equivalents (173, 174, 178, 179, 181, 182,
186, 187), \ttt{CKIN(51) - CKIN(54)} again corresponds
to $\pT$ ranges for scattered partons, but in order of appearance,
i.e. \ttt{CKIN(51) - CKIN(52)} for the parton scattered off the beam
and \ttt{CKIN(53) - CKIN(54)} for the one scattered off the target.
\ttt{CKIN(55)} and \ttt{CKIN(56)} here sets $\pT$ limits for the
third product, the $\H^0$, i.e. the \ttt{CKIN(3)} and \ttt{CKIN(4)}
values have no effect for this process. Since the $\pT$ of the Higgs
is not one of the primary variables selected, any constraints here
may mean reduced Monte Carlo efficiency, while for these processes
\ttt{CKIN(51) - CKIN(54)}
are `hardwired'  and therefore do not cost anything.
 
\end{entry}
 
\subsection{The General Switches and Parameters}
\label{ss:PYswitchpar}
 
The \ttt{PYPARS} common block contains the status code and parameters
that regulate the performance of the program. All of them are
provided with sensible default values, so that a novice user can
neglect them, and only gradually explore the full range of
possibilities. Some of the switches and parameters in \ttt{PYPARS}
will be described later, in the shower and beam remnants sections.

\drawbox{COMMON/PYPARS/MSTP(200),PARP(200),MSTI(200),PARI(200)}%
\label{p:PYPARS1}
\begin{entry}
 
\itemc{Purpose:} to give access to status code and parameters that
regulate the performance of the program. If the default values,
denoted below by (D=\ldots), are not satisfactory, they must in
general be changed before the \ttt{PYINIT} call. Exceptions, i.e.
variables that can be changed for each new event, are denoted by
(C).
 
\iteme{MSTP(1) :}\label{p:MSTP} (D=3) maximum number of generations. 
Automatically set $\leq 4$.
 
\iteme{MSTP(2) :} (D=1) calculation of $\alphas$ at hard interaction,
in the routine \ttt{ULALPS}.
\begin{subentry}
\iteme{= 0 :} $\alphas$ is fixed at value \ttt{PARU(111)}.
\iteme{= 1 :} first-order running $\alphas$.
\iteme{= 2 :} second-order running $\alphas$.
\end{subentry}
 
\iteme{MSTP(3) :} (D=2) selection of $\Lambda$ value in $\alphas$
for \ttt{MSTP(2)}$\geq 1$.
\begin{subentry}
\iteme{= 1 :} $\Lambda$ is given by \ttt{PARP(1)} for hard
interactions, by \ttt{PARP(61)} for space-like showers,
by \ttt{PARP(72)} for time-like showers not from a resonance decay,
and by \ttt{PARJ(81)} for time-like ones from a resonance decay
(including e.g. $\gamma/\Z^0 \to \q\qbar$ decays, i.e. conventional
$\ee$ physics). This $\Lambda$ is assumed
to be valid for 5 flavours; for the hard interaction the number of
flavours assumed can be changed by \ttt{MSTU(112)}.
\iteme{= 2 :} $\Lambda$ value is chosen according to the 
parton-distribution-function para\-metri\-zations, 
i.e. $\Lambda$ = 0.20 GeV for EHLQ1, = 0.29 GeV for EHLQ2, 
= 0.20 GeV for DO1, = 0.40 GeV for DO2, = 0.213 GeV for CTEQ2M, 
= 0.208 GeV for CTEQ2MS, = 0.208 GeV for CTEQ2MF,
= 0.322 GeV for CTEQ2ML, = 0.190 GeV for CTEQ2L, 
= 0.235 GeV for CTEQ2D, = 0.25 GeV for GRV LO,
and similarly for parton-distribution functions in the \tsc{Pdflib}
library (cf. (\ttt{MSTP(51)}, \ttt{MSTP(52)}). The choice is
always based on the proton parton-distribution set selected, i.e.
is unaffected by pion and photon parton-distribution selection.
All the $\Lambda$ values above
are assumed to refer to 4 flavours, and \ttt{MSTU(112)} is set
accordingly. This $\Lambda$ value is used both for the hard
scattering and the initial- and final-state radiation. The
ambiguity in the choice of the $Q^2$ argument still remains (see
\ttt{MSTP(32)}, \ttt{MSTP(64)} and \ttt{MSTJ(44)}). This $\Lambda$
value is used also for \ttt{MSTP(57)=0}, but the sensible choice
here would be to use \ttt{MSTP(2)=0} and have no initial- or
final-state radiation. This option does \textit{not} change the
\ttt{PARJ(81)} value of timelike parton showers in resonance decays,
so that LEP experience on this specific parameter is not overwritten 
unwittingly. Therefore \ttt{PARJ(81)} can be updated completely 
independently.
\iteme{= 3 :} as \ttt{=2}, except that here also \ttt{PARJ(81)} is
overwritten in accordance with the $\Lambda$ value of the proton
parton-distribution-function set. 
\end{subentry}
 
\iteme{MSTP(4) :} (D=0) treatment of the Higgs sector,
predominantly the neutral one.
\begin{subentry}
\iteme{= 0 :} the $\H^0$ is given the Standard Model Higgs
couplings, while $\H'^0$ and $\A^0$ couplings should be set by
the user in \ttt{PARU(171) - PARU(175)} and
\ttt{PARU(181) - PARU(185)}, respectively.
\iteme{= 1 :} the user should set couplings for all three Higgses,
for the $\H^0$ in \ttt{PARU(161) - PARU(165)}, and for the
$\H'^0$ and $\A^0$ as above.
\iteme{= 2 :} the mass of $\H^0$ in \ttt{PMAS(25,1)} and the
$\tan\beta$ value in \ttt{PARU(141)} are used to derive
$\H'^0$, $\A^0$ and $\H^{\pm}$ masses, and $\H^0$, $\H'^0$,
$\A^0$ and $\H^{\pm}$ couplings, using the relations of the Minimal
Supersymmetric extension of the Standard Model at Born level
\cite{Gun90}. Existing masses and couplings are overwritten by the
derived values. See section \ref{sss:extneutHclass} for discussion 
on parameter constraints.
\iteme{= 3:} as \ttt{=2}, but using relations at the one-loop level.
This option is not yet implemented.
\end{subentry}
 
\iteme{MSTP(5) :} (D=0) presence of anomalous couplings in processes.
\begin{subentry}
\iteme{= 0 :} absent.
\iteme{$\geq$1 :} present, wherever implemented. See section
\ref{sss:ancoupclass} for further details.
\end{subentry}
 
\iteme{MSTP(6) :} (D=0) usage of the fourth-generation fermions to
simulate other fermion kinds.
\begin{subentry}
\iteme{= 0 :} none, i.e. can be used as a standard fourth
generation.
\iteme{= 1 :} excited fermions, as present in compositeness
scenarios; see section \ref{sss:qlstarclass}.
\end{subentry}
 
\iteme{MSTP(7) :} (D=0) choice of heavy flavour in subprocesses
81--85. Does not apply for \ttt{MSEL=4-8}, where the MSEL value
always takes precedence. 
\begin{subentry}
\iteme{= 0 :} for processes 81--84 (85) the `heaviest' flavour 
allowed for gluon (photon) splitting into a quark--antiquark 
(fermion--antifermion) pair, as set in the \ttt{MDME} array. 
Note that `heavy' is defined as the one with largest KF code, 
so that leptons take precedence if they are allowed. 
\iteme{= 1 - 8 :} pick this particular quark flavour; e.g., 
\ttt{MSTP(7)=6} means that top will be produced.
\iteme{= 11 - 18 :} pick this particular lepton flavour. Note that
neutrinos are not possible, i.e. only 11, 13, 15 and 17 are 
meaningful alternatives. Lepton pair production can only occur in
process 85, so if any of the other processes have been switched on
they are generated with the same flavour as would be obtained in 
the option \ttt{MSTP(7)=0}.
\end{subentry}

\iteme{MSTP(8) :} (D=0) choice of electroweak parameters to use 
in the decay widths of resonances ($\W$, $\Z$, $\H$, \ldots) and 
cross sections (production of $\W$'s, $\Z$'s, $\H$'s, \ldots).
\begin{subentry} 
\iteme{= 0 :} everything is expressed in terms of a running 
$\alphaem(Q^2)$ and a fixed $\ssintw$, i.e. $G_{\F}$ is nowhere 
used.
\iteme{= 1 :} a replacement is made according to 
$\alphaem(Q^2) \to \sqrt{2} G_{\F} m_{\W}^2 \ssintw / \pi$
in all widths and cross sections. If $G_{\F}$ and $m_{\Z}$ are 
considered as given, this means that $\ssintw$ and $m_{\W}$ are the 
only free electroweak parameter.
\iteme{= 2 :} a replacement is made as for \ttt{=1}, but additionally 
$\ssintw$ is constrained by the relation 
$\ssintw = 1 - m_{\W}^2/m_{\Z}^2$.
This means that $m_{\W}$ remains as a free parameter, but that the
$\ssintw$ value in \ttt{PARU(102)} is never used, \textit{except} in
the vector couplings in the combination $v = a - 4 \ssintw e$.
This latter degree of freedom enters e.g. for forward-backward
asymmetries in $\Z^0$ decays.
\itemc{Note:} This option does not affect the emission of real photons 
in the initial and final state, where $\alphaem$ is always used. However, 
it does affect also purely electromagnetic hard processes, such as 
$\q \qbar \to \gamma \gamma$.
\end{subentry}
 
\iteme{MSTP(11) :} (D=1) use of electron parton distribution in
$\ee$ and $\ep$ interactions.
\begin{subentry}
\iteme{= 0 :} no, i.e. electron carries the whole beam energy.
\iteme{= 1 :} yes, i.e. electron carries only a fraction of beam
energy in agreement with next-to-leading electron parton-distribution
function, thereby including the effects of initial-state
bremsstrahlung.
\end{subentry}
 
\iteme{MSTP(12) :} (D=0) use of $\e^-$ (`sea', i.e. from
$\e \to \gamma \to \e$), $\e^+$, quark and gluon distribution
functions inside an electron.
\begin{subentry}
\iteme{= 0 :} off.
\iteme{= 1 :} on, provided that \ttt{MSTP(11)}$\geq 1$.
Quark and gluons distributions
are obtained by numerical convolution of the photon content
inside an electron (as given by the bremsstrahlung spectrum of
\ttt{MSTP(11)=1}) with the quark and gluon content inside a
photon. The required numerical precision is set by \ttt{PARP(14)}.
Since the need for numerical integration makes this option
somewhat more time-consuming than ordinary parton-distribution
evaluation, one should only use it when studying processes
where it is needed.
\end{subentry}
 
\iteme{MSTP(13) :} (D=1) choice of $Q^2$ range over which electrons
are assumed to radiate photons; affects normalization of $\e^-$
(sea), $\e^+$, $\gamma$, quark and gluon distributions inside an
electron.
\begin{subentry}
\iteme{= 1 :} range set by $Q^2$ argument of 
parton-distribution-function call, i.e. by $Q^2$ scale of the hard 
interaction. Therefore 
parton distributions are proportional to $\ln(Q^2/m_e^2)$. This is 
normally most appropriate for $\ee$ annihilation.
\iteme{= 2 :} range set by the user-determined $Q_{\mmax}^2$, given in
\ttt{PARP(13)}. Parton distributions are assumed to be  proportional to
$\ln((Q_{\mmax}^2/m_e^2)(1-x)/x^2)$. This is normally most appropriate
for photoproduction, where the electron is supposed to go
undetected, i.e. scatter less than $Q_{\mmax}^2$.
\itemc{Note:} the choice of effective range is especially touchy for
the quark and gluon distributions. An (almost) on-the-mass-shell 
photon has a VMD piece that dies away for a virtual photon. A simple
folding of distribution functions does not take this into account
properly. Therefore the contribution from $Q$ values above the 
$\rho$ mass should be suppressed. A choice of 
$Q_{\mmax} \approx 1$~GeV is then appropriate for a 
photoproduction limit description of physics.
\end{subentry}
 
\iteme{MSTP(14) :} (D=0) structure of incoming photon beam or
target (does not affect photon inside electron, only photons
appearing as argument in the \ttt{PYINIT} call).
\begin{subentry}
\iteme{= 0 :} a photon is assumed to be point-like (a  direct photon), 
i.e. can only interact in processes which explicitly contain the 
incoming photon, such as $\f_i \gamma \to \f_i \g$ for $\gamma\p$ 
interactions. In $\gamma\gamma$ interactions both photons are
direct, i.e the main process is $\gamma \gamma \to \f_i \fbar_i$.
\iteme{= 1 :} a photon is assumed to be resolved, i.e. can only interact
through its constituent quarks and gluons, giving either high-$\pT$
parton--parton scatterings or low-$\pT$ events. Hard processes are 
calculated with the use of the full photon parton distributions. 
In $\gamma\gamma$ interactions both photons are resolved.
\iteme{= 2 :} a photon is assumed resolved, but only the VMD piece is 
included in the parton distributions, which therefore mainly 
are scaled-down versions of the $\rho^0 / \pi^0$ ones. Both high-$\pT$
parton--parton scatterings and low-pT events are allowed. In 
$\gamma\gamma$ interactions both photons are VMD-like.
\iteme{= 3 :} a photon is assumed resolved, but only the anomalous 
piece of the photon parton distributions is included. Only high-$\pT$ 
parton--parton scatterings are allowed. In $\gamma\gamma$ 
interactions both photons are anomalous.
\iteme {= 4 :} in $\gamma\gamma$ interactions one photon is direct 
and the other resolved. A typical process is thus 
$\f_i \gamma \to \f_i \g$. Hard processes are calculated with the use 
of the full photon parton distributions for the resolved photon. 
Both possibilities of which photon is direct are included, in event 
topologies and in cross sections. This option cannot be used in 
configurations with only one incoming photon.
\iteme{= 5 :} in $\gamma\gamma$ interactions one photon is direct 
and the other VMD-like. Both possibilities of which photon is direct 
are included, in event topologies and in cross sections. This option 
cannot be used in configurations with only one incoming photon.
\iteme{= 6 :} in $\gamma\gamma$ interactions one photon is direct 
and the other anomalous. Both possibilities of which photon is direct 
are included, in event topologies and in cross sections. This option 
cannot be used in configurations with only one incoming photon.
\iteme{= 7 :} in $\gamma\gamma$ interactions one photon is VMD-like 
and the other anomalous. Only high-$\pT$ parton--parton scatterings 
are allowed. Both possibilities of which photon is VMD-like are 
included, in event topologies and in cross sections. This option 
cannot be used in configurations with only one incoming photon.
\iteme{= 10 :} the VMD, direct and anomalous components of the photon
are automatically mixed. For $\gamma\p$ interactions, this means an
automatic mixture of the three classes 0, 2 and 3 above 
\cite{Sch93,Sch93a}, for $\gamma\gamma$ ones a mixture of the 
six classes 0, 2, 3, 5, 6 and 7 above \cite{Sch94a}. Various 
restrictions exist for this option, as discussed in section 
\ref{sss:photoprodclass}.  
\itemc{Note:} our best understanding of how to mix event classes is
provided by the option 10 above, which also can be obtained by 
combining three (for $\gamma\p$) or six (for $\gamma\gamma$) separate
runs. In a simpler alternative the VMD and anomalous classes are joined
into a single resolved class. Then $\gamma\p$ physics only requires two
separate runs, with 0 and 1, and $\gamma\gamma$ physics requires three, 
with 0, 1 and 4.
\end{subentry}
 
\iteme{MSTP(15) :} (D=5) possibility to modify the nature of the 
anomalous photon component (as used with the appropriate \ttt{MSTP(14)}
options), in particular with respect to the scale choices and
cut-offs of hard processes. This option is mainly intended for 
comparative studies and should not normally be touched.
\begin{subentry}
\iteme{= 0 :} none, i.e. the same treatment as for the VMD component.
\iteme{= 1 :} evaluate the anomalous parton distributions at a scale
$Q^2/$\ttt{PARP(17)}$^2$.
\iteme{= 2 :} as \ttt{=1}, but instead of \ttt{PARP(17)} use either
\ttt{PARP(81)/PARP(15)} or \ttt{PARP(82)/PARP(15)}, depending on 
\ttt{MSTP(82)} value.
\iteme{= 3 :} evaluate the anomalous parton distribution functions 
of the photon as $f^{\gamma,\mrm{anom}}(x, Q^2, p_0^2) - 
f^{\gamma,\mrm{anom}}(x, Q^2, r^2 Q^2)$ with $r = $\ttt{PARP(17)}.
\iteme{= 4 :} as \ttt{=3}, but instead of \ttt{PARP(17)} use either
\ttt{PARP(81)/PARP(15)} or \ttt{PARP(82)/PARP(15)}, depending on 
\ttt{MSTP(82)} value.
\iteme{= 5 :} use larger $\pTmin$ for the anomalous component than 
for the VMD one, but otherwise no difference.
\end{subentry}

\iteme{MSTP(21) :} (D=1) nature of fermion--fermion scatterings
simulated in process 10 by $t$-channel exchange.
\begin{subentry}
\iteme{= 0 :} all off.
\iteme{= 1 :} full mixture of $\gammaZ$ neutral current and
$\W^{\pm}$ charged current.
\iteme{= 2 :} $\gamma$ neutral current only.
\iteme{= 3 :} $\Z^0$ neutral current only.
\iteme{= 4 :} $\gammaZ$ neutral current only.
\iteme{= 5 :} $\W^{\pm}$ charged current only.
\end{subentry}
 
\iteme{MSTP(22) :} (D=0) special override of normal $Q^2$ definition
used for maximum of parton-shower evolution, intended for deep 
inelastic scattering) in lepton--hadron events, see section
\ref{ss:showrout}.
 
\iteme{MSTP(23) :} (D=1) for deep inelastic scattering processes
(10 and 83), this option allows the $x$ and $Q^2$ of the original
hard scattering to be retained by the final electron.
\begin{subentry}
\iteme{= 0 :} no correction procedure, i.e. $x$ and $Q^2$ of the
scattered electron differ from the originally generated $x$ and
$Q^2$.
\iteme{= 1 :} post facto correction, i.e. the change of electron
momentum, by initial and final QCD radiation, primordial $k_{\perp}$
and beam remnant treatment, is corrected for by a shuffling of
momentum between the electron and hadron side in the final state.
Only process 10 is corrected, while process 83 is not.
\iteme{= 2 :} as \ttt{=1}, except that both process 10 and 83 are
treated. This option is dangerous, especially for top, since it
may well be impossible to `correct' in process 83: the standard DIS 
kinematics definitions are based on the assumption of massless quarks.
Therefore infinite loops are not excluded.
\itemc{Note:} the correction procedure will fail for a fraction of
the events, which are thus rejected (and new ones generated in their
place). The correction option is not unambiguous, and should
not be taken too seriously. For very small $Q^2$ values, the $x$
is not exactly preserved even after this procedure.
\end{subentry}
 
\iteme{MSTP(31) :} (D=1) parametrization of total, elastic and
diffractive cross sections.
\begin{subentry}
\iteme{= 0 :} everything is to be set by you yourself in 
the \ttt{PYINT7} common block. For photoproduction, additionally
you need to set \ttt{VINT(281)}. Normally you would set these
values once and for all before the \ttt{PYINIT} call, but if
you run with variable energies (see \ttt{MSTP(171)}) you can
also set it before each new \ttt{PYEVNT} call. 
\iteme{= 1 :} Donnachie--Landshoff for total cross section 
\cite{Don92}, and Schuler--Sj\"ostrand for elastic and diffractive
cross sections \cite{Sch94,Sch93a}.
\end{subentry}
 
\iteme{MSTP(32) :} (D=2) $Q^2$ definition in hard scattering for
$2 \to 2$ processes. For resonance production $Q^2$ is always chosen
to be $\hat{s} = m_R^2$, where $m_R$ is the mass of the resonance.
For gauge boson scattering processes $VV \to VV$ the $\W$ or $\Z^0$
squared mass is used as scale in parton distributions. See 
\ttt{PARP(34)} for a possibility to modify the choice below by 
a multiplicative factor.
\begin{subentry}
\iteme{= 1 :} $Q^2 = 2 \hat{s} \hat{t} \hat{u} / (\hat{s}^2 +
\hat{t}^2 + \hat{u}^2)$.
\iteme{= 2 :} $Q^2 = (m_{\perp 1}^2 + m_{\perp 2}^2)/2$.
\iteme{= 3 :} $Q^2 = \min(-\hat{t}, -\hat{u})$.
\iteme{= 4 :} $Q^2 = \hat{s}$.
\iteme{= 5 :} $Q^2 = -\hat{t}$.
\end{subentry}
 
\iteme{MSTP(33) :} (D=0) inclusion of $K$ factors in hard
cross sections for parton--parton interactions (i.e. for
incoming quarks and gluons).
\begin{subentry}
\iteme{= 0 :} none, i.e. $K = 1$.
\iteme{= 1 :} a common $K$ factor is used, as stored in
\ttt{PARP(31)}.
\iteme{= 2 :} separate factors are used for ordinary
(\ttt{PARP(31)}) and colour annihilation graphs (\ttt{PARP(32)}).
\iteme{= 3 :} A $K$ factor is introduced by a shift in the
$\alphas$ $Q^2$ argument,
$\alphas = \alphas($\ttt{PARP(33)}$Q^2)$.
\end{subentry}
 
\iteme{MSTP(34) :} (D=1) use of interference term in matrix
elements for QCD processes, see section \ref{sss:QCDjetclass}.
\begin{subentry}
\iteme{= 0 :} excluded (i.e. string-inspired matrix elements).
\iteme{= 1 :} included (i.e. conventional QCD matrix elements).
\itemc{Note:} for the option \ttt{MSTP(34)=1}, i.e. interference
terms included, these terms are divided between the different
possible colour configurations according to the pole structure
of the (string-inspired) matrix elements for the different
colour configurations.
\end{subentry}
 
\iteme{MSTP(35) :} (D=0) threshold behaviour for heavy-flavour
production, i.e. ISUB = 81, 82, 84, 85, and also for $\Z$ and $\Z'$
decays. The non-standard options are mainly intended for top, but
can be used, with less theoretical reliability, also for charm and
bottom (for $\Z$ and $\Z'$ only top and heavier flavours
are affected). The threshold factors are given in 
eqs.~(\ref{pp:threshenh}) and (\ref{pp:threshsup}).
\begin{subentry}
\iteme{= 0 :} na\"{\i}ve lowest-order matrix-element behaviour.
\iteme{= 1 :} enhancement or suppression close to threshold,
according to the colour structure of the process. The
$\alphas$ value appearing in the threshold factor (which is not
the same as the $\alphas$ of the lowest-order $2 \to 2$ process)
is taken to be fixed at the value given in \ttt{PARP(35)}. The
threshold factor used in an event is stored in \ttt{PARI(81)}.
\iteme{= 2 :} as \ttt{=1}, but the $\alphas$ value appearing in
the threshold factor is taken to be running, with argument
$Q^2 = m_{\Q} \sqrt{ (\hat{m} - 2m_{\Q})^2 + \Gamma_{\Q}^2}$.
Here $m_{\Q}$ is the nominal heavy-quark mass, $\Gamma_{\Q}$ is
the width of the heavy-quark-mass distribution, and $\hat{m}$ is
the invariant mass of the heavy-quark pair. The $\Gamma_{\Q}$ value
has to be stored by the user in \ttt{PARP(35)}. The regularization
of $\alphas$ at low $Q^2$ is given by \ttt{MSTP(36)}.
\end{subentry}
 
\iteme{MSTP(36) :} (D=2) regularization of $\alphas$ in the
limit $Q^2 \to 0$ for the threshold factor obtainable in the
\ttt{MSTP(35)=2} option; see \ttt{MSTU(115)} for a list of
the possibilities.
 
\iteme{MSTP(37) :} (D=1) inclusion of running quark masses in
Higgs production ($\q \qbar \to \H^0$) and decay
($\H^0 \to \q \qbar$) couplings. Also included for charged Higgs
production and decay, but there only for the down-type quark,
since the up-type one normally is a top quark, with 
$m_{\t} \approx m_{\H}$.
\begin{subentry}
\iteme{= 0 :} not included, i.e. fixed quark masses are used
according to the values in the \ttt{PMAS} array.
\iteme{= 1 :} included, with running starting from the
value given in the \ttt{PMAS} array, at a $Q_0$ 
scale of \ttt{PARP(37)} times the quark mass itself,
up to a $Q$ scale given by the Higgs mass.
This option only works when $\alphas$ is allowed to run (so one can 
define a $\Lambda$ value). Therefore it is only applied if additionally
\ttt{MSTP(2)}$\geq 1$.
\end{subentry}
 
\iteme{MSTP(38) :} (D=5) handling of quark loop masses in the box
graphs $\g \g \to \gamma \gamma$ and $\g \g \to \g \gamma$.
\begin{subentry}
\iteme{= 0 :} the program will for each flavour automatically
choose the massless approximation for light quarks and the full
massive formulae for heavy quarks, with a dividing line
between light and heavy quarks that depends on the actual
$\hat{s}$ scale.
\iteme{$\geq$1 :} the program will use the massless approximation
throughout, assuming the presence of \ttt{MSTP(38)} effectively
massless quark species (however, at most 8). Normally one would use
\ttt{=5} for the inclusion of all quarks up to bottom, and
\ttt{=6} to include top as well.
\itemc{Warning:} for \ttt{=0}, numerical instabilities may arise
for scattering at small angles. Users are therefore recommended in
this case to set \ttt{CKIN(27)} and \ttt{CKIN(28)} so as to exclude
the range of scattering angles that are not of interest anyway.
\end{subentry}

\iteme{MSTP(39) :} (D=2) choice of $Q^2$ scale for parton distributions 
and initial state parton showers in processes $\g\g$ or 
$\q \qbar \to \Q \Qbar \H$.
\begin{subentry}
\iteme{= 1 :} $m_{\Q}^2$.
\iteme{= 2 :} $\max(m_{\perp\Q}^2,m_{\perp\Qbar}^2 ) = 
        m_{\Q}^2 + \max(p_{\perp\Q}^2 , p_{\perp\Qbar}^2)$.
\iteme{= 3 :} $m_{\H}^2$.
\iteme{= 4 :} $\hat{s} = (p_{\H} + p_{\Q} + p_{\Qbar})^2$.
\end{subentry}
 
\iteme{MSTP(41) :} (D=1) master switch for all resonance decays
($\Z^0$, $\W^{\pm}$, $\H^0$, $\Z'^0$, $\W'^{\pm}$, $\H'^0$,
$\A^0$, $\H^{\pm}$, $\L_{\Q}$, $\R^0$, $\d^*$, $\u^*$, \ldots).
\begin{subentry}
\iteme{= 0 :} off.
\iteme{= 1 :} on.
\itemc{Note:} also for \ttt{MSTP(41)=1} it is possible to switch
off the decays of specific resonances by using the \ttt{MDCY(KC,1)}
switches in {\Je}. However, since the \ttt{MDCY} values are
overwritten in the \ttt{PYINIT} call, individual resonances should
be switched off after the \ttt{PYINIT} call.
\itemc{Warning:} leptoquark decays must not be switched off if one
later on intends to let leptoquarks decay (with \ttt{LUEXEC}); see
section \ref{sss:LQclass}.
\end{subentry}
 
\iteme{MSTP(42) :} (D=1) mass treatment in $2 \to 2$ processes, where
the final-state resonances have finite width (see \ttt{PARP(41)}).
(Does not apply for the production of a single $s$-channel resonance,
where the mass appears explicitly in the cross section of the
process, and thus is always selected with width.)
\begin{subentry}
\iteme{= 0 :} particles are put on the mass shell.
\iteme{= 1 :} mass generated according to a Breit--Wigner.
\end{subentry}
 
\iteme{MSTP(43) :} (D=3) treatment of $\Z^0/\gamma^*$ interference
in matrix elements. So far implemented in subprocesses 1, 15, 19, 22,
30 and 35; in other processes what is called a $\Z^0$ is really a 
$\Z^0$ only, without the $\gamma^*$ piece.
\begin{subentry}
\iteme{= 1 :} only $\gamma^*$ included.
\iteme{= 2 :} only $\Z^0$ included.
\iteme{= 3 :} complete $\Z^0/\gamma^*$ structure (with
interference) included.
\end{subentry}
 
\iteme{MSTP(44) :} (D=7) treatment of $\Z'^0/\Z^0/\gamma^*$
interference in matrix elements.
\begin{subentry}
\iteme{= 1 :} only $\gamma^*$ included.
\iteme{= 2 :} only $\Z^0$ included.
\iteme{= 3 :} only $\Z'^0$ included.
\iteme{= 4 :} only $\Z^0/\gamma^*$ (with interference) included.
\iteme{= 5 :} only $\Z'^0/\gamma^*$ (with interference) included.
\iteme{= 6 :} only $\Z'^0/\Z^0$ (with interference) included.
\iteme{= 7 :} complete $\Z'^0/\Z^0/\gamma^*$ structure
(with interference) included.
\end{subentry}
 
\iteme{MSTP(45) :} (D=3) treatment of $\W\W \to \W\W$ structure
(ISUB = 77).
\begin{subentry}
\iteme{= 1 :} only $\W^+\W^+ \to \W^+\W^+$ and
$\W^-\W^- \to \W^-\W^-$ included.
\iteme{= 2 :} only $\W^+\W^- \to \W^+\W^-$ included.
\iteme{= 3 :} all charge combinations $\W\W \to \W\W$ included.
\end{subentry}
 
\iteme{MSTP(46) :} (D=1) treatment of $VV \to V'V'$ structures
(ISUB = 71--77), where $V$ represents a longitudinal gauge boson.
\begin{subentry}
\iteme{= 0 :} only $s$-channel Higgs exchange included (where
existing). With this option, subprocesses 71--72 and 76--77
will essentially be equivalent to subprocesses 5 and 8,
respectively, with the proper decay channels (i.e. only $\Z^0\Z^0$
or $\W^+\W^-$) set via \ttt{MDME}.
The description obtained for subprocesses 5 and 8 in this case
is more sophisticated, however; see section \ref{sss:heavySMHclass}.
\iteme{= 1 :} all graphs contributing to $VV \to V'V'$
processes are included.
\iteme{= 2 :} only graphs not involving Higgs exchange
(either in $s$, $t$ or $u$ channel) are included; this option
then gives the na\"{\i}ve behaviour one would expect if no Higgs 
exists, including unphysical unitarity violations at high energies.
\iteme{= 3 :} the strongly interacting Higgs-like model of
Dobado, Herrero and Terron \cite{Dob91} with Pad\'e unitarization.
Note that to use this option it is necessary to set the Higgs mass
to a large number like 20 TeV (i.e. \ttt{PMAS(25,1)=20000}). The
parameter $\nu$ is stored in \ttt{PARP(44)}, but should not have
to be changed.
\iteme{= 4 :} as \ttt{=3}, but with K-matrix unitarization.
\iteme{= 5 :} the strongly interacting QCD-like model of
Dobado, Herrero and Terron \cite{Dob91} with Pad\'e unitarization.
The parameter $\nu$ is stored in \ttt{PARP(44)}, but should not
have to be changed. The effective techni-$\rho$ mass in this model
is stored in \ttt{PARP(45)}; by default it is 2054 GeV, which is
the expected value for three technicolors, based on scaling up
the ordinary $\rho$ mass appropriately.
\iteme{= 6 :} as \ttt{=5}, but with K-matrix unitarization.
While \ttt{PARP(45)} still is a parameter of the model, this type
of unitarization does not give rise to a resonance at a mass of
\ttt{PARP(45)}.
\end{subentry}
 
\iteme{MSTP(47) :} (D=1) (C) angular orientation of decay products
of resonances ($\Z^0$, $\W^{\pm}$, $\H^0$, $\Z'^0$, $\W'^{\pm}$,
etc.), either when produced singly or in pairs (also
from an $\H^0$ decay), or in combination with a single quark,
gluon or photon.
\begin{subentry}
\iteme{= 0 :} independent decay of each resonance, isotropic in c.m.
frame of the resonance.
\iteme{= 1 :} correlated decay angular distributions according to
proper matrix elements, to the extent these are known.
\end{subentry}

\iteme{MSTP(48) :} (D=2) (C) possibility to switch between top decay 
before or after fragmentation. As a rule of thumb, option 0 is
recommendable for top masses below 120 GeV and option 1 above that,
but clearly there is a gradual transition between the two.
\begin{subentry}
\iteme{= 0 :} top quarks fragment to top hadrons, which subsequently 
decay. The $\b$ quark may be allowed to shower, see \ttt{MSTJ(27)}.
The $\W$ produced in the decay is code 89.
\iteme{= 1 :} top quarks decay, $\t \to \b \W^+$, and thereafter the 
$\b$ quark fragments. Parton showering of the $\b$ is automatically 
included, but can be switched off with \ttt{MSTP(71)}. The $\W$
has the ordinary code 24, and is allowed to decay isotropically. 
\iteme{= 2 :} as \ttt{=1}, except that the $\W$ decay is anisotropic,
as expected from $\W$ polarization in the top decay.
\itemc{Note:} in options 1 and 2 the cross section is reduced to take 
into account restrictions on allowed decay modes, while no such 
reduction occurs for option 0. See further section 
\ref{sss:heavflavclass}. 
\end{subentry} 

\iteme{MSTP(49) :} (D=2) (C) possibility to switch between fourth 
generation decay before or after fragmentation. For the quarks $\hrm$ 
and $\lrm$ the meaning is exactly as \ttt{MSTP(48)} is for the $\t$ 
quark. For the lepton $\chi$ the difference is whether decay is 
handled as part of the {\Py} resonance machinery or as part of the 
{\Je} particle decay one. The $\nu_{\chi}$ is assumed stable, so the 
option above would currently make no difference.
\begin{subentry}
\iteme{= 0 :} hadrons are first produced, which subsequently 
decay (in \ttt{LUDECY}). The new quark may be allowed to shower, 
see \ttt{MSTJ(27)}. The $\W$ produced in the decay is code $\pm89$.
\iteme{= 1 :} the heavy quark first decays (in \ttt{PYRESD}) to a 
light one, and thereafter the light quark fragments. Parton showering 
in the decay is automatically included, but can be switched off with 
\ttt{MSTP(71)}. The $\W$ has the ordinary code $\pm24$, and is allowed 
to decay isotropically. 
\iteme{= 2 :} as \ttt{=1}, except that the $\W$ decay is anisotropic,
as expected from $\W$ polarization in the heavy flavour decay.
\itemc{Note:} in options 1 and 2 the cross section is reduced to take 
into account restrictions on allowed decay modes, while no such 
reduction occurs for option 0. See further section 
\ref{sss:heavflavclass}. 
\end{subentry} 

\iteme{MSTP(51) :} (D=9) choice of proton parton-distribution set; 
see also \ttt{MSTP(52)}.
\begin{subentry}
\iteme{= 1 :} EHLQ set 1 (1986 updated version).
\iteme{= 2 :} EHLQ set 2 (1986 updated version).
\iteme{= 3 :} Duke--Owens set 1.
\iteme{= 4 :} Duke--Owens set 2.
\iteme{= 5 :} CTEQ2M (best $\br{\mrm{MS}}$ fit).
\iteme{= 6 :} CTEQ2MS (singular at small $x$).
\iteme{= 7 :} CTEQ2MF (flat at small $x$).
\iteme{= 8 :} CTEQ2ML (large $\Lambda$).
\iteme{= 9 :} CTEQ2L (best leading order fit).
\iteme{= 10 :} CTEQ2D (best DIS fit).
\iteme{= 11 :} GRV LO (1992 updated version). 
\itemc{Note:} since all parametrizations have some region of
applicability, the parton distributions are assumed frozen below
the lowest $Q^2$ covered by the parametrizations; the CTEQ2 ones
have been allowed to extend down to $Q_{\mmin} = 1$~GeV. For the
former four, evolution is also frozed above the maximum $Q^2$.
The extrapolation of EHLQ to low $x$ is covered by \ttt{PARP(51)}.
\end{subentry}
 
\iteme{MSTP(52) :} (D=1) choice of proton 
parton-distribution-function library.
\begin{subentry}
\iteme{= 1 :} the internal {\Py} one, with parton distributions
according to the \ttt{MSTP(51)} above.
\iteme{= 2 :} the \tsc{Pdflib} one \cite{Plo93}, with the
\tsc{Pdflib} (version 4) \ttt{NGROUP} and \ttt{NSET} numbers to be 
given as \ttt{MSTP(51) = 1000}$\times$\ttt{NGROUP + NSET}.
\itemc{Note:} to make use of option 2, it is necessary to link 
\tsc{Pdflib}. Additionally, on most computers, the two dummy routines
\ttt{PDFSET} and \ttt{STRUCTM} at the end of the {\Py} file should be
removed or commented out.
\itemc{Warning:} For external parton distribution libraries,
{\Py} does not check whether \ttt{MSTP(51)} corresponds to a
valid code, or if special $x$ and $Q^2$ restrictions exist
for a given set, such that crazy values could be returned.
This puts an extra responsibility on you.
\end{subentry}
 
\iteme{MSTP(53) :} (D=1) choice of pion parton-distribution set;
see also \ttt{MSTP(54)}.
\begin{subentry}
\iteme{= 1 :} Owens set 1.
\iteme{= 2 :} Owens set 2.
\iteme{= 3 :} GRV LO (updated version).
\end{subentry}
 
\iteme{MSTP(54) :} (D=1) choice of pion parton-distribution-function
library.
\begin{subentry}
\iteme{= 1 :} the internal {\Py} one, with parton distributions
according to the \ttt{MSTP(53)} above.
\iteme{= 2 :} the \tsc{Pdflib} one \cite{Plo93}, with the
\tsc{Pdflib} (version 4) \ttt{NGROUP} and \ttt{NSET} numbers to be 
given as \ttt{MSTP(53) = 1000}$\times$\ttt{NGROUP + NSET}.
\itemc{Note:} to make use of option 2, it is necessary to link 
\tsc{Pdflib}. Additionally, on most computers, the two dummy routines
\ttt{PDFSET} and \ttt{STRUCTM} at the end of the {\Py} file should be
removed or commented out.
\itemc{Warning:} For external parton distribution libraries,
{\Py} does not check whether \ttt{MSTP(53)} corresponds to a valid
code, or if special $x$ and $Q^2$ restrictions exist for a given
set, such that crazy values could be returned. This puts an extra
responsibility on you.
\end{subentry}

\iteme{MSTP(55)} : (D=5) choice of the parton-distribution 
set of the photon; see also \ttt{MSTP(56)}.
\begin{subentry}
\iteme{= 1 :} Drees--Grassie.
\iteme{= 5 :} SaS 1D (in DIS scheme, with $Q_0=0.6$~GeV).
\iteme{= 6 :} SaS 1M (in {\MSbar} scheme, with $Q_0=0.6$~GeV).
\iteme{= 7 :} SaS 2D (in DIS scheme, with $Q_0=2$~GeV).
\iteme{= 8 :} SaS 2M (in {\MSbar} scheme, with $Q_0=2$~GeV).
\iteme{= 9 :} SaS 1D (in DIS scheme, with $Q_0=0.6$~GeV).
\iteme{= 10 :} SaS 1M (in {\MSbar} scheme, with $Q_0=0.6$~GeV).
\iteme{= 11 :} SaS 2D (in DIS scheme, with $Q_0=2$~GeV).
\iteme{= 12 :} SaS 2M (in {\MSbar} scheme, with $Q_0=2$~GeV).
\itemc{Note 1:} sets 5--8 use the parton distributions of the respective
set, and nothing else. These are appropriate for most applications, e.g. 
jet production in $\gamma\p$ and $\gamma\gamma$ collisions. Sets 9--12 
instead are appropriate for $\gamma^*\gamma$ processes, i.e. DIS 
scattering on a photon, as measured in $F_2^{\gamma}$. Here the anomalous 
contribution for $\c$ and $\b$ quarks are handled by the Bethe-Heitler 
formulae, and the direct term is artificially lumped with the anomalous
one, so that the event simulation more closely agrees with what will be 
experimentally observed in these processes. The agreement with the 
$F_2^{\gamma}$ parametrization is still not perfect, e.g. in the treatment 
of heavy flavours close to threshold. 
\itemc{Note 2:} Sets 5--12 contain both VMD pieces and anomalous pieces,
separately parametrized. Therefore the respective piece is automatically 
called, whatever \ttt{MSTP(14)} value is used to select only a part of the
allowed photon interactions. For other sets (set 1 above or \tsc{Pdflib} 
sets), usually there is no corresponding subdivision. Then an option like 
\ttt{MSTP(14)=2} (VMD part of photon only) is based on a rescaling of the 
pion distributions, while \ttt{MSTP(14)=3} gives the SaS anomalous 
parametrization.     
\itemc{Note 3:} Formally speaking, the $k_0$ (or $p_0$) cut-off in 
\ttt{PARP(15)} need not be set in any relation to the $Q_0$ cut-off 
scales used by the various parametrizations. Indeed, due to the 
familiar scale choice ambiguity problem, there could well be some offset 
between the two. However, unless you know what you are doing, it is 
strongly recommended that you let the two agree, i.e. set
\ttt{PARP(15)=0.6} for the SaS 1 sets and \ttt{=2.} for the SaS 2 sets.
\end{subentry}
 
\iteme{MSTP(56) :} (D=1) choice of photon parton-distribution-function
library. 
\begin{subentry}
\iteme{= 1 :} the internal {\Py} one, with parton distributions
according to the \ttt{MSTP(55)} above.
\iteme{= 2 :} the \tsc{Pdflib} one \cite{Plo93}, with the
\tsc{Pdflib} (version 4) \ttt{NGROUP} and \ttt{NSET} numbers to be 
given as \ttt{MSTP(55) = 1000}$\times$\ttt{NGROUP + NSET}.
When the VMD and anomalous parts of the photon are split,
like for \ttt{MSTP(14)=10}, it is necessary to specify pion set to be
used for the VMD component, in \ttt{MSTP(53)} and \ttt{MSTP(54)},
while \ttt{MSTP(55)} here is irrelevant.
\iteme{= 3 :} when the parton distributions of the anomalous photon
are requested, the homogeneous solution is provided, evolved from a 
starting value \ttt{PARP(15)} to the requested $Q$ scale. The homogeneous 
solution is normalized so that the net momentum is unity,
i.e. any factors of $\alphaem/2\pi$ and charge have been left out. 
The flavour of the original $\q$ is given in \ttt{MSTP(55)} (1, 2, 3, 4 
or 5 for $\d$, $\u$, $\s$, $\c$ or $\b$); the value 0 gives a mixture 
according to squared charge, with the exception that $\c$ and $\b$ 
are only allowed above the respective mass threshold ($Q > m_{\q}$).
The four-flavour $\Lambda$ value is assumed given in \ttt{PARP(1)};
it is automatically recalculated for 3 or 5 flavours at 
thresholds. This option is not intended for standard event 
generation, but is useful for some theoretical studies.
\itemc{Note:} to make use of option 2, it is necessary to link 
\tsc{Pdflib}. Additionally, on most computers, the two dummy routines
\ttt{PDFSET} and \ttt{STRUCTM} at the end of the {\Py} file should be
removed or commented out.
\itemc{Warning:} For external parton-distribution libraries, {\Py}
does not check whether \ttt{MSTP(55)} corresponds to a valid code,
or if special $x$ and $Q^2$ restrictions exist for a given set,
such that crazy values could be returned. This puts an extra
responsibility on you.
\end{subentry}
 
\iteme{MSTP(57) :} (D=1) choice of $Q^2$ dependence in 
parton-distribution functions. 
\begin{subentry}
\iteme{= 0 :} parton distributions are evaluated at nominal lower
cut-off value $Q_0^2$, i.e. are made $Q^2$-independent.
\iteme{= 1 :} the parametrized $Q^2$ dependence is used.
\iteme{= 2 :} the parametrized parton-distribution behaviour is kept
at large $Q^2$ and $x$, but modified at small $Q^2$ and/or $x$,
so that parton distributions vanish in the limit $Q^2 \to 0$ and
have a theoretically motivated small-$x$ shape \cite{Sch93a}.
This option is only valid for the $\p$ and $\n$.
\iteme{= 3 :} as \ttt{=2}, except that also the $\pi^{\pm}$ and the
VMD component of a photon is modified in a corresponding manner.
\end{subentry}
 
\iteme{MSTP(58) :} (D=min(6, 2$\times$\ttt{MSTP(1)})) maximum number of
quark flavours used in parton distributions, and thus also for 
initial-state space-like showers. If some distributions (notably $\t$) 
are absent in the parametrization selected in \ttt{MSTP(51)}, these 
are obviously automatically excluded.
 
\iteme{MSTP(61) :} (D=1) (C) master switch for initial-state QCD and
QED radiation.
\begin{subentry}
\iteme{= 0 :} off.
\iteme{= 1 :} on.
\end{subentry}
 
\iteme{MSTP(62) - MSTP(67) :} (C) further switches for initial-state 
radiation, see section \ref{ss:showrout}.
 
\iteme{MSTP(71) :} (D=1) (C) master switch for final-state QCD and
QED radiation.
\begin{subentry}
\iteme{= 0 :} off.
\iteme{= 1 :} on.
\itemc{Note:} additional switches (e.g. for conventional/coherent
showers) are available in \ttt{MSTJ(40) - MSTJ(50)} and
\ttt{PARJ(81) - PARJ(89)}, see section \ref{ss:showrout}.
\end{subentry}
 
\iteme{MSTP(81) :} (D=1) master switch for multiple interactions.
\begin{subentry}
\iteme{= 0 :} off.
\iteme{= 1 :} on.
\end{subentry}

\iteme{MSTP(82) - MSTP(83) :} further switches for multiple
interactions, see section \ref{ss:multintpar}.

\iteme{MSTP(85) :} possibility to dampen hard scattering matrix 
elements in the limit $\pT \to 0$. It parellels some of the 
multiple interactions options for QCD processes, but can be 
used for any $2 \to 2$ process.
\begin{subentry}
\iteme{= 0 :} off, i.e. standard matrix elements are kept.
\iteme{= 1 :} on, i.e. matrix elements are multiplied by a 
factor $\pT^4/(\pT^2 + p_{\perp 0}^2)^2$, where $p_{\perp 0}$
is given by \ttt{PARP(82)}. Additionally $\alphas$ is evaluated 
at a scale $\pT^2 + p_{\perp 0}^2$ rather than just $\pT^2$. 
\end{subentry}

\iteme{MSTP(91) - MSTP(94) :} switches for beam remnant treatment,
see section \ref{ss:multintpar}.
 
\iteme{MSTP(101) :} (D=3) (C) structure of diffractive system.
\begin{subentry}
\iteme{= 1 :} forward moving diquark + interacting quark.
\iteme{= 2 :} forward moving diquark + quark joined via interacting
gluon (`hairpin' configuration).
\iteme{= 3 :} a mixture of the two options above, with a fraction
\ttt{PARP(101)} of the former type.
\end{subentry}

\iteme{MSTP(102) :} (D=1) (C) decay of a $\rho^0$ meson produced by
`elastic' scattering of an incoming $\gamma$, as in
$\gamma \p \to \rho^0 \p$, or the same with the hadron diffractively 
excited.
\begin{subentry}
\iteme{= 0 :} the $\rho^0$ is allowed to decay isotropically, like 
any other $\rho^0$.
\iteme{= 1 :} the decay $\rho^0 \to \pi^+ \pi^-$ is done with an
angular distribution proportional to $\sin^2 \theta$ in its rest frame,
where the $z$ axis is given by the direction of motion of the 
$\rho^0$. The $\rho^0$ decay is then done as part of the hard process, 
i.e. also when \ttt{MSTP(111)=0}.
\end{subentry}
 
\iteme{MSTP(111) :} (D=1) (C) master switch for fragmentation
and decay, as obtained with a \ttt{LUEXEC} call.
\begin{subentry}
\iteme{= 0 :} off.
\iteme{= 1 :} on.
\iteme{= -1 :} only choose kinematical variables for hard scattering,
i.e. no jets are defined. This is useful, for instance, to calculate
cross sections (by Monte Carlo integration) without wanting
to simulate events; information obtained with \ttt{PYSTAT(1)}
will be correct.
\end{subentry}
 
\iteme{MSTP(112) :} (D=1) (C) cuts on partonic events; only affects
an exceedingly tiny fraction of events.
\begin{subentry}
\iteme{= 0 :} no cuts (can be used only with independent
fragmentation, at least in principle).
\iteme{= 1 :} string cuts (as normally required for fragmentation).
\end{subentry}
 
\iteme{MSTP(113) :} (D=1) (C) recalculation of energies of partons
from their momenta and masses, to be done immediately before
and after fragmentation, to partly compensate for some numerical 
problems appearing at high energies.
\begin{subentry}
\iteme{= 0 :} not performed.
\iteme{= 1 :} performed.
\end{subentry}
 
\iteme{MSTP(121) :} (D=0) calculation of kinematics selection
coefficients and differential cross section maxima for
included (by user or default) subprocesses.
\begin{subentry}
\iteme{= 0 :} not known; to be calculated at initialization.
\iteme{= 1 :} not known; to be calculated at initialization;
however, the maximum value then obtained is to be multiplied by
\ttt{PARP(121)} (this may be useful if a violation factor has
been observed in a previous run of the same kind).
\iteme{= 2 :} known; kinematics selection coefficients stored
by user in \ttt{COEF(ISUB,J)} (\ttt{J} = 1--20) in common block
\ttt{PYINT2} and maximum of the corresponding differential
cross section times Jacobians in \ttt{XSEC(ISUB,1)} in
common block \ttt{PYINT5}. This is to be done for each included
subprocess ISUB before initialization, with the sum of all
\ttt{XSEC(ISUB,1)} values, except for ISUB = 95, stored in
\ttt{XSEC(0,1)}.
\end{subentry}
 
\iteme{MSTP(122) :} (D=1) initialization and differential
cross section maximization print-out.
\begin{subentry}
\iteme{= 0 :} none.
\iteme{= 1 :} short message.
\iteme{= 2 :} detailed message, including full maximization.
\end{subentry}
 
\iteme{MSTP(123) :} (D=2) reaction to violation of maximum
differential cross section.
\begin{subentry}
\iteme{= 0 :} stop generation, print message.
\iteme{= 1 :} continue generation, print message for each
subsequently larger violation.
\iteme{= 2 :} as \ttt{=1}, but also increase value of maximum.
\end{subentry}
 
\iteme{MSTP(124) :} (D=1) (C) frame for presentation of event.
\begin{subentry}
\iteme{= 1 :} as specified in \ttt{PYINIT}.
\iteme{= 2 :} c.m. frame of incoming particles.
\end{subentry}
 
\iteme{MSTP(125) :} (D=1) (C) documentation of partonic process,
see section \ref{sss:PYrecord} for details.
\begin{subentry}
\iteme{= 0 :} only list ultimate string/particle configuration.
\iteme{= 1 :} additionally list short summary of the hard process.
\iteme{= 2 :} list complete documentation of intermediate steps of
parton-shower evolution.
\end{subentry}
 
\iteme{MSTP(126) :} (D=20) number of lines at the beginning of event
record that are reserved for event-history information; see section
\ref{sss:PYrecord}. This value should never be reduced, but may be
increased at a later date if more complicated processes are included.
 
\iteme{MSTP(128) :} (D=0) storing of copy of resonance decay
products in the documentation section of the event record, and
mother pointer (\ttt{K(I,3)}) relation of the actual resonance
decay products (stored in the main section of the event record)
to the documentation copy.
\begin{subentry}
\iteme{= 0 :} products are stored also in the documentation section,
and each product stored in the main section points back
to the corresponding entry in the documentation section.
\iteme{= 1 :} products are stored also in the documentation section,
but the products stored in the main section point back to
the decaying resonance copy in the main section.
\iteme{= 2 :} products are not stored in the documentation section;
the products stored in the main section point back to the
decaying resonance copy in the main section.
\end{subentry}
 
\iteme{MSTP(129) :} (D=10) for the maximization of $2 \to 3$ processes
(\ttt{ISET(ISUB)=5}) each phase-space point in $\tau$, $y$ and $\tau'$
is tested \ttt{MSTP(129)} times in the other dimensions (at randomly
selected points) to determine the effective maximum in the
($\tau$, $y$, $\tau'$) point.
 
\iteme{MSTP(131) :} (D=0) master switch for pile-up events, i.e. several
independent hadron--hadron interactions generated in the same
bunch--bunch crossing, with the events following one after the
other in the event record.
\begin{subentry}
\iteme{= 0 :} off, i.e. only one event is generated at a time.
\iteme{= 1 :} on, i.e. several events are allowed in the same event
record. Information on the processes generated may be found in
\ttt{MSTI(41) - MSTI(50)}.
\end{subentry}

\iteme{MSTP(132) - MSTP(134) :} further switches for pile-up events,
see section \ref{ss:multintpar}.
 
\iteme{MSTP(141) :} (D=0) calling of \ttt{PYKCUT} in the 
event-generation chain, for inclusion of user-specified cuts.
\begin{subentry}
\iteme{= 0 :} not called.
\iteme{= 1 :} called.
\end{subentry}
 
\iteme{MSTP(142) :} (D=0) calling of \ttt{PYEVWT} in the 
event-generation chain, either to give weighted events or to modify
standard cross sections. See \ttt{PYEVWT} description in section
\ref{ss:PYTmainroutines} for further details.
\begin{subentry}
\iteme{= 0 :} not called.
\iteme{= 1 :} called; the distribution of events among subprocesses
and in kinematics variables is modified by the factor \ttt{WTXS},
set by the user in the \ttt{PYEVWT} call, but events come with a
compensating weight \ttt{PARI(10)=1./WTXS}, such that total
cross sections are unchanged.
\iteme{= 2 :} called; the cross section itself is modified by the
factor \ttt{WTXS}, set by the user in the \ttt{PYEVWT} call.
\end{subentry}

\iteme{MSTP(151) :} (D=0) introduce smeared position of primary vertex
of events.
\begin{subentry}
\iteme{= 0 :} no, i.e. the primary vertex of each event is at the 
origin.
\iteme{= 1 :} yes, with Gaussian distributions separately in $x$, $y$,
$z$ and $t$. The respective widths of the Gaussians have to be given 
in \ttt{PARP(151) - PARP(154)}. Also pile-up events obtain separate
primary vertices. No provisions are made for more complicated 
beam-spot shapes, e.g. with a spread in $z$ that varies as a 
function of $t$. Note that a large beam spot combined with some of the
\ttt{MSTJ(22)} options may lead to many particles not being allowed to
decay at all.
\end{subentry}

\iteme{MSTP(171) :} (D=0) possibility of variable energies from one 
event to the next. For further details see subsection 
\ref{ss:PYvaren}.
\begin{subentry}
\iteme{= 0 :} no; i.e. the energy is fixed at the initialization call.
\iteme{= 1 :} yes; i.e. a new energy has to be given for each new 
event.
\end{subentry}

\iteme{MSTP(172) :} (D=2) options for generation of events with 
variable energies, applicable when \ttt{MSTP(171)=1}. 
\begin{subentry}
\iteme{= 1 :} an event is generated at the requested energy, i.e. 
internally a loop is performed over possible event configurations 
until one is accepted. If the requested c.m. energy of an event 
is below \ttt{PARP(2)} the run is aborted. Cross-section information 
can not be trusted with this option, since it depends on how you 
decided to pick the requested energies.
\iteme{= 2 :} only one event configuration is tried. If that is 
accepted, the event is generated in full. If not, no event is
generated, and the status code \ttt{MSTI(61)=1} is returned.
You are then expected to give a new energy, looping until an
acceptable event is found. No event is generated if the 
requested c.m. energy is below \ttt{PARP(2)}, instead
\ttt{MSTI(61)=1} is set to signal the failure. In principle,
cross sections should come out correctly with this option. 
\end{subentry}

\iteme{MSTP(173) :} (D=0) possibility for user to give in an event 
weight to compensate for a biased choice of beam spectrum.
\begin{subentry}
\iteme{= 0 :} no, i.e. event weight is unity.
\iteme{= 1 :} yes; weight to be given for each event in 
\ttt{PARP(173)}, with maximum weight given at initialization
in \ttt{PARP(174)}.
\end{subentry}
 
\iteme{MSTP(181) :} (R) {\Py} version number.
 
\iteme{MSTP(182) :} (R) {\Py} subversion number.
 
\iteme{MSTP(183) :} (R) last year of change for {\Py}.
 
\iteme{MSTP(184) :} (R) last month of change for {\Py}.
 
\iteme{MSTP(185) :} (R) last day of change for {\Py}.

\iteme{MSTP(186) :} (R) earliest subversion of {\Je} version 7
with which this {\Py} subversion can be run. 

\boxsep
 
\iteme{PARP(1) :}\label{p:PARP} (D=0.25 GeV) nominal 
$\Lambda_{\mrm{QCD}}$ used in running $\alphas$ for hard 
scattering (see \ttt{MSTP(3)}).
 
\iteme{PARP(2) :} (D=10. GeV) lowest c.m. energy for the
event as a whole that the program will accept to simulate.
 
\iteme{PARP(13) :} (D=1. GeV$^2$) $Q_{\mmax}^2$ scale, to be set by
user for defining maximum scale allowed for photoproduction when
using the option \ttt{MSTP(13)=2}.
 
\iteme{PARP(14) :} (D=0.01) in the numerical integration of quark
and gluon parton distributions inside an electron, the successive
halvings of evaluation-point spacing is interrupted when two values
agree in relative size, $|$new$-$old$|$/(new$+$old), to better than
\ttt{PARP(14)}. There are hardwired lower and upper limits of 2 and
8 halvings, respectively.

\iteme{PARP(15) :} (D=0.5 GeV) lower cut-off $p_0$ used to define
minimum transverse momentum in branchings $\gamma \to \q\qbar$ in
the anomalous event class of $\gamma\p$ interactions. 

\iteme{PARP(16) :} (D=1.) the anomalous parton-distribution functions 
of the photon are taken to have the charm and bottom flavour thresholds 
at virtuality \ttt{PARP(16)}$\times m_{\q}^2$.

\iteme{PARP(17) :} (D=1.) rescaling factor used for the $Q$ argument 
of the anomalous parton distributions of the photon, see 
\ttt{MSTP(15)}.
 
\iteme{PARP(31) :} (D=1.5) common $K$ factor multiplying the
differential cross section for hard parton--parton processes
when \ttt{MSTP(33)=1} or \ttt{2}, with the exception of colour
annihilation graphs in the latter case.
 
\iteme{PARP(32) :} (D=2.0) special $K$ factor multiplying the
differential cross section in hard colour annihilation graphs,
including resonance production, when \ttt{MSTP(33)=2}.
 
\iteme{PARP(33) :} (D=0.075) this factor is used to multiply the
ordinary $Q^2$ scale in $\alphas$ at the hard interaction for
\ttt{MSTP(33)=3}. The effective $K$ factor thus obtained is in
accordance with the results in \cite{Ell86}.

\iteme{PARP(34) :} (D=1.) the $Q^2$ scale defined by \ttt{MSTP(32)} is 
multiplied by \ttt{PARP(34)} when it is used as argument for parton
distributions and $\alphas$ at the hard interaction. It does not affect 
$\alphas$ when \ttt{MSTP(33)=3}, nor does it change the $Q^2$ argument 
of parton showers.
 
\iteme{PARP(35) :} (D=0.20) fix $\alphas$ value that is used in 
the heavy-flavour threshold factor when \ttt{MSTP(35)=1}.
 
\iteme{PARP(36) :} (D=0. GeV) the width $\Gamma_{\Q}$ for the heavy
flavour studied in processes ISUB = 81 or 82; to be used for the
threshold factor when \ttt{MSTP(35)=2}.
 
\iteme{PARP(37) :} (D=2.) for \ttt{MSTP(37)=1} this regulates the
point at which the reference on-shell quark mass in Higgs couplings is
assumed defined; specifically the running quark mass is assumed
to coincide with the fix one at an energy scale \ttt{PARP(37)} times 
the fix quark mass,
i.e. $m_{\mrm{running}}($\ttt{PARP(37)}$\times m_{\mrm{fix}}) = 
m_{\mrm{fix}}$.
 
\iteme{PARP(38) :} (D=0.70 GeV$^3$) the squared wave function at the
origin, $|R(0)|^2$, of the $\Jpsi$ wave function. Used for process
86. See ref. \cite{Glo88}.
 
\iteme{PARP(39) :} (D=0.006 GeV$^3$) the squared derivative of the
wave function at the origin, $|R'(0)|^2/m^2$, of the $\chi_{\c}$
wave functions. Used for the processes 87, 88 and 89. See ref.
\cite{Glo88}.
 
\iteme{PARP(41) :} (D=0.020 GeV) in the process of generating mass
for resonances, and optionally to force that mass to be in a given
range, only resonances with a total width in excess of \ttt{PARP(41)}
are generated according to a Breit--Wigner shape (if allowed by
\ttt{MSTP(42)}), while narrower resonances are put on the mass
shell.
 
\iteme{PARP(42) :} (D=2. GeV) minimum mass of resonances assumed
to be allowed when evaluating total width of $\H^0$ to $\Z^0 \Z^0$ or
$\W^+ \W^-$ for cases when the $\H^0$ is so light that (at least)
one $\Z/\W$ is forced to be off the mass shell. Also generally used
as safety check on minimum mass of resonance. Note that some 
\ttt{CKIN} values may provide additional constraints. 
 
\iteme{PARP(43) :} (D=0.10) precision parameter used in numerical
integration of width into channel with at least one daughter off
the mass shell.
 
\iteme{PARP(44) :} (D=1000.) the $\nu$ parameter of the strongly
interacting $\Z/\W$ model of Dobado, Herrero and Terron \cite{Dob91}.
 
\iteme{PARP(45) :} (D=2054. GeV) the effective techni-$\rho$ mass
parameter of the strongly interacting model of Dobado, Herrero and
Terron \cite{Dob91}; see \ttt{MSTP(46)=5}. On physical grounds it
should not be chosen smaller than about 1 TeV or larger than
about the default value.

\iteme{PARP(46) :} (D=123. GeV) the $F_{\pi}$ decay constant that
appears inversely quadratically in all techni-$\eta$ partial decay
widths \cite{Eic84,App92}.

\iteme{PARP(47) :} (D=246. GeV) vacuum expectation value $v$ used 
in the DHT scenario \cite{Dob91} to define the width of the 
techni-$\rho$; this width is inversely proportional $v^2$.
 
\iteme{PARP(51) :} (D=1.) if parton distributions for light flavours
have to be extrapolated to $x$ values lower than covered by the
parametrizations, an $x^{-b}$ behaviour, with $b =$\ttt{PARP(51)},
is assumed in that region. This option only applies for the EHLQ
proton parton distributions that are internal to {\Py}.
 
\iteme{PARP(61) - PARP(65) :} (C) parameters for initial-state 
radiation, see section \ref{ss:showrout}.
 
\iteme{PARP(71) - PARP(72) :} (C) parameter for final-state 
radiation, see section \ref{ss:showrout}.

\iteme{PARP(81) - PARP(88) :} parameters for multiple interactions,
see section \ref{ss:multintpar}.

\iteme{PARP(91) - PARP(100) :} parameters for beam remnant
treatment, see section \ref{ss:multintpar}.

\iteme{PARP(101) :} (D=0.50) fraction of diffractive systems in which
a quark is assumed kicked out by the pomeron rather than a gluon;
applicable for option \ttt{MSTP(101)=3}.
 
\iteme{PARP(102) :} (D=0.28 GeV) the mass spectrum of diffractive 
states (in single and double diffractive scattering) is assumed to
start \ttt{PARP(102)} above the mass of the particle that is 
diffractively excited. In this connection, an incoming $\gamma$
is taken to have the selected VMD meson mass, i.e. $m_{\rho}$,
$m_{\omega}$, $m_{\phi}$ or $m_{\Jpsi}$.

\iteme{PARP(103) :} (D=1.0 GeV) if the mass of a diffractive state 
is less than \ttt{PARP(103)} above the mass of the particle that is
diffractively excited, the state is forced to decay isotropically
into a two-body channel. In this connection, an incoming $\gamma$
is taken to have the selected VMD meson mass, i.e. $m_{\rho}$,
$m_{\omega}$, $m_{\phi}$ or $m_{\Jpsi}$. If the mass is higher than 
this threshold, the standard string fragmentation machinery is used. 
The forced two-body decay is always carried out, also when 
\ttt{MSTP(111)=0}. 
 
\iteme{PARP(111) :} (D=2. GeV) used to define the minimum invariant
mass of the remnant hadronic system (i.e. when interacting partons
have been taken away), together with original hadron masses and
extra parton masses.
 
\iteme{PARP(121) :} (D=1.) the maxima obtained at initial
maximization are multiplied by this factor if \ttt{MSTP(121)=1};
typically \ttt{PARP(121)} would be given as the product of the
violation factors observed (i.e. the ratio of final maximum value
to initial maximum value) for the given process(es).
 
\iteme{PARP(122) :} (D=0.4) fraction of total probability that is
shared democratically between the \ttt{COEF} coefficients open for
the given variable, with the remaining fraction distributed according
to the optimization results of \ttt{PYMAXI}.

\iteme{PARP(131) :} parameter for pile-up events, see section
\ref{ss:multintpar}. 

\iteme{PARP(151) - PARP(154) :} (D=4*0.) (C) regulate the assumed
beam-spot size. For \ttt{MSTP(151)=1} the $x$, $y$, $z$ and $t$ 
coordinates of the primary vertex of each event are selected 
according to four independent Gaussians. The widths of these 
Gaussians are given by the four parameters, where the first three 
are in units of mm and the fourth in mm/$c$. 

\iteme{PARP(161) - PARP(164) :} (D=2.20, 23.6, 18.4, 11.5) couplings 
$f_V^2/4\pi$ of the photon to the $\rho^0$, $\omega$, $\phi$ and
$\Jpsi$ vector mesons.

\iteme{PARP(171) :} to be set, event-by-event, when variable 
energies are allowed, i.e. when \ttt{MSTP(171)=1}. If \ttt{PYINIT} is 
called with \ttt{FRAME='CMS'} (\ttt{='FIXT'}), \ttt{PARP(171)} 
multiplies the c.m. energy (beam energy) used at initialization. 
For the options \ttt{'USER'}, \ttt{'FOUR'} and \ttt{'FIVE'}, 
\ttt{PARP(171)} is dummy, since there the momenta are set in the
\ttt{P} array.

\iteme{PARP(173) :} event weight to be given by user when 
\ttt{MSTP(173)=1}.  

\iteme{PARP(174) :} (D=1.) maximum event weight that will be 
encountered in \ttt{PARP(173)} during the course of a run with 
\ttt{MSTP(173)=1}; to be used to optimize the efficiency of the 
event generation. It is always allowed to use a larger bound than 
the true one, but with a corresponding loss in efficiency.
 
\end{entry}
 
\subsection{General Event Information}
 
When an event is generated with \ttt{PYEVNT}, some information on
it is stored in the \ttt{MSTI} and \ttt{PARI} arrays of the
\ttt{PYPARS} common block (often copied directly from the internal
\ttt{MINT} and \ttt{VINT} variables). Further information is stored
in the complete event record; see section \ref{ss:evrec}.
 
Part of the information is only relevant for some subprocesses; by
default everything irrelevant is set to 0. Kindly note that, like the
\ttt{CKIN} constraints described in section \ref{ss:PYswitchkin},
kinematical variables normally (i.e. where it is not explicitly stated
otherwise) refer to the na\"{\i}ve hard scattering, before initial- and
final-state radiation effects have been included.
 
\drawbox{COMMON/PYPARS/MSTP(200),PARP(200),MSTI(200),PARI(200)}%
\label{p:PYPARS2}
\begin{entry}
\itemc{Purpose:} to provide information on latest event generated or,
in a few cases, on statistics accumulated during the run.
 
\iteme{MSTI(1) :}\label{p:MSTI} specifies the general type of 
subprocess that has occurred, according to the ISUB code given 
in section \ref{ss:ISUBcode}.
 
\iteme{MSTI(2) :} whenever \ttt{MSTI(1)} (together with \ttt{MSTI(15)}
and \ttt{MSTI(16)}) are not enough to specify the type of process
uniquely, \ttt{MSTI(2)} provides an ordering of the different
possibilities. This is particularly relevant for the different
colour-flow topologies possible in QCD $2 \to 2$ processes. With
$i = $\ttt{MSTI(15)}, $j = $\ttt{MSTI(16)} and $k = $\ttt{MSTI(2)},
the QCD possibilities are, in the classification scheme of
\cite{Ben84} (cf. section \ref{sss:QCDjetclass}):
\begin{subentry}
\itemn{ISUB = 11,} $i = j$, $\q_i \q_i \to \q_i \q_i$; \\
$k = 1$ : colour configuration $A$. \\
$k = 2$ : colour configuration $B$.
\itemn{ISUB = 11,} $i \neq j$, $\q_i \q_j \to \q_i \q_j$; \\
$k = 1$ : only possibility.
\itemn{ISUB = 12,} $\q_i \qbar_i \to \q_l \qbar_l$; \\
$k = 1$ : only possibility.
\itemn{ISUB = 13,} $\q_i \qbar_i \to \g \g$; \\
$k = 1$ : colour configuration $A$. \\
$k = 2$ : colour configuration $B$.
\itemn{ISUB = 28,} $\q_i \g \to \q_i \g$; \\
$k = 1$ : colour configuration $A$. \\
$k = 2$ : colour configuration $B$.
\itemn{ISUB = 53,} $\g \g \to \q_l \qbar_l$; \\
$k = 1$ : colour configuration $A$. \\
$k = 2$ : colour configuration $B$.
\itemn{ISUB = 68,} $\g \g \to \g \g$; \\
$k = 1$ : colour configuration $A$. \\
$k = 2$ : colour configuration $B$. \\
$k = 3$ : colour configuration $C$.
\itemn{ISUB = 83,} $\f \q \to \f' \Q$ (by $t$-channel $\W$ exchange;
does not distinguish colour flows but result of user selection); \\
$k = 1$ : heavy flavour $\Q$ is produced on side 1. \\
$k = 2$ : heavy flavour $\Q$ is produced on side 2.
\end{subentry}
 
\iteme{MSTI(3) :} the number of partons produced in the hard 
interactions, i.e. the number $n$ of the $2 \to n$ matrix elements 
used; it is sometimes 3 or 4 when a basic $2 \to 1$ or $2 \to 2$ 
process has been folded with two $1 \to 2$ initial branchings (like
$\q \q' \to \q'' \q''' \H^0$).
 
\iteme{MSTI(4) :} number of documentation lines at the beginning of
the common block \ttt{LUJETS} that are given with \ttt{K(I,1)=21};
0 for \ttt{MSTP(125)=0}.
 
\iteme{MSTI(5) :} number of events generated to date in current
run. In runs with the variable-energy option, \ttt{MSTP(171)=1}
and \ttt{MSTP(172)=2}, only those events that survive (i.e. that
do not have \ttt{MSTI(61)=1}) are counted in this number. That
is, \ttt{MSTI(5)} may be less than the total number of \ttt{PYEVNT}
calls.
 
\iteme{MSTI(6) :} current frame of event, cf. \ttt{MSTP(124)}.
 
\iteme{MSTI(7), MSTI(8) :} line number for documentation of
outgoing partons/particles from hard scattering for $2 \to 2$ or
$2 \to 1 \to 2$ processes (else = 0).

\iteme{MSTI(9) :} event class used in current event for $\gamma\p$ or 
$\gamma\gamma$ events generated with the \ttt{MSTP(14)=10} option.
\begin{subentry}
\iteme{= 0 :} for other processes than the ones listed above.
\iteme{= 1 :} VMD (for $\gamma\p$) or VMD*VMD (for $\gamma\gamma$).
\iteme{= 2 :} direct (for $\gamma\p$) or VMD*direct (for 
$\gamma\gamma$).
\iteme{= 3 :} anomalous (for $\gamma\p$) or VMD*anomalous (for 
$\gamma\gamma$).
\iteme{= 4 :} direct*direct (for $\gamma\gamma$).
\iteme{= 5 :} direct*anomalous (for $\gamma\gamma$).
\iteme{= 6 :} anomalous*anomalous (for $\gamma\gamma$).
\end{subentry}
 
\iteme{MSTI(10) :} is 1 if cross section maximum was violated
in current event, and 0 if not.
 
\iteme{MSTI(11) :} KF flavour code for beam (side 1) particle.
 
\iteme{MSTI(12) :} KF flavour code for target (side 2) particle.
 
\iteme{MSTI(13), MSTI(14) :} KF flavour codes for side 1 and side 2
initial-state shower initiators.
 
\iteme{MSTI(15), MSTI(16) :} KF flavour codes for side 1 and side 2
incoming partons to the hard interaction.
 
\iteme{MSTI(17), MSTI(18) :} flag to signal if particle on side
1 or side 2 has been scattered diffractively; 0 if no, 1 if yes.
 
\iteme{MSTI(21) - MSTI(24) :} KF flavour codes for outgoing partons
from the hard interaction. The number of positions actually used is
process-dependent, see \ttt{MSTI(3)}; trailing positions not used
are set = 0.
 
\iteme{MSTI(25), MSTI(26) :} KF flavour codes of the products in the
decay of a single $s$-channel resonance formed in the hard interaction.
Are thus only used when \ttt{MSTI(3)=1} and the resonance is allowed
to decay.
 
\iteme{MSTI(31) :} number of hard or semi-hard scatterings that occurred
in the current event in the multiple-interaction scenario; is = 0 for a
low-$\pT$ event.
 
\iteme{MSTI(41) :} the number of pile-up events generated in the latest
\ttt{PYEVNT} call (including the first, `hard' event).
 
\iteme{MSTI(42) - MSTI(50) :} ISUB codes for the events 2--10
generated in the pile-up-events scenario. The first event ISUB code is
stored in \ttt{MSTI(1)}. If \ttt{MSTI(41)} is less than 10, only as
many positions are filled as there are pile-up events. If MSTI(41) is
above 10, some ISUB codes will not appear anywhere.

\iteme{MSTI(51) :} normally 0 but set to 1 if a \ttt{PYUPEV} call
did not return an event, such that \ttt{PYEVNT} could not generate
an event. For further details, see end of section \ref{ss:PYnewproc}.

\iteme{MSTI(52) :} counter for the number of times the current event 
configuration failed in the generation machinery. For accepted events
this is always 0, but the counter can be used inside \ttt{PYUPEV}
to check on anomalous occurrences. For further 
details, see end of section \ref{ss:PYnewproc}.

\iteme{MSTI(61) :} status flag set when events are generated. It is 
only of interest for runs with variable energies, \ttt{MSTP(171)=1}, 
with the option \ttt{MSTP(172)=2}.
\begin{subentry}
\iteme{= 0 :} an event has been generated.
\iteme{= 1 :} no event was generated, either because the c.m. energy 
was too low or because the Monte Carlo phase space point selection 
machinery rejected the trial point. A new energy is to be picked by 
the user.
\end{subentry}

\boxsep
 
\iteme{PARI(1) :}\label{p:PARI} total integrated cross section for 
the processes
under study, in mb. This number is obtained as a by-product of the
selection of hard-process kinematics, and is thus known with better
accuracy when more events have been generated. The value stored here
is based on all events until the latest one generated.
 
\iteme{PARI(2) :} is the ratio \ttt{PARI(1)/MSTI(5)}, i.e. the ratio
of total integrated cross section and number of events generated.
Histograms filled with unit event weight have to be
multiplied by this factor, at the end of the run, to convert
results to mb. For \ttt{MSTP(142)=1}, \ttt{MSTI(5)} is replaced by
the sum of \ttt{PARI(10)} values. Histograms are then filled with
weight \ttt{PARI(10)} for each event and multiplied by \ttt{PARI(2)}
at the end. In runs with the variable-energy option, \ttt{MSTP(171)=1}
and \ttt{MSTP(172)=2}, only those events that survive (i.e. that
do not have \ttt{MSTI(61)=1}) are counted. 
calls.
 
\iteme{PARI(9) :} is weight \ttt{WTXS} returned from \ttt{PYEVWT}
call when \ttt{MSTP(142)}$\geq 1$, otherwise is 1.
 
\iteme{PARI(10) :} is compensating weight \ttt{1./WTXS} that should
be associated to events when \ttt{MSTP(142)=1}, else is 1.
 
\iteme{PARI(11) :} $E_{\mrm{cm}}$, i.e. total c.m. energy.
 
\iteme{PARI(12) :} $s$, i.e. squared total c.m. energy.
 
\iteme{PARI(13) :} $\hat{m} = \sqrt{\hat{s}}$, i.e. mass of the 
hard-scattering subsystem.
 
\iteme{PARI(14) :} $\hat{s}$ of the hard subprocess
($2 \to 2$ or $2 \to 1$).
 
\iteme{PARI(15) :} $\hat{t}$ of the hard subprocess
($2 \to 2$ or $2 \to 1 \to 2$).
 
\iteme{PARI(16) :} $\hat{u}$ of the hard subprocess
($2 \to 2$ or $2 \to 1 \to 2$).
 
\iteme{PARI(17) :} $\hat{p}_{\perp}$ of the hard subprocess
($2 \to 2$ or $2 \to 1 \to 2$),
evaluated in the rest frame of the hard interaction.
 
\iteme{PARI(18) :} $\hat{p}_{\perp}^2$ of the hard subprocess;
see \ttt{PARI(17)}.
 
\iteme{PARI(19) :} $\hat{m}'$, the mass of the complete three- or
four-body final state in $2 \to 3$ or $2 \to 4$ processes (while
$\hat{m}$, given in \ttt{PARI(13)}, here corresponds to the one-
or two-body central system).
Kinematically $\hat{m} \leq \hat{m}' \leq E_{\mrm{cm}}$.
 
\iteme{PARI(20) :} $\hat{s}' = \hat{m}'^2$; see \ttt{PARI(19)}.
 
\iteme{PARI(21) :} $Q$ of the hard-scattering subprocess. The exact 
definition is process-dependent, see \ttt{MSTP(32)}.
 
\iteme{PARI(22) :} $Q^2$ of the hard-scattering subprocess; see 
\ttt{PARI(21)}.
 
\iteme{PARI(23) :} $Q$ of the outer hard-scattering subprocess.
Agrees with \ttt{PARI(21)} for a $2 \to 1$ or $2 \to 2$ process.
For a $2 \to 3$ or $2 \to 4$ $\W/\Z$ fusion process, it is set by
the $\W/\Z$ mass scale, and for subprocesses 121 and 122 by the
heavy-quark mass.
 
\iteme{PARI(24) :} $Q^2$ of the outer hard-scattering subprocess;
see \ttt{PARI(23)}.
 
\iteme{PARI(25) :} $Q$ scale used as maximum virtuality in parton
showers. Is equal to \ttt{PARI(23)}, except for 
deep-inelastic-scattering processes when \ttt{MSTP(22)}$\geq 1$.
 
\iteme{PARI(26) :} $Q^2$ scale in parton showers; see \ttt{PARI(25)}.
 
\iteme{PARI(31), PARI(32) :} the momentum fractions $x$ of the
initial-state parton-shower initiators on side 1 and 2, respectively.
 
\iteme{PARI(33), PARI(34) :} the momentum fractions $x$ taken by the
partons at the hard interaction, as used e.g. in the 
parton-distribution functions.
 
\iteme{PARI(35) :} Feynman-$x$,
$x_{\mrm{F}} = x_1 - x_2 = $\ttt{PARI(33)}$-$\ttt{PARI(34)}.
 
\iteme{PARI(36) :}
$\tau = \hat{s}/s = x_1 \, x_2 = $\ttt{PARI(33)}$\times$\ttt{PARI(34)}.
 
\iteme{PARI(37) :} $y = (1/2) \ln(x_1/x_2)$, i.e. rapidity of the
hard-interaction subsystem in the c.m. frame of the event as a whole.
 
\iteme{PARI(38)  :} $\tau' = \hat{s}'/s = $\ttt{PARI(20)/PARI(12)}.
 
\iteme{PARI(39), PARI(40) :} the primordial $k_{\perp}$ values
selected in the two beam remnants.
 
\iteme{PARI(41) :} $\cos\hat{\theta}$, where $\hat{\theta}$ is the
scattering angle of a $2 \to 2$ (or $2 \to 1 \to 2$) interaction,
defined in the rest frame of the hard-scattering subsystem.
 
\iteme{PARI(42) :} $x_{\perp}$, i.e. scaled transverse momentum of the
hard-scattering subprocess, 
$x_{\perp} = 2 \hat{p}_{\perp}/E_{\mrm{cm}}$.
 
\iteme{PARI(43), PARI(44) :} $x_{L3}$ and $x_{L4}$, i.e. longitudinal
momentum fractions of the two scattered partons, in the range
$-1 < x_{\mrm{L}} < 1$, in the c.m. frame of the event as a whole.
 
\iteme{PARI(45), PARI(46) :} $x_3$ and $x_4$, i.e. scaled energy
fractions of the two scattered partons, in the c.m. frame of the
event as a whole.
 
\iteme{PARI(47), PARI(48) :} $y^*_3$ and $y^*_4$, i.e. rapidities
of the two scattered partons in the c.m. frame of the
event as a whole.
 
\iteme{PARI(49), PARI(50) :} $\eta^*_3$ and $\eta^*_4$, i.e.
pseudorapidities of the two scattered partons in the c.m. frame
of the event as a whole.
 
\iteme{PARI(51), PARI(52) :} $\cos\theta^*_3$ and $\cos\theta^*_4$,
i.e. cosines of the polar angles of the two scattered partons in
the c.m. frame of the event as a whole.
 
\iteme{PARI(53), PARI(54) :} $\theta^*_3$ and $\theta^*_4$, i.e.
polar angles of the two scattered partons, defined in the range
$0 < \theta^* < \pi$, in the c.m. frame of the event as a whole.
 
\iteme{PARI(55), PARI(56) :} azimuthal angles $\phi^*_3$ and
$\phi^*_4$ of the two scattered partons, defined in the range
$-\pi < \phi^* < \pi$, in the c.m. frame of the event as a whole.
 
\iteme{PARI(61) :} multiple interaction enhancement factor for
current event. A large value corresponds to a central collision
and a small value to a peripheral one.
 
\iteme{PARI(65) :} sum of the transverse momenta of partons
generated at the hardest interaction of the event, excluding
initial- and final-state radiation, i.e. $2 \times$\ttt{PARI(17)}.
 
\iteme{PARI(66) :} sum of the transverse momenta of all partons
generated at the hardest interaction, including initial- and 
final-state radiation, resonance decay products, and primordial
$k_{\perp}$.
 
\iteme{PARI(67) :} sum of transverse momenta of partons generated
at hard interactions, excluding the hardest one (see \ttt{PARI(65)}),
and also excluding initial- and final-state radiation. Is non-vanishing
only in the multiple-interaction scenario.
 
\iteme{PARI(68) :} sum of transverse momenta of all partons generated
at hard interactions, excluding the hardest one (see \ttt{PARI(66)}),
but including initial- and final-state radiation. Is non-vanishing only
in the multiple-interaction scenario.
 
\iteme{PARI(69) :} sum of transverse momenta of all partons generated
in hard interactions (\ttt{PARI(66) + PARI(68)}) and, additionally,
of all beam remnant partons.
 
\iteme{PARI(71), PARI(72) :} sum of the momentum fractions $x$ taken
by initial-state parton-shower initiators on side 1 and and side 2,
excluding those of the hardest interaction. Is non-vanishing only in
the multiple-interaction scenario.
 
\iteme{PARI(73), PARI(74) :} sum of the momentum fractions $x$ taken
by the partons at the hard interaction on side 1 and side 2, excluding
those of the hardest interaction. Is non-vanishing only in the
multiple-interaction scenario.
 
\iteme{PARI(75), PARI(76) :} the $x$ value of a photon that branches
into quarks or gluons, i.e. $x$ at interface between initial-state QED
and QCD cascades.
 
\iteme{PARI(77), PARI(78) :} the $\chi$ values selected for beam
remnants that are split into two objects, describing how the energy
is shared (see \ttt{MSTP(92)} and \ttt{MSTP(94)}); is vanishing if 
no splitting is needed.
 
\iteme{PARI(81) :} size of the threshold factor (enhancement or
suppression) in the latest event with heavy-flavour production;
see \ttt{MSTP(35)}.
 
\iteme{PARI(91) :} average multiplicity $\br{n}$ of pile-up events,
see \ttt{MSTP(133)}. Only relevant for \ttt{MSTP(133)=} 1 or 2.
 
\iteme{PARI(92) :} average multiplicity $\langle n \rangle$ of pile-up 
events as actually simulated, i.e. with multiplicity = 0 events 
removed and the high-end tail truncated. Only relevant for 
\ttt{MSTP(133)=} 1 or 2.
 
\iteme{PARI(93) :} for \ttt{MSTP(133)=1} it is the probability that
a beam crossing will produce a pile-up event at all, i.e. that there
will be at least one hadron--hadron interaction; for
\ttt{MSTP(133)=2} the probability that a beam crossing will produce
a pile-up event with one hadron--hadron interaction of the desired rare
type.
\end{entry}
 
\subsection{How to include external processes in PYTHIA}
\label{ss:PYnewproc}
 
Despite a large repertory of processes in {\Py}, the number
of missing ones clearly is even larger, and with time this discrepancy 
is likely to increase. There are several reasons why it is not
practicable to imagine a {\Py} which has `everything'. One is
the amount of time it takes to implement a process for the single
{\Py} author, compared with the rate of new cross section results
produced by the rather larger matrix-element calculations community.
Another is the length of currently produced matrix-element expressions,
which would make the program very bulky. A third argument is that,
whereas the phase space of $2 \to 1$ and $2 \to 2$ processes can be set
up once and for all according to a reasonably flexible machinery,
processes with more final-state particles are less easy to generate.
To achieve a reasonable efficiency, it is necessary to tailor the
phase-space selection procedure to the dynamics of the given process,
and to the desired experimental cuts.
 
If the desired subprocess is missing, it can be included into {\Py}
as an `external' subprocess. In this section we will describe how it
is possible to
specify the partonic state of some hard-scattering process in an
interface common block. {\Py} will read this common block, and add
initial- and final-state showers, beam remnants and underlying events,
fragmentation and decays, to build up an event in as much detail as an
ordinary {\Py} one. You may also use {\Py} to mix events of different
kinds, and to keep track of cross section statistics. You have to
provide the matrix elements, the phase-space generator, and the storage
of event information in the common block.
 
First a minor comment, however. Some processes may be seen just as
trivial modifications of already existing ones. For instance, you might
want to add some extra term, corresponding to contact interactions,
to the matrix elements of a {\Py} $2 \to 2$ process. In that case it is
not necessary to go through the machinery below, but instead you can use
the \ttt{PYEVWT} routine to introduce an additional weight for the 
event, defined as the ratio of the modified to the unmodified 
differential cross sections. If you use the option \ttt{MSTP(142)=2}, 
this weight is considered as part of the `true' cross section of the 
process, and the generation is changed accordingly.
 
The more generic facility for including an external process is a bit
more complicated, and involves two routines and one common block. All
names contain \ttt{UP}, which is short for User Process.
 
If you want to include a new process, first you have to pick an unused
subprocess number ISUB (see tables in section \ref{ss:ISUBcode}). For 
instance, the numbers 191--200 are currently unused, so this might be a 
logical place to put a new process. This number and the `title' of the 
process (plus \ttt{SIGMAX}, to be described below) have to be given in 
to {\Py} in a subroutine call
 
\drawbox{CALL PYUPIN(ISUB,TITLE,SIGMAX)}\label{p:PYUPIN}
\boxsep
 
\noindent
before the call to \ttt{PYINIT}. The \ttt{TITLE} can be any character
string up to 28 characters, e.g.
\begin{verbatim}
      CALL PYUPIN(191,'g + g -> t + tbar + gamma',SIGMAX)
\end{verbatim}
The call to \ttt{PYUPIN} tells the program that a process ISUB exists,
but not that you want to generate it. This is done, as with normal
processes, by setting \ttt{MSUB(ISUB)=1} before the \ttt{PYINIT}
call.
 
Once the event generation chain has been started and \ttt{PYEVNT} is
called to generate an event, this routine may in its turn call the
routine \ttt{PYUPEV}, which is the routine you must supply, in which
the next event is selected. (A dummy copy of \ttt{PYUPEV} has been
included at the end of {\Py}; depending on the machine you may have to
comment out this copy when you link your own.) The call arguments are
 
\drawbox{CALL PYUPEV(ISUB,SIGEV)}\label{p:PYUPEV}
\boxsep
 
\noindent
where \ttt{ISUB} is given by \ttt{PYEVNT}, while \ttt{SIGEV} is to be
calculated (see below) and returned to \ttt{PYEVNT}. If there is only
one user-defined process, then the \ttt{ISUB} input is superfluous;
otherwise it is necessary to branch to the relevant process.
 
The \ttt{SIGEV} variable is supposed to give the differential
cross section ot the current event, times the phase-space volume
within which events are generated, expressed in millibarns. This means
that, in the limit that many events are generated, the average value
of \ttt{SIGEV} gives the total cross section of the simulated process.
The \ttt{SIGMAX} value, handed to {\Py} in the \ttt{PYUPIN} call, is
assumed to be the maximum value that \ttt{SIGEV} will reach. Events
will be accepted with a probability \ttt{SIGEV/SIGMAX}, i.e. the
acceptance/rejection of events according to
differential cross section is done by \ttt{PYEVNT}, not by the user.
This means that the events that come out in the end all have unit
weight, i.e. the user does not have to worry about events with
different weights. It also allows several subprocesses to  be
generated together, in the proper mixture.
 
Of course, the tricky part is that the differential cross section
usually is strongly peaked in a few regions of the phase space, such
that the average probability to accept an event,
$\langle$\ttt{SIGEV}$\rangle /$\ttt{SIGMAX} is
small. It may then be necessary to find a suitable set of transformed
phase-space coordinates, for which the correspondingly transformed
differential cross section is better behaved.
 
To avoid unclarities, here is a more formal version of the two above
paragraphs. Call $\d X$ the differential phase space, e.g. for a 
$2 \to 2$
process $\d X = \d x_1 \, \d x_2 \, \d \hat{t}$, where $x_1$ and $x_2$ 
are the momentum fractions carried by the two incoming partons and
$\hat{t}$ the Mandelstam variable of the scattering. Call
$\d \sigma / \d X$ the differential cross section
of the process, e.g. for $2 \to 2$: $\d \sigma / \d X = \sum_{ij}
f_i(x_1,Q^2) \, f_j(x_2,Q^2) \, \d \hat{\sigma}_{ij} / \d \hat{t}$,
i.e. the product of parton distributions and hard-scattering matrix
elements, summed over all allowed incoming flavours $i$ and $j$.
The physical cross section that one then wants to generate is
$\sigma = \int (\d \sigma / \d X) \, \d X$, where the integral is over 
the allowed phase-space volume. The event generation procedure consists
of selecting an $X$ uniformly in $\d X$ and then evaluating the weight
$\d \sigma / \d X$ at this point. \ttt{SIGEV} is now simply
\ttt{SIGEV}$ = \d \sigma / \d X \, \int \d X$, i.e. the differential
cross section times the considered volume of phase space. Clearly,
when averaged over many events, \ttt{SIGEV} will correctly estimate
the desired cross section. If \ttt{SIGEV} fluctuates too much, one
may try to transform to new variables $X'$,
where events are now picked accordingly to $\d X'$ and
\ttt{SIGEV}$ = \d \sigma / \d X' \, \int \d X'$.

A warning. It is important that $X$ is indeed uniformly picked within
the allowed phase space, alternatively that any Jacobians are properly 
taken into account. For instance, in the case above, one approach 
would be to pick $x_1$, $x_2$ and $\hat{t}$ uniformly in the ranges
$0 < x_1 < 1$, $0 < x_2 < 1$, and $-s < \hat{t} < 0$, with full phase 
space
volume  $\int \d X = s$. The cross section would only be non-vanishing
inside the physical region given by $-s x_1 x_2 < \hat{t}$ (in the
massless case), i.e. Monte Carlo efficiency is likely to be low.
However, if one were to choose $\hat{t}$ values only in the range 
$-\hat{s} < \hat{t} < 0$, small $\hat{s}$ values would be favoured, 
since the density of selected $\hat{t}$ values would be larger there. 
Without the use of a compensating Jacobian $\hat{s}/s$, an incorrect 
answer would be obtained. Alternatively, one could start out with a
phase space like $\d X = \d x_1 \, \d x_2 \, \d (\cos\hat{\theta})$, 
where the limits decouple. Of course, the $\cos\hat{\theta}$ variable 
can be translated back into a $\hat{t}$, which will then always be in 
the desired range $-\hat{s} < \hat{t} < 0$. The transformation itself 
here gives the necessary Jacobian.
 
If you do not know how big \ttt{SIGMAX} is, you can put it to some
very small value (but larger than zero, however) and do an
exploratory run. When the program encounters events with
\ttt{SIGEV}$ > $\ttt{SIGMAX}, a warning message is printed, which
gives the new \ttt{SIGMAX} that the program will use from then
on. Hopefully such maximum violations only appear at the beginning of
the run, and later \ttt{SIGMAX} stabilizes to a level that can then  
be used as input for a second, correct run.
 
If you want to do the event rejection yourself, simply put \ttt{SIGEV}
equal to \ttt{SIGMAX}. In that case events will not be rejected by
{\Py} (except if there is something else wrong with them). If
\ttt{SIGMAX} is the correct total cross section of the process, event
mixing with other processes will still work fine. You could also decide
not to reject any events, but to use weighted ones. In that case you
can only have one ISUB switched on in a run, since the program will not
know how to mix different kinds of events, and you cannot use {\Py} to
do cross section statistics for you. Therefore you could, for instance, 
put \ttt{SIGMAX = SIGEV = 1}, and use a common block to transfer event 
weight and other information from your \ttt{PYUPEV} routine to your 
main program.
 
In addition to the \ttt{SIGEV} value returned for each event, it is also
necessary to return the event itself. This is done via the common block
 
\drawbox{COMMON/PYUPPR/NUP,KUP(20,7),PUP(20,5),NFUP,IFUP(10,2),%
Q2UP(0:10)}\label{p:PYUPPR}
\boxsep
 
The first part closely parallels the standard event record in the
\ttt{LUJETS} common block, see section \ref{ss:evrec}, although with 
a few simplifications. The number \ttt{NUP}\label{p:NUP} gives the
number of particles involved in the process, where a particle may be
a quark, a lepton, a gauge boson, or anything else. The first two
are simply the two incoming particles that initiate the hard
scattering, while the remaining \ttt{NUP-2} are the outgoing
particles from the hard process. For each particle \ttt{I}, with
$1 \leq $\ttt{I}$ \leq $\ttt{NUP}, the following information is stored:
\begin{entry}
 
\iteme{KUP(I,1) :}\label{p:KUP} is = 1 normally. However, if you put 
it = 2 that signifies intermediate states that are not to be treated 
by {\Py}, but are included only to make the event record easier to 
read.
 
\iteme{KUP(I,2) :} is the flavour code of a particle, i.e. the two
incoming partons for \ttt{I} = 1 and 2, and the outgoing particles
for \ttt{I}$ \geq 3$. The flavour codes are the standard KF ones,
as used elsewhere in the program.
 
\iteme{KUP(I,3) :} may be used to indicate the position of a mother.
Such information may again make the record more readable, but is not
really needed, and so one may well put all \ttt{KUP(I,3)=0}.
 
\iteme{KUP(I,4) :} for a final-state parton which carries colour,
\ttt{KUP(I,4)} gives the position of the parton from which the colour
comes; otherwise it must be 0.
 
\iteme{KUP(I,5) :} for a final-state parton that carries anticolour,
\ttt{KUP(I,5)} gives the position of the parton from which the
anticolour comes; else it must be 0.
 
\iteme{KUP(I,6) :} for an initial-state parton that carries colour,
\ttt{KUP(I,6)} gives the position of the parton to which the colour
goes; else it must be 0.
 
\iteme{KUP(I,7) :} for an initial-state parton which carries
anticolour, \ttt{KUP(I,7)} gives the position of the parton to
which the anticolour goes; else it must be 0.
 
\iteme{PUP(I,1) :}\label{p:PUP} $p_x$, i.e. $x$ momentum.
 
\iteme{PUP(I,2) :} $p_y$, i.e. $y$ momentum.
 
\iteme{PUP(I,3) :} $p_z$, i.e. $z$ momentum.
 
\iteme{PUP(I,4) :} $E$, i.e. energy.
 
\iteme{PUP(I,5) :} $m$, i.e. mass.
\end{entry}
After this brief summary, we proceed with more details and examples.
 
To illustrate the issue of documentation in \ttt{KUP(I,1)} and
\ttt{KUP(I,3)}, consider the case of $\W^+$ production and decay to
$\u \dbar$, maybe as part of a more complex process. The final-state
particles clearly are $\u$ and $\dbar$, so the $\W^+$ need not be
given at all, but it would make the event listing easier to read.
Therefore one should add the $\W^+$, but with \ttt{KUP(I,1)=2}.
(If the $\W^+$ would have been added with \ttt{KUP(I,1)=1}, it would
later have been treated by {\Py}/{\Je}, which means it would have
been allowed to decay once more.) If the $\W^+$ is in line 3, the
$\u$ in 4 and the $\dbar$ in 5, one could further put \ttt{KUP(4,3)=3}
and \ttt{KUP(5,3)=3} to indicate that the $\u$ and $\dbar$ in lines
4 and 5 come from the $\W^+$ stored in line 3.

The switch \ttt{MSTP(128)} works in the same way for user-defined 
processes as for ordinary ones, i.e. decay products of
resonances can optionally be omitted from the documentation section
of the event record, and history pointers can be set slightly 
differently. The information the program has at its disposal 
for this purpose is in \ttt{KUP(I,3)}; an entry with this
value non-zero is considered as a resonance decay product.
 
The colour-flow information for coloured particles (quarks, gluons,
leptoquarks, \ldots) is needed to set up parton showers and
fragmentation properly. Sometimes many different colour flows are
possible for one and the same process, as discussed in section
\ref{sss:QCDjetclass}.
It is up to you whether or not you will include all possible colour
flows in the appropriate mixture, but at least you must pick some
representative colour configuration. Consider e.g. the case of
$\g(1) + \g(2) \to \q(3) + \qbar(4)$, where the numbers give the
position in the array. It is clear the $\q$ must get its colour from
either of the two gluons, which means there are (at least) two
possibilities. Picking the $\q$ colour
to come from gluon 1, one would thus write
\ttt{KUP(3,4)=1}, to be read `the colour of parton 3 comes from
parton 1'. By implication therefore also \ttt{KUP(1,6)=3}, i.e.
`the colour of parton 1 goes to parton 3', i.e. the colour flow is
bookkept doubly. The anticolour now must flow from parton 2 to parton
4, i.e. \ttt{KUP(2,7)=4} and \ttt{KUP(4,5)=2}. This completely
specifies the colours of the $\q$ and $\qbar$, but not of the
two gluons. In fact, one colour in the initial state `annihilates'
between the $\g(1)$ and $\g(2)$, i.e. the anticolour of gluon 1 and the
colour of gluon 2 match, which may be expressed by \ttt{KUP(1,7)=2}
and \ttt{KUP(2,6)=1}. In other words colour/anticolour of an 
initial-state parton may either go to a final-state parton or to 
another
initial-state parton. Correspondingly, the colour/anticolour of a
final-state parton may come either from an initial-state parton or
from another final-state parton. An example of the latter possibility
is $\W$ decays, or generically the decay of any colour-singlet
particle. (Thus a third colour flow above is represented by
$\g \g \to \H^0 \to \q \qbar$, where no colour passes through the
Higgs, and therefore colour flows between the two gluons and,
separately, between the $\q$ and $\qbar$.)
 
Storing of momenta should be straightforward, but a few comments must
be made. Even if you ask, in the \ttt{PYINIT} call, to have events
generated in a fixed target or a user-specified frame, at
intermediate stages {\Py} will still work in the c.m. frame of the
two incoming beam particles, with the first beam moving in the $+z$
direction and the second in the $-z$ one. This c.m. frame must also
be used when giving the momenta of the process. In addition, the two
incoming partons in lines 1 and 2 are assumed massless. Therefore the
initial-state partons are characterized only by the two energies
\ttt{P(1,4)} and \ttt{P(2,4)}, with \ttt{P(1,3) = P(1,4)},
\ttt{P(2,3) = -P(2,4)}, and everything else is zero. In the
final state, energies, momenta and masses are free, but must add up to
give the same four-momentum as that of the initial state. All momenta 
are given in GeV, with the speed of light $c = 1$.
 
The second part of the \ttt{PYUPPR} common block is used to regulate the
initial- and final-state showering, as follows:
\begin{entry}
\iteme{Q2UP(0) :}\label{p:Q2UP} $Q^2$ scale of initial-state showers.
\iteme{NFUP :}\label{p:NFUP} number of parton pairs that undergo 
final-state showers.
\iteme{IFUP(IF,1), IFUP(IF,2) :}\label{p:IFUP} positions of the two 
partons of a final-state showering pair, where the index \ttt{IF} 
runs between 1 and \ttt{NFUP}.
\iteme{Q2UP(IF) :} the $Q^2$ scale of the final-state shower between
parton pair \ttt{IF} above.
\end{entry}
 
If you do not want any showering at all, you can put \ttt{MSTP(61)=0}
and \ttt{MSTP(71)=0}, and then you do not have to give the above 
quantities. In general the scale choices $Q^2$ are not unique, which 
means that some guesswork is involved. Since the showers add extra 
partonic activity at mass scales below the mentioned $Q^2$ choices, the
\ttt{Q2UP} should be of the order of the phase-space cut-offs, so as to
provide a reasonably smooth joining between partonic activity from
matrix elements and that from showers. There are a few cases where
choices are rather easy. In the decay of any $s$-channel colour
neutral state, such as a $\W^{\pm}$, the $Q^2$ scale of final-state
showers is just set by the squared mass of the resonance. For 
initial-state radiation, \ttt{Q2UP(0)} should be about the same as 
the $Q^2$
scale used for the evaluation of parton distributions for the hard
process, up to some factor of order unity. (One frequent choice for
this factor would be 4, if your parton-distribution scale is something
like the squared transverse momentum, simply because $m^2$ is of order
$4 \pT^2$.)
 
The `parton'-shower evolution actually also can include photon emission
off quarks and leptons, if the shower switches are properly set. It is
not possible to define only one particle in the above arrays, since
it would then not be possible to conserve energy and momentum in the
shower. You can very well have a pair where only one of the two can
branch, however. For instance, in a $\g \gamma$ final state, only
the gluon can shower, but the photon can lose energy to the gluon
in such a way that the gluon branchings becomes possible.
 
Currently, it is not possible to do showering where three or more 
final-state particles are involved at the same time. This may be added 
at a
later stage. It is therefore necessary to subdivide suitably into pairs,
and maybe leave some (especially colour-neutral) particles unshowered.
 
You are free to make use of whatever tools you want in your \ttt{PYUPEV}
routine, and normally there would be no contact with the rest of
{\Py}, except as described above. However, you may want to use
some of the tools already available. One attractive possibility
is to use \ttt{PYSTFU} for parton-distribution-function evaluation.
Other possible tools could be \ttt{RLU} for random-number 
generation, \ttt{ULALPS} for $\alphas$ evaluation,
\ttt{ULALEM} for evaluation of a running $\alphaem$, and 
maybe a few more.

We end with a few comments on anomalous situations. In some cases
one may want to decide, inside \ttt{PYUPEV}, when to stop
the event-generation loop. This is the case, for instance, if event 
configurations are read in from a file, and the end of the file is 
reached. One might be tempted just to put \ttt{SIGEV=0} when this 
happens. Then \ttt{PYEVNT} will discard the event, as part of the 
matrix-element-weighting procedure. However, next \ttt{PYEVNT} will
generate another event, which normally means a new request to 
\ttt{PYUPEV}, so one does not really get out of the loop. Instead
you should put \ttt{NUP=0}. If the program encounters this value at a
return from \ttt{PYUPEV}, then it will also exit from \ttt{PYEVNT},
without incrementing the counters for the number of events generated.
It is then up to you to have a check on this condition in your main
event-generation loop. This you do either by looking at \ttt{NUP} or
at \ttt{MSTI(51)}; the latter is set to 1 if no event was generated.

It may also happen that a user-defined configuration fails elsewhere
in the \ttt{PYEVNT} call. For instance, the beam-remnant treatment
occasionally encounters situations it cannot handle, wherefore the
hard interaction is rejected and a new one generated. This happens also
with ordinary (not user-defined) events, and usually comes about as a
consequence of the initial-state radiation description leaving too
little energy for the remnant. If the same hard scattering were to
be used as input for a new initial-state radiation and beam-remnant
attempt, it could then work fine. There is a possibility to give events
that chance, as follows. \ttt{MSTI(52)} counts the number of times a
hard-scattering configuration has failed to date. If you come in to
\ttt{PYUPEV} with \ttt{MSTI(52)} non-vanishing, this means that the 
latest configuration failed. So long as the contents of the \ttt{PYUPPR} 
common block are not changed, such an event may be given another try.
For instance, a line 
\begin{verbatim}
      IF(MSTI(52).GE.1.AND.MSTI(52).LE.4) RETURN
\end{verbatim}
at the beginning of \ttt{PYUPEV} will give each event up to five 
tries; thereafter a new one would be generated as usual. Note that the
counter for the number of events is updated at each new try. The
fraction of failed configurations is given in the bottom line of
the \ttt{PYSTAT(1)} table.

The above comment only refers to very rare occurrences (less than 
one in a hundred), which are not errors in a strict sense; for instance, 
they do not produce any error messages on output. If you get
warnings and error messages that the program does not understand the 
flavour codes or cannot reconstruct the colour flows, it is due to 
faults of yours, and giving such events more tries is not going to 
help. 
 
\subsection{How to run PYTHIA with varying energies}
\label{ss:PYvaren}

It is possible to use {\Py} in a mode where the energy can be
varied from one event to the next, without the need to reinitialize
with a new \ttt{PYINIT} call. This allows a significant speed-up of
execution, although it is not as fast as running at a fixed
energy. It can not be used for everything --- we will come to
the fine print at the end --- but it should be applicable for most 
tasks.

The master switch to access this possibility is in \ttt{MSTP(171)}.
By default it is off, so you must set \ttt{MSTP(171)=1} before 
initialization. There are two submodes of running, with 
\ttt{MSTP(172)} being 1 or 2. In the former mode, {\Py} will generate 
an event at the requested energy. This means that you have to know 
which energy you want beforehand. In the latter mode, {\Py}
will often return without having generated an event --- with flag
\ttt{MSTI(61)=1} to signal that --- and you are then requested to give 
a new energy. The energy spectrum of accepted events will then,
in the end, be your naive input spectrum weighted with the 
cross-section of the processes you study. We will come back to 
this. 

The energy can be varied, whichever frame is given in the \ttt{PYINIT}
call. When the frame is \ttt{'CMS'}, \ttt{PARP(171)} should be filled
with the fractional energy of each event, i.e. 
$E_{\mrm{cm}} =$\ttt{PARP(171)}$\times$\ttt{WIN}, where \ttt{WIN} is 
the nominal c.m. energy of the \ttt{PYINIT} call. Here \ttt{PARP(171)} 
should normally be smaller than unity, i.e. initialization should
be done at the maximum energy to be encountered. For the \ttt{'FIXT'}
frame, \ttt{PARP(171)} should be filled by the fractional beam energy 
of that one, i.e. $E_{\mrm{beam}} = $\ttt{PARP(171)}$\times$\ttt{WIN}. 
For the \ttt{'USER'}, \ttt{'FOUR'} and \ttt{'FIVE'} options, the
two four-momenta are given in for each event in the same format as
used for the \ttt{PYINIT} call. Note that there is a minimum c.m.
energy allowed, \ttt{PARP(2)}. If you give in values below this, 
the program will stop for \ttt{MSTP(172)=1}, and will return with
\ttt{MSTI(61)=1} for \ttt{MSTP(172)=1}. 

To illustrate the use of the \ttt{MSTP(172)=2} facility, consider the 
case of beamstrahlung in $\ee$ linear colliders. This is just for 
convenience; what is said here can be translated easily into other 
situations. Assume that the beam spectrum is given by $D(z)$, where 
$z$ is the fraction retained by the original $\e$ after beamstrahlung. 
Therefore $0 \leq z \leq 1$ and the integral of $D(z)$ is unity. 
This is not perfectly general; one could imagine branchings 
$\e^- \to \e^- \gamma \to \e^-\e^+\e^-$, which gives a multiplication 
in the number of beam particles. This could either be expressed in 
terms of a $D(z)$ with integral larger than unity or in terms of an 
increased luminosity. We will assume the latter, and use $D(z)$ 
properly normalized. Given a nominal $s = 4E_{\mrm{beam}}^2$, 
the actual $s'$ after beamstrahlung is given by $s' = z_1 z_2 s$. 
For a process with a cross section $\sigma(s)$ the total 
cross section is then
\begin{equation}
\sigma_{\mrm{tot}} = \int_0^1 \int_0^1 D(z_1) \, D(z_2)   
\sigma(z_1 z_2 s) \, \d z_1  \, \d z_2~.
\end{equation} 
The cross section $\sigma$ may in itself be an integral over a number
of additional phase space variables. If the maximum of the differential
cross section is known, a correct procedure to generate events is 
\begin{Enumerate}
\item pick $z_1$ and $z_2$ according to $D(z_1) \, \d z_1$ and 
$D(z_2) \, \d z_2$, respectively;
\item pick a set of phase space variables of the process, for the 
given $s'$ of the event;
\item evaluate $\sigma(s')$ and compare with $\sigma_{\mmax}$;
\item if event is rejected, then return to step 1 to generate new
variables;
\item else continue the generation to give a complete event.
\end{Enumerate}
You as a user are assumed to take care of step 1, and present the
resulting kinematics with incoming $\e^+$ and $\e^-$ of varying energy. 
Thereafter {\Py} will do steps 2--5, and either return an event or
put \ttt{MSTI(61)=1} to signal failure in step 4.

The maximization procedure does search in phase space to find 
$\sigma_{\mmax}$, but it does not vary the $s'$ energy in this process.
Therefore the maximum search in the \ttt{PYINIT} call should be 
performed where the cross section is largest. For processes with 
increasing cross section as a function of energy this means at the 
largest energy that will ever be encountered, i.e. $s' = s$ in the 
case above. This is the `standard' case, but often one encounters 
other behaviours, where more complicated procedures are needed. One 
such case would be the process $\ee \to \Z^{*0} \to \Z^0 \H^0$, 
which is known to have a cross section that increases near the 
threshold but is decreasing asymptotically. If one already knows
that the maximum, for a given Higgs mass, appears at 300 GeV, say, 
then the \ttt{PYINIT} call should be made with that energy, even if 
subsequently one will be generating events for a 500 GeV collider. 

In general, it may be necessary to modify the selection of $z_1$ and
$z_2$ and assign a compensating event weight. For instance, consider
a process with a cross section behaving roughly like $1/s$. Then the 
$\sigma_{\mrm{tot}}$ expression above may be rewritten as
\begin{equation}
\sigma_{\mrm{tot}} = \int_0^1 \int_0^1 \frac{D(z_1)}{z_1} \, 
\frac{D(z_2)}{z_2} \, z_1 z_2 \sigma(z_1 z_2 s) \, \d z_1 \, \d z_2 ~. 
\end{equation}
The expression $z_1 z_2 \sigma(s')$ is now essentially flat in $s'$, 
i.e. not only can $\sigma_{\mmax}$ be found at a convenient energy
such as the maximum one, but additionally the {\Py} generation 
efficiency (the likelihood of surviving step 4) is greatly 
enhanced. The price to be paid is that $z$ has to be selected
according to $D(z)/z$ rather than according to $D(z)$. Note that
$D(z)/z$ is not normalized to unity. One therefore needs to define
\begin{equation}
{\cal I}_D = \int_0^1 \frac{D(z)}{z} \, \d z ~, 
\end{equation}
and a properly normalized 
\begin{equation}
D'(z) = \frac{1}{{\cal I}_D} \, \frac{D(z)}{z} ~.
\end{equation}
Then  
\begin{equation}
\sigma_{\mrm{tot}} = \int_0^1 \int_0^1 D'(z_1) \, D'(z_2) \, 
{\cal I}_D^2 \, z_1 z_2 \sigma(z_1 z_2 s) \, \d z_1 \, \d z_2 ~.
\end{equation}
Therefore the proper event weight is ${\cal I}_D^2 \, z_1 z_2$.
This weight should be stored, for each event, in \ttt{PARP(173)}.
The maximum weight that will be encountered should be stored in
\ttt{PARP(174)} before the \ttt{PYINIT} call, and not changed
afterwards. It is not necessary to know the precise maximum;
any value larger than the true maximum will do, but the inefficiency
will be larger the cruder the approximation.
Additionally you must put \ttt{MSTP(173)=1} for the program to 
make use of weights at all. Often $D(z)$ are not known analytically;
therefore ${\cal I}_D$ is also not known beforehand, but may have to 
be evaluated (by you) during the course of the run. Then you should
just use the weight $z_1 z_2$ in \ttt{PARP(173)} and do the overall 
normalization yourself in the end. Since \ttt{PARP(174)=1.} by
default, in this case you need not set this variable specially. 
Only the cross sections are affected by the procedure selected for
overall normalization, the events themselves still are properly 
distributed in $s'$ and internal phase space.  

Above it has been assumed tacitly that $D(z) \to 0$ for $z \to 0$. 
If not, $D(z)/z$ is divergent, and it is not possible to define a 
properly normalized $D'(z) = D(z)/z$. If the cross section is truly 
diverging like $1/s$, then a $D(z)$ which is nonvanishing for 
$z \to 0$ does imply an infinite total cross section, whichever way 
things are considered. In cases like that, it is necessary to impose
a lower cut on $z$, based on some physics or detector consideration. 
Some such cut is anyway needed to keep away from the minimum c.m. 
energy required for {\Py} events, see above.

The most difficult cases are those with a very narrow and high
peak, such as the $\Z^0$. One could initialize at the energy of 
maximum cross section and use $D(z)$ as is, but efficiency might 
turn out to be very low. One might then be tempted to do more 
complicated transforms of the kind illustrated above. As a rule 
it is then convenient to work in the variables $\tau_z = z_1 z_2$ 
and $y_z = (1/2) \ln (z_1/z_2)$, cf. subsection \ref{ss:kinemtwo}.

Clearly, the better the behaviour of the cross section can be 
modelled in the choice of $z_1$ and $z_2$, the better the overall
event generation efficieny. Even under the best of circumstances,
the efficiency will still be lower than for runs with fix energy.
There is also a non-negligible time overhead for using variable
energies in the first place, from kinematics reconstruction
and (in part) from the phase space selection. One should therefore
not use variable energies when not needed, and not use a large
range of energies $\sqrt{s'}$ if in the end only a smaller range 
is of experimental interest. 

This facility may be combined with most other aspects of the program.
For instance, it is possible to simulate beamstrahlung as above and
still include bremsstrahlung with \ttt{MSTP(11)=1}. Further, one may
multiply the overall event weight of \ttt{PARP(173)} with a 
kinematics-dependent weight given by \ttt{PYEVWT}, although it is not
recommended (since the chances of making a mistake are also 
multiplied). However, a few things do \textit{not} work.
\begin{Itemize}
\item It is not possible to use pile-up events, i.e. you must have 
\ttt{MSTP(131)=0}.
\item The possibility of giving in your own cross-section optimization
coefficients, option \ttt{MSTP(121)=2}, would require more input
than with fixed energies, and this option should therefore not be 
used. You can still use \ttt{MSTP(121)=1}, however.
\item The multiple interactions scenario with \ttt{MSTP(82)}$\geq 2$ 
only works approximately for energies different from the initialization 
one. If the c.m. energy spread is smaller than a factor 2, say, the
approximation should be reasonable, but if the spread is larger
one may have to subdivide into subruns of different energy bins.
The initialization should be made at the largest energy to be 
encountered --- whenever multiple interactions are possible (i.e. for 
incoming hadrons and resolved photons) this is where the cross sections 
are largest anyway, and so this is no further constraint. There is no 
simple possibility to change \ttt{PARP(82)} during the course of the 
run, i.e. an energy-independent $p_{\perp 0}$ must be assumed. The 
default option \ttt{MSTP(82)=1} works fine, i.e. does not suffer
from the constraints above. If so desired,
$\pTmin=$\ttt{PARP(81)} can be set differently for each event, 
as a function of c.m. energy. Initialization should then be done with 
\ttt{PARP(81)} as low as it is ever supposed to become.
\end{Itemize}
  
\subsection{Other Routines and Common Blocks}
 
The subroutines and common blocks that you will come in direct
contact with have already been described. A number of other routines
and common blocks exist, and those not described elsewhere are here 
briefly listed for the sake of completeness. The \ttt{PYG***} 
routines are slightly modified versions of the \ttt{SAS***} ones
of the \tsc{SaSgam} library. The common block \ttt{SASCOM} is 
renamed \ttt{PYINT8}. If you want to use the parton distributions 
for standalone purposes, you are encouraged to use the original 
\tsc{SaSgam} routines rather than going the way via the
{\Py} adaptations. 
 
\begin{entry}
 
\iteme{SUBROUTINE PYINRE :}\label{p:PYINRE}
to initialize the widths and effective widths of resonances.

\iteme{SUBROUTINE PYINBM(CHFRAM,CHBEAM,CHTARG,WIN) :}\label{p:PYINBM}
to read in and identify the beam (\ttt{CHBEAM}) and target 
(\ttt{CTTARG}) particles and the frame (\ttt{CHFRAM}) as given in 
the \ttt{PYINIT} call; also to save the original energy (\ttt{WIN}).

\iteme{SUBROUTINE PYINKI(MODKI) :}\label{p:PYINKI}
to set up the event kinematics, either at initialization 
(\ttt{MODKI=0}) or for each separate event, the latter when the 
program is run with varying kinematics (\ttt{MODKI=1}). 

\iteme{SUBROUTINE PYINPR :}\label{p:PYINPR} 
to set up the partonic subprocesses selected with \ttt{MSEL}. 
For $\gamma\p$ and $\gamma\gamma$, also the \ttt{MSTP(14)} value
affects the choice of processes. In particular, the option 
\ttt{MSTP(14)=10} sets up the three or six different processes that
need to be mixed, with separate cuts for each.

\iteme{SUBROUTINE PYXTOT :}\label{p:PYXTOT}
to give the parametrized total, double diffractive, single
diffractive and elastic cross sections for different energies and
colliding hadrons.
 
\iteme{SUBROUTINE PYMAXI :}\label{p:PYMAXI}
to find optimal coefficients \ttt{COEF} for the selection of
kinematical variables, and to find the related maxima for
the differential cross section times Jacobian factors,
for each of the subprocesses included.
 
\iteme{SUBROUTINE PYPILE(MPILE) :}\label{p:PYPILE}
to determine the number of pile-up events, i.e. events
appearing in the same beam--beam crossing.

\iteme{SUBROUTINE PYSAVE(ISAVE,IGA) :}\label{p:PYSAVE}  
saves and restores parameters and cross section values between the 
three $\gamma\p$ and the six $\gamma\gamma$ components
of \ttt{MSTP(14)=10}. The options for \ttt{ISAVE} are (1)
a complete save of all parameters specific to a given component,
(2) a partial save of cross-section information, (3) a restoration
of all parameters specific to a given component, (4) as 3 but
preceded by a random selection of component, and (5) a summation of
component cross sections (for \ttt{PYSTAT}). The subprocess code
in \ttt{IGA} is the one described for \ttt{MSTI(9)}; it is input
for options 1, 2 and 3 above, output for 4 and dummy for 5.
 
\iteme{SUBROUTINE PYRAND :}\label{p:PYRAND}
to generate the quantities characterizing a hard scattering
on the parton level, according to the relevant matrix elements.
 
\iteme{SUBROUTINE PYSCAT :}\label{p:PYSCAT}
to find outgoing flavours and to set up the kinematics and
colour flow of the hard scattering.
 
\iteme{SUBROUTINE PYRESD :}\label{p:PYRESD}
to allow resonances to decay,
including chains of successive decays and parton showers.
 
\iteme{SUBROUTINE PYMULT(MMUL) :}\label{p:PYMULT}
to generate semi-hard interactions according to the
multiple interaction formalism.
 
\iteme{SUBROUTINE PYREMN(IPU1,IPU2) :}\label{p:PYREMN}
to add on target remnants and include primordial $k_{\perp}$.
 
\iteme{SUBROUTINE PYDIFF :}\label{p:PYDIFF}
to handle diffractive and elastic scattering events.
 
\iteme{SUBROUTINE PYDOCU :}\label{p:PYDOCU}
to compute cross sections of processes, based on current
Monte Carlo statistics, and to store event information in the
\ttt{MSTI} and \ttt{PARI} arrays.
 
\iteme{SUBROUTINE PYWIDT(KFLR,SH,WDTP,WDTE) :}\label{p:PYWIDT}
to calculate widths and effective widths of resonances.
 
\iteme{SUBROUTINE PYOFSH(MOFSH,KFMO,KFD1,KFD2,PMMO,RET1,RET2) :}%
\label{p:PYOFSH}
to calculate partial widths into channels off the mass shell,
and to select correlated masses of resonance pairs.
 
\iteme{SUBROUTINE PYKLIM(ILIM) :}\label{p:PYKLIM}
to calculate allowed kinematical limits.
 
\iteme{SUBROUTINE PYKMAP(IVAR,MVAR,VVAR) :}\label{p:PYKMAP}
to calculate the value of a kinematical variable when this
is selected according to one of the simple pieces.
 
\iteme{SUBROUTINE PYSIGH(NCHN,SIGS) :}\label{p:PYSIGH}
to give the differential cross section (multiplied by the
relevant Jacobians) for a given subprocess and kinematical
setup.

\iteme{SUBROUTINE PYSTFL(KF,X,Q2,XPQ) :}\label{p:PYSTFL}
to give parton distributions for $\p$ and $\n$ in the option
with modified behaviour at small $Q^2$ and $x$, see
\ttt{MSTP(57)}.
 
\iteme{SUBROUTINE PYSTFU(KF,X,Q2,XPQ) :}\label{p:PYSTFU}
to give parton-distribution functions (multiplied by $x$, i.e. 
$x f_i(x,Q^2)$) for an arbitrary particle (of those recognized by 
{\Py}). Generic driver routine for the following, specialized ones.
\begin{subentry}
\iteme{KF :} flavour of probed particle, according to KF code.
\iteme{X :} $x$ value at which to evaluate parton distributions.
\iteme{Q2 :} $Q^2$ scale at which to evaluate parton distributions.
\iteme{XPQ :} array of dimensions \ttt{XPQ(-25:25)}, which contains
the evaluated parton distributions $x f_i(x,Q^2)$. Components $i$
ordered according to standard KF code; additionally the gluon is
found in position 0 as well as 21 (for historical reasons).
\end{subentry}
 
\iteme{SUBROUTINE PYSTEL(X,Q2,XPEL) :}\label{p:PYSTEL}
to give electron parton distributions.

\iteme{SUBROUTINE PYSTGA(X,Q2,XPGA) :}\label{p:PYSTGA}
to give the photon parton distributions for sets other than the SaS 
ones.

\iteme{SUBROUTINE PYGGAM(ISET,X,Q2,P2,F2GM,XPDFGM) :}\label{p:PYGGAM}
to construct the SaS $F_2$ and parton distributions of the photon
by summing homogeneous (VMD) and inhomogeneous (anomalous) terms.
For $F_2$, $\c$ and $\b$ are included by the Bethe-Heitler formula;
in the `{\MSbar}' scheme additionally a $C^{\gamma}$ term 
is added. Calls \ttt{PYGVMD}, \ttt{PYGANO}, \ttt{PYGBEH},
and \ttt{PYGDIR}.
 
\iteme{SUBROUTINE PYGVMD(ISET,KF,X,Q2,P2,ALAM,XPGA) :}\label{p:PYGVMD}
to evaluate the VMD parton distributions of a photon,
evolved homogeneously from an initial scale $P^2$ to $Q^2$.
 
\iteme{SUBROUTINE PYGANO(KF,X,Q2,P2,ALAM,XPGA) :}\label{p:PYGANO}
to evaluate the parton distributions of the anomalous
photon, inhomogeneously evolved from a scale $P^2$ (where it vanishes)
to $Q^2$.
 
\iteme{SUBROUTINE PYGBEH(KF,X,Q2,P2,PM2,XPBH) :}\label{p:PYGBEH}
to evaluate the Bethe-Heitler cross section for
heavy flavour production.
 
\iteme{SUBROUTINE PYGDIR(X,Q2,P2,AK0,XPGA) :}\label{p:PYGDIR}
to evaluate the direct contribution, i.e. the $C^{\gamma}$ term,
as needed in {\MSbar} parametrizations.
 
\iteme{SUBROUTINE PYSTPI(X,Q2,XPPI) :}\label{p:PYSTPI}
to give pion parton distributions.
 
\iteme{SUBROUTINE PYSTPR(X,Q2,XPPR) :}\label{p:PYSTPR}
to give proton parton distributions.
 
\iteme{FUNCTION PYCTQ2(ISET,IPRT,X,Q) :}\label{p:PYCTQ2}
to give the CTEQ2 proton parton distributions.
 
\iteme{FUNCTION PYHFTH(SH,SQM,FRATT) :}\label{p:PYHFTH}
to give heavy-flavour threshold factor in matrix elements.
 
\iteme{SUBROUTINE PYSPLI(KF,KFLIN,KFLCH,KFLSP) :}\label{p:PYSPLI}
to give hadron remnant or remnants left behind when the reacting
parton is kicked out.
 
\iteme{FUNCTION PYGAMM(X) :}\label{p:PYGAMM}
to give the value of the ordinary $\Gamma (x)$ function
(used in some parton-distribution parametrizations).
 
\iteme{SUBROUTINE PYWAUX(IAUX,EPS,WRE,WIM) :}\label{p:PYWAUX}
to evaluate the two auxiliary functions $W_1$ and $W_2$ appearing
in the cross section expressions in \ttt{PYSIGH}.
 
\iteme{SUBROUTINE PYI3AU(EPS,RAT,Y3RE,Y3IM) :}\label{p:PYI3AU}
to evaluate the auxiliary function $I_3$ appearing
in the cross section expressions in \ttt{PYSIGH}.
 
\iteme{FUNCTION PYSPEN(XREIN,XIMIN,IREIM) :}\label{p:PYSPEN}
to calculate the real and imaginary part of the Spence
function.
 
\iteme{SUBROUTINE PYQQBH(WTQQBH) :}\label{p:PYQQBH}
to calculate matrix elements for the two processes 
$\g \g \to \Q \Qbar \H^0$ and $\q \qbar \to \Q \Qbar \H^0$.
 
\iteme{BLOCK DATA PYDATA :}\label{p:PYDATA}
to give sensible default values to all status codes and
parameters.
 
\end{entry}
 
\drawbox{COMMON/PYINT1/MINT(400),VINT(400)}\label{p:PYINT1}
\begin{entry}
 
\itemc{Purpose:} to collect a host of integer- and real-valued
variables used internally in the program during the initialization
and/or event generation stage. These variables must not be changed
by you.
 
\iteme{MINT(1) :}\label{p:MINT} specifies the general type of 
subprocess that has occurred, according to the ISUB code given in 
section \ref{ss:ISUBcode}.

\iteme{MINT(2) :} whenever \ttt{MINT(1)} (together with \ttt{MINT(15)}
and \ttt{MINT(16)}) are not sufficient to specify the type of process
uniquely, \ttt{MINT(2)} provides an ordering of the different
possibilities, see \ttt{MSTI(2)}.
 
\iteme{MINT(3) :} number of partons produced in the hard
interactions, i.e. the number $n$ of the $2 \to n$ matrix elements
used; is sometimes 3 or 4 when a basic $2 \to 1$ or $2 \to 2$ process
has been folded with two $1 \to 2$ initial branchings
(like $\q \q' \to \q'' \q''' \H^0$).
 
\iteme{MINT(4) :} number of documentation lines at the beginning of
the common block \ttt{LUJETS} that are given with \ttt{K(I,1)=21};
0 for \ttt{MSTP(125)=0}.
 
\iteme{MINT(5) :} number of events generated to date in current run.
In runs with the variable-energy option, \ttt{MSTP(171)=1}
and \ttt{MSTP(172)=2}, only those events that survive (i.e. that
do not have \ttt{MSTI(61)=1}) are counted in this number. That
is, \ttt{MINT(5)} may be less than the total number of \ttt{PYEVNT}
calls.
 
\iteme{MINT(6) :} current frame of event (see \ttt{MSTP(124)} for
possible values).
 
\iteme{MINT(7), MINT(8) :} line number for documentation of outgoing
partons/particles from hard scattering for $2 \to 2$ or
$2 \to 1 \to 2$ processes (else = 0).
\iteme{MINT(10) :} is 1 if cross section maximum was violated in
current event, and 0 if not.
 
\iteme{MINT(11) :} KF flavour code for beam (side 1) particle.
 
\iteme{MINT(12) :} KF flavour code for target (side 2) particle.
 
\iteme{MINT(13), MINT(14) :} KF flavour codes for side 1 and side 2
initial-state shower initiators.
 
\iteme{MINT(15), MINT(16) :} KF flavour codes for side 1 and side 2
incoming partons to the hard interaction.
 
\iteme{MINT(17), MINT(18) :} flag to signal if particle on side 1 or
side 2 has been scattered diffractively; 0 if no, 1 if yes.
 
\iteme{MINT(19), MINT(20) :} flag to signal initial-state structure
with parton inside photon inside electron on side 1 or side 2;
0 if no, 1 if yes.
 
\iteme{MINT(21) - MINT(24) :} KF flavour codes for outgoing partons
from the hard interaction. The number of positions actually used is
process-dependent, see \ttt{MINT(3)}; trailing positions not used are
set = 0.
 
\iteme{MINT(25), MINT(26) :} KF flavour codes of the products in the
decay of a single $s$-channel resonance formed in the hard interaction.
Are thus only used when \ttt{MINT(3)=1} and the resonance is allowed
to decay.
 
\iteme{MINT(31) :} number of hard or semi-hard scatterings that occurred
in the current event in the multiple-interaction scenario; is = 0 for a
low-$\pT$ event.
 
\iteme{MINT(35) :} in a true $2 \to 3$ process, where one particle is
a resonance with decay channel selected already before the
\ttt{PYRESD} call, the decay channel number (in the \ttt{/LUDAT3/}
numbering) is stored here.
 
\iteme{MINT(41), MINT(42) :} type of incoming beam or target particle;
1 for lepton and 2 for hadron. A photon counts as a lepton if it
is not resolved (\ttt{MSTP(14)=0}) and as a hadron if it is resolved
(\ttt{MSTP(14)}$\geq 1$).
 
\iteme{MINT(43) :} combination of incoming beam and target particles.
A photon counts as a hadron.
\begin{subentry}
\iteme{= 1 :} lepton on lepton.
\iteme{= 2 :} lepton on hadron.
\iteme{= 3 :} hadron on lepton.
\iteme{= 4 :} hadron on hadron.
\end{subentry}
 
\iteme{MINT(44) :} as \ttt{MINT(43)}, but a photon counts as a lepton.
 
\iteme{MINT(45), MINT(46) :} structure of incoming beam and target
particles.
\begin{subentry}
\iteme{= 1 :} no internal structure, i.e. an electron or photon
carrying the full beam energy.
\iteme{= 2 :} defined with parton distributions that are not peaked
at $x = 1$, i.e. a hadron or a resolved photon.
\iteme{= 3 :} defined with parton distributions that are peaked at
$x = 1$, i.e. a resolved electron.
\end{subentry}
 
\iteme{MINT(47) :} combination of incoming beam- and target-particle
parton-distribution function types.
\begin{subentry}
\iteme{= 1 :} no parton distribution either for beam or target.
\iteme{= 2 :} parton distributions for target but not for beam.
\iteme{= 3 :} parton distributions for beam but not for target.
\iteme{= 4 :} parton distributions for both beam and target, but not
both peaked at $x = 1$.
\iteme{= 5 :} parton distributions for both beam and target, with both
peaked at $x = 1$.
\end{subentry}
 
\iteme{MINT(48) :} total number of subprocesses switched on.
 
\iteme{MINT(49) :} number of subprocesses that are switched on, apart
from elastic scattering and single, double and central diffractive.

\iteme{MINT(50) :} combination of incoming particles from a multiple
interactions point of view.
\begin{subentry}
\iteme{= 0 :} the total cross section is not known; therefore no 
multiple interactions are possible.
\iteme{= 1 :} the total cross section is known; therefore multiple
interactions are possible if switched on.
\end{subentry}
 
\iteme{MINT(51) :} internal flag that event failed cuts.
\begin{subentry}
\iteme{= 0 :} no problem.
\iteme{= 1 :} event failed; new one to be generated.
\iteme{= 2 :} event failed; no new event is to be generated but
instead control is to be given back to used. Is intended for
user-defined processes, when \ttt{NUP=0}.
\end{subentry}
 
\iteme{MINT(52) :} internal counter for number of lines used
(in \ttt{/LUJETS/}) before multiple interactions are considered.
 
\iteme{MINT(53) :} internal counter for number of lines used
(in \ttt{/LUJETS/}) before beam remnants are considered.
 
\iteme{MINT(55) :} the heaviest new flavour switched on for QCD
processes, specifically the flavour to be generated for ISUB = 81,
82, 83 or 84.
 
\iteme{MINT(56) :} the heaviest new flavour switched on for QED
processes, specifically for ISUB = 85. Note that, unlike
\ttt{MINT(55)}, the heaviest flavour may here be a lepton, and
that heavy means the one with largest KF code.

\iteme{MINT(57) :} number of times the beam remnant treatment has 
failed, and the same basic kinematical setup is used to produce 
a new parton shower evolution and beam remnant set. Mainly used
in leptoproduction, for the option when $x$ and $Q^2$ are to be
preserved.   
 
\iteme{MINT(61) :} internal switch for the mode of operation of
resonance width calculations in \ttt{PYWIDT} for $\gammaZ$ or
$\gamma^*/\Z^0/\Z'^0$.
\begin{subentry}
\iteme{= 0 :} without reference to initial-state flavours.
\iteme{= 1 :} with reference to given initial-state flavours.
\iteme{= 2 :} for given final-state flavours.
\end{subentry}
 
\iteme{MINT(62) :} internal switch for use at initialization of
$\H^0$ width.
\begin{subentry}
\iteme{= 0 :} use widths into $\Z \Z^*$ or $\W \W^*$ calculated
before.
\iteme{= 1 :} evaluate widths into $\Z \Z^*$ or $\W \W^*$ for
current Higgs mass.
\end{subentry}
 
\iteme{MINT(65) :} internal switch to indicate initialization
without specified reaction.
\begin{subentry}
\iteme{= 0 :} normal initialization.
\iteme{= 1 :} initialization with argument \ttt{'none'} in
\ttt{PYINIT} call.
\end{subentry}
 
\iteme{MINT(71) :} switch to tell whether current process is singular 
for $\pT \to 0$ or not.
\begin{subentry}
\iteme{= 0 :} non-singular process, i.e. proceeding via an
$s$-channel resonance or with both products having a mass above
\ttt{CKIN(6)}.
\iteme{= 1 :} singular process.
\end{subentry}

\iteme{MINT(72) :} number of $s$-channel resonances that may
contribute to the cross section.
 
\iteme{MINT(73) :} KF code of first $s$-channel resonance;
0 if there is none.
 
\iteme{MINT(74) :} KF code of second $s$-channel resonance;
0 if there is none.
 
\iteme{MINT(81) :} number of selected pile-up events.
 
\iteme{MINT(82) :} sequence number of currently considered pile-up
event.
 
\iteme{MINT(83) :} number of lines in the event record already
filled by previously considered pile-up events.
 
\iteme{MINT(84) :} \ttt{MINT(83) + MSTP(126)}, i.e. number of lines
already filled by previously considered events plus number of lines
to be kept free for event documentation.

\iteme{MINT(91) :} is 1 for a lepton--hadron event and 0
else. Used to determine whether a \ttt{PYFRAM(3)} call is possible.

\iteme{MINT(92) :} is used to denote region in $(x,Q^2)$ plane when
\ttt{MSTP(57)=2}, according to numbering in \cite{Sch93a}. Simply put,
0 means that the modified proton parton distributions were not used,
1 large $x$ and $Q^2$, 2 small $Q^2$ but large $x$,
3 small $x$ but large $Q^2$ and 4 small $x$ and $Q^2$.

\iteme{MINT(93) :} is used to keep track of parton distribution set
used in the latest \ttt{STRUCTM} call to \tsc{Pdflib}. The code for this
set is stored in the form
\ttt{MINT(93) = 1000000}$\times$\ttt{NPTYPE + 1000}$\times$%
\ttt{NGROUP + NSET}.
The stored previous value is compared with the current new value
to decide whether a \ttt{PDFSET} call is needed to switch to another
set.

\iteme{MINT(101), MINT(102) :} is normally 1, but is 4 when a resolved
photon (appearing on side 1 or 2) can be represented by either of the 
four vector mesons $\rho^0$, $\omega$, $\phi$ and $\Jpsi$.

\iteme{MINT(103), MINT(104) :} KF flavour code for the two incoming
particles, i.e. the same as \ttt{MINT(11)} and \ttt{MINT(12)}. The 
exception is when a resolved photon is represented by a vector meson 
(a $\rho^0$, $\omega$, $\phi$ or $\Jpsi$). Then the code of the vector 
meson is given.

\iteme{MINT(105) :} is either \ttt{MINT(103)} or \ttt{MINT(104)}, 
depending on which side of the event currently is being studied. 

\iteme{MINT(107), MINT(108) :} if either or both of the two incoming 
particles is a photon, then the respective value gives the nature 
assumed for that photon. The code follows the one used for 
\ttt{MSTP(14)}:
\begin{subentry}
\iteme{= 0 :} direct photon.
\iteme{= 1 :} resolved photon.
\iteme{= 2 :} VMD-like photon.
\iteme{= 3 :} anomalous photon.
\end{subentry}

\iteme{MINT(109) :} is either \ttt{MINT(107)} or \ttt{MINT(108)}, 
depending on which side of the event currently is being studied.
 
\iteme{MINT(111) :} the frame given in \ttt{PYINIT} call, 0--5 for
\ttt{'NONE'}, \ttt{'CMS'}, \ttt{'FIXT'}, \ttt{'USER'}, \ttt{'FOUR'} 
and \ttt{'FIVE'}, respectively.

\iteme{MINT(121) :} number of separate event classes to initialize 
and mix.
\begin{subentry}
\iteme{= 1 :} the normal value.
\iteme{= 3 :} for a $\gamma\p$ interaction when \ttt{MSTP(14)=10}.
\iteme{= 6 :} for a $\gamma\gamma$ interaction when 
\ttt{MSTP(14)=10}.
\end{subentry}

\iteme{MINT(122) :} event class used in current event for $\gamma\p$ or 
$\gamma\gamma$ events generated with the \ttt{MSTP(14)=10} option;
code as described for \ttt{MSTI(9)}.

\iteme{MINT(123) :} event class used in the current event, with the 
same list of possibilities as for \ttt{MSTP(14)}, except that
options 1, 4 or 10 do not appear. Apart from a different coding, this
is exactly the same information as is available in \ttt{MINT(122)}. 
 
\boxsep
 
\iteme{VINT(1) :}\label{p:VINT} $E_{\mrm{cm}}$, c.m. energy.
 
\iteme{VINT(2) :} $s$ ($=E_{\mrm{cm}}^2$) squared mass of 
complete system.
 
\iteme{VINT(3) :} mass of beam particle.
 
\iteme{VINT(4) :} mass of target particle.
 
\iteme{VINT(5) :} momentum of beam (and target) particle in c.m.
frame.
 
\iteme{VINT(6) - VINT(10) :} $\theta$, $\varphi$ and
\mbox{{\boldmath $\beta$}} for rotation and boost
from c.m. frame to user-specified frame.
 
\iteme{VINT(11) :} $\tau_{\mmin}$.
 
\iteme{VINT(12) :} $y_{\mmin}$.
 
\iteme{VINT(13) :} $\cos\hat{\theta}_{\mmin}$ for
$\cos\hat{\theta} \leq 0$.
 
\iteme{VINT(14) :} $\cos\hat{\theta}_{\mmin}$ for
$\cos\hat{\theta} \geq 0$.
 
\iteme{VINT(15) :} $x^2_{\perp \mmin}$.
 
\iteme{VINT(16) :} $\tau'_{\mmin}$.
 
\iteme{VINT(21) :} $\tau$.
 
\iteme{VINT(22) :} $y$.
 
\iteme{VINT(23) :} $\cos\hat{\theta}$.
 
\iteme{VINT(24) :} $\varphi$ (azimuthal angle).
 
\iteme{VINT(25) :} $x_{\perp}^2$.
 
\iteme{VINT(26) :} $\tau'$.
 
\iteme{VINT(31) :} $\tau_{\mmax}$.
 
\iteme{VINT(32) :} $y_{\mmax}$.
 
\iteme{VINT(33) :} $\cos\hat{\theta}_{\mmax}$ for
$\cos\hat{\theta} \leq 0$.
 
\iteme{VINT(34) :} $\cos\hat{\theta}_{\mmax}$ for
$\cos\hat{\theta} \geq 0$.
 
\iteme{VINT(35) :} $x^2_{\perp \mmax}$.
 
\iteme{VINT(36) :} $\tau'_{\mmax}$.
 
\iteme{VINT(41), VINT(42) :} the momentum fractions $x$ taken by the
partons at the hard interaction, as used e.g. in the 
parton-distribution functions.
 
\iteme{VINT(43) :} $\hat{m} = \sqrt{\hat{s}}$, mass of 
hard-scattering subsystem.
 
\iteme{VINT(44) :} $\hat{s}$ of the hard subprocess ($2 \to 2$ or
$2 \to 1$).
 
\iteme{VINT(45) :} $\hat{t}$ of the hard subprocess ($2 \to 2$ or
$2 \to 1 \to 2$).
 
\iteme{VINT(46) :} $\hat{u}$ of the hard subprocess ($2 \to 2$ or
$2 \to 1 \to 2$).
 
\iteme{VINT(47) :} $\hat{p}_{\perp}$ of the hard subprocess
($2 \to 2$ or $2 \to 1 \to 2$), i.e. transverse momentum evaluated
in the rest frame of the scattering.
 
\iteme{VINT(48) :} $\hat{p}_{\perp}^2$ of the hard subprocess;
see \ttt{VINT(47)}.
 
\iteme{VINT(49) :} $\hat{m}'$, the mass of the complete three- or
four-body final state in $2 \to 3$ or $2 \to 4$ processes.
 
\iteme{VINT(50) :} $\hat{s}' = \hat{m}'^2$; see \ttt{VINT(49)}.
 
\iteme{VINT(51) :} $Q$ of the hard subprocess. The exact definition is
process-dependent, see \ttt{MSTP(32)}.
 
\iteme{VINT(52) :} $Q^2$ of the hard subprocess; see \ttt{VINT(51)}.
 
\iteme{VINT(53) :} $Q$ of the outer hard-scattering subprocess.
Agrees with \ttt{VINT(51)} for a $2 \to 1$ or $2 \to 2$ process.
For a $2 \to 3$ or $2 \to 4$ $\W/\Z$ fusion process, it is set by
the $\W/\Z$ mass scale, and for
subprocesses 121 and 122 by the heavy-quark mass.
 
\iteme{VINT(54) :} $Q^2$ of the outer hard-scattering subprocess;
see \ttt{VINT(53)}.
 
\iteme{VINT(55) :} $Q$ scale used as maximum virtuality in parton
showers. Is equal to \ttt{VINT(53)}, except for 
deep-inelastic-scattering processes when \ttt{MSTP(22)}$ > 0$.
 
\iteme{VINT(56) :} $Q^2$ scale in parton showers; see \ttt{VINT(55)}.
 
\iteme{VINT(57) :} $\alphaem$ value of hard process.
 
\iteme{VINT(58) :} $\alphas$ value of hard process.
 
\iteme{VINT(59) :} $\sin\hat{\theta}$ (cf. \ttt{VINT(23)}); used for
improved numerical precision in elastic and diffractive scattering.
 
\iteme{VINT(61), VINT(62) :} nominal $m^2$ values, i.e. without
initial-state radiation effects, for the two partons entering
the hard interaction.
 
\iteme{VINT(63), VINT(64) :} nominal $m^2$ values, i.e. without 
final-state radiation effects, for the two (or one) partons/particles
leaving the hard interaction.
 
\iteme{VINT(65) :} $\hat{p}_{\mrm{init}}$, i.e. common nominal absolute
momentum of the two partons entering the hard interaction, in their
rest frame.
 
\iteme{VINT(66) :} $\hat{p}_{\mrm{fin}}$, i.e. common nominal absolute
momentum of the two partons leaving the hard interaction, in their
rest frame.

\iteme{VINT(67), VINT(68) :} mass of beam and target particle, as 
\ttt{VINT(3)} and \ttt{VINT(4)}, except that an incoming $\gamma$ is
assigned the $\rho^0$, $\omega$ or $\phi$ mass. Used for elastic 
scattering $\gamma \p \to \rho^0 \p$ and other similar processes.
 
\iteme{VINT(71) :} $\pTmin$ of process, i.e. \ttt{CKIN(3)} or
\ttt{CKIN(5)}, depending on which is larger, and whether the process
is singular in $\pT \to 0$ or not.
 
\iteme{VINT(73) :} $\tau = m^2/s$ value of first resonance, if any;
see \ttt{MINT(73)}.
 
\iteme{VINT(74) :} $m \Gamma/s$ value of first resonance, if any;
see \ttt{MINT(73)}.
 
\iteme{VINT(75) :} $\tau = m^2/s$ value of second resonance, if any;
see \ttt{MINT(74)}.
 
\iteme{VINT(76) :} $m \Gamma/s$ value of second resonance, if any;
see \ttt{MINT(74)}.
 
\iteme{VINT(80) :} correction factor (evaluated in \ttt{PYOFSH}) for
the cross section of resonances produced in $2 \to 2$ processes, if
only some mass range of the full Breit--Wigner shape is allowed by
user-set mass cuts (\ttt{CKIN(2)}, \ttt{CKIN(45) - CKIN(48)}).
 
\iteme{VINT(81) - VINT(84) :} the $\cos\theta$ and $\varphi$ variables
of a true $2 \to 3$ process, where one product is a resonance,
effectively giving $2 \to 4$. The first two are $\cos\theta$ and
$\varphi$ for the resonance decay, the other two ditto for the
effective system formed by the other two particles.
 
\iteme{VINT(85), VINT(86) :} transverse momenta in a true $2 \to 3$
process; one is stored in \ttt{VINT(47)} (that of the
$\Z^0$ in $\g \g \to \Z^0 \Q \Qbar$), while the smaller of the other
two is stored in \ttt{VINT(85)} and the larger in \ttt{VINT(86)}.
 
\iteme{VINT(91), VINT(92) :} gives a dimensionless suppression factor,
to take into account reduction in cross section due to the allowed
channels for a $\W^+\W^+$ or $\W^-\W^-$ pair, respectively, in the
same sense as \ttt{WIDS(24,1)} gives it for a $\W^+\W^-$ pair.
 
\iteme{VINT(98) :} is sum of \ttt{VINT(100)} values for current run.
 
\iteme{VINT(99) :} is weight \ttt{WTXS} returned from \ttt{PYEVWT} call
when \ttt{MSTP(142)}$\geq 1$, otherwise is 1.
 
\iteme{VINT(100) :} is compensating weight \ttt{1./WTXS} that should be
associated with events when \ttt{MSTP(142)=1}, otherwise is 1.
 
\iteme{VINT(108) :} ratio of maximum differential cross section
observed to maximum differential cross section assumed for the
generation; cf. \ttt{MSTP(123)}.
 
\iteme{VINT(109) :} ratio of minimal (negative!) cross section
observed to maximum differential cross section assumed for the
generation; could only become negative if cross sections are
incorrectly included.
 
\iteme{VINT(111) - VINT(116) :} for \ttt{MINT(61)=1} gives
kinematical factors for the different pieces contributing to
$\gammaZ$ or $\gamma^*/\Z^0/\Z'^0$ production, for \ttt{MINT(61)=2}
gives sum of final-state weights for the same; coefficients are
given in the order pure $\gamma^*$, $\gamma^*$--$\Z^0$ interference,
$\gamma^*$--$\Z'^0$ interference, pure $\Z^0$, $\Z^0$--$\Z'^0$
interference and pure $\Z'^0$.
 
\iteme{VINT(117) :} width of $\Z^0$; needed in
$\gamma^*/\Z^0/\Z'^0$ production.
 
\iteme{VINT(131) :} total cross section (in mb) for subprocesses
allowed in the pile-up events scenario according to the
\ttt{MSTP(132)} value.
 
\iteme{VINT(132) :} $\br{n} = $\ttt{VINT(131)}$\times$\ttt{PARP(131)},
cf. \ttt{PARI(91)}.
 
\iteme{VINT(133) :} $\langle n \rangle = \sum_i i \, {\cal P}_i  /
\sum_i {\cal P}_i$ as actually simulated, i.e. $1 \leq i \leq 200$
(or smaller), see \ttt{PARI(92)}.
 
\iteme{VINT(134) :} number related to probability to have event in
beam--beam crossing; is $\exp(-\br{n}) \sum_i \br{n}^i/i!$ for
\ttt{MSTP(133)=1} and $\exp(-\br{n}) \sum_i \br{n}^i/(i-1)!$
for \ttt{MSTP(133)=2}, cf. \ttt{PARI(93)}.
 
\iteme{VINT(138) :} size of the threshold factor (enhancement or
suppression) in the latest event with heavy-flavour production;
see \ttt{MSTP(35)}.
 
\iteme{VINT(141), VINT(142) :} $x$ values for the parton-shower
initiators of the hardest interaction; used to find what is left
for multiple interactions.
 
\iteme{VINT(143), VINT(144) :} $1 - \sum_i x_i$ for all scatterings;
used for rescaling each new $x$-value in the multiple-interaction
parton-distribution-function evaluation.
 
\iteme{VINT(145) :} estimate of total parton--parton cross section for
multiple interactions; used for \ttt{MSTP(82)}$\geq 2$.
 
\iteme{VINT(146) :} common correction factor $f_c$ in the 
multiple-interaction probability; used for \ttt{MSTP(82)}$\geq 2$ 
(part of $e(b)$, see eq.~(\ref{mi:ebenh})).
 
\iteme{VINT(147) :} average hadronic matter overlap; used for
\ttt{MSTP(82)}$\geq 2$ (needed in evaluation of $e(b)$, see 
eq.~(\ref{mi:ebenh})).
 
\iteme{VINT(148) :} enhancement factor for current event in the
multiple-interaction probability, defined as the actual overlap 
divided by the average one; used for \ttt{MSTP(82)}$\geq 2$ 
(is $e(b)$ of eq.~(\ref{mi:ebenh})).
 
\iteme{VINT(149) :} $x_{\perp}^2$ cut-off or turn-off for multiple
interactions. For \ttt{MSTP(82)}$\leq 1$ it is $4 \pTmin^2/s$,
for \ttt{MSTP(82)}$\geq 2$ it is $4 p_{\perp 0}^2/s$.
 
\iteme{VINT(150) :} probability to keep the given event in the
multiple-interaction scenario, as given by the `Sudakov' form factor.
 
\iteme{VINT(151), VINT(152) :} sum of $x$ values for all the
multiple-interaction partons.
 
\iteme{VINT(153) :} current differential cross section value
obtained from \ttt{PYSIGH}; used in multiple interactions only.
 
\iteme{VINT(155), VINT(156) :} the $x$ value of a photon that branches
into quarks or gluons, i.e. $x$ at interface between initial-state QED
and QCD cascades.
 
\iteme{VINT(157), VINT(158) :} the primordial $k_{\perp}$ values
selected in the two beam remnants.
 
\iteme{VINT(159), VINT(160) :} the $\chi$ values selected for beam
remnants that are split into two objects, describing how the energy
is shared (see \ttt{MSTP(92)} and \ttt{MSTP(94)}); is 0 if no 
splitting is needed.
 
\iteme{VINT(161) - VINT(200) :} sum of Cabibbo--Kobayashi--Maskawa
squared matrix elements that a given flavour is allowed to couple to.
Results are stored in format \ttt{VINT(180+KF)} for quark and lepton
flavours and antiflavours (which need not be the same; see
\ttt{MDME(IDC,2)}). For leptons, these factors are normally unity.
 
\iteme{VINT(201) - VINT(220) :} additional variables needed in 
phase-space selection for $2 \to 3$ processes with \ttt{ISET(ISUB)=5}.
Below indices 1, 2 and 3 refer to scattered partons 1, 2 and 3, except
that the $q$ four-momentum variables are $q_1 + q_2 \to q_1' q_2' q_3'$.
All kinematical variables refer to the internal kinematics of
the 3-body final state --- the kinematics of the system as a whole
is described by $\tau'$ and $y$, and the mass distribution of
particle 3 (a resonance) by $\tau$.
\begin{subentry}
\iteme{VINT(201) :} $m_1$.
\iteme{VINT(202) :} $p_{\perp 1}^2$.
\iteme{VINT(203) :} $\varphi_1$.
\iteme{VINT(204) :} $M_1$ (mass of propagator particle).
\iteme{VINT(205) :} weight for the $p_{\perp 1}^2$ choice.
\iteme{VINT(206) :} $m_2$.
\iteme{VINT(207) :} $p_{\perp 2}^2$.
\iteme{VINT(208) :} $\varphi_2$.
\iteme{VINT(209) :} $M_2$ (mass of propagator particle).
\iteme{VINT(210) :} weight for the $p_{\perp 2}^2$ choice.
\iteme{VINT(211) :} $y_3$.
\iteme{VINT(212) :} $y_{3 \mmax}$.
\iteme{VINT(213) :} $\epsilon = \pm 1$; choice between two mirror
solutions $1 \leftrightarrow 2$.
\iteme{VINT(214) :} weight associated to $\epsilon$-choice.
\iteme{VINT(215) :} $t_1 = (q_1 - q_1')^2$.
\iteme{VINT(216) :} $t_2 = (q_2 - q_2')^2$.
\iteme{VINT(217) :} $q_1 q_2'$ four-product.
\iteme{VINT(218) :} $q_2 q_1'$ four-product.
\iteme{VINT(219) :} $q_1' q_2'$ four-product.
\iteme{VINT(220) :} $\sqrt{(m_{\perp 12}^2 - m_{\perp 1}^2 -
m_{\perp 2}^2)^2 - 4 m_{\perp 1}^2 m_{\perp 2}^2}$, where
$m_{\perp 12}$ is the transverse mass of the $q'_1 q'_2$ system.
\end{subentry}
 
\iteme{VINT(221) - VINT(225) :} $\theta$, $\varphi$ and
\mbox{{\boldmath $\beta$}} for rotation and boost
from c.m. frame to hadronic c.m. frame of a lepton--hadron
event.

\iteme{VINT(231) :} $Q^2_{\mmin}$ scale for current 
parton-distribution function set.

\iteme{VINT(281) :} for resolved photon events, it gives the 
ratio between the total $\gamma X$ cross section and the total 
$\pi^0 X$ cross section, where $X$ represents the target particle.

\iteme{VINT(283), VINT(284) :} virtuality scale at which an anomalous 
photon on the beam or target side of the event is being resolved.
More precisely, it gives the $\pT^2$ of the $\gamma \to \q\qbar$
vertex.

\iteme{VINT(285) :} the \ttt{CKIN(3)} value provided by the user 
at initialization; subsequently \ttt{CKIN(3)} may be overwritten 
(for \ttt{MSTP(14)=10}) but \ttt{VINT(285)} stays.

\iteme{VINT(289) :} squared c.m. energy found in \ttt{PYINIT} call.

\iteme{VINT(290) :} the \ttt{WIN} argument of a \ttt{PYINIT} call.

\iteme{VINT(291) - VINT(300) :} the two five-vectors of the two 
incoming particles, as reconstructed in \ttt{PYINKI}.
These may vary from one event to the next.
 
\end{entry}
 
\drawbox{COMMON/PYINT2/ISET(200),KFPR(200,2),COEF(200,20),ICOL(40,4,2)}%
\label{p:PYINT2}
\begin{entry}
\itemc{Purpose:} to store information necessary for efficient
generation of the different subprocesses, specifically type of
generation scheme and coefficients of the Jacobian. Also to store
allowed colour-flow configurations. These variables must not be
changed by you.
 
\iteme{ISET(ISUB) :}\label{p:ISET} gives the type of 
kinematical-variable selection scheme used for subprocess ISUB.
\begin{subentry}
\iteme{= 0 :} elastic, diffractive and low-$\pT$ processes.
\iteme{= 1 :} $2 \to 1$ processes (irrespective of subsequent decays).
\iteme{= 2 :} $2 \to 2$ processes (i.e. the bulk of processes).
\iteme{= 3 :} $2 \to 3$ processes
(like $\q \q' \to \q'' \q''' \H^0$).
\iteme{= 4 :} $2 \to 4$ processes
(like $\q \q' \to \q'' \q''' \W^+ \W^-$).
\iteme{= 5 :} `true' $2 \to 3$ processes, one method.
\iteme{= 6 :} `true' $2 \to 3$ processes, another method;
currently only $\g  \g \to \Z^0 \Q \Qbar$.
\iteme{= 9 :} $2 \to 2$ in multiple interactions ($\pT$ as kinematics
variable).
\iteme{= 11 :} a user-defined process.
\iteme{= -1 :} legitimate process which has not yet been implemented.
\iteme{= -2 :} ISUB is an undefined process code.
\end{subentry}
 
\iteme{KFPR(ISUB,J) :}\label{p:KFPR} give the KF flavour codes for the 
products produced in subprocess ISUB. If there is only one product, the
\ttt{J=2} position is left blank. Also, quarks and leptons assumed
massless in the matrix elements are denoted by 0. The main
application is thus to identify resonances produced
($\Z^0$, $\W^{\pm}$, $\H^0$, etc.).
 
\iteme{COEF(ISUB,J) :}\label{p:COEF} factors used in the Jacobians in 
order to speed
up the selection of kinematical variables. More precisely, the shape
of the cross section is given as the sum of terms with different
behaviour, where the integral over the allowed phase space is
unity for each term. \ttt{COEF} gives the relative strength of these
terms, normalized so that the sum of coefficients for each variable
used is unity. Note that which coefficients are indeed used is
process-dependent.
\begin{subentry}
\iteme{ISUB :} standard subprocess code.
\iteme{J = 1 :} $\tau$ selected according $1/\tau$.
\iteme{J = 2 :} $\tau$ selected according to $1/\tau^2$.
\iteme{J = 3 :} $\tau$ selected according to $1/(\tau(\tau+\tau_R))$,
where $\tau_R = m_R^2/s$ is $\tau$ value of resonance; only used for
resonance production.
\iteme{J = 4 :} $\tau$ selected according to Breit--Wigner of form
$1/((\tau-\tau_R)^2+\gamma_R^2)$, where $\tau_R = m_R^2/s$ is $\tau$
value of resonance and $\gamma_R = m_R \Gamma_R/s$ is its scaled mass
times width; only used for resonance production.
\iteme{J = 5 :} $\tau$ selected according to
$1/(\tau(\tau+\tau_{R'}))$, where $\tau_{R'} = m_{R'}^2/s$ is $\tau$
value of second resonance; only used for simultaneous production of
two resonances.
\iteme{J = 6 :} $\tau$ selected according to second Breit--Wigner of
form $1/((\tau-\tau_{R'})^2+\gamma_{R'}^2)$, where
$\tau_{R'} = m_{R'}^2/s$ is $\tau$ value of second resonance and
$\gamma_{R'} = m_{R'} \Gamma_{R'}/s$ is its scaled
mass times width; is used only for simultaneous production
of two resonances, like $\gamma^*/\Z^0/\Z'^0$.
\iteme{J = 7 :} $\tau$ selected according to $1/(1-\tau)$; only used
when both parton distributions are peaked at $x = 1$.
\iteme{J = 8 :} $y$ selected according to $y - y_{\mmin}$.
\iteme{J = 9 :} $y$ selected according to $y_{\mmax} - y$.
\iteme{J = 10 :} $y$ selected according to $1/\cosh(y)$.
\iteme{J = 11 :} $y$ selected according to $1/(1-\exp(y-y_{\mmax}))$;
only used when beam parton distribution is peaked close to $x = 1$.
\iteme{J = 12 :} $y$ selected according to $1/(1-\exp(y_{\mmin}-y))$;
only used when target parton distribution is peaked close to $x = 1$.
\iteme{J = 13 :} $z = \cos\hat{\theta}$ selected evenly between limits.
\iteme{J = 14 :} $z = \cos\hat{\theta}$ selected according to $1/(a-z)$,
where $a = 1 + 2 m_3^2 m_4^2/\hat{s}^2$, $m_3$ and $m_4$ being the
masses of the two final-state particles.
\iteme{J = 15 :} $z = \cos\hat{\theta}$ selected according to $1/(a+z)$,
with $a$ as above.
\iteme{J = 16 :} $z = \cos\hat{\theta}$ selected according to
$1/(a-z)^2$, with $a$ as above.
\iteme{J = 17 :} $z = \cos\hat{\theta}$ selected according to
$1/(a+z)^2$, with $a$ as above.
\iteme{J = 18 :} $\tau'$ selected according to $1/\tau'$.
\iteme{J = 19 :} $\tau'$ selected according to
$(1 - \tau/\tau')^3/\tau'^2$.
\iteme{J = 20 :} $\tau'$ selected according to $1/(1-\tau')$; only
used when both parton distributions are peaked close to $x = 1$.
\end{subentry}
 
\iteme{ICOL :}\label{p:ICOL} contains information on different 
colour-flow topologies in hard $2 \to 2$ processes.
 
\end{entry}
 
\drawbox{COMMON/PYINT3/XSFX(2,-40:40),ISIG(1000,3),SIGH(1000)}%
\label{p:PYINT3}
\begin{entry}
 
\itemc{Purpose:} to store information on parton distributions,
subprocess cross sections and different final-state relative
weights. These variables must not be changed by you.
 
\iteme{XSFX :}\label{p:XSFX} current values of parton-distribution
functions $xf(x)$ on beam and target side.
 
\iteme{ISIG(ICHN,1) :}\label{p:ISIG} incoming parton/particle on the 
beam side to
the hard interaction for allowed channel number \ttt{ICHN}. The number
of channels filled with relevant information is given by \ttt{NCHN},
one of the arguments returned in a \ttt{PYSIGH} call. Thus only
$1 \leq $\ttt{ICHN}$ \leq $\ttt{NCHN} is filled with relevant
information.
 
\iteme{ISIG(ICHN,2) :} incoming parton/particle on the target side
to the hard interaction for allowed channel number \ttt{ICHN}. See also
comment above.
 
\iteme{ISIG(ICHN,3) :} colour-flow type for allowed channel number
\ttt{ICHN}; see \ttt{MSTI(2)} list. See also above comment. For
`subprocess' 96 uniquely, \ttt{ISIG(ICHN,3)} is also used to
translate information on what is the correct subprocess number
(11, 12, 13, 28, 53 or 68); this is used for reassigning subprocess
96 to either of these.
 
\iteme{SIGH(ICHN) :}\label{p:SIGH} evaluated differential 
cross section for allowed
channel number \ttt{ICHN}, i.e. matrix-element value times
parton distributions, for current kinematical setup (in addition,
Jacobian factors are included in the figures, as used to speed up
generation). See also comment for \ttt{ISIG(ICHN,1)}.
 
\end{entry}
 
\drawbox{COMMON/PYINT4/WIDP(21:40,0:40),WIDE(21:40,0:40),WIDS(21:40,3)}%
\label{p:PYINT4}
\begin{entry}
 
\itemc{Purpose:} to store partial and effective decay widths for the
different resonances. These variables must not be changed by you.
 
\iteme{WIDP(KF,J) :}\label{p:WIDP} gives partial decay widths of 
resonances into different channels (in GeV), given that all physically 
allowed final states are included.
\begin{subentry}
\iteme{KF :} standard KF code for resonance considered. When top is
treated like a resonance (see \ttt{MSTP(48)}) it is stored in position
26. When the fourth generation fermions $\lrm$, $\hrm$, $\chi^-$ and
$\nu_{\chi}$ are treated like resonances (see \ttt{MSTP(49)}) they
are stored in positions 27, 28, 29 and 30, respectively. 
\iteme{J :} enumerates the different decay channels possible for
resonance KF, as stored in the {\Je} \ttt{LUDAT3} common block,
with the first channel in \ttt{J=1}, etc.
\end{subentry}
 
\iteme{WIDE(KF,J) :}\label{p:WIDE} gives effective decay widths of 
resonances into
different channels (in GeV), given the decay modes actually left open
in the current run. The on/off status of decay modes is set by the
\ttt{MDME} switches in {\Je}; see section \ref{ss:parapartdat}.
\begin{subentry}
\iteme{KF :} standard KF code for resonance considered. For comment
about top and fourth generation see \ttt{WIDP} above.
\iteme{J :} enumerates the different decay channels possible for
resonance KF, as stored in the {\Je} \ttt{LUDAT3} common block,
with the first channel in \ttt{J=1}, etc.
\end{subentry}
 
\iteme{WIDS(KF,J) :}\label{p:WIDS} gives a dimensionless suppression 
factor, which
is defined as the ratio of the total width of channels switched on to
the total width of all possible channels (replace width by squared
width for a pair of resonances). The on/off status of channels is set
by the \ttt{MDME} switches in {\Je}; see section \ref{ss:parapartdat}.
The information in \ttt{WIDS} is used e.g. in cross-section
calculations.
\begin{subentry}
\iteme{KF :} standard KF code for resonance considered. For comment
about top and fourth generation see \ttt{WIDP} above.
\iteme{J = 1 :} suppression when a pair of resonances of type KF are
produced together. When an antiparticle exists, the
particle--antiparticle pair (such as $\W^+ \W^-$) is the relevant
combination, else the particle--particle one (such as $\Z^0 \Z^0$).
\iteme{J = 2 :} suppression for a particle of type KF when produced
on its own, or together with a particle of another type.
\iteme{J = 3 :} suppression for an antiparticle of type KF when produced
on its own, or together with a particle of another type.
\end{subentry}
 
\end{entry}
 
\drawbox{COMMON/PYINT5/NGEN(0:200,3),XSEC(0:200,3)}\label{p:PYINT5}
\begin{entry}
 
\itemc{Purpose:} to store information necessary for cross-section
calculation and differential cross-section maximum violation.
These variables must not be changed by you.
 
\iteme{NGEN(ISUB,1) :}\label{p:NGEN} gives the number of times that 
the differential cross section (times Jacobian factors) has been
evaluated for subprocess ISUB, with \ttt{NGEN(0,1)} the sum of
these.
 
\iteme{NGEN(ISUB,2) :} gives the number of times that a kinematical
setup for subprocess ISUB is accepted in the generation procedure,
with \ttt{NGEN(0,2)} the sum of these.
 
\iteme{NGEN(ISUB,3) :} gives the number of times an event of
subprocess type ISUB is generated, with \ttt{NGEN(0,3)} the sum of
these. Usually \ttt{NGEN(ISUB,3) = NGEN(ISUB,2)}, i.e. an accepted
kinematical configuration can normally be used to produce an event.
 
\iteme{XSEC(ISUB,1) :}\label{p:XSEC} estimated maximum differential 
cross section
(times the Jacobian factors used to speed up the generation process)
for the different subprocesses in use, with \ttt{XSEC(0,1)} the sum
of these (except low-$\pT$, i.e. ISUB = 95).
 
\iteme{XSEC(ISUB,2) :} gives the sum of differential cross sections
(times Jacobian factors) for the \ttt{NGEN(ISUB,1)} phase-space
points evaluated so far.
 
\iteme{XSEC(ISUB,3) :} gives the estimated integrated cross section
for subprocess ISUB, based on the statistics accumulated so far,
with \ttt{XSEC(0,3)} the estimated total cross section for all
subprocesses included (all in mb). This is exactly the
information obtainable by a \ttt{PYSTAT(1)} call.
 
\end{entry}
 
\drawboxtwo{COMMON/PYINT6/PROC(0:200)}{CHARACTER PROC*28}%
\label{p:PYINT6}
\begin{entry}
 
\itemc{Purpose:} to store character strings for the different
possible subprocesses; used when printing tables.
 
\iteme{PROC(ISUB) :}\label{p:PROC} name for the different 
subprocesses, according to ISUB code. \ttt{PROC(0)} denotes 
all processes.
 
\end{entry}
 
\drawbox{COMMON/PYINT7/SIGT(0:6,0:6,0:5)}\label{p:PYINT7}
\begin{entry}
 
\itemc{Purpose:} to store information on total, elastic and 
diffractive cross sections. These variables should only be set 
by you for the option \ttt{MSTP(31)=0}; else they should not be
touched. All numbers are given in mb. 
 
\iteme{SIGT(I1,I2,J) :}\label{p:SIGT} the cross section, both total
and subdivided by class (elastic, diffractive etc.). For a photon
to be considered as a VMD meson the cross sections are additionally
split into the contributions from the various meson states. 
\begin{subentry}
\iteme{I1, I2 :} allowed states for the incoming particle on side 
1 and 2, respectively.
\begin{subentry}
\iteme{= 0 :} sum of all allowed states. Except for a photon to
be considered as a VMD meson this is the only nonvanishing entry.
\iteme{= 1 :} the contribution from the $\rho^0$ VMD state.
\iteme{= 2 :} the contribution from the $\omega$ VMD state.
\iteme{= 3 :} the contribution from the $\phi$ VMD state.
\iteme{= 4 :} the contribution from the $\Jpsi$ VMD state.
\iteme{= 5, 6 :} reserved for future use.
\end{subentry}
\iteme{J :} the total and partial cross sections. 
\begin{subentry}
\iteme{= 0 :} the total cross section.
\iteme{= 1 :} the elastic cross section.
\iteme{= 2 :} the single diffractive cross section $AB \to XB$.
\iteme{= 3 :} the single diffractive cross section $AB \to AX$.
\iteme{= 4 :} the double diffractive cross section.
\iteme{= 5 :} the inelastic, non-diffractive cross section.
\end{subentry} 
\itemc{Warning:} If you set these values yourself, it is important
that they are internally consistent, since this is not explicitly
checked by the program. Thus the contributions \ttt{J=}1--5 should
add up to the \ttt{J=}0 one and, for VMD photons, the contributions
\ttt{I=}1--4 should add up to the \ttt{I=}0 one.
\end{subentry}

\end{entry}
 
\drawboxtwo{~COMMON/PYINT8/XPVMD(-6:6),XPANL(-6:6),XPANH(-6:6),%
XPBEH(-6:6),}{\&XPDIR(-6:6)}\label{p:PYINT8}
\begin{entry}

\itemc{Purpose:} to store the various components of the photon 
parton distributions when the \ttt{PYGGAM} routine is called.

\iteme{XPVMD(KFL) :}\label{p:XPVMD} gives distributions of the VMD part 
($\rho^0$, $\omega$ and $\phi$). 
     
\iteme{XPANL(KFL) :}\label{p:XPANL} gives distributions of the anomalous 
part of light quarks ($\d$, $\u$ and $\s$).

\iteme{XPANH(KFL) :}\label{p:XPANH} gives distributions of the anomalous 
part of heavy quarks ($\c$ and $\b$).

\iteme{XPBEH(KFL) :}\label{p:XPBEH} gives Bethe-Heitler distributions of 
heavy quarks ($\c$ and $\b$). This provides an alternative to \ttt{XPANH}, 
i.e. both should not be used at the same time.

\iteme{XPDIR(KFL) :}\label{p:XPDIR} gives direct correction to the 
production of light quarks ($\d$, $\u$ and $\s$). This term is 
nonvanishing only in the {\MSbar} scheme, and is applicable for 
$F_2^{\gamma}$ rather than for the parton distributions themselves.

\end{entry}

\boxsep
 
Finally, in addition a number of routines and common blocks with names
beginning with \ttt{RK} come with the program. These contain the
matrix-element evaluation for the process
$\g \g \to \Z \q \qbar$, based on a
program of Ronald Kleiss, with only minor modifications.
 
\subsection{Examples}
 
The program is built as a slave system, i.e. you supply the main
program, which calls on the {\Py} and {\Je} routines to perform
specific tasks and then resumes control.
 
A typical program for the analysis of collider events at 630 GeV c.m.
energy with a minimum $\pT$ of 10 GeV/c at the hard scattering
(because of initial-state radiation, fragmentation effects, etc.,
the actual $\pT$ cut-off will be smeared around this value) might look
like
\begin{verbatim}
      COMMON/LUJETS/N,K(4000,5),P(4000,5),V(4000,5)
      COMMON/PYSUBS/MSEL,MSUB(200),KFIN(2,-40:40),CKIN(200)
      COMMON/PYPARS/MSTP(200),PARP(200),MSTI(200),PARI(200)
      ...                   ! set all common block variables that
      ...                   ! did not have desired default values
      CKIN(3)=10.           ! lower p_T cut-off
      CALL PYINIT('CMS','p','pbar',630.)         ! initialize
      ...                   ! initialize analysis statistics
      DO 100 IEVENT=1,1000                ! loop over events
      CALL PYEVNT                         ! generate event
      IF(IEVENT.EQ.1) CALL LULIST(1)      ! list first event
      ...                   ! insert desired analysis chain for
      ...                   ! each event
  100 CONTINUE
      CALL PYSTAT(1)        ! print cross sections
      ...                   ! user output
      END
\end{verbatim}
 
\clearpage
 
\section{Initial- and Final-State Radiation}
\label{s:showinfi}
 
Starting from the hard interaction, initial- and final-state
radiation corrections may be added. This is normally done by
making use of the parton-shower language --- only for the
$\ee \to \q \qbar$ process does {\Je} offer a matrix-element
option (described in section \ref{ss:eematrix}).
The algorithms used to generate initial- and final-state showers
are rather different, and are therefore described separately
below, starting with the conceptually easier final-state one.
Before that, some common elements are introduced.
 
The main reference for final-state showers is ref. \cite{Ben87a} and
for initial-state ones ref. \cite{Sjo85}.
 
\subsection{Shower Evolution}
 
In the leading log picture, a shower may be viewed as a sequence
of $1 \to 2$ branchings $a \to bc$. Here $a$ is called the mother
and $b$ and $c$ the two daughters. Each daughter is free to branch
in its turn, so that a tree-like stucture can evolve. We will use
the work `parton' for all the objects $a$, $b$ and $c$ involved
in the branching process, i.e. not only for quarks and gluons
but also for leptons and photons. The branchings included in the
program are $\q \to \q \g$, $\g \to \g \g$, $\g \to \q \qbar$,
$\q \to \q \gamma$ and $\ell \to \ell \gamma$. Photon branchings,
i.e. $\gamma \to \q \qbar$ and $\gamma \to \ell \br{\ell}$, have not
been included so far, since they are reasonably rare and since no
urgent need for them has been perceived.
 
\subsubsection{The evolution equations}
 
In the shower formulation, the kinematics of each branching is
given in terms of two variables, $Q^2$ and $z$. Slightly different
interpretations may be given to these variables, and indeed this is
one main area where the various programs on the market differ.
$Q^2$ has dimensions of squared mass, and is related to the
mass or transverse momentum scale of the branching. $z$ gives the
sharing of the $a$ energy and momentum between the two daughters,
with parton $b$ taking a fraction $z$ and parton $c$ a fraction
$1-z$. To specify the kinematics, an azimuthal angle
$\varphi$ of the $b$ around the $a$ direction is needed in addition; 
normally $\varphi$ is chosen to be isotropically distributed, 
although options for non-isotropic distributions exist.
 
The probability for a parton to branch is given by the evolution
equations (also called DGLAP or Altarelli--Parisi \cite{Gri72,Alt77}).
It is convenient to introduce
\begin{equation}
t = \ln(Q^2/\Lambda^2) ~~~ \Rightarrow ~~~
\d t = \d \ln(Q^2) = \frac{\d Q^2}{Q^2} ~,
\end{equation}
where $\Lambda$ is the QCD $\Lambda$ scale in $\alphas$. Of course,
this choice is more directed towards the QCD parts of the shower,
but it can be used just as well for the QED ones. In terms of the two
variables $t$ and $z$, the differential probability $\d {\cal P}$ for
parton $a$ to branch is now
\begin{equation}
\d {\cal P}_a = \sum_{b,c} \frac{\alpha_{abc}}{2 \pi} \,
P_{a \to bc}(z) \, \d t \, \d z ~.
\label{sh:Patobc}
\end{equation}
Here the sum is supposed to run over all allowed branchings,
for a quark $\q \to \q \g$ and $\q \to \q \gamma$, and so on. The
$\alpha_{abc}$ factor is $\alphaem$ for QED branchings and
$\alphas$ for QCD ones (to be evaluated at some suitable scale,
see below).
 
The splitting kernels $P_{a \to bc}(z)$ are
\begin{eqnarray}
 P_{\q \to \q\g}(z)&=&C_F \, \frac{1+z^2}{1-z} ~,
                     \nonumber \\
 P_{\g \to \g\g}(z)&=&N_C \, \frac{(1-z(1-z))^2}{z(1-z)} ~,
                     \nonumber \\
 P_{\g \to \q\qbar}(z)&=&T_R \, (z^2 + (1-z)^2) ~,
                     \nonumber \\
 P_{\q \to \q\gamma}(z)&=&e_{\q}^2 \, \frac{1+z^2}{1-z} ~,
                     \nonumber \\
 P_{\ell \to \ell\gamma}(z)&=&e_{\ell}^2 \, \frac{1+z^2}{1-z} ~,
\label{sh:APker}
\end{eqnarray}
with $C_F = 4/3$, $N_C = 3$, $T_R = n_f/2$ (i.e. $T_R$ receives
a contribution of $1/2$ for each allowed $\q\qbar$ flavour),
and $e_{\q}^2$ and $e_{\ell}^2$ the squared electric charge
($4/9$ for $\u$-type quarks, $1/9$ for $\d$-type ones, and 1 for
leptons).
 
Persons familiar with analytical calculations may wonder why the
`+ prescriptions' and $\delta(1-z)$ terms of the splitting kernels
in eq.~(\ref{sh:APker}) are missing. These complications fulfil
the task of ensuring flavour and energy conservation in the analytical
equations. The corresponding problem is solved trivially in
Monte Carlo programs, where the shower evolution is traced in detail,
and flavour and four-momentum are conserved at each branching.
The legacy left is the need to introduce a cut-off on the allowed range
of $z$ in splittings, so as to avoid the singular regions corresponding
to excessive production of very soft gluons.
 
Also note that $P_{\g \to \g\g}(z)$ is given here with a factor
$N_C$ in front, while it is sometimes shown with $2 N_C$. The
confusion arises because the final state contains two identical
partons. With the normalization above, $P_{a \to bc}(z)$
is interpreted as the branching probability for the original parton
$a$. On the other hand, one could also write down the probability
that a parton $b$ is produced with a fractional energy $z$. Almost
all the above kernels can be used unchanged also for this purpose,
with the obvious symmetry $P_{a \to bc}(z) = P_{a \to cb}(1-z)$.
For $\g \to \g\g$, however, the total probability to find a gluon
with energy fraction $z$ is the sum of the probability to find either
the first or the second daughter there, and that gives the factor of
2 enhancement.
 
\subsubsection{The Sudakov form factor}
\label{ss:sudakov}
 
The $t$ variable fills the function of a kind of time for the shower
evolution. In final-state showers, $t$ is constrained to be gradually
decreasing away from the hard scattering, in initial-state ones to
be gradually increasing towards the hard scattering.  This does not
mean that an individual parton runs through a range of $t$ values: 
in the end, each parton is associated with a fixed $t$ value, and the
evolution procedure is just a way of picking that value. It is only
the ensemble of partons in many events that evolves continuously with
$t$, cf. the concept of parton distributions.
 
For a given $t$ value we define the integral of the branching
probability over all allowed $z$ values,
\begin{equation}
{\cal I}_{a \to bc}(t) = \int_{z_{-}(t)}^{z_{+}(t)} \d z \,
\frac{\alpha_{abc}}{2 \pi} \, P_{a \to bc}(z) ~.
\end{equation}
The na\"{\i}ve probability that a branching occurs during a small range
of $t$ values, $\delta t$, is given by
$\sum_{b,c} {\cal I}_{a \to bc}(t) \, \delta t$, and thus the
probability for no emission by
$1 - \sum_{b,c} {\cal I}_{a \to bc}(t) \, \delta t$.
 
If the evolution of parton $a$ starts at a `time' $t_0$, the
probability that the parton has not yet branched at a `later time'
$t > t_0$ is given by the product of the probabilities that it did not
branch in any of the small intervals $\delta t$ between $t_0$ and $t$.
In other words, letting $\delta t \to 0$, the no-branching probability
exponentiates:
\begin{equation}
{\cal P}_{\mrm{no-branching}}(t_0,t) =
\exp \left\{ - \int_{t_0}^t \d t' \, \sum_{b,c}
{\cal I}_{a \to bc}(t') \right\} = S_a(t) ~.
\label{sh:Pnobranch}
\end{equation}
Thus the actual probability that a branching of $a$ occurs at $t$
is given by
\begin{equation}
\frac{\d {\cal P}_a}{\d t} =
- \frac{\d {\cal P}_{\mrm{no-branching}}(t_0,t)}{\d t} =
\left( \sum_{b,c} {\cal I}_{a \to bc}(t) \right)
\exp \left\{ - \int_{t_0}^t \d t' \, \sum_{b,c}
{\cal I}_{a \to bc}(t') \right\} ~.
\label{sh:Pbranch}
\end{equation}
 
The first factor is the na\"{\i}ve branching probability, the second
the suppression due to the conservation of total
probability: if a parton has already branched at a `time' $t' < t$,
it can no longer branch at $t$. This is nothing but the exponential
factor that is familiar from radioactive decay. In parton-shower
language the exponential factor
$S_a(t) = {\cal P}_{\mrm{no-branching}}(t_0,t)$ is referred to as
the Sudakov form factor \cite{Sud56}.
 
The ordering in terms of increasing $t$ above
is the appropriate one for initial-state showers. In 
final-state showers the evolution is from an initial $t_{\mmax}$ 
(set by the hard scattering) and towards smaller $t$. In that case 
the integral from $t_0$ to $t$ in eqs.~(\ref{sh:Pnobranch}) 
and (\ref{sh:Pbranch}) is replaced by an integral from $t$ 
to $t_{\mmax}$. Since, by convention,
the Sudakov factor is still defined from the lower cut-off $t_0$,
i.e. gives the probability that a parton starting at scale $t$ will
not have branched by the lower cut-off scale $t_0$, the no-branching
factor is actually 
${\cal P}_{\mrm{no-branching}}(t_{\mmax},t) = 
S_a(t_{\mmax})/S_a(t)$.
 
We note that the above structure is exactly of the kind discussed
in section \ref{ss:vetoalg}. The veto algorithm is therefore
extensively used in the Monte Carlo simulation of parton showers.
 
\subsubsection{Matching to the hard scattering}
\label{sss:showermatching}
 
The evolution in $Q^2$ is begun from some maximum scale
$Q_{\mmax}^2$ for final-state parton showers, and is terminated
at (a possibly different) $Q_{\mmax}^2$ for initial-state showers.
In general $Q_{\mmax}^2$ is not known. Indeed, since
the parton-shower language does not guarantee agreement with
higher-order matrix-element results, neither in absolute
shape nor normalization, there is no unique prescription for
a `best' choice. Generically $Q_{\mmax}$ should be of the order
of the hard-scattering scale, i.e. the largest virtuality
should be associated with the hard scattering, and initial- and
final-state parton showers should only involve virtualities
smaller than that. This may be viewed just as a matter of sound
bookkeeping: in a $2 \to n$ graph, a $2 \to 2$ hard-scattering
subgraph could be chosen several different ways, but if all 
the possibilities were to be generated then the cross section
would be double-counted. Therefore one should define the $2 \to 2$
`hard' piece of a $2 \to n$ graph as the one that involves the 
largest virtuality. 
 
Of course, the issue of double-counting depends a bit on what
processes are actually generated in the program. If one
considers a $\q \qbar \g$ final state in hadron colliders,
this could come either as final-state radiation off a
$\q \qbar$ pair, or by a gluon splitting in a $\q \qbar$ pair,
or many other ways, so that the danger of double-counting is
very real. On the other hand, consider the production of a
low-$\pT$, low-mass Drell--Yan pair of leptons, together with two
quark jets. Such a process in principle could proceed by having
a $\gamma^*$ emitted off a quark leg, with a quark--quark
scattering as hard interaction. However, since this process is
not included in the program, there is no actual danger of
(this particular) double-counting, and so the scale of evolution
could be picked larger than the mass of the Drell--Yan pair, at 
least by some amount.
 
For most $2 \to 2$ scattering processes in {\Py}, the
$Q^2$ scale of the hard scattering is chosen to be
$Q_{\mrm{hard}}^2 = \pT^2$ (when the final-state particles are
massless, otherwise masses are added). In final-state showers, where
$Q$ is associated with the mass of the branching parton,
transverse momenta generated in the shower are constrained by
$\pT < Q/2$. An ordering that shower $\pT$ be smaller than
the hard-scattering $\pT$ therefore corresponds roughly
to $Q_{\mmax}^2 = 4 Q_{\mrm{hard}}^2$, which is the default 
assumption. In principle, the constraints are slightly different 
for initial-state showers, but not enough to warrant a separate 
$Q_{\mmax}$ choice.
 
The situation is rather better for the final-state showers in the
decay of any colour-singlet particles, such as the $\Z^0$ or the
$\H^0$, either as part of a hard $2 \to 1 \to 2$ process, or
anywhere else in the final state. Then we know that $Q_{\mmax}$
has to be put equal to the particle mass. It is also possible
to match the parton-shower evolution to the first-order
matrix-element results. In the program this is done under the
assumption that the resonance has spin one, and this approach is known
to work very well for $\gammaZ$. The machinery is not fully correct
for the spin-zero $\H^0$, but should also there provide a rather good
description.
 
QCD processes such as $\q\g \to \q\g$ pose a special problem when 
the scattering angle is small. Coherence effects (see below) may
then restrict the emission further than what is just given by 
the $Q_{\mmax}$ scale introduced above. This is most easily viewed
in the rest frame of the $2 \to 2$ hard scattering subprocess.
Some colours flow from the initial to the final state. The radiation
associated with such a colour flow should be restricted to a cone 
with opening angle given by the difference between the original and 
the final colour directions; there is one such cone around the 
incoming parton for initial state radiation and one around the
outgoing parton for final state radiation. Colours that are 
annihilated or created in the process effectively correspond to an
opening angle of 180$^{\circ}$ and therefore the emission is not
constrained for these. For a gluon, which have two colours and 
therefore two different cones, a random choice is made between the
two for the first branching. Further, coherence effects also imply
azimuthal anisotropies of the emission inside the allowed cones.

\subsection{Final-State Showers}
 
Final-state showers are time-like, i.e. all virtualities
$m^2 = E^2 - \mbf{p}^2 \geq 0$. The maximum allowed virtuality
scale $Q^2_{\mmax}$ is set by the hard-scattering process, and
thereafter the virtuality is decreased in each subsequent
branching, down to the cut-off scale $Q_0^2$. This cut-off scale
is used to regulate both soft and collinear divergences in the
emission probabilities.
 
The main points of the {\Je} showering algorithm are as follows.
\begin{Itemize}
\item It is a leading-log algorithm, of the improved, coherent kind,
i.e. with angular ordering.
\item It can be used for an arbitrary initial pair of partons
or, in fact, for any one, two or three given entities (including
hadrons and gauge bosons) although only quarks, gluons and
leptons can initiate a shower.
\item The pair of showering partons may be given in any frame, but
the evolution is carried out in the c.m. frame of the showering
partons.
\item Energy and momentum are conserved exactly at each step of the
showering process.
\item If the initial pair is $\q\qbar$ or $\ell^+ \ell^-$ (coming
from a resonance decay) an additional rejection technique is used
in the first branching of each of the two original partons, so as
to reproduce the lowest-order differential 3-jet cross section.
\item In subsequent branchings, angular
ordering (coherence effects) is imposed.
\item Gluon helicity effects, i.e. correlations between the
production plane and the decay plane of a gluon, can be included.
\item The first-order
$\alphas$ expression is used, with the $Q^2$ scale given by (an
approximation to) the squared transverse momentum of a branching.
The default $\Lambda_{\mrm{QCD}}$,
which should not be regarded as a proper 
$\Lambda_{\br{\mrm{MS}}}$, is 0.4 GeV.
\item The parton shower is by default
cut off at a mass scale of 1 GeV.
\end{Itemize}
Let us now proceed with a more detailed description.
 
\subsubsection{The choice of evolution variable}
 
In the {\Je} shower algorithm, the evolution variable $Q^2$ is
associated with the squared mass of the branching parton,
$Q^2 = m_a^2$ for a branching $a \to bc$. As a consequence,
$t = \ln(Q^2/\Lambda^2) = \ln(m_a^2/\Lambda^2)$.
This $Q^2$ choice is not unique, and indeed other programs have
other definitions: \tsc{Herwig} uses $Q^2 \approx m^2/(2z(1-z))$
\cite{Mar88} and \tsc{Ariadne}
$Q^2 = \pT^2 \approx z(1-z)m^2$ \cite{Pet88}.
 
With $Q$ a mass scale, the lower cut-off $Q_0$ is one in mass.
To be more precise, in a QCD shower, the $Q_0$ parameter is used
to derive effective masses
\begin{eqnarray}
m_{\mrm{eff},\g} & = & \frac{1}{2} Q_0 ~, \nonumber \\
m_{\mrm{eff},\q} & = & \sqrt{ m_{\q}^2 + \frac{1}{4} Q_0^2 } ~,
\end{eqnarray}
where the $m_{\q}$ have been chosen as typical current-algebra quark
masses. A parton cannot branch unless its mass is at least the sum
of the lightest pair of allowed decay products, i.e. the minimum
mass scale at which a branching is possible is
\begin{eqnarray}
m_{\mmin,\g} & = & 2 \, m_{\mrm{eff},\g} = Q_0 ~, \nonumber \\
m_{\mmin,\q} & = & m_{\mrm{eff},\q} + m_{\mrm{eff},\g} \geq Q_0 ~.
\end{eqnarray}
The above masses are used to constrain the allowed range of
$Q^2$ and $z$ values. However, once it has been decided that a
parton cannot branch any further, that parton is put on the
mass shell, i.e. `final-state' gluons are massless.
 
When also photon emission is included, a separate $Q_0$ scale is
introduced for the QED part of the shower, exactly reproducing
the QCD one above \cite{Sjo92c}. By default the two $Q_0$ scales 
are chosen equal, and have the value 1 GeV. If anything, one would 
be inclined to allow a cut-off lower for photon emission than for
gluon one. In that case the allowed $z$ range of photon emission
would be larger than that of gluon emission, and at the end of
the shower evolution only photon emission would be allowed.
 
Photon and gluon emission differ fundamentally in that photons appear
as physical particles in the final state, while gluons are confined.
For photon emission off quarks, however, the confinement forces
acting on the quark may provide an effective photon emission
cut-off at larger scales than the bare quark mass. Soft and collinear
photons could also be emitted by the final-state charged hadrons;
the matching between emission off quarks and off hadrons is a
delicate issue, and we therefore do not attempt to address the 
soft-photon region.
 
For photon emission off leptons, there is no need to introduce any
collinear emission cut-off beyond what is given by the lepton mass,
but we keep the same cut-off approach as for quarks: firstly,
the program is not aimed at high-precision studies of lepton
pairs (where interference terms between initial- and final-state
radiation also would have to be included); secondly, most experimental
procedures would include the energy of collinear photons into the
effective energy of a final-state lepton.
 
\subsubsection{The choice of energy splitting variable}
 
The final-state radiation
machinery is always applied in the c.m. frame of the hard scattering,
from which normally emerges a pair of evolving partons.
Occasionally there may be one evolving parton recoiling against a
non-evolving one, as in $\q \qbar \to \g \gamma$, where only the
gluon evolves in the final state, but where the energy of the photon
is modifed by the branching activity of the gluon. (With only one
evolving parton and nothing else, it would not be possible to
conserve energy and momentum when the parton is assigned a mass.)
Thus, before the evolution is performed, the parton pair is boosted
to their common c.m. frame, and rotated to sit along the $z$ axis.
After the evolution, the full parton shower is rotated and boosted
back to the original frame of the parton pair.
 
The interpretation of the energy and momentum splitting variable
$z$ is not unique, and in fact the program allows the possibility
to switch between four different alternatives \cite{Ben87a},
`local' and `global' $z$ definition combined with `constrained'
or `unconstrained' evolution. In all
four of them, the $z$ variable is interpreted as an energy fraction,
i.e. $E_b = z E_a$ and $E_c = (1-z) E_a$. In the `local' choice of
$z$ definition, energy fractions are defined in the rest frame
of the grandmother, i.e. the mother of parton $a$. The preferred
choice is the `global' one, in which energies are always evaluated
in the c.m. frame of the hard scattering. The two definitions agree
for the branchings of the partons that emerge directly from the hard
scattering, since the hard scattering itself is considered to be the
`mother' of the first generation of partons. For instance, in
$\Z^0 \to \q\qbar$ the $\Z^0$ is considered the mother of the $\q$
and $\qbar$, even though the branching is not handled by the 
parton-showering machinery. The `local' and `global' definitions 
diverge for subsequent branchings, where the `global' tends to allow 
more shower evolution.
 
In a branching $a \to bc$ the kinematically allowed range of
$z = z_a$ values, $z_{-} < z < z_{+}$, is given by
\begin{equation}
z_{\pm} = \frac{1}{2} \left\{ 1 + \frac{m_b^2 - m_c^2}{m_a^2} \pm
\frac{|\mbf{p}_a|}{E_a} \, \frac{\sqrt{ (m_a^2 - m_b^2 - m_c^2)^2
- 4 m_b^2 m_c^2}}{m_a^2} \right\} ~.
\label{sh:zbounds}
\end{equation}
With `constrained' evolution, these bounds are respected in the
evolution. The cut-off masses $m_{\mrm{eff},b}$ and $m_{\mrm{eff},c}$ 
are used to
define the maximum allowed $z$ range, within which $z_a$ is chosen,
together with the $m_a$ value. In the subsequent evolution of $b$ and
$c$, only pairs of $m_b$ and $m_c$ are allowed for which the
already selected $z_a$ fulfils the constraints in 
eq.~(\ref{sh:zbounds}).
 
For `unconstrained' evolution, which is the preferred alternative,
one may start off by assuming the daughters to be massless, so that the
allowed $z$ range is
\begin{equation}
z_{\pm} = \frac{1}{2} \left\{ 1 \pm \frac{|\mbf{p}_a|}{E_a}
\theta(m_a - m_{\mmin,a}) \right\} ~,
\label{sh:zrange}
\end{equation}
where $\theta(x)$ is the step function, $\theta(x) = 1$ for $x > 0$
and $\theta(x) = 0$ for $x < 0$. The decay kinematics into two
massless four-vectors $p_b^{(0)}$ and $p_c^{(0)}$ is now
straightforward. Once $m_b$ and $m_c$ have been found from the
subsequent evolution, subject only to the constraints
$m_b < z_a E_a$, $m_c < (1-z_a) E_a$ and $m_b + m_c < m_a$,
the actual massive four-vectors may be defined as
\begin{equation}
p_{b,c} = p_{b,c}^{(0)} \pm (r_c p_c^{(0)} - r_b p_b^{(0)}) ~,
\label{sh:pshowershift}
\end{equation}
where
\begin{equation}
r_{b,c} = \frac{m_a^2 \pm (m_c^2 -m_b^2) -
\sqrt{ (m_a^2 - m_b^2 - m_c^2)^2 - 4 m_b^2 m_c^2}}{2 m_a^2} ~.
\end{equation}
In other words, the meaning of $z_a$ is somewhat reinterpreted
{\it post facto}. Needless to say, the `unconstrained' option allows
more branchings to take place than the `constrained' one.
In the following discussion we will only refer to the
`global, unconstrained' $z$ choice.
 
\subsubsection{First branchings and matrix-element matching}
 
The final-state evolution is normally started from some initial
parton pair $1 + 2$, at a $Q_{\mmax}^2$ scale determined by
deliberations already discussed. When the evolution of parton 1
is considered, it is assumed that parton 2 is massless,
so that the parton 1 energy and momentum are simple
functions of its mass (and of the c.m. energy of the pair, which is
fixed), and hence also the allowed $z_1$ range for splittings is a
function of this mass,
eq.~(\ref{sh:zrange}). Correspondingly, parton 2 is evolved under the
assumption that parton 1 is massless. After both partons have been
assigned masses, their correct energies may be found, which are
smaller than originally assumed. Therefore the allowed $z$ ranges
have shrunk, and it may happen that a branching has been assigned
a $z$ value outside this range. If so, the parton is evolved
downwards in mass from the rejected mass value; if both $z$ values
are rejected, the parton with largest mass is evolved further.
It may also happen that the sum of $m_1$ and $m_2$ is larger
than the c.m. energy, in which case the one with the larger mass
is evolved downwards. The checking and evolution steps are
iterated until an acceptable set of $m_1$, $m_2$, $z_1$
and $z_2$ has been found.
 
The procedure is an extension of the veto
algorithm, where an initial overestimation of the allowed $z$
range is compensated by rejection of some branchings. One should
note, however, that the veto algorithm is not strictly
applicable for the coupled evolution in two variables ($m_1$
and $m_2$), and that therefore some arbitrariness is involved.
This is manifest in the choice of which parton will be evolved
further if both $z$ values are unacceptable, or if the mass sum
is too large.
 
For quark and lepton pairs which come from the decay of a
colour-singlet particle, the first branchings are matched to
the explicit first-order matrix elements for gauge boson decays.
This is also done, e.g. in $\H^0$ decays, which has spin 0 rather
than 1, and for which in principle therefore the matrix elements
are slightly different.
 
The matching is based on a mapping of the parton-shower variables
on to the 3-jet phase space. To produce a 3-jet event,
$\gammaZ \to \q(p_1) \qbar(p_2) \g(p_3)$, in the shower language,
one will pass through an intermediate
state, where either the $\q$ or the $\qbar$ is off the mass shell.
If the former is the case then
\begin{eqnarray}
m^2 & = & (p_1 + p_3)^2 = E_{\mrm{cm}}^2 (1 - x_2) ~, \nonumber \\
z   & = & \frac{E_1}{E_1 + E_3} = \frac{x_1}{x_1 + x_3} =
\frac{x_1}{2-x_2} ~,
\label{sh:MEfrPS}
\end{eqnarray}
where $x_i = 2 E_i/E_{\mrm{cm}}$. The $\qbar$ emission case is obtained
with $1 \leftrightarrow 2$. The parton-shower splitting expression
in terms of $m^2$ and $z$, eq.~(\ref{sh:Patobc}), can therefore be
translated into the following differential 3-jet rate:
\begin{eqnarray}
\frac{1}{\sigma} \, \frac{\d \sigma_{\mrm{PS}}}{\d x_1 \, \d x_2} 
& = & \frac{\alphas}{2 \pi} \, C_F \, \frac{1}{(1-x_1)(1-x_2)}
\times
\nonumber \\
& \times &
\left\{ \frac{1-x_1}{x_3} \left( 1 + \left( \frac{x_1}{2-x_2}
\right)^2 \right) + \frac{1-x_2}{x_3} \left( 1 +
\left( \frac{x_2}{2-x_1} \right)^2 \right) \right\} ~,
\label{sh:PSwt}
\end{eqnarray}
where the first term inside the curly bracket comes from emission
off the quark and the second term from emission off the antiquark.
The corresponding expression in matrix-element language is
\begin{equation}
\frac{1}{\sigma} \, \frac{\d \sigma_{\mrm{ME}}}{\d x_1 \, \d x_2} =
\frac{\alphas}{2 \pi} \, C_F \, \frac{1}{(1-x_1)(1-x_2)}
\left\{ x_1^2 +x_2^2 \right\} ~.
\label{sh:MEwt}
\end{equation}
With the kinematics choice of \tsc{Jetset},
the matrix-element expression is always smaller than the parton-shower
one. It is therefore possible to run the shower as usual,
but to impose an extra weight factor 
$\d \sigma_{\mrm{ME}} / \d \sigma_{\mrm{PS}}$,
which is just the ratio of the expressions in curly brackets.
If a branching is rejected, the evolution is continued from the
rejected $Q^2$ value onwards (the veto algorithm). The weighting
procedure is applied to the first branching of both the $\q$ and the
$\qbar$, in each case with the (nominal) assumption that none of
the other partons branch (neither the sister nor the daughters),
so that the relations of eq.~(\ref{sh:MEfrPS}) are applicable.
 
If a photon is emitted instead of a gluon, the emission rate in
parton showers is given by
\begin{eqnarray}
\frac{1}{\sigma} \, \frac{\d \sigma_{\mrm{PS}}}{\d x_1 \, \d x_2} 
& = & \frac{\alphaem}{2 \pi} \, \frac{1}{(1-x_1)(1-x_2)} 
\times \nonumber \\
& \times &
\left\{ e_{\q}^2 \, \frac{1-x_1}{x_3} \left( 1 + \left(
\frac{x_1}{2-x_2} \right)^2 \right) + e_{\qbar}^2 \, \frac{1-x_2}{x_3}
\left( 1 + \left( \frac{x_2}{2-x_1} \right)^2 \right) \right\} ~,
\end{eqnarray}
and in matrix elements by \cite{Gro81}
\begin{equation}
\frac{1}{\sigma} \, \frac{\d \sigma_{\mrm{ME}}}{\d x_1 \, \d x_2} =
\frac{\alphaem}{2 \pi} \, \frac{1}{(1-x_1)(1-x_2)}
\left\{ \left( e_{\q} \, \frac{1-x_1}{x_3} - e_{\qbar} \,
\frac{1-x_2}{x_3} \right)^2 \left( x_1^2 +x_2^2 \right) \right\} ~.
\end{equation}
As in the gluon emission case, a weighting factor
$\d \sigma_{\mrm{ME}} / \d \sigma_{\mrm{PS}}$ can therefore be applied 
when either the original $\q$ ($\ell$) or the original $\qbar$
($\br{\ell}$) emits a photon. For a
neutral resonance, such as $\Z^0$, where $e_{\qbar} = - e_{\q}$,
the above expressions simplify and one recovers
exactly the same ratio $\d \sigma_{\mrm{ME}} / \d \sigma_{\mrm{PS}}$ 
as for gluon emission.
 
Compared with the standard matrix-element treatment, a few
differences remain. The shower one automatically contains the
Sudakov form factor and an $\alphas$ running as a function
of the $\pT^2$ scale of the branching.
The shower also allows all partons to evolve
further, which means that the na\"{\i}ve kinematics assumed for a
comparison with matrix elements is modified by subsequent
branchings, e.g. that the energy of parton 1 is reduced when
parton 2 is assigned a mass. All these effects are formally of
higher order, and so do not affect a first-order comparison.
This does not mean that the corrections need be small, but experimental
results are encouraging: the approach outlined does every bit as
good as explicit second-order matrix elements for the description
of 4-jet production.
 
\subsubsection{Subsequent branches and angular ordering}
 
The shower evolution is (almost) always done on a pair of partons,
so that energy and momentum can be conserved. In the first
step of the evolution, the two original partons thus undergo
branchings $1 \to 3 + 4$ and $2 \to 5 + 6$. As described
above, the allowed $m_1$, $m_2$, $z_1$ and $z_2$ ranges
are coupled by kinematical constraints. In the second step, 
the pair $3 + 4$ is evolved and,
separately, the pair $5 + 6$. Considering only the former (the
latter is trivially obtained by symmetry), the partons thus have
nominal initial energies $E_3^{(0)} = z_1 E_1$ and
$E_4^{(0)} = (1-z_1) E_1$, and maximum allowed virtualities
$m_{\mmax,3} = \min(m_1,E_3^{(0)})$ and
$m_{\mmax,4} = \min(m_1,E_4^{(0)})$. Initially partons 3 and 4 are
evolved separately, giving masses $m_3$ and $m_4$ and splitting
variables $z_3$ and $z_4$. If $m_3 + m_4 > m_1$,
the parton of 3 and 4 that has the largest ratio of 
$m_i/m_{\mmax,i}$ is
evolved further. Thereafter eq.~(\ref{sh:pshowershift}) is used to
construct corrected energies $E_3$ and $E_4$, and the $z$
values are checked for consistency. If a branching has to be
rejected because the change of parton energy puts $z$ outside the
allowed range, the parton is evolved further.
 
This procedure can then be iterated for the evolution of the two
daughters of parton 3 and for the two of parton 4, etc., until each
parton reaches the cut-off mass $m_{\mmin}$. Then the parton is put on
the mass shell.
 
The model, as described so far, produces so-called conventional
showers, wherein masses are strictly decreasing in the shower
evolution. Emission angles are decreasing only in an average sense,
however, which means that also fairly `late' branchings can give
partons at large angles. Theoretical studies beyond the leading-log
level show that this is not correct \cite{Mue81}, but that destructive
interference effects are large in the region of non-ordered
emission angles. To a very good first approximation, these so-called
coherence effects can be taken into account in parton shower
programs by requiring a strict ordering in terms of decreasing
emission angles.
 
The coherence phenomenon is known already from QED. One
manifestation is the Chudakov effect \cite{Chu55}, discovered
in the study of high-energy cosmic $\gamma$ rays impinging on a
nuclear target. If a $\gamma$ is
converted into a highly collinear $\ee$ pair inside the
emulsion, the $\e^+$ and $\e^-$ in their travel through the
emulsion ionize atoms and thereby produce blackening.
However, near the conversion point the blackening is small:
the $\e^+$ and $\e^-$ then are still close together, so that
an atom traversed by the pair does not resolve the individual
charges of the $\e^+$ and the $\e^-$, but only feels a net
charge close to zero. Only later,
when the $\e^+$ and $\e^-$ are separated by more than a typical
atomic radius, are the two able to ionize independently of
each other.
 
The situation is similar in QCD, but is further extended, since
now also gluons carry colour. For example, in a branching
$\q_0 \to \q\g$ the $\q$ and $\g$ share a colour--anticolour
pair, and therefore the $\q$ and $\g$ cannot emit subsequent gluons
incoherently. Again the net effect is to reduce the amount of soft
gluon emission: since a soft gluon (emitted at large angles)
corresponds to a large (transverse) wavelength,
the soft gluon is unable to resolve the separate colour charges
of the $\q$ and the $\g$, and only feels the net charge carried by
the $\q_0$. Such a soft gluon $\g'$ (in the region
$\theta_{\q_0 \g'} > \theta_{\q \g}$)
could therefore be thought of as being
emitted by the $\q_0$ rather than by the $\q$--$\g$ system.
If one considers only emission that should be associated with the
$\q$ or the $\g$, to a good approximation, there is
a complete destructive interference in the
regions of non-decreasing opening angles, while partons
radiate independently of each other inside the regions
of decreasing opening angles ($\theta_{q g'} < \theta_{q g}$ and
$\theta_{g g'} < \theta_{q g}$), once azimuthal angles are averaged
over. The details of the colour interference pattern are
reflected in non-uniform azimuthal emission probabilities.
 
The first branchings of the shower are not affected by the
angular-ordering requirement --- since the evolution is performed
in the c.m. frame of the original parton pair, where the original
opening angle is  180$^{\circ}$, any angle would anyway be smaller
than this --- but here instead the matrix-element matching procedure
is used, where applicable. Subsequently, each opening angle is
compared with that of the preceding branching in the shower.
 
For a branching $a \to bc$ the kinematical approximation
\begin{equation}
\theta_a \approx \frac{p_{\perp b}}{E_b} + \frac{p_{\perp c}}{E_c}
\approx \sqrt{z_a (1-z_a)} m_a \left( \frac{1}{z_a E_a} +
\frac{1}{(1-z_a) E_a} \right) = \frac{1}{\sqrt{z_a(1-z_a)}}
\frac{m_a}{E_a}
\end{equation}
is used to derive the opening angle (this is anyway to the same level
of approximation as the one in which angular ordering is derived).
With $\theta_b$ of the $b$ branching calculated similarly, the
requirement $\theta_b < \theta_a$ can be reduced to
\begin{equation}
\frac{z_b (1-z_b)}{m_b^2} > \frac{1-z_a}{z_a m_a^2} ~.
\end{equation}
 
Since photons do not obey angular ordering, the check on angular
ordering is not performed when a photon is emitted.
When a gluon is emitted in the branching after a photon, its emission
angle is restricted by that of the preceding QCD branching in the
shower, i.e. the photon emission angle does not enter.
 
\subsubsection{Other final-state shower aspects}
 
The electromagnetic coupling constant for the emission of photons
on the mass shell is
$\alphaem = \alphaem(Q^2 = 0) \approx 1/137$. For 
the strong coupling constant several alternatives are available, the 
default being the first-order expression $\alphas(\pT^2)$, where
$\pT^2$ is defined by the approximate expression
$\pT^2 \approx z(1-z) m^2$. Studies of next-to-leading-order
corrections favour this choice \cite{Ama80}. The other alternatives
are a fixed $\alphas$ and an $\alphas(m^2)$.
 
With the default choice of $\pT^2$ as scale in $\alphas$,
a further cut-off is introduced on the allowed phase space
of gluon emission, not present in the options with fixed
$\alphas$ or with $\alphas(m^2)$, nor in the QED shower.
A minimum requirement, to ensure a well-defined $\alphas$,
is that $\pT / \Lambda > 1.1$, but additionally
{\Je} requires that $\pT > Q_0/2$. This latter
requirement is not a necessity, but it makes sense when
$\pT$ is taken to be the preferred scale of the branching
process, rather than e.g. $m$. It reduces the allowed $z$ range,
compared with the purely kinematical constraints.
Since the $\pT$ cut is not
present for photon emission, the relative ratio of photon to gluon
emission off a quark is enhanced at small virtualities compared with
na\"{\i}ve expectations; in actual fact this enhancement is largely
compensated by the running of $\alphas$, which acts in the
opposite direction. The main consequence, however, is that the
gluon energy spectrum is peaked at around $Q_0$ and rapidly
vanishes for energies below that, whilst the photon spectum
extends all the way to zero energy.
 
Previously it was said that azimuthal angles in branchings are
chosen isotropically. In fact, as an option, it is possible to
include some effects of gluon polarization, which correlate the
production and the decay planes of a gluon, such that a
$\g \to \g\g$ branching tends to take place in the production
plane of the gluon, while a decay out of the plane is favoured
for $\g \to \q\qbar$. The formulae are given e.g. in ref.
\cite{Web86}, as simple functions of the $z$ value at the
vertex where the gluon is produced and of the $z$ value when
it branches. Also coherence phenomena lead to non-isotropic
azimuthal distributions \cite{Web86}, which are included as a
further option. In either case the $\varphi$ azimuthal variable
is first chosen isotropically, then the weight factor due to
polarization times coherence is evaluated, and the $\varphi$ value
is accepted or rejected. In case of rejection, a new $\varphi$ is
generated, and so on.
 
While the rule is to have an initial pair of partons, there are
a few examples where one or three partons have to be allowed to
shower. If only one parton is given, it is not possible to
conserve both energy and momentum. The choice has been made
to conserve energy and jet direction, but the momentum vector is
scaled down when the radiating parton acquires a mass. The `rest
frame of the system', used e.g. in the $z$ definition, is taken to
be whatever frame the jet is given in.
 
In $\Upsilon \to \g\g\g$ decays and other primary three-parton
configurations, one is left with the issue how the energy
sharing variables $x_1$ and $x_2$ from the massless matrix elements
should be reinterpreted for a massive three-parton configuration.
We have made the arbitrary choice of preserving the energy of each
parton, which means that relative angles between the original partons
is changed. Mass triplets outside the allowed phase space are
rejected and the evolution continued.
 
Finally, it should be noted that two toy shower models are
included as options. One is a scalar gluon model, in which the
$\q \to \q\g$ branching kernel is replaced by
$P_{\q \to \q\g}(z) = \frac{2}{3} (1-z)$. The couplings of the
gluon, $\g \to \g\g$ and $\g \to \q\qbar$, have been left as free
parameters, since they depend on the colour structure assumed in the
model. The spectra are flat in $z$ for a spin 0 gluon.
Higher-order couplings of the type $\g \to \g\g\g$ could well
contribute significantly, but are not included.
The second toy model is an Abelian vector one. In this
option $\g \to \g \g$ branchings are absent, and $\g \to \q \qbar$
ones enhanced. More precisely, in the splitting kernels, eq.
(\ref{sh:APker}), the Casimir factors are changed as follows:
$C_F = 4/3 \to 1$, $N_C = 3 \to 0$, $T_R = n_f/2 \to 3n_f$.
When using either of these options, one should be aware that also
a number of other components in principle should be changed, from
the running of $\alphas$ to the whole concept of fragmentation.
One should therefore not take them too seriously.
 
\subsection{Initial-State Showers}
 
The initial-state showe algorithm in {\Py} is not quite as
sophisticated as the final-state one. This is partly because
initial-state radiation is less well understood theoretically,
partly because the programming task is
more complicated and ambiguous. Still, the program at disposal
is known to do a reasonably good job of describing existing data,
such as $\Z^0$ production properties at hadron colliders
\cite{Sjo85}.
 
\subsubsection{The shower structure}
\label{sss:initshowstruc}
 
A fast hadron may be viewed as a cloud of quasireal partons.
Similarly a fast lepton may be viewed as surrounded by a cloud
of photons and partons; in the program the two situations are on
an equal footing, but here we choose the hadron as example. At
each instant, an individual parton can initiate a virtual cascade,
branching into a number of partons. This cascade can be described
in terms of a tree-like structure, composed of many subsequent
branchings $a \to bc$. Each branching involves some relative
transverse momentum between the two daughters. In a language where
four-momentum is conserved at each vertex, this implies that at
least one of the $b$ and $c$ partons must have a space-like
virtuality, $m^2 < 0$. Since the partons are not on the mass
shell, the cascade only lives a finite time before reassembling, with
those parts of the cascade that are most off the mass shell living the
shortest time.
 
A hard scattering, e.g. in deep inelastic leptoproduction, will
probe the hadron at a given instant. The probe, i.e. the virtual
photon in the leptoproduction case, is able to resolve fluctuations
in the hadron up to the $Q^2$ scale of the hard scattering. Thus
probes at different $Q^2$ values will seem to see different parton
compositions in the hadron. The change in parton composition with
$t = \ln(Q^2/\Lambda^2)$ is given by the evolution equations
\begin{equation}
\frac{\d f_b(x,t)}{\d t} = \sum_{a,c} \int \frac{\d x'}{x'} \,
f_a(x',t) \, \frac{\alpha_{abc}}{2 \pi} \,
P_{a \to bc} \left( \frac{x}{x'} \right) ~.
\label{sh:sfevol}
\end{equation}
Here the $f_i(x,t)$ are the parton-distribution functions, expressing
the probability of finding a parton $i$ carrying a fraction $x$
of the total momentum if the hadron is probed at virtuality $Q^2$.
The $P_{a \to bc}(z)$ are given in eq.~(\ref{sh:APker}). As before,
$\alpha_{abc}$ is $\alphas$ for QCD shower and $\alphaem$
for QED ones.
 
Eq.~(\ref{sh:sfevol}) is closely related to eq.~(\ref{sh:Patobc}):
$\d {\cal P}_a$ describes the probability that a given parton $a$ will
branch (into partons $b$ and $c$), $\d f_b$ the influx of partons
$b$ from the branchings of partons $a$. (The expression $\d f_b$ in
principle also should contain a loss term for partons $b$ that
branch; this term is important for parton-distribution evolution,
but does not appear explicitly in what we shall be using eq.
(\ref{sh:sfevol}) for.) The absolute form of
hadron parton distributions cannot be predicted in perturbative
QCD, but rather have to be parametrized at some $Q_0$ scale, with
the $Q^2$ dependence thereafter given by eq.~(\ref{sh:sfevol}).
Available parametrizations are discussed in section \ref{ss:structfun}.
The lepton and photon parton distributions inside a lepton can be
fully predicted, but here for simplicity are treated on equal footing
with hadron parton distributions.
 
If a hard interaction scatters a parton out of the incoming hadron,
the `coherence' \cite{Gri83} of the cascade is broken: the partons can
no longer reassemble completely back to the cascade-initiating parton.
In this semiclassical picture, the partons on the `main chain' of
consecutive branchings that lead directly from the initiating parton
to the scattered parton can no longer reassemble, whereas fluctuations
on the `side branches' to this chain may still disappear. A
convenient description is obtained by assigning a space-like
virtuality to the partons on the main chain, in such a way that
the partons on the side branches may still be on the mass shell.
Since the momentum transfer of the hard process can put the scattered
parton on the mass shell (or even give it a time-like virtuality, so
that it can initiate a final-state shower), one is then guaranteed
that no partons have a space-like virtuality in the final state. (In
real life, confinement effects obviously imply that partons need not
be quite on the mass shell.) If no hard scattering had taken place,
the virtuality of the space-like parton line would still force the
complete cascade to reassemble. Since the virtuality of the cascade
probed is carried by one single parton, it is possible to equate the
space-like virtuality of this parton with the $Q^2$ scale of the
cascade, to be used e.g. in the evolution equations. Further,
coherence effects \cite{Gri83,Bas83} guarantee that the $Q^2$ vaules
of the partons along the main chain are strictly ordered, with the
largest $Q^2$ values close to the hard scattering.
 
In recent years, further coherence effects have been studied
\cite{Cia87}, with particular implications for the
structure of parton showers at small $x$. None of these additional
complications are implemented in the current algorithm, with the
exception of a few rather primitive options that do not address
the full complexity of the problem.
 
Instead of having a tree-like structure, where all legs are treated
democratically, the cascade is reduced to a single sequence of
branchings $a \to bc$, where the $a$ and $b$ partons are on the
main chain of space-like virtuality, $m_{a,b}^2 < 0$, while the $c$
partons are on the mass shell and do not branch. (Later we will
include the possibility that the $c$ partons may have positive
virtualities, $m_c^2 > 0$, which leads to the appearance of time-like
`final-state' parton showers on the side branches.) This truncation
of the cascade is only possible when it is known which parton
actually partakes in the hard scattering: of all the possible
cascades that exist virtually in the incoming hadron, the hard
scattering will select one.
 
To obtain the correct $Q^2$ evolution of parton distributions,
e.g., it is essential that all branches of the cascade be treated
democratically. In Monte Carlo simulation of space-like showers
this is a major problem. If indeed the evolution of the complete
cascade is to be followed from some small $Q_0^2$ up to the
$Q^2$ scale of the hard scattering, it is no possible at the
same time to handle kinematics exactly, since the virtuality of the
various partons cannot be found until after the hard scattering
has been selected. This kind of `forward evolution' scheme therefore
requires a number of extra tricks to be made to work. Further, in
this approach it is not known e.g. what the $\hat{s}$ of the
hard scattering subsystem will be until the evolution has been
carried out, which means that the initial-state evolution
and the hard scattering have to be selected jointly, a not so
trivial task.
 
Instead we use the `backwards evolution' approach \cite{Sjo85},
in which the hard scattering is first selected, and the parton
shower that preceded it is subsequently reconstructed. This
reconstruction is started at the hard interaction, at the
$Q_{\mmax}^2$ scale, and thereafter step by step one moves
`backwards' in  `time', towards smaller $Q^2$, all the way back
to the parton-shower initiator at the cut-off scale $Q_0^2$.
This procedure is possible if evolved parton distributions are
used to select the hard scattering, since the $f_i(x,Q^2)$
contain the inclusive summation of all initial-state parton-shower 
histories that can lead to the appearance of an interacting
parton $i$ at the hard scale. What remains is thus to select an
exclusive history from the set of inclusive ones.
 
\subsubsection{Longitudinal evolution}
 
The evolution equations, eq.~(\ref{sh:sfevol}), express that, during
a small increase $\d t$ there is a probability for parton $a$ with
momentum fraction $x'$ to become resolved into parton $b$ at
$x = z x'$ and another parton $c$ at $x' - x = (1-z) x'$.
Correspondingly, in backwards evolution, during a decrease $\d t$
a parton $b$ may be `unresolved' into parton $a$. The relative
probability $\d {\cal P}_b$ for this to happen is given by
the ratio $\d f_b / f_b$. Using eq.~(\ref{sh:sfevol}) one obtains
\begin{equation}
\d {\cal P}_b = \frac{\d f_b(x,t)}{f_b(x,t)} = |\d t| \, \sum_{a,c}
\int \frac{\d x'}{x'} \, \frac{f_a(x',t)}{f_b(x,t)} \,
\frac{\alpha_{abc}}{2 \pi} \,
P_{a \to bc} \left( \frac{x}{x'} \right) ~.
\end{equation}
Summing up the cumulative effect of many small changes $\d t$, the
probability for no radiation exponentiates. Therefore one may
define a form factor
\begin{eqnarray}
S_b(x,t_{\mmax},t) & = &
\exp \left\{ - \int_t^{t_{\mmax}} \d t' \, \sum_{a,c} \int
\frac{\d x'}{x'} \, \frac{f_a(x',t')}{f_b(x,t')} \,
\frac{\alpha_{abc}(t')}{2\pi} \, P_{a \to bc} \left(
\frac{x}{x'} \right) \right\} \nonumber \\
& = & \exp \left\{ - \int_t^{t_{\mmax}} \d t' \, \sum_{a,c} 
\int \d z \, \frac{\alpha_{abc}(t')}{2\pi} \, P_{a \to bc}(z) \,
\frac{x'f_a(x',t')}{xf_b(x,t')} \right\} ~,
\end{eqnarray}
giving the probability that a parton $b$ remains at $x$ from
$t_{\mmax}$ to a $t < t_{\mmax}$.
 
It may be useful to compare this with the corresponding expression
for forward evolution, i.e. with $S_a(t)$ in eq.~(\ref{sh:Pnobranch}).
The most obvious difference is the appearance of parton distributions
in $S_b$. Parton distributions are absent in $S_a$: the probability
for a given parton $a$ to branch, once it exists, is independent of
the density of partons $a$ or $b$. The parton distributions in $S_b$,
on the other hand, express the fact that the probability for a parton
$b$ to come from the branching of a parton $a$ is proportional to
the number of partons $a$ there are in the hadron, and inversely
proportional to the number of partons $b$. Thus the numerator
$f_a$ in the exponential of $S_b$ ensures that the parton composition
of the hadron
is properly reflected. As an example, when a gluon is chosen at the
hard scattering and evolved backwards, this gluon is more likely to
have been emitted by a $\u$ than by a $\d$ if the incoming hadron is
a proton. Similarly, if a heavy flavour is chosen at the hard
scattering, the denominator $f_b$ will vanish at the $Q^2$ threshold
of the heavy-flavour production, which means that the integrand
diverges and $S_b$ itself vanishes, so that no heavy flavour remain
below threshold.
 
Another difference between $S_b$ and $S_a$, already touched upon, is
that the $P_{\g \to \g\g}(z)$ splitting kernel appears with a
normalization $2 N_C$ in $S_b$ but only with $N_C$ in $S_a$, since
two gluons are produced but only one decays in a branching.
 
A knowledge of $S_b$ is enough to reconstruct the parton shower
backwards. At each branching $a \to bc$, three quantities have to
be found: the $t$ value of the branching (which defines the
space-like virtuality $Q_b^2$ of parton $b$), the parton flavour $a$
and the splitting variable $z$. This information may be extracted as
follows:
\begin{Enumerate}
\item If parton $b$ partook in the hard scattering or branched into
other partons at a scale $t_{\mmax}$, the probability that $b$ was
produced in a branching $a \to bc$ at a lower scale $t$ is
\begin{equation}
\frac{\d {\cal P}_b}{\d t} = - \frac{\d S_b(x,t_{\mmax},t)}{\d t} =
\left( \sum_{a,c} \int dz \, \frac{\alpha_{abc}(t')}{2\pi} \,
P_{a \to bc}(z) \, \frac{x'f_a(x',t')}{xf_b(x,t')} \right)
S_b(x,t_{\mmax},t) ~.
\end{equation}
If no branching is found above the cut-off scale $t_0$ the
iteration is stopped and parton $b$ is assumed to be massless.
\item Given the $t$ of a branching, the relative probabilities
for the different allowed branchings $a \to bc$ are given by the
$z$ integrals above, i.e. by
\begin{equation}
\int \d z \, \frac{\alpha_{abc}(t)}{2\pi} \, P_{a \to bc}(z) \, 
\frac{x'f_a(x',t)}{xf_b(x,t)} ~.
\label{sh:Pbspace}
\end{equation}
\item Finally, with $t$ and $a$ known, the probability distribution
in the splitting variable $z = x/x' = x_b/x_a$ is given by the
integrand in eq.~(\ref{sh:Pbspace}).
\end{Enumerate}
In addition, the azimuthal angle $\varphi$ of the branching is
selected isotropically, i.e. no spin or coherence effects are
included in this distribution.
 
The selection of $t$, $a$ and $z$ is then a standard task of the
kind than can be performed with the help of the veto algorithm.
Specifically, upper and lower bounds for parton distributions are
used to find simple functions that are everywhere larger than the
integrands in eq.~(\ref{sh:Pbspace}). Based on these simple expressions,
the integration over $z$ may be carried out, and $t$, $a$ and $z$
values selected. This set is then accepted with a weight given
by a ratio of the correct integrand in eq.~(\ref{sh:Pbspace}) to
the simple approximation used, both evaluated for the given set.
Since parton distributions, as a rule, are not in a simple
analytical form, it may be tricky to find reasonably good bounds to
parton distributions. It is necessary to make different assumptions
for valence and sea quarks, and be especially attentive close to a
flavour threshold (\cite{Sjo85}). An electron distribution
inside an electron behaves differently from parton distributions
encountered in hadrons, and has to be considered separately.
 
A comment on soft gluon emission. Nominally the range of the $z$
integral in $S_b$ is $x \leq z \leq 1$. The lower limit corresponds
to $x' = x/z = 1$, and parton distributions vanish in this limit,
wherefore no problems are encountered here. At the upper cut-off
$z=1$ the splitting kernels $P_{\q \to \q\g}(z)$ and
$P_{\g \to \g\g}$ diverge. This is the soft gluon singularity:
the energy carried by the emitted gluon is vanishing,
$x_{\g} = x' - x = (1-z) x' = (1-z) x/z \to 0$ for $z \to 1$.
In order to calculate the integral over $z$ in $S_b$, an upper
cut-off $z_{\mmax} = x/(x + x_{\epsilon})$ is introduced, i.e. only
branchings with $z \leq z_{\mmax}$ are included in $S_b$. Here
$x_{\epsilon}$ is a small number, typically chosen so that the
gluon energy $x_{\g} \sqrt{s}/2 \geq x_{\epsilon} \sqrt{s}/2 = 2$
GeV. The average amount of energy carried away by gluons in the
range $x_{g} < x_{\epsilon}$, over the given range of $t$ values
from $t_a$ to $t_b$, may be estimated \cite{Sjo85}. The finally
selected $z$ value may thus be picked as
$z = z_{\mrm{hard}} \langle z_{\mrm{soft}}(t_a, t_b) \rangle$, 
where $z_{\mrm{hard}}$
is the originally selected $z$ value and $z_{\mrm{soft}}$ is the
correction factor for soft gluon emission.
 
In QED showers, the smallness of $\alphaem$ means that one can
use rather smaller cut-off values without obtaining large amounts of
emission. A fixed small cut-off $x_{\gamma} > 10^{-6}$ is therefore
used to avoid the region of very soft photons. As has been discussed
in section \ref{sss:estructfun}, the electron distribution
inside the electron is cut off at $x_{\e} < 1 - 10^{-6}$, for
numerical reasons, so the two cuts are closely matched.
 
The cut-off scale $Q_0$ may be chosen separately for QCD and QED
showers, just as in final-state radiation. The defaults are
1 GeV and 0.001 GeV, respectively. The former is the typical hadronic
mass scale, below which radiation is not expected resolvable; the
latter is of the order of the electron mass.
 
Normally QED and QCD showers do not appear mixed. The most notable
exception is resolved photoproduction (in $\ep$) and resolved
2$\gamma$ events
(in $\ee$), i.e. shower histories of the type $\e \to \gamma \to \q$.
Here the $Q^2$ scales need not be ordered at the interface, i.e.
the last $\e \to \e\gamma$ branching may well have a larger $Q^2$
than the first $\q \to \q \g$ one, and the branching $\gamma \to \q$
does not even have a strict parton-shower interpretation for the
vector dominance model part of the photon parton distribution. 
These issues are currently
not addressed in full. Rather, based on the $x$ selected for the
parton (quark or gluon) at the hard scattering, the $x_{\gamma}$
is selected once and for all in the range $x < x_{\gamma} <1$,
according to the distribution implied by eq.~(\ref{pg:foldqgine}).
The QCD parton shower is then traced backwards from the hard
scattering to the QCD shower initiator at $t_0$. No attempt is
made to perform the full QED shower, but rather the beam remnant
treatment (see section \ref{ss:beamrem}) is used to find the
$\qbar$ (or $\g$) remnant that matches the $\q$ (or $\g$) QCD shower
initiator, with the electron itself considered as a second beam
remnant.
 
\subsubsection{Transverse evolution}
\label{sss:initshowtrans}
 
We have above seen that two parton lines may be defined, stretching
back from the hard scattering to the initial incoming hadron
wavefunctions at small $Q^2$. Specifically, all parton flavours
$i$, virtualities $Q^2$ and energy fractions $x$ may be found.
The exact kinematical interpretation of the $x$ variable is not
unique, however. For partons with small virtualities and transverse
momenta, essentially all definitions agree, but differences may
appear for branchings close to the hard scattering.
 
In first-order QED \cite{Ber85} and in some simple QCD toy models
\cite{Got86}, one may show that the `correct' choice is the
`$\hat{s}$ approach'. Here one requires that $\hat{s} = x_1 x_2 s$,
both at the hard scattering scale and at any lower scale, i.e.
$\hat{s}(Q^2) = x_1(Q^2) \, x_2(Q^2) \, s$, where $x_1$ and $x_2$ are
the $x$ values of the two resolved partons (one from each incoming
beam particle) at the given $Q^2$ scale. In practice this means
that, at a branching with the splitting variable $z$, the total
$\hat{s}$ has to be increased by a factor $1/z$ in the backwards
evolution. It also means that branchings on the two incoming legs
have to be interleaved in a single monotonic sequence of $Q^2$
values of branchings.
 
For a reconstruction of the complete kinematics in this approach,
one should start with the hard scattering, for which $\hat{s}$
has been chosen according to the hard scattering matrix element.
By backwards evolution, the virtualities $Q_1^2 = -m_1^2$ and 
$Q_2^2 = -m_2^2$ of
the two interacting partons are reconstructed. Initially the two
partons are considered in their common c.m. frame, coming in along
the $\pm z$ directions. Then the four-momentum vectors have the
non-vanishing components
\begin{eqnarray}
E_{1,2} & = & \frac{ \hat{s} \pm (Q_2^2 - Q_1^2)}{2 \sqrt{\hat{s}}} ~,
\nonumber \\
p_{z1} = - p_{z2} & = & \sqrt{ \frac{ (\hat{s} + Q_1^2 + Q_2^2 )^2
- 4 Q_1^2 Q_2^2 }{ 4 \hat{s} } } ~,
\end{eqnarray}
with $(p_1 + p_2)^2 = \hat{s}$.
 
If, say, $Q_1^2 > Q_2^2$, then the branching $3 \to 1 + 4$, which
produced parton 1, is the one that took place closest to the hard
scattering, and the one to be reconstructed first. With the
four-momentum $p_3$ known, $p_4 = p_3 - p_1$ is automatically
known, so there are four degrees of freedom. One corresponds to
a trivial azimuthal angle around the $z$ axis. The $z$ splitting
variable for the $3 \to 1 + 4$ vertex is found as the same time as
$Q_1^2$, and provides the constraint $(p_3 + p_2)^2 = \hat{s}/z$.
The virtuality $Q_3^2$ is given by backwards evolution of parton 3.
 
One degree of freedom remains to be specified, and this is related
to the possibility that parton 4 initiates a time-like parton shower,
i.e. may have a non-zero mass. The maximum allowed squared mass
$m_{\mmax,4}^2$ is found for a collinear branching $3 \to 1 + 4$.
In terms of the combinations
\begin{eqnarray}
s_1 & = & \hat{s} + Q_2^2 + Q_1^2 ~,    \nonumber \\
s_3 & = & \frac{\hat{s}}{z} + Q_2^2 + Q_3^2 ~,  \nonumber \\
r_1 & = & \sqrt{s_1^2 - 4 Q_2^2 Q_1^2} ~, \nonumber \\
r_3 & = & \sqrt{s_3^2 - 4 Q_2^2 Q_3^2} ~,
\end{eqnarray}
one obtains
\begin{equation}
m_{\mmax,4}^2 = \frac{s_1 s_3 - r_1 r_3}{2 Q_2^2} - Q_1^2 - Q_3^2 ~,
\end{equation}
which, for the special case of $Q_2^2 = 0$, reduces to
\begin{equation}
m_{\mmax,4}^2 = \left\{ \frac{Q_1^2}{z} - Q_3^2 \right\}
\left\{ \frac{\hat{s}}{\hat{s} + Q_1^2} -
\frac{\hat{s}}{\hat{s}/z + Q_3^2} \right\} ~.
\label{sh:zrangespace}
\end{equation}
These constraints on $m_4$ are only the kinematical ones, in
addition coherence phenomena could constrain the $m_{\mmax,4}$
values further. Some options of this kind are available; the
default one is to require additionally that $m_4^2 \leq Q_1^2$,
i.e. lesser than the space-like virtuality of the sister parton.
 
With the maximum virtuality given, the final-state showering
machinery may be used to give the development of the subsequent
cascade, including the actual mass $m_4^2$, with
$0 \leq m_4^2 \leq m_{\mmax,4}^2$. The evolution is performed in
the c.m. frame of the two `resolved' partons, i.e. that of
partons 1 and 2 for the
branching $3 \to 1 + 4$, and parton 4 is assumed to have a nominal
energy $E_{\mrm{nom},4} = (1/z - 1) \sqrt{\hat{s}}/2$. (Slight
modifications appear if parton 4 has a non-vanishing mass
$m_{\q}$ or $m_{\ell}$.)
 
Using the relation $m_4^2 = (p_3 - p_1)^2$, the momentum of parton
3 may now be found as
\begin{eqnarray}
E_3 & = & \frac{1}{2 \sqrt{\hat{s}} } \left\{ \frac{\hat{s}}{z}
+ Q_2^2 - Q_1^2 - m_4^2 \right\} ~,  \nonumber \\
p_{z3} & = & \frac{1}{2 p_{z1}} \left\{ s_3 - 2 E_2 E_3 \right\} ~,
\nonumber \\
p_{\perp ,3}^2 & = & \left\{ m_{\mmax,4}^2 - m_4^2 \right\} \,
\frac{ (s_1 s_3 + r_1 r_3)/2 - Q_2^2 (Q_1^2 + Q_3^2 + m_4^2)}{r_1^2} ~.
\end{eqnarray}
 
The requirement that $m_4^2 \geq 0$ (or $\geq m_f^2$ for heavy
flavours) imposes a constraint on allowed $z$ values. This constraint
cannot be included in the choice of $Q_1^2$, where it logically
belongs, since it also depends on $Q_2^2$ and $Q_3^2$, which are
unknown at this point. It is fairly rare (in the order of 10\% of all
events) that an unallowed $z$ value is generated, and when it happens
it is almost always for one of the two branchings closest to the
hard interaction: for $Q_2^2 = 0$ eq.~(\ref{sh:zrangespace}) may be
solved to yield $z \leq \hat{s}/(\hat{s} + Q_1^2 - Q_3^2)$, which is
a more severe cut for $\hat{s}$ small and $Q_1^2$ large. Therefore
an essentially bias-free way of coping is to redo completely any
initial-state cascade for which this problem appears.
 
This completes the reconstruction of the $3 \to 1 + 4$ vertex.
The subsystem made out of partons 3 and 2 may now be boosted to its
rest frame and rotated to bring partons 3 and 2 along the $\pm z$
directions. The partons 1 and 4 now have opposite and compensating
transverse momenta with respect to the event axis. When the next
vertex is considered, either the one that produces parton 3 or the
one that produces parton 2, the 3--2 subsystem will fill the function
the 1--2 system did above, e.g. the r\^ole of 
$\hat{s} = \hat{s}_{12}$ in the formulae above is now played by 
$\hat{s}_{32} = \hat{s}_{12}/z$. The internal structure of the 
3--2 system, i.e. the branching $3 \to 1 + 4$, appears nowhere 
in the continued description, but has become
`unresolved'. It is only reflected in the successive rotations and
boosts performed to bring back the new endpoints to their common
rest frame. Thereby the hard scattering subsystem 1--2 builds up
a net transverse momentum and also an overall rotation of the
hard scattering subsystem.
 
After a number of steps, the two outermost partons have virtualities
$Q^2 < Q_0^2$ and then the shower is terminated and the endpoints
assigned $Q^2 = 0$. Up to small corrections from primordial
$k_{\perp}$, discussed in section \ref{ss:beamrem}, a final boost
will bring the partons from their c.m. frame to the overall c.m.
frame, where the $x$ values of the outermost partons agree also
with the light-cone definition.
 
\subsubsection{Other initial-state shower aspects}
 
In the formulae above, $Q^2$ has been used as argument for
$\alphas$, and not only as the space-like virtuality of partons.
This is one possibility, but in fact loop calculations tend to
indicate that the proper argument for $\alphas$ is not $Q^2$
but $\pT^2 = (1-z) Q^2$ \cite{Bas83}. The variable $\pT$ does
have the interpretation of transverse momentum, although it
is only exactly so for a branching $a \to bc$
with $a$ and $c$ massless and $Q^2 = - m_b^2$, and with $z$
interpreted as light-cone fraction of energy and momentum.
The use of $\alphas((1-z)Q^2)$ is default in the program.
Indeed, if one wanted to, the complete shower might be
interpreted as an evolution in $\pT^2$ rather than in $Q^2$.
 
As we see, the initial-state showering algorithm leads to a
net boost and rotation of the hard scattering subsystems. The
overall final state is made even more complex by the additional
final-state radiation. In principle, the complexity is very
physical, but it may still have undesirable side effects. One
such, discussed further in section \ref{ss:PYswitchkin}, is that 
it is very difficult to generate events that fulfill specific 
kinematics conditions, since kinematics is smeared and even, 
at times, ambiguous.
 
A special case is encountered in deep inelastic scattering in
$\ep$ collisions. Here the DIS $x$ and $Q^2$ values are defined
in terms of the scattered electron direction and energy, and
therefore are unambiguous (except for issues of final-state 
photon radiation close to the electron direction).
Neither initial- nor final-state showers preserve the kinematics
of the scattered electron, however, and hence the DIS $x$ and
$Q^2$ are changed. In principle, this is perfectly legitimate,
with the caveat that one then also should use different sets
of parton distributions than ones derived from DIS, since these
are based on the kinematics of the scattered lepton and nothing
else. Alternatively, one might consider showering schemes that
leave $x$ and $Q^2$ unchanged. In \cite{Ben88} detailed
modifications are presented that make a preservation
possible when radiation off the incoming and outgoing electron is
neglected, but these are not included in the current version of
{\Py}.
 
What is available, as an option, is a simple machinery
which preserves $x$ and $Q^2$ from the effects of QCD radiation,
and also from those of primordial $k_{\perp}$ and the beam remnant
treatment, as follows. After the showers have been generated,
the four-momentum of the scattered lepton is changed to the
expected one, based on the nominal $x$ and $Q^2$ values.
The azimuthal angle of the lepton is maintained when the transverse
momentum is adjusted. Photon radiation off the lepton leg is not fully
accounted for, i.e. it is assumed that the energy of final-state
photons is added to that of the scattered electron for the definition
of $x$ and $Q^2$ (this is the normal procedure for parton-distribution
definitions). 

The change of three-momentum on the lepton side of the event is 
balanced by the final state partons on the hadron side, excluding 
the beam remnant but including all the partons both from initial-
and final-state showering. The fraction of three-momentum shift taken 
by each parton is proportional to its original light-cone momentum
in the direction of the incoming lepton, i.e. to $E \mp p_z$ for a
hadron moving in the $\pm$ direction. This procedure guarantees 
momentum but not energy conservation. For the latter, one additional 
degree of freedom is needed, which is taken to be the longitudinal 
momentum of the initial state shower initiator. As this momentum is 
modified, the change is shared by the final state partons on the 
hadron side, according to the same light-cone fractions as before
(based on the original momenta). Energy conservation 
requires that the total change in final state parton energies plus 
the change in lepton side energy equals the change in initiator 
energy. This condition can be turned into an iterative procedure to 
find the initiator momentum shift.

Sometimes the procedure may break down. For instance, an initiator
with $x > 1$ may be reconstructed. If this should happen, the $x$ and 
$Q^2$ values of the event are preserved, but new initial and final
state showers are generated. After five such failures, the event is 
completely discared in favour of a new kinematical setup.

Kindly note that the four-momentum of intermediate partons in the
shower history are not being adjusted. In a listing of the complete
event history, energy and momentum need then not be conserved in 
shower branchings. This mismatch could be fixed up, if need be.

The scheme presented above should not be taken too
literally, but is rather intended as a contrast to the more
sophisticated schemes already on the market, if one would like to
understand whether the kind of conservation scheme chosen does affect
the observable physics.
 
\subsection{Routines and Common Block Variables}
\label{ss:showrout}
 
In this section we collect information on how to use the initial-
and final-state showering routines. Of these \ttt{LUSHOW} for 
final-state radiation is the more generally interesting, since it 
can be called to let a user-defined parton configuration shower. 
\ttt{PYSSPA}, on the other hand, is so intertwined with the general
structure of a {\Py} event that it is of little use as a 
stand-alone product.
  
\drawbox{CALL LUSHOW(IP1,IP2,QMAX)}\label{p:LUSHOW}
\begin{entry}
\itemc{Purpose:} to generate time-like parton showers, conventional
or coherent. The performance of the program is regulated by the
switches \ttt{MSTJ(40) - MSTJ(50)} and parameters
\ttt{PARJ(81) - PARJ(89)}. In order to keep track of the colour
flow information, the positions \ttt{K(I,4)} and \ttt{K(I,5)} have
to be organized properly for showering partons. Inside the {\JePy}
programs, this is done automatically, but for external use
proper care must be taken.
\iteme{IP1 > 0, IP2 = 0 :} generate a time-like parton shower for the
parton in line \ttt{IP1} in common block \ttt{LUJETS}, with maximum
allowed mass \ttt{QMAX}. With only one parton at hand, one cannot
simultaneously conserve both energy and momentum: we here choose to
conserve energy and jet direction, while longitudinal momentum (along
the jet axis) is not conserved.
\iteme{IP1 > 0, IP2 > 0 :} generate time-like parton showers for the
two partons in lines \ttt{IP1} and \ttt{IP2} in the common block
\ttt{LUJETS}, with maximum allowed mass for each parton \ttt{QMAX}.
For shower evolution, the two partons are boosted to their c.m. frame.
Energy and momentum is conserved for the pair of partons, although
not for each individually. One of the two partons may be
replaced by a nonradiating particle, such as a photon or a
diquark; the energy and momentum of this particle will then be
modified to conserve the total energy and momentum.
\iteme{IP1 > 0, IP2 < 0 :} generate time-like parton showers for the
\ttt{-IP2} (at most 3) partons in lines \ttt{IP1}, \ttt{IP1+1}, \ldots
\ttt{IPI-IP2-1} in the common block \ttt{LUJETS}, with maximum allowed
mass for each parton \ttt{QMAX}. The actions for \ttt{IP2=-1} and
\ttt{IP2=-2} correspond to what is described above, but additionally
\ttt{IP2=-3} may be used to generate the evolution starting from
three given partons (e.g. in $\Upsilon \to \g\g\g$). Then the three
partons are boosted to their c.m. frame, energy is conserved for each
parton individually and momentum for the system as a whole.
\iteme{QMAX :} the maximum allowed mass of a radiating parton, i.e.
the starting value for the subsequent evolution. (In addition, the
mass of a single parton may not exceed its energy, the mass of a
parton in a system may not exceed the invariant mass of the
system.)
\end{entry}
 
\boxsep
\begin{entry}
\iteme{SUBROUTINE PYSSPA(IPU1,IPU2) :}\label{p:PYSSPA} to generate 
the space-like showers of the initial-state radiation.
\end{entry}
 
\drawbox{COMMON/LUDAT1/MSTU(200),PARU(200),MSTJ(200),PARJ(200)}
\begin{entry}
\itemc{Purpose:} to give access to a number of status codes and
parameters which regulate the performance of {\Je}.
Most parameters are described in section \ref{ss:JETswitch};
here only those related to \ttt{LUSHOW} are described.

\boxsep

\iteme{MSTJ(40) :}\label{p:MSTJ40} (D=0) possibility to suppress the 
branching probability for a branching $\q \to \q\g$ (or 
$\q \to \q\gamma$) of a quark produced in the decay of an unstable 
particle with width $\Gamma$, where this width has to
be specified by the user in \ttt{PARJ(89)}. The algorithm used is not 
exact, but still gives some impression of potential effects. This 
switch ought to have appeared at the end of the current list of shower 
switches (after \ttt{MSTJ(50)}), but because of lack of 
space it appears immediately before.  
\begin{subentry}
\iteme{= 0 :} no suppression, i.e. the standard parton-shower machinery.
\iteme{= 1 :} suppress radiation by a factor 
$\chi(\omega) = \Gamma^2 / (\Gamma^2 + \omega^2)$, where $\omega$ is 
the energy of the gluon (or photon) in the rest frame of the radiating 
dipole. Essentially this means that hard radiation with 
$\omega > \Gamma$ is removed. 
\iteme{= 2 :} suppress radiation by a factor 
$1 - \chi(\omega) = \omega^2 / (\Gamma^2 + \omega^2)$, where $\omega$
is the energy of the gluon (or photon) in the rest frame of the
radiating dipole. Essentially this means that soft radiation with 
$\omega < \Gamma$ is removed. 
\end{subentry}

\iteme{MSTJ(41) :} (D=2) type of branchings allowed in shower.
\begin{subentry}
\iteme{= 0 :} no branchings at all, i.e. shower is switched off.
\iteme{= 1 :} QCD type branchings of quarks and gluons.
\iteme{= 2 :} also emission of photons off quarks and leptons; the
photons are assumed on the mass shell.
\iteme{= 10 :} as \ttt{=2}, but enhance photon emission by a factor
\ttt{PARJ(84)}. This option is unphysical, but for moderate values,
\ttt{PARJ(84)}$\leq 10$, it may be used to enhance the prompt photon
signal in $\q\qbar$ events. The normalization of the prompt photon 
rate should then be scaled down by the same factor. The dangers
of an improper use are significant, so do not use this option if you
do not know what you are doing.
\end{subentry}
 
\iteme{MSTJ(42) :} (D=2) branching mode for time-like showers.
\begin{subentry}
\iteme{= 1 :} conventional branching, i.e. without angular ordering.
\iteme{= 2 :} coherent branching, i.e. with angular ordering.
\end{subentry}
 
\iteme{MSTJ(43) :} (D=4) choice of $z$ definition in branching.
\begin{subentry}
\iteme{= 1 :} energy fraction in grandmother's rest frame (`local,
constrained').
\iteme{= 2 :} energy fraction in grandmother's rest frame assuming
massless daughters, with energy and momentum reshuffled for massive
ones (`local, unconstrained').
\iteme{= 3 :} energy fraction in c.m. frame of the showering partons
(`global, constrained').
\iteme{= 4 :} energy fraction in c.m. frame of the showering partons
assuming massless daughters, with energy and momentum reshuffled for
massive ones (`global, unconstrained').
\end{subentry}
 
\iteme{MSTJ(44) :} (D=2) choice of $\alphas$ scale for shower.
\begin{subentry}
\iteme{= 0 :} fixed at \ttt{PARU(111)} value.
\iteme{= 1 :} running with $Q^2 = m^2/4$, $m$ mass of decaying
parton, $\Lambda$ as stored in \ttt{PARJ(81)} (natural choice for
conventional showers).
\iteme{= 2 :} running with $Q^2 = z(1-z)m^2$, i.e. roughly $\pT^2$
of branching, $\Lambda$ as stored in \ttt{PARJ(81)} (natural choice
for coherent showers).
\end{subentry}
 
\iteme{MSTJ(45) :} (D=5) maximum flavour that can be produced in
shower by $\g \to \q \qbar$; also used to determine the maximum
number of active flavours in the $\alphas$ factor in parton showers
(here with a minimum of 3).
 
\iteme{MSTJ(46) :} (D=3) nonhomogeneous azimuthal distributions in
a shower branching.
\begin{subentry}
\iteme{= 0 :} azimuthal angle is chosen uniformly.
\iteme{= 1 :} nonhomogeneous azimuthal angle in gluon decays due to
a kinematics-dependent effective gluon polarization.
Not meaningful for scalar model, i.e. then same as \ttt{=0}.
\iteme{= 2 :} nonhomogeneous azimuthal angle in gluon decay due to
interference with nearest neighbour (in colour).
Not meaningful for Abelian model, i.e. then same as \ttt{=0}.
\iteme{= 3 :} nonhomogeneous azimuthal angle in gluon decay due to
both polarization (\ttt{=1}) and interference (\ttt{=2}).
Not meaningful for Abelian model, i.e. then same as \ttt{=1}.
Not meaningful for scalar model, i.e. then same as \ttt{=2}.
\end{subentry}
 
\iteme{MSTJ(47) :} (D=3) corrections to the lowest-order $\q\qbar\g$,
$\q\qbar\gamma$, $\ell^+\ell^-\gamma$ or $\ell\nu_{\ell}\gamma$
3-parton matrix element at the first branching of either initial
parton in a shower.
\begin{subentry}
\iteme{= 0 :} no corrections.
\iteme{= 1 :} included whenever scattered partons are $\q\qbar$,
$\ell^+\ell^-$ or $\ell\nu_{\ell}$.
\iteme{= 2 :} always included when shower starts from two partons.
\iteme{= 3 :} as \ttt{=1} except that for massive quarks also the
massive matrix element expression is used, eq.~(\ref{ee:threejMEmass}),
 while \ttt{=1} is always based on massless matrix elements.
\iteme{= 4 :} as \ttt{=2} except that for massive quarks also the
massive matrix element expression is used, while \ttt{=2} is always
based on massless matrix elements.
\end{subentry}
 
\iteme{MSTJ(48) :} (D=0) possibility to impose maximum angle for the 
first branching in a shower.
\begin{subentry}
\iteme{= 0 :} no explicit maximum angle.
\iteme{= 1 :} maximum angle given by \ttt{PARJ(85)} for single
showering parton, by \ttt{PARJ(85)} and \ttt{PARJ(86)} for pair
of showering partons.
\end{subentry}
 
\iteme{MSTJ(49) :} (D=0) possibility to change the branching
probabilities according to some alternative toy models (note that
the $Q^2$ evolution of $\alphas$ may well be different in these
models, but that only the \ttt{MSTJ(44)} options are at the
disposal of the user).
\begin{subentry}
\iteme{= 0 :} standard QCD branchings.
\iteme{= 1 :} branchings according to a scalar gluon theory, i.e. the
splitting kernels in the evolution equations are,
with a common factor $\alphas/(2\pi)$ omitted,
$P_{\q \to \q\g} = (2/3) (1-z)$, $P_{\g \to \g\g} =$ \ttt{PARJ(87)},
$P_{\g \to \q\qbar} =$ \ttt{PARJ(88)} (for each separate flavour).
The couplings of the gluon have been left as free parameters,
since they depend on the colour structure assumed. Note that,
since a spin 0 object decays isotropically, the gluon splitting
kernels contain no $z$ dependence.
\iteme{= 2 :} branchings according to an Abelian vector gluon theory,
i.e. the colour factors are changed (compared with QCD) according to
$C_F = 4/3 \to 1$, $N_C = 3 \to 0$, $T_R = 1/2 \to 3$. Note that an
Abelian model is not expected to contain any coherence effects
between gluons, so that one should normally use \ttt{MSTJ(42)=1} and
\ttt{MSTJ(46)=} 0 or 1. Also, $\alphas$ is expected to increase
with increasing $Q^2$ scale, rather than decrease. No such
$\alphas$ option is available; the one that comes closest
is \ttt{MSTJ(44)=0}, i.e. a fix value.
\end{subentry}

\iteme{MSTJ(50) :} (D=3) possibility to introduce colour coherence 
effects in the first branching of a final state shower; mainly 
of relevance for QCD parton--parton scattering processes.
\begin{subentry}
\iteme{= 0 :} none.
\iteme{= 1 :} impose an azimuthal anisotropy.
\iteme{= 2 :} restrict the polar ange of a branching to be smaller 
than the scattering angle of the relevant colour flow.
\iteme{= 3 :} both azimuthal anisotropy and restricted polar angles.
\itemc{Note:} for subsequent branchings the (polar) angular ordering 
is automatic (\ttt{MSTP(42)=2}) and \ttt{MSTJ(46)=3}).
\end{subentry}

\boxsep 

\iteme{PARJ(81) :}\label{p:PARJ81} (D=0.29 GeV) $\Lambda$ value
in running $\alphas$ for parton showers (see \ttt{MSTJ(44)}). This is
used in all user calls to \ttt{LUSHOW}, in the $\ee$ routines of {\Je},
and in a {\Py} (or {\Je}) resonance decay. It is not intended for 
other timelike showers in {\Py}, however, for which \ttt{PARP(72)}
is used.
 
\iteme{PARJ(82) :} (D=1.0 GeV) invariant mass cut-off $m_{\mmin}$ of
parton showers, below which partons are not assumed to radiate.
For $Q^2 = \pT^2$ (\ttt{MSTJ(44)=2}) \ttt{PARJ(82)}/2
additionally gives the minimum $\pT$ of a branching. To avoid
infinite $\alphas$ values, one must have
\ttt{PARJ(82)}$ > 2 \times$\ttt{PARJ(81)} for \ttt{MSTJ(44)}$\geq 1$
(this is automatically checked in the program, with
$2.2 \times$\ttt{PARJ(81)} as the lowest value attainable).
 
\iteme{PARJ(83) :} (D=1.0 GeV) invariant mass cut-off $m_{\mmin}$ used
for photon emission in parton showers, below which quarks and leptons
are not assumed to radiate. The function of \ttt{PARJ(83)} closely
parallels that of \ttt{PARJ(82)} for QCD branchings, but there is a
priori no requirement that the two be equal.

\iteme{PARJ(84) :} (D=1.) used for option \ttt{MSTJ(41)=10} as a 
multiplicative factor in the promt photon emission rate in final
state parton showers. Unphysical but useful technical trick, so
beware!
 
\iteme{PARJ(85), PARJ(86) :} (D=10.,10.) maximum opening angles
allowed in the first branching of parton showers; see \ttt{MSTJ(48)}.
 
\iteme{PARJ(87) :} (D=0.) coupling of $\g \to \g\g$ in scalar gluon
shower, see \ttt{MSTJ(49)=1}.
 
\iteme{PARJ(88) :} (D=0.) coupling of $\g \to \q\qbar$ in scalar
gluon shower (per quark species), see \ttt{MSTJ(49)=1}.
 
\iteme{PARJ(89) :} (D=0. GeV) the width of the unstable particle studied 
for the \ttt{MSTJ(40) > 0} options; to be set by the user (separately 
for each \ttt{LUSHOW} call, if need be). 

\end{entry}

\drawbox{COMMON/PYPARS/MSTP(200),PARP(200),MSTI(200),PARI(200)}
\begin{entry}
 
\itemc{Purpose:} to give access to status code and parameters which
regulate the performance of {\Py}.
Most parameters are described in section \ref{ss:PYswitchpar};
here only those related to \ttt{PYSSPA} and \ttt{LUSHOW} are 
described.
 
\iteme{MSTP(22) :}\label{p:MSTP22} (D=0) special override of normal 
$Q^2$ definition used for maximum of parton-shower evolution. This 
option only affects processes 10 and 83 (deep inelastic scattering) 
and only in lepton--hadron events.
\begin{subentry}
\iteme{= 0 :} use the scale as given in \ttt{MSTP(32)}.
\iteme{= 1 :} use the DIS $Q^2$ scale, i.e. $-\hat{t}$.
\iteme{= 2 :} use the DIS $W^2$ scale, i.e. $(-\hat{t})(1-x)/x$.
\iteme{= 3 :} use the DIS $Q \times W$ scale, i.e.
$(-\hat{t}) \sqrt{(1-x)/x}$.
\iteme{= 4 :} use the scale $Q^2 (1-x) \max(1, \ln(1/x))$, as 
motivated by first order matrix elements \cite{Ing80,Alt78}.
\itemc{Note:} in all of these alternatives, a multiplicative factor is
introduced by \ttt{PARP(67)} and \ttt{PARP(71)}, as usual.
\end{subentry}
 
\iteme{MSTP(61) :}\label{p:MSTP61} (D=1) master switch for 
initial-state QCD and QED radiation.
\begin{subentry}
\iteme{= 0 :} off.
\iteme{= 1 :} on.
\end{subentry}

\iteme{MSTP(62) :} (D=3) level of coherence imposed on the
space-like parton-shower evolution.
\begin{subentry}
\iteme{= 1 :} none, i.e. neither $Q^2$ values nor angles need be
ordered.
\iteme{= 2 :} $Q^2$ values at branches are strictly ordered,
increasing towards the hard interaction.
\iteme{= 3 :} $Q^2$ values and opening angles of emitted
(on-mass-shell or time-like) partons are both strictly ordered,
increasing towards the hard interaction.
\end{subentry}
 
\iteme{MSTP(63) :} (D=2) structure of associated time-like
showers, i.e. showers initiated by emission off the incoming
space-like partons.
\begin{subentry}
\iteme{= 0 :} no associated showers are allowed, i.e. emitted
partons are put on the mass shell.
\iteme{= 1 :} a shower may evolve, with maximum allowed time-like
virtuality set by the phase space only.
\iteme{= 2 :} a shower may evolve, with maximum allowed time-like
virtuality set by phase space or by \ttt{PARP(71)} times the $Q^2$
value of the space-like parton created in the same vertex, whichever
is the stronger constraint.
\end{subentry}
 
\iteme{MSTP(64) :} (D=2) choice of $\alphas$ and $Q^2$ scale
in space-like parton showers.
\begin{subentry}
\iteme{= 0 :} $\alphas$ is taken to be fix at the value
\ttt{PARU(111)}.
\iteme{= 1 :} first-order running $\alphas$ with argument
\ttt{PARP(63)}$Q^2$.
\iteme{= 2 :} first-order running $\alphas$ with argument
\ttt{PARP(64)}$k_{\perp}^2 = $\ttt{PARP(64)}$(1-z)Q^2$.
\end{subentry}
 
\iteme{MSTP(65) :} (D=1) treatment of soft gluon emission in
space-like parton-shower evolution.
\begin{subentry}
\iteme{= 0 :} soft gluons are entirely neglected.
\iteme{= 1 :} soft gluon emission is resummed and included
together with the hard radiation as an effective $z$ shift.
\end{subentry}

\iteme{MSTP(66) :} (D=1) choice of lower cut-off for initial-state
QCD radiation in anomalous photoproduction events 
(see \ttt{MSTP(14)=3}).
\begin{subentry}
\iteme{= 0 :} the lower $Q^2$ cutoff is the standard one in 
\ttt{PARP(62)}$^2$.
\iteme{= 1 :} the lower cutoff is the larger of \ttt{PARP(62)}$^2$
and \ttt{VINT(283)} or \ttt{VINT(284)}, where the latter is the 
virtuality scale of the $\gamma \to \q\qbar$ vertex on the appropriate
side of the event.
\end{subentry}
 
\iteme{MSTP(67) :} (D=2) possibility to introduce colour coherence 
effects in the first branching of the backwards evolution of an 
initial state shower; mainly of relevance for QCD parton--parton 
scattering processes.
\begin{subentry}
\iteme{= 0 :} none.
\iteme{= 2 :} restrict the polar angle of a branching to be smaller 
than the scattering angle of the relevant colour flow.
\itemc{Note 1:} azimuthal anisotropies have not yet been included.
\itemc{Note 2:} for subsequent branchings, \ttt{MSTP(62)=3} is
used to restrict the (polar) angular range of branchings.
\end{subentry}

\iteme{MSTP(71) :} (D=1) master switch for final-state QCD and
QED radiation.
\begin{subentry}
\iteme{= 0 :} off.
\iteme{= 1 :} on.
\end{subentry}

\boxsep
 
\iteme{PARP(61) :}\label{p:PARP61} (D=0.25 GeV) $\Lambda$ value 
used in space-like parton shower (see \ttt{MSTP(64)}). This value 
may be overwritten, see \ttt{MSTP(3)}.
 
\iteme{PARP(62) :} (D=1. GeV) effective cut-off $Q$ or
$k_{\perp}$ value (see \ttt{MSTP(64)}), below which space-like
parton showers are not evolved.
 
\iteme{PARP(63) :} (D=0.25) in space-like shower evolution the
virtuality $Q^2$ of a parton is multiplied by \ttt{PARP(63)} for use
as a scale in $\alphas$ and parton distributions when
\ttt{MSTP(64)=1}.
 
\iteme{PARP(64) :} (D=1.) in space-like parton-shower evolution
the squared transverse momentum evolution scale $k_{\perp}^2$ is
multiplied by \ttt{PARP(64)} for use as a scale in $\alphas$ and
parton distributions when \ttt{MSTP(64)=2}.
 
\iteme{PARP(65) :} (D=2. GeV) effective minimum energy (in c.m.
frame) of time-like or on-shell parton emitted in space-like shower;
see also \ttt{PARP(66)}.
 
\iteme{PARP(66) :} (D=0.001) effective lower cut-off on $1-z$ in
space-like showers, in addition to the cut implied by \ttt{PARP(65)}.
 
\iteme{PARP(67) :} (D=4.) the $Q^2$ scale of the hard scattering
(see \ttt{MSTP(32)}) is multiplied by \ttt{PARP(67)} to define the
maximum parton virtuality allowed in space-like showers. This does not
apply to $s$-channel resonances, where the maximum virtuality is set
by $m^2$.
 
\iteme{PARP(68) :} (D=1E-3) lower $Q$ cut-off for QED space-like
showers.
 
\iteme{PARP(71) :} (D=4.) the $Q^2$ scale of the hard scattering
(see \ttt{MSTP(32)}) is multiplied by \ttt{PARP(71)} to define the
maximum parton virtuality allowed in time-like showers. This does not
apply to $s$-channel resonances, where the maximum virtuality is set
by $m^2$.

\iteme{PARP(72) :} (D=0.25 GeV) $\Lambda$ value used in running 
$\alphas$ for timelike parton showers, except for showers in the
decay of a resonance. (Resonance decay, e.g. $\gamma^*/Z^0$ decay,
is instead set by \ttt{PARJ(81)}.) 

\end{entry}

\clearpage
 
\section{Beam Remnants and Underlying Events}
 
Each incoming beam particle may leave behind a beam remnant, which
does not take part in the initial-state radiation or hard scattering
process. If nothing else, the remnants need be reconstructed and
connected to the rest of the event. In hadron--hadron collisions,
the composite nature of the two incoming beam particles implies the
additional possibility that several parton pairs undergo separate
hard or semi-hard scatterings, `multiple interactions'.
This may give a non-negligible
contribution to the `underlying event' structure, and thus to the
total multiplicity. Finally, in high-luminosity colliders, it is
possible to have several collisions between beam particles in one
and the same beam crossing, i.e. pile-up events, which further act
to build up the
general particle production activity that is to be observed by
detectors. These three aspects are described in turn, with emphasis
on the middle one, that of multiple interactions within a single
hadron--hadron collision.
 
The main reference on the multiple interactions model is
\cite{Sjo87a}.
 
\subsection{Beam Remnants}
\label{ss:beamrem}
 
The initial-state radiation algorithm reconstructs one shower
initiator in each beam. (If initial-state radiation is not included,
the initiator is nothing but the incoming parton to the hard
interaction.) Together the two initiators delineate an
interaction subsystem, which contains all the partons that
participate in the initial-state showers, in the hard interaction,
and in the final-state showers. Left behind are two beam remnants
which, to first approximation, just sail through, unaffected by the
hard process. (The issue of additional interactions is covered in
the next section.)
 
A description of the beam remnant structure contains a few components.
First, given the flavour content of a (colour-singlet) beam particle,
and the flavour and colour of the initiator parton, it is possible
to reconstruct the flavour and colour of the
beam remnant. Sometimes the remnant may be
represented by just a single parton or diquark, but often the
remnant has to be subdivided into two separate objects. In the latter
case it is necessary to share the remnant energy and momentum between
the two. Due to Fermi motion inside hadron beams, the initiator parton
may have a `primordial $k_{\perp}$' transverse momentum motion,
which has to be compensated by the beam remnant. If the remnant is
subdivided, there may also be a relative transverse momentum.
In the end, total energy and momentum has to be conserved.
To first approximation, this is ensured within each remnant
separately, but some final global adjustments are necessary to
compensate for the
primordial $k_{\perp}$ and any effective beam remnant mass.
 
Consider first a proton (or, with trivial modifications, any other
baryon or antibaryon).
\begin{Itemize}
\item If the initiator parton is a $\u$ or $\d$ quark, it is
assumed to be a valence quark, and therefore leaves behind a
diquark beam remnant, i.e. either a $\u\d$ or a $\u\u$ diquark,
in a colour antitriplet state.
Relative probabilities for different diquark spins are derived within
the context of the non-relativistic {\bf SU(6)} model, i.e. flavour
{\bf SU(3)} times spin {\bf SU(2)}. Thus a $\u\d$ is $3/4$ $\u\d_0$
and $1/4$ $\u\d_1$, while a $\u\u$ is always $\u\u_1$.
\item An initiator gluon leaves behind a colour octet $\u\u\d$
state, which is subdivided into a colour triplet quark and a colour
antitriplet diquark. {\bf SU(6)} gives the appropriate subdivision,
$1/2$ of the time into $\u + \u\d_0$, $1/6$ into $\u + \u\d_1$ and
$1/3$ into $\d + \u\u_1$.
\item A sea quark initiator, such as an $\s$, leaves behind a
$\u\u\d\sbar$ four-quark state. The PDG flavour coding scheme and
the fragmentation routines do not foresee such a state, so therefore
it is subdivided into a meson plus a diquark, i.e. $1/2$ into
$\u\sbar + \u\d_0$, $1/6$ into $\u\sbar +  \u\d_1$ and $1/3$ into
$\d\sbar + \u\u_1$. Once the flavours of the meson are determined,
the choice of meson multiplet is performed as in the standard
fragmentation description.
\item Finally, an antiquark initiator, such as an $\sbar$, leaves
behind a $\u\u\d\s$ four-quark state, which is subdivided into a
baryon plus a quark. Since, to first approximation, the $\s\sbar$
pair comes from the branching $\g \to \s\sbar$ of a colour octet
gluon, the subdivision $\u\u\d + \s$ is not allowed, since it would
correspond to a colour-singlet $\s\sbar$. Therefore the subdivision
is $1/2$ into $\u\d_0\s + \u$, $1/6$ into $\u\d_1\s + \u$ and $1/3$ 
into $\u\u_1\s + \d$. A baryon is formed among the ones possible for 
the given flavour content and diquark spin, according to the relative
probabilities used in the fragmentation. One could argue for an
additional weighting to count the number of baryon states available
for a given diquark plus quark combination, but this has not been
included.
\end{Itemize}
 
One may note that any $\u$ or $\d$ quark taken out of the proton is
automatically assumed to be a valence quark. Clearly this is
unrealistic,
but not quite as bad as it might seem. In particular, one should
remember that the beam remnant scenario is applied to the 
initial-state shower initiators at a scale of $Q_0 \approx 1$ GeV 
and at an
$x$ value usually much larger than the $x$ at the hard scattering.
The sea quark contribution therefore normally is negligible.
 
For a meson beam remnant, the rules are in the same spirit, but
somewhat easier, since no diquark or baryons need be taken into
account. Thus a valence quark (antiquark) initiator leaves
behind a valence antiquark (quark), a gluon initiator leaves
behind a valence quark plus a valence antiquark, and a sea quark
(antiquark) leaves behind a meson (which contains the partner to
the sea parton) plus a valence antiquark (quark).
 
A resolved photon is even simpler than a meson,
since one does not have to make
the distinction between valence and sea flavour. Thus any quark
(antiquark) initiator leaves behind the matching antiquark (quark),
and a gluon leaves behind a quark + antiquark pair. The relative
quark flavour composition in the latter case is assumed proportional
to $e_{\q}^2$ among light flavours, i.e. $2/3$ into $\u + \ubar$,
$1/6$ into $\d + \dbar$, and $1/6$ into $\s + \sbar$. If one wanted
to, one could also have chosen to represent the remnant by a single
gluon.
 
If no initial-state radiation is assumed, an electron (or, in general,
a lepton or a neutrino) leaves behind no beam remnant. Also when
radiation is included, one would expect to recover a single electron
with the full beam energy when the shower initiator is reconstructed.
This does not have to happen, e.g. if the initial-state shower is cut
off at a non-vanishing scale, such that some of the emission at low
$Q^2$ values is not simulated. Further, for purely technical reasons,
the distribution of an electron inside an electron,
$f_{\e}^{\e}(x,Q^2)$, is cut off at $x = 1 - 10^{-6}$. This means that
always, when initial-state radiation is included, a fraction
of at least $10^{-6}$ of the beam energy has to be put into one single
photon along the beam direction, to represent this not simulated
radiation. The physics is here slightly different from the standard
beam remnant concept, but it is handled with the same machinery.
Beam remnants can also apper when the electron is resolved with the
use of parton distributions, but initial-state radiation is switched
off. Conceptually, this is a contradiction, since it is the 
initial-state radiation that builds up the parton distributions, 
but sometimes
the combination is still useful. Finally, since QED radiation has not
yet been included in events with resolved photons inside electrons,
also in this case effective beam remnants have to be assigned by the
program.
 
The beam remnant assignments inside an electron, in either of the cases
above, is as follows.
\begin{Itemize}
\item An $\e^-$ initiator leaves behind a $\gamma$ remnant.
\item A $\gamma$ initiator leaves behind an $\e^-$ remnant.
\item An $\e^+$ initiator leaves behind an $\e^- + \e^-$ remnant.
\item A $\q$ ($\qbar$) initiator leaves behind a
$\qbar + \e^-$ ($\q + \e^-$) remnant.
\item A $\g$ initiator leaves behind a $\g + \e^-$ remnant.
One could argue that, in agreement with the treatment of photon
beams above, the remnant should be $\q + \qbar + \e^-$. The program
currently does not allow for three beam remnant objects, however.
\end{Itemize}
 
By the hard scattering and initial-state radiation machinery,
the shower initiator has been assigned some fraction $x$ of the
four-momentum of the beam particle, leaving behind $1-x$ to the
remnant. If the remnant consists of two objects, this energy
and momentum has to be shared, somehow. For an electron, the
sharing is given from first principles: if, e.g., the initiator
is a $\q$, then that $\q$ was produced in the sequence of branchings
$\e \to \gamma \to \q$, where $x_{\gamma}$ is distributed according
to the convolution in eq.~(\ref{pg:foldqgine}). Therefore the $\qbar$
remnant takes a fraction $\chi = (x_{\gamma} -x)/(1-x)$ of the total
remnant energy, and the $\e$ takes $1 - \chi$.
 
For the other beam remnants, the relative energy-sharing variable
$\chi$ is not known from first principles, but picked according to
some suitable parametrization. Normally several different options are
available, that can be set separately for baryon and meson beams, and
for hadron + quark and quark + diquark (or antiquark) remnants. In one
extreme are shapes in agreement with na\"{\i}ve counting rules, i.e.
where energy is shared evenly between `valence' partons. For instance,
${\cal P}(\chi) = 2 \, (1-\chi)$ for the energy fraction taken by the
$\q$ in a $\q + \q\q$ remnant. In the other extreme, an uneven 
distribution could be used, like in parton distributions, where the 
quark only takes a small fraction and most is retained by the diquark. 
The default for a $\q + \q\q$ remnant is of this type,
\begin{equation}
{\cal P}(\chi) \propto \frac{(1 - \chi)^3}
{\sqrt{\chi^2 + c_{\mmin}^2}} ~,
\end{equation}
with $c_{\mmin} = 2 \langle m_{\q} \rangle / E_{\mrm{cm}} = (0.6$
GeV)$/ E_{\mrm{cm}}$ providing a lower cut-off. In general, the more 
uneven the sharing of the energy, the less the total multiplicity in the
beam remnant fragmentation. If no multiple interactions are allowed,
a rather even sharing is needed to come close to the experimental
multiplicity (and yet one does not quite make it). With an
uneven sharing there is room to generate more of the total multiplicity
by multiple interactions \cite{Sjo87a}.
 
In a photon beam, with a remnant $\q + \qbar$, the $\chi$ variable is
chosen the same way it would have been in a corresponding meson 
remnant. 
 
Before the $\chi$ variable is used to assign remnant momenta, it
is also necessary to consider the issue of primordial $k_{\perp}$.
The initiator partons are thus assigned each a $k_{\perp}$ value,
vanishing for an electron or photon inside an electron, distributed
either according to a Gaussian or an exponential shape for a hadron,
and according to either of these shapes or a power-like shape
for a quark or gluon inside a photon (which may in its turn be inside
an electron). The interaction subsystem is boosted and rotated to bring
it from the frame assumed so far, with each initiator along the $\pm z$
axis, to one where the initiators have the required primordial
$k_{\perp}$ values.
 
The $\pT$ recoil is taken by the remnant. If the remnant is composite,
the recoil is all taken by one of the two, namely the one that,
in some imagined perturbative splitting language, is the sister of
the initiator parton. For instance, when a gluon is taken out of a
proton, the recoil is all taken by the lone quark
(i.e. nothing by the diquark), since one could
have imagined an earlier branching $\q_0 \to \q \g$, below the shower
cut-off scale $Q_0$, with $p_{\perp q_0} = 0$.
In addition, however, two beam remnants may be
given a relative $\pT$, which is then always chosen as for
$\q_i \qbar_i$ pairs in the fragmentation description.
 
The $\chi$ variable is interpreted as a sharing of light-cone energy and
momentum, i.e. $E + p_z$ for the beam moving in the $+z$ direction and
$E - p_z$ for the other one. When the two transverse masses
$m_{\perp 1}$ and $m_{\perp 2}$ of a composite remnant have been
constructed, the total transverse mass can therefore be found as
\begin{equation}
m_{\perp}^2 = \frac{m_{\perp 1}^2}{\chi} +
\frac{m_{\perp 2}^2}{1 - \chi}   ~,
\end{equation}
if remnant 1 is the one that takes the fraction $\chi$. The choice of
a light-cone interpretation to $\chi$ means the definition is invariant
under longitudinal boosts, and therefore does not depend on the beam
energy itself. A $\chi$ value close to the na\"{\i}ve borders 0 or 1 can
lead to an unreasonably large remnant $m_{\perp}$.
Therefore an additional check is introduced, that the remnant
$m_{\perp}$ be smaller than the na\"{\i}ve c.m. frame remnant energy,
$(1-x) E_{\mrm{cm}}/2$. If this is not the case, a new $\chi$ and a 
new relative transverse momentum is selected.
 
Whether there is one remnant parton or two, the transverse mass of
the remnant is not likely to agree with $1-x$ times the mass of the
beam particle, i.e. it is not going to be possible to preserve
the energy and momentum
in each remnant separately. One therefore allows a
shuffling of energy and momentum between the beam remnants from
each of the two incoming beams. This may be achieved by performing a
(small) longitudinal boost of each remnant system. Since there are
two boost degrees of freedom, one for each remnant, and two constraints,
one for energy and one for longitudinal momentum, a solution may be
found.
 
Under some circumstances, one beam remnant may be absent or of very
low energy, while the other one is more complicated. One example is
deep inelastic scattering in $\ep$ collisions, where the electron
leaves no remnant, or maybe only a low-energy photon.
It is clearly then not possible to balance the two beam remnants
against each other. Therefore, if one beam remnant has an energy
below 0.2 of the beam energy, i.e. if the initiator parton has
$x > 0.8$, then the two boosts needed to ensure energy and momentum
conservation are instead performed on the other remnant and on the
interaction subsystem. If there is a low-energy remnant at all then,
before that, energy and momentum are assigned to the remnant
constituent(s) so that the appropriate light-cone combination
$E \pm p_z$ is conserved, but not energy or momentum separately.
If both beam remnants have low energy, but both still exist, then
the one with lower $m_{\perp} / E$ is the one that will not be
boosted.
 
\subsection{Multiple Interactions}
\label{ss:multint}
 
In this section we present the model used in {\Py} to describe the
possibility that several parton pairs undergo hard interactions in a
hadron--hadron collision, and thereby contribute to the overall event
activity, in particular at low $\pT$. The same model is also used
to describe the VMD $\gamma \p$ events, where the photon interacts 
like a hadron. It should from the onset be made
clear that this is not an easy topic. In fact, in the full event
generation process, probably no other area is as poorly understood
as this one. The whole concept of multiple interactions is very
controversial, with contraditory experimental conclusions
\cite{AFS87}.
 
The multiple interactions scenario presented here \cite{Sjo87a} was
the first detailed model for this kind of physics, and is still one
of the very few available. We will present two related but separate
scenarios, one `simple' model and one somewhat more sophisticated.
In fact, neither of them are all that simple, which may make the
models look unattractive. However, the world of hadron physics
{\it is} complicated, and if we err, it is most likely in being
too unsophisticated. The experience gained with the model(s), in
failures as well as successes, could be used as a guideline in
the evolution of yet more detailed models.
 
Our basic philosophy will be as follows. The total rate of
parton--parton interactions, as a function of the transverse
momentum scale $\pT$, is assumed to be given by perturbative
QCD. This is certainly true for reasonably large $\pT$ values, but
here we shall also extend the perturbative parton--parton
scattering framework into the low-$\pT$ region. A regularization of
the divergence in the cross section for $\pT \to 0$ has to be
introduced, however, which will provide us with the main free
parameter of the model. Since each incoming hadron is a composite
object, consisting of many partons, there should exist the possibility
of several parton pairs interacting when two hadrons collide. It is
not unreasonable to assume that the different pairwise interactions
take place essentially independently of each other, and that therefore
the number of interactions in an event is given by a Poissonian
distribution. This is the strategy of the `simple' scenario.
 
Furthermore, hadrons are not only composite but also extended
objects, meaning that collisions range from very central to rather
peripheral ones. Reasonably, the average number of interactions
should be larger in the former than in the latter case. Whereas
the assumption of a Poissonian distribution should hold for each
impact parameter separately, the distribution
in number of interactions should be widened by the spread of impact
parameters. The amount of widening depends on the assumed
matter distribution inside the colliding hadrons. In the `complex'
scenario, different matter distributions are therefore introduced.
 
\subsubsection{The basic cross sections}
 
The QCD cross section for hard $2 \to 2$ processes, as a function
of the $\pT^2$ scale, is given by
\begin{equation}
\frac{\d \sigma}{\d \pT^2} = \sum_{i,j,k} \int \d x_1 \int \d x_2 
\int \d \hat{t} \, f_i(x_1,Q^2) \, f_j(x_2,Q^2)
\, \frac{\d \hat{\sigma}_{ij}^k}{\d \hat{t}} \,
\delta \! \left( \pT^2 - \frac{\hat{t}\hat{u}}{\hat{s}} \right) ~,
\label{mi:sigmapt}
\end{equation}
cf. section \ref{ss:kinemtwo}. Implicitly, from now on we are assuming
that the `hardness' of processes is given by the $\pT$ scale of the
scattering. For an application of the formula above to small $\pT$
values, a number of caveats could be made. At low $\pT$, the
integrals receive major contributions from the small-$x$ region,
where parton distributions are poorly understood theoretically
(Regge limit behaviour, dense packing problems etc. \cite{Lev90})
and not yet measured. Different sets of parton distributions can
therefore give numerically rather different results for the
phenomenology of interest. One may also worry about higher-order
corrections to the jet rates ($K$ factors), beyond what is given
by parton-shower corrections --- one simple option we allow here
is to evaluate $\alphas$ of the hard scattering process at an
optimized scale, $\alphas(0.075\pT^2)$ \cite{Ell86}.
 
The hard scattering cross section above some given $\pTmin$
is given by
\begin{equation}
\sigma_{\mrm{hard}} (\pTmin) = \int_{\pTmin^2}^{s/4}
\frac{\d \sigma}{\d \pT^2} \, \d \pT^2 ~.
\label{mi:sigmahard}
\end{equation}
Since the differential cross section diverges roughly like
$\d \pT^2 / \pT^4$, $\sigma_{\mrm{hard}}$ is also divergent for
$\pTmin \to 0$. We may compare this with the total
inelastic, non-diffractive cross section $\sigma_{\mrm{nd}}(s)$
--- elastic and diffractive events are not the topic of this
section. At current collider energies
$\sigma_{\mrm{hard}} (\pTmin)$ 
becomes comparable with $\sigma_{\mrm{nd}}$
for $\pTmin \approx$1.5--2 GeV. This need not lead to
contradictions: $\sigma_{\mrm{hard}}$ does not give the hadron--hadron
cross section but the parton--parton one. Each of the incoming
hadrons may be viewed as a beam of partons, with the possibility of
having several parton--parton interactions when the hadrons pass
through each other. In this language,
$\sigma_{\mrm{hard}} (\pTmin) / \sigma_{\mrm{nd}}(s)$ 
is simply the
average number of parton--parton scatterings above $\pTmin$
in an event, and this number may well be larger than unity.
 
While the introduction of several interactions per event is the
natural consequence of allowing small $\pTmin$ values
and hence large $\sigma_{\mrm{hard}}$ ones, it is not the solution of
$\sigma_{\mrm{hard}} (\pTmin)$ being divergent for
$\pTmin \to 0$: the average $\hat{s}$ of a scattering
decreases slower with $\pTmin$ than the number of
interactions increases, so na\"{\i}vely the total amount of scattered
partonic energy becomes infinite. One cut-off is therefore
obtained via the need to introduce proper multi-parton correlated
parton distributions inside a hadron. This is not a part of the
standard perturbative QCD formalism and is therefore not built into
eq.~(\ref{mi:sigmahard}). In practice, even correlated 
parton-distribution functions seems to provide too weak a cut, 
i.e. one is lead to a
picture with too little of the incoming energy remaining in the
small-angle beam jet region.
 
A more credible reason for an effective cut-off is that the incoming
hadrons are colour neutral objects. Therefore, when the $\pT$ of an
exchanged gluon is made small and the transverse wavelength
correspondingly large, the gluon can no longer resolve the individual
colour charges, and the effective coupling is decreased. This
mechanism is not in contradiction to perturbative
QCD calculations, which are always performed assuming scattering
of free partons (rather than partons inside hadrons), but neither does
present knowledge of QCD provide an understanding of how such a
decoupling mechanism would work in detail. In the simple model one
makes use of a sharp cut-off at some scale $\pTmin$, while
a more smooth dampening is assumed for the complex scenario.
 
\subsubsection{The simple model}
 
In an event with several interactions, it is convenient to impose an
ordering. The logical choice is to arrange the scatterings in falling
sequence of $x_{\perp} = 2 \pT / E_{\mrm{cm}}$. The `first' scattering 
is thus the hardest one, with the `subsequent' (`second', `third', etc.)
successively softer. It is important to remember that this terminology
is in no way related to any picture in physical time; we do not know
anything about the latter. In principle, all the scatterings that occur
in an event must be correlated somehow, na\"{\i}vely by momentum 
and flavour conservation for the partons from each incoming hadron, 
less na\"{\i}vely by
various quantum mechanical effects. When averaging over all
configurations of soft partons, however, one should effectively obtain
the standard QCD phenomenology for a hard scattering, e.g. in terms of
parton distributions. Correlation effects, known or estimated, can be
introduced in the choice of subsequent scatterings, given that the
`preceding' (harder) ones are already known.
 
With a total cross section of hard interactions
$\sigma_{\mrm{hard}} (\pTmin)$ to be distributed among
$\sigma_{\mrm{nd}}(s)$ (non-diffractive, inelastic) events, the average
number of interactions per event is just the ratio
$\br{n} = \sigma_{\mrm{hard}} (\pTmin) / \sigma_{\mrm{nd}}(s)$. 
As a starting point we will assume that all hadron collisions are
equivalent (no impact parameter dependence), and that the different
parton--parton interactions take place completely independently of
each other. The number of scatterings per event is then distributed
according to a Poissonian with mean $\br{n}$. A fit to collider
multiplicity data gives $\pTmin \approx 1.6$ GeV, which
corresponds to $\br{n} \approx 1$. For Monte Carlo
generation of these interactions it is useful to define
\begin{equation}
f(x_{\perp}) = \frac{1}{\sigma_{\mrm{nd}}(s)} \,
\frac{\d \sigma}{\d x_{\perp}}   ~,
\end{equation}
with $\d \sigma / \d x_{\perp}$ obtained by analogy with eq.
(\ref{mi:sigmapt}). Then $f(x_{\perp})$ is simply the probability to
have a parton--parton interaction at $x_{\perp}$, given that the two
hadrons undergo a non-diffractive, inelastic collision.
 
The probability that the hardest interaction, i.e. the one with
highest $x_{\perp}$, is at $x_{\perp 1}$, is now given by
\begin{equation}
f(x_{\perp 1}) \exp \left\{ - \int_{x_{\perp 1}}^1
f(x'_{\perp}) \, \d x'_{\perp} \right\}   ~,
\label{mi:hardest}
\end{equation}
i.e. the na\"{\i}ve probability to have a scattering at $x_{\perp 1}$
multiplied by the probability that there was no scattering with
$x_{\perp}$ larger than $x_{\perp 1}$. This is the familiar
exponential dampening in radioactive decays, encountered e.g. in
parton showers in section \ref{ss:sudakov}. Using the same technique
as in the proof of the veto algorithm, section \ref{ss:vetoalg},
the probability to have an $i$:th scattering at an
$x_{\perp i} < x_{\perp i-1} < \cdots < x_{\perp 1} < 1$ is found
to be
\begin{equation}
f(x_{\perp i}) \, \frac{1}{(i-1)!} \left( \int_{x_{\perp i}}^1
f(x'_{\perp}) \, \d x'_{\perp} \right)^{i-1} \exp \left\{
- \int_{x_{\perp i}}^1 f(x'_{\perp}) \, \d x'_{\perp} \right\}  ~.
\end{equation}
The total probability to have a scattering at a given $x_{\perp}$,
irrespectively of it being the first, the second or whatever,
obviously adds up to give back $f(x_{\perp})$. The multiple
interaction formalism thus retains the correct perturbative
QCD expression for the scattering probability at any given
$x_{\perp}$.
 
With the help of the integral
\begin{equation}
F(x_{\perp}) = \int_{x_{\perp}}^1 f(x'_{\perp}) \, \d x'_{\perp}
= \frac{1}{\sigma_{\mrm{nd}}(s)} \, \int_{s x_{\perp}^2/4}^{s/4}
\frac{\d \sigma}{\d \pT^2} \, \d \pT^2
\end{equation}
(where we assume $F(x_{\perp}) \to \infty$ for $x_{\perp} \to 0$)
and its inverse $F^{-1}$, the iterative procedure to generate
a chain of scatterings
$1 > x_{\perp 1} > x_{\perp 2} > \cdots > x_{\perp i}$
is given by
\begin{equation}
x_{\perp i} = F^{-1}(F(x_{\perp i-1}) - \ln R_i) ~.
\label{mi:iter}
\end{equation}
Here the $R_i$ are random numbers evenly distributed between 0 and 1.
The iterative chain is started with a fictitious $x_{\perp 0} = 1 $
and is terminated when $x_{\perp i}$ is smaller than
$x_{\perp \mmin} = 2 \pTmin / E_{\mrm{cm}}$. Since $F$ and 
$F^{-1}$ are not known analytically, the standard veto algorithm is 
used to generate a much denser set of $x_{\perp}$ values, whereof 
only some are retained in the end. In addition to the $\pT^2$ of an
interaction, it is also necessary to generate the other flavour
and kinematics variables according to the relevant matrix elements.
 
Whereas the ordinary parton distributions should be used for the
hardest scattering, in order to reproduce standard QCD
phenomenology, the parton distributions to be used for subsequent
scatterings must depend on all preceding $x$ values and flavours
chosen. We do not know enough about the hadron wave function to
write down such joint probability distributions. To take
into account the energy `already' used in harder scatterings, a
conservative approach is to evaluate the parton distributions, not at
$x_i$ for the $i$:th scattered parton from hadron, but at
the rescaled value
\begin{equation}
x'_i = \frac{x_i}{\sum_{j=1}^{i-1} x_j} ~.
\end{equation}
This is our standard procedure in the program; we have
tried a few alternatives without finding any significantly
different behaviour in the final physics.
 
In a fraction $\exp(-F(x_{\perp \mmin}))$ of the events studied, there
will be no hard scattering above $x_{\perp \mmin}$ when the iterative
procedure in eq.~(\ref{mi:iter}) is applied. It is therefore also
necessary to have a model for what happens in events with no
(semi)hard interactions. The simplest possible way to produce an event
is to have an exchange of a very soft gluon between the two colliding
hadrons. Without (initially) affecting the momentum distribution of
partons, the `hadrons' become colour octet objects rather than colour
singlet ones. If only valence quarks are considered, the colour
octet state of a baryon can be decomposed into a colour triplet quark
and an antitriplet diquark. In a baryon-baryon collision, one would
then obtain a two-string picture, with each string stretched from the
quark of one baryon to the diquark of the other. A baryon-antibaryon
collision would give one string between a quark and an antiquark and
another one between a diquark and an antidiquark.
 
In a hard interaction, the number of possible string drawings are many
more, and the overall situation can become quite complex
when several hard scatterings are present in an event.
Specifically, the string drawing now depends on the relative colour
arrangement, in each hadron individually, of the partons that are about
to scatter. This is a subject about which nothing is known.
To make matters worse, the standard string fragmentation description
would have to be extended, to handle events where two or more valence
quarks have been kicked out of an incoming hadron by separate
interactions. In particular, the position of the baryon number would
be unclear. We therefore here assume that, following the hardest
interaction, all subsequent interactions belong to one of three
classes.
\begin{Itemize}
\item Scatterings of the $\g\g \to \g\g $ type, with
the two gluons in a colour-singlet state, such that a double string is
stretched directly between the two outgoing gluons, decoupled from the
rest of the system.
\item Scatterings $\g\g \to \g\g$, but colour correlations
assumed to be such that each of the gluons is connected to one
of the strings `already' present. Among the different possibilities of
connecting the colours of the gluons, the one which minimizes the total
increase in string length is chosen. This is in contrast to the
previous alternative, which roughly corresponds to a maximization of
the extra string length.
\item Scatterings $\g\g \to \q\qbar$, with the final pair again
in a colour-singlet state, such that a single string is stretched
between the outgoing $\q$ and $\qbar$.
\end{Itemize}
By default, the three possibilities are assumed equally probable.
Note that the total jet rate is maintained at its nominal value, i.e.
scatterings such as $\q\g \to \q\g$ are included in the cross section,
but are replaced by a mixture of $\g\g$ and $\q\qbar$ events for string
drawing issues. Only the hardest interaction is guaranteed to give
strings coupled to the beam remnants. One should not take this approach
to colour flow too seriously --- clearly
it is a simplification --- but the overall picture does not tend to be
very dependent on the particular choice you make.
 
Since a $\g\g \to \g\g$ or $\q\qbar$ scattering need not remain that if
initial- and final-state showers were to be included, radiation is only
included for the hardest interaction. In practice, there is no
problem: except for the hardest interaction, which can be hard because
of experimental trigger conditions, it is unlikely for a parton
scattering to be so hard that radiation plays a significant r\^ole.
 
In events with multiple interactions, the beam remnant treatment is
slightly modified. First the hard scattering is generated, with its
associated initial- and final-state radiation, and next any additional
multiple interactions. Only thereafter are beam remnants attached to
the initator partons of the hardest scattering, using the same
machinery as before, except that the energy and momentum already
taken away from the beam remnants also include that of the
subsequent interactions.
 
\subsubsection{A model with varying impact parameters}
 
Up to this point, it has been assumed that the initial state is the
same for all hadron collisions, whereas in fact each collision also
is characterized by a varying impact parameter $b$. Within the
classical framework of this paper, $b$ is to be thought of as a
distance of closest approach, not as the Fourier transform of the
momentum transfer. A small $b$ value corresponds to a large
overlap between the two colliding hadrons, and hence an enhanced
probability for multiple interactions. A large $b$, on the other
hand, corresponds to a grazing collision, with a large probability
that no parton--parton interactions at all take place.
 
In order to quantify the concept of hadronic matter overlap, one may
assume a spherically symmetric distribution of matter inside the
hadron, $\rho(\mbf{x}) \, \d^3 x = \rho(r) \, \d^3 x$.
For simplicity, the same spatial distribution is taken to apply
for all parton species and momenta. Several different matter
distributions have been tried, and are available. We will here
concentrate on the most extreme one, a double Gaussian
\begin{equation}
\rho(r) \propto \frac{1 - \beta}{a_1^3} \, \exp \left\{
- \frac{r^2}{a_1^2} \right\} + \frac{\beta}{a_2^3} \,
\exp \left\{ - \frac{r^2}{a_2^2} \right\} ~.
\label{mi:doubleGauss}
\end{equation}
This corresponds to a distribution with a small core region, of radius
$a_2$ and containing a fraction $\beta$ of the total hadronic matter,
embedded in a larger hadron of radius $a_1$. While it is mathematically
convenient to have the origin of the two Gaussians coinciding, the
physics could well correspond to having three disjoint core regions,
reflecting the presence of three valence quarks, together carrying
the fraction $\beta$ of the proton momentum. One could alternatively
imagine a hard hadronic core surrounded by a pion cloud.
Such details would affect e.g. the predictions for the $t$ distribution
in elastic scattering, but are not of any consequence for the
current topics. To be
specific, the values $\beta = 0.5$ and $a_2/a_1 = 0.2$ have been
picked as default values. It should be noted that the overall distance
scale $a_1$ never enters in the subsequent calculations, since the
inelastic, non-diffractive cross section $\sigma_{\mrm{nd}}(s)$ is taken
from literature rather than calculated from the $\rho(r)$.
 
Compared to other shapes, like a simple Gaussian, the double Gaussian
tends to give larger fluctuations, e.g. in the multiplicity
distribution of minimum bias events: a collision in which the two
cores overlap tends to have a strongly increased activity, while
ones where they do not are rather less active. One also has a
biasing effect: hard processes are more likely
when the cores overlap, thus hard scatterings are associated with
an enhanced multiple interaction rate. This provides one possible
explanation for the experimental `pedestal effect'.
 
For a collision with impact parameter $b$, the time-integrated
overlap ${\cal O}(b)$ between the matter distributions of the
colliding hadrons is given by
\begin{eqnarray}
& & {\cal O}(b) \propto \int \d t \int \d^3 x \, \rho(x,y,z) \,
\rho(x+b,y,z+t)      \nonumber \\
& & \propto \frac{(1 - \beta)^2}{2a_1^2} \exp \left\{
- \frac{b^2}{2a_1^2} \right\} +
\frac{2 \beta (1-\beta)}{a_1^2+a_2^2} \exp \left\{
- \frac{b^2}{a_1^2+ a_2^2} \right\} +
\frac{\beta^2}{2a_2^2} \exp \left\{ - \frac{b^2}{2a_2^2} \right\} ~.
\end{eqnarray}
The necessity to use boosted $\rho(\mbf{x})$ distributions
has been circumvented by a suitable scale transformation of the $z$
and $t$ coordinates.
 
The overlap ${\cal O}(b)$ is obviously strongly related to the
eikonal $\Omega(b)$ of optical models. We have kept a separate
notation, since the physics context of the two is slightly
different: $\Omega(b)$ is based on the quantum mechanical
scattering of waves in a potential, and is normally used to
describe the elastic scattering of a hadron-as-a-whole, while
${\cal O}(b)$ comes from a purely classical picture of point-like
partons distributed inside the two colliding hadrons. Furthermore,
the normalization and energy dependence is differently realized
in the two formalisms.

The larger the overlap ${\cal O}(b)$ is, the more likely it is to
have interactions between partons in the two colliding hadrons.
In fact, there should be a linear relationship
\begin{equation}
\langle \tilde{n}(b) \rangle = k {\cal O}(b) ~,
\end{equation}
where $\tilde{n} = 0, 1, 2, \ldots$ counts the number of interactions
when two hadrons pass each other with an impact parameter $b$.
The constant of proportionality, $k$, is related to the
parton--parton cross section and hence increases with c.m. energy.
 
For each given impact parameter, the number of interactions is
assumed to be distributed according to a Poissonian. If the matter
distribution has a tail to infinity (as the double Gaussian does),
events may be obtained with arbitrarily large $b$ values. In order
to obtain finite total cross sections, it is necessary to assume
that each event contains at least one semi-hard interaction. The
probability that two hadrons, passing each other with an impact
parameter $b$, will actually undergo a collision is then given by
\begin{equation}
{\cal P}_{\mrm{int}}(b) = 1 - \exp ( - \langle \tilde{n}(b) \rangle )
= 1 - \exp (- k {\cal O}(b) ) ~,
\end{equation}
according to Poissonian statistics. The average number of
interactions per event at impact parameter $b$ is now
\begin{equation}
\langle n(b) \rangle =
\frac{ \langle \tilde{n}(b) \rangle }{{\cal P}_{\mrm{int}}(b)} =
\frac{ k {\cal O}(b) }{ 1 - \exp (- k {\cal O}(b) ) } ~,
\label{mi:nofb}
\end{equation}
where the denominator comes from the removal of hadron pairs which
pass without colliding, i.e. with $\tilde{n} =  0$.
 
The relationship 
$\langle n \rangle = \sigma_{\mrm{hard}}/\sigma_{\mrm{nd}}$
was earlier introduced for the average number of interactions per
non-diffractive, inelastic event. When averaged over all
impact parameters, this relation must still hold true: the
introduction of variable impact parameters may give more interactions
in some events and less in others, but it does not affect either
$\sigma_{\mrm{hard}}$ or $\sigma_{\mrm{nd}}$. 
For the former this is because the
perturbative QCD calculations only depend on the total parton flux,
for the latter by construction. Integrating eq.~(\ref{mi:nofb}) over
$b$, one then obtains
\begin{equation}
\langle n \rangle =
\frac{ \int \langle n(b) \rangle \, {\cal P}_{\mrm{int}}(b) \, \d^2 b}
{ \int {\cal P}_{\mrm{int}}(b) \, \d^2 b} =
\frac{ \int k {\cal O}(b) \, \d^2 b}
{ \int \left( 1 - \exp (- k {\cal O}(b) ) \right) \, \d^2 b} =
\frac{\sigma_{\mrm{hard}}}{\sigma_{\mrm{nd}}}   ~.
\label{mi:kfromsigma}
\end{equation}
For ${\cal O}(b)$, $\sigma_{\mrm{hard}}$ and $\sigma_{\mrm{nd}}$ given, 
with $\sigma_{\mrm{hard}} / \sigma_{\mrm{nd}} > 1$, $k$ can thus always 
be found (numerically) by solving the last equality.
 
The absolute normalization of ${\cal O}(b)$ is not interesting
in itself, but only the relative variation with impact parameter.
It is therefore useful to introduce an `enhancement factor'
$e(b)$, which gauges how the interaction probability for a passage
with impact parameter $b$ compares with the average, i.e.
\begin{equation}
\langle \tilde{n}(b) \rangle = k{\cal O}(b) =
e(b) \, \langle k{\cal O}(b) \rangle  ~.
\label{mi:ebenh}
\end{equation}
The definition of the average $\langle k{\cal O}(b) \rangle$ is a
bit delicate, since the average number of interactions per event
is pushed up by the requirement that each event contain at least
one interaction. However, an exact meaning can be given \cite{Sjo87a}.
 
With the knowledge of $e(b)$, the $f(x_{\perp})$ function of the
simple model generalizes to
\begin{equation}
f(x_{\perp},b) = e(b) \, f(x_{\perp})     ~.
\end{equation}
The na\"{\i}ve generation procedure is thus to pick a $b$ according 
to the phase space $\d^2 b$, find the relevant $e(b)$ and plug in the
resulting $f(x_{\perp},b)$ in the formalism of the simple model.
If at least one hard interaction is generated, the event is retained,
else a new $b$ is to be found. This algorithm would work fine for
hadronic matter distributions which vanish outside some radius, so
that the $\d^2 b$ phase space which needs to be probed is finite.
Since this is not true for the distributions under study, it is
necessary to do better.
 
By analogy with eq.~(\ref{mi:hardest}), it is possible to ask what
the probability is to find the hardest scattering of an event at
$x_{\perp 1}$. For each impact parameter separately, the probability
to have an interaction at $x_{\perp 1}$ is given by $f(x_{\perp},b)$,
and this should be multiplied by the probability that the event
contains no interactions at a scale $x'_{\perp} > x_{\perp 1}$,
to yield the total probability distribution
\begin{eqnarray}
\frac{\d {\cal P}_{\mrm{hardest}}}{\d^2 b \, \d x_{\perp 1}} & = &
f(x_{\perp 1},b) \, \exp \left\{ - \int_{x_{\perp 1}}^1
f(x'_{\perp},b) \, \d x'_{\perp} \right\}  \nonumber \\
& = & e(b) \, f(x_{\perp 1}) \, \exp \left\{ - e(b)
\int_{x_{\perp 1}}^1 f(x'_{\perp}) \, \d x'_{\perp} \right\} ~.
\end{eqnarray}
If the treatment of the exponential is deferred for a moment,
the distribution in $b$ and $x_{\perp 1}$ appears in factorized form,
so that the two can be chosen independently of each other. In
particular, a high-$\pT$ QCD scattering or any other hard scattering
can be selected with whatever kinematics desired for that process,
and thereafter assigned some suitable `hardness' $x_{\perp 1}$.
With the $b$ chosen according to $e(b) \, \d^2 b$, the neglected
exponential can now be evaluated, and the event retained with a
probability proportional to it. From the $x_{\perp 1}$ scale of the
selected interaction, a sequence of softer $x_{\perp i}$ values may
again be generated as in the simple model, using the known
$f(x_{\perp},b)$. This sequence may be empty, i.e. the
event need not contain any further interactions.
 
It is interesting to understand how the algorithm above works.
By selecting $b$ according to $e(b) \, \d^2 b$, i.e.
${\cal O}(b) \, \d^2 b$, the primary $b$ distribution is
maximally biased towards small impact parameters. If the first
interaction is hard, by choice or by chance, the integral of
the cross section above $x_{\perp 1}$ is small, and the exponential
close to unity. The rejection procedure is therefore very efficient
for all standard hard processes in the program --- one may even
safely drop the weighting with the exponential completely. The large
$e(b)$ value is also likely to lead to the generation of many further,
softer interactions. If, on the other hand, the first interaction is
not hard, the exponential is no longer close to unity, and many
events are rejected. This pulls down the efficiency for `minimum bias'
event generation. Since the exponent is proportional to $e(b)$,
a large $e(b)$ leads to an enhanced probability for rejection,
whereas the chance of acceptance is larger with a small $e(b)$.
Among events where the hardest interaction is soft, the $b$
distribution is therefore biased towards larger values
(smaller $e(b)$), and there is a small probability for yet softer
interactions.
 
To evaluate the exponential factor, the program pretabulates the
integral of $f(x_{\perp})$ at the initialization stage, and further
increases the Monte Carlo statistics of this tabulation as the run
proceeds. The $x_{\perp}$ grid is concentrated towards small
$x_{\perp}$, where the integral is large. For a selected
$x_{\perp 1}$ value, the $f(x_{\perp})$ integral is obtained by
interpolation. After multiplication by the known $e(b)$ factor,
the exponential factor may be found.
 
In this section, nothing has yet been assumed about the form of the
$\d \sigma / \d \pT$ spectrum. Like in the impact parameter independent
case, it is possible to use a sharp cut-off at some given
$\pTmin$ value. However, now each event is required to have
at least one interaction, whereas before events without interactions
were retained and put at $\pT = 0$. It is therefore aesthetically
more appealing to assume a gradual turn-off, so that a (semi)hard
interaction can be rather soft part of the time. The matrix elements
roughly diverge like $\alphas(\pT^2) \, \d \pT^2 / \pT^4$ for
$\pT \to 0$. They could therefore be regularized as follows. Firstly,
to remove the $1/\pT^4$ behaviour, multiply by a factor
$\pT^4 / (\pT^2 + p_{\perp 0}^2)^2$. Secondly, replace the $\pT^2$
argument in $\alphas$ by $\pT^2 + p_{\perp 0}^2$ or, with the
inclusion of the $K$ factor procedure introduced earlier, replace
$0.075 \, \pT^2$ by $0.075 \, (\pT^2 + p_{\perp 0}^2)$.
 
With these substitutions, a continuous $\pT$ spectrum is obtained,
stretching from $\pT = 0$ to $E_{\mrm{cm}}/2$. For 
$\pT \gg p_{\perp 0}$
the standard perturbative QCD cross section is recovered, while
values $\pT \ll p_{\perp 0}$ are strongly damped. The $p_{\perp 0}$
scale, which now is the main free parameter of the model, in
practice comes out to be of the same order of magnitude as the sharp
cut-off $\pTmin$ did, i.e. 1.5--2 GeV.
 
If gluons with large transverse wavelength decouple because of the
colour-singlet nature of hadrons, and if the transverse structure of
hadrons is assumed to be energy-independent, it is natural to assume
that also $\pTmin$ and $p_{\perp 0}$ are independent of the
c.m. energy of the hadron collision. For the impact parameter
independent picture this works out fine, with all events being reduced
to low-$\pT$ two-string ones when the c.m. energy is reduced. In the
variable impact parameter picture, the whole formalism only makes sense
if $\sigma_{\mrm{hard}} > \sigma_{\mrm{nd}}$, see e.g. 
eq.~(\ref{mi:kfromsigma}). Since $\sigma_{\mrm{nd}}$ does not vanish 
with decreasing energy, but $\sigma_{\mrm{hard}}$ 
would do that for a fixed $p_{\perp 0}$, this means
that $p_{\perp 0}$ has to be reduced when the energy is decreased
below some given threshold. The more `sophisticated' model of this
section therefore makes sense at collider energies, whereas it is not
well suited for applications at lower energies.
 
\subsection{Pile-up Events}
\label{ss:pileup}
 
In high luminosity colliders, there is a non-negligible probability
that one single bunch crossing may produce several separate events,
so-called pile-up events.
This in particular applies to future $\p\p$ colliders like LHC,
but one could also consider e.g. $\ee$ colliders with high rates of
$\gamma\gamma$ collisions. The program therefore contains an option,
currently only applicable to hadron--hadron collisions,
wherein several events may be generated and put one after the other
in the event record, to simulate the full amount of particle
production a detector might be facing.
 
The program needs to know the assumed luminosity per bunch--bunch
crossing, expressed in mb$^{-1}$. Multiplied by the cross section
for pile-up processes studied, $\sigma_{\mrm{pile}}$, this gives the 
average number of collisions per beam crossing, $\br{n}$. These pile-up
events are taken to be of the minimum bias type, with diffractive
and elastic events included or not (and a further subdivision of
diffractive events into single and double). This means that
$\sigma_{\mrm{pile}}$ may be either $\sigma_{\mrm{tot}}$,
$\sigma_{\mrm{tot}} - \sigma_{\mrm{el}}$ or
$\sigma_{\mrm{tot}} - \sigma_{\mrm{el}} - \sigma_{\mrm{diffr}}$.
Which option to choose depends on the detector: most detectors
would not be able to observe elastic $\p\p$ scattering, and therefore
it would be superfluous to generate that kind of events.
In addition, we allow for the possibility that one interaction may
be of a rare kind, selected freely by the user.
There is no option to generate two `rare' events in the same crossing;
normally the likelihood for that kind of occurences should be small.
 
If only minimum bias type events are generated, i.e. if only one
cross section is involved in the problem, then the number of
events in a crossing is distributed according to a Poissonian
with the average number $\br{n}$ as calculated above.
The program actually will simulate only those beam crossings
where at least one event occurs, i.e. not consider the fraction
$\exp(-\br{n})$ of zero-event crossings. Therefore the actually
generated average number of pile-up events is
$\langle n \rangle = \br{n}/(1-\exp(-\br{n}))$.
 
Now instead consider the other extreme, where one event is supposed
be rare, with a cross section $\sigma_{\mrm{rare}}$ much smaller 
than $\sigma_{\mrm{pile}}$, i.e. 
$f \equiv \sigma_{\mrm{rare}}/\sigma_{\mrm{pile}} \ll 1$.
The probability that a bunch crossing will give $i$ events, whereof
one of the rare kind, now is
\begin{equation}
{\cal P}_i = f \, i \, \exp(-\br{n}) \, \frac{\br{n}^i}{i!} =
f \, \br{n} \exp(-\br{n}) \, \frac{\br{n}^{i-1}}{(i-1)!} ~.
\end{equation}
The na\"{\i}ve Poissonian is suppressed by a factor $f$, since one of 
the events is rare rather than of the normal kind, but enhanced by a
factor $i$, since any one of the $i$ events may be the rare one.
As the equality shows, the probability distribution is now a
Poissonian in $i-1$:
in a beam crossing which produces one rare event, the multiplicity of
additional pile-up events is distributed according to a Poissonian
with average number $\br{n}$. The total average number of events
thus is $\langle n \rangle = \br{n} + 1$.
 
Clearly, for processes with intermediate cross sections,
$\br{n} \, \sigma_{\mrm{rare}}/\sigma_{\mrm{pile}} \simeq 1$, 
also the average number of events will be intermediate, and it is not 
allowed to assume only one event to be of the `rare' type. We do not 
consider that kind of situations.
 
Kindly note that, in the current implementation, all events are
supposed to be produced at the same vertex (the origin). To simulate
the spatial extent of the colliding beams, you would have to assign
interaction points yourself, and then shift each event separately
by the required amount in space and time.
 
When the pile-up option is used, one main limitation is that event
records may become very large when several events are put one after
the other, so that the space limit in the \ttt{LUJETS} common block
is reached. It is possible to expand the dimension of the common block,
see \ttt{MSTU(4)} and \ttt{MSTU(5)}, but only up to about 
20\,000 entries, which might not always be enough.
 
For practical reasons, the program will only allow a $\br{n}$ up to
120. The multiplicity distribution is truncated above 200,
or when the probability for a multiplicity has fallen below
$10^{-6}$, whichever occurs sooner. Also low multiplicities with
probabilities below $10^{-6}$ are truncated.

\subsection{Common Block Variables}
\label{ss:multintpar}

Of the routines used to generate beam remnants, multiple interactions 
and pile-up events, none are intended to be used in standalone mode.
The only way to regulate these aspects is therefore via the 
variables in the \ttt{PYPARS} common block. 

\drawbox{COMMON/PYPARS/MSTP(200),PARP(200),MSTI(200),PARI(200)}
\begin{entry}
 
\itemc{Purpose:} to give access to a number of status codes and
parameters which regulate the performance of {\Py}.
Most parameters are described in section \ref{ss:PYswitchpar};
here only those related to beam remnants, multiple interactions
and pile-up events are described. If the default values,
below denoted by (D=\ldots), are not satisfactory, they must in
general be changed before the \ttt{PYINIT} call. Exceptions, i.e.
variables which can be changed for each new event, are denoted by
(C).

\iteme{MSTP(81) :}\label{p:MSTP81} (D=1) master switch for 
multiple interactions.
\begin{subentry}
\iteme{= 0 :} off.
\iteme{= 1 :} on.
\end{subentry}

\iteme{MSTP(82) :} (D=1) structure of multiple interactions. For QCD
processes, used down to $\pT$ values below $\pTmin$, it also
affects the choice of structure for the one hard/semi-hard interaction.
\begin{subentry}
\iteme{= 0 :} simple two-string model without any hard interactions.
\iteme{= 1 :} multiple interactions assuming the same probability in
all events, with an abrupt $\pTmin$ cut-off at \ttt{PARP(81)}.
\iteme{= 2 :} multiple interactions assuming the same probability in
all events, with a continuous turn-off of the cross section at
$p_{\perp 0} = $\ttt{PARP(82)}.
\iteme{= 3 :} multiple interactions assuming a varying impact
parameter and a hadronic matter overlap consistent with a Gaussian
matter distribution, with a continuous turn-off of the cross section
at $p_{\perp 0} = $\ttt{PARP(82)}.
\iteme{= 4 :} multiple interactions assuming a varying impact
parameter and a hadronic matter overlap consistent with a double
Gaussian matter distribution given by \ttt{PARP(83)} and
\ttt{PARP(84)}, with a continuous turn-off of the cross section at
$p_{\perp 0} = $\ttt{PARP(82)}.
\itemc{Note 1:} For \ttt{MSTP(82)}$\geq 2$ and
\ttt{CKIN(3)}$ > $\ttt{PARP(82)}, cross sections
given with \ttt{PYSTAT(1)} may be somewhat too large, since (for
reasons of efficiency) the probability factor that the hard
interaction is indeed the hardest in the event is not
included in the cross sections. It is included in the event
selection, however, so the events generated are correctly
distributed. For \ttt{CKIN(3)} values a couple of times larger than
\ttt{PARP(82)} this ceases to be a problem.
\itemc{Note 2:} The \ttt{PARP(81)} and, in particular, \ttt{PARP(82)}
values are sensitive to the choice of parton distributions,
$\Lambda_{\mrm{QCD}}$,
etc., in the sense that a change in the latter variables
leads to a net change in the multiple interaction rate, which has
to be compensated by a retuning of \ttt{PARP(81)} or \ttt{PARP(82)}
if one wants to keep the net multiple interaction structure the
same. The default \ttt{PARP(81)} value is consistent with the other
default values give, i.e. CTEQ 2L parton distributions etc.
When options \ttt{MSTP(82)=} 2--4 are used, the default
\ttt{PARP(82)} value is to be used in conjunction with
\ttt{MSTP(2)=2} and \ttt{MSTP(33)=3}. These switches
should be set by you.
\end{subentry}
 
\iteme{MSTP(83) :} (D=100) number of Monte Carlo generated phase-space 
points per bin (whereof there are 20) in the initialization
(in \ttt{PYMULT}) of multiple interactions for
\ttt{MSTP(82)}$\geq 2$.
 
\iteme{MSTP(91) :} (D=1) (C) primordial $k_{\perp}$ distribution
in hadron. See \ttt{MSTP(93)} for photon.
\begin{subentry}
\iteme{= 0 :} no primordial $k_{\perp}$.
\iteme{= 1 :} Gaussian, width given in \ttt{PARP(91)}, upper cut-off
in \ttt{PARP(93)}.
\iteme{= 2 :} exponential, width given in \ttt{PARP(92)}, upper
cut-off in \ttt{PARP(93)}.
\end{subentry}
 
\iteme{MSTP(92) :} (D=4) (C) energy partitioning in hadron or 
resolved photon remnant, when this remnant is split into two jets.
(For a splitting into a hadron plus a jet, see \ttt{MSTP(94)}.)
The energy fraction $\chi$ taken by one of the two objects, with
conventions as described for \ttt{PARP(94)} and \ttt{PARP(96)}, 
is chosen according to the different distributions below. Here
$c_{\mmin} = 2 \langle m_{\q} \rangle/E_{\mrm{cm}} = 0.6$ 
GeV$/E_{\mrm{cm}}$.
\begin{subentry}
\iteme{= 1 :} 1 for meson or resolved photon, $2(1-\chi)$ for 
baryon, i.e. simple counting rules.
\iteme{= 2 :} $(k+1)(1-\chi)^k$, with $k$ given by
\ttt{PARP(94)} or \ttt{PARP(96)}.
\iteme{= 3 :} proportional to 
$(1-\chi)^k/\sqrt[4]{\chi^2+c_{\mmin}^2}$, with $k$ given by
\ttt{PARP(94)} or \ttt{PARP(96)}.
\iteme{= 4 :} proportional to $(1-\chi)^k/\sqrt{\chi^2+c_{\mmin}^2}$, 
with $k$ given by \ttt{PARP(94)} or \ttt{PARP(96)}.
\iteme{= 5 :} proportional to $(1-\chi)^k/(\chi^2+c_{\mmin}^2)^{b/2}$, 
with $k$ given by \ttt{PARP(94)} or \ttt{PARP(96)}, and 
$b$ by \ttt{PARP(98)}.
\end{subentry}
 
\iteme{MSTP(93) :} (D=1) (C) primordial $k_{\perp}$ distribution
in photon, either it is one of the incoming particles or inside
an electron.
\begin{subentry}
\iteme{= 0 :} no primordial $k_{\perp}$.
\iteme{= 1 :} Gaussian, width given in \ttt{PARP(99)}, upper
cut-off in \ttt{PARP(100)}.
\iteme{= 2 :} exponential, width given in \ttt{PARP(99)}, upper
cut-off in \ttt{PARP(100)}.
\iteme{= 3 :} power-like of the type
$\d k_{\perp}^2/(k_{\perp 0}^2 + k_{\perp}^2)^2$, with
$k_{\perp 0}$ in \ttt{PARP(99)} and upper $k_{\perp}$ cut-off in
\ttt{PARP(100)}.
\iteme{= 4 :} power-like of the type
$\d k_{\perp}^2/(k_{\perp 0}^2 + k_{\perp}^2)$, with
$k_{\perp 0}$ in \ttt{PARP(99)} and upper $k_{\perp}$ cut-off in
\ttt{PARP(100)}.
\iteme{= 5 :} power-like of the type
$\d k_{\perp}^2/(k_{\perp 0}^2 + k_{\perp}^2)$, with
$k_{\perp 0}$ in \ttt{PARP(99)} and upper $k_{\perp}$ cut-off
given by the $\pT$ of the hard process or by
\ttt{PARP(100)}, whichever is smaller.
\itemc{Note:} for options 1 and 2 the \ttt{PARP(100)} value is of
minor importance, once \ttt{PARP(100)}$\gg$\ttt{PARP(99)}. However,
options 3 and 4 correspond to distributions with infinite
$\langle k_{\perp}^2 \rangle$ if the $k_{\perp}$ spectrum is not
cut off, and therefore the \ttt{PARP(100)} value is as important
for the overall distribution as is \ttt{PARP(99)}.
\end{subentry}

\iteme{MSTP(94) :} (D=2) (C) energy partitioning in hadron or 
resolved photon remnant, when this remnant is split into a hadron 
plus a remainder-jet. The energy fraction chi is taken by one of the 
two objects, with conventions as described below or for \ttt{PARP(95)} 
and \ttt{PARP(97)}.
\begin{subentry}
\iteme{= 1 :} 1 for meson or resolved photon, $2(1-\chi)$ for 
baryon, i.e. simple counting rules.
\iteme{= 2 :} $(k+1)(1-\chi)^k$, with $k$ given by
\ttt{PARP(95)} or \ttt{PARP(97)}.
\iteme{= 3 :} the $\chi$ of the hadron is selected according to the
normal fragmentation function used for the hadron in jet 
fragmentation, see \ttt{MSTJ(11)}. The possibility of a changed 
fragmentation function shape in diquark fragmentation 
(see \ttt{PARJ(45)}) is not included.  
\iteme{= 4 :} as \ttt{=3}, but the shape is changed as allowed in 
diquark fragmentation (see \ttt{PARJ(45)}); this change is here also 
allowed for meson production. (This option is not so natural
for mesons, but has been added to provide the same amount of freedom
as for baryons).  
\end{subentry}

\iteme{MSTP(131) :}\label{p:MSTP131} (D=0) master switch for pile-up 
events, i.e. several
independent hadron--hadron interactions generated in the same
bunch--bunch crossing, with the events following one after the
other in the event record.
\begin{subentry}
\iteme{= 0 :} off, i.e. only one event is generated at a time.
\iteme{= 1 :} on, i.e. several events are allowed in the same event
record. Information on the processes generated may be found in
\ttt{MSTI(41) - MSTI(50)}.
\end{subentry}
 
\iteme{MSTP(132) :} (D=4) the processes that are switched on for
pile-up events. The first event may be set up completely arbitrarily,
using the switches in the \ttt{PYSUBS} common block, while all the
subsequent events have to be of one of the `inclusive' processes
which dominate the cross section, according to the options below.
It is thus not possible to generate two rare events in the pile-up
option.
\begin{subentry}
\iteme{= 1 :} low-$\pT$ processes (ISUB = 95) only. The
low-$\pT$ model actually used, both in the hard event and in
the pile-up events, is the one set by \ttt{MSTP(81)} etc. This means
that implicitly also high-$\pT$ jets can be generated in the pile-up
events.
\iteme{= 2 :} low-$\pT$ + double diffractive processes
(ISUB = 95 and 94).
\iteme{= 3 :} low-$\pT$ + double diffractive + single diffractive
processes (ISUB = 95, 94, 93 and 92).
\iteme{= 4 :} low-$\pT$ + double diffractive + single diffractive
+ elastic processes, together corresponding to the full
hadron--hadron cross section (ISUB = 95, 94, 93, 92 and 91).
\end{subentry}
 
\iteme{MSTP(133) :} (D=0) multiplicity distribution of pile-up events.
\begin{subentry}
\iteme{= 0 :} selected by user, before each \ttt{PYEVNT} call, by
giving the \ttt{MSTP(134)} value.
\iteme{= 1 :} a Poissonian multiplicity distribution in the total
number of pile-up events. This is the relevant distribution if the
switches set for the first event in \ttt{PYSUBS} give the same
subprocesses as are implied by \ttt{MSTP(132)}. In that case the
mean number of events per beam crossing is
$\br{n} = \sigma_{\mrm{pile}} \times$\ttt{PARP(31)}, where 
$\sigma_{\mrm{pile}}$ is the sum of the cross section for allowed
processes. Since bunch crossing which do not give any events at all
(probability $\exp(-\br{n})$) are not simulated, the actual average
number per \ttt{PYEVNT} call is
$\langle n \rangle = \br{n}/(1-\exp(-\br{n}))$.
\iteme{= 2 :} a biased distribution, as is relevant when one of the
events to be generated is assumed to belong to an event class
with a cross section much smaller than the total hadronic
cross section. If $\sigma_{\mrm{rare}}$ is the cross section for this
rare process (or the sum of the cross sections of several rare
processes) and $\sigma_{\mrm{pile}}$ the cross section for the 
processes allowed by \ttt{MSTP(132)}, then define
$\br{n} = \sigma_{\mrm{pile}} \times$\ttt{PARP(131)}
and $f = \sigma_{\mrm{rare}}/\sigma_{\mrm{pile}}$. The probability 
that a bunch crossing will give $i$ events is then
${\cal P}_i = f \, i \, \exp(-\br{n}) \, \br{n}^i/i!$,
i.e. the na\"{\i}ve Poissonian is suppressed by a factor $f$ since
one of the events will be rare rather than frequent, but
enhanced by a factor $i$ since any of the $i$ events may be the
rare one. Only beam crossings which give at least one event
of the required rare type are simulated, and the distribution
above normalized accordingly.
\itemc{Note:} for practical reasons, it is required that
$\br{n} < 120$, i.e. that an average beam crossing does not contain
more than 120 pile-up events. The multiplicity distribution is
truncated above 200, or when the probability for a multiplicity
has fallen below $10^{-6}$, whichever occurs sooner. Also low
multiplicities with probabilities below $10^{-6}$ are truncated.
See also \ttt{PARI(91) - PARI(93)}.
\end{subentry}
 
\iteme{MSTP(134) :} (D=1) a user selected multiplicity, i.e. total
number of pile-up events, to be generated in the next \ttt{PYEVNT}
call. May be reset for each new event, but must be in the range
$1 \leq$\ttt{MSTP(134)}$\leq 200$.
 
\boxsep
 
\iteme{PARP(81) :}\label{p:PARP81} (D=1.40 GeV/$c$) effective 
minimum transverse momentum $\pTmin$ for multiple 
interactions with \ttt{MSTP(82)=1}.
 
\iteme{PARP(82) :} (D=1.55 GeV/$c$) regularization scale $p_{\perp 0}$
of the transverse momentum spectrum for multiple interactions with
\ttt{MSTP(82)}$\geq 2$.
 
\iteme{PARP(83), PARP(84) :} (D=0.5, 0.2) parameters of an assumed
double Gaussian matter distribution inside the colliding hadrons for
\ttt{MSTP(82)=4}, of the form given in eq.~(\ref{mi:doubleGauss}),
i.e. with a core of radius \ttt{PARP(84)} of the main radius and
containing a fraction \ttt{PARP(83)} of the total hadronic matter.
 
\iteme{PARP(85) :} (D=0.33) probability that an additional
interaction in the multiple interaction formalism gives two gluons,
with colour connections to `nearest neighbours' in momentum space.
 
\iteme{PARP(86) :} (D=0.66) probability that an additional
interaction in the multiple interaction formalism gives two gluons,
either as described in \ttt{PARP(85)} or as a closed gluon loop.
Remaining fraction is supposed to consist of quark--antiquark pairs.
 
\iteme{PARP(87), PARP(88) :} (D=0.7, 0.5) in order to account for an
assumed dominance of valence quarks at low transverse momentum scales,
a probability is introduced that a $\g\g$-scattering according to
na\"{\i}ve cross section is replaced by a $\q\qbar$ one; this is used 
only for \ttt{MSTP(82)}$\geq 2$. The probability is parametrized as
${\cal P} = a (1 - (\pT^2/(\pT^2 + b^2)^2)$, where
$a =$\ttt{PARP(87)} and $b =$\ttt{PARP(88)}$\times$\ttt{PARP(82)}.
 
\iteme{PARP(91) :} (D=0.44 GeV/$c$) (C) width of Gaussian primordial
$k_{\perp}$ distribution inside hadron for \ttt{MSTP(91)=1}, i.e.
$\exp(-k_{\perp}^2/\sigma^2) \, k_{\perp} \, \d k_{\perp}$ with
$\sigma =$\ttt{PARP(91)} and
$\langle k_{\perp}^2 \rangle = $\ttt{PARP(91)}$^2$.
 
\iteme{PARP(92) :} (D=0.20 GeV/$c$) (C) width parameter of 
exponential primordial $k_{\perp}$ distribution inside hadron for
\ttt{MSTP(91)=2}, i.e.
$\exp(-k_{\perp}/\sigma) \, k_{\perp} \, \d k_{\perp}$ with
$\sigma =$\ttt{PARP(92)} and 
$\langle k_{\perp}^2 \rangle = 6 \times$\ttt{PARP(92)}$^2$.
Thus one should put \ttt{PARP(92)}$\approx$\ttt{PARP(91)}$/\sqrt{6}$
to have continuity with the option above.  
 
\iteme{PARP(93) :} (D=2. GeV/$c$) (C) upper cut-off for primordial
$k_{\perp}$ distribution inside hadron.
 
\iteme{PARP(94) :} (D=1.) (C) for \ttt{MSTP(92)}$\geq 2$ this gives
the value of the parameter $k$ for the case when a meson or
resolved photon remnant is split into two fragments (which is which 
is chosen at random).
 
\iteme{PARP(95) :} (D=0.) (C) for \ttt{MSTP(94)=2} this gives
the value of the parameter $k$ for the case when a meson or
resolved photon remnant is split into a meson and a spectator 
fragment jet, with $\chi$ giving the energy fraction taken by the 
meson.
 
\iteme{PARP(96) :} (D=3.) (C) for \ttt{MSTP(92)}$\geq 2$ this gives
the value of the parameter $k$ for the case when a nucleon remnant
is split into a diquark and a quark fragment, with $\chi$ giving
the energy fraction taken by the quark jet.
 
\iteme{PARP(97) :} (D=1.) (C) for \ttt{MSTP(94)=2} this gives
the value of the parameter $k$ for the case when a nucleon remnant
is split into a baryon and a quark jet or a meson and a diquark jet,
with $\chi$ giving the energy fraction taken by the quark jet or
meson, respectively.
 
\iteme{PARP(98) :} (D=0.75) (C) for \ttt{MSTP(92)=5} this gives
the power of an assumed basic $1/\chi^b$ behaviour in the splitting
distribution, with $b =$\ttt{PARP(98)}.
 
\iteme{PARP(99) :} (D=0.44 GeV/$c$) (C) width parameter of primordial
$k_{\perp}$ distribution inside photon; exact meaning depends on
\ttt{MSTP(93)} value chosen (cf. \ttt{PARP(91)} and \ttt{PARP(92)}
above).
 
\iteme{PARP(100) :} (D=2. GeV/$c$) (C) upper cut-off for primordial
$k_{\perp}$ distribution inside photon.
 
\iteme{PARP(131) :}\label{p:PARP131} (D=0.01 mb$^{-1}$) in the 
pile-up events scenario, \ttt{PARP(131)}
gives the assumed luminosity per bunch--bunch crossing, i.e.
if a subprocess has a cross section $\sigma$, the average number
of events of this type per bunch--bunch crossing is
$\br{n} = \sigma \times$\ttt{PARP(131)}. \ttt{PARP(131)} may be
obtained by dividing the integrated luminosity over a given time
(1 s, say) by the number of bunch--bunch crossings that this
corresponds to. Since the program will not generate more than
200 pile-up events, the initialization procedure will crash if
$\br{n}$ is above 120. 

\end{entry}
 
\clearpage
 
\section{Fragmentation}
 
The main fragmentation option in {\JePy}
is the Lund string scheme, but independent fragmentation
options are also available. These latter options should not be
taken too seriously, since we know that independent fragmentation
does not provide a consistent alternative, but occasionally one
may like to compare string fragmentation with something else.
 
The subsequent four subsections give further details;
the first one on flavour selection, which is common to the two
approaches, the second on string fragmentation, the third on
independent fragmentation, while the fourth and final contains
information on a few other minor issues.
 
The Lund fragmentation model is described in \cite{And83}, where
all the basic ideas are presented and earlier papers
\cite{And79,And80,And82,And82a} summarized.
The details given there on how a multiparton jet system is allowed
to fragment are out of date, however, and for this one should turn
to \cite{Sjo84}. Also the `popcorn' baryon production mechanism
is not covered, see \cite{And85}. Reviews of fragmentation models
in general may be found in \cite{Sjo88,Sjo89}.
 
\subsection{Flavour Selection}
\label{ss:flavoursel}
 
In either string or independent fragmentation, an iterative
approach is used to describe the fragmentation process.
Given an initial quark $\q = \q_0$, it is assumed that a new
$\q_1 \qbar_1$
pair may be created, such that a meson $\q_0 \qbar_1$ is
formed, and a $\q_1$ is left behind. This $\q_1$ may at a later
stage pair off with a $\qbar_2$, and so on. What need be
given is thus the relative probabilities to produce the various
possible $\q_i \qbar_i$ pairs, $\u \ubar$, $\d \dbar$,
$\s \sbar$, etc., and the relative probilities that a given
$\q_{i-1}\qbar_i$ quark pair combination forms a specific meson,
e.g. for $\u \ubar$ either $\pi^+$, $\rho^+$ or some higher state.
 
In {\Je}, it is assumed that the two aspects can be factorized,
i.e. that it is possible first to select a $\q_i \qbar_i$ pair,
without any reference to allowed physical meson states, and
that, once the $\q_{i-1} \qbar_i$ flavour combination is given,
it can be assigned to a given meson state with total probability
unity.
 
\subsubsection{Quark flavours and transverse momenta}
 
In order to generate the quark--antiquark pairs $\q_i \qbar_i$ which
lead to string breakups, the Lund model invokes the idea of
quantum mechanical tunnelling, as follows. If the $\q_i$ and
$\qbar_i$ have no (common) mass or transverse momentum, the pair can
classically be created at one point and then be pulled apart
by the field. If the quarks have mass and/or transverse momentum,
however, the $\q_i$ and $\qbar_i$ must classically be produced at a
certain distance so that the field energy between them can be
transformed into the sum of the two transverse masses $m_{\perp}$.
Quantum mechanically, the quarks may be created in one point
(so as to keep the concept of local flavour conservation) and then
tunnel out to the classically allowed region. In terms of a
common transverse mass $m_{\perp}$ of the $\q_i$ and the $\qbar_i$,
the tunnelling probability is given by
\begin{equation}
   \exp \left( -\frac{\pi m_{\perp}^2}{\kappa} \right) =
   \exp \left( -\frac{\pi m^2}{\kappa} \right)
   \exp \left( -\frac{\pi \pT^2}{\kappa} \right) ~.
\end{equation}
 
The factorization of the transverse momentum and the mass terms leads
to a flavour-independent Gaussian spectrum for the $p_x$ and $p_y$
components of $\q_i \qbar_i$ pairs. Since the string is assumed to 
have no transverse excitations, this $\pT$ is locally compensated 
between the quark and the antiquark of the pair. The $\pT$ of a 
meson $\q_{i-1} \qbar_i$ is given by the vector sum of the $\pT$:s 
of the $\q_{i-1}$ and $\qbar_i$ constituents, which implies 
Gaussians in $p_x$ and $p_y$ with a width $\sqrt{2}$ that of the 
quarks themselves. The assumption of a Gaussian shape may be a good 
first approximation, but there remains the possibility of non-Gaussian
tails, that can be important in some situations.
 
In a perturbative QCD framework, a hard scattering is
associated with gluon radiation, and further contributions to what is
na\"{\i}vely called fragmentation $\pT$ comes from unresolved radiation.
This is used as an explanation
why the experimental $\left\langle \pT \right\rangle$ is somewhat
higher than obtained with the formula above.
 
The formula also implies a suppression of heavy quark production
$u : d : s : c \approx$ \mbox{$1 : 1 : 0.3 : 10^{-11}$}. Charm and
heavier quarks
are hence not expected to be produced in the soft fragmentation.
Since the predicted flavour suppressions are in terms of quark masses,
which are notoriously difficult to assign (should it be current algebra,
or constituent, or maybe something in between?), the
suppression of $\s \sbar$ production is left as a free parameter in
the program: $\u \ubar$ : $\d \dbar$ : $\s \sbar$ = 1 : 1 : $\gamma_s$,
where by default $\gamma_s = 0.3$. At least qualitatively, the
experimental value agrees with theoretical prejudice.
There is no production at all of heavier flavours in the fragmentation
process, but only as part of the shower evolution.
 
\subsubsection{Meson production}
\label{sss:mesonprod}
 
Once the flavours $\q_{i-1}$ and $\qbar_i$ have been selected, a choice 
is made between the possible multiplets. The relative composition of
different multiplets is not given from first principles, but must
depend on the details of the fragmentation process. To some
approximation one would expect a negligible fraction of states with
radial excitations or non-vanishing orbital angular momentum. Spin
counting arguments would then suggest a 3:1 mixture between the
lowest lying vector and pseudoscalar multiplets. Wave function
overlap arguments lead to a relative enhancement of the lighter
pseudoscalar states, which is more pronounced the larger the mass
splitting is \cite{And82a}.
 
In the program, six meson multiplets are
included. If the nonrelativistic classification scheme is used, i.e.
mesons are assigned a valence quark spin $S$ and an internal orbital
angular momentum $L$, with the physical spin $s$ denoted $J$,
$\mbf{J} = \mbf{L} + \mbf{S}$, then the multiplets are:
\begin{Itemize}
\item $L = 0$, $S = 0$, $J = 0$: the ordinary pseudoscalar meson
   multiplet;
\item $L = 0$, $S = 1$, $J = 1$: the ordinary vector meson multiplet;
\item $L = 1$, $S = 0$, $J = 1$: an axial vector meson multiplet;
\item $L = 1$, $S = 1$, $J = 0$: the scalar meson multiplet;
\item $L = 1$, $S = 1$, $J = 1$: another axial vector meson multiplet;
    and
\item $L = 1$, $S = 1$, $J = 2$: the tensor meson multiplet.
\end{Itemize}
Each multiplet has the full four-generation setup of $8 \times 8$
states included in the program, although many could never actually be
produced. Some simplifications have been made; thus
there is no mixing included between the two axial vector multiplets.
 
In the program, the spin $S$ is first chosen to be either 0 or 1.
This is done according to parametrized relative probabilities,
where the probability for spin 1 by default is taken to be 0.5 for
a meson consisting only of $\u$ and $\d$ quark, 0.6 for one which
contains $\s$ as well, and $0.75$ for quarks with $\c$ or heavier
quark, in accordance with the deliberations above.
 
By default, it is assumed that $L = 0$, such that only pseudoscalar
and vector mesons are produced. For inclusion of $L = 1$ production,
four parameters can be used, one to give the probability that a $S = 0$
state also has $L =1$, the other three for the probability that a
$S = 1$ state has $L = 1$ and $J$ either 0, 1, or 2.
 
For the
flavour-diagonal meson states $\u \ubar$, $\d \dbar$ and $\s \sbar$,
it is also necessary to include mixing into the physical mesons.
This is done according to a parametrization, based on the mixing angles
given in the Review of Particle Properties \cite{PDG88}. In particular,
the default choices correspond to
\begin{eqnarray}
\eta & = & \frac{1}{2} (\u\ubar + \d\dbar) - \frac{1}{\sqrt{2}}
\s\sbar ~;   \nonumber \\
\eta' & = & \frac{1}{2} (\u\ubar + \d\dbar) + \frac{1}{\sqrt{2}}
\s\sbar ~;   \nonumber \\
\omega & = & \frac{1}{\sqrt{2}} (\u\ubar + \d\dbar)    \nonumber \\
\phi & = & \s\sbar ~.
\end{eqnarray}
In the $\pi^0 - \eta - \eta'$ system, no account is therefore
taken of the difference in masses, an approximation which seems to
lead to an overestimate of $\eta'$ rates \cite{ALE92}. Recently,
parameters have been introduced to allow an additional `brute force'
suppression of $\eta$ and $\eta'$ states.
 
\subsubsection{Baryon production}
 
Baryon production may, in its simplest form, be obtained by assuming
that any flavour $\q_i$ given above could represent either a quark or
an antidiquark in a colour triplet state. Then the same basic
machinery can be run through as above, supplemented with the probability
to produce various diquark pairs. In principle, there is one
parameter for each diquark, but if tunnelling is still assumed to give
an effective description, mass relations can be used to reduce the
effective number of parameters. There are three main ones
appearing in the program: 
\begin{Itemize}
\item the relative probability to pick a $\qbar\qbar$ diquark
rather than a $\q$; 
\item the extra suppression associated with a diquark
containing a strange quark (over and above the ordinary $\s / \u$
suppression factor $\gamma_s$); and 
\item the suppression of spin 1 diquarks relative to spin 0 ones 
(apart from the factor of 3 enhancement of the former based on 
counting the number of spin states). 
\end{Itemize}
The extra strange diquark suppression factor comes about since what 
appears in the exponent of the tunnelling formula is $m^2$ and not 
$m$, so that the diquark and the strange quark suppressions do not 
factorize.
 
Only two baryon multiplets are included, i.e. there are no $L=1$
excited states.  The two multiplets are:
\begin{Itemize}
\item $S = J = 1/2$: the `octet' multiplet of SU(3) (in the full
four-generation scenario in the program 168 states are available);
\item $S = J = 3/2$: the `decuplet' multiplet of SU(3) (120 states
in the program).
\end{Itemize}
In contrast to the meson case, different flavour combinations have
different numbers of states available: for $\u \u \u$ only
$\Delta^{++}$, whereas $\u \d \s$ may become either $\Lambda$,
$\Sigma^0$ or $\Sigma^{*0}$.
 
An important constraint is that a baryon is a
symmetric state of three quarks, neglecting the colour degree of
freedom. When a diquark and a quark are joined to form a baryon,
the combination is therefore weighted with the probability that
they form a symmetric three-quark state. The program implementation
of this principle is to first select a diquark at random, with
the strangeness and spin 1 suppression factors above included,
but then to accept the selected diquark with a weight proportional to
the number of states available for the quark-diquark combination. This
means that, were it not for the tunnelling suppression factors, all
states in the {\bf SU(6)} (flavour {\bf SU(3)} times spin {\bf SU(2)})
56-multiplet would become equally populated. Of course also heavier
baryons may come from the fragmentation of e.g. $\c$ quark jets, but
although the particle classification scheme used in the program is
{\bf SU(16)}, i.e. with eight flavours, all possible quark-diquark
combinations can be related to {\bf SU(6)} by symmetry arguments.
As in the case for mesons, one could imagine an explicit further
suppression of the heavier spin 3/2 baryons. We do not expect it to
be an important effect, since baryon mass splittings are much smaller
than in the meson case.
 
In case of rejection, a new diquark is selected and tested, etc.
A corresponding procedure is used for the quark selection when
a diquark has already been formed in the previous step. Properly
speaking both the quark and the diquark flavour should be chosen anew.
This would become a tedious process, since also the hadron produced
in the step before would have to be rejected. In practice only the
last produced pair, be that the quark or diquark one, is rejected.
The error introduced by this is small.
 
A more general framework for baryon production is the `popcorn' one
\cite{And85}, in which diquarks as such are never produced, but
rather baryons appear from the successive production of several
$\q_i \qbar_i$ pairs. The picture is the following. Assume that the
original $\q$ is red $r$ and the $\qbar$ is $\br{r}$. Normally a
new $\q_1 \qbar_1$ pair produced in the field would also be
$r \br{r}$, so that the $\qbar_1$ is pulled towards the $\q$
end and vice versa, and two separate colour-singlet systems
$\q \qbar_1$ and $\q_1 \qbar$ are formed. Occasionally, the
$\q_1 \qbar_1$ pair may be e.g. $g \br{g}$ ($g$ = green), in which
case there is no net colour charge acting on either $\q_1$ or
$\qbar_1$. Therefore, the pair cannot gain energy from the field,
and normally would exist only as a fluctuation. If $\q_1$ moves
towards $\q$ and $\qbar_1$ towards $\qbar$, the net field remaining
between $\q_1$ and $\qbar_1$ is $\br{b} b$ ($b$ = blue;
$g + r = \br{b}$ if only colour triplets are assumed). In this central
field, an additional $\q_2 \qbar_2$ pair can be created, where
$\q_2$ now is pulled towards $\q \q_1$ and $\qbar_2$
towards $\qbar \qbar_1$, with no net colour field between $\q_2$
and $\qbar_2$. If this is all that happens, the baryon $B$ will be
made up out of $\q_1$, $\q_2$ and some $\q_4$ produced between
$\q$ and $\q_1$, and $\br{B}$ of $\qbar_1$, $\qbar_2$ and some
$\qbar_5$, i.e. the $B$ and $\br{B}$ will be nearest neighbours in
rank and share two quark pairs. Specifically, $\q_1$ will gain
energy from $\q_2$ in order to end up on mass shell, and the
tunnelling formula for an effective $\q_1 \q_2$ diquark is recovered.
 
Part of the time, several $b \br{b}$ colour pair productions
may take place between the $\q_1$ and $\qbar_1$, however. With two
production vertices $\q_2 \qbar_2$ and $\q_3 \qbar_3$, a central
meson $\qbar_2 \q_3$ may be formed, surrounded by a baryon
$\q_4 \q_1 \q_2$ and an antibaryon $\qbar_3 \qbar_1 \qbar_5$.
We call this a $BM\br{B}$ configuration to distinguish it from the
$\q_4 \q_1 \q_2$ + $\qbar_2 \qbar_1 \qbar_5$ $B\br{B}$ configuration
above. For $BM\br{B}$ the $B$ and $\br{B}$ only share one
quark--antiquark pair, as opposed to two for $B\br{B}$
configurations. The relative probability for a $BM\br{B}$
configuration is given by the uncertainty relation suppression for
having the $\q_1$ and $\qbar_1$ sufficiently far apart that a meson
may be formed in between. Strictly speaking, also configurations
like $BMM\br{B}$, $BMMM\br{B}$, etc. should be possible, but the
probability for this is small in our model. Further, since larger
masses corresponds to longer string pieces, the production of
pseudoscalar mesons is favoured over that of vector ones. If only
$B\br{B}$ and $BM\br{B}$ states are included, and if the probability
for having a vector meson $M$ is not suppressed extra, two partly
compensating errors are made (since a vector meson typically
decays into two or more pseudoscalar ones).
 
In total, the flavour iteration procedure therefore contains the
following possible subprocesses (plus, of course, their charge
conjugates):
\begin{Itemize}
\item $\q_1 \to \q_2 + (\q_1 \qbar_2)$ meson;
\item $\q_1 \to \qbar_2\qbar_3 + (\q_1 \q_2 \q_3)$ baryon;
\item $\q_1 \q_2 \to \qbar_3 + (\q_1 \q_2 \q_3)$ baryon;
\item $\q_1 \q_2 \to \q_1 \q_3 + (\q_2 \qbar_3)$ meson;
\end{Itemize}
with the constraint that the last process cannot be iterated to
obtain several mesons in between the baryon and the antibaryon.
 
Unfortunately, the resulting baryon production model has a fair
number of parameters, which would be given by the model only if
quark and diquark masses were known unambiguously.
We have already mentioned the $\s / \u$ ratio and the $\q\q / \q$
one; the latter has to be increased from 0.09 to 0.10 for the
popcorn model, since the total number of possible baryon
production configurations is lower in this case (the particle
produced between the $B$ and $\br{B}$ is constrained to be a
meson). For the popcorn model, exactly the same parameters as
already found in the diquark model are needed to describe the
$B\br{B}$ configurations. For $BM\br{B}$ configurations, the square
root of a suppression factor should be applied if the factor is
relevant only for one of the $B$ and $\br{B}$, e.g. if the $B$
is formed with a spin 1 `diquark' $\q_1 \q_2$ but the $\br{B}$
with a spin 0 diquark $\qbar_1 \qbar_3$. Additional parameters
include the relative probability for $BM\br{B}$ configurations,
which is assumed to be roughly 0.5 (with the remaining 0.5 being
$B\br{B}$), a suppression factor for having a strange meson $M$
between the $B$ and $\br{B}$ (as opposed to having a lighter
nonstrange one) and a suppression factor for having a $\s \sbar$
pair (rather than a $\u \ubar$ one) shared between the $B$ and
$\br{B}$ of a $BM\br{B}$ configuration. The default parameter
values are based on a combination of experimental observation
and internal model predictions.
 
In the diquark model, a diquark is expected to have exactly the
same transverse momentum distribution as a quark. For $BM\br{B}$
configurations the situation is somewhat more unclear, but we
have checked that various possibilities give very similar results.
The option implemented in the program is to assume no transverse
momentum at all for the $\q_1 \qbar_1$ pair shared by the $B$ and
$\br{B}$, with all other pairs having the standard Gaussian
spectrum with local momentum conservation. This means that the
$B$ and $\br{B}$ $\pT$:s are uncorrelated in a $BM\br{B}$
configuration and (partially) anticorrelated in the $B\br{B}$
configurations, with the same mean transverse momentum for primary
baryons as for primary mesons.
 
Occasionally, the endpoint of a string is not a single parton,
but a diquark or antidiquark, e.g. when a quark has been kicked
out of a proton beam particle. One could consider fairly complex
schemes for the resulting fragmentation. One such \cite{And81}
was available in {\Je} version 6 but is no longer found in version
7. Instead the same basic scheme is used as for diquark pair
production above. Thus a $\q\q$ diquark endpoint is let to fragment
just as would a $\q\q$ produced in the field behind a matching
$\qbar\qbar$ flavour, i.e. either the two quarks of the diquark enter
into the same leading baryon, or else a meson is first produced,
containing one of the quarks, while the other is contained in the
baryon produced in the next step.
 
\subsection{String Fragmentation}
 
An iterative procedure can also be used for other aspects of the
fragmentation. This is possible because, in the string picture,
the various points where the string break by the production of
$\q \qbar$ pairs are causally disconnected. Whereas the space--time
picture in the c.m. frame is such that slow particles
(in the middle of the system) are formed first, this ordering is
Lorentz frame dependent and hence irrelevant. One may therefore
make the convenient choice of starting an iteration process at
the ends of the string and proceeding towards the middle.
 
The string fragmentation scheme is rather complicated for a generic
multiparton state. In order to simplify the discussion, we will
therefore start with the simple $\q \qbar$ process, and only later
survey the complications that appear when additional gluons
are present. (This distinction is made for pedagogical reasons,
in the program there is only one general-purpose algorithm).
 
\subsubsection{Fragmentation functions}
 
Assume a $\q \qbar$ jet system, in its c.m. frame, with the quark moving
out in the $+z$ direction and the antiquark in the $-z$ one. We have
discussed how it is possible to start the flavour iteration from the
$\q$ end, i.e. pick a $\q_1 \qbar_1$ pair, form a hadron $\q \qbar_1$,
etc. It has also been noted that the tunnelling mechanism
is assumed to give a transverse momentum $\pT$ for each new
$\q_i\qbar_i$ pair created, with the $\pT$ locally compensated between
the $\q_i$ and the $\qbar_i$ member of the pair, and with a Gaussian
distribution in $p_x$ and $p_y$ separately. In the program, this is
regulated by one parameter, which gives the root-mean-square $\pT$
of a quark. Hadron transverse momenta are obtained as the sum of
$\pT$:s of the constituent $\q_i$ and $\qbar_{i+1}$, where a diquark
is considered just as a single quark.
 
What remains to be determined is the energy and longitudinal
momentum of the hadron. In fact, only one variable can be selected
independently, since the momentum of the hadron is constrained by
the already determined hadron transverse mass $m_{\perp}$,
\begin{equation}
(E+p_z)(E-p_z) = E^2 - p_z^2 = m_{\perp}^2 = m^2 + p_x^2 + p_y^2 ~.
\label{fr:massconstr}
\end{equation}
In an iteration from the quark end, one is led (by the desire for
longitudinal boost invariance and other considerations)
to select the $z$ variable as the fraction of
$E+p_z$ taken by the hadron, out of the available $E+p_z$.
As hadrons are split off, the $E+p_z$ (and $E-p_z$) left for
subsequent steps is reduced accordingly:
\begin{eqnarray}
(E+p_z)_{\mrm{new}} & = & (1-z) (E+p_z)_{\mrm{old}} ~, \nonumber \\
(E-p_z)_{\mrm{new}} & = & (E-p_z)_{\mrm{old}} -
\frac{m_{\perp}^2}{z (E+p_z)_{\mrm{old}}} ~.
\end{eqnarray}

The fragmentation function $f(z)$, which expresses the probability
that a given $z$ is picked, could in principle be arbitrary --- indeed,
several such choices can be used inside the program, see below.
 
If one, in addition, requires that the fragmentation
process as a whole should look the same,
irrespectively of whether the iterative procedure is performed
from the $\q$ end or the $\qbar$ one, `left--right symmetry',
the choice is essentially unique \cite{And83a}: the `Lund symmetric
fragmentation function',
\begin{equation}
f(z) \propto \frac{1}{z} z^{a_{\alpha}} \left( \frac{1-z}{z}
   \right)^{a_{\beta}} \exp \left( - \frac{bm_{\perp}^2}{z} 
   \right) ~.
\label{fr:LSFFlong}
\end{equation}
There is one separate parameter $a$ for each flavour,
with the index $\alpha$ corresponding to the `old' flavour in the
iteration process, and $\beta$ to the `new' flavour.
It is customary to put all $a_{\alpha,\beta}$ the same, and thus
arrive at the simplified expression
\begin{equation}
 f(z) \propto z^{-1} (1-z)^a \exp (-bm_{\perp}^2/z) ~.
 \label{fr:LSFF}
\end{equation}
In the program, only two separate $a$ values can be given, that
for quark pair production and that for diquark one; by default the
two are taken to be the same. In addition, there is the $b$ parameter,
which is universal.
 
It should be noted that the explicit mass dependence in $f(z)$ implies
a harder fragmentation function for heavier hadrons; the
asymptotic behaviour of the mean $z$ value for heavy hadrons is
\begin{equation}
\langle z \rangle \approx 1 - \frac{1+a}{bm_{\perp}^2} ~.
\end{equation}
Unfortunately it seems this predicts a somewhat harder spectrum
for $\B$ mesons than observed in data.
 
For future reference we note that the derivation of $f(z)$ as a
by-product also gives the probability distribution in proper
time $\tau$ of $\q_i \qbar_i$ breakup vertices. In terms of
$\Gamma = (\kappa \tau)^2$, this distribution is
\begin{equation}
{\cal P}(\Gamma) \, \d \Gamma \propto \Gamma^a \, \exp(-b \Gamma) \,
\d \Gamma ~,
\end{equation}
with the same $a$ and $b$ as above.
 
Many different other fragmentation functions have been proposed,
and a few are available as options in the program.
\begin{Itemize}
\item The Field-Feynman parametrization \cite{Fie78},
\begin{equation}
  f(z) = 1 - a + 3a(1-z)^2 ~,
\end{equation}
with default value $a = 0.77$, is intended to be used only for
ordinary hadrons made out of $\u$, $\d$ and $\s$ quarks.
\item Since there are indications that the shape above is too
strongly peaked at $z = 0$, instead a shape like
\begin{equation}
  f(z) = (1+c) (1-z)^c
\end{equation}
may be used.
\item Charm and bottom data clearly indicate the need for a
harder fragmentation function for heavy flavours.
The best known of these is the Peterson et al.~formula \cite{Pet83}
\begin{equation}
  f(z) \propto \frac{1}{ z \left( 1 - 
      \frac{\displaystyle 1}{\displaystyle z} -
      \frac{\displaystyle \epsilon_Q}{\displaystyle 1-z} 
      \right)^2 } ~,
  \label{fr:PetHF}
\end{equation}
where $\epsilon_Q$ is a free parameter, expected to scale between
flavours like $\epsilon_Q \propto 1/m_Q^2$.
\item As a crude alternative, that is also peaked at $z=1$, one may
use
\begin{equation}
  f(z) = (1+c) z^c ~.
\end{equation}
\item Bowler \cite{Bow81} has shown, within the framework of the
Artru--Mennessier model \cite{Art74}, that a massive endpoint quark
with mass $m_Q$ leads to a modification of the symmetric
fragmentation function, due to
the fact that the string area swept out is reduced for massive endpoint
quarks, compared with massless ditto. The Artru--Mennessier model in
principle only applies for clusters with a continuous mass spectrum,
and does not allow an $a$ term (i.e. $a \equiv 0$); however, it has
been shown \cite{Mor89} that, for a discrete mass spectrum, one may
still retain an effective $a$ term. In the program an approximate
form with an $a$ term has therefore been used:
\begin{equation}
f(z) \propto \frac{1}{z^{1 + r_Q b m_Q^2}}
   z^{a_{\alpha}} \left( \frac{1-z}{z}
   \right)^{a_{\beta}} \exp \left( - \frac{b m_{\perp}^2}{z} \right) ~.
\label{fr:LSFFBowler}
\end{equation}
In principle the prediction is that $r_Q \equiv 1$, but so as to
be able to extrapolate smoothly between this form and the Lund
symmetric one, it is possible to pick $r_Q$ separately for $\c$, $\b$
and $\t$ hadrons.
\end{Itemize}
 
\subsubsection{Joining the jets}
 
The $f(z)$ formula above is only valid, for the breakup of a jet
system into a hadron plus a remainder-system, when the remainder
mass is large. If the fragmentation algorithm were to be used all the
way from the $\q$ end to the $\qbar$ one, the mass of the last hadron to
be formed at the $\qbar$ end would be completely constrained by energy
and momentum conservation, and could not be on its mass shell. In theory
it is known how to take such effects into account, but the resulting
formulae are wholly unsuitable for Monte Carlo implementation.
 
The practical solution to this problem is to carry out the
fragmentation both from the $\q$ and the $\qbar$ end, such that for
each new step in the fragmentation process, a random
choice is made as to from what side the step is to be taken.
If the step is on the $\q$ side, then $z$ is interpreted as fraction
of the remaining $E+p_z$ of the system, while $z$ is interpreted as
$E-p_z$ fraction for a step from the $\qbar$ end. At some point, when
the remaining mass of the system has dropped below a given value,
it is decided that the next breakup will produce two final hadrons,
rather than a hadron and a remainder-system.
Since the momenta of two hadrons are to be selected, rather than
that of one only, there are enough degrees of freedom to have both
total energy and total momentum completely conserved.
 
The mass at which the normal fragmentation
process is stopped and the final two hadrons formed is not actually
a free parameter of the model: it is given by the requirement that
the string everywhere looks the same, i.e. that the rapidity spacing
of the final two hadrons, internally and with respect to surrounding
hadrons, is the same as elsewhere in the fragmentation process.
The stopping mass, for a given setup of fragmentation parameters,
has therefore been determined in separate runs. If the fragmentation
parameters are changed, some retuning should be done but, in practice,
reasonable changes can be made without any special arrangements.
 
Consider a fragmentation process which has already split off a number
of hadrons from the $\q$ and $\qbar$ sides, leaving behind a
a $\q_i \qbar_j$ remainder system. When this system breaks by the
production of a $\q_n \qbar_n$ pair, it is decided to make this pair
the final one, and produce the last two hadrons $\q_i\qbar_n$ and
$\q_n\qbar_j$, if
\begin{equation}
( (E+p_z)(E-p_z) )_{\mrm{remaining}} = W_{\mrm{rem}}^2 < 
W_{\mmin}^2 ~.
\end{equation}
The $W_{\mmin}$ is calculated according to
\begin{equation}
W_{\mmin} = ( W_{\mmin 0} + m_{\q i} + m_{\q j} + k \, m_{\q n} )
\, (1 \pm \delta) ~.
\end{equation}
Here $W_{\mmin 0}$ is the main free parameter, typically around 
1 GeV, determined to give a flat
rapidity plateau (separately for each particle species), while the
default $k = 2$ corresponds to the mass of the final pair being taken
fully into account. Smaller values may also be considered, depending
on what criteria are used to define the `best' joining of the $\q$
and the $\qbar$ chain. The factor $1 \pm \delta$, by default evenly
distributed between 0.8 and 1.2, signifies a smearing of the 
$W_{\mmin}$ value, to avoid an abrupt and unphysical cut-off in
the invariant mass distribution of the final two hadrons. Still,
this distribution will be somewhat different from that of any two
adjacent hadrons elsewhere. Due to the cut there will be no tail up to
very high masses; there are also fewer events close to the lower limit,
where the two hadrons are formed at rest with respect to each other.
 
This procedure does not work all that well for heavy flavours, since it
does not fully take into account the harder fragmentation function
encountered. Therefore, in addition to the check above, one further
test is performed for charm and heavier flavours, as follows. If the
check above allows more particle production,
a heavy hadron $\q_i \qbar_n$ is formed, leaving a
remainder $\q_n \qbar_j$.  The range of allowed $z$ values, i.e. the
fraction of remaining $E+p_z$ that may be taken by the
$\q_i \qbar_n$ hadron, is constrained away from 0 and 1 by the
$\q_i \qbar_n$ mass and minimal mass of the $\q_n \qbar_j$ system.
The limits of the physical $z$ range is obtained when the
$\q_n \qbar_j$ system only consists of one single particle, which then
has a well-determined
transverse mass $m_{\perp}^{(0)}$. From the $z$ value obtained with the
infinite-energy fragmentation function formulae, a rescaled $z'$ value
between these limits is given by
\begin{equation}
z' = \frac{1}{2} \left\{ 1 + \frac{m_{\perp i n}^2}{W_{\mrm{rem}}^2} - 
\frac{m_{\perp n j}^{(0)2}}{W_{\mrm{rem}}^2}  + 
\sqrt{ \left( 1 - \frac{m_{\perp i n}^2}{W_{\mrm{rem}}^2} - 
\frac{m_{\perp n j}^{(0)2}}{W_{\mrm{rem}}^2} \right)^2 
- 4 \frac{m_{\perp i n}^2}{W_{\mrm{rem}}^2}
\frac{m_{\perp n j}^{(0)2}}{W_{\mrm{rem}}^2} }
\, (2 z -1) \right\} ~.
\end{equation}
From the $z'$ value, the actual transverse mass
$m_{\perp n j} \geq m_{\perp n j}^{(0)}$ of the $\q_n \qbar_j$
system may be calculated. For more than one particle
to be produced out of this system, the requirement
\begin{equation}
m_{\perp n j}^2 = (1-z') \, \left( W_{\mrm{rem}}^2 -
\frac{m_{\perp i n}^2}{z'} \right) > (m_{q j} + W_{\mmin 0})^2 
+ \pT^2
\end{equation}
has to be fulfilled. If not, the $\q_n \qbar_j$ system is assumed to
collapse to one single particle.
 
The consequence of the procedure above is that, the more the infinite
energy fragmentation function $f(z)$ is peaked close to $z=1$, the more
likely it is that only two particles are produced. In particular, for
$\t \tbar$ systems, where very large $\langle z \rangle$ values are
predicted, the
expectation is that two particle final states will dominate far above
the threshold region. The procedure above has been constructed so that
the two particle fraction can be calculated directly from the shape of
$f(z)$ and the (approximate) mass spectrum, but it is not unique.
For the symmetric Lund fragmentation function, a number of
alternatives tried all give essentially the same result, whereas other
fragmentation functions may be more sensitive to details.
 
Assume now that two final hadrons have been picked. If the
transverse mass of the remainder-system is smaller than the sum of
transverse masses of the final two hadrons, the whole fragmentation
chain is rejected, and started over from the $\q$ and $\qbar$
endpoints. This does not introduce any significant bias, since the
decision to reject a fragmentation chain only depends on what happens
in the very last step, specifically that the next-to-last step took
away too much energy, and not on what happened in the steps before
that.
 
If, on the other hand,  the remainder-mass is large enough, there
are two kinematically allowed solutions for the final two hadrons:
the two mirror images in the rest frame of the remainder-system. Also 
the choice between these two solutions is given by the consistency
requirements, and can be derived from studies of infinite energy jets.
The probability for the reverse ordering, i.e. where
the rapidity and the flavour orderings disagree, is parametrized by
\begin{equation}
{\cal P}_{\mrm{reverse}} = \frac{1}{2} \left(
\frac{m_{\perp i n} + m_{\perp n j}}{W_{\mrm{rem}}} \right)^d ~.
\label{fr:revord}
\end{equation}
For symmetric fragmentation, the ordering is expected to be
increasingly strict when the particles involved are more massive.
In the program it is therefore assumed that $d$ is a function of the
masses, $d = d_0 (m_{\perp i n} + m_{\perp n j})^2$, where $d_0$ is a
free parameter.
 
When baryon production is included, some particular problems arise.
First consider $B\br{B}$ situations. In the na\"{\i}ve iterative scheme,
away from the middle of the event, one
already has a quark and is to chose a matching diquark flavour or the
other way around. In either case the choice of the new flavour can
be done taking into account the number of {\bf SU(6)} states available
for the quark-diquark combination. For a
case where the final $\q_n \qbar_n$ breakup is an
antidiquark-diquark one, the weights for forming $\q_i \qbar_n$ and
$\q_n \qbar_i$ enter at the same time, however. We do not know how to
handle this problem; what is done is to use weights as usual for the
$\q_i \qbar_n$ baryon to select $\q_n$, but then consider
$\q_n \qbar_i$ as given (or the other way around with equal
probability). If $\q_n \qbar_i$ turns out to be
an antidiquark-diquark combination, the whole fragmentation chain is
rejected, since we do not know how to form corresponding hadrons.
A similar problem arises, and is solved in the same spirit, for a
$BM\br{B}$ configuration in which the $B$ (or $\br{B}$) was chosen
as third-last particle. When only two particles remain to be
generated, it is obviously too late to consider having a $BM\br{B}$
configuration. This is as it should, however, as can be found by
looking at all possible ways a hadron of given rank can be a baryon.
 
While some practical compromises have to be accepted in the
joining procedure, the fact that the joining takes place in
different parts of the string in different events means that,
in the end, essentially no visible effects remain.
 
\subsubsection{String motion and infrared stability}
 
We have now discussed the SF scheme for the fragmentation of a simple
$\q \qbar$ jet system. In order to understand how these results
generalize
to arbitrary jet systems, it is first necessary to understand the
string motion for the case when no fragmentation takes place. In the
following we will assume that quarks as well as gluons are massless,
but all arguments can be generalized to massive quarks without too
much problem.
 
For a $\q \qbar$ event viewed in the c.m. frame, with total energy $W$,
the partons start moving out back-to-back, carrying half the energy
each. As they move apart, energy and momentum is lost to the string.
When the partons are a distance $W / \kappa$ apart, all the energy
is stored in the string. The partons now turn around and come
together again with the original momentum vectors reversed. This
corresponds to half a period of the full string motion; the second
half the process is repeated, mirror-imaged. For further
generalizations to multiparton systems, a convenient description of
the energy and momentum flow is given in terms of `genes'
\cite{Art83}, infinitesimal packets of the four-momentum given up
by the partons to the string. Genes with $p_z = E$, emitted from the
$\q$ end in the initial stages of the string motion above,
will move in the $\qbar$ direction with the speed of light, whereas
genes with $p_z = -E$ given up by the $\qbar$ will move in the $\q$
direction. Thus, in this simple case, the direction of motion for a
gene is just opposite to that of a free particle with the same
four-momentum. This is due to the string tension. If the system is
not viewed in the c.m. frame, the rules are that any parton gives up
genes with four-momentum proportional to its own four-momentum, but
the direction of motion of any gene is given by the momentum direction
of the genes it meets, i.e. that were emitted by the parton at the
other end of that particular string piece. When the $\q$ has lost
all its energy, the $\qbar$ genes, which before could not catch up
with $\q$, start impinging on it, and the $\q$ is pulled back,
accreting $\qbar$ genes in the process. When the $\q$ and $\qbar$
meet in the origin again, they have completely traded genes with
respect to the initial situation.
 
A 3-jet $\q \qbar \g$ event initially corresponds to having a
string piece stretched between $\q$ and $\g$ and another between
$\g$ and $\qbar$. Gluon four-momentum genes are thus flowing towards
the $\q$ and $\qbar$. Correspondingly, $\q$ and $\qbar$ genes are
flowing towards the $\g$. When the gluon has lost all its energy,
the $\g$ genes continue moving apart, and instead a third
string region is formed in the `middle' of the total string,
consisting of overlapping $\q$ and $\qbar$ genes. The two `corners'
on the string, separating the three string regions, are not of the
gluon-kink type: they do not carry any momentum.
 
If this third region would only appear at a time later than the typical
time scale for fragmentation, it could not affect the sharing of energy
between different particles. This is true in the limit of high energy,
well separated partons. For a small gluon energy, on the other hand, the
third string region appears early, and the overall drawing of the string
becomes fairly 2-jet-like, since the third string region consists of
$\q$ and $\qbar$ genes and therefore behaves exactly as a sting pulled
out directly between the $\q$ and $\qbar$.
In the limit of vanishing gluon energy,
the two initial string regions collapse to naught, and the ordinary
2-jet event is recovered. Also for a collinear gluon, i.e.
$\theta_{\q \g}$ (or $\theta_{\qbar \g}$) small, the stretching
becomes 2-jet-like. In particular, the $\q$ string endpoint first
moves out a distance $\mbf{p}_{\q} / \kappa$ losing genes to the
string, and then a further distance $\mbf{p}_{\g} / \kappa$, a first
half accreting genes from the $\g$ and the second half re-emitting
them. (This latter half actually includes yet another
string piece; a corresponding piece appears at the $\qbar$ end, such
that half a period of the system involves five different string
regions.) The end result is, approximately, that a string is drawn out
as if there had only been a single parton with energy
$|\mbf{p}_{\q} + \mbf{p}_{\g}|$, such that the simple 2-jet event
again is recovered in the limit $\theta_{\q \g} \to 0$. These
properties of the string motion are the reason why
the string fragmentation scheme is `infrared safe' with respect to
soft or collinear gluon emission.
 
The discussions for the 3-jet case can be generalized to the motion
of a string with $\q$ and $\qbar$ endpoints and an arbitrary number of
intermediate gluons. For $n$ partons, whereof $n-2$
gluons, the original string contains $n-1$ pieces. Anytime one of the
original gluons has lost its energy, a new string region is formed,
delineated by a pair of `corners'.
As the extra `corners' meet each other, old string regions vanish
and new are created, so that half a period of the string contains
$2n^2 - 6n + 5$ different string regions. Each of these regions can
be understood simply as built up from the overlap of (opposite-moving)
genes from two of the original partons, according to well specified
rules.
 
\subsubsection{Fragmentation of multiparton systems}
 
The full machinery needed for a multiparton system is very
complicated, and is described in detail in \cite{Sjo84}. The
following outline is far from complete, and is complicated
nonetheless. The main message to be conveyed is that a
Lorentz covariant algorithm exists for handling an arbitrary parton
configuration, but that the necessary machinery is more complex
than in either cluster or independent fragmentation.
 
Assume $n$ partons, with ordering along the string, and related
four-momenta, given by
$\q(p_1) \g(p_2) \g(p_3) \cdots \g(p_{n-1}) \qbar(p_n)$.
The initial string then contains $n-1$ separate pieces.
The string piece between the quark and its neigbouring gluon is, in
four-momentum space, spanned by one side with four-momentum
$p_+^{(1)} = p_1$ and another with $p_-^{(1)} = p_2/2$. The factor of
1/2 in the second expression
comes from the fact that the gluon shares its energy between two
string pieces. The indices `$+$' and `$-$' denotes direction towards
the $\q$ and $\qbar$ end, respectively. The next string piece,
counted from the quark end, is spanned by $p_+^{(2)} = p_2/2$ and
$p_-^{(2)} = p_3/2$, and so on, with the last one being
$p_+^{(n-1)} = p_{n-1}/2$ and $p_-^{(n-1)} = p_n$.
 
For the algorithm to work, it is important
that all $p_{\pm}^{(i)}$ be light-cone-like, i.e. $p_{\pm}^{(i)2} = 0$.
Since gluons are massless, it is only the two endpoint quarks which
can cause problems. The procedure here is to create new $p_{\pm}$
vectors for each of the two endpoint regions, defined to be linear
combinations of the old $p_{\pm}$ ones for the same region, with
coefficients determined so that the
new vectors are light-cone-like. De facto, this corresponds to
replacing a massive quark at the end of a string piece with a massless
quark at the end of a somewhat longer string piece. With the exception
of the added fictitious piece, which anyway ends up entirely within
the heavy hadron produced from the heavy quark, the string motion
remains unchanged by this.
 
In the continued string motion, when new string regions appear as
time goes by, the first such string regions that appear can be
represented as being spanned by one $p_+^{(j)}$ and another
$p_-^{(k)}$ four-vector, with $j$ and $k$ not necessarily
adjacent. For instance, in the $\q \g \qbar$
case, the `third' string region is spanned by $p_+^{(1)}$ and
$p_-^{(3)}$. Later on in the string evolution history,
it is also possible to have regions made up of two $p_+$ or two
$p_-$ momenta. These appear when an endpoint quark has
lost all its original momentum, has accreted the momentum of an
gluon, and is now re-emitting this momentum. In practice, these
regions may be neglected. Therefore only pieces made up by a
$(p_+^{(j)},p_-^{(k)})$ pair of momenta are considered in the
program.
 
The allowes string regions may be ordered in an abstract parameter
plane, where the $(j,k)$ indices of the four-momentum pairs define
the position of each region along the two (parameter plane)
coordinate axes. In this plane the fragmentation procedure can be
described as a sequence of steps, starting at the quark end,
where $(j,k) = (1,1)$, and ending at the antiquark one,
$(j,k) = (n-1,n-1)$. Each step is taken from an `old' 
$\q_{i-1} \qbar_{i-1}$ pair production vertex, to the 
production vertex of a `new' $\q_i \qbar_i$
pair, and the string piece between these two string
breaks represent a hadron. Some steps may be taken within one and
the same region, while others may have one vertex in one region
and the other vertex in another region. Consistency requirements,
like energy-momentum conservation, dictates that vertex $j$ and
$k$ region values be ordered in a monotonic sequence, and that
the vertex positions are monotonically ordered inside each region.
The four-momentum of each hadron can be read off, for $p_+$ ($p_-$)
momenta, by projecting the separation between the old and the new
vertex on to the $j$ ($k$) axis. If the four-momentum fraction of
$p_{\pm}^{(i)}$ taken by a hadron is denoted $x_{\pm}^{(i)}$,
then the total hadron four-momentum is given by
\begin{equation}
p = \sum_{j=j_1}^{j_2} x_+^{(j)} p_+^{(j)} +
      \sum_{k=k_1}^{k_2} x_-^{(k)} p_-^{(k)} +
      p_{x1} \hat{e}_x^{(j_1 k_1)} + p_{y1} \hat{e}_y^{(j_1 k_1)} +
      p_{x2} \hat{e}_x^{(j_2 k_2)} + p_{y2} \hat{e}_y^{(j_2 k_2)} ~,
\label{fr:fourmom}
\end{equation}
for a step from region $(j_1,k_1)$ to region $(j_2,k_2)$.
By necessity, $x_+^{(j)}$ is unity for a $j_1 < j < j_2$,
and correspondingly for $x_-^{(k)}$.
 
The $(p_x,p_y)$ pairs are the transverse momenta produced at
the two string breaks, and the $(\hat{e}_x,\hat{e}_y)$ pairs 
four-vectors transverse to the string directions in the regions 
of the respective string breaks:
\begin{eqnarray}
 & & \hat{e}_x^{(jk)2} = \hat{e}_y^{(jk)2} = -1 ~,        \nonumber \\
 & & \hat{e}_x^{(jk)} \hat{e}_y^{(jk)} = \hat{e}_{x,y}^{(jk)}
  p_+^{(j)} = \hat{e}_{x,y}^{(jk)} p_-^{(k)} = 0 ~.
\end{eqnarray}
 
The fact that the hadron should be on mass shell, $p^2 = m^2$, puts
one constraint on where a new breakup may be, given that the old
one is already known, just as eq.~(\ref{fr:massconstr}) did in the
simple 2-jet case. The remaining degree of freedom is, as before, 
to be given by
the fragmentation function $f(z)$. The interpretation of the $z$
is only well-defined for a step entirely constrained to one of the
initial string regions, however, which is not enough. In the
2-jet case, the $z$ values can be related to the proper times
of string breaks, as follows. The variable
$\Gamma = (\kappa \tau)^2$, with $\kappa$ the string tension and
$\tau$ the proper time between the production vertex of the
partons and the breakup point, obeys an iterative relation of the
kind
\begin{eqnarray}
\Gamma_0 & = & 0 ~,              \nonumber   \\
\Gamma_i & = & (1-z_i) \left( \Gamma_{i-1} + \frac{m_{\perp i}^2}{z_i}
   \right) ~.
\end{eqnarray}
Here $\Gamma_0$ represents the value at the $\q$ and $\qbar$ endpoints,
and $\Gamma_{i-1}$ and $\Gamma_i$ the values
at the old and new breakup vertices needed to produce
a hadron with transverse mass $m_{\perp i}$, and with the $z_i$ of the
step chosen according to $f(z_i)$. The proper time can be
defined in an unambiguous way, also over boundaries between the
different string regions, so for multijet events the $z$ variable may
be interpreted just as
an auxiliary variable needed to determine the next $\Gamma$ value.
(In the Lund symmetric fragmentation function derivation, the
$\Gamma$ variable actually does appear naturally, so the choice
is not as arbitrary as it may seem here.)
The mass and $\Gamma$ constraints together are sufficient to determine
where the next string breakup is to be chosen, given the preceding
one in the iteration scheme. Actually, several ambiguities remain,
but are of no importance for the overall picture.
 
The algorithm for finding the next breakup then works something
like follows. Pick a hadron, $\pT$, and $z$, and calculate the next
$\Gamma$. If the old breakup is in the region $(j,k)$, and if the
new breakup is also assumed to be in the same region, then the
$m^2$ and $\Gamma$ constraints can be reformulated in terms of
the fractions $x_+^{(j)}$ and $x_-^{(k)}$ the hadron must take of the
total four-vectors $p_+^{(j)}$ and $p_-^{(k)}$:
\begin{eqnarray}
 m^2 &=&
 c_1 + c_2 x_+^{(j)} + c_3 x_-^{(k)} + c_4 x_+^{(j)} x_-^{(k)} ~,
       \nonumber  \\
 \Gamma &=&
 d_1 + d_2 x_+^{(j)} + d_3 x_-^{(k)} + d_4 x_+^{(j)} x_-^{(k)} ~.
\label{fr:mGa}
\end{eqnarray}
Here the coefficients
$c_n$ are fairly simple expressions, obtainable by squaring
eq.~(\ref{fr:fourmom}), while $d_n$ are slightly more complicated 
in that they depend on the position of the old string break,
but both the $c_n$ and the $d_n$ are explicitly calculable. What
remains is an equation system with two unknowns, $x_+^{(j)}$
and $x_-^{(k)}$.
The absence of any quadratic terms is due to
the fact that all $p_{\pm}^{(i)2} = 0$, i.e. to the choice of a
formulation based on light-cone-like longitudinal vectors.
Of the two possible solutions to the equation system (elimination of
one variable gives a second degree equation in the other), one is
unphysical and can be discarded outright. The other solution is checked
for
whether the $x_{\pm}$ values are actually inside the physically allowed
region, i.e. whether the $x_{\pm}$ values of the current step, plus
whatever has already been used up in previous steps, are less than
unity. If yes, a solution has been found. If no, it is because the
breakup could not take place inside the region studied, i.e. because
the equation system was solved for the wrong region. One therefore
has to change either index $j$ or index $k$ above by one step, i.e.
go to the next nearest string region. In this new region, a new equation
system of the type in eq.~(\ref{fr:mGa}) may be written down, with new
coefficients. A new solution is found and tested, and so on until a
physically
acceptable solution is found. The hadron four-momentum is now
given by an expression of the type (\ref{fr:fourmom}). The breakup
found forms the starting point for the new step in the fragmentation
chain, and so on. The final joining in the middle is done as in the
2-jet case, with minor extensions.
 
\subsection{Independent Fragmentation}
 
The independent fragmentation (IF) approach dates back to the early
seventies \cite{Krz72}, and gained widespread popularity with
the Field-Feynman paper \cite{Fie78}. Subsequently, IF was the basis
for two programs widely used in the early PETRA/PEP days, the
Hoyer et al.~\cite{Hoy79} and the Ali et al.~\cite{Ali80} programs.
JETSET has as (non-default) options a wide selection of independent
fragmentation algorithms.
 
\subsubsection{Fragmentation of a single jet}
 
In the IF approach, it is assumed that the fragmentation of any
system of partons can be described as an incoherent sum of
independent fragmentation procedures for each parton separately.
The process is to be carried out in the overall c.m. frame of the
jet system, with each jet fragmentation axis given by the direction
of motion of the corresponding parton in that frame.
 
Exactly as in string fragmentation, an iterative ansatz can be used
to describe the sucessive production of one hadron after the next.
Assume that a quark is kicked out by some hard interaction, carrying a
well-defined amount of energy and momentum. This quark jet $\q$ is
split into a hadron $\q \qbar_1$ and a remainder-jet $\q_1$, essentially
collinear with each other. New quark and hadron flavours are picked
as already described. The sharing of energy and momentum is given
by some probability distribution $f(z)$, where $z$ is the fraction
taken by the hadron, leaving $1-z$ for the remainder-jet. The
remainder-jet is assumed to be just a scaled-down version of the
original jet, in an average sense. The process of splitting off a
hadron can therefore be iterated, to yield a sequence of hadrons.
In particular, the function $f(z)$ is assumed to be the
same at each step, i.e. independent of remaining energy. If $z$ is
interpreted as the fraction of the jet $E+p_{\mrm{L}}$, i.e. energy plus
longitudinal momentum with respect to the jet axis, this leads to a
flat central rapidity plateau $dn/dy$ for a large initial energy.
 
Fragmentation functions can be chosen among those listed above for
string fragmentation, but also here the default is the Lund symmetric
fragmentation function.
 
The normal $z$ interpretation means that
a choice of a $z$ value close to $0$ corresponds to a particle
moving backwards, i.e. with $p_{\mrm{L}} < 0$.
It makes sense to allow only the production of particles with 
$p_{\mrm{L}} > 0$, but to explicitly constrain $z$ accordingly
would destroy longitudinal invariance. The most
straightforward way out is to allow all $z$ values but discard
hadrons with $p_{\mrm{L}} < 0$. Flavour, transverse momentum and 
$E + p_{\mrm{L}}$
carried by these hadrons are `lost' for the forward jet. The average
energy of the final jet comes out roughly right this way, with a spread
of 1--2 GeV around the mean. The jet longitudinal
momentum is decreased, however, since the jet acquires an effective
mass during the fragmentation procedure. For a 2-jet event this is
as it should be, at least on average, because also the momentum of the
compensating opposite-side parton is decreased.
 
In addition to local flavour conservation in $\q_i \qbar_i$ splittings,
it is also assumed that transverse momentum is locally conserved, i.e.
the net $\pT$ of the $\q_i \qbar_i$ pair as a whole is assumed to be
vanishing. The $\pT$ of the $\q$ is taken to be a Gaussian in the two
transverse degrees of freedom separately, with the transverse momentum
of a hadron obtained by the sum of constituent quark transverse momenta.
 
Within the IF framework, there is no unique recipe for how gluon jet
fragmentation should be handled. One possibility is to treat it exactly
like a quark jet, with the initial quark flavour chosen
at random among $\u$, $\ubar$, $\d$, $\dbar$, $\s$ and
$\sbar$, including the ordinary $\s$ quark suppression factor.
Since the gluon is supposed to fragment more softly
than a quark jet, the fragmentation fuction may be chosen
independently. Another common option is to split the $\g$ jet into
a pair of parallel $\q$ and $\qbar$ ones, sharing the energy,
e.g. as in a perturbative branching $\g \to \q \qbar$, i.e.
$f(z) \propto z^2 + (1-z)^2$.
The fragmentation function could still be chosen
independently, if so desired. Further, in either case the fragmentation
$\pT$ could be chosen to have a different mean.
 
\subsubsection{Fragmentation of a jet system}
 
In a system of many jets, each jet is fragmented independently. Since
each jet by itself does not conserves the flavour, energy and momentum,
then neither does a system of jets.
At the end of the generation, special algorithms are
therefore used to patch this up. The choice of approach has major
consequences, e.g. for event shapes and $\alphas$ determinations
\cite{Sjo84a}.
 
Little attention is usually given to flavour conservation, and we only
offer one scheme. When the fragmentation of all jets has been performed,
independently of each other, the net initial flavour composition, i.e.
number of $\u$ quarks minus number of $\ubar$ quarks etc., is compared
with the net final flavour composition. In case of an imbalance,
the flavours of the hadron with lowest three-momentum are removed, and
the imbalance is re-evaluated. If the remaining imbalance could be
compensated by a suitable choice of new flavours for this hadron,
flavours are so chosen, a new mass is found and the new energy can be
evaluated, keeping the three-momentum of the original hadron. If
the removal of flavours from the hadron with lowest momentum is not
enough, flavours are removed from the one with next-lowest momentum,
and so on until enough freedom is obtained, whereafter the necessary
flavours are recombined at random to form the new hadrons. Occasionally
one extra $\q_i \qbar_i$ pair must be created, which is then done
according to the customary probabilities.
 
Several different schemes for energy and momentum conservation have
been devised. One \cite{Hoy79} is to conserve transverse momentum
locally within each jet, so that the final momentum vector of a jet 
is always parallel with that of the corresponding parton. Then 
longitudinal momenta
may be rescaled separately for particles within each jet, such that
the ratio of rescaled jet momentum to initial parton momentum is the
same in all jets. Since the initial partons had net vanishing
three-momentum, so do now the hadrons. The rescaling factors may
be chosen such that also energy comes out right. Another common
approach \cite{Ali80} is to boost the event to the frame
where the total hadronic momentum is vanishing. After that, energy
conservation can be obtained by rescaling all particle three-momenta
by a common factor.
 
The number of possible schemes is infinite.
Two further options are available in the program. One is to
shift all particle three-momenta by a common amount to give
net vanishing momentum, and then rescale as before. Another is
to shift all particle three-momenta, for each particle by an
amount proportional to the longitudinal mass with respect to the
imbalance direction, and with overall magnitude selected to give
momentum conservation, and then rescale as before.
In addition, there is a choice of whether to treat separate
colour singlets (like $\q \qbar'$ and $\q' \qbar$ in a
$\q \qbar \q' \qbar'$ event) separately or as one single big system.
 
A serious conceptual weakness of the IF framework is the issue of
Lorentz invariance. The outcome of the fragmentation procedure
depends on the coordinate frame chosen, a problem circumvented by
requiring fragmentation always to be carried out in the c.m. frame.
This is a consistent procedure for 2-jet events, but only a
technical trick for multijets.
 
It should be noted, however, that a Lorentz covariant generalization 
of the independent fragmentation model exists, in which separate 
`gluon-type' and `quark-type' strings are used,
the Montvay scheme \cite{Mon79}.
The `quark string' is characterized  by the ordinary string constant
$\kappa$, whereas a `gluon string' is taken to have a string constant
$\kappa_{\g}$. If $\kappa_{\g} > 2 \kappa$ it is always energetically
favourable to split a gluon string into two quark ones, and the
ordinary Lund string model is recovered. Otherwise, for a 3-jet
$\q \qbar \g$ event the three different string pieces are joined at
a junction. The motion of this junction is given by the composant of
string tensions acting on it. In particular, it is always possible
to boost an event to a frame where this junction is at rest. In this
frame, much of the standard na\"{\i}ve IF picture holds for the
fragmentation of the three jets; additionally, a correct treatment
would automatically give flavour, momentum and energy conservation.
Unfortunately, the simplicity is lost when studying events with
several gluon jets. In general, each event will contain a number of
different junctions, resulting in a polypod shape with a number of
quark and gluons strings sticking out from a skeleton of gluon
strings. With the shift of emphasis from three-parton to
multi-parton configurations, the simple option existing in {\Je}~6.3 
therefore is no longer included.
 
A second conceptual weakness of IF is the issue of collinear
divergences. In a parton-shower picture, where a quark or gluon is
expected to branch into several reasonably collimated partons,
the independent fragmentation of one single parton or of a bunch of
collinear ones gives quite different outcomes, e.g. with a much
larger hadron multiplicity in the latter case. It is conceivable that
a different set of fragmentation functions could be constructed in the
shower case in order to circumvent this problem
(local parton--hadron duality \cite{Dok89} would correspond to having
$f(z) = \delta(z-1)$).
 
\subsection{Other Fragmentation Aspects}
 
Here two aspects are considered, which are applicable
regardless of whether string or
independent fragmentation is used.
 
\subsubsection{Small mass systems}
 
Occasionally, a jet system may have too small an invariant mass for the
ordinary jet fragmentation schemes. This is particularly a problem when
showers are used, since two nearby $\g \to \q' \qbar'$ branchings 
may give rise to an intermediate low-mass colour-singlet system. Before 
the ordinary fragmentation, one includes an optional additional step,
to catch situations of this kind.
First the jet system with lowest invariant mass, minus endpoint quark
masses, is found. If this is too low for jet fragmentation, an attempt
is made to split the system into two hadrons by producing a new
$\q_n \qbar_n$ pair (with $\q_n$ chosen according to the standard
fragmentation scheme, so that e.g. also diquarks are allowed)
to go with the existing endpoint flavours. If the sum of the two
thus constructed
hadron masses is smaller than the total invariant mass, a simple
isotropic two-particle decay is performed. If not, the endpoint flavours
are combined to give one single hadron. Next, the parton (or hadron) is
found which, when taken together with the jet system, has the largest
invariant mass. A minimal transfer of four-momentum is then performed,
which puts the hadron on mass shell while keeping the mass of the
parton unchanged. With this done, one may again search for a low-mass
jet system, and iterate the procedure above, if need be.
The procedure may be seen as a `poor man's cluster fragmentation',
i.e. a cluster and a low-mass string are considered to be more or
less the same thing.
 
\subsubsection{Bose--Einstein effects}
 
A crude option for the simulation of Bose--Einstein effects is
included, but is turned off by default. Here the detailed physics is
not that well understood, see e.g. \cite{Lor89}. What is
offered is an algorithm, more than just a parametrization (since very
specific assumptions and choices have been made), and yet less than a
true model (since the underlying physics picture is rather fuzzy).
In this scheme, the
fragmentation is allowed to proceed as usual, and so is the decay of
short-lived particles like $\rho$. Then pairs of identical particles,
$\pi^+$ say, are considered one by one. The $Q_{ij}$ value of a pair
$i$ and $j$ is evaluated,
\begin{equation}
 Q_{ij} = \sqrt{ (p_i + p_j)^2 - 4m^2} ~,
\end{equation}
where $m$ is the common particle mass. A shifted (smaller) $Q'_{ij}$
is then to be found such that the (infinite statistics) ratio
$C_2(Q)$ of shifted to unshifted $Q$ distributions is given by the
requested parametrization. The shape may be chosen either
exponential or Gaussian,
\begin{equation}
  C_2(Q) = 1 + \lambda  \exp \left( - (Q/d)^r \right),
  ~~~~r = 1~\mrm{or}~2 ~.
\end{equation}
(In fact, the distribution has to dip slightly below unity at $Q$
values outside the Bose enhancement region, from conservation of
total multiplicity.) If the inclusive distribution of $Q_{ij}$
values is assumed given
just by phase space, at least at small relative momentum then,
with $\d^3 p / E \propto Q^2 \, \d Q / \sqrt{Q^2 + 4m^2}$, 
then $Q'_{ij}$ is found as the solution to the equation
\begin{equation}
\int_0^{Q_{ij}} \frac{Q^2 \, \d Q}{\sqrt{Q^2 + 4 m^2}} =
\int_0^{Q'_{ij}} C_2(Q) \, \frac{Q^2 \, \d Q}{\sqrt{Q^2 + 4 m^2}} ~.
\end{equation}
The change of $Q_{ij}$ can be translated into an effective shift of
the three-momenta of the two particles, if one uses
as extra constraint that the total three-momentum of each pair be
conserved in the c.m. frame of the event. Only after all pairwise
momentum shifts have been evaluated, with respect to the original
momenta, are these momenta actually shifted, for each particle by the
sum of evaluated shifts. The total energy of the event is slightly
reduced in the process, which is compensated by an overall rescaling of
all c.m. frame momentum vectors. It can be discussed which are the 
particles to involve in this rescaling. Currently the only
exceptions to using everything are leptons and neutrinos coming from
resonance decays (such as $\W$'s) and photons radiated by leptons
(also in initial state radiation). Finally, the decay chain is resumed 
with more long-lived particles like $\pi^0$.
 
Two comments can be made.
The Bose--Einstein effect is here interpreted almost as a
classical force acting on the `final state', rather than
as a quantum mechanical phenomenon on the production amplitude. This
is not a credo, but just an ansatz to make things manageable.
Also, since only pairwise interactions
are considered, the effects associated with three or more nearby
particles tend to get overestimated. (More exact, but also more
time-consuming methods may be found in
\cite{Zaj87}.) Thus the input $\lambda$
may have to be chosen smaller than what one wants to get out.
(On the other hand, many of the pairs of an event contains at least
one particle produced in some secondary vertex, like a $\D$ decay.
This reduces the fraction of pairs which may contribute to the
Bose--Einstein effects, and thus reduces the potential signal.)
This option should therefore be used with caution, and only as a
first approximation to what Bose--Einstein effects can mean.
 
\clearpage
 
\section{Particles and Their Decays}
 
Particles are the building blocks from which events are constructed.
We here use the word `particle' in its broadest sense, i.e. including
partons, resonances, hadrons, and so on, subgroups we will describe
in the following. Each particle is characterized by some quantities,
such as charge and mass. In addition, many of the particles are
unstable and subsequently decay. This section contains a survey of
the particle content of the programs, and the particle properties
assumed. In particular, the decay treatment is discussed.  Some
particle and decay properties form part already of the hard
subprocess description, and are therefore described in sections
\ref{s:JETSETproc}, \ref{s:PYTprocgen} and \ref{s:pytproc}.
 
\subsection{The Particle Content}
\label{ss:decpartcont}
 
In order to describe both current and potential future physics,
a number of different particles are needed. A list of some
particles, along with their codes, is given in section \ref{ss:codes}.
Here we therefore emphasize the generality rather than the details.
 
Four full generations of quarks and leptons are included in the
program, although indications from LEP strongly suggest that
only three exist in Nature. There is no standard terminology for the
fourth generation; we use $\lrm$ for the down type quark
($\lrm$ for low), $\hrm$ for the up type quark
($\hrm$ for high), $\chi$ for the lepton
and $\nu_{\chi}$ for the neutrino. Quarks may appear either singly
or in pairs; the latter are called diquarks and are characterized
by their flavour content and their spin. A diquark is always assumed
to be in a colour antitriplet state.
 
From the coloured quarks (and diquarks), the colour neutral hadrons
may be build up. Six full meson multiplets are included and two
baryon ones, see section \ref{ss:flavoursel}. In addition, 
$\K_{\mrm{S}}^0$ and $\K_{\mrm{L}}^0$ are considered as separate 
particles coming from the `decay' of $\K^0$ and $\br{\K}^0$ (or, 
occasionally, produced directly).
 
Other particles from the Standard Model include the gluon $\g$, the
photon $\gamma$, the intermediate gauge bosons $\Z^0$ and $\W^{\pm}$,
and the standard Higgs $\H^0$. Non-standard particles include
additional gauge bosons, $\Z'^0$ and $\W'^{\pm}$, additional
Higgs bosons $\H'^0$, $\A^0$ and $\H^{\pm}$, a leptoquark $\L_{\Q}$
and a horizontal gauge boson $\R^0$. It is also possible
to use the particle codes of the current fourth generation
fermions to represent excited quarks and leptons, $\q^*$ and
$\ell^*$.
 
From the point of view of usage inside the programs, particles
may be subdivided into three classes, partly overlapping.
\begin{Enumerate}
\item A parton is generically any object which may be found in the
wave function of the incoming beams, and may participate in initial-
or final-state showers. This includes what is normally meant by
partons, i.e. quarks and gluons, but here also leptons
and photons. In a few cases other particles may be
classified as partons in this sense.
\item A resonance is an unstable particle produced as part of the
hard process, and where the decay treatment normally is also part
of the hard process. Resonance partial widths are perturbatively
calculable, and therefore it is possible to dynamically recalculate
branching ratios as a function of the mass assigned to a resonance.
Resonances includes particles like the $\Z^0$ and other massive
gauge bosons and Higgs particles. It does not include hadrons
with non-vanishing width, like the $\rho$, which are just called
`unstable hadrons'.
\item Hadrons, i.e. mesons and baryons produced either in the
fragmentation process, in secondary decays or as part of the beam
remnant treatment, but not directly as part of the hard process
(except in a few special cases). Hadrons may be stable or unstable.
Branching ratios are not assumed perturbatively calculable, and can
therefore be set freely. Also leptons and photons produced in decays
belong to this class.
\end{Enumerate}
 
Usually the subdivision above is easy to understand and gives you
the control you would expect. However, the classification of top
and the fourth generation fermions may lead to some confusion, as
already mentioned, section \ref{sss:heavflavclass}. The problem is 
that the top did not use to be
treated as a resonance, but was rather allowed to fragment
to hadrons, which subsequently decayed. This approach was a reasonable
choice in the days when the top mass was assumed quite light by
today's standards. However, given current top limits, the fragmentation
and the decay of the top quark is being played out on comparable time
scales, and the treatment becomes much more difficult (see e.g. ref.
\cite{Sjo92a} for a toy model description). Starting at masses of
around 120 GeV the top decay time is so short that no top hadrons at
all are formed, and then a true resonance description is
appropriate, but still with some complications due to the net colour
charge of the top quark. Such an option is now default, wherein the top
quark is assumed to decay immediately, but intermediate scenarios can
not be modelled. The appearance of intermediate top hadrons
in the description has little influence on event shapes, even for a
very heavy top. 
 
\subsection{Masses, Widths and Lifetimes}
 
\subsubsection{Masses}
 
Quark masses are not particularly well defined. In the program it is
necessary to make use of two kinds of masses, current algebra ones
and constituent ones. The former are relevant for the kinematics in
hard processes (e.g. in $\g\g \to \c\cbar$) and for couplings to
Higgs particles, and therefore directly affect cross sections.
These values are the ones stored in the standard mass array
\ttt{PMAS}. Constituent masses are used to derive the masses of hadrons,
and are stored separately in the \ttt{PARF} array. We maintain
this distinction for the five first flavours, using the following
values by default: \\
\begin{tabular}{ccc@{\protect\rule{0mm}{\tablinsep}}}
quark & current algebra mass & constituent mass \\
d & 0.0099 GeV & 0.325 GeV \\
u & 0.0056 GeV & 0.325 GeV \\
s & 0.199 GeV & 0.5 GeV \\
c & 1.35 GeV & 1.6 GeV \\
b & 5.0 GeV & 5.0 GeV. \\
\end{tabular} \\
For top and fourth generation quarks the distinction is not as
important, so only one set of mass values is used, namely the one in
\ttt{PMAS}. The default top mass is 160 GeV. Constituent masses for
diquarks are defined as the sum of the respective quark masses. The
gluon is always assumed massless.
 
Particle masses, when known, are taken from ref. \cite{PDG92}.
Hypothesized particles, such as fourth generation fermions and
Higgs bosons, are assigned some not unreasonable set of default
values, in the sense of where you want to search for them in the
not too distant future. Here it is understood that you will go in
and change the default values according to your own opinions at
the beginning of a run.
 
The total number of hadrons in the program is very large, whereof
many are not yet discovered (like charm and bottom baryons) and
other may or may not exist (top and fourth generation hadrons).
In particular for the latter, it would be messy for the user to have
to recalculate the masses of hadron each time the assumed quark mass
is changed. Therefore the masses of yet undiscovered mesons and
baryons are built up, when needed, from the constituent masses.
For this purpose one uses formulae of the type
\cite{DeR75}
\begin{equation}
m = m_0 + \sum_i m_i + k \, m_{\d}^2 \sum_{i<j} \frac{\langle
\mbox{\boldmath $\sigma$}_i \cdot
\mbox{\boldmath $\sigma$}_j \rangle}{m_i \, m_j} ~,
\end{equation}
i.e. one constant term, a sum over constituent masses
and a spin-spin interaction term for each quark pair in the hadron.
The constants $m_0$ and $k$ are fitted from known masses,
treating mesons and baryons separately. For mesons with orbital
angular momentum $L=1$ the spin-spin coupling is assumed vanishing,
and only $m_0$ is fitted.
One may also define `constituent diquarks masses' using the formula
above, with a $k$ value $2/3$ that of baryons. The default values
are: \\
\begin{tabular}{ccc@{\protect\rule{0mm}{\tablinsep}}}
multiplet & $m_0$ & $k$ \\
pseudoscalars and vectors & 0. & 0.16 GeV \\
axial vectors ($S=0$) & 0.50 GeV & 0. \\
scalars & 0.45 GeV & 0. \\
axial vectors ($S=1$) & 0.55 GeV & 0. \\
tensors & 0.60 GeV & 0. \\
baryons  & 0.11 GeV & 0.048 GeV \\
diquarks & 0.077 GeV & 0.048 GeV.\\
\end{tabular} \\
There is one exception to the rule above, and that is flavour neutral
mesons, i.e. the onia states of a heavy quark--antiquark pair. These 
are defined individually, to allow more flexibility.
 
\subsubsection{Widths}
 
A width is calculated perturbatively for those resonances which
appear in the {\Py} hard process generation machinery. The width is
used to select masses in hard processes according to a relativistic
Breit--Wigner shape. In many processes the width is allowed to be
$\hat{s}$-dependent, see section \ref{ss:kinemreson}.
 
Other particle masses, as discussed so far, have been fixed at their
nominal value, i.e. with no mass broadening for short-lived particles
such as $\rho$, $\K^*$ or $\Delta$. Compared to the $\Z^0$, it is
much more difficult to describe the $\rho$ resonance shape, since
nonperturbative and threshold effects act to distort the na\"{\i}ve
shape. Thus the $\rho$ mass is limited from below by its decay
$\rho \to \pi\pi$, but also from above, e.g. in the decay
$\phi \to \rho \pi$. In some decay chains, several mass choices are
coupled, like in $\a_2 \to \rho \pi$, where also the $\a_2$ has a
non-negligible width. Finally, there are some extreme cases, like
the $\f_0$, which has a nominal mass below the $\K\K$ threshold, but
a tail extending beyond that threshold, and therefore a
non-negligible branching ratio to the $\K\K$ channel.
 
In view of examples like these, no attempt is made to provide a
full description. Instead a simplified description is used, which
should be enough to give the general smearing of events due to
mass broadening, but maybe not sufficient for detailed studies of
a specific resonance. By default, hadrons are therefore given a
mass distribution according to a non-relativistic Breit--Wigner
\begin{equation}
{\cal P}(m) \, \d m \propto \frac{1}{(m - m_0)^2 + \Gamma^2/4}
\, \d m  ~.
\label{dec:BWlin}
\end{equation}
Leptons and resonances not taken care of by the hard
process machinery are distributed according to a relativistic
Breit--Wigner
\begin{equation}
{\cal P}(m^2) \, \d m^2 \propto \frac{1}{(m^2 - m_0^2)^2 +
m_0^2 \Gamma^2} \, \d m^2 ~.
\label{dec:BWtwo}
\end{equation}
Here $m_0$ and $\Gamma$ are the nominal mass and width of the
particle. The Breit--Wigner shape is truncated symmetrically,
$|m - m_0| < \delta$, with $\delta$ arbitrarily chosen for each
particle so that no problems are encountered in the decay chains.
It is possible to switch off the mass broadening, or to use either
a non-relativistic or a relativistic Breit--Wigners everywhere.
 
The $\f_0$ problem has been `solved' by shifting the $\f_0$ mass to
be slightly above the $\K\K$ threshold and have vanishing width.
Then kinematics in decays $\f_0 \to \K\K$ is reasonably well
modelled. The $\f_0$ mass is too large in the
$\f_0 \to \pi\pi$ channel, but this does not really matter, since
one anyway is far above threshold here.
 
\subsubsection{Lifetimes}
 
Clearly the lifetime and the width of a particle are inversely
related. For practical applications, however, any particle with
a non-negligible width decays too close to its production vertex
for the lifetime to be of any interest. In the program, the two
aspects are therefore considered separately. Particles with a
non-vanishing nominal proper lifetime
$\tau_0 = \langle \tau \rangle$ are
assigned an actual lifetime according to
\begin{equation}
{\cal P}(\tau) \, \d \tau \propto \exp(- \tau / \tau_0 ) \, \d \tau ~,
\end{equation}
i.e. a simple exponential decay is assumed. Since the program
uses dimensions where the speed of light $c \equiv 1$, and
space dimensions are in mm, then actually the unit of $c \tau_0$ is
mm and of $\tau_0$ itself mm$/c \approx 3.33\times10^{-12}$ s.
 
If a particle is produced at a vertex $v = (\mbf{x}, t)$ with a
momentum $p = (\mbf{p}, E)$ and a lifetime $\tau$, the
decay vertex position is assumed to be
\begin{equation}
v' = v + \tau \, \frac{p}{m} ~,
\label{dec:newvertex}
\end{equation}
where $m$ is the mass of the particle. With the primary
interaction (normally) in the origin, it is therefore possible to
construct all secondary vertices in parallel with the ordinary
decay treatment.
 
The formula above does not take into account any detector effects,
such as a magnetic field. It is therefore possible to stop the
decay chains at some suitable point, and leave any subsequent
decay treatment to the detector simulation program. One may
select that particles are only allowed to decay if they have a
nominal lifetime $\tau_0$ shorter than some given value or,
alternatively, if their decay vertices $\mbf{x}'$ are inside some
spherical or cylindrical volume around the origin.
 
\subsection{Decays}
\label{ss:partdecays}
 
Several different kinds of decay treatment are used in the program,
depending on the nature of the decay. Not discussed here are the
decays of resonances which are handled as part of the hard process.
 
\subsubsection{Strong and electromagnetic decays}
 
The decays of hadrons containing the `ordinary' $\u$, $\d$ and $\s$
quarks into two or three particles are known, and branching ratios
may be found in \cite{PDG92}. We normally assume that the momentum
distributions are given by phase space. There are a few exceptions,
where the phase space is weighted by a matrix-element expression,
as follows.
 
In $\omega$ and $\phi$ decays to $\pi^+ \pi^- \pi^0$, a matrix element
of the form
\begin{equation}
|{\cal M}|^2 \propto | \mbf{p}_{\pi^+} \times \mbf{p}_{\pi^-} |^2
\label{dec:omegphi}
\end{equation}
is used, with the $\mbf{p}_{\pi}$ the pion momenta in the rest frame
of the decay. (Actually, what is coded is the somewhat more lengthy
Lorentz invariant form of the expression above.)
 
Consider the decay chain $P_0 \to P_1 + V \to P_1 + P_2 + P_3$,
with $P$ representing pseudoscalar mesons and $V$ a vector one. Here
the decay angular distribution of the $V$ in its rest frame is
\begin{equation}
|{\cal M}|^2 \propto \cos^2 \theta_{02} ~,
\label{dec:psvpsps}
\end{equation}
where $\theta_{02}$ is the angle between $P_0$ and $P_2$.
The classical example is $\D \to \K^* \pi \to \K \pi \pi$.
If the $P_1$ is replaced by a $\gamma$, the angular distribution in
the $V$ decay is instead $\propto \sin^2 \theta_{02}$. 
 
In Dalitz decays, $\pi^0$ or $\eta \to \e^+\e^- \gamma$, the mass $m^*$
of the $\ee$ pair is selected according to
\begin{equation}
{\cal P}(m^{*2}) \, \d m^{*2} \propto \frac{\d m^{*2}}{m^{*2}} \,
\left( 1 + \frac{2m_{\e}^2}{m^{*2}} \right) \,
\sqrt{ 1 - \frac{4m_{\e}^2}{m^{*2}} } \,
\left( 1 - \frac{m^{*2}}{m_{\pi,\eta}^2} \right)^3 \,
\frac{1}{ (m_{\rho}^2 - m^{*2})^2 + m_{\rho}^2 \Gamma_{\rho}^2 } ~.
\label{dec:Dalitz}
\end{equation}
The last factor, the VMD-inspired $\rho^0$ propagator, is negligible
for $\pi^0$ decay. Once the $m^*$ has been selected, the angular
distribution of the $\ee$ pair is given by
\begin{equation}
|{\cal M}|^2 \propto (m^{*2} - 2 m_{\e}^2) \left\{
(p_{\gamma} p_{\e^+})^2 + (p_{\gamma} p_{\e^-})^2 \right\} +
4m_{\e}^2 \left\{ (p_{\gamma} p_{\e^+}) (p_{\gamma} p_{\e^-}) +
(p_{\gamma} p_{\e^+})^2 + (p_{\gamma} p_{\e^-})^2 \right\} ~.
\end{equation}
 
Also a number of simple decays involving resonances of heavier
hadrons, e.g. $\Sigma_{\c}^0 \to \Lambda_{\c}^+ \pi^-$ or
$\B^{*-} \to \B^- \gamma$ are treated in the same way as the
other two-particle decays.
 
\subsubsection{Weak decays of charm hadrons}
 
The charm hadrons have a mass in an intermediate range, where the
effects of the na\"{\i}ve $V-A$ weak decay matrix element is partly but
not fully reflected in the kinematics of final-state particles.
Therefore different decay strategies ar combined. We start with
hadronic decays, and subseqently consider semileptonic ones.
 
For the four `main' charm hadrons, $\D^+$, $\D^0$, $\D_{\s}^+$ and
$\Lambda_{\c}^+$, a number of branching ratios are already known.
The known braching ratios have been combined with reasonable
guesses, to construct more or less complete tables of all channels.
For hadronic decays of $\D^0$ and $\D^+$, where rather much is known, 
all channels have an explicitly listed particle content. 
However, only for the two-body decays is resonance production 
properly taken into account. It means that the experimentally measured 
branching ratio for a $\K \pi \pi$ decay channel, say, is represented 
by contributions from a direct $\K \pi \pi$ channel as well as from
indirect ones, such as $\K^* \pi$ and $\K \rho$. For a channel like
$\K \pi \pi \pi$, on the other hand, only the $\K^* \rho$ appears 
separately, while the rest is lumped into one entry in the decay 
tables. This is more or less in agreement with the philosophy adopted
in the PDG tables \cite{PDG92}. For $\D_{\s}^+$ and $\Lambda_{\c}^+$ 
knowledge is rather incomplete, and only two-body decay channels are
listed. Final states with three or more hadron are only listed in
terms of a flavour content. 
 
The way the program works, it is important to include all the
allowed decay channels up to a given multiplicity. Channels with
multiplicity higher than this may then be generated according to
a simple flavour combination scheme. For instance, in a $\D_{\s}^+$
decay, the normal quark content is $\s\sbar\u\dbar$, where one
$\sbar$ is the spectator quark and the others come from the weak
decay of the $\c$ quark. The spectator quark may also be annihilated,
like in $\D_{\s}^+ \to \u\dbar$. The flavour content to make up
one or two hadrons is therefore present from the onset.
If one decides to generate more hadrons,
this means new flavour-antiflavour pairs have to be generated
and combined with the existing flavours. This is done using the
same flavour approach as in fragmentation.
 
In more detail, the following scheme is used.
\begin{Enumerate}
\item The multiplicity is first selected. The $\D_{\s}^+$ and 
$\Lambda_{\c}^+$ multiplicity is selected according to a distribution 
described further below. The program can also be asked to 
generate events of a predetermined multiplicity.
\item One of the non-spectator flavours is selected at random.
This flavour is allowed to `fragment' into a hadron plus a new
remaining flavour, using exactly the same flavour generation
algorithm as in the standard jet fragmentation, section
\ref{ss:flavoursel}.
\item Step 2 is iterated until only one or two hadrons remain to
be generated, depending on whether the original number of flavours
is two or four. In each step one `unpaired' flavour is replaced by
another one as a hadron is `peeled off', so the number of unpaired
flavours is preserved.
\item If there are two flavours, these are combined to form the last
hadron. If there are four, then one of the two possible pairings
into two final hadrons is selected at random. To find the hadron
species, the same flavour rules are used as when final flavours are 
combined in the joining of two jets.
\item If the sum of decay product masses is larger than the mass of
the decaying particle, the flavour selection is rejected and the
process is started over at step 1. Normally a new multiplicity is
picked, but for $\D^0$ and $\D^+$ the old multiplicity is retained.
\item Once an acceptable set of hadrons has been found, these are
distributed according to phase space.
\end{Enumerate}
The picture then is one of a number of partons moving apart,
fragmenting almost like jets, but with momenta so low that phase-space 
considerations are enough to give the average behaviour of
the momentum distribution. Like in jet fragmentation, endpoint
flavours are not likely to recombine with each other. Instead
new flavour pairs are created in between them. One should also note
that, while vector and pseudoscalar mesons are produced at their
ordinary relative rates, events with many vectors are likely to
fail in step 5. Effectively, there is therefore a shift towards
lighter particles, especially at large multiplicities.
 
When a multiplicity is to be picked, this is done according to a
Gaussian distribution, centered at $c + n_{\q}/4$ and with a
width $\sqrt{c}$, with the final number rounded off to the nearest
integer. The value for the number of quarks $n_{\q}$ is 2 or 4,
as described above, and
\begin{equation}
c = c_1 \, \ln \left( \frac{m - \sum m_{\q}}{c_2} \right) ~,
\label{dec:multsel}
\end{equation}
where $m$ is the hadron mass and $c_1$ and $c_2$ have been tuned
to give a reasonable description of multiplicities. There is always
some lower limit for the allowed multiplicity; if a number
smaller than this is picked the choice is repeated. Since two-body
decays are explicitly enumerated for $\D_{\s}^+$ and
$\Lambda_{\c}^+$, there the mimimum multiplicity  is three.
 
Semileptonic branching ratios are explicitly given in the program
for all the four particles discussed here, i.e. it is never
necessary to generate the flavour content using the fragmentation
description. This does not mean that all branching ratios are known;
a fair amount of guesswork is involved for
the channels with higher multiplicities, based
on a knowledge of the inclusive semileptonic branching ratio and
the exclusive branching ratios for low multiplicities.
 
In semileptonic decays it is not appropriate to distribute the
lepton and neutrino momenta according to phase space. Instead the
simple $V-A$
matrix element is used, in the limit that decay product masses may
be neglected and that quark momenta can be replaced by hadron
momenta. Specifically, in the decay $H \to \ell^+ \nu_{\ell} h$,
where $H$ is a charm hadron and $h$ and ordinary hadron, the matrix
element
\begin{equation}
|{\cal M}|^2 = (p_H p_{\ell}) (p_{\nu} p_h)
\end{equation}
is used to distribute the products. It is not clear how to
generalize this formula when several hadrons are present in the final
state. In the program, the same matrix element is used as above,
with $p_h$ replaced by the total four-momentum of all the hadrons.
This tends to favour a low invariant mass for the hadronic system
compared with na\"{\i}ve phase space.

There are a few charm hadrons, such as $\Xi_c$ and $\Omega_c$, which
decay weakly but are so rare that little is known about them. For
these a simplified generic charm decay treatment is used. For
hadronic decays only the quark content is given, and then a
multiplicity and a flavour composition is picked at random, as
already described. Semileptonic decays are assumed to produce only
one hadron, so that $V-A$ matrix element can be simply applied.
 
\subsubsection{Weak decays of the $\tau$ lepton}
 
For the $\tau$ lepton, an explicit list of decay channels has been
put together, which includes channels with up to five final-state
particles, some of which may be unstable and subsequently decay to
produce even larger total multiplicities. Because of the well-known
`$\tau$ puzzle', i.e. that experimentally the sum of branching
ratios for exclusive one-prong decays is lower than the inclusive
one-prong branching ratio, such a table cannot be constructed in
full agreement with the PDG data. (The problem is nowadays less severe 
than it used to be, but still not fully resolved.)
 
The leptonic decays $\tau^- \to \nu_{\tau} \ell^- \br{\nu}_{\ell}$,
where $\ell$ is $\e$ or $\mu$, are distributed according to the
standard $V-A$ matrix element
\begin{equation}
|{\cal M}|^2 = (p_{\tau} p_{\br{\nu}_{\ell}})
(p_{\ell} p_{\nu_{\tau}}) ~.
\end{equation}
(The corresponding matrix element is also used in $\mu$ decays, but
normally the $\mu$ is assumed stable.)
 
In $\tau$ decays to hadrons, the hadrons and the $\nu_{\tau}$ are
distributed according to phase space times the factor
$x_{\nu} \, (3 - x_{\nu})$, where $x_{\nu} = 2E_{\nu}/m_{\tau}$
in the rest frame of the $\tau$. The latter factor is the
$\nu_{\tau}$ spectrum predicted by the parton level $V-A$ matrix
element, and therefore represents an attempt to take into account
that the $\nu_{\tau}$ should take a larger momentum fraction than
given by phase space alone.
 
The probably largest shortcoming of the $\tau$ decay treatment is
that no polarization effects are included, i.e. the $\tau$ is
always assumed to decay isotropically. Usually this is not correct,
since a $\tau$ is produced polarized in $\Z^0$ and $\W^{\pm}$ decays.
The \ttt{LUTAUD} routine provides a generic interface to an external
$\tau$ decay library, where such effects could be handled (see also
\ttt{MSTJ(28)}). 
 
\subsubsection{Weak decays of bottom hadrons}
 
Some exclusive branching ratios now start to be known for $\B$
decays. In this version, the $\B^0$, $\B^+$, $\B_{\s}^0$ and  
$\Lambda_{\b}^0$ therefore appear in a similar vein to the one
outlined above for $\D_{\s}^+$ and $\Lambda_{\c}^+$ above. That 
is, all leptonic channels and all hadronic two-body decay channels 
are explicitly listed, while hadronic channels with three or more 
particles are only given in terms of a quark content. The $\B_{\c}$
is exceptional, in that either the bottom or the charm quark may
decay first, and in that annihilation graphs may be non-negligible.
Leptonic and semileptonic channels are here given in full, while 
hadronic channels are only listed in terms of a quark content,
with a relative composition as given in \cite{Lus91}. No separate
branching ratios are set for any of the other weakly decaying
bottom hadrons, but instead a pure `spectator quark' model is assumed, 
where the decay of the $\b$ quark is the same in all hadrons and the 
only difference in final flavour content comes from the spectator quark. 
Compared to the charm decays, the weak decay matrix elements are given
somewhat larger importance in the hadronic decay channels.
 
In semileptonic decays $\b \to \c \ell^- \br{\nu}_{\ell}$ the $\c$
quark is combined with the spectator antiquark or diquark to form
one single hadron. This hadron may be either a pseudoscalar, a vector
or a higher resonance (tensor etc.). The relative fraction of the
higher resonances has been picked to be about 30\%, in order to give
a leptonic spectrum in reasonable experiment with data. (This only 
applies to the main particles  $\B^0$, $\B^+$, $\B_{\s}^0$ and  
$\Lambda_{\b}^0$; for the rest the choice is according to the standard 
composition in the fragmentation.) The overall process is therefore
$H \to h \ell^- \br{\nu}_{\ell}$, where $H$ is a bottom antimeson
or a bottom baryon (remember that $\br{\B}$ is the one that contains
a $\b$ quark), and the matrix element used to distribute momenta is
\begin{equation}
|{\cal M}|^2 = (p_H p_{\nu}) (p_{\ell} p_h)  ~.
\end{equation}
Again decay product masses have been neglected in the matrix element,
but in the branching ratios the $\tau^- \br{\nu}_{\tau}$ channel has
been reduced in rate, compared with $\e^- \br{\nu}_{\e}$ and
$\mu^- \br{\nu}_{\mu}$ ones, according to the expected mass effects.
No CKM-suppressed decays $\b \to \u \ell^- \br{\nu}_{\ell}$ are
currently included.
 
In most multibody hadronic decays, e.g. 
$\b \to \c \d \ubar$, the $\c$ quark is
again combined with the spectator flavour to form one single hadron,
and thereafter the hadron and the two quark momenta are distributed
according to the same matrix element as above, with
$\ell^- \leftrightarrow \d$ and
$\br{\nu}_{\ell} \leftrightarrow \ubar$.
The invariant mass of the two quarks is calculated next. If this mass
is so low that two hadrons cannot be formed from the system, the
two quarks are combined into one single hadron. Else the same kind of
approach as in hadronic charm decays is adopted, wherein a
multiplicity  is selected, a number of hadrons are formed and
thereafter momenta are distributed according to phase space. The
difference is that here the charm decay product is distributed according
to the $V-A$ matrix element, and only the rest of the system is
assumed isotropic in its rest frame, while in charm decays all hadrons
are distributed isotropically.
 
Note that the $\c$ quark and the spectator are assumed to form
one colour singlet and the $\d \ubar$ another, separate one.
It is thus assumed that the original colour assignments of the
basic hard process are better retained than in charm decays.
However, sometimes this will not be true, and with about 20\%
probability the colour assignment is flipped around so that
$\c \ubar$ forms one singlet. (In the program, this is achieved by
changing the order in which decay products are given.) In particular,
the decay $\b \to \c \s \cbar$ is allowed to give a $\c\cbar$
colour-singlet state part of the time, and this state may collapse
to a single $\Jpsi$. Two-body decays of this type are explicitly 
listed for $\B^0$, $\B^+$, $\B_{\s}^0$ and $\Lambda_{\b}^0$;
while other $\Jpsi$ production channels appear from the flavour
content specification.
 
The $\B^0$--$\br{\B}^0$ and $\B_{\s}^0$--$\br{\B}_{\s}^0$ systems
mix before decay. This is optionally included. With a probability
\begin{equation}
{\cal P}_{\mrm{flip}} = \sin^2 \left( \frac{x \, \tau}
{ 2\, \langle \tau \rangle} \right)
\end{equation}
a $\B$ is therefore allowed to decay like a $\br{\B}$, and vice versa.
The mixing parameters are by default $x_{\d} = 0.7$ in the
$\B^0$--$\br{\B}^0$ system and $x_{\s} = 10$ in the
$\B_{\s}^0$--$\br{\B}_{\s}^0$ one.

The generic $\B$ meson and baryon decay properties are stored for 
`particle' 85. This particle contains a description of the free $\b$ 
quark decay, with an instruction to find the spectator flavour 
according to the particle code of the actual decaying hadron. 
Currently baryons other than $\Lambda_{\b}^0$ are treated this 
way. If so desired, each hadron could be given a separate decay 
channel list, or all $\B$ hadrons could be mapped to particle 85,
as used to be the case..
 
\subsubsection{Weak decays of top and fourth generation}
 
As already explained in section \ref{ss:decpartcont}, 
heavy quarks are normally assumed to decay before they fragment. 
Optionally, they may be allowed to fragment before they decay.
In either case, the decay itself is handled as
if the heavy flavour is free. 

For a hadron, some of the hadron energy is reserved for the spectator
quark. The decay matrix element used for $\Q \to \q \fbar \f$ is
\begin{equation}
|{\cal M}|^2 \propto \frac{ (p_{\Q} p_{\fbar}) (p_{\f} p_{\q})}
{\left( (p_{\f} + p_{\fbar})^2 - m_{\W}^2 \right)^2 +
m_{\W}^2 \Gamma_{\W}^2} ~.
\end{equation}
Here $\Q$ may represent the $\t$ or any of the fourth generation
quarks, $\lrm$ and $\hrm$. With trivial change of
notation, the lepton $\chi$ obeys the same formula. The $\f\fbar$
pair are the fermions from the $\W$ decay, either quarks or leptons.
The program takes care of the effects of the $\W$ propagator,
whatever the mass difference $m_{\Q} - m_{\q}$, with one proviso:
the selection of the $\q$ flavour is done according to fixed branching
ratios, and does thus not take into account the relative enhancement
of a CKM-suppressed $\q$ due to mass effects. This would play a
r\^ole around thresholds, e.g., with $m_{\t} \approx m_{\W}$, the
$\t \to \s$ would be enhanced compared with $\t \to \b$.
On the other hand, threshold factors are included for the
choice of the $\f\fbar$ fermion pair from the $\W$ decay.

For the alternative with a rapidly decaying top quark, so that no 
hadron is formed, one is not close to threshold. The composition of
the light flavour produced in the decay is then calculated according
to the respective phase space times CKM weight. By default the $\W$
decays with the spin information implicit in the matrix element above,
but isotropic $\W$ decay is an option.

The $\b$ quark produced in the decay $\t \to \b\W^+$ may be allowed 
to radiate. It thereby acquires an effective mass, which means that
the kinematics of the decay is changed, with energy shuffled from the
$\W$ to the $\b$. 
 
The system containing the spectator quark will often have a mass
too small to allow it to fragment like a jet system. In these cases
a single particle is formed from the flavour content, with a momentum
vector given by the sum of the two quark momenta. Since the energy of
this particle then will come out wrong, the momenta of the other jets
or leptons in the decay are modified slightly to obtain total energy
conservation. (Of course, for $\chi$ decay, there is no spectator and
thus no treatment of this kind.)
 
The $\f \fbar$ pair from the $\W$ decay is allowed to shower, i.e.
emit gluons and photons according to the standard final-state
radiation algorithm, including matching to first-order matrix
elements. The resulting jet system is fragmented with
ordinary string fragmentation --- the mass is here so high that
a fragmentation description is quite appropriate. Only very rarely
would the $\W$ mass be below the threshold for the production of a
pair of particles; such kinematical configurations are rejected.
 
\subsubsection{Other decays}
 
For onia spin 1 resonances, decay channels into a pair of leptons
are explicitly given. Hadronic decays of the $\Jpsi$ are simulated
using the flavour generation model introduced for charm. For
$\Upsilon$ a fraction of the hadronic decays is into $\q\qbar$
pairs, while the rest is into $\g\g\g$ or $\g\g\gamma$, using the
matrix elements of eq.~(\ref{ee:Upsilondec}). The $\eta_c$ and
$\eta_b$ are both allowed to decay into a $\g\g$ pair, which then
subsequently fragments. In $\Upsilon$ and $\eta_b$ decays the partons
are allowed to shower before fragmentation, but energies are too low
for showering to have any impact.
 
With current bounds on the top mass, one does not expect the
formation of well-defined toponium states. A complete description of 
the resonance structure in the threshold region is beyond the scope
of the program. The approach taken for the toponium states that
have been defined is to let either the $\t$ or the $\tbar$ decay
weakly first, then do the fragmentation, and subsequently let the
produced antitop or top hadron decay. A better description is provided
by the {\Py} machinery for resonance decays.
 
Default branching ratios are given for resonances like the $\Z^0$,
the $\W^{\pm}$ or the $\H^0$. When {\Py} is initialized, these
numbers are replaced by branching ratios evaluated from the given
masses. For $\Z^0$ and $\W^{\pm}$ the branching ratios depend only
marginally on the masses assumed, while effects are large e.g. for
the $\H^0$. In fact, branching ratios may vary over the
Breit--Wigner resonance shape, something which is also taken into
account in {\Py}. Therefore the default resonance treatment of {\Je}
is normally not so useful, and should be avoided (except, of course,
the standard $\ee \to \gammaZ \to \q\qbar$ description). When it is
used, a channel is selected according to the given fixed branching
ratios. If the decay is into a $\q\qbar$ pair, the quarks are allowed
to shower and subsequently the parton system is fragmented.
 
\clearpage
 
\section{The JETSET Program Elements}

In this section we collect information on most of the routines and 
common block variables found in {\Je}. A few parts are discussed 
elsewhere; this includes the $\ee$ routines, parton showers
and event-analysis routines. In this section
the emphasis is on the fragmentation and decay package, and on 
generic utilities for things like event listing.
 
\subsection{Definition of Initial Configuration or Variables}
 
With the use of the conventions described for the event record, it
is possible to specify any initial jet/particle configuration. This
task is simplified for a number of often occuring situations by
the existence of the filling routines below. It should be noted that 
many users do not come in direct contact with these routines, since 
that is taken care of by higher-level routines for specific 
processes, particularly \ttt{LUEEVT} and \ttt{PYEVNT}.

Several calls to the routines can be combined in the specification. 
In case one call is enough, the complete
fragmentation/decay chain may be simulated at the same time. At each
call, the value of \ttt{N} is updated to the last line used for
information in the call, so if several calls are used, they should be
made with increasing \ttt{IP} number, or else \ttt{N} should be
redefined by hand afterwards.  

The routine \ttt{LUJOIN} is very useful
to define the colour flow in more complicated parton configurations;
thereby one can bypass the not so trivial rules for how to set the
\ttt{K(I,4)} and \ttt{K(I,5)} colour-flow information. 

As an experiment, the routine \ttt{LUGIVE} contains a facility to set
various comonblock variables in a controlled and documented fashion.
 
\drawbox{CALL LU1ENT(IP,KF,PE,THE,PHI)}\label{p:LU1ENT}
\begin{entry}
\itemc{Purpose:} to add one entry to the event record, i.e. either
a jet or a particle.
\iteme{IP :} normally line number for the jet/particle. There are two
exceptions. \\
If \ttt{IP=0}, line number 1 is used and \ttt{LUEXEC} is called. \\
If \ttt{IP<0}, line \ttt{-IP} is used, with status code 
\ttt{K(-IP,2)=2} rather than 1; thus a jet system may be built up by 
filling all but the last jet of the system with \ttt{IP<0}.
\iteme{KF :} jet/particle flavour code.
\iteme{PE :} jet/particle energy. If \ttt{PE} is smaller than the mass,
the jet/particle is taken to be at rest.
\iteme{THE, PHI :} polar and azimuthal angle for the momentum vector
of the jet/particle.
\end{entry}
 
\drawbox{CALL LU2ENT(IP,KF1,KF2,PECM)}\label{p:LU2ENT}
\begin{entry}
\itemc{Purpose:} to add two entries to the event record, i.e. either a
2-jet system or two separate particles.
\iteme{IP :} normally line number for the first jet/particle, with
second in line \ttt{IP+1}. There are two exceptions. \\
If \ttt{IP=0}, lines 1 and 2 are used and \ttt{LUEXEC} is called. \\
If \ttt{IP<0}, lines \ttt{-IP} and \ttt{-IP+1} are used, with status
code \ttt{K(I,1)=3}, i.e. with special colour connection information,
so that a parton shower can be generated by a \ttt{LUSHOW} call,
followed by a \ttt{LUEXEC} call, if so desired (only relevant for jets).
\iteme{KF1, KF2 :} flavour codes for the two jets/particles.
\iteme{PECM :} ($=E_{\mrm{cm}}$) the total energy of the system.
\itemc{Remark:} the system is given in the c.m. frame, with the
first jet/particle going out in the $+z$ direction.
\end{entry}
 
\drawbox{CALL LU3ENT(IP,KF1,KF2,KF3,PECM,X1,X3)}\label{p:LU3ENT}
\begin{entry}
\itemc{Purpose:} to add three entries to the event record, i.e. either
a 3-jet system or three separate particles.
\iteme{IP :} normally line number for the first jet/particle, with
other two in \ttt{IP+1} and \ttt{IP+2}. There are two exceptions. \\
If \ttt{IP=0}, lines 1, 2 and 3 are used and \ttt{LUEXEC} is
called. \\
If \ttt{IP<0}, lines \ttt{-IP} through \ttt{-IP+2} are used, with
status code \ttt{K(I,1)=3}, i.e. with special colour connection
information, so that a parton shower can be generated by a \ttt{LUSHOW}
call, followed by a \ttt{LUEXEC} call, if so desired (only relevant
for jets).
\iteme{KF1, KF2, KF3:} flavour codes for the three jets/particles.
\iteme{PECM :} ($E_{\mrm{cm}}$) the total energy of the system.
\iteme{X1, X3 :} $x_i = 2E_i/E_{\mrm{cm}}$, i.e. twice the energy 
fraction taken by the $i$'th jet. Thus $x_2 = 2 - x_1 - x_3$, and 
need not be given. Note that not all combinations of $x_i$ are 
inside the physically allowed region.
\itemc{Remark :} the system is given in the c.m. frame, in the
$xz$-plane, with the first jet going out in the $+z$ direction and the
third one having $p_x > 0$.
\end{entry}
 
\drawbox{CALL LU4ENT(IP,KF1,KF2,KF3,KF4,PECM,X1,X2,X4,X12,X14)}%
\label{p:LU4ENT}
\begin{entry}
\itemc{Purpose:} to add four entries to the event record, i.e. either a
4-jet system or four separate particles (or, for $\q\qbar\q'\qbar'$
events, two 2-jet systems).
\iteme{IP :} normally line number for the first jet/particle, with
other three in lines \ttt{IP+1}, \ttt{IP+2} and \ttt{IP+3}.
There are two exceptions. \\
If \ttt{IP=0}, lines 1, 2, 3 and 4 are used and \ttt{LUEXEC} is
called. \\
If \ttt{IP<0}, lines \ttt{-IP} through \ttt{-IP+3} are used, with
status code \ttt{K(I,1)=3}, i.e. with special colour connection
information, so that a parton shower can be generated by a
\ttt{LUSHOW} call, followed by a \ttt{LUEXEC} call, if so desired
(only relevant for jets).
\iteme{KF1,KF2,KF3,KF4 :} flavour codes for the four jets/particles.
\iteme{PECM :} ($=E_{\mrm{cm}}$) the total energy of the system.
\iteme{X1,X2,X4 :} $x_i = 2E_i/E_{\mrm{cm}}$, i.e. twice the energy
fraction taken by the $i$'th jet. Thus $x_3 = 2 - x_1 - x_2 - x_4$,
and need not be given.
\iteme{X12,X14 :} $x_{ij} = 2 p_i p_j/E_{\mrm{cm}}^2$, i.e. twice the
four-vector product of the momenta for jets $i$ and $j$, properly
normalized. With the masses known, other $x_{ij}$ may be constructed
from the $x_i$ and $x_{ij}$ given. Note that not all combinations of
$x_i$ and $x_{ij}$ are inside the physically allowed region.
\itemc{Remark:} the system is given in the c.m. frame, with the first
jet going out in the $+z$ direction and the fourth jet lying in the
$xz$-plane with $p_x > 0$. The second jet will have $p_y > 0$ and
$p_y < 0$ with equal probability, with the third jet balancing this
$p_y$ (this corresponds to a random choice between the two possible
stereoisomers).
\end{entry}
 
\drawbox{CALL LUJOIN(NJOIN,IJOIN)}\label{p:LUJOIN}
\begin{entry}
\itemc{Purpose:} to connect a number of previously defined partons
into a string configuration. Initially the partons must be given with
status codes \ttt{K(I,1)=} 1, 2 or 3. Afterwards the partons
all have status code 3, i.e. are given with full colour-flow
information. Compared to the normal way of defining a parton
system, the partons need therefore not appear in the same sequence
in the event record as they are assumed to do along the string. It
is also possible to call \ttt{LUSHOW} for all or some of the
entries making up the string formed by \ttt{LUJOIN}.
\iteme{NJOIN:} the number of entries that are to be joined by one
string.
\iteme{IJOIN:} an one-dimensional array, of size at least \ttt{NJOIN}.
The \ttt{NJOIN} first numbers are the positions of the partons that
are to be joined, given in the order the partons are assumed to appear
along the string. If the system consists entirely of gluons,
the string is closed by connecting back the last to the first
entry.
\itemc{Remarks:} only one string (i.e. one colour singlet) may be
defined per call, but one is at liberty to use any number of
\ttt{LUJOIN} calls for a given event. The program will check that the
parton configuration specified makes sense, and not take any action
unless it does. Note, however, that an initially sensible parton
configuration may become nonsensical, if only some of the partons
are reconnected, while the others are left unchanged.
\end{entry}
 
\drawbox{CALL LUGIVE(CHIN)}\label{p:LUGIVE}
\begin{entry}
\itemc{Purpose:} to set the value of any variable residing in the
commmonblocks \ttt{LUJETS}, \ttt{LUDAT1}, \ttt{LUDAT2}, \ttt{LUDAT3},
\ttt{LUDAT4}, \ttt{LUDATR}, \ttt{PYSUBS}, \ttt{PYPARS}, \ttt{PYINT1},
\ttt{PYINT2}, \ttt{PYINT3}, \ttt{PYINT4}, \ttt{PYINT5}, \ttt{PYINT6},
or \ttt{PYINT7}. This is done in a more controlled fashion than by
directly including the common blocks in the user program, in that
array bounds are checked and the old and new values for the variable
changed are written to the output for reference.
\iteme{CHIN :} character expression of length at most 100 characters,
with requests for variables to be changed, stored in the form \\
\ttt{variable1=value1;variable2=value2;variable3=value3}\ldots~. \\
Note that an arbitrary number of instructions can be stored in
one call if separated by semicolons, and that blanks may be included
anyplace. The variable$_i$ may be any single variable in the {\JePy}
common blocks, and the value$_i$ must be of the correct integer, real
or character (without extra quotes) type. Array indices and values
must be given explicitly, i.e. cannot be variables in their own
right. The exception is that the first index can be preceded by
a \ttt{C}, signifying that the index should be translated from normal
KF to compressed KC code with a \ttt{LUCOMP} call; this is allowed for
the \ttt{KCHG}, \ttt{PMAS}, \ttt{MDCY} and \ttt{CHAF} arrays. If a
value$_i$ is omitted, i.e. with the construction \ttt{variable=},
the current value is written to the output, but the variable itself
is not changed.
\itemc{Remark :} The checks on array bounds are hardwired into this
routine. Therefore, if some user changes array dimensions and
\ttt{MSTU(3)}, \ttt{MSTU(6)} and/or \ttt{MSTU(7)}, as allowed by
other considerations, these changes will not be known to \ttt{LUGIVE}.
Normally this should not be a problem, however.
\end{entry}
 
\subsection{The JETSET Physics Routines}
\label{ss:JETphysrout}
 
The physics routines form the major part of {\Je}, but once
the initial jet/particle configuration has been specified and
default parameter values changed, if so desired, only a \ttt{LUEXEC}
call is necessary to simulate the whole fragmentation and decay chain.
Therefore a normal user will not directly see any of the other routines
in this section. Some of them could be called directly, but the danger
of faulty usage is then non-negligible.

The \ttt{LUTAUD} routine provides an optional interface to an external 
$\tau$ decay library, where polarization effects could be included. 
It is up to the user to write the appropriate calls, as explained at 
the end of this section. 
 
\drawbox{CALL LUEXEC}\label{p:LUEXEC}
\begin{entry}
\itemc{Purpose:} to administrate the fragmentation and decay chain.
\ttt{LUEXEC} may be called several times, but only entries which have
not yet been treated (more precisely, which have 
$1 \leq$\ttt{K(I,1)}$\leq 10$)
can be affected by further calls. This may apply if more jets/particles
have been added by the user, or if particles previously considered
stable are now allowed to decay. The actions that will be taken during
a \ttt{LUEXEC} call can be tailored extensively via the
\ttt{LUDAT1}--\ttt{LUDAT3} common blocks, in particular by setting the
\ttt{MSTJ} values suitably.
\end{entry}
 
\boxsep
\begin{entry}
\iteme{SUBROUTINE LUPREP(IP) :}\label{p:LUPREP}
to rearrange parton shower end products (marked with \ttt{K(I,1)=3})
sequentially along strings; also to (optionally) allow small jet
systems to collapse into two particles or one only, in the latter
case with energy and momentum to be shuffled elsewhere in the event;
also to perform checks that e.g. flavours of colour-singlet systems
make sense.
 
\iteme{SUBROUTINE LUSTRF(IP) :}\label{p:LUSTRF}
to generate the fragmentation of an arbitrary colour-singlet
jet system according to the Lund string fragmentation model. In many
respects, this routine is the very heart and soul of {\Je}.
 
\iteme{SUBROUTINE LUINDF(IP) :}\label{p:LUINDF}
to handle the fragmentation of a jet system according to
independent fragmentation models, and implement energy, momentum
and flavour conservation, if so desired. Also the fragmentation of
a single jet, not belonging to a jet system, is considered here
(this is of course physical nonsense, but may sometimes be
convenient for specific tasks).
 
\iteme{SUBROUTINE LUDECY(IP) :}\label{p:LUDECY}
to perform a particle decay, according to known
branching ratios or different kinds of models, depending on our level
of knowledge. Various matrix elements are included for specific
processes.
 
\iteme{SUBROUTINE LUKFDI(KFL1,KFL2,KFL3,KF) :}\label{p:LUKFDI}
to generate a new quark or diquark flavour and to
combine it with an existing flavour to give a hadron.
\begin{subentry}
\iteme{KFL1:} incoming flavour.
\iteme{KFL2:} extra incoming flavour, e.g. for formation of final
particle, where the flavours are completely specified. Is normally 0.
\iteme{KFL3:} newly created flavour; is 0 if \ttt{KFL2} is non-zero.
\iteme{KF:} produced hadron. Is 0 if something went wrong (e.g.
inconsistent combination of incoming flavours).
\end{subentry}
 
\iteme{SUBROUTINE LUPTDI(KFL,PX,PY) :}\label{p:LUPTDI}
to give transverse momentum, e.g. for a $\q\qbar$ pair created in the
colour field, according to independent Gaussian distributions in
$p_x$ and $p_y$.
 
\iteme{SUBROUTINE LUZDIS(KFL1,KFL3,PR,Z) :}\label{p:LUZDIS}
to generate the longitudinal scaling variable $z$ in jet
fragmentation, either according to the Lund symmetric fragmentation
function, or according to a choice of other shapes.
 
\iteme{SUBROUTINE LUBOEI :}\label{p:LUBOEI}
to include Bose--Einstein effects according to a simple
para\-metri\-zation. By default, this routine is not called. 
If called from \ttt{LUEXEC}, this is done after the decay of 
short-lived resonances, but before the decay of long-lived ones.
This means the routine should never be called directly by you,
nor would effects be correctly simulated if decays are switched off.
See \ttt{MSTJ(51) - MSTJ(52)} for switching on the routine.
 
\iteme{FUNCTION ULMASS(KF) :}\label{p:ULMASS}
to give the mass for a parton/particle.
 
\iteme{SUBROUTINE LUNAME(KF,CHAU) :}\label{p:LUNAME}
to give the parton/particle name (as a string of type
\ttt{CHARACTER CHAU*16}).
 
\iteme{FUNCTION LUCHGE(KF) :}\label{p:LUCHGE}
to give three times the charge for a parton/particle.
 
\iteme{FUNCTION LUCOMP(KF) :}\label{p:LUCOMP}
to give the compressed parton/particle code KC for a given
KF code, as required to find entry into mass and decay data tables.
Also checks whether the given KF code is actually an allowed one
(i.e. known by the program), and returns 0 if not. Note that KF
may be positive or negative, while the resulting KC code is never
negative.
 
\iteme{SUBROUTINE LUERRM(MERR,MESSAG) :}\label{p:LUERRM}
to keep track of the number of errors and warnings encountered,
write out information on them, and abort the program in case of too
many errors.
 
\iteme{FUNCTION ULANGL(X,Y) :}\label{p:ULANGL}
to calculate the angle from the $x$ and $y$ coordinates.

\iteme{SUBROUTINE LULOGO :}\label{p:LULOGO}
to write a titlepage for the {\JePy} programs. Called by LULIST(0).
 
\iteme{BLOCK DATA LUDATA :}\label{p:LUDATA}
to give default values for variables in the \ttt{LUDAT1}, \ttt{LUDAT2},
\ttt{LUDAT3}, \ttt{LUDAT4} and \ttt{LUDATR} common blocks.
 
\end{entry}
 
\drawbox{CALL LUTAUD(ITAU,IORIG,KFORIG,NDECAY)}\label{p:LUTAUD}
\begin{entry}
\itemc{Purpose:} to  act as an interface between the standard decay 
routine \ttt{LUDECY} and a user-supplied $\tau$ lepton decay library. 
The latter library would normally know how to handle polarized $\tau$'s, 
given the $\tau$ polarization as input, so one task of the interface 
routine is to construct the $\tau$ polarization/helicity from the 
information available. Input to the routine (from \ttt{LUDECY}) is 
provided in the first three arguments, while the last argument and 
some event record information have to be set before return. 
To use this facility you have to set the switch \ttt{MSTJ(28)},
include your own interface routine \ttt{LUTAUD} and see to it that 
the dummy routine \ttt{LUTAUD} in {\Je} is not linked. The dummy 
routine is there only to avoid unresolved external references when 
no user-supplied interface is linked.
\iteme{ITAU :} line number in the event record where the $\tau$ is 
stored. The four-momentum of this $\tau$ has first been boosted back 
to the rest frame of the decaying mother and thereafter rotated to move 
out along the $+z$ axis. It would have been possible to also perform
a final boost to the rest frame of the $\tau$ itself, but this has been
avoided so as not to suppress the kinematics aspect of 
close-to-threshold production (e.g. in $\B$ decays) vs. 
high-energy production (e.g. in real $\W$ decays). The choice of frame 
should help the calculation of the helicity configuration. After the 
\ttt{LUTAUD} call the $\tau$ and its decay products will automatically
be rotated and boosted back. However, seemingly, the event record does 
not conserve momentum at this intermediate stage.  
\iteme{IORIG :} line number where the mother particle to the $\tau$ 
is stored. Is 0 if the mother is not stored. This does not have 
to mean the mother is unknown. For instance, in semileptonic $\B$ decays 
the mother is a $\W^{\pm}$ with known four-momentum
$p_{\W} = p_{\tau} + p_{\nu_{\tau}}$, but there is no $\W$ line in the 
event record. When several copies of the mother is stored (e.g. one in
the documentation section of the event record and one in the main
section), \ttt{IORIG} points to the last. If a branchings like 
$\tau \to \tau\gamma$ occurs, the `grandmother' is given, i.e. the
mother of the direct $\tau$ before branching.
\iteme{KFORIG :} flavour code for the mother particle. Is 0 if the 
mother is unknown. The mother would typically be a resonance such as 
$\gammaZ$ (23), $\W^{\pm}$ ($\pm 24$), $\H^0$ (25), or $\H^{\pm}$ 
($\pm 37$). Often the helicity choice would be clear just by the 
knowledge of this mother species, e.g., $\W^{\pm}$ vs. $\H^{\pm}$. 
However, sometimes further complications may exist. For instance, 
the KF code 23 represents a mixture of $\gamma^*$ and $\Z^0$; a 
knowledge of the mother mass (in \ttt{P(IORIG,5)}) would here be 
required to make the choice of helicities. Further, a $\W^{\pm}$ or 
$\Z^0$ may either be (predominantly) transverse or longitudinal, 
depending on the production process under study.
\iteme{NDECAY :} the number of decay products of the $\tau$; to be 
given by the user. You must also store the KF flavour codes of those 
decay products in the positions \ttt{K(I,2)}, 
\ttt{N+1}$\leq$\ttt{I}$\leq$\ttt{N+NDECAY}, of the event record. 
The corresponding five-momentum (momentum, energy and mass) should be 
stored in the associated \ttt{P(I,J)} positions, 1$\leq$\ttt{J}$\leq$5. 
The four-momenta are expected to add up to the four-momentum of the 
$\tau$ in position \ttt{ITAU}. You should not change the \ttt{N} value 
or any of the other \ttt{K} or \ttt{V} values (neither for the 
$\tau$ nor for its decay products) since this is automatically done 
in \ttt{LUDECY}.  
\end{entry}
 
\subsection{Event Study and Data Listing Routines}
 
After an \ttt{LUEXEC} call, the event generated is stored in the
\ttt{LUJETS} common block, and whatever physical variable is desired
may be constructed from this record. An event may be rotated, boosted
or listed, and particle data may be listed or modified. Via the
functions \ttt{KLU} and \ttt{PLU} the values of some frequently
appearing variables may be obtained more easily. As described in
section \ref{ss:evanrout}, also more detailed event shape analyses
may be performed simply.
 
\drawbox{CALL LUROBO(THE,PHI,BEX,BEY,BEZ)}\label{p:LUROBO}
\begin{entry}
\itemc{Purpose:} to perform rotations and Lorentz boosts (in that
order, if both in the same call) of jet/particle momenta and vertex
position variables.
\iteme{THE, PHI :} standard polar coordinates $\theta, \varphi$,
giving the rotated direction of a momentum vector initially along
the $+z$ axis.
\iteme{BEX, BEY, BEZ :} gives the direction and size
\mbox{\boldmath $\beta$} of a Lorentz boost,
such that a particle initially at rest will have
$\mbf{p}/E = $\mbox{\boldmath $\beta$} afterwards.
\itemc{Remark:} all entries 1 through \ttt{N} are affected by the
transformation, unless lower and upper bounds are explicitly given
by \ttt{MSTU(1)} and \ttt{MSTU(2)}, or if status code 
\ttt{K(I,1)}$\leq 0$.
\end{entry}
 
\drawbox{ENTRY LUDBRB(IMI,IMA,THE,PHI,DBEX,DBEY,DBEZ)}\label{p:LUDBRB}
\begin{entry}
\itemc{Purpose:} to perform rotations and Lorentz boosts (in that
order, if both in the same call) of jet/particle momenta and vertex
position variables, for a specific range of entries, and with the
boost vector given in double precision. Is entry to \ttt{LUROBO},
mainly intended for internal use.
\iteme{IMI, IMA :} range of entries affected by transformation,
\ttt{IMI}$\leq$\ttt{I}$\leq$\ttt{IMA}.
\iteme{THE, PHI :} standard polar coordinates $\theta, \varphi$,
giving the rotated direction of a momentum vector initially along
the $+z$ axis.
\iteme{DBEX, DBEY, DBEZ :} gives the direction and size
\mbox{\boldmath $\beta$} of a Lorentz boost,
such that a particle initially at rest will have
$\mbf{p}/E = $\mbox{\boldmath $\beta$} afterwards.
Is to be given in double precision.
\itemc{Remark:} all entries with status codes \ttt{K(I,1)>0} in 
the requested range are affected by the transformation.
\end{entry}
 
\drawbox{CALL LUEDIT(MEDIT)}\label{p:LUEDIT}
\begin{entry}
\itemc{Purpose:} to exclude unstable or undetectable jets/particles
from the event record. One may also use \ttt{LUEDIT} to store spare
copies of events (specifically initial parton configuration) that can
be recalled to allow e.g. different fragmentation schemes to be run
through with one and the same parton configuration. Finally, an
event which has been analyzed with \ttt{LUSPHE}, \ttt{LUTHRU} or
\ttt{LUCLUS} (see section \ref{ss:evanrout}) may be rotated to align 
the event axis with the $z$ direction.
\iteme{MEDIT :} tells which action is to be taken.
\begin{subentry}
\iteme{= 0 :} empty (\ttt{K(I,1)=0}) and documentation 
(\ttt{K(I,1)>20}) lines
are removed. The jets/particles remaining are compressed in the
beginning of the \ttt{LUJETS} common block and the \ttt{N} value is
updated accordingly. The event history is lost, so that information
stored in \ttt{K(I,3)}, \ttt{K(I,4)} and \ttt{K(I,5)} is no longer
relevant.
\iteme{= 1 :} as \ttt{=0}, but in addition all jets/particles that have
fragmented/decayed (\ttt{K(I,1)>10}) are removed.
\iteme{= 2 :} as \ttt{=1}, but also all neutrinos and unknown particles
(i.e. compressed code KC$ = 0$) are removed.
\iteme{= 3 :} as \ttt{=2}, but also all uncharged, colour neutral
particles are removed, leaving only charged, stable particles (and
unfragmented partons, if fragmentation has not been performed).
\iteme{= 5 :} as \ttt{=0}, but also all partons which have branched or
been rearranged in a parton shower and all particles which have
decayed are removed, leaving only the fragmenting parton
configuration and the final-state particles.
\iteme{= 11 :} remove lines with \ttt{K(I,1)<0}. Update event
history information (in \ttt{K(I,3) - K(I,5)}) to refer to remaining
entries.
\iteme{= 12 :} remove lines with \ttt{K(I,1)=0}. Update event
history information (in \ttt{K(I,3) - K(I,5)}) to refer to remaining
entries.
\iteme{= 13 :} remove lines with \ttt{K(I,1)}$ = 11$, 12 or 15, except
for any line with \ttt{K(I,2)=94}. Update event history information
(in \ttt{K(I,3) - K(I,5)}) to refer to remaining entries. In
particular, try to trace origin of daughters, for which the mother
is decayed, back to entries not deleted.
\iteme{= 14 :} remove lines with \ttt{K(I,1)}$ = 13$ or 14, and also
any line with \ttt{K(I,2)=94}. Update event history information
(in \ttt{K(I,3) - K(I,5)}) to refer to remaining entries. In
particular, try to trace origin of rearranged jets back through
the parton-shower history to the shower initiator.
\iteme{= 15 :} remove lines with \ttt{K(I,1)>20}. Update event
history information (in \ttt{K(I,3) - K(I,5)}) to refer to remaining
entries.
\iteme{= 16 :} try to reconstruct missing daughter pointers of 
decayed particles from the mother pointers of decay products.
These missing pointers typically come from the need to use
\ttt{K(I,4)} and \ttt{K(I,5)} also for colour flow information.
\iteme{= 21 :} all partons/particles in current event record are
stored (as a spare copy) in bottom of common block \ttt{LUJETS}
(is e.g. done to save original partons before calling \ttt{LUEXEC}).
\iteme{= 22 :} partons/particles stored in bottom of event record with
\ttt{=21} are placed in beginning of record again, overwriting previous
information there (so that e.g. a different fragmentation scheme
can be used on the same partons). Since the copy at bottom is
unaffected, repeated calls with \ttt{=22} can be made.
\iteme{= 23 :} primary partons/particles in the beginning of event
record are marked as not fragmented or decayed, and number of entries
\ttt{N} is updated accordingly. Is simpe substitute for \ttt{=21} plus
\ttt{=22} when no fragmentation/decay products precede any of the
original partons/particles.
\iteme{= 31 :} rotate largest axis, determined by \ttt{LUSPHE},
\ttt{LUTHRU} or \ttt{LUCLUS}, to sit along the $z$ direction, and the
second largest axis into the $xz$ plane. For \ttt{LUCLUS} it can be
further specified to $+z$ axis and $xz$ plane with $x > 0$,
respectively. Requires that one of these routines has been called
before.
\iteme{= 32 :} mainly intended for \ttt{LUSPHE} and \ttt{LUTHRU}, this
gives a further alignment of the event, in addition to the one implied
by \ttt{=31}. The `slim' jet, defined as the side ($z > 0$ or $z < 0$)
with the smallest summed $\pT$ over square root of number of particles,
is rotated into the $+z$ hemisphere. In the opposite hemisphere
(now $z < 0$), the side of $x > 0$ and $x < 0$ which has the largest
summed $|p_z|$ is rotated into the $z < 0, x > 0$ quadrant.
Requires that \ttt{LUSPHE} or \ttt{LUTHRU} has been called before.
\end{subentry}
\itemc{Remark:} all entries 1 through \ttt{N} are affected by the
editing. For options 0--5 lower and upper bounds can be explicitly 
given by \ttt{MSTU(1)} and \ttt{MSTU(2)}.
\end{entry}
 
\drawbox{CALL LULIST(MLIST)}\label{p:LULIST}
\begin{entry}
\itemc{Purpose:} to list an event, jet or particle data, or current
parameter values.
\iteme{MLIST :} determines what is to be listed.
\begin{subentry}
\iteme{= 0 :} writes a title page, common for {\Je} and {\Py}, with 
program version numbers and last dates of change; is mostly for 
internal use.
\iteme{= 1 :} gives a simple list of current event record, in an 80
column format suitable for viewing directly on the computer terminal.
For each entry, the following information is given: the entry
number \ttt{I}, the parton/particle name (see below), the status code
(\ttt{K(I,1)}), the flavour code KF (\ttt{K(I,2)}), the line number
of the mother (\ttt{K(I,3)}), and the three-momentum, energy and mass
(\ttt{P(I,1) - P(I,5)}). If \ttt{MSTU(3)} is non-zero, lines immediately
after the event record proper are also listed. A final line
contains information on total charge, momentum, energy and
invariant mass. \\
The particle name is given by a call to the routine \ttt{LUNAME}.
For an entry which has decayed/fragmented (\ttt{K(I,1)=} 11--20), 
this particle name is given within parentheses. Similarly, a
documentation line (\ttt{K(I,1)=} 21--30) has the name enclosed in
expression signs (!\ldots!) and an event/jet axis information line
the name within inequality signs ($<$\ldots$>$). If the last
character of the name is a `?', it is a signal that the complete name
has been truncated to fit in, and can therefore not be trusted;
this is very rare. For partons which have been arranged along
strings (\ttt{K(I,1)=} 1, 2, 11 or 12), the end of the parton name
column contains information about the colour string arrangement:
an \ttt{A} for the first entry of a string, an \ttt{I} for all
intermediate ones, and a \ttt{V} for the final one (a poor man's
rendering of a vertical doublesided arrow, $\updownarrow$). \\
It is possible to insert lines just consisting of sequences
of \ttt{======} to separate different sections of the event record,
see \ttt{MSTU(70) - MSTU(80)}.
\iteme{= 2 :} gives a more extensive list of the current event record,
in a 132 column format, suitable for printers or workstations.
For each entry, the following information is given: the entry
number \ttt{I}, the parton/particle name (with padding as described
for \ttt{=1}), the status code (\ttt{K(I,1)}), the flavour code
KF (\ttt{K(I,2)}), the line number of the mother (\ttt{K(I,3)}), the
decay product/colour-flow pointers (\ttt{K(I,4), K(I,5)}), and the
three-momentum, energy and mass (\ttt{P(I,1) - P(I,5)}). If
\ttt{MSTU(3)} is non-zero, lines immediately after the event record
proper are also listed. A final line contains information on total
charge, momentum, energy and invariant mass. Lines with only
\ttt{======} may be inserted as for \ttt{=1}.
\iteme{= 3 :} gives the same basic listing as \ttt{=2}, but with an
additional line for each entry containing information on production
vertex position and time (\ttt{V(I,1) - V(I,4)}) and, for unstable
particles, proper lifetime (\ttt{V(I,5)}).
\iteme{= 11 :} provides a simple list of all parton/particle codes
defined in the program, with KF code and corresponding particle name.
The list is grouped by particle kind, and only within each group
in ascending order.
\iteme{= 12 :} provides a list of all parton/particle and decay data
used in the program. Each parton/particle code is represented by one
line containing KF flavour code, KC compressed code, particle
name, antiparticle name (where appropriate), electrical and
colour charge (stored in \ttt{KCHG}), mass, resonance width and
maximum broadening, average proper lifetime (in \ttt{PMAS}) and
whether the particle is considered stable or not (in \ttt{MDCY}).
Immediately after a particle, each decay channel gets one line,
containing decay channel number (\ttt{IDC} read from \ttt{MDCY}),
on/off switch for the channel, matrix element type (\ttt{MDME}),
branching ratio (\ttt{BRAT}), and decay products (\ttt{KFDP}).
The \ttt{MSTU(14)} flag can be used to set the maximum flavour
for which particles are listed, with the default (= 0)
corresponding to separately defined ones (KC$ > 100$ if KF$ > 0$).
In order to keep the size down, decay modes of heavy hadrons
collectively defined are never listed; these have KC codes
84--88, where the relevant information may be found.
\iteme{= 13 :} gives a list of current parameter values for
\ttt{MSTU}, \ttt{PARU}, \ttt{MSTJ} and \ttt{PARJ}, and the first
200 entries of \ttt{PARF}. This is useful to keep check of which
default values were changed in a given run.
\end{subentry}
\itemc{Remark:} for options 1--3 and 12 lower and upper bounds of the
listing can be explicitly given by \ttt{MSTU(1)} and \ttt{MSTU(2)}.
\end{entry}
 
\drawbox{CALL LUUPDA(MUPDA,LFN)}\label{p:LUUPDA}
\begin{entry}
\itemc{Purpose:} to give you the ability to update particle data,
or to keep several versions of modified particle data for special
purposes (e.g. charm studies).
\iteme{MUPDA :} gives the type of action to be taken.
\begin{subentry}
\iteme{= 1 :} write a table of particle data, that you then can
edit at leisure. For ordinary listing of decay data, \ttt{LULIST(12)}
should be used, but that listing could not be read back in
by the program. \\
For each compressed flavour code KC = 1--500, one line is
written containing KC (\ttt{I5}), the basic particle name (i.e.
excluding charge etc.) (\ttt{2X,A8}) in \ttt{CHAF}, the electric
(\ttt{I3}), colour charge (\ttt{I3}) and particle/antiparticle
distinction (\ttt{I3}) codes in \ttt{KCHG}, the mass (\ttt{F12.5}),
the mass width (\ttt{F12.5}), maximum broadening (\ttt{F12.5}) and
average proper lifetime (\ttt{2X,F12.5}) in \ttt{PMAS}, and the
on/off decay switch (I3) in \ttt{MDCY(KC,1).} \\
After a KC line follows one line for each possible decay
channel, containing the \ttt{MDME} codes (\ttt{5X,2I5}), the branching
ratio (\ttt{5X,F12.5}) in \ttt{BRAT}, and the \ttt{KFDP} codes for the
decay products (\ttt{5I8}), with trailing 0's if the number of decay
products is smaller than 5.
\iteme{= 2 :} read in particle data, as written with \ttt{=1} and
thereafter edited by you, and use this data subsequently in
the current run. Reading is done with fixed format,
which means that you have to preserve the format codes
described for \ttt{=1} during the editing. A number of checks will
be made to see if input looks reasonable, with warnings if not.
If some decay channel is said not to conserve charge, it should
be taken seriously. Warnings that decay is kinematically
unallowed need not be as serious, since that particular decay
mode may not be switched on unless the particle mass is
increased.
\iteme{= 3 :} write current particle data as data lines, which
can be edited into \ttt{BLOCK DATA LUDATA} for a permanent
replacement of the particle data. This option is intended for the
program author only, not for you.
\end{subentry}
\iteme{LFN :} the file number which the data should be written to or
read from. You must see to it that this file is properly
opened for read or write (since the definition of file names
is machine dependent).
\end{entry}
 
\drawbox{KK = KLU(I,J)}\label{p:KLU}
\begin{entry}
\itemc{Purpose:} to provide various integer-valued event data. Note
that many of the options available (in particular \ttt{I}$ > 0$,
\ttt{J}$\geq 14$) which refer to event history will not work after a
\ttt{LUEDIT} call. Further, the options 14--18 depend on the way the
event history has been set up, so with the explosion of different
allowed formats these options are no longer as safe as they may have
been. For instance, option 16 can only work if \ttt{MSTU(16)=2}. 
\iteme{I=0, J= :} properties referring to the complete event.
\begin{subentry}
\iteme{= 1 :} \ttt{N}, total number of lines in event record.
\iteme{= 2 :} total number of partons/particles remaining after
fragmentation and decay.
\iteme{= 6 :} three times the total charge of remaining (stable)
partons and particles.
\end{subentry}
\iteme{I>0, J= :} properties referring to the entry in line no.
\ttt{I} of the event record.
\begin{subentry}
\iteme{= 1 - 5 :} \ttt{K(I,1) - K(I,5)}, i.e. parton/particle status 
KS, flavour code KF and origin/decay product/colour-flow information.
\iteme{= 6 :} three times parton/particle charge.
\iteme{= 7 :} 1 for a remaining entry, 0 for a decayed, fragmented or
documentation entry.
\iteme{= 8 :} KF code (\ttt{K(I,2)}) for a remaining entry, 0 for a
decayed, fragmented or documentation entry.
\iteme{= 9 :} KF code (\ttt{K(I,2)}) for a parton (i.e. not colour
neutral entry), 0 for a particle.
\iteme{= 10 :} KF code (\ttt{K(I,2)}) for a particle (i.e. colour
neutral entry), 0 for a parton.
\iteme{= 11 :} compressed flavour code KC.
\iteme{= 12 :} colour information code, i.e. 0 for colour neutral,
1 for colour triplet, -1 for antitriplet and 2 for octet.
\iteme{= 13 :} flavour of `heaviest' quark or antiquark (i.e. with
largest code) in hadron or diquark (including sign for antiquark),
0 else.
\iteme{= 14 :} generation number. Beam particles or virtual exchange
particles are generation 0, original jets/particles generation
1 and then 1 is added for each step in the fragmentation/decay
chain.
\iteme{= 15 :} line number of ancestor, i.e. predecessor in first
generation (generation 0 entries are disregarded).
\iteme{= 16 :} rank of a hadron in the jet it belongs to. Rank denotes
the ordering in flavour space, with hadrons containing the original
flavour of the jet having rank 1, increasing by 1 for each step
away in flavour ordering. All decay products inherit the rank
of their parent. Whereas the meaning of a first-rank hadron
in a quark jet is always well-defined, the definition of higher
ranks is only meaningful for independently fragmenting quark
jets. In other cases, rank refers to the ordering in the actual
simulation, which may be of little interest.
\iteme{= 17 :} generation number after a collapse of a jet system into
one particle, with 0 for an entry not coming from a collapse, and
-1 for entry with unknown history. A particle formed in a
collapse is generation 1, and then one is added in each decay
step.
\iteme{= 18 :} number of decay/fragmentation products (only defined
in a collective sense for fragmentation).
\iteme{= 19 :} origin of colour for showering parton, 0 else.
\iteme{= 20 :} origin of anticolour for showering parton, 0 else.
\iteme{= 21 :} position of colour daughter for showering parton,
0 else.
\iteme{= 22 :} position of anticolour daughter for showering parton,
0 else.
\end{subentry}
\end{entry}
 
\drawbox{PP = PLU(I,J)}\label{p:PLU}
\begin{entry}
\itemc{Purpose:} to provide various real-valued event data. Note that
some of the options available (\ttt{I}$ > 0$, \ttt{J}$ = 20$--25),
which are primarily intended for studies of systems in their
respective c.m. frame, requires that a \ttt{LUEXEC} call has been
made for the current initial parton/particle configuration, but that
the latest \ttt{LUEXEC} call has not been followed by a \ttt{LUROBO}
one.
\iteme{I=0, J= :} properties referring to the complete event.
\begin{subentry}
\iteme{= 1 - 4 :} sum of $p_x$, $p_y$, $p_z$ and $E$, respectively,
for the stable remaining entries.
\iteme{= 5 :} invariant mass of the stable remaining entries.
\iteme{= 6 :} sum of electric charge of the stable remaining entries.
\end{subentry}
\iteme{I>0, J= :} properties referring to the entry in line no.
\ttt{I} of the event record.
\begin{subentry}
\iteme{= 1 - 5 :} \ttt{P(I,1) - P(I,5)}, i.e. normally $p_x$, $p_y$,
$p_z$, $E$ and $m$ for jet/particle.
\iteme{= 6 :} electric charge $e$.
\iteme{= 7 :} squared momentum $|\mbf{p}|^2 = p_x^2 + p_y^2 + p_z^2$.
\iteme{= 8 :} absolute momentum $|\mbf{p}|$.
\iteme{= 9 :} squared transverse momentum $\pT^2 = p_x^2 + p_y^2$.
\iteme{= 10 :} transverse momentum $\pT$.
\iteme{= 11 :} squared transverse mass
$m_{\perp}^2 = m^2 + p_x^2 + p_y^2$.
\iteme{= 12 :} transverse mass $m_{\perp}$.
\iteme{= 13 - 14 :} polar angle $\theta$ in radians (between 0 and
$\pi$) or degrees, respectively.
\iteme{= 15 - 16 :} azimuthal angle $\varphi$ in radians (between
$-\pi$ and $\pi$)  or degrees, respectively.
\iteme{= 17 :} true rapidity $y = (1/2) \, \ln((E+p_z)/(E-p_z))$.
\iteme{= 18 :} rapidity $y_{\pi}$ obtained by assuming that the
particle is a pion when calculating the energy $E$, to be used in the
formula above, from the (assumed known) momentum $\mbf{p}$.
\iteme{= 19 :} pseudorapidity $\eta = (1/2) \, \ln((p+p_z)/(p-p_z))$.
\iteme{= 20 :} momentum fraction $x_p = 2|\mbf{p}|/W$, where $W$
is the total energy of initial jet/particle configuration.
\iteme{= 21 :} $x_{\mrm{F}} = 2p_z/W$ (Feynman-$x$ if system is 
studied in the c.m. frame).
\iteme{= 22 :} $x_{\perp} = 2\pT/W$.
\iteme{= 23 :} $x_E = 2E/W$.
\iteme{= 24 :} $z_+ = (E+p_z)/W$.
\iteme{= 25 :} $z_- = (E-p_z)/W$.
\end{subentry}
\end{entry}
  
\subsection{The General Switches and Parameters}
\label{ss:JETswitch}
 
The common block \ttt{LUDAT1} is, next to \ttt{LUJETS}, the one a {\Je}
user is most likely to access. Here one may control in detail what the
program is to do, if the default mode of operation is not satisfactory.
 
\drawbox{COMMON/LUDAT1/MSTU(200),PARU(200),MSTJ(200),PARJ(200)}%
\label{p:LUDAT1}
\begin{entry}
\itemc{Purpose:} to give access to a number of status codes and
parameters which regulate the performance of the program as a whole.
Here \ttt{MSTU} and \ttt{PARU} are related to utility functions, as
well as a few parameters of the Standard Model, while \ttt{MSTJ} and
\ttt{PARJ} affect the underlying physics assumptions. Some of the
variables in \ttt{LUDAT1} are described elsewhere, and are therefore 
here only reproduced as references to the relevant sections. This 
in particular applies to many coupling constants mainly used by
{\Py}, which are found just after this, in section 
\ref{ss:coupcons}.
 
\boxsep
 
\iteme{MSTU(1),MSTU(2) :}\label{p:MSTU} (D=0,0) can be used to replace 
the ordinary
lower and upper limits (normally 1 and \ttt{N}) for the action of
\ttt{LUROBO}, and most \ttt{LUEDIT} and \ttt{LULIST} calls. Are reset
to 0 in a \ttt{LUEXEC} call.
 
\iteme{MSTU(3) :} (D=0) number of lines with extra information added
after line \ttt{N}. Is reset to 0 in a \ttt{LUEXEC} call, or in an
\ttt{LUEDIT} call when particles are removed.
 
\iteme{MSTU(4) :} (D=4000) number of lines available in the
common block \ttt{LUJETS}. Should always be changed if the
dimensions of the \ttt{K} and \ttt{P} arrays are changed by the user,
but should otherwise never be touched. Maximum allowed value is 10000,
unless \ttt{MSTU(5)} is also changed.
 
\iteme{MSTU(5) :} (D=10000) is used in building up the special
colour-flow information stored in \ttt{K(I,4)} and \ttt{K(I,5)}
for \ttt{K(I,3)=} 3, 13 or 14. The generic form for \ttt{j=} 4 or 5
is \\
\ttt{K(I,j)}$ = 2 \times$\ttt{MSTU(5)}$^2 \times$MCFR$ + 
$\ttt{MSTU(5)}$^2 \times$MCTO$ + $\ttt{MSTU(5)}$\times$ICFR$ + $ICTO, \\
with notation as in section \ref{ss:evrec}.
One should always have \ttt{MSTU(5)}$\geq$\ttt{MSTU(4)}. On a 32 bit
machine, values \ttt{MSTU(5)}$> 20000$ may lead to overflow problems,
and should be avoided.
 
\iteme{MSTU(6) :} (D=500) number of KC codes available in the
\ttt{KCHG}, \ttt{PMAS}, \ttt{MDCY}, and \ttt{CHAF} arrays; should be
changed if these dimensions are changed.
 
\iteme{MSTU(7) :} (D=2000) number of decay channels available in the
\ttt{MDME}, \ttt{BRAT} and \ttt{KFDP} arrays; should be changed if
these dimensions are changed.
 
\iteme{MSTU(10) :} (D=2) use of parton/particle masses in filling
routines (\ttt{LU1ENT}, \ttt{LU2ENT}, \ttt{LU3ENT}, \ttt{LU4ENT}).
\begin{subentry}
\iteme{= 0 :} assume the mass to be zero.
\iteme{= 1 :} keep the mass value stored in \ttt{P(I,5)}, whatever it
is. (This may be used e.g. to describe kinematics with
off-mass-shell partons).
\iteme{= 2 :} find masses according to mass tables as usual.
\end{subentry}
 
\iteme{MSTU(11) :} (D=6) file number to which all program output is
directed. It is your responsibility to see to it that the
corresponding file is also opened for output.
 
\iteme{MSTU(12) :} (D=1) writing of title page (version number and last
date of change for {\Py} and {\Je}) on output file.
\begin{subentry}
\iteme{= 0 :} not done.
\iteme{= 1 :} title page is written at first occasion, at which time
\ttt{MSTU(12)} is set =0.
\end{subentry}
 
\iteme{MSTU(13) :} (D=1) writing of information on variable values
changed by a \ttt{LUGIVE} call.
\begin{subentry}
\iteme{= 0 :} no information is provided.
\iteme{= 1 :} information is written to standard output.
\end{subentry}
 
\iteme{MSTU(14) :} (D=0) if non-zero, this gives the maximum flavour
for which a \ttt{LULIST(12)} call will give particle data on possible
hadrons. With \ttt{MSTU(14)=5} only known hadrons, i.e. up to bottom,
are listed. If \ttt{=0}, only separately specified particles are
listed (i.e. either KF$\leq 100$ or else both KF$> 100$ and
KC$> 100$).
 
\iteme{MSTU(15) :} (D=1) selection for characters used in particle
names to denote an antiparticle; appear in \ttt{LULIST} listings
or other \ttt{LUNAME} applications.
\begin{subentry}
\iteme{= 1 :} the tilde character `$\sim$'.
\iteme{= 2 :} the characters `bar'.
\end{subentry}
 
\iteme{MSTU(16) :} (D=1) choice of mother pointers for the particles
produced by a fragmenting parton system.
\begin{subentry}
\iteme{= 1 :} all primary particles of a system point to a line with
KF = 92 or 93, for string or independent fragmentation, respectively,
or to a line with KF = 91 if a jet system has so small a mass
that it is forced to decay into one or two particles. The two
(or more) shower initiators of a showering parton system point
to a line with KF = 94. The entries with KF = 91--94 in their
turn point back to the predecessor partons, so that the
KF = 91--94 entries form a part of the event history proper.
\iteme{= 2 :} although the lines with KF = 91--94 are present, and
contain the correct mother and daughter pointers, they are not part
of the event history proper, in that particles produced in string
fragmentation point directly to either of the two endpoint
partons of the string (depending on the side they were generated
from), particles produced in independent fragmentation point
to the respective parton they were generated from, particles in
small mass systems point to either endpoint parton, and
shower initiators point to the original on-mass-shell
counterparts. Also the daugher pointers bypass the KF = 91--94
entries. In independent fragmentation, a parton need not produce
any particles at all, and then have daughter pointers 0.
\itemc{Note :} \ttt{MSTU(16)} should not be changed between the
generation of an event and the translation of this event record with
a \ttt{LUHEPC} call, since this may give an erroneous translation of
the event history.
\end{subentry}
 
\iteme{MSTU(17) :} (D=0) storage option for \ttt{MSTU(90)} and
associated information on $z$ values for heavy-flavour production.
\begin{subentry}
\iteme{= 0 :} \ttt{MSTU(90)} is reset to zero at each \ttt{LUEXEC}
call. This is the appropriate course if \ttt{LUEXEC} is only called
once per event, as is normally the case when you do not
yourself call \ttt{LUEXEC}.
\iteme{= 1 :} you have to reset \ttt{MSTU(90)} to zero yourself
before each new event. This is the appropriate course if several
\ttt{LUEXEC} calls may appear for one event, i.e. if you
call \ttt{LUEXEC} directly.
\end{subentry}
 
\iteme{MSTU(19) :} (D=0) advisory warning for unphysical flavour
setups in \ttt{LU2ENT}, \ttt{LU3ENT} or \ttt{LU4ENT} calls.
\begin{subentry}
\iteme{= 0 :} yes.
\iteme{= 1 :} no; \ttt{MSTU(19)} is reset to 0 in such a call.
\end{subentry}
 
\iteme{MSTU(21) :} (D=2) check on possible errors during program
execution. Obviously no guarantee is given that all errors will be
caught, but some of the most trivial user-caused errors may be found.
\begin{subentry}
\iteme{= 0 :} errors do not cause any immediate action, rather the
program will try to cope, which may mean e.g. that it
runs into an infinite loop.
\iteme{= 1 :} parton/particle configurations are checked for possible
errors. In case of problem, an exit is made from the
misbehaving subprogram, but the generation of the event is
continued from there on. For the first \ttt{MSTU(22)} errors a
a message is printed; after that no messages appear.
\iteme{= 2 :} parton/particle configurations are checked for possible
errors. In case of problem, an exit is made from the
misbehaving subprogram, and subsequently from \ttt{LUEXEC}.
You may then choose to correct the error, and continue the
execution by another \ttt{LUEXEC} call. For the first \ttt{MSTU(22)}
errors a message is printed, after that the last event is
printed and execution is stopped.
\end{subentry}
 
\iteme{MSTU(22) :} (D=10) maximum number of errors that are printed.
 
\iteme{MSTU(23) :} (I) count of number of errors experienced to date.
 
\iteme{MSTU(24) :} (R) type of latest error experienced; reason that
event was not generated in full. Is reset at each \ttt{LUEXEC} call.
\begin{subentry}
\iteme{= 0 :} no error experienced.
\iteme{= 1 :} have reached end of or are writing outside
\ttt{LUJETS} memory.
\iteme{= 2 :} unknown flavour code or unphysical combination of codes;
may also be caused by erroneous string connection information.
\iteme{= 3 :} energy or mass too small or unphysical kinematical
variable setup.
\iteme{= 4 :} program is caught in an infinite loop.
\iteme{= 5 :} momentum, energy or charge was not conserved (even
allowing for machine precision errors, see \ttt{PARU(11)}); is
evaluated only after event has been generated in full, and does not
apply when independent fragmentation without momentum conservation
was used.
\iteme{= 6 :} error call from outside the fragmentation/decay package
(e.g. the $\ee$ routines).
\iteme{= 7 :} inconsistent particle data input in \ttt{LUUPDA}
(\ttt{MUPDA = 2}) or other \ttt{LUUPDA}-related problem.
\iteme{= 8 :} problems in more peripheral service routines.
\iteme{= 9 :} various other problems.
\end{subentry}
 
\iteme{MSTU(25) :} (D=1) printing of warning messages.
\begin{subentry}
\iteme{= 0 :} no warnings are written.
\iteme{= 1 :} first \ttt{MSTU(26)} warnings are printed, thereafter
no warnings appear.
\end{subentry}
 
\iteme{MSTU(26) :} (D=10) maximum number of warnings that are printed.
 
\iteme{MSTU(27) :} (I) count of number of warnings experienced to
date.
 
\iteme{MSTU(28) :} (R) type of latest warning given, with codes
paralleling those for \ttt{MSTU(24)}, but of a less serious nature.
 
\iteme{MSTU(31) :} (I) number of \ttt{LUEXEC} calls in present run.
 
\iteme{MSTU(32) :} (I) number of entries stored with
\ttt{LUEDIT(-1)} call.
 
\iteme{MSTU(33) :} (I) if set 1 before a \ttt{LUDBRB} call, the
\ttt{V} vectors (in the particle range to be rotated/boosted) are
set 0 before the rotation/boost. \ttt{MSTU(33)} is set back to 0 in
the \ttt{LUDBRB} call. Is inactive in a \ttt{LUROBO} call.
 
\iteme{MSTU(41) - MSTU(63) :} switches for event-analysis routines,
see section \ref{ss:evanrout}.
 
\iteme{MSTU(70) :} (D=0) the number of lines consisting only of equal
signs (\ttt{======}) that are inserted in the event listing obtained
with \ttt{LULIST(1)}, \ttt{LULIST(2)} or \ttt{LULIST(3)}, so as to
distinguish different sections of the event record on output. At most
10 such lines can be inserted; see \ttt{MSTU(71) - MSTU(80)}. Is reset
at \ttt{LUEDIT} calls with arguments 0--5.
 
\iteme{MSTU(71) - MSTU(80) :} line numbers below which lines
consisting only of equal signs (\ttt{======}) are inserted in event
listings. Only the first \ttt{MSTU(70)} of the 10 allowed positions
are enabled.
 
\iteme{MSTU(90) :} number of heavy-flavour hadrons (i.e. hadrons
containing charm or heavier flavours) produced in current event, for
which the positions in the event record are stored in
\ttt{MSTU(91) - MSTU(98)} and the $z$ values in the fragmentation in
\ttt{PARU(91) - PARU(98)}. At most eight values will be stored
(normally this is no problem). No $z$ values can be stored for those
heavy hadrons produced when a string has so small mass that it
collapses to one or two particles, nor for those produced as one of
the final two particles in the fragmentation of a string. If
\ttt{MSTU(17)=1}, \ttt{MSTU(90)} should be reset to zero by you
before each new event, else this is done automatically.
 
\iteme{MSTU(91) - MSTU(98) :} the first \ttt{MSTU(90)} positions will
be filled with the line numbers of the heavy-flavour hadrons produced
in the current event. See \ttt{MSTU(90)} for additional comments. Note
that the information is corrupted by calls to \ttt{LUEDIT} with options
0--5 and 21--23; calls with options 11--15 work, however.
 
\iteme{MSTU(101) - MSTU(118) :} switches related to couplings, see
section \ref{ss:coupcons}.
 
\iteme{MSTU(161), MSTU(162) :} information used by event-analysis 
routines, see section \ref{ss:evanrout}.
 
\iteme{MSTU(181) :} (R) {\Je} version number.
 
\iteme{MSTU(182) :} (R) {\Je} subversion number.
 
\iteme{MSTU(183) :} (R) last year of change for {\Je}.
 
\iteme{MSTU(184) :} (R) last month of change for {\Je}.
 
\iteme{MSTU(185) :} (R) last day of change for {\Je}.

\iteme{MSTU(186) :} (R) earliest subversion of {\Py} version 5
with which this {\Je} subversion can be run. 
 
\boxsep
 
\iteme{PARU(1) :}\label{p:PARU} (R) $\pi \approx 3.1415927$.
 
\iteme{PARU(2) :} (R) $2\pi \approx 6.2831854$.
 
\iteme{PARU(3) :} (D=0.1973) conversion factor for GeV$^{-1} \to$ fm
or fm$^{-1} \to$ GeV.
 
\iteme{PARU(4) :} (D=5.068) conversion factor for fm $\to$ GeV$^{-1}$
or GeV $\to$ fm$^{-1}$.
 
\iteme{PARU(5) :} (D=0.3894) conversion factor for GeV$^{-2} \to$ mb
or mb$^{-1} \to$ GeV$^2$.
 
\iteme{PARU(6) :} (D=2.568) conversion factor for mb $\to$ GeV$^{-2}$
or GeV$^2 \to$ mb$^{-1}$.
 
\iteme{PARU(11) :} (D=0.001) relative error, i.e. nonconservation of
momentum and energy divided by total energy, that may be attributable
to machine precision problems before a physics error is suspected
(see \ttt{MSTU(24)=5}).
 
\iteme{PARU(12) :} (D=0.09 GeV$^2$) effective cut-off in squared mass,
below which partons may be recombined to simplify (machine precision
limited) kinematics of string fragmentation.
 
\iteme{PARU(13) :} (D=0.01) effective angular cut-off in radians for
recombination of partons, used in conjunction with \ttt{PARU(12)}.
 
\iteme{PARU(21) :} (I) contains the total energy $W$ of all first
generation jets/particles after a \ttt{LUEXEC} call; to be used by the
\ttt{PLU} function for \ttt{I>0}, \ttt{J=} 20--25.

\iteme{PARU(41) - PARU(63) :} parameters for event-analysis routines,
see section \ref{ss:evanrout}.
 
\iteme{PARU(91) - PARU(98) :} the first \ttt{MSTU(90)} positions will
be filled with the fragmentation $z$ values used internally in the
generation of heavy-flavour hadrons --- how these are translated into
the actual energies and momenta of the observed hadrons is a
complicated function of the string configuration. The particle with
$z$ value stored in \ttt{PARU(i)} is to be found in line \ttt{MSTU(i)}
of the event record. See \ttt{MSTU(90)} and \ttt{MSTU(91) - MSTU(98)}
for additional comments.
 
\iteme{PARU(101) - PARU(195) :} various coupling constants and 
parameters related to couplings, see section \ref{ss:coupcons}.
 
\boxsep
 
\iteme{MSTJ(1) :}\label{p:MSTJ} (D=1) choice of fragmentation scheme.
\begin{subentry}
\iteme{= 0 :} no jet fragmentation at all.
\iteme{= 1 :} string fragmentation according to the Lund model.
\iteme{= 2 :} independent fragmentation, according to specification
in \ttt{MSTJ(2)} and \ttt{MSTJ(3)}.
\end{subentry}
 
\iteme{MSTJ(2) :} (D=3) gluon jet fragmentation scheme in independent
fragmentation.
\begin{subentry}
\iteme{= 1 :} a gluon is assumed to fragment like a random $\d$, $\u$
or $\s$ quark or antiquark.
\iteme{= 2 :} as \ttt{=1}, but longitudinal (see \ttt{PARJ(43)},
\ttt{PARJ(44)} and \ttt{PARJ(59)}) and transverse (see \ttt{PARJ(22)})
momentum properties of quark or antiquark substituting for gluon may
be separately specified.
\iteme{= 3 :} a gluon is assumed to fragment like a pair of a
$\d$, $\u$ or $\s$ quark and its antiquark, sharing the gluon energy
according to the Altarelli-Parisi splitting function.
\iteme{= 4 :} as \ttt{=3}, but longitudinal (see \ttt{PARJ(43)},
\ttt{PARJ(44)} and \ttt{PARJ(59)}) and transverse (see \ttt{PARJ(22)})
momentum properties of quark and antiquark substituting for gluon may
be separately specified.
\end{subentry}
 
\iteme{MSTJ(3) :} (D=0) energy, momentum and flavour conservation
options in independent fragmentation. Whenever momentum conservation
is described below, energy and flavour conservation is also
implicitly assumed.
\begin{subentry}
\iteme{= 0 :} no explicit conservation of any kind.
\iteme{= 1 :} particles share momentum imbalance compensation according
to their energy (roughly equivalent to boosting event to c.m.
frame). This is similar to the approach in the Ali et al.
program \cite{Ali80}.
\iteme{= 2 :} particles share momentum imbalance compensation according
to their longitudinal mass with respect to the imbalance
direction.
\iteme{= 3 :} particles share momentum imbalance compensation equally.
\iteme{= 4 :} transverse momenta are compensated separately within
each jet, longitudinal momenta are rescaled so that ratio
of final jet to initial parton momentum is the same for
all the jets of the event. This is similar to the approach in
the Hoyer et al. program \cite{Hoy79}.
\iteme{= 5 :} only flavour is explicitly conserved.
\iteme{= 6 - 10 :} as \ttt{=1 - 5}, except that above several colour
singlet systems that followed immediately after each other
in the event listing (e.g. $\q\qbar\q\qbar$) were treated as
one single system, whereas here they are treated as
separate systems.
\iteme{= -1 :} independent fragmentation, where also particles moving
backwards with respect to the jet direction are kept, and
thus the amount of energy and momentum mismatch may be large.
\end{subentry}
 
\iteme{MSTJ(11) :} (D=4) choice of longitudinal fragmentation
function, i.e. how large a fraction of the energy available a
newly-created hadron takes.
\begin{subentry}
\iteme{= 1 :} the Lund symmetric fragmentation function, see
\ttt{PARJ(41) - PARJ(45)}.
\iteme{= 2 :} choice of some different forms for each flavour
separately, see \ttt{PARJ(51) - PARJ(59)}.
\iteme{= 3 :} hybrid scheme, where light flavours are treated with
symmetric Lund (\ttt{=1}), but charm and heavier can be separately
chosen, e.g. according to the SLAC function (\ttt{=2}).
\iteme{= 4 :} the Lund symmetric fragmentation function (\ttt{=1}),
for heavy endpoint quarks modified according to the Bowler
(Artru--Mennessier, Morris) space--time picture of string evolution,
see \ttt{PARJ(46)}.
\iteme{= 5 :} as \ttt{=4}, but with possibility to interpolate
between Bowler and Lund separately for $\c$, $\b$ and $\t$; see
\ttt{PARJ(46) - PARJ(48)}.
\end{subentry}
 
\iteme{MSTJ(12) :} (D=2) choice of baryon production model.
\begin{subentry}
\iteme{= 0 :} no baryon-antibaryon pair production at all; initial
diquark treated as a unit.
\iteme{= 1 :} diquark-antidiquark pair production allowed; diquark
treated as a unit.
\iteme{= 2 :} diquark-antidiquark pair production allowed, with
possibility for diquark to be split according to the `popcorn'
scheme.
\iteme{= 3 :} as \ttt{=2}, but additionally the production of first
rank baryons may be suppressed by a factor \ttt{PARJ(19)}.
\end{subentry}
 
\iteme{MSTJ(13) :} (D=0) generation of transverse momentum for
endpoint quark(s) of single quark jet or $\q\qbar$ jet system (in
multijet events no endpoint transverse momentum is ever allowed for).
\begin{subentry}
\iteme{= 0 :} no transverse momentum for endpoint quarks.
\iteme{= 1 :} endpoint quarks obtain transverse momenta like ordinary
$\q\qbar$ pairs produced in the field (see \ttt{PARJ(21)}); for
2-jet systems the endpoints obtain balancing transverse momenta.
\end{subentry}
 
\iteme{MSTJ(14) :} (D=1) treatment of a colour-singlet jet system
with a low invariant mass.
\begin{subentry}
\iteme{= 0 :} no precautions are taken, meaning that problems may
occur in \ttt{LUSTRF} (or \ttt{LUINDF}) later on.
\iteme{= 1 :} small jet systems are allowed to collapse into two
particles or, failing that, one single particle. Normally
all small systems are treated this way, starting with the
smallest one, but some systems would require more work and
are left untreated; they include diquark-antidiquark pairs
below the two-particle threshold.
\iteme{= -1 :} special option for \ttt{LUPREP} calls, where no
precautions are taken (as for \ttt{=0}), but, in addition, no checks
are made on the presence of small-mass systems; i.e. \ttt{LUPREP}
only rearranges colour strings.
\end{subentry}
 
\iteme{MSTJ(15) :} (D=0) production probability for new flavours.
\begin{subentry}
\iteme{= 0 :} according to standard Lund parametrization, as given
by \ttt{PARJ(1) - PARJ(20)}.
\iteme{= 1 :} according to probabilities stored in
\ttt{PARF(201) - PARF(1960)}; note that no default values exist here,
i.e. \ttt{PARF} must be set by you. The \ttt{MSTJ(12)} switch
can still be used to set baryon production mode, with the
modification that \ttt{MSTJ(12)=2} here allows an arbitrary number
of mesons to be produced between a baryon and an antibaryon (since
the probability for diquark $\to$ meson $+$ new diquark is assumed
independent of prehistory).
\end{subentry}
 
\iteme{MSTJ(21) :} (D=2) form of particle decays.
\begin{subentry}
\iteme{= 0 :} all particle decays are inhibited.
\iteme{= 1 :} a particle declared unstable in the \ttt{MDCY} vector,
and with decay channels defined, may decay within the region given
by \ttt{MSTJ(22)}. A particle may decay into jets, which then fragment
further according to the \ttt{MSTJ(1)} value.
\iteme{= 2 :} as \ttt{=1}, except that a $\q\qbar$ jet system produced 
in a decay (e.g. of a $\B$ meson) is always allowed to fragment 
according to string fragmentation, rather than according to the 
\ttt{MSTJ(1)} value (this means that momentum, energy and charge 
are conserved in the decay).
\end{subentry}
 
\iteme{MSTJ(22) :} (D=1) cut-off on decay length for a particle that is
allowed to decay according to \ttt{MSTJ(21)} and the \ttt{MDCY} value.
\begin{subentry}
\iteme{= 1 :} a particle declared unstable is also forced to decay.
\iteme{= 2 :} a particle is decayed only if its average proper
lifetime is larger than \ttt{PARJ(71)}.
\iteme{= 3 :} a particle is decayed only if the decay vertex is
within a distance \ttt{PARJ(72)} of the origin.
\iteme{= 4 :} a particle is decayed only if the decay vertex is
within a cylindrical volume with radius \ttt{PARJ(73)} in the
$xy$-plane and extent to $\pm$\ttt{PARJ(74)} in the $z$ direction.
\end{subentry}
 
\iteme{MSTJ(23) :} (D=1) possibility of having a shower evolving from
a $\q\qbar$ pair created as decay products.
\begin{subentry}
\iteme{= 0 :} never.
\iteme{= 1 :} whenever the decay channel matrix-element code is
\ttt{MDME(IDC,2)=} 4, 32, 33, 44 or 46, the two first decay products
(if they are partons) are allowed to shower, like a colour-singlet
subsystem, with maximum virtuality given by the invariant mass
of the pair.
\end{subentry}
 
\iteme{MSTJ(24) :} (D=2) particle masses.
\begin{subentry}
\iteme{= 0 :} discrete mass values are used.
\iteme{= 1 :} particles registered as having a mass width in the
\ttt{PMAS} vector are given a mass according to a truncated
Breit--Wigner shape, linear in $m$, eq.~(\ref{dec:BWlin}).
\iteme{= 2 :} as \ttt{=1}, but gauge bosons (actually all particles
with $|$KF$| \leq 100$) are distributed according to a Breit--Wigner
quadratic in $m$, as obtained from propagators.
\iteme{= 3 :} as \ttt{=1}, but Breit--Wigner shape is always
quadratic in $m$, eq.~(\ref{dec:BWtwo}).
 
\end{subentry}
 
\iteme{MSTJ(25) :} (D=1) inclusion of the $\W^{\pm}$ propagator, in
addition to the standard, `infinitely heavy' weak $V-A$ matrix
element, in the decay of a $\t$, $\lrm$ or $\hrm$ quark, or
$\chi$ lepton.
\begin{subentry}
\iteme{= 0 :} not included.
\iteme{= 1 :} included.
\end{subentry}
 
\iteme{MSTJ(26) :} (D=2) inclusion of $\B$--$\br{\B}$ mixing in
decays.
\begin{subentry}
\iteme{= 0 :} no.
\iteme{= 1 :} yes, with mixing parameters given by \ttt{PARJ(76)}
and \ttt{PARJ(77)}. Mixing decays are not specially marked.
\iteme{= 2 :} yes, as \ttt{=1}, but a $\B$ ($\br{\B}$) that decays
as a $\br{\B}$ ($\B$) is marked as \ttt{K(I,1)=12} rather than the
normal \ttt{K(I,1)=11}.
\end{subentry}

\iteme{MSTJ(27) :} (D=2) possibility for the $\b$ quark to develop
a shower in the decay of a top hadron, i.e. $\T \to \W^+ \b \qbar$,
where $\qbar$ is a spectator quark.
\begin{subentry}
\iteme{= 0 :} no, i.e. $\b$ jet is narrow, low-multiplicity.
\iteme{= 1 :} the $\b$ is allowed to shower and the $\W$ momentum
(in the rest frame of the $\T$) is reduced acccordingly. The $\W$
is therafter assumed to decay isotropically.
\iteme{= 2 :} the $\b$ is allowed to shower, as in \ttt{=1}, but 
the $\W$ decays anisotropically, with the same polarization as in
the standard weak decay of option \ttt{=0}. In principle this is 
better than option \ttt{=1}, but in practice there is no big 
difference.
\end{subentry}

\iteme{MSTJ(28) :} (D=0) call to an external $\tau$ decay library.
For this option to be meaningful, it is up to you to write the 
appropriate interface and include that in the routine \ttt{LUTAUD},
as explained in section \ref{ss:JETphysrout}. 
\begin{subentry}
\iteme{= 0 :} not done, i.e. the internal \ttt{LUDECY} treatment is 
used.
\iteme{= 1 :} done whenever the $\tau$ mother particle species can 
be identified, else the internal \ttt{LUDECY} treatment is used. 
Normally the mother particle should always be identified, but it is 
possible for a user to remove event history information or to add 
extra $\tau$'s directly to the event record, and then the mother is 
not known.
\iteme{= 2 :} always done. 
\end{subentry}
 
\iteme{MSTJ(40) - MSTJ(50) :} switches for time-like parton showers,
see section \ref{ss:showrout}.
 
\iteme{MSTJ(51) :} (D=0) inclusion of Bose--Einstein effects.
\begin{subentry}
\iteme{= 0 :} no effects included.
\iteme{= 1 :} effects included according to an exponential
parametrization
$C_2(Q) = 1 + $\ttt{PARJ(92)}$\times \exp(-Q/$\ttt{PARJ(93)}$)$,
where $C_2(Q)$ represents the ratio of particle production at
$Q$ with Bose--Einstein effects to that without, and the relative
momentum $Q$ is defined by
$Q^2(p_1,p_2) = -(p_1 - p_2)^2 = (p_1 + p_2)^2 - 4m^2$. Particles
with width broader than \ttt{PARJ(91)} are assumed to have time to
decay before Bose--Einstein effects are to be considered.
\iteme{= 2 :} effects included according to a Gaussian
parametrization
$C_2(Q) = 1 + $\ttt{PARJ(92)}$\times \exp(- (Q/$\ttt{PARJ(93)}$)^2 )$,
with notation and comments as above.
\end{subentry}
 
\iteme{MSTJ(52) :} (D=3) number of particle species for which
Bose--Einstein correlations are to be included, ranged along the
chain $\pi^+$, $\pi^-$, $\pi^0$, $\K^+$, $\K^-$, $\K^0_{\mrm{L}}$,
$\K^0_{\mrm{S}}$, $\eta$ and $\eta'$. Default corresponds to
including all pions ($\pi^+$, $\pi^-$, $\pi^0$), 7 to including all
Kaons as well, and 9 is maximum.
 
\iteme{MSTJ(91) :} (I) flag when generating gluon jet with options
\ttt{MSTJ(2)=} 2 or 4 (then \ttt{=1}, else \ttt{=0}).
 
\iteme{MSTJ(92) :} (I) flag that a $\q\qbar$ or $\g\g$ pair or a
$\g\g\g$ triplet created in \ttt{LUDECY} should be allowed to shower,
is 0 if no pair or triplet, is the entry number of the first parton
if a pair indeed exists, is the entry number of the first parton,
with a $-$ sign, if a triplet indeed exists.
 
\iteme{MSTJ(93) :} (I) switch for \ttt{ULMASS} action. Is reset to 0
in \ttt{ULMASS} call.
\begin{subentry}
\iteme{= 0 :} ordinary action.
\iteme{= 1 :} light ($\d$, $\u$, $\s$, $\c$, $\b$) quark masses are
taken from \ttt{PARF(101) - PARF(105)} rather than
\ttt{PMAS(1,1) - PMAS(5,1)}. Diquark masses are
given as sum of quark masses, without spin splitting term.
\iteme{= 2 :} as \ttt{=1}. Additionally the constant terms
\ttt{PARF(121)} and \ttt{PARF(122)} are subtracted from quark and
diquark masses, respectively.
\end{subentry}

\iteme{MSTJ(101) - MSTJ(121) :} switches for $\ee$ event generation,
see section \ref{ss:eeroutines}.
 
\boxsep
 
\iteme{PARJ(1) :}\label{p:PARJ} (D=0.10) is 
${\cal P}(\q\q)/{\cal P}(\q)$, the
suppression of diquark-antidiquark pair production in the colour
field, compared with quark--antiquark production.
 
\iteme{PARJ(2) :} (D=0.30) is ${\cal P}(\s)/{\cal P}(u)$, the
suppression of $\s$ quark pair production in the field compared with
$\u$ or $\d$ pair production.
 
\iteme{PARJ(3) :} (D=0.4) is
$({\cal P}(\u\s)/{\cal P}(\u\d))/({\cal P}(\s)/{\cal P}(\d))$,
the extra suppression of strange diquark production  compared with
the normal suppression of strange quarks.
 
\iteme{PARJ(4) :} (D=0.05) is $(1/3){\cal P}(\u\d_1)/{\cal P}(\u\d_0)$,
the suppression of spin 1 diquarks compared with spin 0 ones
(excluding the factor 3 coming from spin counting).
 
\iteme{PARJ(5) :} (D=0.5) parameter determining relative occurence of
baryon production by $BM\br{B}$ and by $B\br{B}$ configurations in the
popcorn baryon production model, roughly
${\cal P}(BM\br{B})/({\cal P}(B\br{B})+{\cal P}(BM\br{B})) =$
\ttt{PARJ(5)}$/(0.5+$\ttt{PARJ(5)}$)$.
 
\iteme{PARJ(6) :} (D=0.5) extra suppression for having a $\s\sbar$
pair shared by the $B$ and $\br{B}$ of a $BM\br{B}$ situation.
 
\iteme{PARJ(7) :} (D=0.5) extra suppression for having a strange
meson $M$ in a $BM\br{B}$ configuration.
 
\iteme{PARJ(11) - PARJ(17) :} parameters that determine the spin of
mesons.
\begin{subentry}
\iteme{PARJ(11) :} (D=0.5) is the probability that a light meson
(containing $\u$ and $\d$ quarks only) has spin 1 (with
\ttt{1-PARJ(11)} the probability for spin 0) when formed in
fragmentation.
\iteme{PARJ(12) :} (D=0.6) is the probability that a strange meson
has spin 1.
\iteme{PARJ(13) :} (D=0.75) is the probability that a charm or
heavier meson has spin 1.
\iteme{PARJ(14) :} (D=0.) is the probability that a spin = 0
meson is produced with an orbital angular momentum 1, for a
total spin = 1.
\iteme{PARJ(15) :} (D=0.) is the probability that a spin = 1
meson is produced with an orbital angular momentum 1, for a
total spin = 0.
\iteme{PARJ(16) :} (D=0.) is the probability that a spin = 1
meson is produced with an orbital angular momentum 1, for a
total spin = 1.
\iteme{PARJ(17) :} (D=0.) is the probability that a spin = 1
meson is produced with an orbital angular momentum 1, for a
total spin = 2.
\itemc{Note :} the end result of the numbers above is
that, with \ttt{i} = 11, 12 or 13, depending on flavour content, \\
${\cal P}(S = 0, L = 0, J = 0) = (1 - \mtt{PARJ(i)}) \times 
(1 - \mtt{PARJ(14)})$, \\
${\cal P}(S = 0, L = 1, J = 1) = (1 - \mtt{PARJ(i)}) \times 
\mtt{PARJ(14)}$, \\
${\cal P}(S = 1, L = 0, J = 1) = $ \\
\mbox{~}\hfill$\mtt{PARJ(i)} \times (1 - \mtt{PARJ(15)} - 
\mtt{PARJ(16)} - \mtt{PARJ(17)})$, \\
${\cal P}(S = 1, L = 1, J = 0) = \mtt{PARJ(i)} \times 
\mtt{PARJ(15)}$, \\
${\cal P}(S = 1, L = 1, J = 1) = \mtt{PARJ(i)} \times 
\mtt{PARJ(16)}$, \\
${\cal P}(S = 1, L = 1, J = 2) = \mtt{PARJ(i)} \times 
\mtt{PARJ(17)}$, \\
where $S$ is the quark `true' spin and $J$ is the total spin, usually
called the spin $s$ of the meson.
\end{subentry}
 
\iteme{PARJ(18) :} (D=1.) is an extra suppression factor multiplying
the ordinary {\bf SU(6)} weight for spin $3$/2 baryons, and hence a
means to break {\bf SU(6)} in addition to the dynamic breaking implied
by \ttt{PARJ(2)}, \ttt{PARJ(3)}, \ttt{PARJ(4)}, \ttt{PARJ(6)} and
\ttt{PARJ(7)}.
 
\iteme{PARJ(19) :} (D=1.) extra baryon suppression factor, which
multiplies the ordinary diquark-antidiquark production probability
for the breakup closest to the endpoint of a string, but leaves other
breaks unaffected. Is only used for \ttt{MSTJ(12)=3}.
 
\iteme{PARJ(21) :} (D=0.36 GeV) corresponds to the width $\sigma$ in
the Gaussian $p_x$ and $p_y$ transverse momentum distributions for
primary hadrons. See also \ttt{PARJ(22) - PARJ(24)}.
 
\iteme{PARJ(22) :} (D=1.) relative increase in transverse momentum in
a gluon jet generated with \ttt{MSTJ(2)=} 2 or 4.

\iteme{PARJ(23), PARJ(24) :} (D=0.01, 2.) a fraction \ttt{PARJ(23)}
of the Gaussian transverse momentum distribution is taken to be a
factor \ttt{PARJ(24)} larger than input in \ttt{PARJ(21)}. This
gives a simple parametrization of non-Gaussian tails to the Gaussian
shape assumed above. 

\iteme{PARJ(25) :} (D=1.) extra suppression factor for $\eta$ 
production in fragmentation; if an $\eta$ is rejected a new flavour 
pair is generated and a new hadron formed.

\iteme{PARJ(26) :} (D=0.4) extra suppression factor for $\eta'$ 
production in fragmentation; if an $\eta'$ is rejected a new flavour 
pair is generated and a new hadron formed.
 
\iteme{PARJ(31) :} (D=0.1 GeV) gives the remaining $W_+$ below which
the generation of a single jet is stopped (it is chosen smaller than a
pion mass, so that no hadrons moving in the forward direction are
missed).
 
\iteme{PARJ(32) :} (D=1. GeV) is, with quark masses added, used to
define the minimum allowable energy of a colour-singlet jet system.
 
\iteme{PARJ(33) - PARJ(34) :} (D=0.8 GeV, 1.5 GeV) are, together with
quark masses, used to define the remaining energy below which the
fragmentation of a jet system is stopped and two final hadrons
formed. \ttt{PARJ(33)} is normally used, except for \ttt{MSTJ(11)=2},
when \ttt{PARJ(34)} is used.
 
\iteme{PARJ(36) :} (D=2.) represents the dependence on the mass of the
final quark pair for defining the stopping point of the fragmentation.
Is strongly correlated to the choice of \ttt{PARJ(33) - PARJ(35)}.
 
\iteme{PARJ(37) :} (D=0.2) relative width of the smearing of the
stopping point energy.
 
\iteme{PARJ(38) - PARJ(39) :} (D=2.5, 0.6) refers to the probability
for reverse rapidity ordering of the final two hadrons,
according to eq.~(\ref{fr:revord}), where
$d_0 = $\ttt{PARJ(39)} for \ttt{MSTJ(11)}$\neq 2$, and
$d = $\ttt{PARJ(39)} for \ttt{MSTJ(11)=2}.
 
\iteme{PARJ(41), PARJ(42) :} (D=0.3, 0.58 GeV$^{-2}$) give the $a$ and
$b$ parameters of the symmetric Lund fragmentation function for
\ttt{MSTJ(11)=}1, 4 and 5 (and \ttt{MSTJ(11)=3} for ordinary hadrons).
 
\iteme{PARJ(43), PARJ(44) :} (D=0.5, 0.9 GeV$^{-2}$) give the $a$ and
$b$ parameters as above for the special case of a gluon jet generated
with IF and \ttt{MSTJ(2)=} 2 or 4.
 
\iteme{PARJ(45) :} (D=0.5) the amount by which the effective $a$
parameter in the Lund flavour dependent symmetric fragmentation
function is assumed to be larger than the normal $a$ when diquarks
are produced. More specifically, referring to eq.~(\ref{fr:LSFFlong}),
$a_{\alpha} = $\ttt{PARJ(41)} when considering the
fragmentation of a quark and = \ttt{PARJ(41) + PARJ(45)} for
the fragmentation of a diquark, with corresponding expression for
$a_{\beta}$ depending on whether the newly created object is a quark
or diquark (for an independent gluon jet generated with
\ttt{MSTJ(2)=} 2 or 4, replace \ttt{PARJ(41)} by \ttt{PARJ(43)}).
In the popcorn model, a meson created in between the baryon and
antibaryon has $a_{\alpha} = a_{\beta} = $\ttt{PARJ(41) + PARJ(45)}.
 
\iteme{PARJ(46) - PARJ(48) :} (D=3*1.) modification of the Lund
symmetric fragmentation for heavy endpoint quarks according to the
recipe by Bowler, available when
\ttt{MSTJ(11)=} 4 or 5 is selected. The shape is given
by eq.~(\ref{fr:LSFFBowler}). If \ttt{MSTJ(11)=4} then
$r_{\Q} = $\ttt{PARJ(46)} for all flavours, while if
\ttt{MSTJ(11)=5} then $r_{\c} = $\ttt{PARJ(46)},
$r_{\b} = $\ttt{PARJ(47)} and $r_{\Q} = $\ttt{PARJ(48)} for
$\t$ and heavier. \ttt{PARJ(46) - PARJ(48)} thus provide a
possibility to interpolate between the `pure' Bowler shape,
$r = 1$, and the normal Lund one, $r = 0$. The additional
modifications made in \ttt{PARJ(43) - PARJ(45)}
are automatically taken into account, if necessary.
 
\iteme{PARJ(51) - PARJ(58) :} (D=3*0.77, $-0.05$, $-0.005$, 
3*$-0.00001$) give a choice of four possible ways to parametrize the 
fragmentation function for \ttt{MSTJ(11)=2} (and \ttt{MSTJ(11)=3} 
for charm and heavier). The fragmentation of each flavour KF may 
be chosen separately; for a diquark the flavour of the heaviest 
quark is used. With $c = $\ttt{PARJ(50+KF)}, the parametrizations 
are: \\
$0 \leq c \leq 1$ : Field-Feynman, $f(z) = 1 - c + 3c(1-z)^2$; \\
$-1 \leq c < 0$ : SLAC, $f(z) = 1/(z(1-1/z-(-c)/(1-z))^2)$; \\
$c > 1$ : power peaked at $z=0$, $f(z) = (1-z)^{c-1}$; \\
$c < -1$ : power peaked at $z=1$, $f(z) = z^{-c-1}$.
 
\iteme{PARJ(59) :} (D=1.) replaces \ttt{PARJ(51) - PARJ(53)} for
gluon jet generated with \ttt{MSTJ(2)=} 2 or 4.
 
\iteme{PARJ(61) - PARJ(63) :} (D=4.5, 0.7, 0.) parametrizes the
energy dependence of the primary multiplicity distribution in
phase-space decays. The former two correspond to $c_1$ and $c_2$ 
of eq.~(\ref{dec:multsel}), while the latter allows a further
additive term in the multiplicity specifically for onium decays.
 
\iteme{PARJ(64) :} (0.003 GeV) minimum kinetic energy in decays
(safety margin for numerical precision errors).
 
\iteme{PARJ(65) :} (D=0.5 GeV) mass which, in addition to the
spectator quark ordiquark mass, is not assumed to partake in the
weak decay of a heavy quark in a hadron.
 
\iteme{PARJ(66) :} (D=0.5) relative probability that colour is
rearranged when two singlets are to be formed from decay products.
Only applies for \ttt{MDME(IDC,2)=} 11--30, i.e. low-mass 
phase-space decays.
 
\iteme{PARJ(71) :} (D=10 mm) maximum average proper lifetime for
particles allowed to decay in the \ttt{MSTJ(22)=2} option. With
the default value, $\K_{\mrm{S}}^0$, $\Lambda$, $\Sigma^-$, $\Sigma^+$,
$\Xi^-$, $\Xi^0$ and $\Omega^-$ are stable (in addition to those
normally taken to be stable), but charm and bottom do still decay.
 
\iteme{PARJ(72) :} (D=1000 mm) maximum distance from the origin at
which a decay is allowed to take place in the \ttt{MSTJ(22)=3}
option.
 
\iteme{PARJ(73) :} (D=100 mm) maximum cylindrical distance
$\rho = \sqrt{x^2 + y^2}$ from the origin at which a decay is allowed
to take place in the \ttt{MSTJ(22)=4} option.
 
\iteme{PARJ(74) :} (D=1000 mm) maximum z distance from the origin
at which a decay is allowed to take place in the \ttt{MSTJ(22)=4}
option.
 
\iteme{PARJ(76) :} (D=0.7) mixing parameter $x_d = \Delta M/\Gamma$
in $\B^0$--$\br{\B}^0$ system.
 
\iteme{PARJ(77) :} (D=10.) mixing parameter $x_s = \Delta M/\Gamma$
in $\B_s^0$--$\br{\B}_s^0$ system.
 
\iteme{PARJ(81) - PARJ(89) :} parameters for time-like parton showers,
see section \ref{ss:showrout}.
 
\iteme{PARJ(91) :} (D=0.020 GeV) minimum particle width in
\ttt{PMAS(KC,2)}, above which particle decays are assumed to take
place before the stage where Bose--Einstein effects are introduced.
 
\iteme{PARJ(92) :} (D=1.) nominal strength of Bose--Einstein effects
for $Q = 0$, see \ttt{MSTJ(51)}. This parameter, often denoted
$\lambda$, expresses the amount of incoherence in particle
production. Due to the simplified picture used for the
Bose--Einstein effects, in particular for effects from three
nearby identical particles, the actual $\lambda$
of the simulated events may be larger than the input value.
 
\iteme{PARJ(93) :} (D=0.20 GeV) size of the Bose--Einstein effect
region in terms of the $Q$ variable, see \ttt{MSTJ(51)}. The more
conventional measure, in terms of the radius $R$ of the production
volume, is given by
$R = \hbar/$\ttt{PARJ(93)}$ \approx 0.2$
fm$\times$GeV/\ttt{PARJ(93)}$ = $\ttt{PARU(3)}/\ttt{PARJ(93)}.

\iteme{PARJ(121) - PARJ(171) :} parameters for $\ee$ event generation,
see section \ref{ss:eeroutines}.
 
\end{entry}

\subsection{Couplings}
\label{ss:coupcons}

In this section we collect information on the two routines for 
running $\alphas$ and $\alphaem$, and on other couplings
of standard and non-standard particles. Although originally begun
for {\Je} applications, this section has rapidly expanded towards
the non-standard aspects, and is thus more of interest for {\Py}
applications than for {\Je} itself. It could therefore equally
well have been put somewhere else in this manual. A few couplings
indeed appear in the \ttt{PARP} array, see section 
\ref{ss:PYswitchpar}.

\drawbox{ALEM = ULALEM(Q2)}\label{p:ULALEM}
\begin{entry}
\itemc{Purpose:} to calculate the running electromagnetic coupling
constant $\alphaem$. Expressions used are described in
ref. \cite{Kle89}. See \ttt{MSTU(101)}, \ttt{PARU(101)},
\ttt{PARU(103)} and \ttt{PARU(104)}.
\iteme{Q2 :} the momentum transfer scale $Q^2$ at which to evaluate
$\alphaem$.
\end{entry}

\drawbox{ALPS = ULALPS(Q2)}\label{p:ULALPS}
\begin{entry}
\itemc{Purpose:} to calculate the running strong coupling constant 
$\alphas$. The first- and second-order expressions are given by 
eqs.~(\ref{ee:aS3j}) and (\ref{ee:aS4j}). See 
\ttt{MSTU(111) - MSTU(118)} 
and \ttt{PARU(111) - PARU(118)} for options.
\iteme{Q2 :} the momentum transfer scale $Q^2$ at which to evaluate
$\alphas$.
\end{entry}
 
\drawbox{COMMON/LUDAT1/MSTU(200),PARU(200),MSTJ(200),PARJ(200)}
\begin{entry}
\itemc{Purpose:} to give access to a number of status codes and
parameters which regulate the performance of the program as a whole.
Here only those related to couplings are described; the main
description is found in section \ref{ss:JETswitch}.

\boxsep
 
\iteme{MSTU(101) :}\label{p:MSTU101} (D=1) procedure for 
$\alphaem$ evaluation in the \ttt{ULALEM} function.
\begin{subentry}
\iteme{= 0 :} $\alphaem$ is taken fixed at the value 
\ttt{PARU(101)}.
\iteme{= 1 :} $\alphaem$ is running with the $Q^2$ scale, 
taking into account corrections from fermion loops ($\e$, $\mu$, 
$\tau$, $\d$, $\u$, $\s$, $\c$, $\b$).
\iteme{= 2 :} $\alphaem$ is fixed, but with separate values at low 
and high $Q^2$. For $Q^2$ below (above) \ttt{PARU(104)} the value 
\ttt{PARU(101)} (\ttt{PARU(103)}) is used. The former value is
then intended for real photon emission, the latter for 
electroweak physics, e.g. of the $\W/\Z$ gauge bosons.

\end{subentry}
 
\iteme{MSTU(111) :} (D=1) order of $\alphas$ evaluation in the
\ttt{ULALPS} function. Is overwritten in \ttt{LUEEVT}, \ttt{LUONIA} or
\ttt{PYINIT} calls with the value desired for the process under study.
\begin{subentry}
\iteme{= 0 :} $\alphas$ is fixed at the value \ttt{PARU(111)}.
\iteme{= 1 :} first-order running $\alphas$ is used.
\iteme{= 2 :} second-order running $\alphas$ is used.
\end{subentry}
 
\iteme{MSTU(112) :} (D=5) the nominal number of flavours assumed in
the $\alphas$ expression, with respect to which $\Lambda$ is defined.
 
\iteme{MSTU(113) :} (D=3) minimum number of flavours that may be
assumed in $\alphas$ expression, see \ttt{MSTU(112)}.
 
\iteme{MSTU(114) :} (D=5) maximum number of flavours that may be
assumed in $\alphas$ expression, see \ttt{MSTU(112)}.
 
\iteme{MSTU(115) :} (D=0) treatment of $\alphas$ singularity for
$Q^2 \to 0$.
\begin{subentry}
\iteme{= 0 :} allow it to diverge like $1/\ln(Q^2/\Lambda^2)$.
\iteme{= 1 :} soften the divergence to $1/\ln(1 + Q^2/\Lambda^2)$.
\iteme{= 2 :} freeze $Q^2$ evolution below \ttt{PARU(114)}, i.e.
the effective argument is $\max(Q^2, $\ttt{PARU(114)}$)$.
\end{subentry}
 
\iteme{MSTU(118) :} (I) number of flavours $n_f$ found and used in
latest \ttt{ULALPS} call.

\boxsep

\iteme{PARU(101) :}\label{p:PARU101} (D=0.00729735=1/137.04) 
$\alphaem$, the electromagnetic fine structure constant at 
vanishing momentum transfer.
 
\iteme{PARU(102) :} (D=0.232) $\ssintw$, the weak mixing angle of the
standard electroweak model.

\iteme{PARU(103) :} (D=0.007764=1/128.8) typical $\alphaem$ in
electroweak processes; used for $Q^2 >$\ttt{PARU(104)} in the
option \ttt{MSTU(101)=2} of \ttt{ULALEM}.

\iteme{PARU(104) :} (D=1 GeV$^2$) dividing line between `low' and
`high' $Q^2$ values in the option \ttt{MSTU(101)=2} of \ttt{ULALEM}.

\iteme{PARU(105) :} (D=1.16639E-5 GeV$^{-2}$) $G_{\F}$, the Fermi 
constant of weak interactions.

\iteme{PARU(108) :} (I) the $\alphaem$ value obtained in the 
latest call to the \ttt{ULALEM} function.
 
\iteme{PARU(111) :} (D=0.20) fix $\alphas$ value assumed in
\ttt{ULALPS} when \ttt{MSTU(111)=0} (and also in parton showers
when $\alphas$ is assumed fix there).
 
\iteme{PARU(112) :} (D=0.25 GeV) $\Lambda$ used in running $\alphas$
expression in \ttt{ULALPS}. Like \ttt{MSTU(111)}, this value is
overwritten by the calling physics routines, and is therefore purely
nominal.
 
\iteme{PARU(113) :} (D=1.) the flavour thresholds, for the effective
number of flavours $n_f$ to use in the $\alphas$ expression, are
assumed to sit at $Q^2 = $\ttt{PARU(113)}$\times m_{\q}^2$, where
$m_{\q}$ is the quark mass. May be overwritten from the calling
physics routine.
 
\iteme{PARU(114) :} (D=4 GeV$^2$) $Q^2$ value below which the
$\alphas$ value is assumed constant for \ttt{MSTU(115)=2}.
 
\iteme{PARU(115) :} (D=10.) maximum $\alphas$ value that \ttt{ULALPS}
will ever return; is used as a last resort to avoid singularities.
 
\iteme{PARU(117) :} (I) $\Lambda$ value (associated with
\ttt{MSTU(118)} effective flavours) obtained in latest \ttt{ULALPS}
call.
 
\iteme{PARU(118) :} (I) $\alphas$ value obtained in latest
\ttt{ULALPS} call.
 
\iteme{PARU(121) - PARU(130) :} couplings of a new $\Z'^0$; for
fermion default values are given by the Standard Model $\Z^0$ values,
assuming $\ssintw = 0.23$. Note that e.g. the $\Z'^0$ width contains
squared couplings, and thus depends quadratically on the values below.
\begin{subentry}
\iteme{PARU(121), PARU(122) :} (D=-0.693,-1.) vector and axial
couplings of down type quarks to $\Z'^0$.
\iteme{PARU(123), PARU(124) :} (D=0.387,1.) vector and axial
couplings of up type quarks to $\Z'^0$.
\iteme{PARU(125), PARU(126) :} (D=-0.08,-1.) vector and axial
couplings of leptons to $\Z'^0$.
\iteme{PARU(127), PARU(128) :} (D=1.,1.) vector and axial
couplings of neutrinos to $\Z'^0$.
\iteme{PARU(129) :} (D=1.) the coupling $Z'^0 \to \W^+ \W^-$ is
taken to be \ttt{PARU(129)}$\times$(the Standard Model
$\Z^0 \to \W^+ \W^-$ coupling)$\times (m_{\W}/m_{\Z'})^2$.
This gives a $\Z'^0 \to \W^+ \W^-$ partial width that
increases proportionately to the $\Z'^0$ mass.
\iteme{PARU(130) :} (D=0.) in the decay chain
$\Z'^0 \to \W^+ \W^- \to 4$ fermions, the angular distribution in
the $\W$ decays is supposed to be a mixture, with fraction
\ttt{1-PARU(130)} corresponding to the same angular distribution
between the four final fermions as in $\Z^0 \to \W^+ \W^-$ (mixture
of transverse and longitudinal $\W$'s), and fraction \ttt{PARU(130)}
corresponding to $\H^0 \to \W^+ \W^-$ the same way (longitudinal
$\W$'s).
\end{subentry}
 
\iteme{PARU(131) - PARU(136) :} couplings of a new $\W'^{\pm}$;
for fermions default values are given by the Standard Model
$\W^{\pm}$ values (i.e. $V-A$). Note that e.g. the $\W'^{\pm}$
width contains squared couplings, and
thus depends quadratically on the values below.
\begin{subentry}
\iteme{PARU(131), PARU(132) :} (D=1.,-1.) vector and axial couplings
of a quark--antiquark pair to $\W'^{\pm}$; is further multiplied by the
ordinary CKM factors.
\iteme{PARU(133), PARU(134) :} (D=1.,-1.) vector and axial couplings
of a lepton-neutrino pair to $\W'^{\pm}$.
\iteme{PARU(135) :} (D=1.) the coupling $\W'^{\pm} \to \Z^0 \W^{\pm}$
is taken to be \ttt{PARU(135)}$\times$(the Standard Model
$\W^{\pm} \to \Z^0 \W^{\pm}$ coupling)$\times (m_{\W}/m_{W'})^2$.
This gives a $\W'^{\pm} \to \Z^0 \W^{\pm}$ partial width that
increases proportionately to the $\W'$ mass.
\iteme{PARU(136) :} (D=0.) in the decay chain
$\W'^{\pm} \to \Z^0 \W^{\pm} \to 4$ fermions,
the angular distribution in the $\W/\Z$ decays is supposed to be a
mixture, with fraction \ttt{1-PARU(130)} corresponding to the same
angular distribution between the four final fermions as in
$\W^{\pm} \to \Z^0 \W^{\pm}$ (mixture of transverse and longitudinal
$\W/\Z$'s), and fraction \ttt{PARU(130)} corresponding to
$\H^{\pm} \to \Z^0 \W^{\pm}$ the same way (longitudinal $\W/\Z$'s).
\end{subentry}
 
\iteme{PARU(141) :} (D=5.) $\tan\beta$ parameter of a two Higgs
doublet scenario, i.e. the ratio of vacuum expectation values.
This affects mass relations and couplings in the Higgs sector.
 
\iteme{PARU(142) :} (D=1.) the $\Z^0 \to \H^+ \H^-$ coupling is
taken to be \ttt{PARU(142)}$\times$(the MSSM $\Z^0 \to \H^+ \H^-$
coupling).
 
\iteme{PARU(143) :} (D=1.) the $\Z'^0 \to \H^+ \H^-$ coupling is
taken to be \ttt{PARU(143)}$\times$(the MSSM $\Z^0 \to \H^+ \H^-$
coupling).
 
\iteme{PARU(145) :} (D=1.) quadratically multiplicative factor in the
$\Z'^0 \to \Z^0 \H^0$ partial width in left--right-symmetric models,
expected to be unity (see \cite{Coc91}).
 
\iteme{PARU(146) :} (D=1.) $\sin(2\alpha)$ parameter, enters
quadratically as multiplicative factor in the
$\W'^{\pm} \to \W^{\pm} \H^0$ partial width in
left--right-symmetric models (see \cite{Coc91}).
 
\iteme{PARU(151) :} (D=1.) multiplicative factor in the
$\L_{\Q} \to \q \ell$ squared Yukawa coupling, and thereby in the
$\L_{\Q}$ partial width and the $\q \ell \to \L_{\Q}$ and other
cross sections. Specifically,
$\lambda^2/(4\pi) = $\ttt{PARU(151)}$\times \alphaem$, i.e. it
corresponds to the $k$ factor of \cite{Hew88}.
 
\iteme{PARU(153) :} (D=0.) anomalous magnetic moment of the
$\W^{\pm}$; $\eta = \kappa - 1$, where $\eta = 0$ ($\kappa = 1$)
is the Standard Model value.
 
\iteme{PARU(155) :} (D=1000. GeV) compositeness scale $\Lambda$.
 
\iteme{PARU(156) :} (D=1.) sign of interference term between
standard cross section and composite term ($\eta$ parameter);
should be $\pm 1$.
 
\iteme{PARU(157) - PARU(159) :} (D=3*1.) strength of {\bf SU(2)},
{\bf U(1)} and {\bf SU(3)} couplings, respectively, in an excited
fermion scenario; cf. $f$, $f'$ and $f_s$ of \cite{Bau90}.
 
\iteme{PARU(161) - PARU(168) :} (D=5*1.,3*0.) multiplicative factors
that can be used to modify the default couplings of the $\H^0$
particle in {\Py}. Note that the factors enter quadratically in the
partial widths. The default values correspond to the couplings given
in the minimal one-Higgs-doublet Standard Model.
\begin{subentry}
\iteme{PARU(161) :} $\H^0$ coupling to down type quarks.
\iteme{PARU(162) :} $\H^0$ coupling to up type quarks.
\iteme{PARU(163) :} $\H^0$ coupling to leptons.
\iteme{PARU(164) :} $\H^0$ coupling to $\Z^0$.
\iteme{PARU(165) :} $\H^0$ coupling to $\W^{\pm}$.
\iteme{PARU(168) :} $\H^0$ coupling to $\H^{\pm}$ in
$\gamma\gamma \to \H^0$ loops, in MSSM
$\sin(\beta-\alpha)+\cos(2\beta)\sin(\beta+\alpha) /
(2\scostw)$.
\end{subentry}
 
\iteme{PARU(171) - PARU(178) :} (D=7*1.,0.) multiplicative factors
that can be used to modify the default couplings of the $\H'^0$
particle in {\Py}. Note that the factors enter quadratically in
partial widths. The default values for \ttt{PARU(171) - PARU(175)}
correspond to the couplings given to $\H^0$ in the minimal
one-Higgs-doublet Standard Model, and are therefore not realistic
in a two-Higgs-doublet scenario. The default values should
be changed appropriately by you. Also the last two default
values should be changed; for these the expressions of the
minimal supersymmetric Standard Model (MSSM) are given to show
parameter normalization.
\begin{subentry}
\iteme{PARU(171) :} $\H'^0$ coupling to down type quarks.
\iteme{PARU(172) :} $\H'^0$ coupling to up type quarks.
\iteme{PARU(173) :} $\H'^0$ coupling to leptons.
\iteme{PARU(174) :} $\H'^0$ coupling to $\Z^0$.
\iteme{PARU(175) :} $\H'^0$ coupling to $W^{\pm}$.
\iteme{PARU(176) :} $\H'^0$ coupling to $\H^0 \H^0$, in MSSM
$\cos(2\alpha) \cos(\beta+\alpha) - 2 \sin(2\alpha)
\sin(\beta+\alpha)$.
\iteme{PARU(177) :} $\H'^0$ coupling to $\A^0 \A^0$, in MSSM
$\cos(2\beta) \cos(\beta+\alpha)$.
\iteme{PARU(178) :} $\H'^0$ coupling to $\H^{\pm}$ in
$\gamma \gamma \to \H'^0$ loops, in MSSM
$\cos(\beta-\alpha) - \cos(2\beta)\cos(\beta+\alpha) /
(2\scostw)$.
\end{subentry}
 
\iteme{PARU(181) - PARU(190) :} (D=3*1.,2*0.,2*1.,3*0.)
multiplicative factors that can be used to modify the default
couplings of the $\A^0$ particle in PYTHIA. Note that the factors 
enter quadratically in partial widths. The default values for 
\ttt{PARU(181) - PARU(183)} correspond
to the couplings given to $\H^0$ in the minimal one-Higgs-doublet
Standard Model, and are therefore not realistic in a
two-Higgs-doublet scenario. The default values should
be changed appropriately by you. \ttt{PARU(184)} and \ttt{PARU(185)}
should be vanishing at the tree level, and are so set;
normalization of these couplings agrees with what is used for
H and $\H'^0$. Also the other default values should be changed; for
these the expressions of the Minimal Supersymmetric Standard
Model (MSSM) are given to show parameter normalization.
\begin{subentry}
\iteme{PARU(181) :} $\A^0$ coupling to down type quarks.
\iteme{PARU(182) :} $\A^0$ coupling to up type quarks.
\iteme{PARU(183) :} $\A^0$ coupling to leptons.
\iteme{PARU(184) :} $\A^0$ coupling to $\Z^0$.
\iteme{PARU(185) :} $\A^0$ coupling to $\W^{\pm}$.
\iteme{PARU(186) :} $\A^0$ coupling to $\Z^0 \H^0$ (or
$\Z^*$ to $\A^0 \H^0$), in MSSM $\cos(\beta-\alpha)$.
\iteme{PARU(187) :} $\A^0$ coupling to $\Z^0 \H'^0$ (or $\Z^*$ to
$\A^0 \H'^0$), in MSSM $\sin(\beta-\alpha)$.
\iteme{PARU(188) :} As \ttt{PARU(186)}, but coupling to $\Z'^0$
rather than $\Z^0$.
\iteme{PARU(189) :} As \ttt{PARU(187)}, but coupling to $\Z'^0$
rather than $\Z^0$.
\iteme{PARU(190) :} $\A^0$ coupling to $\H^{\pm}$ in
$\gamma \gamma \to \A^0$ loops, 0 in MSSM.
\end{subentry}
 
\iteme{PARU(191) - PARU(195) :} (D=4*0.,1.) multiplicative factors
that can be used to modify the couplings of the $\H^{\pm}$ particle
in {\Py}. Currently only \ttt{PARU(195)} is in use. See above for
related comments.
\begin{subentry}
\iteme{PARU(195) :} $\H^{\pm}$ coupling to $\W^{\pm} \H^0$ (or
$\W^{* \pm}$ to $\H^{\pm} \H^0$), in MSSM $\cos(\beta-\alpha)$.
\end{subentry}

\end{entry}
 
\subsection{Further Parameters and Particle Data}
\label{ss:parapartdat}
 
The following common blocks are maybe of a more peripheral interest,
with the exception of the \ttt{MDCY} array, which allows a selective
inhibiting of particle decays, and masses of not yet discovered
particles, such as \ttt{PMAS(6,1)}, the top quark mass.
 
\drawbox{COMMON/LUDAT2/KCHG(500,3),PMAS(500,4),PARF(2000),VCKM(4,4)}%
\label{p:LUDAT2}
\begin{entry}
\itemc{Purpose:} to give access to a number of flavour treatment
constants or parameters and particle/parton data. Particle data is
stored by compressed code KC rather than by the full KF code.
You are reminded that the way to know the KC value is to use the 
\ttt{LUCOMP} function, i.e. \ttt{KC = LUCOMP(KF)}.
 
\boxsep
 
\iteme{KCHG(KC,1) :}\label{p:KCHG} three times particle/parton charge 
for compressed code KC.
 
\iteme{KCHG(KC,2) :} colour information for compressed code KC.
\begin{subentry}
\iteme{= 0 :} colour-singlet particle.
\iteme{= 1 :} quark or antidiquark.
\iteme{= -1 :} antiquark or diquark.
\iteme{= 2 :} gluon.
\end{subentry}
 
\iteme{KCHG(KC,3) :} particle/antiparticle distinction for
compressed code KC.
\begin{subentry}
\iteme{= 0 :} the particle is its own antiparticle.
\iteme{= 1 :} a nonidentical antiparticle exists.
\end{subentry}
 
\boxsep
 
\iteme{PMAS(KC,1) :}\label{p:PMAS} particle/parton mass $m$ (in GeV) 
for compressed code KC.
 
\iteme{PMAS(KC,2) :} the total width $\Gamma$ (in GeV) of an assumed
symmetric Breit--Wigner mass shape for compressed particle code KC.
 
\iteme{PMAS(KC,3) :} the maximum deviation (in GeV) from the
\ttt{PMAS(KC,1)} value at which the Breit--Wigner shape above is
truncated. (Is used in particle decays, but not in the {\Py} resonance
treatment; cf. the \ttt{CKIN} variables.)
 
\iteme{PMAS(KC,4) :} the average lifetime $\tau$ for compressed
particle code KC, with $c \tau$ in mm, i.e. $\tau$ in units of
about $3.33 \times 10^{-12}$ s.
 
\boxsep
 
\iteme{PARF(1) - PARF(60) :}\label{p:PARF} give a parametrization of 
the $\d\dbar$--$\u\ubar$--$\s\sbar$ flavour mixing in production of
flavour-diagonal mesons. Numbers are stored in groups of 10, for
the six multiplets pseudoscalar, vector, axial vector ($S=0$),
scalar, axial vector ($S=1$) and tensor, in this order; see
section \ref{sss:mesonprod}. Within each group, the first two 
numbers determine the fate of a $\d\dbar$ flavour state, the 
second two that of a $\u\ubar$ one, the next two that of an 
$\s\sbar$ one, while the last four are unused. Call the numbers 
of a pair $p_1$ and $p_2$. Then the probability to produce the 
state with smallest KF code is $1-p_1$, the probability for the 
middle one is $p_1 - p_2$ and the probability for the one with 
largest code is $p_2$, i.e. $p_1$ is the probability to produce 
either of the two `heavier' ones.
 
\iteme{PARF(61) - PARF(80) :} give flavour {\bf SU(6)} weights for
the production of a spin 1/2 or spin 3/2 baryon from a given
diquark--quark combination. Should not be changed.
 
\iteme{PARF(101) - PARF(108) :} first five contain  $\d$, $\u$, $\s$,
$\c$ and $\b$ constituent masses, as to be used in mass formulae, and
should not be changed. For $\t$, $\lrm$ and $\hrm$ masses the current
values stored in \ttt{PMAS(6,1) - PMAS(8,1)} are copied in.
 
\iteme{PARF(111), PARF(112) :} (D=0.0, 0.11 GeV) constant terms in the
mass formulae for heavy mesons and baryons, respectively (with diquark
getting 2/3 of baryon).
 
\iteme{PARF(113), PARF(114) :} (D=0.16,0.048 GeV) factors which,
together with Clebsch-Gordan coefficients and quark constituent
masses, determine the mass splitting due to spin-spin interactions
for heavy mesons and baryons, respectively. The latter factor is
also used for the splitting between spin 0 and spin 1 diquarks.
 
\iteme{PARF(115) - PARF(118) :} (D=0.50, 0.45, 0.55, 0.60 GeV),
constant mass terms, added to the constituent masses, to get the
mass of heavy mesons with orbital angular momentum $L = 1$. The four
numbers are for pseudovector mesons with quark spin 0, and for scalar,
pseudovector and tensor mesons with quark spin 1, respectively.
 
\iteme{PARF(121), PARF(122) :} (D=0.1, 0.2 GeV) constant terms, which
are subtracted for quark and diquark masses, respectively, in defining
the allowed phase space in particle decays into partons.
 
\iteme{PARF(201) - PARF(1960) :} (D=1760*0) relative probabilities
for flavour production in the \ttt{MSTJ(15)=1} option; to be
defined by you before any {\Je} calls. \\
The index in \ttt{PARF} is of the compressed form \\
$120 + 80 \times$KTAB1$ + 25 \times$KTABS$ + $KTAB3. \\
Here KTAB1 is the
old flavour, fixed by preceding fragmentation history, while KTAB3
is the new flavour, to be selected according to the relevant relative
probabilities (except for the very last particle, produced when
joining two jets, where both KTAB1 and KTAB3 are known).
Only the most frequently appearing quarks/diquarks are defined,
according to the code $1 = \d$, $2 = \u$, $3 = \s$, $4 = \c$,
$5 = \b$, $6 = \t$, $7 = \d\d_1$, $8 = \u\d_0$, $9 = \u\d_1$,
$10 = \u\u_1$, $11 = \s\d_0$, $12 = \s\d_1$, $13 = \s\u_0$,
$14 = \s\u_1$, $15 = \s\s_1$, $16 = \c\d_0$, $17 = \c\d_1$,
$18 = \c\u_0$, $19 = \c\u_1$, $20 = \c\s_0$, $21 = \c\s_1$,
$22 = \c\c_1$.
These are thus the only possibilities for the new flavour to be
produced; for an occasional old flavour not on this list, the
ordinary relative flavour production probabilities will be used. \\
Given the initial and final flavour, the intermediate hadron that
is produced is almost fixed. (Initial and final diquark here
corresponds to `popcorn' production of mesons intermediate between
a baryon and an antibaryon). The additional index KTABS gives the
spin type of this hadron, with \\
0 = pseudoscalar meson or $\Lambda$-like spin 1/2 baryon, \\
1 = vector meson or $\Sigma$-like spin 1/2 baryon, \\
2 = tensor meson or spin 3/2 baryon. \\
(Some meson multiplets, not frequently produced, are not accessible
by this parametrization.) \\
Note that some combinations of KTAB1, KTAB3 and KTABS do not
correspond to a physical particle (a $\Lambda$-like baryon must
contain three different quark flavours, a $\Sigma$-like one at least
two), and that you must see to it that the corresponding
\ttt{PARF} entries are vanishing. One additional complication exist
when KTAB3 and KTAB1 denote the same flavour content (normally
KTAB3$ = $KTAB1, but for diquarks the spin freedom may give
KTAB3$ = $KTAB1$\pm 1$): then a flavour neutral meson is to be
produced, and here $\d\dbar$, $\u\ubar$ and $\s\sbar$ states mix
(heavier flavour states do not, and these are therefore no problem).
For these cases the ordinary KTAB3 value gives the total probability
to produce either of the mesons possible, while KTAB3$ = $23 gives the
relative probability to produce the lightest meson state ($\pi^0$,
$\rho^0$, $\a_2^0$), KTAB3$ = $24 relative probability for the middle
meson ($\eta$, $\omega$, $\f_2^0$), and KTAB3 = 25 relative
probability for the heaviest one ($\eta'$, $\phi$, $f'^0_2$). Note
that, for simplicity, these relative probabilities are assumed the
same whether initial and final diquark have the same spin or not; the
total probability may well be assumed different, however. \\
As a general comment, the sum of \ttt{PARF} values for a given KTAB1
need not be normalized to unity, but rather the program will find the
sum of relevant weights and normalize to that. The same goes for
the KTAB3$ = $23--25 weights. This makes it straightforward to use
one common setup of \ttt{PARF} values and still switch between
different \ttt{MSTJ(12)} baryon production modes.
 
\boxsep
 
\iteme{VCKM(I,J) :}\label{p:VCKM} squared matrix elements of the
Cabibbo-Kobayashi-Maskawa flavour mixing matrix.
\begin{subentry}
\iteme{I :} up type generation index, i.e. $1 = \u$, $2 = \c$,
$3 = \t$ and $4 = \hrm$.
\iteme{J :} down type generation index, i.e. $1 = \d$, $2 = \s$,
$3 = \b$ and $4 = \lrm$.
\end{subentry}
 
\end{entry}
 
\drawbox{COMMON/LUDAT3/MDCY(500,3),MDME(2000,2),BRAT(2000),%
KFDP(2000,5)}\label{p:LUDAT3}
\begin{entry}
\itemc{Purpose:} to give access to particle decay data and
parameters. In particular, the \ttt{MDCY(KC,1)} variables may be 
used to switch on or off the decay of a given particle species, 
and the \ttt{MDME(IDC,1)} ones
to switch on or off an individual decay channel of a particle.
For quarks, leptons and gauge bosons, a number of decay channels
are included that are not allowed for on-mass-shell particles, see
\ttt{MDME(IDC,2)=102}. These channels are not currently used in
{\Je}, but instead find applications in {\Py}. Particle data is
stored by compressed code KC rather than by the full KF code.
You are reminded that the way to know the KC value is to use the 
\ttt{LUCOMP} function, i.e. \ttt{KC = LUCOMP(KF)}.
 
\boxsep
 
\iteme{MDCY(KC,1) :}\label{p:MDCY} switch to tell whether a particle 
with compressed code KC may be allowed to decay or not. 
\begin{subentry}
\iteme{= 0 :} the particle is not allowed to decay.
\iteme{= 1 :} the particle is allowed to decay (if decay information
is defined below for the particle).
\end{subentry} 
 
\iteme{MDCY(KC,2) :} gives the entry point into the decay channel
table for compressed particle code KC. Is 0 if no decay channels 
have been defined.
 
\iteme{MDCY(KC,3) :} gives the total number of decay channels defined
for compressed particle code KC, independently of whether they have
been assigned a non-vanishing branching ratio or not. Thus the
decay channels are found in positions \ttt{MDCY(KC,2)} to
\ttt{MDCY(KC,2)+MDCY(KC,3)-1}.
 
\boxsep
 
\iteme{MDME(IDC,1) :}\label{p:MDME} on/off switch for individual 
decay channel IDC. In addition, a channel may be left selectively 
open; this has some special applications in {\Py} which are not 
currently used in {\Je}. Effective branching ratios are 
automatically recalculated for the decay channels left open. 
Also process cross sections are affected; see section 
\ref{sss:resdecaycross}. If a particle is allowed to decay by the 
\ttt{MDCY(KC,1)} value, at least one channel must be left open
by you. A list of decay channels with current IDC numbers
may be obtained with \ttt{LULIST(12)}.
\begin{subentry}
\iteme{= -1 :} this is a non-Standard Model decay mode, which by
default is assumed not to exist. Normally, this option is used for
decays involving fourth generation or $\H^{\pm}$ particles.
\iteme{= 0 :} channel is switched off.
\iteme{= 1 :} channel is switched on.
\iteme{= 2 :} channel is switched on for a particle but off for an
antiparticle. It is also on for a particle its own antiparticle,
i.e. here it means the same as \ttt{=1}.
\iteme{= 3 :} channel is switched on for an antiparticle but off for
a particle. It is off for a particle its own antiparticle.
\iteme{= 4 :} in the production of a pair of equal or charge
conjugate resonances in {\Py}, say $\H^0 \to \W^+ \W^-$, either one
of the resonances is allowed to decay according to this group of
channels, but not both. If the two particles of the pair
are different, the channel is on.
Within {\Je}, this option only means that the channel is
switched off.
\iteme{= 5 :} as \ttt{=4}, but an independent group of channels, such
that in a pair of equal or charge conjugate resonances the decay of
either resonance may be specified independently. If the two
particles in the pair are different, the channel is off.
Within {\Je}, this option only means that the channel is
switched off.
\itemc{Warning:} the two values -1 and 0 may look similar, but in fact
are quite different. In neither case the channel so set is
generated, but in the latter case the channel still contributes
to the total width of a resonance, and thus affects both
simulated line shape and the generated cross section when
{\Py} is run. The value 0 is appropriate to a channel
we assume exists, even if we are not currently simulating it,
while -1 should be used for channels we believe do not exist.
In particular, you are warned unwittingly to set fourth
generation channels 0 (rather than -1), since by now the
support for a fourth generation is small.
\itemc{Remark:} all the options above may be freely mixed. The
difference, for those cases where both make sense, between using
values 2 and 3 and using 4 and 5 is that the latter automatically
include charge conjugate states, e.g.
$\H^0 \to \W^+ \W^- \to \e^+ \nu_e \d \ubar$ or
$\dbar \u \e^- \br{\nu}_e$, but the former only one
of them. In calculations of the joint branching ratio, this
makes a factor 2 difference.
\itemc{Example:} to illustrate the above options, consider the case
of a $\W^+ \W^-$ pair. One might then set the following combination
of switches for the $\W$:\\
\begin{tabular}{ccl@{\protect\rule{0mm}{\tablinsep}}}
channel & value & comment \\
$\u\dbar$ & 1 & allowed for $\W^+$ and $\W^-$ in any combination, \\
$\u\sbar$ & 0 & never produced but contributes to $\W$ width, \\
$\c\dbar$ & 2 & allowed for $\W^+$ only, \\
$\c\sbar$ & 3 & allowed for $\W^-$ only, i.e. properly 
$\W^- \to \cbar\s$, \\
$\t\bbar$ & 0 & never produced but contributes to $\W$ width \\
  &  & if the channel is kinematically allowed, \\
$\nu_{\e}\e^+$ & 4 & allowed for one of $\W^+$ or $\W^-$, but not 
both, \\
$\nu_{\mu}\mu^+$ & 4 & allowed for one of $\W^+$ or $\W^-$, but not 
both, \\
  &  & and not in combination with $\nu_{\e}\e^+$, \\
$\nu_{\tau}\tau^+$ & 5 & allowed for the other $\W$, but not both, \\
$\nu_{\chi}\chi^-$ & $-1$ & not produced and does not contribute to 
$\W$ width.\\
\end{tabular}\\
A $\W^+\W^-$ final state $\u\dbar + \cbar\s$  is allowed, but not its
charge conjugate $\ubar\d + \c\sbar$, since the latter decay mode is 
not allowed for a $\W^+$. The combination 
$\nu_{\e}\e^+ + \bar{\nu}_{\tau}\tau^-$ is allowed, since the two 
channels belong to different groups, but not 
$\nu_{\e}\e^+ + \bar{\nu}_{\mu}\mu^-$, where both belong to the same.
Both $\u\dbar + \bar{\nu}_{\tau}\tau^-$ and 
$\ubar\d + \nu_{\tau}\tau^+$ are allowed, since there is no clash.
The full rulebook, for this case, is given by 
eq.~(\ref{eq:WWallchancombgen}). A term $r_i^2$ means channel 
$i$ is allowed
for $\W^+$ and $\W^-$ simultaneously, a term $r_i r_j$ that channels
$i$ and $j$ may be combined, and a term $2 r_i r_j$ that channels 
$i$ and $j$ may be combined two ways, i.e. that also a charge 
conjugate combination is allowed. 
\end{subentry}
 
\iteme{MDME(IDC,2) :} information on special matrix-element treatment
for decay channel IDC. In addition to the outline below, special rules
apply for the order in which decay products should be given, so that
matrix elements and colour flow is properly treated. One such example
is the weak matrix elements, which only will be correct if decay
products are given in the right order. The program does not police 
this, so if you introduce channels of your own and use these codes,
you should be guided by the existing particle data. 
\begin{subentry}
\iteme{= 0 :} no special matrix-element treatment; partons and
particles are copied directly to the event record, with momentum
distributed according to phase space.
\iteme{= 1 :} $\omega$ and $\phi$ decays into three pions, eq.
(\ref{dec:omegphi}).
\iteme{= 2 :} $\pi^0$ or $\eta$ Dalitz decay to $\gamma \e^+ \e^-$,
eq.~(\ref{dec:Dalitz}).
\iteme{= 3 :} used for vector meson decays into two pseudoscalars,
to signal non-isotropic decay angle according to eq.
(\ref{dec:psvpsps}), where relevant.
\iteme{= 4 :} decay of a spin 1 onium resonance to three gluons or
to a photon and two gluons, eq.~(\ref{ee:Upsilondec}). The gluons may
subsequently develop a shower if \ttt{MSTJ(23)=1}.
\iteme{= 11 :} phase-space production of hadrons from the quarks
available.
\iteme{= 12 :} as \ttt{=11}, but for onia resonances, with the option
of modifying the multiplicity distribution separately.
\iteme{= 13 :} as \ttt{=11}, but at least three hadrons to be produced
(useful when the two-body decays are given explicitly).
\iteme{= 14 :} as \ttt{=11}, but at least four hadrons to be produced.
\iteme{= 15 :} as \ttt{=11}, but at least five hadrons to be produced.
\iteme{= 22 - 30 :} phase-space production of hadrons from the quarks
available, with the multiplicity fixed to be \ttt{MDME(IDC,2)-20},
i.e. 2--10.
\iteme{= 31 :} two or more quarks and particles are distributed
according to phase space. If three or more products, the last product
is a spectator quark, i.e. sitting at rest with respect to the
decaying hadron.
\iteme{= 32 :} a $\q\qbar$ or $\g\g$ pair, distributed according to
phase space (in angle), and allowed to develop a shower if
\ttt{MSTJ(23)=1}.
\iteme{= 33 :} a triplet $\q X \qbar$, where $X$ is either a gluon or
a colour-singlet particle; the final particle ($\qbar$) is assumed to
sit at rest with respect to the decaying hadron, and the
two first particles ($\q$ and $X$) are allowed to develop a shower if
\ttt{MSTJ(23)=1}.
\iteme{= 41 :} weak decay, where particles are distributed according
to phase space, multiplied by a factor from the expected shape of
the momentum spectrum of the direct product of the weak decay
(the $\nu_{\tau}$ in $\tau$ decay).
\iteme{= 42 :} weak decay matrix element for quarks and leptons.
Products may be given either in terms of quarks or hadrons, or
leptons for some channels. If the spectator system is given in
terms of quarks, it is assumed to collapse into one particle from
the onset. If the virtual $\W$ decays into quarks, these quarks
are converted to particles, according to phase space in the
$\W$ rest frame, as in \ttt{=11}. Is intended for $\tau$, charm and
bottom.
\iteme{= 43 :} as \ttt{=42}, but if the $\W$ decays into quarks, these
will either appear as jets or, for small masses, collapse into a
one- or two-body system.
\iteme{= 44 :} weak decay matrix element for quarks and leptons, where
the spectator system may collapse into one particle for a small
invariant mass. If the first two decay products are a $\q\qbar'$
pair, they may develop a parton shower if \ttt{MSTJ(23)=1}.
Is intended for top and beyond, but largely superseded by
the following option.
\iteme{= 45 :} weak decay $\q \to \W \q'$ or $\ell \to \W \nu_{\ell}$,
where the $\W$ is registered as a decay product and subsequently
treated with \ttt{MDME=46}. To distinguish from ordinary $\W$'s
on the mass shell, code KF$ = \pm 89$ is used. The virtual $\W$ mass
is selected according to the standard weak decay matrix element,
times the $\W$ propagator (for \ttt{MSTJ(25)=1}). There may be two
or three decay products; if a third this is a spectator taken to sit
at rest. The spectator system may collapse into one particle. Is
intended for top and beyond.
\iteme{= 46 :} $\W$ (KF = 89) decay into $\q\qbar'$ or
$\ell \nu_{\ell}$ according to relative probabilities given
by couplings (as stored in the \ttt{BRAT} vector)
times a dynamical phase-space factor given by the current $\W$
mass. In the decay, the correct $V-A$ angular distribution is
generated if the $\W$ origin is known (heavy quark or lepton).
This is therefore the second step of a decay with \ttt{MDME=45}.
A $\q\qbar'$ pair may subsequently develop a shower if 
\ttt{MSTJ(23)=1}.
\iteme{= 84 - 88 :} map the decay of this particle onto the generic
$\c$, $\b$, $\t$, $\lrm$ or $\hrm$ decay modes defined for
KC$ = $84--88.
\iteme{= 48 :} as \ttt{=42}, but require at least three decay
products.
\iteme{= 101 :} this is not a proper decay channel, but only to be
considered as a continuation line for the decay product listing
of the immediately preceding channel. Since the \ttt{KFDP} array can
contain five decay products per channel, with this code it is
possible to define channels with up to ten decay products. It is
not allowed to have several continuation lines after each other.
\iteme{= 102 :} this is not a proper decay channel for a decaying
particle on the mass shell (or nearly so), and is therefore assigned
branching ratio 0. For a particle off the mass shell, this decay
mode is allowed, however. By including this channel among the
others, the switches \ttt{MDME(IDC,1)} may be used to allow or
forbid these channels in hard processes, with cross sections
to be calculated separately. As an example, $\gamma \to \u \ubar$
is not possible for a massless photon, but is an allowed
channel in $\ee$ annihilation.
\end{subentry}
 
\boxsep
 
\iteme{BRAT(IDC) :}\label{p:BRAT} give branching ratios for the 
different decay
channels. In principle, the sum of branching ratios for a given
particle should be unity. Since the program anyway has to calculate
the sum of branching ratios left open by the \ttt{MDME(IDC,1)} values
and normalize to that, you need not explicitly ensure this
normalization, however. (Warnings are printed in \ttt{LUUPDA(2)} calls
if the sum is not unity, but this is entirely intended as a help
for finding user mistypings.) For decay channels with
\ttt{MDME(IDC,2)}$> 80$ the \ttt{BRAT} values are dummy.
 
\boxsep
 
\iteme{KFDP(IDC,J) :}\label{p:KFDP} contain the decay products in the 
different channels, with five positions \ttt{J=} 1--5 reserved for each
channel IDC. The decay products are given following the standard KF
code for jets and particles, with 0 for trailing empty positions. Note
that the \ttt{MDME(IDC+1,2)=101} option allows you to double the
maximum number of decay product in a given channel from 5 to 10,
with the five latter products stored \ttt{KFDP(IDC+1,J)}.
 
\end{entry}
 
\drawboxtwo{COMMON/LUDAT4/CHAF(500)}{CHARACTER CHAF*8}\label{p:LUDAT4}
\begin{entry}
\itemc{Purpose:} to give access to character type variables.
 
\boxsep
 
\iteme{CHAF :}\label{p:CHAF} particle names (excluding charge) 
according to KC code.
\end{entry}
 
\subsection{Miscellaneous Comments}
 
The previous sections have dealt with the subroutine options and
variables one at a time. This is certainly important, but for a full
use of the capabilities of the program, it is also necessary
to understand how to make different pieces work together. This is
something that cannot be explained fully in a manual, but must also
be learnt by trial and error. This section contains some examples of
relationships between subroutines, common blocks and parameters.
It also contains comments on issues that did not fit in naturally
anywhere else, but still might be useful to have on record.
 
\subsubsection{Interfacing to detector simulation}
 
Very often, the output of the program is to be fed into a subsequent
detector simulation program. It therefore becomes necessary to set up
an interface between the \ttt{LUJETS} common block and the detector
model. Preferrably this should be done via the \ttt{HEPEVT} standard
common block, see section \ref{ss:HEPEVT}, but sometimes this may not
be convenient. If a \ttt{LUEDIT(2)} call is made, the remaining
entries exactly correspond to those an ideal detector could see: all
non-decayed particles, with the exception of neutrinos. The translation
of momenta should be trivial (if need be, a \ttt{LUROBO} call can be
made to rotate the `preferred' $z$ direction to whatever is the
longitudinal direction of the detector), and so should the translation
of particle codes. In particular, if the
detector simulation program also uses the standard Particle Data Group
codes, no conversion at all is needed. The problem then is to select
which particles are allowed to decay, and how decay vertex information
should be used.
 
Several switches regulate which particles are allowed to decay. First,
the master switch \ttt{MSTJ(21)} can be used to switch on/off all
decays (and it also contains a choice of how fragmentation should
be interfaced). Second, a particle must have decay modes defined for
it, i.e. the corresponding \ttt{MDCY(KC,2)} and \ttt{MDCY(KC,3)}
entries must be non-zero for compressed code \ttt{KC = LUCOMP(KF)}.
This is true for all colour neutral particles except the neutrinos,
the photon, the proton and the neutron. (This statement is actually
not fully correct, since irrelevant `decay modes' with
\ttt{MDME(IDC,2)=102} exist in some cases.) Third, the
individual switch in \ttt{MDCY(KC,1)} must be on. Of all the particles
with decay modes defined, only $\mu^{\pm}$, $\pi^{\pm}$, $\K^{\pm}$
and $\K_{\mrm{L}}^0$ are by default considered stable.
 
Finally, if \ttt{MSTJ(22)} does not have its default value 1, checks
are also made on the lifetime of a particle before it is allowed to
decay. In the simplest alternative, \ttt{MSTJ(22)=2}, the comparison
is based on the average lifetime, or rather $c \tau$, measured in mm.
Thus if the limit \ttt{PARJ(71)} is (the default) 10 mm, then decays
of $\K_{\mrm{S}}^0$, $\Lambda$, $\Sigma^-$, $\Sigma^+$, $\Xi^-$, 
$\Xi^0$ and $\Omega^-$ are all switched off, but charm and bottom 
still decay. No $c \tau$ values below 1 micron are defined. With the 
two options \ttt{MSTJ(22)=} 3 or 4, a spherical or cylindrical volume 
is defined around the origin, and all decays taking place inside this
volume are ignored.
 
Whenever a particle is in principle allowed to decay, i.e.
\ttt{MSTJ(21)} and \ttt{MDCY} on, an proper lifetime is selected
once and for all and stored in \ttt{V(I,5)}. The \ttt{K(I,1)} is then
also changed to 4. For \ttt{MSTJ(22)=1}, such a particle will also
decay, but else it could remain in the event record. It is then
possible, at a later stage, to expand the volume inside which decays
are allowed, and do a new \ttt{LUEXEC} call to have particles
fulfilling the new conditions (but not the old) decay. As a further
option, the \ttt{K(I,1)} code may be put to 5, signalling that the
particle will definitely decay in the next \ttt{LUEXEC} call, at the
vertex position given (by the user) in the \ttt{V} vector.
 
This then allows the {\Je} decay routines to be used inside a
detector simulation program, as follows. For a particle which did not
decay before entering the detector, its point of decay is still well
defined (in the absence of deflections by electric or magnetic fields),
eq.~(\ref{dec:newvertex}). If it interacts before that point, the 
detector simulation program is left to handle things. If not, the 
\ttt{V} vector is updated according to the formula above, \ttt{K(I,1)} 
is set to 5, and \ttt{LUEXEC} is called, to give a set of decay 
products, that can again be tracked.
 
A further possibility is to force particles to decay into specific
decay channels; this may be particularly interesting for charm or
bottom physics. The choice of channels left open is determined by the
values of the switches \ttt{MDME(IDC,1)} for decay channel IDC
(use \ttt{LULIST(12)} to obtain the full listing). One or several
channels may be left open;
in the latter case effective branching ratios are automatically
recalculated without the need for your intervention. It is also
possible to differentiate between which channels are left open for
particles and which for antiparticles. Lifetimes are not affected by
the exclusion of some decay channels. Note that, whereas forced decays
can enhance the efficiency for several kinds of studies, it can
also introduce unexpected biases, in particular when events may contain
several particles with forced decays, cf. section 
\ref{sss:resdecaycross}.
 
\subsubsection{Parameter values}
 
A non-trivial question is to know which parameter values to use. The
default values stored in the program are based on comparisons with 
LEP $\ee \to \Z^0$ data at around 91 GeV \cite{LEP90}, using a 
parton-shower picture followed by string fragmentation. If 
fragmentation is indeed an universal phenomenon, as
we would like to think, then the same parameters should also apply at
other energies and in other processes. The former aspect, at least, 
seems to be borne out by comparisons with lower-energy PETRA/PEP
data. Note, however, that the choice of parameters is intertwined with 
the choice of perturbative QCD description. If instead matrix elements 
are used, a best fit to 30 GeV data would require the values 
\ttt{PARJ(21)=0.40}, \ttt{PARJ(41)=1.0} and \ttt{PARJ(42)=0.7}. 
With matrix elements one does not expect an energy
independence of the parameters, since the effective minimum invariant
mass cut-off is then energy dependent, i.e. so is the amount of soft
gluon emission effects lumped together with the fragmentation
parameters. This is indeed confirmed by the LEP data.
A mismatch in the perturbative QCD treatment could also
lead to small differences between different processes.
 
It is often said that the string fragmentation model contains a wealth
of parameters. This is certainly true, but it must be remembered that
most of these deal with flavour properties, and to a large extent
factorize from the treatment of the general event shape. In a fit to
the latter it is therefore usually enough to consider the parameters of
the perturbative QCD treatment, like $\Lambda$ in $\alphas$ and a
shower cut-off $Q_0$ (or $\alphas$ itself and $y_{\mmin}$, if matrix
elements are used), the $a$ and $b$ parameter of the Lund symmetric
fragmentation function (\ttt{PARJ(41)} and \ttt{PARJ(42)}) and the
width of the transverse momentum distribution
($\sigma = $\ttt{PARJ(21)}). In addition, the $a$ and $b$ parameters
are very strongly correlated by the requirement of having the correct
average multiplicity, such that in a typical $\chi^2$ plot, the
allowed region corresponds to a very narrow but very long valley,
stretched diagonally from small ($a$,$b$) pairs to large ones.
As to the flavour parameters, these are certainly many more, but most
of them are understood qualitiatively within one single framework, that
of tunnelling pair production of flavours.
 
Since the use of independent fragmentation has fallen in disrespect, it
should be pointed out that the default parameters here are not
particularly well tuned to the data. This especially applies if one,
in addition to asking for independent fragmentation, also asks for
another setup of fragmentation functions, i.e. other than the standard
Lund symmetric one. In particular, note that most fits to the popular
Peterson et al. (SLAC) heavy-flavour fragmentation function are based
on the actual observed spectrum. In a Monte Carlo simulation, one must
then start out with something harder, to compensate for the energy lost
by initial-state photon radiation and gluon bremsstrahlung. Since
independent fragmentation is not collinear safe (i.e, the emission of
a collinear gluon changes the properties of the final event), the tuning
is strongly dependent on the perturbative QCD treatment chosen. All the
parameters needed for a tuning of independent fragmentation are
available, however.
 
\subsubsection{Particle properties}
 
The masses of most frequently used particles are taken from tables.
For some rare charm and bottom hadrons, and for heavier flavour
hadrons, this would be unwieldy, and instead mass formulae are used,
based on the quark content. For the known quarks $\d$, $\u$, $\s$,
$\c$ and $\b$, the masses used for this purpose are actually the ones
stored in positions 101--105 in the \ttt{PARF} vector, rather than
the ones found in \ttt{PMAS}. This means that the \ttt{PMAS} masses
can be freely changed by you, to modify the masses that appear
in the event record, without courting disaster elsewhere (since mass
formulae typically contain $1/m$ terms from spin-spin splittings, it
is necessary to have the non-zero `constituent' masses here). Thus you
should never touch the mass values stored in \ttt{PARF}. For the
heavier flavours top, low and high, the current \ttt{PMAS}
values are always used. For these flavours, the only individually
defined hadrons are the flavour neutral $\eta$, $\Theta$,
$\hrm_1$, $\chi_0$, $\chi_1$ and $\chi_2$ states. A complete change 
of top mass in the program thus requires changing \ttt{PMAS(6,1)}, 
\ttt{PMAS(LUCOMP(661),1)}, \ttt{PMAS(LUCOMP(663),1)}, 
\ttt{PMAS(LUCOMP(665),1)}, \ttt{PMAS(LUCOMP(10661),1)},
\ttt{PMAS(LUCOMP(10663),1)} and \ttt{PMAS(LUCOMP(20663),1)}.
Since the latter heavy-flavour-diagonal states
are not normally produced in fragmentation, it would be no
disaster to forget changing their masses.

Most particles have separately defined decay channels. However, 
there are so many heavy-flavour hadrons with common decay desciptions,
that a few `pseudoparticles' have been introduced for generic
decays. The most frequently used ones are 84 for charm decays, 
85 for bottom decays and 86 for top decays. Instead of a long list 
of decay channels, several bottom and 
charm baryons and all top hadrons therefore only have one 
`decay channel', which is the instruction to make use of the decay 
data for particle 84/85/86. The spectator quark of the generic
decay channels is found as the light quark/diquark of the hadron 
considered. All the mesons in the bottom and charm sectors are 
individually defined, as are the $\Lambda_{\c}$ and $\Lambda_{\b}$ 
states. For top and heavier hadrons, the decay is likely to be so 
fast that no hadrons at all are produced, but if they are, the generic
pseudoparticle approach is a good approximation. 
 
The program contains space so that additional new particles may be
introduced. Although not completely trivial, this should not be
beyond the ability of an ordinary user. Basically, three steps are
involved. First, a mechanism of production has to be introduced.
This production may well take place in another program, like
{\Py} or some user-written correspondence, where matrix elements are
used to select the hard process. In this case the new particle already
exists in the \ttt{LUJETS} common block when {\Je} is called. A new
particle, meson, baryon or glueball, may also be a part of the
fragmentation process, in which case \ttt{LUKFDI} would have to be
suitably modified. The particle might also appear as a decay product
from some already existing particle, and then the decay data in
\ttt{/LUDAT3/} would have to be expanded; conceivably also
\ttt{LUDECY} would be affected.
 
The second step is to teach to program to recognize the new particle.
If a KF code in the range 41 to 80 is used, this is automatically taken
care of, and in particular the compressed code KC coincides with KF.
If a whole sequence of particles is to be introduced, with KF codes
paralleling that of ordinary mesons/baryons (a supersymmetric `meson'
multiplet, made of a squark plus an antiquark, say), then \ttt{LUCOMP}
must be modified to include a mapping from these KF values to currently
unused KC ones, like the range 401 - 500. It is the presence of such a
mapping that the program uses to accept a given KF code as bona fide.
 
The third and final step is to define the properties of this new
particle. Thus particle charge information must be given in \ttt{KCHG},
mass, width and lifetime in \ttt{PMAS}, particle name in \ttt{CHAF},
and decay data in the \ttt{MDCY}, \ttt{MDME}, \ttt{BRAT} and \ttt{KFDP}
arrays. This process is most conveniently carried out by using
\ttt{LUUPDA(1)} to produce a table of particle data, which can then be
modified, and afterwards read back in with \ttt{LUUPDA(2)}. Note that
the particle data is to be introduced for the compressed code KC, not
for KF proper.
 
\subsection{Examples}
 
A 10 GeV $\u$ quark jet going out along the $+z$ axis is generated
with
\begin{verbatim}
      CALL LU1ENT(0,2,10.,0.,0.)
\end{verbatim}
Note that such a single jet is not required to conserve energy,
momentum or flavour. In the generation scheme, particles with
negative $p_z$ are produced as well, but these are automatically
rejected unless \ttt{MSTJ(3)=-1}. While frequently used in former
days, the one-jet generation option is not of much current interest.
 
In e.g. a leptoproduction event a typical situation could be a $\u$
quark going out in the $+z$ direction and a $\u\d_0$ target remnant
essentially at rest. (Such a process can be simulated by {\Py}, but
here we illustrate how to do part of it yourself.) The simplest
procedure is probably to treat the process in the c.m. frame
and boost it to the lab frame afterwards. Hence, if the c.m. energy is
20 GeV and the boost $\beta_z = 0.996$ (corresponding to
$x_B = 0.045$), then
\begin{verbatim}
      CALL LU2ENT(0,2,2101,20.)
      CALL LUROBO(0.,0.,0.,0.,0.996)
\end{verbatim}
The jets could of course also be defined and allowed to fragment in the
lab frame with
\begin{verbatim}
      CALL LU1ENT(-1,2,223.15,0.,0.)
      CALL LU1ENT(2,12,0.6837,3.1416,0.)
      CALL LUEXEC
\end{verbatim}
Note here that the target diquark is required to move in the backwards
direction with $E-p_z = m_{\p}(1-x_B)$ to obtain the correct
invariant mass for the system. This is, however, only an artefact of
using a fixed diquark mass to represent a varying target remnant mass,
and is of no importance for the fragmentation. If one wants a
nicer-looking event record, it is possible to use the following
\begin{verbatim}
      CALL LU1ENT(-1,2,223.15,0.,0.)
      MSTU(10)=1
      P(2,5)=0.938*(1.-0.045)
      CALL LU1ENT(2,2101,0.,0.,0.)
      MSTU(10)=2
      CALL LUEXEC
\end{verbatim}
 
A 30 GeV $\u\ubar\g$ event with $E_{\u} = 8$ GeV and
$E_{\ubar} = 14$ GeV is simulated with
\begin{verbatim}
      CALL LU3ENT(0,2,21,-2,30.,2.*8./30.,2.*14./30.)
\end{verbatim}
The event will be given in a standard orientation with the $\u$
quark along the $+z$ axis and the $\ubar$ in the $-z, +x$ quadrant.
Note that the flavours of the three partons have to be given in the
order they are found along a string, if string fragmentation options
are to work. Also note that, for 3-jet events, and particularly
4-jet ones, not all setups of kinematical variables $x$ lie within
the kinematically allowed regions of phase space.
 
All common block variables can obviously be changed by including the
corresponding common block in the user-written main program.
Alternatively, the routine \ttt{LUGIVE} can be used to feed in
values, with some additional checks on array bounds then performed.
A call
\begin{verbatim}
      CALL LUGIVE('MSTJ(21)=3;PMAS(C663,1)=210.;CHAF(401)=funnyino;'//
     &'PMAS(21,4)=')
\end{verbatim}
will thus change the value of \ttt{MSTJ(21)} to 3, the value of
\ttt{PMAS(LUCOMP(663),1) = PMAS(136,1)} to 210., the value of
\ttt{CHAF(401)} to 'funnyino', and print the current value of
\ttt{PMAS(21,4)}. Since old and new values of parameters changed are
written to output, this may offer a convenient way of documenting
non-default values used in a given run. On the other hand, if a
variable is changed back and forth frequently, the resulting
voluminous output may be undesirable, and a direct usage
of the common blocks is then to be recommended (the output can also be
switched off, see \ttt{MSTU(13)}).
 
A general rule of thumb is that none of the physics routines
(\ttt{LUSTRF}, \ttt{LUINDF}, \ttt{LUDECY}, etc.) should ever be
called directly, but only via \ttt{LUEXEC}. This routine may be
called repeatedly for one single event. At each call only those
entries that are allowed to fragment or decay, and have not yet
done so, are treated. Thus
\begin{verbatim}
      CALL LU2ENT(1,1,-1,20.)        ! fill 2 jets without fragmenting
      MSTJ(1)=0                      ! inhibit jet fragmentation
      MSTJ(21)=0                     ! inhibit particle decay
      MDCY(LUCOMP(111),1)=0          ! inhibit pi0 decay
      CALL LUEXEC                    ! will not do anything
      MSTJ(1)=1                      !
      CALL LUEXEC                    ! jets will fragment, but no decays
      MSTJ(21)=2                     !
      CALL LUEXEC                    ! particles decay, except pi0
      CALL LUEXEC                    ! nothing new can happen
      MDCY(LUCOMP(111),1)=1          !
      CALL LUEXEC                    ! pi0:s decay
\end{verbatim}
 
A partial exception to the rule above is \ttt{LUSHOW}. Its main
application is for internal use by \ttt{LUEEVT}, \ttt{LUDECY}, and
\ttt{PYEVNT}, but it can also be directly called by you. Note that a
special format for storing colour-flow information in \ttt{K(I,4)}
and \ttt{K(I,5)} must then be used. For simple cases, the \ttt{LU2ENT}
can be made to take care of that automatically, by calling with the
first argument negative.
\begin{verbatim}
      CALL LU2ENT(-1,1,-2,80.)       ! store d ubar with colour flow
      CALL LUSHOW(1,2,80.)           ! shower partons
      CALL LUEXEC                    ! subsequent fragmentation/decay
\end{verbatim}
For more complicated configurations, \ttt{LUJOIN} should be used.
 
It is always good practice to list one or a few events during a run to
check that the program is working as intended. With
\begin{verbatim}
      CALL LULIST(1)
\end{verbatim}
all particles will be listed and in addition total charge, momentum and
energy of stable entries will be given. For string fragmentation these
quantities should be conserved exactly (up to machine precision errors),
and the same goes when running independent fragmentation with one of
the momentum conservation options. \ttt{LULIST(1)} gives a format that
comfortably fits on an 80 column screen, at the price of not giving
the complete story. With \ttt{LULIST(2)} a more extensive listing is
obtained, and \ttt{LULIST(3)} also gives vertex information. Further
options are available, like \ttt{LULIST(12)}, which gives a list of
particle data.
 
An event, as stored in the \ttt{LUJETS} common block, will contain the
original jets and the whole decay chain, i.e. also particles which 
subsequently decayed. If parton showers are used, the amount of parton 
information is also considerable: first the on-shell partons before 
showers have been considered, then a \ttt{K(I,1)=22} line with total 
energy of the showering
subsystem, after that the complete shower history tree-like structure,
starting off with the same initial partons (now off-shell), and finally
the end products of the shower rearranged along the string directions.
This detailed record is useful in many connections, but if one only
wants to retain the final particles, superfluous information may be
removed with \ttt{LUEDIT}. Thus e.g.
\begin{verbatim}
      CALL LUEDIT(2)
\end{verbatim}
will leave you with the final charged and neutral particles, except
for neutrinos.
 
The information in \ttt{LUJETS} may be used directly to study an event.
Some useful additional quantities derived from these, such as charge
and rapidity, may easily be found via the \ttt{KLU} and \ttt{PLU}
functions. Thus electric charge \ttt{=PLU(I,6)} (as integer,
three times charge \ttt{=KLU(I,6)}) and true rapidity $y$ with respect
to the $z$ axis \ttt{= PLU(I,17)}.
 
A number of utility (\ttt{MSTU}, \ttt{PARU}) and physics (\ttt{MSTJ},
\ttt{PARJ}) switches and parameters are available in common block
\ttt{LUDAT1}. All of these have sensible default values. Particle data
is stored in common blocks \ttt{LUDAT2}, \ttt{LUDAT3} and \ttt{LUDAT4}.
Note that the data in the arrays \ttt{KCHG}, \ttt{PMAS}, \ttt{MDCY}
and \ttt{CHAF} is not stored by KF code, but by the compressed
code KC. This code is not to be learnt by heart, but instead accessed
via the conversion function \ttt{LUCOMP}, \ttt{KC = LUCOMP(KF)}.
 
In the particle tables, the following particles are considered stable:
the photon, $\e^{\pm}$, $\mu^{\pm}$, $\pi^{\pm}$, $\K^{\pm}$, 
$\K_{\mrm{L}}^0$,
$\p$, $\pbar$, $\n$, $\br{\n}$ and all the neutrinos. It is, however,
always possible to inhibit the decay of any given particle by putting
the corresponding \ttt{MDCY} value zero or negative, e.g.
\ttt{MDCY(LUCOMP(310),1)=0} makes $\K_{\mrm{S}}^0$ and
\ttt{MDCY(LUCOMP(3122),1)=0} $\Lambda$ stable. It is also possible
to select stability based on the average lifetime (see \ttt{MSTJ(22)}),
or based on whether the decay takes place within a given spherical
or cylindrical volume around the origin.
 
The Field-Feynman jet model \cite{Fie78} is available in the program
by changing the following values: \ttt{MSTJ(1)=2} (independent
fragmentation), \ttt{MSTJ(3)=-1} (retain particles with $p_z < 0$;
is not mandatory), \ttt{MSTJ(11)=2} (choice of longitudinal
fragmentation function, with the $a$ parameter stored in
\ttt{PARJ(51) - PARJ(53)}), \ttt{MSTJ(12)=0} (no baryon production),
\ttt{MSTJ(13)=1} (give endpoint quarks $\pT$ as quarks created
in the field), \ttt{MSTJ(24)=0} (no mass broadening of resonances),
\ttt{PARJ(2)=0.5} ($\s/\u$ ratio for the production of new $\q\qbar$
pairs), \ttt{PARJ(11)=PARJ(12)=0.5}  (probability for mesons to
have spin 1) and \ttt{PARJ(21)=0.35} (width of Gaussian transverse
momentum distribution). In addition only $\d$, $\u$ and $\s$ single
quark jets may be generated following the FF recipe. Today the FF
`standard jet' concept is probably dead and buried, so the numbers
above should more be taken as an example of the flexibility of the
program, than as something to apply in practice.
 
A wide range of independent fragmentation options are implemented,
to be accessed with the master switch \ttt{MSTJ(1)=2}. In particular,
with \ttt{MSTJ(2)=1} a gluon jet is assumed to fragment like a
random $\d$, $\dbar$, $\u$, $\ubar$, $\s$ or $\sbar$ jet, while with
\ttt{MSTJ(2)=3} the gluon is split into a $\d\dbar$, $\u\ubar$ or
$\s\sbar$ pair of jets sharing the energy according to the
Altarelli-Parisi splitting function. Whereas energy, momentum and
flavour is not explicitly conserved in independent fragmentation, a
number of options are available in \ttt{MSTJ(3)} to ensure this
`post facto', e.g. \ttt{MSTJ(3)=1} will boost the event to ensure
momentum conservation and then (in the c.m. frame) rescale momenta by
a common factor to obtain energy conservation, whereas
\ttt{MSTJ(3)=4} rather uses a method of stretching the jets in
longitudinal momentum along the respective jet axis to keep angles
between jets fixed.
 
\clearpage
 
\section{Event Analysis Routines}
 
To describe the complicated geometries encountered in multihadronic
events, a number of event measures have been introduced. These
measures are intended to provide a global view of the properties
of a given event, wherein the full information content of the event
is condensed into one or a few numbers. A steady stream of such
measures are proposed for different purposes. Many are rather
specialized or never catch on, but a few become standards, and are
useful to have easy access to. \tsc{Jetset} therefore contains a
number of routines that can be called for any event, and that will
directly access the event record to extract the required information.
 
In the presentation below, measures have been grouped in three kinds.
The first contains simple event shape quantities, such as sphericity
and thrust. The second is jet finding algorithms.
The third is a mixed bag of particle multiplicities and compositions,
factorial moments and energy--energy correlations, put together
in a small statistics package.
 
None of the measures presented here are Lorentz invariant. The
analysis will be performed in whatever frame the event happens to
be given in. It it therefore up to you to decide whether the
frame in which events were generated is the right one, or whether
events beforehand should be boosted, e.g. to the c.m. frame. You
can also decide which particles you want to have affected by the
analysis.
 
\subsection{Event Shapes}
\label{ss:evshape}
 
In this section we study general event shape variables: sphericity,
thrust, Fox-Wolfram moments, and jet masses. These measures are
implemented in the routines \ttt{LUSPHE}, \ttt{LUTHRU}, \ttt{LUFOWO}
and \ttt{LUJMAS}, respectively.
 
Each event is assumed characterized by the particle four-momentum
vectors $p_i = (\mbf{p}_i, E_i)$, with $i = 1, 2, \cdots , n$ an
index running over the particles of the event.
 
\subsubsection{Sphericity}
 
The sphericity tensor is defined as \cite{Bjo70}
\begin{equation}
S^{\alpha \beta} = \frac{\displaystyle \sum_i p^{\alpha}_{i} \,
   p^{\beta}_{i} } {\displaystyle \sum_i |\mbf{p}_i|^2 }  ~,
\end{equation}
where $\alpha, \beta = 1, 2, 3$ corresponds to the $x$, $y$ and $z$
components.
By standard diagonalization of $S^{\alpha \beta}$ one may find three
eigenvalues $\lambda_1 \geq \lambda_2 \geq \lambda_3$, with
$\lambda_1 + \lambda_2 + \lambda_3 = 1$. The sphericity of the event
is then defined as
\begin{equation}
S = \frac{3}{2} \, (\lambda_2 + \lambda_3) ~,
\end{equation}
so that $0 \leq S \leq 1$. Sphericity is essentially a measure of the
summed $\pT^2$ with respect to the event axis; a 2-jet event
corresponds to $S \approx 0$ and an isotropic event to $S \approx 1$.
 
The aplanarity $A$, with definition $A = \frac{3}{2} \lambda_3$, is
constrained to the range $0 \leq A \leq \frac{1}{2}$. It measures the
transverse momentum component out of the event plane: a planar event
has $A \approx 0$ and an isotropic one $A \approx \frac{1}{2}$.
 
Eigenvectors $\mbf{v}_j$ can be found that correspond to the three
eigenvalues $\lambda_j$ of the sphericity tensor. The $\mbf{v}_1$
one is called the sphericity axis (or event axis, if it is clear
from the context that sphericity has been used), while the sphericity
event plane is spanned by $\mbf{v}_1$ and $\mbf{v}_2$.
 
The sphericity tensor is quadratic in particle momenta. This means
that the sphericity
value is changed if one particle is split up into two collinear ones
which share the original momentum. Thus sphericity is
not an infrared safe quantity in QCD perturbation theory. A useful
generalization of the sphericity tensor is
\begin{equation}
S^{(r) \alpha \beta} = \frac{\displaystyle \sum_i |\mbf{p}_i|^{r-2} \,
    p^{\alpha}_{i} \, p^{\beta}_{i} }{\displaystyle
    \sum_i |\mbf{p}_i|^r }  ~,
\end{equation}
where $r$ is the power of the momentum dependence. While $r = 2$
thus corresponds to sphericity, $r = 1$ corresponds to linear
measures calculable in perturbation theory \cite{Par78}:
\begin{equation}
S^{(1) \alpha \beta} = \frac{\displaystyle \sum_i \frac{\displaystyle
    p^{\alpha}_{i} \, p^{\beta}_{i}}{\displaystyle |\mbf{p}_i|} }
    {\displaystyle \sum_i |\mbf{p}_i| } ~.
\end{equation}
 
Eigenvalues and eigenvectors may be defined exactly as before, and
therefore also equivalents of $S$ and $A$. These have no standard
names; I tend to call them linearized sphericity $S_{\mrm{lin}}$ and
linearized aplanarity $A_{\mrm{lin}}$. Quantities that are standard in
the literature are instead the combinations \cite{Ell81}
\begin{eqnarray}
   C & = & 3 ( \lambda_1 \lambda_2 + \lambda_1 \lambda_3 +
   \lambda_2 \lambda_3 ) ~,    \\
   D & = & 27 \lambda_1 \lambda_2 \lambda_3 ~.
\end{eqnarray}
Each of these is constrained to be in the range between 0 and 1.
Typically, $C$ is used to measure the 3-jet structure and $D$
the 4-jet one, since $C$ is vanishing for a perfect 2-jet event
and $D$ is vanishing for a planar event. The $C$ measure is related
to the second Fow-Wolfram moment (see below), $C = 1 - H_2$.
 
Noninteger $r$ values may also be used, and corresponding generalized
sphericity and aplanarity measures calculated. While perturbative
arguments favour $r = 1$, we know that the fragmentation `noise', e.g.
from transverse momentum fluctuations, is
proportionately larger for low momentum particles, and so $r > 1$
should be better for experimental event axis determinations. The use
of too large an $r$ value, on the other hand, puts all the emphasis
on a few high-momentum particles, and therefore involves a loss of
information. It should then come as no surprise that intermediate
$r$ values, of around 1.5, gives the best performance for event
axis determinations in 2-jet events, where the theoretical
meaning of the event axis is well-defined. The gain in accuracy
compared with the more conventional choices $r=2$ or $r=1$ is
rather modest, however.
 
\subsubsection{Thrust}
 
The quantity thrust $T$ is defined by \cite{Bra64}
\begin{equation}
T = \max_{|\mbf{n}| = 1} \,
\frac{\displaystyle \sum_i |\mbf{n} \cdot \mbf{p}_i|}
{\displaystyle \sum_i |\mbf{p}_i|} ~,
\label{em:thrust}
\end{equation}
and the thrust axis $\mbf{v}_1$ is given by the $\mbf{n}$ vector
for which maximum is attained. The allowed range is
$1/2 \leq T \leq 1$, with a 2-jet event corresponding
to $T \approx 1$ and an isotropic event to $T \approx 1/2$.
 
In passing, we note that this is not the only definition found in
the literature. The definitions agree for events studied in the c.m.
frame and where all particles are detected. However, a definition
like
\begin{equation}
T = 2 \, \max_{|\mbf{n}| = 1} \,
\frac{\displaystyle \left| \sum_i \theta(\mbf{n} \cdot \mbf{p}_i) \,
\mbf{p}_i \right| }{\displaystyle \sum_i |\mbf{p}_i|} =
2 \, \max_{\theta_i = 0,1} \,
\frac{\displaystyle \left| \sum_i \theta_i \, \mbf{p}_i \right| }
{\displaystyle \sum_i |\mbf{p}_i|}
\label{em:thrusttwo}
\end{equation}
(where $\theta(x)$ is the step function, $\theta(x) = 1 $ if $x > 0$,
else $\theta(x) = 0$) gives different results than the one above
if e.g. only charged particles are detected. It would even be possible
to have $T > 1$; to avoid such problems, often an extra fictitious
particle is introduced to balance the total momentum \cite{Bra79}.
 
Eq.~(\ref{em:thrust}) may be rewritten as
\begin{equation}
T = \max_{\epsilon_i = \pm 1} \,
\frac{\displaystyle \left| \sum_i \epsilon_i \, \mbf{p}_i \right| }
{\displaystyle \sum_i |\mbf{p}_i|} ~.
\label{em:thrustthree}
\end{equation}
(This may also be viewed as applying eq.~(\ref{em:thrusttwo}) to an
event with $2 n$ particles, $n$ carrying the momenta $\mbf{p}_i$
and $n$ the momenta $- \mbf{p}_i$, thus automatically balancing
the momentum.) To find the thrust value and axis this way, $2^{n-1}$
different possibilities would have to be tested. The reduction by a
factor of 2 comes from $T$ being unchanged when all
$\epsilon_i \to - \epsilon_i$. Therefore this approach rapidly becomes
prohibitive. Other exact methods exist, which `only' require about
$4n^2$ combinations to be tried.
 
In the implementation in {\Je}, a faster alternative method is
used, in which the thrust axis is iterated from a starting direction
$\mbf{n}^{(0)}$ according to
\begin{equation}
\mbf{n}^{(j+1)} = \frac{\displaystyle \sum_i \epsilon(
\mbf{n}^{(j)} \cdot \mbf{p}_i) \, \mbf{p}_i }
{\displaystyle \left| \sum_i \epsilon(
\mbf{n}^{(j)} \cdot \mbf{p}_i) \, \mbf{p}_i \right| }
\end{equation}
(where $\epsilon(x) = 1$ for $x > 0$ and $\epsilon(x) = -1$ for
$x < 0$). It is easy to show that the related thrust value will
never decrease, $T^{(j+1)} \geq T^{(j)}$. In fact, the method
normally converges in 2--4 iterations. Unfortunately, this
convergence need not be towards the correct thrust axis but is
occasionally only towards a local maximum of the thrust function
\cite{Bra79}. We know of no foolproof way around this complication,
but the danger of an error may be lowered if several different
starting axes $\mbf{n}^{(0)}$ are tried and found to agree. These
$\mbf{n}^{(0)}$ are suitably constructed from the $n'$ (by default
4) particles with the largest momenta in the event, and the
$2^{n' -1}$ starting directions $\sum_i \epsilon_i \, \mbf{p}_i$
constructed from these are tried in falling order of the
corresponding absolute momentum values. When a predetermined number
of the starting axes have given convergence towards the same
(best) thrust axis this one is accepted.
 
In the plane perpendicular to the thrust axis, a major \cite{MAR79}
axis and value may be defined in just the same fashion as thrust, i.e.
\begin{equation}
M_a = \max_{|\mbf{n}| = 1, \, \mbf{n} \cdot \mbf{v}_1 = 0} \,
\frac{\displaystyle \sum_i |\mbf{n} \cdot \mbf{p}_i|}
{\displaystyle \sum_i |\mbf{p}_i| }    ~.
\end{equation}
In a plane more efficient methods can be used to find an axis than in
three dimensions \cite{Wu79}, but for simplicity we use the same
method as above. Finally, a third axis, the minor axis, is defined
perpendicular to the thrust and major ones, and a minor value $M_i$
is calculated just as thrust and major.
The difference between major and minor is called
oblateness, $O = M_a -M_i$. The upper limit on oblateness depends
on the thrust value in a not so simple way. In general
$O \approx 0$ corresponds to an event symmetrical around the thrust
axis and high $O$ to a planar event.
 
As in the case of sphericity, a generalization to arbitrary momentum
dependence may easily be obtained, here by replacing the $\mbf{p}_i$
in the formulae above by $|\mbf{p}_i|^{r-1} \, \mbf{p}_i$. This
possibility is included, although so far it has not found any
experimental use.
 
\subsubsection{Fox-Wolfram moments}
 
The Fox-Wolfram moments $H_l$, $l = 0, 1, 2, \ldots$, are defined by
\cite{Fox79}
\begin{equation}
H_l = \sum_{i,j} \frac{ |\mbf{p}_i| \, |\mbf{p}_j| }
{E_{\mrm{vis}}^2} \, P_l (\cos \theta_{ij}) ~,
\end{equation}
where $\theta_{ij}$ is the opening angle between hadrons $i$ and
$j$ and $E_{\mrm{vis}}$ the total visible energy of the event. Note
that also autocorrelations, $i = j$, are included. The $P_l(x)$ are 
the Legendre polynomials,
\begin{eqnarray}
P_0(x) & = & 1 ~,   \nonumber \\
P_1(x) & = & x  ~,  \nonumber \\
P_2(x) & = & \frac{1}{2} \, (3x^2 -1) ~,    \nonumber \\
P_3(x) & = & \frac{1}{2} \, (5x^3 - 3x) ~,  \nonumber \\
P_4(x) & = & \frac{1}{8} \, (35x^4 - 30x^2 + 3)  ~.
\end{eqnarray}
To the extent that particle masses may be neglected, $H_0 \equiv 1$.
It is customary to normalize the results to $H_0$, i.e. to give
$H_{l0} = H_l / H_0$. If momentum is balanced then $H_1 \equiv 0$.
2-jet events tend to give $H_l \approx 1$ for $l$ even and
$\approx 0$ for $l$ odd.
 
\subsubsection{Jet masses}
 
The particles of an event may be divided into two classes.
For each class a squared invariant mass may be calculated, $M_1^2$
and $M_2^2$. If the assignment of particles is adjusted such that
the sum $M_1^2 + M_2^2$ is minimized, the two masses thus obtained
are called heavy and light jet mass, $M_{\mrm{H}}$ and $M_{\mrm{L}}$. 
It has been shown that these quantities are well behaved in perturbation
theory \cite{Cla79}. In $\ee$ annihilation, the heavy jet mass
obtains a contribution from $\q\qbar\g$ 3-jet events, whereas
the light mass is non-vanishing only when 4-jet events also
are included. In the c.m. frame of an event one has the limits
$0 \leq M_{\mrm{H}}^2 \leq E_{\mrm{cm}}^2/3$.
 
In general, the subdivision of particles tends to be into two
hemispheres, separated by a plane perpendicular to an event axis.
As with thrust, it is time-consuming to find the exact solution.
Different approximate strategies may therefore be used. In the
program, the sphericity axis is used to perform a fast subdivision
into two hemispheres, and thus into two preliminary jets. Thereafter
one particle at a time is tested to determine whether the sum
$M_1^2 + M_2^2$ would be decreased if that particle were to be
assigned to the other jet. The precedure is stopped when no
further significant change is obtained. Often the original
assignment is retained as it is, i.e. the sphericity axis gives
a good separation. This is not a full guarantee, since the program
might get stuck in a local mimimum which is not the global one.
 
\subsection{Cluster Finding}
\label{ss:clustfind}
 
Global event measures, like sphericity or thrust, can only be used to
determine the jet axes for back-to-back 2-jet events. To determine
the individual jet axes in events with three or more jets,
or with two (main) jets which are not back-to-back, cluster
algorithms are customarily used. In these, nearby particles are
grouped together into a variable number of clusters. Each cluster
has a well-defined direction, given by a suitably weighted average
of the constituent particle directions.
 
The cluster algorithms traditionally used in $\ee$ and in $\pp$
physics differ in several respects. The former tend to be
spherically symmetric, i.e. have no preferred axis in space,
and normally all particles have to be assigned to some jet. The
latter pick the beam axis as preferred direction, and make use
of variables related to this choice, such as rapidity and transverse
momentum; additionally only a fraction of all particles are assigned
to jets.
 
This reflects a difference in the underlying physics: in
$\pp$ collisions, the beam remnants found at low transverse momenta
are not related to any hard processes, and therefore only provide an
unwanted noise to many studies. (Of course, also hard processes may
produce particles at low transverse momenta, but at a rate much
less than that from soft or semi-hard processes.) Further, the
kinematics of hard processes is, to a good approximation, factorized
into the hard subprocess itself, which is boost invariant in rapidity,
and parton-distribution effects, which determine the overall position
of a hard scattering in rapidity. Hence rapidity, azimuthal angle and
transverse momentum is a suitable coordinate frame to describe hard
processes in.
 
In standard $\ee$ annihilation events, on the other hand, the hard
process c.m. frame tends to be almost at rest, and the event
axis is just about randomly distributed in space, i.e. with no 
preferred r\^ole for the axis defined by the incoming $\e^{\pm}$.
All particle production is initiated by and related to the hard
subprocess. Some of the particles may be less easy to associate to
a specific jet, but there is no compelling reason to remove any
of them from consideration.
 
This does not mean that the separation above is always required.
$2\gamma$ events in $\ee$ may have a structure with `beam jets' and
`hard scattering' jets, for which the $\pp$ type algorithms might
be well suited. Conversely, a heavy particle produced in $\pp$
collisions could profitably be studied, in its own rest frame,
with $\ee$ techniques.
 
In the following, particles are only characterized by their
three-momenta or, alternatively, their energy and direction of
motion. No knowledge is therefore assumed of particle types, or
even of mass and charge. Clearly, the more is known, the more
sophisticated clustering algorithms can be used. The procedure
then also becomes more detector-dependent, and therefore less
suitable for general usage.
 
{\Je} contains two cluster finding routines. \ttt{LUCLUS} is of
the $\ee$ type and \ttt{LUCELL} of the $\pp$ one. Each of them
allows some variations of the basic scheme.
 
\subsubsection{Cluster finding in an $\ee$ type of environment}
 
The usage of cluster algorithms for $\ee$ applications started in
the late 1970's. A number of different approaches were proposed
\cite{Dor81}. Of these, we will here only discuss those based on
binary joining. In this kind of
approach, initially each final-state particle is considered to be a
cluster. Using some distance measure, the two nearest clusters are
found. If their distance is smaller than some cut-off value,
the two clusters are joined into one. In this new configuration, the
two clusters that are now nearest are found and joined, and so on until
all clusters are separated by a distance larger than the cut-off.
The clusters remaining at the end are often also called jets. Note that,
in this approach, each single particle belongs to exactly one
cluster. Also note that the resulting jet picture explicitly depends
on the cut-off value used. Normally the number of clusters is allowed to
vary from event to event, but occasionally it is more useful to have
the cluster algorithm find a predetermined number of jets (like 3).
 
The obvious choice for a distance measure is to use squared invariant
mass, i.e. for two clusters $i$ and $j$ to define the
distance to be
\begin{equation}
   m^2_{ij} = (E_i + E_j)^2 - (\mbf{p}_i + \mbf{p}_j)^2 ~.
\end{equation}
(Equivalently, one could have used the invariant mass as measure
rather than its square; this is just a matter of convenience.)
In fact, a number of people (including the author)
tried this measure long ago and gave up on it, since
it turns out to have severe instability problems. The reason is well
understood: in general, particles tend to cluster closer in
invariant mass in the region of small momenta. The clustering process
therefore tends to start in the center of the event, and only
subsequently spread outwards to encompass also the fast particles.
Rather than clustering slow particles around the fast ones (where the
latter na\"{\i}vely should best represent the jet directions), the 
invariant mass measure will tend to cluster fast particles around 
the slow ones.
 
Another instability may be seen by considering the clustering in a
simple 2-jet event. By the time that clustering has reached the
level of three clusters, the `best' the clustering algorithm can
possibly have achieved, in terms of finding three low-mass clusters,
is to have one fast cluster around each jet, plus a third slow cluster
in the middle. In the last step this third cluster would be joined with
one of the fast ones, to produce two final asymmetric clusters:
one cluster would contain all the slow particles, also those that
visually look like belonging to the opposite jet. A simple binary
joining process, with no possiblity to reassign particles between
clusters, is therefore not likely to be optimal.
 
The solution adopted by the author \cite{Sjo83} is to reject invariant
mass as distance measure. Instead a jet is defined as a collection
of particles which have a limited transverse momentum with respect to
a common jet axis, and hence also with respect to each other. This
picture is clearly inspired by the standard fragmentation picture,
e.g. in string fragmentation. A distance measure $d_{ij}$ between two
particles (or clusters) with momenta $\mbf{p}_i$ and $\mbf{p}_j$ should
thus not depend critically on the longitudinal momenta but only on the
relative transverse momentum. A number of such measures were tried,
and the one eventually selected is
\begin{equation}
d_{ij}^2 = \frac{1}{2} \, (|\mbf{p}_i| \, |\mbf{p}_j| -
\mbf{p}_i \cdot \mbf{p}_j) \, \frac{4 \, |\mbf{p}_i| \, |\mbf{p}_j|}
{(|\mbf{p}_i| + |\mbf{p}_j|)^2} =
\frac{4 \, |\mbf{p}_i|^2 \, |\mbf{p}_j|^2 \, \sin^2(\theta_{ij}/2)}
{(|\mbf{p}_i| + |\mbf{p}_j|)^2}   ~.
\end{equation}
 
For small relative angle $\theta_{ij}$, where
$2 \sin(\theta_{ij}/2) \approx \sin\theta_{ij}$ and
$\cos \theta_{ij} \approx 1$, this measure reduces to
\begin{equation}
d_{ij} \approx \frac{|\mbf{p}_i \times \mbf{p}_j|}
{|\mbf{p}_i + \mbf{p}_j|}  ~,
\end{equation}
where `$\times$' represents the cross product. We therefore see that
$d_{ij}$ in this limit has the simple physical interpretation as the
transverse momentum of either particle with respect to the direction
given by the sum of the two particle momenta. Unlike the approximate
expression, however, $d_{ij}$ does not vanish for two back-to-back
particles, but is here more related to the invariant mass
between them.
 
The basic scheme is of the binary joining type, i.e. initially each
particle is assumed to be a cluster by itself. Then the two clusters
with smallest relative distance $d_{ij}$ are found and, if
$d_{ij} < d_{\mrm{join}}$, with $d_{\mrm{join}}$ 
some predetermined distance, the two clusters are joined
to one, i.e. their momenta are added vectorially to give the
momentum of the new cluster. This is repeated until the distance between
any two clusters is $> d_{\mrm{join}}$. The number and momenta of these 
final clusters then represent our reconstruction of the initial jet
configuration, and each particle is assigned to one of the clusters.
 
To make this scheme workable, two further ingredients are necessary,
however. Firstly, after two clusters have been joined, some particles
belonging to the new cluster may actually be closer to
another cluster. Hence, after each joining, all particles in the event
are reassigned to the closest of the clusters. For particle $i$, this
means that the distance $d_{ij}$ to all clusters $j$ in the event has
to be evaluated and compared. After all particles
have been considered, and only then, are cluster momenta recalculated
to take into account any reassignments. To save time, the assignment
procedure is not iterated until a stable configuration is reached,
but, since all particles are reassigned at each step, such an
iteration is effectively taking place in parallel with the cluster
joining. Only at the very end, when all $d_{ij} > d_{\mrm{join}}$, is 
the reassignment procedure iterated to convergence --- still with the
possibility to continue the cluster joining if some $d_{ij}$ should
drop below $d_{\mrm{join}}$ due to the reassignment.
 
Occasionally, it may occur that the reassignment step leads to an empty
cluster, i.e. one to which no particles are assigned. Since such a
cluster has a distance $d_{ij} = 0$ to any other cluster, it is
automatically removed in the next cluster joining. However, it is
possible to run the program in a mode where a minimum number of jets
is to be reconstructed. If this minimum is reached with one cluster
empty, the particle is found which has largest distance to the cluster
it belongs to. That cluster is then split into two, namely the
large-distance particle and a remainder. Thereafter the reassignment
procedure is continued as before.
 
Secondly, the large multiplicities normally encountered means that,
if each particle initially is to be treated as a separate cluster, the
program will become very slow. Therefore a smaller number of clusters,
for a normal $\ee$ event typically 8--12, is constructed as a starting
point for the iteration above, as follows. The particle with the
highest momentum is found, and thereafter all particles within a
distance $d_{ij} < d_{\mrm{init}}$ from it, where 
$d_{\mrm{init}} \ll d_{\mrm{join}}$.
Together these are allowed to form a single cluster. For the
remaining particles, not assigned to this cluster, the procedure is
iterated, until all particles have been used up. Particles in the
central momentum region, $|\mbf{p}| < 2d_{\mrm{init}}$ are treated
separately; if their vectorial momentum sum is above $2d_{\mrm{init}}$
they are allowed to form one cluster, otherwise they are left
unassigned in the initial configuration. The value of $d_{\mrm{init}}$,
as long as reasonably small, has no physical importance, in that the
same final cluster configuration will be found as if each particle
initially is assumed to be a cluster by itself: the particles
clustered at this step are so nearby anyway that they almost
inevitably must enter the same jet; additionally the reassignment
procedure allows any possible `mistake' to be corrected in later
steps of the iteration.
 
Thus the jet reconstruction depends on one single parameter, 
$d_{\mrm{join}}$, with a clearcut physical meaning of a transverse 
momentum `jet-resolution power'. Neglecting smearing from 
fragmentation, $d_{ij}$ between two clusters of equal energy 
corresponds to half the invariant mass of the two original partons. 
If one only wishes to reconstruct well separated jets, a large 
$d_{\mrm{join}}$ should be chosen, while a small $d_{\mrm{join}}$ 
would allow the separation of close jets, at the
cost of sometimes artificially dividing a single jet into two. In
particular, $\b$ quark jets may here be a nuisance. The value of
$d_{\mrm{join}}$ to use for a fixed jet-resolution power in principle
should be independent of the c.m. energy of events, although
fragmentation effects may give a contamination of spurious extra jets
that increases slowly with $E_{\mrm{cm}}$ for fixed $d_{\mrm{join}}$. 
Therefore a $d_{\mrm{join}} = 2.5$ GeV was acceptable at PETRA/PEP, 
while 3--4 GeV may be better for applications at LEP and beyond.
 
This completes the description of the main option of the \ttt{LUCLUS}
routine. Variations are possible. One such is to skip the reassignment
step, i.e. to make use only of the simple binary joining procedure,
without any possibility to reassign particles between jets.
(This option is included mainly as a reference, to check how important
reassignment really is.) The other main alternative is
to replace the distance measure used above with the one used in the
JADE algorithm \cite{JAD86}.
 
The JADE cluster algorithm is an attempt to save the invariant mass
measure. The distance measure is defined to be
\begin{equation}
   y_{ij} = \frac{2 E_i E_j (1-\cos \theta_{ij})}{E^2_{\mrm{vis}}} ~.
\end{equation}
Here $E_{\mrm{vis}}$ is the total visible energy of the event. The
usage of $E^2_{\mrm{vis}}$ in the denominator rather than 
$E_{\mrm{cm}}^2$ tends to make the measure less
sensitive to detector acceptance corrections; in addition the
dimensionless nature of $y_{ij}$ makes it well suited for a comparison
of results at different c.m. energies. For the subsequent discussions,
this normalization will be irrelevant, however.
 
The $y_{ij}$ measure is very closely related to the squared mass
distance measure: the two coincide (up to the difference in
normalization) if $m_i = m_j = 0$. However, consider a pair of
particles or clusters with non-vanishing individual masses and a fixed
pair mass. Then, the larger the net momentum of the pair, the
smaller the $y_{ij}$ measure. This somewhat tends to favour clustering
of fast particles, and makes the algorithm less unstable than the one
based on true invariant mass.
 
The successes of the JADE
algorithm are well known: one obtains a very good
agreement between the number of partons generated on the matrix-element
(or parton-shower) level and the number of clusters reconstructed from
the hadrons, such that QCD aspects like the running of $\alphas$ can
be studied with a minimal dependence on fragmentation effects. Of
course, the insensitivity to fragmentation effects depends on the choice
of fragmentation model. Fragmentation effects are small in the string
model, but not in independent fragmentation scenarios.
Although independent fragmentation in itself is not credible,
this may be seen as a signal for caution.
 
One should note that the JADE measure still suffers from some of the
diseases of the simple mass measure (without reassignments), namely
that particles which go in opposite directions may well be joined
into the same cluster. Therefore, while the JADE algorithm is a good
way to find the number of jets, it is inferior to the standard
$d_{ij}$ measure for a determination of jet directions and energies
\cite{Bet92}. The $d_{ij}$ measure also gives narrower jets, which
agree better with the visual impression of jet structure.
 
Recently, the `Durham algorithm' has been introduced \cite{Cat91},
which works as the JADE one but with a distance measure
\begin{equation}
\tilde{y}_{ij} = \frac{2 \min(E_i^2,E_j^2) (1-\cos \theta_{ij})}
{E^2_{cm}} ~.
\end{equation}
Like the $d_{ij}$ measure, this is a transverse momentum, but
$\tilde{y}_{ij}$ has the geometrical interpretation as the
transverse momentum of the softer particle with respect to the
direction of the harder one, while $d_{ij}$ is the transverse
momentum of either particle with respect to the common direction
given by the momentum vector sum. The two definitions agree when
one cluster is much softer than the other, so the soft gluon
exponentiation proven for the Durham measure also holds for the
$d_{ij}$ one.
 
The main difference therefore is that the standard \ttt{LUCLUS}
option allows reassignments, while the Durham algorithm does not.
The latter is therefore more easily calculable on the perturbative
parton level. This point is sometimes overstressed, and one could
give counterexamples why reassignments in fact may bring better
agreement with the underlying perturbative level. In particular,
without reassignments, one will make the recombination that seems
the `best' in the current step, even when that forces you to make
`worse' choices in subsequent steps. With reassignments, it is
possible to correct for mistakes due to the too local sensitivity of
a simple binary joining scheme.
 
\subsubsection{Cluster finding in a $\pp$ type of environment}
 
The \ttt{LUCELL} cluster finding routines is of the kind pioneered
by UA1 \cite{UA183}, and commonly used in $\pp$ physics. It is based
on a choice of pseudorapidity $\eta$, azimuthal angle $\varphi$ and
transverse momentum $\pT$ as the fundamental coordinates.
This choice is discussed in the introduction to cluster finding
above, with the proviso that the theoretically preferred true
rapidity has to be replaced by pseudorapidity, to make contact with
the real-life detector coordinate system.
 
A fix detector grid is assumed, with the pseudorapidity range
$|\eta| < \eta_{\mmax}$ and the full azimuthal range each divided
into a number of equally large bins, giving a rectangular grid.
The particles of an event impinge on this detector grid. For each
cell in ($\eta$,$\varphi$) space, the transverse momentum which enters
that cell is summed up to give a total cell $E_{\perp}$ flow.
 
Clearly the model remains very primitive in a number of
respects, compared with a real detector. There is no magnetic field
allowed for, i.e. also charged particles move in straight tracks.
The dimensions of the detector are not specified; hence the positions
of the primary vertex and any secondary vertices are neglected when
determining which cell a particle belongs to. The rest mass of
particles is not taken into account, i.e. what is used is really
$\pT = \sqrt{p_x^2 + p_y^2}$, while in a real detector some
particles would decay or annihilate, and then deposit additional
amounts of energy.
 
To take into account the energy resolution of the detector, it is
possible to smear the $E_{\perp}$ contents, bin by bin. This is done
according to a Gaussian, with a width assumed proportional to the
$\sqrt{E_{\perp}}$ of the bin. The Gaussian is cut off at zero and
at some predetermined multiple of the unsmeared $E_{\perp}$, by default
twice it. Alternatively, the smearing may be performed in $E$
rather than in $E_{\perp}$. To find the $E$, it is assumed
that the full energy of a cell is situated at its center, so
that one can translate back and forth with
$E = E_{\perp} \cosh\eta_{\mrm{center}}$.
 
The cell with largest $E_{\perp}$ is taken as a jet initiator if its
$E_{\perp}$ is above some threshold. A candidate jet is defined to
consist of all cells which are within some given radius $R$ in the
($\eta$,$\varphi$) plane, i.e. which have 
$(\eta - \eta_{\mrm{initiator}})^2
 + (\varphi - \varphi_{\mrm{initiator}})^2 < R^2$.
Coordinates are always given with respect to the center of the
cell. If the summed $E_{\perp}$ of the jet is above the
required minimum jet energy, the candidate jet is accepted, and all
its cells are removed from further consideration. If not, the
candidate is rejected. The sequence is now repeated with the
remaining cell of highest $E_{\perp}$, and so on until no single
cell fulfills the jet initiator condition.
 
The number of jets reconstruced can thus vary from none to a maximum
given by purely geometrical considerations, i.e. how many circles of
radius $R$ are needed to cover the allowed ($\eta$,$\varphi$) plane.
Normally only a fraction of the particles are assigned to jets.
 
One could consider to iterate the jet assignment process, using the
$E_{\perp}$-weighted center of a jet to draw a new cirle of radius
$R$. In the current algorithm there is no such iteration step.
For an ideal jet assignment it would also be necessary to improve
the treatment when two jet circles partially overlap.
 
A final technical note. A natural implementation of a cell finding
algorithm is based on having a two-dimensional array of $E_{\perp}$
values, with dimensions to match the detector grid.
Very often most of the cells would then
be empty, in particular for low-multiplicity events in fine-grained
calorimeters. Our implementation is somewhat atypical, since cells are
only reserved space (contents and position) when they are shown to be
non-empty. This means that all non-empty cells have to be looped over
to find which are within the required distance $R$ of a potential jet
initiator. The algorithm is therefore faster than the ordinary kind
if the average cell occupancy is low, but slower if it is high.
 
\subsection{Event Statistics}
 
All the event-analysis routines above are defined on an
event-by-event basis. Once found, the quantities are about equally
often used to define inclusive distributions as to select specific
classes of events for continued study. For instance, the thrust
routine might be used either to find the inclusive $T$ distribution
or to select events with $T < 0.9$. Other measures, although still
defined for the individual event, only make sense to discuss in
terms of averages over many events. A small set of such measures
is found in \ttt{LUTABU}. This routine
has to be called once after each event to accumulate statistics,
and once in the end to print the final tables. Of course, among the
wealth of possibilities imaginable, the ones collected here are only
a small sample, selected because the author at some point has found
a use for them himself.
 
\subsubsection{Multiplicities}
 
Three options are available to collect information on multiplicities
in events. One gives the flavour content of the final state in hard
interaction processes, e.g. the relative composition of
$\d\dbar / \u\ubar / \s\sbar / \c\cbar / \b\bbar$ in $\ee$
annihilation events. Additionally it gives the total parton multiplicity
distribution at the end of parton showering. Another gives the
inclusive rate of all the different particles produced in events,
either as intermediate resonances or as final-state particles.
The number is subdivided into particles produced from fragmentation
(primary particles) and those produced in decays (secondary
particles).
 
The third option tabulates the rate of exclusive final states, after
all allowed decays have occurred. Since only events with up to 8 
final-state particles are analyzed, this is clearly not intended for the
study of complete high-energy events. Rather the main application is
for an analysis of the decay modes of a single particle. For instance,
the decay data for $\D$ mesons is given in terms of channels that
also contain unstable particles, such as $\rho$ and $\eta$, which
decay further. Therefore a given final state may receive
contributions from several tabulated decay channels; e.g.
$\K \pi \pi$ from $\K^* \pi$ and $\K \rho$, and so on.
 
\subsubsection{Energy-Energy Correlation}
 
The Energy-Energy Correlation is defined by \cite{Bas78}
\begin{equation}
\mrm{EEC}(\theta) = \sum_{i < j} 
\frac{2 E_i E_j}{E_{\mrm{vis}}^2} \,
\delta(\theta - \theta_{ij})  ~,
\end{equation}
and its Asymmetry by
\begin{equation}
\mrm{EECA}(\theta) = \mrm{EEC}(\pi - \theta) - \mrm{EEC}(\theta) ~.
\end{equation}
Here $\theta_{ij}$ is the opening angle between the two particles
$i$ and $j$, with energies $E_i$ and $E_j$. In principle,
normalization should be to $E_{\mrm{cm}}$, but if not all particles 
are detected it is convenient to normalize to the total visible
energy $E_{\mrm{vis}}$. Taking into account the autocorrelation term
$i = j$, the total $\mrm{EEC}$ in an event then is unity.
The $\delta$ function peak is smeared out by the
finite bin width $\Delta \theta$ in the histogram, i.e.,
it is replaced by a contribution $1 / \Delta \theta$ to the bin
which contains $\theta_{ij}$.
 
The formulae above refer to an individual event, and are to be
averaged over all events to suppress statistical fluctuations, and
obtain smooth functions of $\theta$.
 
\subsubsection{Factorial moments}
 
Factorial moments may be used to search for intermittency in events
\cite{Bia86}. The whole field has been much studied in recent years,
and a host of different measures have been proposed. We only implement
one of the original prescriptions.
 
To calculate the factorial moments, the full rapidity (or
pseudorapidity) and
azimuthal ranges are subdivided into bins of successively smaller
size, and the multiplicity distributions in bins is studied. The
program calculates pseudorapidity with respect to the $z$ axis;
if desired, one could first find
an event axis, e.g. the sphericity or thrust axis, and subsequently
rotate the event to align this axis with the $z$ direction.
 
The full rapidity range $|y| < y_{\mmax}$ (or pseudorapidity range
$|\eta| < \eta_{\mmax}$) and
azimuthal range $0 < \varphi < 2\pi$ are subdivided into $m_y$ and
$m_{\varphi}$ equally large bins. In fact, the whole analysis is
performed thrice: once with $m_{\varphi}=1$ and the $y$ (or $\eta$)
range gradually
divided into 1, 2, 4, 8, 16, 32, 64, 128, 256 and 512 bins, once
with $m_y = 1$ and the $\varphi$ range subdivided as above, and
finally once with $m_y = m_{\varphi}$ according to the same binary
sequence. Given the multiplicity $n_j$ in bin $j$, the $i$:th
factorial moment is defined by
\begin{equation}
F_i = (m_y m_{\varphi})^{i-1} \, \sum_j
\frac{n_j(n_j-1)\cdots(n_j-i+1)}{n(n-1)\cdots(n-i+1)}  ~.
\end{equation}
Here $n = \sum_j n_j$ is the total multiplicity of the event within
the allowed $y$ (or $\eta$) limits. The calculation is performed for
the second through the fifth moments, i.e. $F_2$ through $F_5$.
 
The $F_i$ as given here are defined for the individual event,
and have to be averaged over many events to give a reasonably smooth
behaviour. If particle production is uniform and uncorrelated
according to Poissonian statistics, one expects
$\langle F_i \rangle \equiv 1$ for all moments and all bin sizes.
If, on the other hand, particles are locally clustered, factorial
moments should increase when bins are made smaller, down to the
characteristic dimensions of the clustering.
 
\subsection{Routines and Common Block Variables}
\label{ss:evanrout}
 
The six routines \ttt{LUSPHE}, \ttt{LUTHRU}, \ttt{LUCLUS},
\ttt{LUCELL}, \ttt{LUJMAS} and \ttt{LUFOWO} give you the possibility
to find some global event shape properties. The routine \ttt{LUTABU}
performs a statistical analysis of a number of different quantities
like particle content, factorial moments and the energy--energy
correlation.
 
Note that, by default, all remaining partons/particles except
neutrinos are used in the analysis. Neutrinos may be included with
\ttt{MSTU(41)=1}. Also note that axes determined are stored in
\ttt{LUJETS}, but are not proper four-vectors and, as a general rule
(with some exceptions), should therefore not be rotated or boosted.
 
\drawbox{CALL LUSPHE(SPH,APL)}\label{p:LUSPHE}
\begin{entry}
\itemc{Purpose:} to diagonalize the momentum tensor, i.e. find the
eigenvalues $\lambda_1 > \lambda_2 > \lambda_3$, with sum unity,
and the corresponding eigenvectors. \\
Momentum power dependence is given by \ttt{PARU(41)}; default
corresponds to sphericity, \ttt{PARU(41)=1.} gives measures linear
in momenta. Which particles (or partons) are used in the analysis
is determined by the \ttt{MSTU(41)} value.
\iteme{SPH :} $\frac{3}{2} (\lambda_2 + \lambda_3)$, i.e. sphericity
(for \ttt{PARU(41)=2.}).
\begin{subentry}
\iteme{= -1. :} analysis not performed because event contained less
than two particles (or two exactly back-to-back particles, in
which case the two transverse directions would be undefined).
\end{subentry}
\iteme{APL :} $\frac{3}{2} \lambda_3$, i.e. aplanarity (for
\ttt{PARU(41)=2.}).
\begin{subentry}
\iteme{= -1. :} as \ttt{SPH=-1.}.
\end{subentry}
\itemc{Remark:} the lines \ttt{N+1} through \ttt{N+3} (\ttt{N-2}
through \ttt{N} for \ttt{MSTU(43)=2}) in \ttt{LUJETS} will, after
a call, contain the following information: \\
\ttt{K(N+i,1) =} 31; \\
\ttt{K(N+i,2) =} 95; \\
\ttt{K(N+i,3) :} $i$, the axis number, $i=1,2,3$; \\
\ttt{K(N+i,4), K(N+i,5) =} 0; \\
\ttt{P(N+i,1) - P(N+i,3) :} the $i$'th eigenvector, $x$, $y$ and $z$
components; \\
\ttt{P(N+i,4) :} $\lambda_i$, the $i$'th eigenvalue; \\
\ttt{P(N+i,5) =} 0; \\
\ttt{V(N+i,1) - V(N+i,5) =} 0. \\
Also, the number of particles used in the analysis is given in
\ttt{MSTU(62)}.
\end{entry}
 
\drawbox{CALL LUTHRU(THR,OBL)}\label{p:LUTHRU}
\begin{entry}
\itemc{Purpose:} to find the thrust, major and minor axes and
corresponding projected momentum quantities, in particular thrust
and oblateness. The performance of the program is affected by
\ttt{MSTU(44)}, \ttt{MSTU(45)}, \ttt{PARU(42)} and \ttt{PARU(48)}.
In particular, \ttt{PARU(42)} gives the momentum dependence, with
the default value \ttt{=1.} corresponding to linear dependence.
Which particles (or partons) are used in the analysis
is determined by the \ttt{MSTU(41)} value.
\iteme{THR :} thrust (for \ttt{PARU(42)=1.}).
\begin{subentry}
\iteme{= -1. :} analysis not performed because event contained
less than two particles.
\iteme{= -2. :} remaining space in \ttt{LUJETS} (partly used as
working area) not large enough to allow analysis.
\end{subentry}
\iteme{OBL :} oblateness (for \ttt{PARU(42)=1.}).
\begin{subentry}
\iteme{= -1., -2. :} as for \ttt{THR}.
\end{subentry}
\itemc{Remark:} the lines \ttt{N+1} through \ttt{N+3} (\ttt{N-2}
through \ttt{N} for \ttt{MSTU(43)=2}) in \ttt{LUJETS} will, after
a call, contain the following information: \\
\ttt{K(N+i,1) =} 31; \\
\ttt{K(N+i,2) =} 96; \\
\ttt{K(N+i,3) :} $i$, the axis number, $i=1,2,3$; \\
\ttt{K(N+i,4), K(N+i,5) =} 0; \\
\ttt{P(N+i,1) - P(N+i,3) :} the thrust, major and minor axis,
respectively, for $i = 1, 2$ and 3; \\
\ttt{P(N+i,4) :} corresponding thrust, major and minor value; \\
\ttt{P(N+i,5) =} 0; \\
\ttt{V(N+i,1) - V(N+i,5) =} 0. \\
Also, the number of particles used in the analysis is given in
\ttt{MSTU(62)}.
\end{entry}
 
\drawbox{CALL LUCLUS(NJET)}\label{p:LUCLUS}
\begin{entry}
\itemc{Purpose:} to reconstruct an arbitrary number of jets using a
cluster analysis method based on particle momenta. \\
Two different distance measures are available, see section
\ref{ss:clustfind}. The choice is controlled by \ttt{MSTU(46)}. The 
distance scale $d_{\mrm{join}}$, above which two clusters may not be
joined, is normally given by \ttt{PARU(44)}. In general, 
$d_{\mrm{join}}$
may be varied to describe different `jet-resolution powers';
the default value, 2.5 GeV, is fairly well suited for $\ee$ physics
at 30--40 GeV. With the alternative mass distance measure,
\ttt{PARU(44)} can be used to set the absolute maximum cluster mass,
or \ttt{PARU(45)} to set the scaled one, i.e. in 
$y = m^2/E_{\mrm{cm}}^2$, where $E_{\mrm{cm}}$ is the total 
invariant mass of the particles being considered. \\
It is possible to continue the cluster search from the configuration
already found, with a new higher $d_{\mrm{join}}$ scale, by selecting
\ttt{MSTU(48)} properly. In \ttt{MSTU(47)} one can also require a
minimum number of jets to be reconstructed; combined with an
artificially large $d_{\mrm{join}}$ this can be used to reconstruct a
predetermined number of jets. \\
Which particles (or partons) are used in the analysis is determined
by the \ttt{MSTU(41)} value, whereas assumptions about particle masses
is given by \ttt{MSTU(42)}. The parameters \ttt{PARU(43)} and
\ttt{PARU(48)} regulate more technical details (for events at high
energies and large multiplicities, however, the choice of a larger
\ttt{PARU(43)} may be necessary to obtain reasonable reconstruction
times).
\iteme{NJET :} the number of clusters reconstructed.
\begin{subentry}
\iteme{= -1 :} analysis not performed because event contained less than
\ttt{MSTU(47)} (normally 1) particles, or analysis failed to
reconstruct the requested number of jets.
\iteme{= -2 :} remaining space in \ttt{LUJETS} (partly used as working
area) not large enough to allow analysis.
\end{subentry}
\itemc{Remark:} if the analysis does not fail, further information is
found in \ttt{MSTU(61) - MSTU(63)} and \ttt{PARU(61) - PARU(63)}.
In particular, \ttt{PARU(61)} contains the invariant mass for the
system analyzed, i.e. the number used in determining the denominator
of $y = m^2/E_{\mrm{cm}}^2$. \ttt{PARU(62)} gives the generalized 
thrust, i.e. the sum of (absolute values of) cluster momenta divided 
by the sum of particle momenta (roughly the same as multicity). 
\ttt{PARU(63)} gives the minimum distance $d$ (in $\pT$ or $m$) 
between two clusters in the final cluster configuration, 0 in case 
of only one cluster. \\
Further, the lines \ttt{N+1} through \ttt{N+NJET} (\ttt{N-NJET+1}
through \ttt{N} for \ttt{MSTU(43)=2}) in \ttt{LUJETS}
will, after a call, contain the following information: \\
\ttt{K(N+i,1) =} 31; \\
\ttt{K(N+i,2) =} 97; \\
\ttt{K(N+i,3) :} $i$, the jet number, with the jets arranged in
falling order of absolute momentum; \\
\ttt{K(N+i,4) :} the number of particles assigned to jet $i$; \\
\ttt{K(N+i,5) =} 0; \\
\ttt{P(N+i,1) - P(N+i,5) :} momentum, energy and invariant mass of
jet $i$; \\
\ttt{V(N+i,1) - V(N+i,5) =} 0. \\
Also, for a particle which was used in the analysis,
\ttt{K(I,4)}$ = i$, where \ttt{I} is the particle number and $i$
the number of the jet it has ben assigned to. Undecayed particles
not used then have \ttt{K(I,4)=0}. An exception is made for lines
with \ttt{K(I,1)=3} (which anyhow are not normally interesting for
cluster search), where the colour-flow information stored in
\ttt{K(I,4)} is left intact.
\end{entry}
 
\drawbox{CALL LUCELL(NJET)}\label{p:LUCELL}
\begin{entry}
\itemc{Purpose:} to provide a simpler cluster routine more in line
with what is currently used in the study of high-$\pT$ collider
events. \\
A detector is assumed to stretch in pseudorapidity between
\ttt{-PARU(51)} and \ttt{+PARU(51)} and be segmented in
\ttt{MSTU(51)} equally large $\eta$ (pseudorapidity) bins and
\ttt{MSTU(52)} $\varphi$ (azimuthal) bins. Transverse
energy $E_{\perp}$ for undecayed entries are summed up in each bin.
For \ttt{MSTU(53)} non-zero, the energy is smeared by calorimetric
resolution effects, cell by cell. This is done according to a Gaussian
distribution; if \ttt{MSTU(53)=1} the standard deviation for the
$E_{\perp}$ is \ttt{PARU(55)}$\times \sqrt{E_{\perp}}$, if
\ttt{MSTU(53)=2} the standard deviation for the $E$ is
\ttt{PARU(55)}$\times \sqrt{E}$, $E_{\perp}$ and $E$ expressed in GeV.
The Gaussian is cut off at 0 and at a factor \ttt{PARU(56)} times the
correct $E_{\perp}$ or $E$. Cells with an  $E_{\perp}$ below a given 
threshold \ttt{PARU(58)} are removed from further consideration;
by default \ttt{PARU(58)=0.} and thus all cells are kept.
\\
All bins with $E_{\perp} > $\ttt{PARU(52)} are taken to be possible
initiators of jets, and are tried in falling $E_{\perp}$ sequence to
check whether the total $E_{\perp}$ summed over cells no more distant
than \ttt{PARU(54)} in $\sqrt{(\Delta\eta)^2 + (\Delta\varphi)^2}$
exceeds \ttt{PARU(53)}. If so, these cells define one jet, and are
removed from further consideration. Contrary to \ttt{LUCLUS}, not all
particles need be assigned to jets. Which particles (or partons) are
used in the analysis is determined by the \ttt{MSTU(41)} value.
\iteme{NJET :} the number of jets reconstructed (may be 0).
\begin{subentry}
\iteme{= -2 :} remaining space in \ttt{LUJETS} (partly used as
working area) not large enough to allow analysis.
\end{subentry}
\itemc{Remark:} the lines \ttt{N+1} through \ttt{N+NJET}
(\ttt{N-NJET+1} through \ttt{N} for \ttt{MSTU(43)=2}) in
\ttt{LUJETS} will, after a call, contain the following information: \\
\ttt{K(N+i,1) =} 31; \\
\ttt{K(N+i,2) =} 98; \\
\ttt{K(N+i,3) :} $i$, the jet number, with the jets arranged in
falling order in $E_{\perp}$; \\
\ttt{K(N+i,4) :} the number of particles assigned to jet $i$; \\
\ttt{K(N+i,5) =} 0; \\
\ttt{V(N+i,1) - V(N+i,5) =} 0. \\
Further, for \ttt{MSTU(54)=1} \\
\ttt{P(N+i,1), P(N+i,2) =} position in $\eta$ and $\varphi$ of the
center of the jet initiator cell, i.e. geometrical center of jet; \\
\ttt{P(N+i,3), P(N+i,4) =} position in $\eta$ and $\varphi$ of the
$E_{\perp}$-weighted center of the jet, i.e. the center of gravity
of the jet; \\
\ttt{P(N+i,5) =} sum $E_{\perp}$ of the jet; \\
while for \ttt{MSTU(54)=2} \\
\ttt{P(N+i,1) - P(N+i,5) :} the jet momentum vector, constructed
from the summed $E_{\perp}$ and the $\eta$ and $\varphi$ of the
$E_{\perp}$-weighted center of the jet as \\
$(p_x, p_y, p_z, E, m)=E_{\perp} (\cos\varphi, \sin\varphi,
\sinh\eta, \cosh\eta, 0)$; \\
and for \ttt{MSTU(54)=3} \\
\ttt{P(N+i,1) - P(N+i,5) :} the jet momentum vector, constructed by
adding vectorially the momentum of each cell assigned to the jet,
assuming that all the $E_{\perp}$ was deposited at the center of the
cell, and with the jet mass in \ttt{P(N+i,5)} calculated from the
summed $E$ and $\mbf{p}$ as $m^2 = E^2 - p_x^2 - p_y^2 - p_z^2$. \\
Also, the number of particles used in the analysis is given in
\ttt{MSTU(62)}, and the number of cells hit in \ttt{MSTU(63)}.
\end{entry}
 
\drawbox{CALL LUJMAS(PMH,PML)}\label{p:LUJMAS}
\begin{entry}
\itemc{Purpose:} to reconstruct high and low jet mass of an event.
A simplified algorithm is used, wherein a preliminary division of
the event into two hemispheres is done transversely to the sphericity
axis. Then one particle at a time is reassigned to the other
hemisphere if that reduces the sum of squares of the two jet masses,
$m_{\mrm{H}}^2 + m_{\mrm{L}}^2$. The procedure is stopped when no 
further significant change (see \ttt{PARU(48)}) is obtained. Often, the
original assignment is retained as it is. Which particles (or partons)
used in the analysis is determined by the \ttt{MSTU(41)} value,
whereas assumptions about particle masses is given by \ttt{MSTU(42)}.
\iteme{PMH :} heavy jet mass (in GeV).
\begin{subentry}
\iteme{= -2. :} remaining space in \ttt{LUJETS} (partly used as
working area) not large enough to allow analysis.
\end{subentry}
\iteme{PML :} light jet mass (in GeV).
\begin{subentry}
\iteme{= -2. :} as for \ttt{PMH=-2.}.
\end{subentry}
\itemc{Remark:} After a successful call, \ttt{MSTU(62)} contains the
number of particles used in the analysis, and \ttt{PARU(61)} the
invariant mass of the system analyzed. The latter number is helpful
in constructing scaled jet masses.
\end{entry}
 
\drawbox{CALL LUFOWO(H10,H20,H30,H40)}\label{p:LUFOWO}
\begin{entry}
\itemc{Purpose:} to do an event analysis in terms of the Fox-Wolfram
moments. The moments $H_i$ are normalized to the lowest one, $H_0$.
Which particles (or partons) are used in the analysis is determined
by the \ttt{MSTU(41)} value.
\iteme{H10 :} $H_1/H_0$. Is $=0$ if momentum is balanced.
\iteme{H20 :} $H_2/H_0$.
\iteme{H30 :} $H_3/H_0$.
\iteme{H40 :} $H_4/H_0$.
\itemc{Remark:} the number of particles used in the analysis is given
in \ttt{MSTU(62)}.
\end{entry}
 
\drawbox{CALL LUTABU(MTABU)}\label{p:LUTABU}
\begin{entry}
\itemc{Purpose:} to provide a number of event-analysis options which
can be be used on each new event, with accumulated statistics to be
written out on request. When errors are quoted, these refer to
the uncertainty in the average value for the event sample as a
whole, rather than to the spread of the individual events, i.e.
errors decrease like one over the square root of the number of
events analyzed. For a correct use of \ttt{LUTABU}, it is not
permissible to freely mix generation and analysis of different
classes of events, since only one set of statistics counters exists.
A single run may still contain sequential `subruns', between
which statistics is reset. Whenever an event is analyzed, the
number of particles/partons used is given in \ttt{MSTU(62)}.
\iteme{MTABU :} determines which action is to be taken. Generally, a
last digit equal to 0 indicates that the statistics counters for this
option is to be reset; since the counters are reset (by \ttt{DATA}
statements) at the beginning of a run, this is not used normally. Last
digit 1 leads to an analysis of current event with respect to the
desired properties. Note that the resulting action may depend on how
the event generated has been rotated, boosted or edited before this
call. The statistics accumulated is output in tabular form with
last digit 2, while it is dumped in the \ttt{LUJETS} common block for
last digit 3. The latter option may be useful for interfacing to
graphics output.
\begin{subentry}
\iteme{= 10 :} statistics on parton multiplicity is reset.
\iteme{= 11 :} the parton content of the current event is analyzed,
classified according to the flavour content of the hard
interaction and the total number of partons. The flavour
content is assumed given in \ttt{MSTU(161)} and \ttt{MSTU(162)};
these are automatically set e.g. in \ttt{LUEEVT} and \ttt{PYEVNT}
calls.
\iteme{= 12 :} gives a table on parton multiplicity distribution.
\iteme{= 13 :} stores the parton multiplicity distribution of events
in \ttt{/LUJETS/}, using the following format: \\
\ttt{N =} total number of different channels found; \\
\ttt{K(I,1) =} 32; \\
\ttt{K(I,2) =} 99; \\
\ttt{K(I,3), K(I,4) =} the two flavours of the flavour content; \\
\ttt{K(I,5) =} total number of events found with flavour content of
\ttt{K(I,3)} and \ttt{K(I,4)}; \\
\ttt{P(I,1) - P(I,5) =} relative probability to find given flavour
content and a total of 1, 2, 3, 4 or 5 partons, respectively; \\
\ttt{V(I,1) - V(I,5) =} relative probability to find given flavour
content and a total of 6--7, 8--10, 11--15, 16--25 or above 25
partons, respectively. \\
In addition, \ttt{MSTU(3)=1} and \\
\ttt{K(N+1,1) =} 32; \\
\ttt{K(N+1,2) =} 99; \\
\ttt{K(N+1,5) =} number of events analyzed.
\iteme{= 20 :} statistics on particle content is reset.
\iteme{= 21 :} the particle/parton content of the current event is
analyzed, also for particles which have subsequently decayed and
partons which have fragmented (unless this has been made impossible
by a preceding \ttt{LUEDIT} call). Particles are subdivided into
primary and secondary ones, the main principle being that primary
particles are those produced in the fragmentation of a string,
while secondary come from decay of other particles. Since
particles (top, say), may decay into partons, the distinction
is not always unique.
\iteme{= 22 :} gives a table of particle content in events.
\iteme{= 23 :} stores particle content in events in \ttt{/LUJETS/},
using the following format: \\
\ttt{N =} number of different particle species found; \\
\ttt{K(I,1) =} 32; \\
\ttt{K(I,2) =} 99; \\
\ttt{K(I,3) =} particle KF code; \\
\ttt{K(I,5) =} total number of particles and antiparticles of this
species; \\
\ttt{P(I,1) =} average number of primary particles per event; \\
\ttt{P(I,2) =} average number of secondary particles per event; \\
\ttt{P(I,3) =} average number of primary antiparticles per event; \\
\ttt{P(I,4) =} average number of secondary antiparticles per event; \\
\ttt{P(I,5) =} average total number of particles or antiparticles
per event. \\
In addition, \ttt{MSTU(3)=1} and \\
\ttt{K(N+1,1) =} 32; \\
\ttt{K(N+1,2) =} 99; \\
\ttt{K(N+1,5) =} number of events analyzed; \\
\ttt{P(N+1,1) =} average primary multiplicity per event; \\
\ttt{P(N+1,2) =} average final multiplicity per event; \\
\ttt{P(N+1,3) =} average charged multiplicity per event.
\iteme{= 30 :} statistics on factorial moments is reset.
\iteme{= 31 :} analyzes the factorial moments of the multiplicity
distribution in different bins of rapidity and azimuth.
Which particles (or partons) are used in the analysis is
determined by the \ttt{MSTU(41)} value. The selection between usage
of true rapidity, pion rapidity or pseudorapidity is regulated
by \ttt{MSTU(42)}. The $z$ axis is assumed to be event axis; if this
is not desirable find an event axis e.g. with \ttt{LUSPHE} or
\ttt{LUTHRU} and use \ttt{LUEDIT(31)}. Maximum (pion-, pseudo-)
rapidity, which sets the limit for the rapidity plateau or the
experimental acceptance, is given by \ttt{PARU(57)}.
\iteme{= 32 :} prints a table of the first four factorial moments
for various bins of pseudorapidity and azimuth. The moments are
properly normalized so that they would be unity (up to
statistical fluctuations) for uniform and uncorrelated particle
production according to Poissonian statistics, but increasing
for decreasing bin size in case of `intermittent' behaviour.
The error on the average value is based on the actual
statistical sample (i.e. does not use any assumptions on the
distribution to relate errors to the average values of higher
moments). Note that for small bin sizes, where the average
multiplicity is small and the factorial moment therefore only
very rarely is non-vanishing, moment values may fluctuate wildly
and the errors given may be too low.
\iteme{= 33 :} stores the factorial moments in \ttt{/LUJETS/},
using the format: \\
\ttt{N =} 30, with \ttt{I = }$i=1$--10 corresponding to results for
slicing the rapidity range in $2^{i-1}$ bins, \ttt{I = }$i = 11$--20
to slicing the azimuth in $2^{i-11}$ bins, and \ttt{I = }$i = 21$--30
to slicing both rapidity and azimuth, each in $2^{i-21}$ bins; \\
\ttt{K(I,1) =} 32; \\
\ttt{K(I,2) =} 99; \\
\ttt{K(I,3) =} number of bins in rapidity; \\
\ttt{K(I,4) =} number of bins in azimuth; \\
\ttt{P(I,1) =} rapidity bin size; \\
\ttt{P(I,2) - P(I,5) =} $\langle F_2 \rangle$--$\langle F_5 \rangle$,
i.e. mean of second, third, fourth and fifth factorial moment; \\
\ttt{V(I,1) =} azimuthal bin size; \\
\ttt{V(I,2) - V(I,5) =} statistical errors on
$\langle F_2 \rangle$--$\langle F_5 \rangle$. \\
In addition, \ttt{MSTU(3) =} 1 and \\
\ttt{K(31,1) =} 32; \\
\ttt{K(31,2) =} 99; \\
\ttt{K(31,5) =} number of events analyzed.
\iteme{= 40 :} statistics on energy--energy correlation is reset.
\iteme{= 41 :} the energy--energy correlation $\mrm{EEC}$ of the 
current
event is analyzed. Which particles (or partons) are used in the
analysis is determined by the \ttt{MSTU(41)} value. Events are
assumed given in their c.m. frame. The weight assigned to a pair
$i$ and $j$ is $2 E_i E_j/E_{\mrm{vis}}^2$, where $E_{\mrm{vis}}$ 
is the sum of energies of
all analyzed particles in the event. Energies are determined from
the momenta of particles, with mass determined according to the
\ttt{MSTU(42)} value. Statistics is accumulated for the relative
angle $\theta_{ij}$, ranging between 0 and 180 degrees, subdivided
into 50 bins.
\iteme{= 42 :} prints a table of the energy--energy correlation 
$\mrm{EEC}$ and its asymmetry $\mrm{EECA}$, with errors. 
The definition of errors is not unique. In our approach each event 
is viewed as one observation, i.e. an $\mrm{EEC}$ and 
$\mrm{EECA}$ distribution is obtained by
summing over all particle pairs of an event, and then the
average and spread of this event-distribution is calculated
in the standard fashion. The quoted error is therefore inversely
proportional to the square root of the number of events. It could
have been possible to view each single particle pair as one
observation, which would have given somewhat lower errors, but then
one would also be forced to do a complicated correction procedure
to account for the pairs in an event not being uncorrelated
(two hard jets separated by a given angle typically corresponds
to several pairs at about that angle). Note, however, that
in our approach the squared error on an $\mrm{EECA}$ bin is smaller
than the sum of the squares of the errors on the corresponding 
$\mrm{EEC}$ bins (as it should be). Also note that it is not possible 
to combine the errors of two nearby bins by hand from the
information given, since nearby bins are correlated (again a
trivial consequence of the presence of jets).
\iteme{= 43 :} stores the $\mrm{EEC}$ and $\mrm{EECA}$ in 
\ttt{/LUJETS/}, using the format: \\
\ttt{N =} 25; \\
\ttt{K(I,1) =} 32; \\
\ttt{K(I,2) =} 99; \\
\ttt{P(I,1) =} $\mrm{EEC}$ for angles between \ttt{I-1} and \ttt{I}, 
in units of $3.6^{\circ}$; \\
\ttt{P(I,2) =} $\mrm{EEC}$ for angles between \ttt{50-I} and 
\ttt{51-I}, in units of $3.6^{\circ}$; \\
\ttt{P(I,3) =} $\mrm{EECA}$ for angles between \ttt{I-1} and 
\ttt{I}, in units of $3.6^{\circ}$; \\
\ttt{P(I,4), P(I,5) :} lower and upper edge of angular range of bin
\ttt{I}, expressed in radians; \\
\ttt{V(I,1) - V(I,3) :} errors on the $\mrm{EEC}$ and $\mrm{EECA}$ 
values stored in \ttt{P(I,1) - P(I,3)} (see \ttt{=42} for comments); \\
\ttt{V(I,4), V(I,5) :} lower and upper edge of angular range of bin
\ttt{I}, expressed in degrees. \\
In addition, \ttt{ MSTU(3)=1} and \\
\ttt{K(26,1) =} 32; \\
\ttt{K(26,2) =} 99; \\
\ttt{K(26,5) =} number of events analyzed.
\iteme{= 50 :} statistics on complete final states is reset.
\iteme{= 51 :} analyzes the particle content of the final state of
the current event record. During the course of the run, statistics
is thus accumulated on how often different final states appear.
Only final states with up to 8 particles are analyzed, and there
is only reserved space for up to 200 different final states.
Most high energy events have multiplicities far above 8, so the
main use for this tool is to study the effective branching
ratios obtained with a given decay model for e.g. charm or bottom
hadrons. Then \ttt{LU1ENT} may be used to generate one decaying
particle at a time, with a subsequent analysis by \ttt{LUTABU}.
Depending on at what level this studied is to be carried out,
some particle decays may be switched off, like $\pi^0$.
\iteme{= 52 :} gives a list of the (at most 200) channels with up
to 8 particles in the final state, with their relative branching
ratio. The ordering is according to multiplicity, and within
each multiplicity according to an ascending order of KF codes.
The KF codes of the particles belonging to a given channel are
given in descending order.
\iteme{= 53 :} stores the final states and branching ratios found in
\ttt{/LUJETS/}, using the format: \\
\ttt{N =} number of different explicit final states found (at most
200); \\
\ttt{K(I,1) =} 32; \\
\ttt{K(I,2) =} 99; \\
\ttt{K(I,5) =} multiplicity of given final state, a number between
1 and 8; \\
\ttt{P(I,1) - P(I,5), V(I,1) - V(I,3) :} the KF codes of the up to 8
particles of the given final state, converted to real
numbers, with trailing zeroes for positions not used; \\
\ttt{V(I,5) :} effective branching ratio for the given final state. \\
In addition, \ttt{MSTU(3)=1} and \\
\ttt{K(N+1,1) =} 32; \\
\ttt{K(N+1,2) =} 99; \\
\ttt{K(N+1,5) =} number of events analyzed; \\
\ttt{V(N+1,5) =} summed branching ratio for finals states not given
above, either because they contained more than 8 particles
or because all 200 channels have been used up.
\end{subentry}
\end{entry}
  
\drawbox{COMMON/LUDAT1/MSTU(200),PARU(200),MSTJ(200),PARJ(200)}
\begin{entry}
\itemc{Purpose:} to give access to a number of status codes and
parameters which regulate the performance of {\Je}.
Most parameters are described in section \ref{ss:JETswitch};
here only those related to the event-analysis routines are described.

\boxsep
 
\iteme{MSTU(41) :}\label{p:MSTU41} (D=2) partons/particles used in 
the event-analysis routines \ttt{LUSPHE}, \ttt{LUTHRU}, \ttt{LUCLUS}, 
\ttt{LUCELL}, \ttt{LUJMAS}, \ttt{LUFOWO} and \ttt{LUTABU} 
(\ttt{LUTABU(11)} excepted).
\begin{subentry}
\iteme{= 1 :} all partons/particles that have not fragmented/decayed.
\iteme{= 2 :} ditto, with the exception of neutrinos and unknown
particles.
\iteme{= 3 :} only charged, stable particles, plus any partons
still not fragmented.
\end{subentry}
 
\iteme{MSTU(42) :} (D=2) assumed particle masses, used in
calculating energies $E^2 = \mbf{p}^2 + m^2$, as subsequently
used in \ttt{LUCLUS}, \ttt{LUJMAS} and \ttt{LUTABU}
(in the latter also for pseudorapidity, pion rapidity or true
rapidity selection).
\begin{subentry}
\iteme{= 0 :} all particles are assumed massless.
\iteme{= 1 :} all particles, except the photon, are assumed to have
the charged pion mass.
\iteme{= 2 :} the true masses are used.
\end{subentry}
 
\iteme{MSTU(43) :} (D=1) storing of event-analysis information (mainly
jet axes), in \ttt{LUSPHE}, \ttt{LUTHRU}, \ttt{LUCLUS} and
\ttt{LUCELL}.
\begin{subentry}
\iteme{= 1 :} stored after the event proper, in positions \ttt{N+1}
through \ttt{N+MSTU(3)}. If several of the routines are used in
succession, all but the latest information is overwritten.
\iteme{= 2 :} stored with the event proper, i.e. at the end of the
event listing, with \ttt{N} updated accordingly. If several of the
routines are used in succession, all the axes determined are
available.
\end{subentry}
 
\iteme{MSTU(44) :} (D=4) is the number of the fastest (i.e. with
largest momentum) particles used to construct the (at most) 10 most
promising starting configurations for the thrust axis determination.
 
\iteme{MSTU(45) :} (D=2) is the number of different starting
configurations above, which have to converge to the same (best) value
before this is accepted as the correct thrust axis.
 
\iteme{MSTU(46) :} (D=1) distance measure used for the joining of
clusters in \ttt{LUCLUS}.
\begin{subentry}
\iteme{= 1 :} $d_{ij}$, i.e. approximately relative transverse 
momentum. Anytime two clusters have been joined, particles are 
reassigned to the cluster they now are closest to. The distance 
cut-off $d_{\mrm{join}}$ is stored in \ttt{PARU(44)}.
\iteme{= 2 :} distance measure as in \ttt{=1}, but particles are
never reassigned to new jets.
\iteme{= 3 :} JADE distance measyre $y_{ij}$, but with dimensions
to correspond approximately to total invariant mass. Particles may
never be reassigned between clusters. The distance cut-off 
$m_{\mmin}$ is stored in \ttt{PARU(44)}.
\iteme{= 4 :} as \ttt{=3}, but a scaled JADE distance $y_{ij}$ is 
used instead of $m_{ij}$. The distance cut-off $y_{\mmin}$ is stored 
in \ttt{PARU(45)}.
\end{subentry}
 
\iteme{MSTU(47) :} (D=1) the minimum number of clusters to be
reconstructed by \ttt{LUCLUS}.
 
\iteme{MSTU(48) :} (D=0) mode of operation of the \ttt{LUCLUS}
routine.
\begin{subentry}
\iteme{= 0 :} the cluster search is started from scratch.
\iteme{= 1 :} the clusters obtained in a previous cluster search on the
same event (with \ttt{MSTU(48)=0}) are to be taken as the starting
point for subsequent cluster joining. For this call to have any effect,
the joining scale in \ttt{PARU(44)} or \ttt{PARU(45)} must have been
changed. If the event record has been modified after the last
\ttt{LUCLUS} call, or if any other cluster search parameter setting
has been changed, the subsequent result is unpredictable.
\end{subentry}
 
\iteme{MSTU(51) :} (D=25) number of pseudorapidity bins that the range
between \ttt{-PARU(51)} and \ttt{+PARU(51)} is divided into to define
cell size for \ttt{LUCELL}.
 
\iteme{MSTU(52) :} (D=24) number of azimuthal bins, used to define the
cell size for \ttt{LUCELL}.
 
\iteme{MSTU(53) :} (D=0) smearing of correct energy, imposed
cell-by-cell in \ttt{LUCELL}, to simulate calorimeter resolution
effects.
\begin{subentry}
\iteme{= 0 :} no smearing.
\iteme{= 1 :} the transverse energy in a cell, $E_{\perp}$, is smeared
according to a Gaussian distribution with standard deviation
\ttt{PARU(55)}$\times \sqrt{E_{\perp}}$, where $E_{\perp}$ is given in
GeV. The Gaussian is cut off so that $0 < E_{\perp \mrm{smeared}} 
< $\ttt{PARU(56)}$\times E_{\perp \mrm{true}}$.
\iteme{= 2 :} as \ttt{=1}, but it is the energy $E$ rather than the
transverse energy $E_{\perp}$ that is smeared.
\end{subentry}
 
\iteme{MSTU(54) :} (D=1) form for presentation of information about
reconstructed clusters in \ttt{LUCELL}, as stored in \ttt{LUJETS}
according to the \ttt{MSTU(43)} value.
\begin{subentry}
\iteme{= 1 :} the \ttt{P} vector in each line contains $\eta$ and
$\varphi$ for the geometric origin of the jet, $\eta$ and $\varphi$
for the weighted center of the jet, and jet $E_{\perp}$, respectively.
\iteme{= 2 :} the \ttt{P} vector in each line contains a massless
four-vector giving the direction of the jet, obtained as \\
$(p_x,p_y,p_z,E,m) = E_{\perp}
(\cos\varphi,\sin\varphi,\sinh\eta,\cosh\eta,0)$, \\
where $\eta$ and $\varphi$ give the weighted center of a jet and
$E_{\perp}$ its transverse energy.
\iteme{= 3 :} the \ttt{P} vector in each line contains a massive
four-vector, obtained by adding the massless four-vectors of all cells
that form part of the jet, and calculating the jet mass from
$m^2 = E^2-p_x^2-p_y^2-p_z^2$. For each cell, the total $E_{\perp}$ is
summed up, and then translated into a massless four-vector
assuming that all the $E_{\perp}$ was deposited in the center of the
cell.
\end{subentry}
 
\iteme{MSTU(61) :} (I) first entry for storage of event-analysis
information in last event analyzed with \ttt{LUSPHE}, \ttt{LUTHRU},
\ttt{LUCLUS} or \ttt{LUCELL}.
 
\iteme{MSTU(62) :} (R) number of particles/partons used in the last
event analysis with \ttt{LUSPHE}, \ttt{LUTHRU}, \ttt{LUCLUS},
\ttt{LUCELL}, \ttt{LUJMAS}, \ttt{LUFOWO} or \ttt{LUTABU}.
 
\iteme{MSTU(63) :} (R) in a \ttt{LUCLUS} call, the number of
preclusters constructed in order to speed up analysis (should be
equal to \ttt{MSTU(62)} if \ttt{PARU(43)=0.}). In a \ttt{LUCELL}
call, the number of cells hit.
 
\iteme{MSTU(161), MSTU(162) :}\label{p:MSTU161} hard flavours involved 
in current event,
as used in an analysis with \ttt{LUTABU(11)}. Either or both may be set
0, to indicate the presence of one or none hard flavours in event.
Is normally set by high-level routines, like \ttt{LUEEVT} or
\ttt{PYEVNT}, but can also be set by you.

\boxsep
 
\iteme{PARU(41) :}\label{p:PARU41} (D=2.) power of momentum-dependence 
in \ttt{LUSPHE}, default corresponds to sphericity, \ttt{=1.} to linear 
event measures.
 
\iteme{PARU(42) :} (D=1.) power of momentum-dependence in \ttt{LUTHRU},
default corresponds to thrust.
 
\iteme{PARU(43) :} (D=0.25 GeV) maximum distance $d_{\mrm{init}}$ 
allowed in
\ttt{LUCLUS} when forming starting clusters used to speed up
reconstruction. The meaning of the parameter is in $\pT$ for
\ttt{MSTU(46)}$\leq 2$ and in $m$ for \ttt{MSTU(46)}$\geq 3$. If
\ttt{=0.}, no preclustering is obtained.
If chosen too large, more joining may be generated at this stage
than is desirable. The main application is at high energies,
where some speedup is imperative, and the small details are not
so important anyway.
 
\iteme{PARU(44) :} (D=2.5 GeV) maximum distance $d_{\mrm{join}}$, 
below
which it is allowed to join two clusters into one in \ttt{LUCLUS}.
Is used for \ttt{MSTU(46)}$\leq 3$, i.e. both for $\pT$ and mass
distance measure.
 
\iteme{PARU(45) :} (D=0.05) maximum distance 
$y_{\mrm{join}} = m^2/E_{\mrm{vis}}^2$,
below which it is allowed to join two clusters into one in
\ttt{LUCLUS} for \ttt{MSTU(46)=4}.
 
\iteme{PARU(48) :} (D=0.0001) convergence criterion for thrust (in
\ttt{LUTHRU}) or generalized thrust (in \ttt{LUCLUS}), or relative
change of $m_{\mrm{H}}^2 + m_{\mrm{L}}^2$ (in \ttt{LUJMAS}), 
i.e. when the value changes by less than this amount between two 
iterations the process is stopped.
 
\iteme{PARU(51) :} (D=2.5) defines maximum absolute pseudorapidity
used for detector assumed in \ttt{LUCELL}.
 
\iteme{PARU(52) :} (D=1.5 GeV) gives minimum $E_{\perp}$ for a cell
to be considered as a potential jet initiator by \ttt{LUCELL}.
 
\iteme{PARU(53) :} (D=7.0 GeV) gives minimum summed $E_{\perp}$ for
a collection of cells to be accepted as a jet.
 
\iteme{PARU(54) :} (D=1.) gives the maximum distance in
$R = \sqrt{(\Delta\eta)^2 + (\Delta\varphi)^2}$ from cell initiator
when grouping cells to check whether they qualify as a jet.
 
\iteme{PARU(55) :} (D=0.5) when smearing the transverse energy 
(or energy, see \ttt{MSTU(53)}) in \ttt{LUCELL}, the calorimeter 
cell resolution is taken to be
\ttt{PARU(55)}$\times \sqrt{E_{\perp}}$ (or
\ttt{PARU(55)}$\times \sqrt{E}$) for $E_{\perp}$ (or $E$) in GeV.
 
\iteme{PARU(56) :} (D=2.) maximum factor of upward fluctuation in
transverse energy or energy in a given cell when calorimeter
resolution is included in \ttt{LUCELL} (see \ttt{MSTU(53)}).
 
\iteme{PARU(57) :} (D=3.2) maximum rapidity (or pseudorapidity or
pion rapidity, depending on \ttt{MSTU(42)}) used in the factorial 
moments analysis in \ttt{LUTABU}.

\iteme{PARU(58) :} (D=0. GeV) in a \ttt{LUCELL} call, cells with a 
transverse energy $E_{\perp}$ below \ttt{PARP(58)} are removed from
further consideration. This may be used to represent a threshold
in an actual calorimeter, or may be chosen just to speed up the 
algorithm in a high-multiplicity environment..
 
\iteme{PARU(61) :} (I) invariant mass $W$ of a system analyzed with
\ttt{LUCLUS} or \ttt{LUJMAS}, with energies calculated according to
the \ttt{MSTU(42)} value.
 
\iteme{PARU(62) :} (R) the generalized thrust obtained after a
successful \ttt{LUCLUS} call, i.e. ratio of summed cluster momenta
and summed particle momenta.
 
\iteme{PARU(63) :} (R) the minimum distance $d$ between two clusters
in the final cluster configuration after a successful \ttt{LUCLUS}
call; is 0 if only one cluster left.

\end{entry}

\clearpage
 
\section{Summary and Outlook}

A complete description of the {\PyJe} programs would have to cover
four aspects:  
\begin{Enumerate}
\item the basic philosophy and principles underlying the programs;
\item the detailed physics scenarios implemented, with all the
necessary compromises and approximations;
\item the structure of the implementation, including program flow,
internal variable names and programming tricks; and  
\item the manual, which describes how to use the programs.
\end{Enumerate}
Of these aspects, the first has been dealt with in reasonable detail. 
The second is unevenly covered: in depth for aspects which are not 
discussed anywhere else, more summarily for areas where separate 
up-to-date papers already exist. The third is not included at all, 
but `left as an exercise' for the reader, to figure out from the code 
itself. The fourth, finally, should be largely covered, although 
many further comments could have been made, in particular about the 
interplay between different parts of the programs. Still, in the 
end, no manual, however complete, can substitute for `hands on' 
experience.

The {\PyJe} programs are continuously being developed. We are aware
of many shortcomings, which hopefully will be addressed in the 
future, such as:
\begin{Itemize}
\item polarization effects should be included in more places,
in particular for $\tau$ production and decay;
\item the photoproduction and $\gamma\gamma$-physics scenarios
should be expanded;
\item many processes of interest are missing; and
\item mass relations and couplings need to be included beyond the
Born level in the MSSM two Higgs doublet scenario.
\end{Itemize}
This list could have been made much longer (I almost certainly missed
your top priority). One other aspect would be to provide more and
longer examples of working main programs for a number of standard
applications. 

Apart from these physics aspects, one may also worry about the
programming ones. For instance, for historical reasons, 
single precision real is used almost everywhere. With the
push to higher energies, this is becoming more and more of a
problem, so it would be logical to move to double precision
throughout.

One should also note that the {\Je} and {\Py} programs these days
are becoming so intertwined, that it would make sense to join
them into one single program. This would e.g. mean that the current
$\ee$ generation routines of {\Je} are made part of the generic
{\Py} process generation machinery --- this is particular affects
the matrix-element options, since $\ee$ events with parton showers
already are available in {\Py}. A joint product would likely adopt
the name {\Py}: although {\Je} is the older of the two programs, it
has a less well developed identity of its own. (It is also often 
referred to as `Lund', which today is more confusing than it was in 
the early days.) In the process of joining the programs, one would 
probably also remove a number of options that are no longer used.

Another possible change on longer time scales would be an introduction
of Fortran 90 programming elements. In particular, derived data types
could be used to define the event record as a one-dimensional array,
where each element represents a particle, with integer and real
components to give flavour, history, momentum and production vertex.

No timetable is set up for future changes. After all, this is not 
a professionally maintained software product, but part of a one-man
physics research project. Very often, developments
of the programs have come about as a direct response to the evolution
of the physics stage, i.e. experimental results and studies for 
future accelerators. Hopefully, the program will keep on evolving
in step with the new challenges opening up.

\clearpage

\section*{Index of Subprograms and Common Block Variables}
\addcontentsline{toc}{section}%
{Index of Subprograms and Common Block Variables}

This index is not intended to be complete, but gives the page where
the main description begins of a subroutine, function, block data, 
common block, variable or array. For common block variables also the
name of the common block is given. When some components of an array 
are described in a separate place, a special reference (indented with
respect to the main one) is given for these components.

\boxsep

\noindent
\begin{minipage}[t]{\halfpagewid}
\begin{tabular*}{\halfpagewid}[t]{@{}l@{\extracolsep{\fill}}r@{}}
\ttt{BRAT} in \ttt{LUDAT3}       & \pageref{p:BRAT} \\
\ttt{CHAF} in \ttt{LUDAT4}       & \pageref{p:CHAF} \\
\ttt{CKIN} in \ttt{PYSUBS}       & \pageref{p:CKIN} \\
\ttt{COEF} in \ttt{PYINT2}       & \pageref{p:COEF} \\
\ttt{HEPEVT} common block        & \pageref{p:HEPEVT} \\
\ttt{ICOL} in \ttt{PYINT2}       & \pageref{p:ICOL} \\
\ttt{IFUP} in \ttt{PYUPPR}       & \pageref{p:IFUP} \\
\ttt{ISET} in \ttt{PYINT2}       & \pageref{p:ISET} \\
\ttt{ISIG} in \ttt{PYINT3}       & \pageref{p:ISIG} \\
\ttt{K} in \ttt{LUJETS}          & \pageref{p:K} \\
\ttt{KCHG} in \ttt{LUDAT2}       & \pageref{p:KCHG} \\
\ttt{KFDP} in \ttt{LUDAT3}       & \pageref{p:KFDP} \\
\ttt{KFIN} in \ttt{PYSUBS}       & \pageref{p:KFIN} \\
\ttt{KFPR} in \ttt{PYINT2}       & \pageref{p:KFPR} \\
\ttt{KLU}    function            & \pageref{p:KLU} \\
\ttt{KUP} in \ttt{PYUPPR}        & \pageref{p:KUP} \\
\ttt{LU1ENT} subroutine          & \pageref{p:LU1ENT} \\
\ttt{LU2ENT} subroutine          & \pageref{p:LU2ENT} \\
\ttt{LU3ENT} subroutine          & \pageref{p:LU3ENT} \\
\ttt{LU4ENT} subroutine          & \pageref{p:LU4ENT} \\
\ttt{LUBOEI} subroutine          & \pageref{p:LUBOEI} \\
\ttt{LUCELL} subroutine          & \pageref{p:LUCELL} \\
\ttt{LUCHGE} function            & \pageref{p:LUCHGE} \\
\ttt{LUCLUS} subroutine          & \pageref{p:LUCLUS} \\
\ttt{LUCOMP} function            & \pageref{p:LUCOMP} \\
\ttt{LUDAT1} common block        & \pageref{p:LUDAT1} \\
\ttt{LUDAT2} common block        & \pageref{p:LUDAT2} \\
\ttt{LUDAT3} common block        & \pageref{p:LUDAT3} \\
\ttt{LUDAT4} common block        & \pageref{p:LUDAT4} \\
\ttt{LUDATA} block data          & \pageref{p:LUDATA} \\
\ttt{LUDATR} common block        & \pageref{p:LUDATR} \\
\ttt{LUDBRB} subroutine          & \pageref{p:LUDBRB} \\
\ttt{LUDECY} subroutine          & \pageref{p:LUDECY} \\
\ttt{LUEDIT} subroutine          & \pageref{p:LUEDIT} \\
\ttt{LUEEVT} subroutine          & \pageref{p:LUEEVT} \\
\ttt{LUERRM} subroutine          & \pageref{p:LUERRM} \\
\ttt{LUEXEC} subroutine          & \pageref{p:LUEXEC} \\
\ttt{LUFOWO} subroutine          & \pageref{p:LUFOWO} \\
\ttt{LUGIVE} subroutine          & \pageref{p:LUGIVE} \\
\ttt{LUHEPC} subroutine          & \pageref{p:LUHEPC} \\
\ttt{LUINDF} subroutine          & \pageref{p:LUINDF} \\
\ttt{LUJETS} common block        & \pageref{p:LUJETS} \\
\ttt{LUJMAS} subroutine          & \pageref{p:LUJMAS} \\
\ttt{LUJOIN} subroutine          & \pageref{p:LUJOIN} \\
\end{tabular*}
\end{minipage}%
\hfill%
\begin{minipage}[t]{\halfpagewid}
\begin{tabular*}{\halfpagewid}[t]{@{}l@{\extracolsep{\fill}}r@{}}
\ttt{LUKFDI} subroutine          & \pageref{p:LUKFDI} \\
\ttt{LULIST} subroutine          & \pageref{p:LULIST} \\
\ttt{LULOGO} subroutine          & \pageref{p:LULOGO} \\
\ttt{LUNAME} subroutine          & \pageref{p:LUNAME} \\
\ttt{LUONIA} subroutine          & \pageref{p:LUONIA} \\
\ttt{LUPREP} subroutine          & \pageref{p:LUPREP} \\
\ttt{LUPTDI} subroutine          & \pageref{p:LUPTDI} \\
\ttt{LURADK} subroutine          & \pageref{p:LURADK} \\
\ttt{LUROBO} subroutine          & \pageref{p:LUROBO} \\
\ttt{LUSHOW} subroutine          & \pageref{p:LUSHOW} \\
\ttt{LUSPHE} subroutine          & \pageref{p:LUSPHE} \\
\ttt{LUSTRF} subroutine          & \pageref{p:LUSTRF} \\
\ttt{LUTABU} subroutine          & \pageref{p:LUTABU} \\
\ttt{LUTAUD} subroutine          & \pageref{p:LUTAUD} \\
\ttt{LUTEST} subroutine          & \pageref{p:LUTEST} \\
\ttt{LUTHRU} subroutine          & \pageref{p:LUTHRU} \\
\ttt{LUUPDA} subroutine          & \pageref{p:LUUPDA} \\
\ttt{LUX3JT} subroutine          & \pageref{p:LUX3JT} \\
\ttt{LUX4JT} subroutine          & \pageref{p:LUX4JT} \\
\ttt{LUXDIF} subroutine          & \pageref{p:LUXDIF} \\
\ttt{LUXJET} subroutine          & \pageref{p:LUXJET} \\
\ttt{LUXKFL} subroutine          & \pageref{p:LUXKFL} \\
\ttt{LUXTOT} subroutine          & \pageref{p:LUXTOT} \\
\ttt{LUZDIS} subroutine          & \pageref{p:LUZDIS} \\
\ttt{MDCY} in \ttt{LUDAT3}       & \pageref{p:MDCY} \\
\ttt{MDME} in \ttt{LUDAT3}       & \pageref{p:MDME} \\
\ttt{MINT} in \ttt{PYINT1}       & \pageref{p:MINT} \\
\ttt{MRLU} in \ttt{LUDATR}       & \pageref{p:MRLU} \\
\ttt{MSEL} in \ttt{PYSUBS}       & \pageref{p:MSEL} \\
\ttt{MSTI} in \ttt{PYPARS}       & \pageref{p:MSTI} \\
\ttt{MSTJ} in \ttt{LUDAT1}, main & \pageref{p:MSTJ} \\
~~~\ttt{MSTJ(40) - MSTJ(50)}     & \pageref{p:MSTJ40} \\
~~~\ttt{MSTJ(101) - MSTJ(121)}   & \pageref{p:MSTJ101} \\
\ttt{MSTP} in \ttt{PYPARS}, main & \pageref{p:MSTP} \\
~~~\ttt{MSTP(22)}                & \pageref{p:MSTP22} \\
~~~\ttt{MSTP(61) - MSTP(71)}     & \pageref{p:MSTP61} \\
~~~\ttt{MSTP(81) - MSTP(94)}     & \pageref{p:MSTP81} \\
~~~\ttt{MSTP(131) - MSTP(134)}   & \pageref{p:MSTP131} \\
\ttt{MSTU} in \ttt{LUDAT1}, main & \pageref{p:MSTU} \\
~~~\ttt{MSTU(41) - MSTU(63)}     & \pageref{p:MSTU41} \\
~~~\ttt{MSTU(101) - MSTU(118)}   & \pageref{p:MSTU101} \\
~~~\ttt{MSTU(161) - MSTU(162)}   & \pageref{p:MSTU161} \\
\ttt{MSUB} in \ttt{PYSUBS}       & \pageref{p:MSUB} \\
\ttt{N} in \ttt{LUJETS}          & \pageref{p:N} \\
\end{tabular*}
\end{minipage} 

\newpage

\noindent
\begin{minipage}[t]{\halfpagewid}
\begin{tabular*}{\halfpagewid}[t]{@{}l@{\extracolsep{\fill}}r@{}}
\ttt{NFUP} in \ttt{PYUPPR}       & \pageref{p:NFUP} \\
\ttt{NGEN} in \ttt{PYINT5}       & \pageref{p:NGEN} \\
\ttt{NUP} in \ttt{PYUPPR}        & \pageref{p:NUP} \\
\ttt{P} in \ttt{LUJETS}          & \pageref{p:P} \\
\ttt{PARF} in \ttt{LUDAT2}       & \pageref{p:PARF} \\
\ttt{PARI} in \ttt{PYPARS}       & \pageref{p:PARI} \\
\ttt{PARJ} in \ttt{LUDAT1}, main & \pageref{p:PARJ} \\
~~~\ttt{PARJ(81) - PARJ(89)}     & \pageref{p:PARJ81} \\
~~~\ttt{PARJ(121) - PARJ(171)}   & \pageref{p:PARJ121} \\
\ttt{PARP} in \ttt{PYPARS}, main & \pageref{p:PARP} \\
~~~\ttt{PARP(61) - PARP(72)}     & \pageref{p:PARP61} \\
~~~\ttt{PARP(81) - PARP(100)}    & \pageref{p:PARP81} \\
~~~\ttt{PARP(131)}               & \pageref{p:PARP131} \\
\ttt{PARU} in \ttt{LUDAT1}, main & \pageref{p:PARU} \\
~~~\ttt{PARU(41) - PARU(63)}     & \pageref{p:PARU41} \\
~~~\ttt{PARU(101) - PARU(195)}   & \pageref{p:PARU101} \\
\ttt{PLU}    function            & \pageref{p:PLU} \\
\ttt{PMAS} in \ttt{LUDAT2}       & \pageref{p:PMAS} \\
\ttt{PROC} in \ttt{PYINT6}       & \pageref{p:PROC} \\
\ttt{PUP} in \ttt{PYUPPR}        & \pageref{p:PUP} \\
\ttt{PYCTQ2} function            & \pageref{p:PYCTQ2} \\
\ttt{PYDATA} block data          & \pageref{p:PYDATA} \\
\ttt{PYDIFF} subroutine          & \pageref{p:PYDIFF} \\
\ttt{PYDOCU} subroutine          & \pageref{p:PYDOCU} \\
\ttt{PYEVNT} subroutine          & \pageref{p:PYEVNT} \\
\ttt{PYEVWT} subroutine          & \pageref{p:PYEVWT} \\
\ttt{PYFRAM} subroutine          & \pageref{p:PYFRAM} \\
\ttt{PYGANO} function            & \pageref{p:PYGANO} \\
\ttt{PYGBEH} function            & \pageref{p:PYGBEH} \\
\ttt{PYGDIR} function            & \pageref{p:PYGDIR} \\
\ttt{PYGAMM} function            & \pageref{p:PYGAMM} \\
\ttt{PYGGAM} function            & \pageref{p:PYGGAM} \\
\ttt{PYGVMD} function            & \pageref{p:PYGVMD} \\
\ttt{PYHFTH} function            & \pageref{p:PYHFTH} \\
\ttt{PYI3AU} subroutine          & \pageref{p:PYI3AU} \\
\ttt{PYINBM} subroutine          & \pageref{p:PYINBM} \\
\ttt{PYINIT} subroutine          & \pageref{p:PYINIT} \\
\ttt{PYINKI} subroutine          & \pageref{p:PYINKI} \\
\ttt{PYINPR} subroutine          & \pageref{p:PYINPR} \\
\ttt{PYINRE} subroutine          & \pageref{p:PYINRE} \\
\ttt{PYINT1} common block        & \pageref{p:PYINT1} \\
\ttt{PYINT2} common block        & \pageref{p:PYINT2} \\
\ttt{PYINT3} common block        & \pageref{p:PYINT3} \\
\ttt{PYINT4} common block        & \pageref{p:PYINT4} \\
\ttt{PYINT5} common block        & \pageref{p:PYINT5} \\
\ttt{PYINT6} common block        & \pageref{p:PYINT6} \\
\ttt{PYINT7} common block        & \pageref{p:PYINT7} \\
\ttt{PYINT8} common block        & \pageref{p:PYINT8} \\
\ttt{PYKCUT} subroutine          & \pageref{p:PYKCUT} \\
\ttt{PYKLIM} subroutine          & \pageref{p:PYKLIM} \\
\ttt{PYKMAP} subroutine          & \pageref{p:PYKMAP} \\
\ttt{PYMAXI} subroutine          & \pageref{p:PYMAXI} \\
\ttt{PYMULT} subroutine          & \pageref{p:PYMULT} \\
\end{tabular*}
\end{minipage}%
\hfill%
\begin{minipage}[t]{\halfpagewid}
\begin{tabular*}{\halfpagewid}[t]{@{}l@{\extracolsep{\fill}}r@{}}
\ttt{PYOFSH} subroutine          & \pageref{p:PYOFSH} \\
\ttt{PYPARS} common block        & \pageref{p:PYPARS1},
\pageref{p:PYPARS2}  \\
\ttt{PYPILE} subroutine          & \pageref{p:PYPILE} \\
\ttt{PYQQBH} subroutine          & \pageref{p:PYQQBH} \\
\ttt{PYRAND} subroutine          & \pageref{p:PYRAND} \\
\ttt{PYREMN} subroutine          & \pageref{p:PYREMN} \\
\ttt{PYRESD} subroutine          & \pageref{p:PYRESD} \\
\ttt{PYSAVE} subroutine          & \pageref{p:PYSAVE} \\
\ttt{PYSCAT} subroutine          & \pageref{p:PYSCAT} \\
\ttt{PYSIGH} subroutine          & \pageref{p:PYSIGH} \\
\ttt{PYSPEN} function            & \pageref{p:PYSPEN} \\
\ttt{PYSPLI} subroutine          & \pageref{p:PYSPLI} \\
\ttt{PYSSPA} subroutine          & \pageref{p:PYSSPA} \\
\ttt{PYSTAT} subroutine          & \pageref{p:PYSTAT} \\
\ttt{PYSTEL} subroutine          & \pageref{p:PYSTEL} \\
\ttt{PYSTFL} subroutine          & \pageref{p:PYSTFL} \\
\ttt{PYSTFU} subroutine          & \pageref{p:PYSTFU} \\
\ttt{PYSTGA} subroutine          & \pageref{p:PYSTGA} \\
\ttt{PYSTPI} subroutine          & \pageref{p:PYSTPI} \\
\ttt{PYSTPR} subroutine          & \pageref{p:PYSTPR} \\
\ttt{PYSUBS} common block        & \pageref{p:PYSUBS} \\
\ttt{PYTEST} subroutine          & \pageref{p:PYTEST} \\
\ttt{PYUPEV} subroutine          & \pageref{p:PYUPEV} \\
\ttt{PYUPIN} subroutine          & \pageref{p:PYUPIN} \\
\ttt{PYUPPR} common block        & \pageref{p:PYUPPR} \\
\ttt{PYWAUX} subroutine          & \pageref{p:PYWAUX} \\
\ttt{PYWIDT} subroutine          & \pageref{p:PYWIDT} \\
\ttt{PYXTOT} subroutine          & \pageref{p:PYXTOT} \\
\ttt{Q2UP} in \ttt{PYUPPR}       & \pageref{p:Q2UP} \\
\ttt{RLU} function               & \pageref{p:RLU} \\
\ttt{RLUGET} subroutine          & \pageref{p:RLUGET} \\
\ttt{RLUSET} subroutine          & \pageref{p:RLUSET} \\
\ttt{RRLU} in \ttt{LUDATR}       & \pageref{p:RRLU} \\
\ttt{SIGH} in \ttt{PYINT3}       & \pageref{p:SIGH} \\
\ttt{SIGT} in \ttt{PYINT7}       & \pageref{p:SIGT} \\
\ttt{ULALEM} function            & \pageref{p:ULALEM} \\
\ttt{ULALPS} function            & \pageref{p:ULALPS} \\
\ttt{ULANGL} function            & \pageref{p:ULANGL} \\
\ttt{ULMASS} function            & \pageref{p:ULMASS} \\
\ttt{V} in \ttt{LUJETS}          & \pageref{p:V} \\
\ttt{VCKM} in \ttt{LUDAT2}       & \pageref{p:VCKM} \\
\ttt{VINT} in \ttt{PYINT1}       & \pageref{p:VINT} \\
\ttt{WIDE} in \ttt{PYINT4}       & \pageref{p:WIDE} \\
\ttt{WIDP} in \ttt{PYINT4}       & \pageref{p:WIDP} \\
\ttt{WIDS} in \ttt{PYINT4}       & \pageref{p:WIDS} \\
\ttt{XPANH} in \ttt{PYINT8}       & \pageref{p:XPANH} \\
\ttt{XPANL} in \ttt{PYINT8}       & \pageref{p:XPANL} \\
\ttt{XPBEH} in \ttt{PYINT8}       & \pageref{p:XPBEH} \\
\ttt{XPDIR} in \ttt{PYINT8}       & \pageref{p:XPDIR} \\
\ttt{XPVMD} in \ttt{PYINT8}       & \pageref{p:XPVMD} \\
\ttt{XSEC} in \ttt{PYINT5}       & \pageref{p:XSEC} \\
\ttt{XSFX} in \ttt{PYINT3}       & \pageref{p:XSFX} \\
\end{tabular*}
\end{minipage}
 
\end{document}